# Foundations of Digital Circuits

## Denotational, Operational, and Algebraic Semantics with Applications to Graph Rewriting

by

George Kaye

A thesis submitted to the
University of Birmingham
for the degree of
Doctor of Philosophy

School of Computer Science
College of Engineering and Physical Sciences
University of Birmingham
July 2024

# Abstract

This thesis details the culmination of a project to define a *fully compositional* theory of synchronous sequential circuits built from primitive components, motivated by applying techniques successfully used in programming languages to hardware.

The first part of the thesis defines the syntactic foundations required to create sequential circuit morphisms, and then builds three different semantic theories on top of this: denotational, operational and algebraic. We characterise the denotational semantics of sequential circuits as certain *causal stream functions*, as well as providing a link to existing circuit methodologies by mapping between circuit morphisms, stream functions and *Mealy machines*. The operational semantics is defined as a strategy for applying some global transformations followed by local reductions in order to demonstrate how a circuit processes a value, leading to a notion of *observational equivalence*. The algebraic semantics consists of equations for bringing circuits into a pseudo-normal form, and then *encoding* between different state sets. This part of the thesis concludes with a discussion of some novel applications, such as those for using *partial evaluation* for digital circuits.

While mathematically rigorous, the categorical string diagram formalism is not suited for reasoning computationally. The second part of this thesis details an extension of existing work on string diagram rewriting with *hypergraphs* so that it is compatible with the *traced comonoid* structure present in the category of digital circuits. We identify the properties that characterise cospans of hypergraphs corresponding to traced comonoid terms, and demonstrate how to identify rewriting contexts valid for rewriting modulo traced comonoid structure. We apply the graph rewriting framework to *fixed point operators* as well as the operational semantics from the first part, and present a new hardware description language based on these theoretical developments.

# Acknowledgements

The hardest part of writing a thesis is determining exactly who to put in the acknowledgements. A good place to start is the people who had the most direct input into my development as an academic, starting of course with the inimitable Dan Ghica, who dragged me kicking and screaming into the world of category theory. Miriam Backens played an instrumental role, by tirelessly reading any paper draft I sent their way and by providing useful observations on problems I was having. The influence of David Sprunger cannot be underestimated, as it was his careful and methodical approach to mathematics that I attempted to emulate in the latter half of my studies, the success of which will be determined by the contents of this thesis. Although the RSMG process is often lambasted by my peers, I found it an excellent way to check how things were going, so thanks must go to Achim Jung and Martín Escardó for providing sage advice during thesis group meetings. And finally, thank you to Timothy Bourke, Eric Finster, and Paul Levy for agreeing to attend the ultimate thesis group meeting: my viva.

The Theory Group at Birmingham is perhaps one of the most vibrant in the field, so I am proud to have been a member over the past five years. I have had some great conversations with the group's PhD students, which have included Nicolas Blanco, Tom de Jong, Alex Rice, Calin Tataru, Paaras Padhiar, and Ayberk Tosun. The students have been supplemented by an army of postdocs, including Chris Barrett, Gianluca Curzi, Abhishek De, Iris van der Giessen, and Sam Speight; and last but not least there have been the permanent staff keeping things (not too) serious at the top: in addition to those already mentioned, Anupam Das and Sonia Marin have done their bit to make the group a friendly and sociable bunch.

Immersing oneself wholly in theory would have an extremely detrimental effect to one's health, so fortunately I have had the pleasure of associating with numerous non-theory PhD students along the way. I started my journey alongside Andrea Basso and Charlotte Weitkämper, and with Alex and Calin we managed to eventually managed to find our feet in the often chaotic world of postgraduate research. Thanks to a minor global event in 2020 this group was sadly splintered, but when there is a void something always springs up to fill the gap. To say that a Mastodon server self-hosted by Jon Freer has had a large impact on me would be an understatement, and thanks to this tight-knit community I have enjoyed countless entertaining shenanigans with Matthew Bowden, Dhurim Cakiqi, Anna Clee, Matthew Hammond, Tobias Schmude, Charlie Street, Yánrong Wang, and Jacob Thomas Whitnell Wilson.

A department is only as strong as its administration: Computer Science is lucky to have had several diligent and competent members of Professional Services staff, including Angeliki Bompetsi, Sarah Brookes, Kate Campbell, Jason Fenemore, Tom Holly, Jamie Hough, Mohammed Idrees, Wenxuan Pan, Eden Whitehead, and Mohammed Zafar. Special mention must go to Lydia Eason for the ease at which she operated as SRSCC secretary while juggling student experience, and also for taking my request for better wine seriously.

If you've been following closely, you might notice that some names are missing. I thrive in a double act and have been fortunate enough to be part of several in my time at Birmingham. For better or for worse, Todd Waugh Ambridge has been there since the beginning, pointing out my most egregious errors and sharing countless balcony chats with tea and biscuits; with him I have engaged in many adventures in academia, including one particularly memorable trip to Edinburgh. Bruno da Rocha Paiva has been a near-constant office companion for the past two years; perhaps his presence has made progress on this thesis slower than it could have been, but the time would not have been nearly as interesting. Jacqueline Henes has acted as an incredibly capable SRSCC chair and provided a much-needed second perspective to the issues PGRs face; together we have battled operational issues galore and without her determination the PGR community would be in a much graver state.

Charlotte was nice to have around too.

# Contents

CHAPTER 1

# Introduction

Today's society has become dependent on digital circuits [KB05], which run our computers, homes, vehicles and much more. These days, digital circuits are so common that one may doubt that there are *any* gaps in our theoretical understanding of them.

But while the *design* of faster and more efficient circuits is a well-trodden area, it is relatively self-contained. We wish to support the existing techniques and procedures with alternatives successfully applied in other fields, such as that of programming languages. To see where the parallels lie between digital circuits and other areas, we need a foundational, mathematically rigorous theory. Although there has been previous work on such a theory [Laf03; GJL17a; GL18], it is the goal of this thesis to bring this project to its ultimate conclusion: a *fully compositional theory of synchronous sequential circuits*. The first point of order is to unpack exactly what this means.

## 1.1 Synchronous sequential circuits

The term '*circuit*' (or '*network*') is often used for any system constructed by connecting wires between primitive components. Two common constructs are the ability to *fork* wires or *join* them together, traditionally drawn using black dots as shown in Figure 1.1. These circuits are not particularly interesting as they are structural in nature; it is the other components that add meaning.

When it comes to picking a set of components, there is a plethora of choices for various applications; even when restricting to *electronic* circuits there are different flavours to consider. One variety is *analog* circuits constructed from components such as resistors, capacitors and inductors, such as in the left of Figure 1.2. Reasoning with analog circuits requires manipulating equations relating quantities such as voltage, current and resistance; in a parallel line of work to our own, analog circuits have already been given a compositional mathematical treatment [BS22].

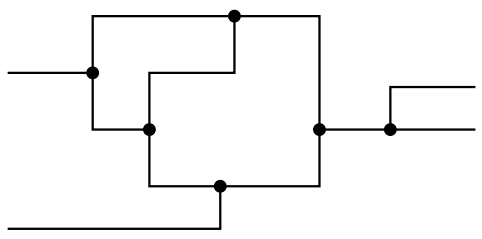

Figure 1.1: A circuit of forks and joins with no other primitives

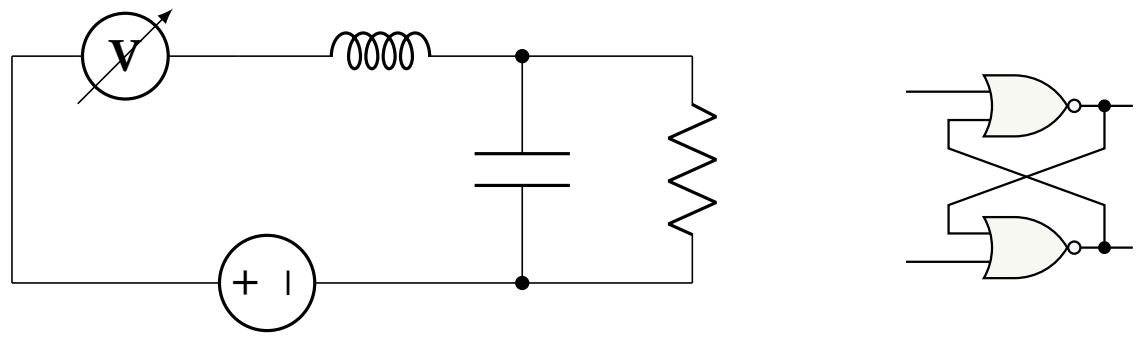

Figure 1.2: An analog circuit, using a *voltmeter*, an *inductor*, a *capacitor*, a *resistor*, and a *voltage source* as primitive components; and a digital circuit, using NAND *gates* as primitive components

We are concerned with *digital* circuits that operate over a finite number of *discrete* values. Here there is a much stricter notion of causality linking input and output; signals provided as input to a digital circuit propagate across the components, producing outputs and updating state. Digital circuits are constructed by connecting *logic gates* together with wires, to the inputs of other logic gates as illustrated in the right of Figure 1.2. A key point to note is that when designing a digital circuit, one may create *cycles*: paths from a component to itself. At a low level the components of a digital circuit are still constructed using analog parts, but the higher-level *abstraction* to digital components and discrete signals makes it far easier to design and reuse them.

Digital circuits can further be divided into classes based on their components. A *combinational* circuit is a circuit that only contains logical operations for computing functions. Such circuits have no memory; the outputs at each tick of the clock only depend on the inputs received at that moment. A far more useful class of circuits is that of *sequential* circuits , which have *delay* and *feedback* in addition to logic gates. Sequential circuits can be divided into two parts: the *combinational logic* for performing logical functions, and the *state registers* for storing data.

A typical digital circuit operates by using the combinational logic to perform some function on the inputs and the current state, and use the results of this computation to produce output signals and update the state with new data. When circuits are used in practice, it is usually desired for them to run as fast as possible: the time between providing the inputs and updating the output and state should be minimised.

With this in mind, there are two different ways one can design a sequential circuit. A *synchronous* circuit is one in which the state only changes in time with some global clock signal, whereas in *asynchronous* circuits the state changes as soon as the inputs do. The latter type of circuits is useful when speed is of the essence, but are harder to design because small differences in how quickly components process inputs can lead to the circuit assuming an unexpected state. For this reason, most practical circuits are restricted to the synchronous kind, and so this is where we are focusing our mathematical theory.

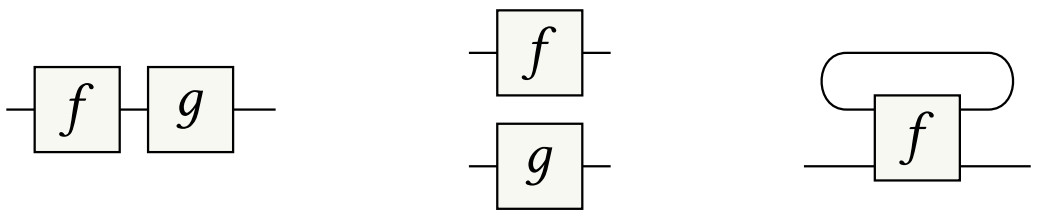

Figure 1.3: Three types of composition: *sequential*, *parallel* (tensor) and using a *trace operator*

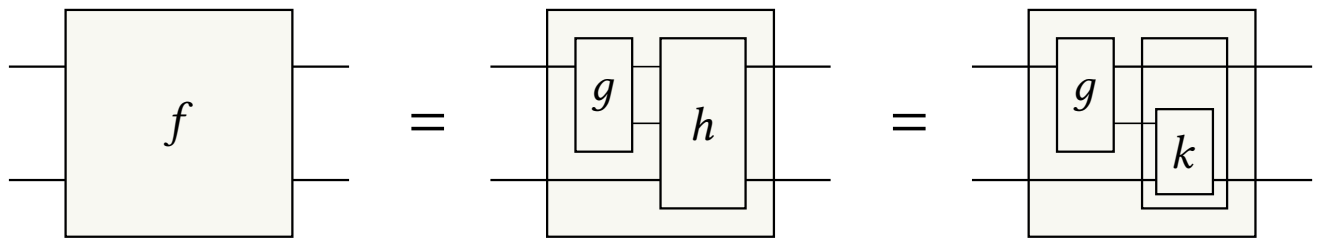

Figure 1.4: Decomposing a large circuit $f$ into smaller circuits $g$ and $k$

## 1.2 Compositionality and category theory

A (synchronous sequential) circuit can be viewed as a component with some input and output wires; these wires can be connected to other components to create bigger and more complex circuits. This is called *composing* circuits, and some ways of doing this are shown in Figure 1.3. We can compose circuits *horizontally* (if the outputs of the first match with the inputs of the second), *vertically*, or even by connecting an output wire to an input wire, creating a *feedback loop*.

Our goal is to define a *fully compositional* theory of synchronous sequential circuits. Here, we take full compositionality to mean that we can compose circuits solely on the basis of their interfaces: we should not have to perform any sort of semantic check or 'peek inside' a circuit to find out if composition is permitted.

Compositionality is an appealing paradigm to follow because it means we can repeatedly split complicated circuits into simpler parts until we reach some indivisible atomic components. If one defines the behaviour of these components and how they interact with composition, it is possible to establish the behaviour of some larger circuit *inductively* by breaking it down into its constituent parts. For example, some circuit $f$ might be decomposed as in Figure 1.4. To prove that $f$ has some property, we can prove it holds for $g$ and $k$ and verify if composition *preserves* these properties.

While it is possible to work with this surface-level notion of compositionality, it is important to establish a mathematical foundation to determine the meaning and properties of composition; for this, we turn to *category theory* [Mac78] . A *category* is made up of *objects* and *morphisms* ('arrows') between them. Any two arrows $f\colon A \to B$ and $g\colon B \to C$ can be *composed* to make a new morphism $g \circ f\colon A \to C$, and every object $A$ has a unique *identity* morphism $\mathsf{id}_A$. Composition is *associative* and *unital*; that is to say, the following equations hold:

$$(h \circ g) \circ f = h \circ (g \circ f) \qquad f \circ \mathsf{id}_A = f \qquad \mathsf{id}_B \circ f = f$$

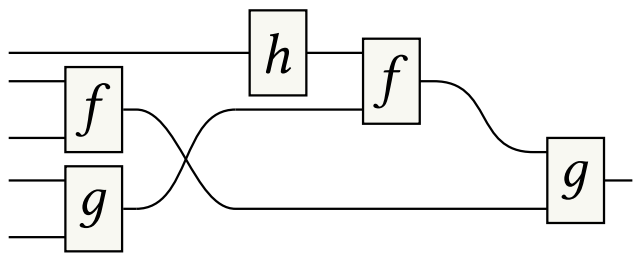

Figure 1.5: Example of a string diagram

It turns out that this simple concept paves the way to an array of theorems that can generalise many areas of computer science, mathematics, and beyond.

It is straightforward to see how a compositional process with inputs of type $A$ and outputs of type $B$ could be modelled as a morphism $f\colon A \to B$. But to model more complex processes one needs to work with a class of categories known as *freely generated symmetric monoidal categories* [Mac63], categories equipped with an additional notion of *parallel* composition $\otimes$ along with a family of morphisms $\sigma_{A,B}\colon A \otimes B \to B \otimes A$ known as *symmetries* to swap over inputs. Using a set of primitive components as *generators*, the two composition operators can be used in combination with the identities and symmetries to build up more complicated terms. These terms can be written as *term strings*, such as the following:

$$\mathsf{id}_1 \otimes (f \otimes g \mathbin{⨟} \sigma_{1,1}) \mathbin{⨟} (h \otimes \mathsf{id}_1 \mathbin{⨟} f) \otimes \mathsf{id}_1 \mathbin{⨟} g$$

Terms described in this way are quite unintuitive, as the two different types of composition are compressed into a one-dimensional text string. Fortunately, symmetric monoidal categories admit an intuitive graphical notation known as *string diagrams*, in which generators are drawn as boxes connected by wires, with the identity depicted as an empty wire and the symmetry as swapping over wires. For example, the term described above can be depicted diagrammatically as in Figure 1.5, which is far easier to comprehend than the original term string.

String diagrams do not add any more computational power to one-dimensional reasoning, but they are immensely beneficial because the categorical axioms of associativity and unitality are 'absorbed' by the notation: two morphisms are equal if and only if their string diagrams share the same connectivity between boxes [KL80; Kis14; Sel11]. This makes proofs far less bureaucratic, as one can focus on the non-trivial steps without having to constantly rearrange the bracketing of a term. String diagrams also make the work more approachable and easier to explain to non-mathematicians; using the diagrammatic approach it is possible to give talks about category theory without mentioning categories at all. There have even been books written with this philosophy [CK18].

## 1.3 Compositionality and sequential circuits

One might argue that composition is already widespread in sequential circuit design, and indeed it is: circuits are constructed by connecting lots of very common primitive components together to make something more complex. But this is done informally, as the behaviour of a circuit is usually tested by *simulating* it and seeing what happens. We can simulate the subcomponents, but what does this mean for their composite? Without a guarantee of full compositionality, we have no reason to believe that connecting two well-behaved circuits together will result in another well-behaved circuit.

Progress towards full compositionality for sequential circuits has been hampered by the presence of the dreaded *non-delay-guarded feedback loop*; a cycle that does not pass through any state registers. Non-delay-guarded feedback can lead to undefined behaviour; for example, if we assume that the rightmost circuit $f$ in Figure 1.3 contains no registers, then it is not immediately obvious how to compute the first input, as it would depend on itself.

Some approaches try to nullify this by considering only some 'safe' subset of circuits which will always be well-behaved [CSBH21], by introducing some sort of 'type system' on wires so that components may only be connected if they are guaranteed to have well-defined behaviour at the same points in time [NAS23], or by only considering certain kinds of composition [Ale14]. While these are useful perspectives, they still shy away from true full compositionality for sequential circuits.

Even though there are indeed times when non-delay-guarded feedback can lead to unwanted behaviour, careful use can still result in useful circuits, so it should not be excluded from our mathematical theory of sequential circuits. Consider the circuit below, where $f$ and $g$ are inverses, i.e. $-\boxed{f}-\boxed{g}- \;=\; ---$ . If we were to enforce that every loop in our circuits is somehow delay-guarded, the line of equational reasoning in Figure 1.6 is forbidden; 'yanking out' the otherwise trivial loop of wires would implicitly delete a delay and alter the outputs of the circuit over time.

Non-delay-guarded feedback can also be used in clever ways to create sequential circuits that exhibit combinational behaviour. The circuit in Figure 1.7 is a classic example [Mal94] in which the feedback is just used to share the two subcircuits $-\boxed{f}-$ and $-\boxed{g}-$ ; the circuit acts as $g \circ f$ or $f \circ g$ depending on the control input c. Such circuits, while not following conventional design methodology, are more efficient in terms of circuit size and power consumption. This once more illustrates how working in a setting where not all loops are delay-guarded may be beneficial to us.

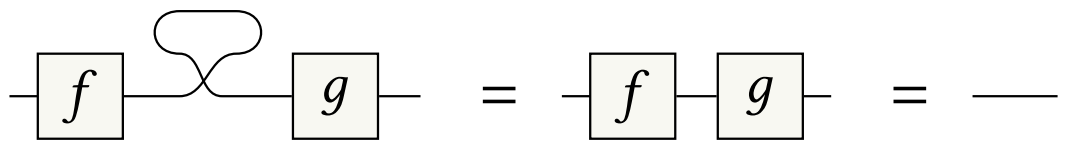

Figure 1.6: Example of diagrammatic equational reasoning by 'yanking' a feedback loop

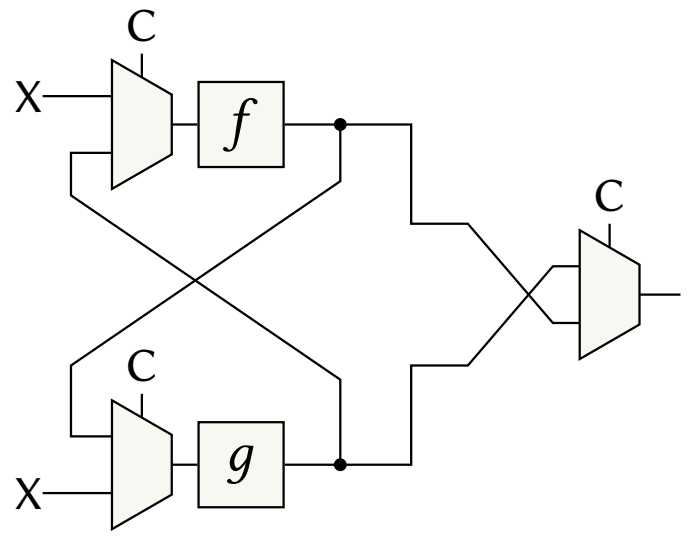

Figure 1.7: Example of a circuit with non-delay-guarded feedback that can exhibit combinational behaviour [Mal94]

## 1.4 Theories of digital circuits

While this thesis details a fully compositional *categorical* theory of digital circuits, this is by no means the first time sequential circuits have been given the mathematical treatment. *Mealy machines* [Mea55] are the *de facto* mathematical structure for specifying the behaviour of sequential circuits, and it is well known how they should be composed [Gin14]. In more recent times Mealy machines have been given a categorical treatment as certain kinds of *coalgebra* [Rut06; BRS08]. However, Mealy machines abstract away from the *components* of the circuit; we are keen to preserve the link between structure and behaviour.

While not explicitly categorical, the idea of representing circuits as mathematical expressions built up from primitive components was studied in the 80s by Gordon, who worked on *denotational semantics for sequential machines* [Gor80] and used this idea to present *a model of register transfer systems* [Gor82]. Gordon subsequently noted that *higher order logic* would make a good fit for a hardware description language [Gor85], and this has become a ubiquitous concept in formal verification of hardware [Gup92].

The first steps towards a categorical theory of digital circuits took place after the turn of the millennium, when Lafont presented an *algebraic theory of Boolean circuits* [Laf03]. This work already bears a great resemblance to the framework presented in this thesis; circuits are presented as morphisms in a symmetric monoidal category freely generated over a set of primitive logic gates. However, Lafont's work only considered circuits for

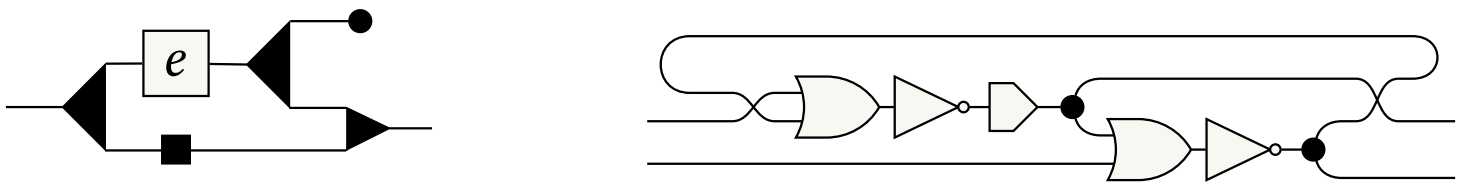

Figure 1.8: Two representations of digital circuits, the first by Lafont [Laf03] and the second by Ghica, Jung, and Lopez [GJ16; GJL17a]

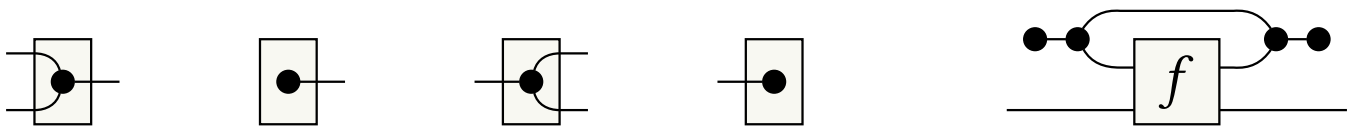

Figure 1.9: The four generators of a Frobenius structure, and the construction of a trace using them

*Boolean functions*; these circuits did not have any notion of delay or feedback. Lafont also made use of a diagrammatic language (shown in Figure 1.8), but the equations of monoidal categories still had to be applied explicitly.

It was not until 2016 that *sequential* circuits were given the categorical treatment by Ghica and Jung [GJ16], who were later joined by Lopez when considering how to use this for a graph-rewriting based operational semantics [GJL17a]. In this line of work, sequential circuits were modelled as morphisms in a *symmetric traced monoidal category*, a symmetric monoidal category extended with a *trace operator*. In the context of sequential circuits, the trace operator models *feedback*: Tr ( 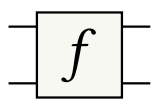 ) = 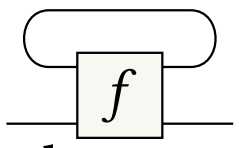 .

This marks a departure from many other recent works on compositional processes such as the work on string diagrammatic *signal flow theory* [BSZ21] or analog circuits [BS22], which operate in a setting with a *Frobenius* structure. In addition to any application-specific components, these settings also contain the four structural components shown in Figure 1.9, forming what is known mathematically as a *commutative monoid* and a *cocommutative comonoid*. These components are used to model the forks and joins alluded to earlier in this chapter; as also illustrated in Figure 1.9, one can use them to create a feedback loop.

So why not opt for such a route for sequential circuits? The issue arises in the form of *copying*: we would like the leftmost equation in Figure 1.10 circuits to hold for any circuit 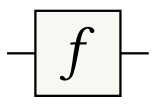 . What this means is that running $f$ and then copying the outputs should be exactly the same as copying the inputs and running two copies of $f$ in parallel. But this seemingly innocent equation causes the construction of the feedback loop using smaller components to fall apart. This is because if we instantiate the circuit 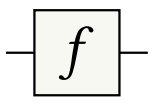 to the 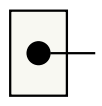 component, it can be propagated over the 'end' of the trace and break the loop, also illustrated in Figure 1.10.

This is known as the *no-cloning theorem*; any setting with a Frobenius structure cannot also admit copying. For this reason, we need to model circuits in a *traced* category in which the feedback loop is built as one piece so it cannot be broken.

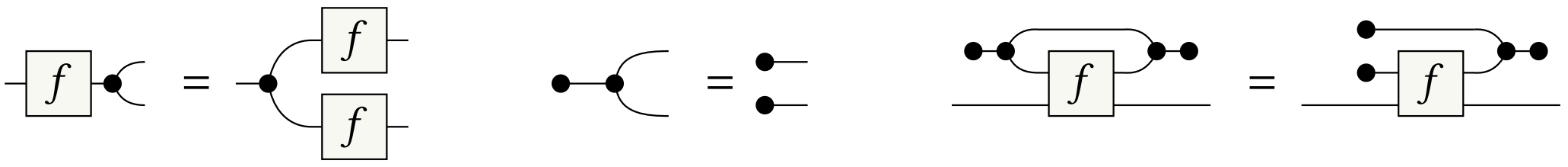

Figure 1.10: The copying equation, and its implications

## 1.5 Contributions

The contributions of this thesis are split into two parts, *Semantics of Digital Circuits*, and *Graph Rewriting for Digital Circuits*. These sections respectively correspond to two papers: *A Fully Compositional Theory of Sequential Digital Circuits: Denotational, Operational and Algebraic Semantics* [GKS24], and *Rewriting Modulo Traced Monoidal Structure* [GK23], which was published in *Formal Structures for Computation and Deduction (FSCD) 2023*.

### 1.5.1 Semantics of Digital Circuits

The first part of this thesis sets about finishing the project on categorical semantics for digital circuits [GJ16; GJL17a] in a methodical, rigorous fashion. We first define the categorical syntax of sequential circuits and follow this up with three sound and complete semantic frameworks: denotational, operational, and algebraic. Each of these frameworks has their own benefits and intended uses; together they form a comprehensive examination of semantics of digital circuits. The framework is sufficiently general to encompass circuits constructed from all manner of components ranging from the level of transistors to the level of logic gates and beyond, but to provide some intuition we include an extended case study into circuits constructed from *Belnap logic gates* [Bel77], an extension of traditional Boolean logic containing the usual AND, OR and NOT gates. An overview of the categories involved can be seen in Figure 1.11.

**Chapter 3: Syntax of sequential circuits**

Previously, the categories of circuits were immediately quotiented by some 'natural laws'; this made it difficult to define maps from circuits to other categories, as equations on circuits also had to be considered. We take a more modular approach, in which we first define the syntactic categories $\mathbf{CCirc}_\Sigma$ of *combinational* circuits and $\mathbf{SCirc}_\Sigma$ of *sequential* circuits. These are categories in which we can construct circuit morphisms; the three semantic theories that we will present next provide different ways of *quotienting* these categories in order to identify circuits with the same behaviour under some interpretation.

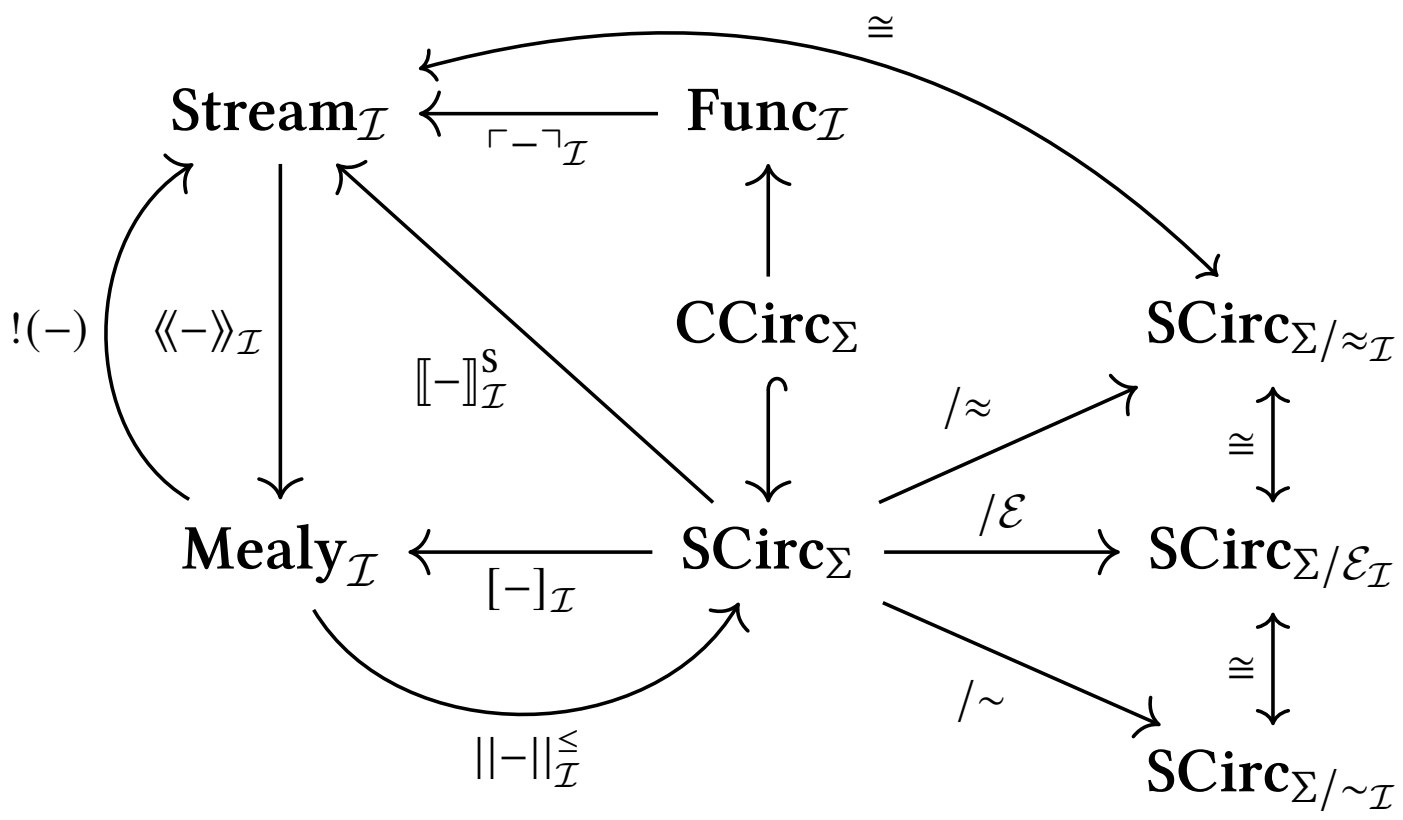

Figure 1.11: Categories of digital circuits

### Chapter 4: Denotational semantics

While the previous circuits work discussed assigning semantics to circuits in terms of streams, this model was not constructed in great detail. It was not even deemed important enough to appear in the conference version of the paper, only being examined in detail in the arXiv preprint [GJL17b]. Here we present a category $\mathbf{Stream}_{\mathcal{I}}$ of *stream functions* with certain properties, which serve as the denotational semantics for sequential circuits. This denotational semantics is sound and complete in that every syntactic circuit can be expressed as one of these stream functions, and every such stream function can be mapped to a syntactic circuit which has the original stream function as its behaviour.

We also define a category $\mathbf{Mealy}_{\mathcal{I}}$ of Mealy machines lifted to work on lattices, as a 'bridge' between circuits and stream functions. As well as being essential for showing the soundness and completeness of the denotational semantics, this category of Mealy machines is nice to have in its own right, as it shows how existing circuit methodologies [KJ09] are compatible with our rigorous mathematic framework.

Circuits that map to the same stream function are called *denotationally equivalent.* Quotienting $\mathbf{SCirc}_{\Sigma}$ by denotational equivalence we obtain a category $\mathbf{SCirc}_{\Sigma/\approx_{\mathcal{I}}}$; this is the category against which we will compare our next two semantic theories.

### Chapter 5: Operational semantics

The original motivation for a categorical theory of circuits was to create an *operational semantics* for digital circuits, bringing techniques from software to hardware. While such a system was presented in [GJL17a], this only worked on *closed* circuits with no *non-delay-guarded feedback.* One of the main contributions of this chapter is to lift this restriction using a novel reduction rule for eliminating non-delay-guarded feedback inspired by the Kleene fixed point theorem. Combined with a generalisation of the previous reduction procedure to work on open circuits, this means that any circuit applied to some inputs can be reduced in order to determine its outputs and next state.

As a result of this, we also present a new formal notion of *observational equivalence* on sequential circuits, and show that it is the correct one using the well-known universal property that it is the largest adequate congruence relation [Gor98]. Quotienting $\mathbf{SCirc}_\Sigma$ by observational equivalence gives us another semantic category of circuits $\mathbf{SCirc}_{\Sigma/\sim_\mathcal{I}}$. By establishing an isomorphism between $\mathbf{SCirc}_{\Sigma/\approx_\mathcal{I}}$ and $\mathbf{SCirc}_{\Sigma/\sim_\mathcal{I}}$ we show that the operational semantics is also sound and complete: two circuits have the same behaviour as stream functions if and only if they reduce to the same outputs for all inputs.

#### Chapter 6: Algebraic semantics

The previous framework of digital circuits was presented as an *algebraic semantics*: the category of circuits was quotiented by certain 'natural laws', which were stated as axioms rather than being derived from any mathematical model. These equations were not actually enough to show the desired results, so additional quotients of 'extensional equivalence' were used to add in the remaining equalities.

In this thesis our equational theory is guided by the stream semantics, building up to an algebraic semantics for circuits without having to add any arbitrary quotients. We try to stick to standard equations on algebraic structures and small 'local' equations detailing the interactions on individual generators, but the nature of digital circuits means that some larger equations including *context* are necessary to include. Ultimately, we define a set of equations $\mathcal{E}_\mathcal{I}$ and show that these equations suffice to bring any circuit to a pseudo-normal form.

Quotienting $\mathbf{SCirc}_\Sigma$ by these equations gives us our last semantic category $\mathbf{SCirc}_{\Sigma/\mathcal{E}_\mathcal{I}}$. Establishing an isomorphism between $\mathbf{SCirc}_{\Sigma/\approx_\mathcal{I}}$ and $\mathbf{SCirc}_{\Sigma/\sim_\mathcal{I}}$ shows that the algebraic semantics is sound and complete: two circuits have the same behaviour as stream functions if and only if they can be translated into each other using the equations.

#### Chapter 7: Potential applications

We conclude the first part of the thesis by examining some potential applications for the categorical framework of digital circuits. In particular, we show how one could use the framework for *partial evaluation* of circuits, and how we can use (in)equational reasoning to develop more efficient circuits.

### 1.5.2 Graph Rewriting for Digital Circuits

While this work marks the first time the semantics of sequential digital circuits have been given a rigorous mathematical treatment, it is not really feasible to apply the techniques to anything more than toy circuits by hand; trying to manually apply the techniques to actual, practical, circuits would quickly become impractical. Instead it is desirable to have a computer deal with all the hard work for us and reason *automatically*. To do this, we need to represent circuits *combinatorially* as graphs.

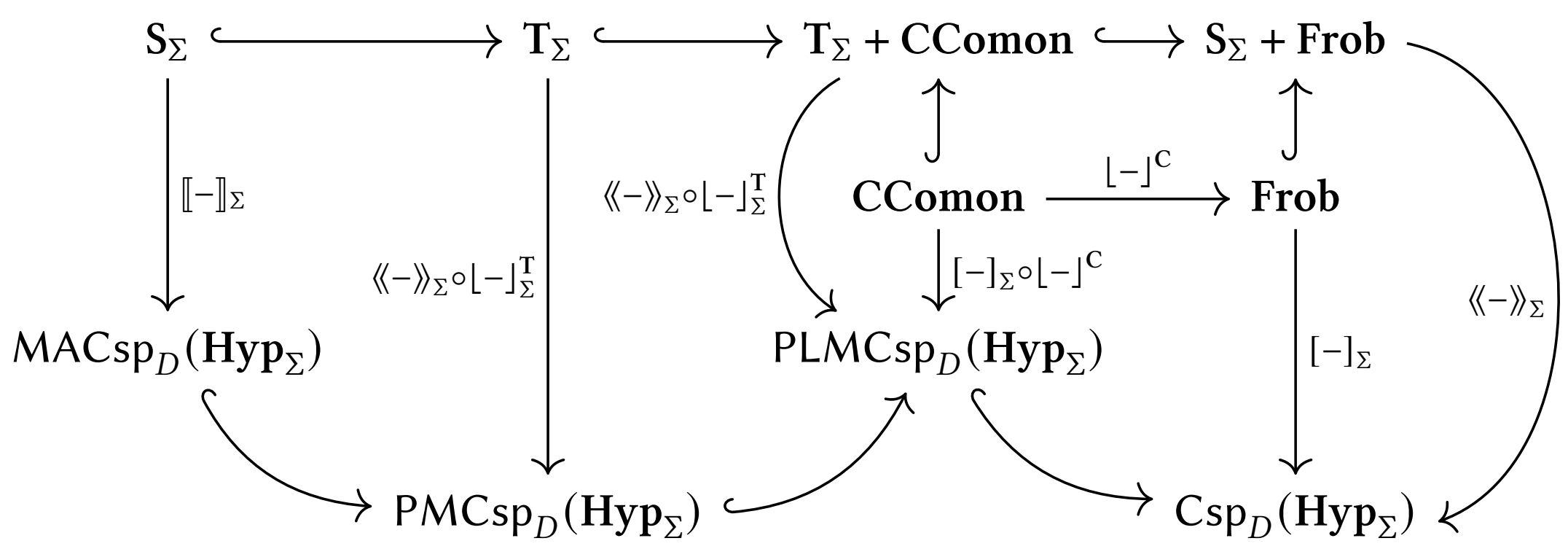

Figure 1.12: Categories of terms and cospans of hypergraphs

Representing the categorical syntax of digital circuits in this way was considered in [GJL17a] using Kissinger's *framed point graphs* [Kis12]. These had several drawbacks: for example, many framed point graphs correspond to the same string diagram term, and the category of such graphs is not fully *adhesive*, a property that provides nice properties for graph rewriting. In the second part of the thesis, we extend more recent work on *hypergraph string diagram rewriting* [BGK+22a; BGK+22b; BGK+22c] so we can apply it to sequential digital circuits. An overview of the categories involved can be seen in Figure 1.12.

### Chapter 8: String diagrams as hypergraphs

Previous work on string diagram rewriting using hypergraphs showed how terms in a freely generated symmetric monoidal category $S_\Sigma$ equipped with a *special commutative Frobenius structure* **Frob** correspond to morphisms in a category of *cospans of hypergraphs* $\mathsf{Csp}_D(\mathbf{Hyp}_\Sigma)$, and terms without a Frobenius structure correspond to morphisms in a category of cospans of *monogamous acyclic* $\mathsf{MACsp}_D(\mathbf{Hyp}_\Sigma)$.

We extend this work to terms in a freely generated symmetric traced monoidal category $\mathbf{T}_\Sigma$; since these terms occupy the space in between regular symmetric monoidal terms and those with a Frobenius structure, the interpretation as cospans accordingly sits between $\mathsf{MACsp}_D(\mathbf{Hyp}_\Sigma)$ and $\mathsf{Csp}_D(\mathbf{Hyp}_\Sigma)$ in the form of the category of *partial monogamous* cospans of hypergraphs $\mathsf{PMCsp}_D(\mathbf{Hyp}_\Sigma)$.

We furthermore extend this to the case where the traced terms are additionally equipped with a cocommutative comonoid structure **CComon**, leading to a category of *partial left-monogamous* cospans of hypergraphs $\mathsf{PLMCsp}_D(\mathbf{Hyp}_\Sigma)$.

### Chapter 9: Graph rewriting

Reasoning on terms interpreted as cospans of hypergraphs is performed using *double pushout (DPO) graph rewriting*. The main computational step during this procedure is identifying a valid *pushout complement*: the context of the rewrite step. Bonchi et al. showed that for Frobenius terms, any pushout complement corresponds to a term rewrite [BGK+22a], and for symmetric monoidal terms exactly one pushout complement corresponds to a term rewrite [BGK+22b]: the *boundary complement*.

For the traced case and the traced comonoid case some pushout complements are valid and some are not. We characterise those that are as *traced boundary complements* for the former and *traced left-boundary complements* for the latter.

### Chapter 10: Applications of graph rewriting

Interpreting circuits as hypergraphs introduces the opportunity to evaluate them *automatically* using graph rewriting. As a first case study, we show how graph rewriting modulo traced comonoid structure can aid reasoning in settings with a *Cartesian* structure, of which the semantic categories of digital circuits are an example.

Subsequently, we illustrate how graph rewriting can be used as a combinatorial implementation of the operational semantics for digital circuits. This culminates in the presentation of a graph-rewriting-based hardware description language based on the work throughout the thesis, with which one can design and (partially) evaluate circuits in a step-by-step manner.

# Chapter 2

# A crash course in category theory

When setting about developing a mathematical foundation for something, the first step is to decide exactly what foundation to pick. As a basis, we will model circuits as *terms* constructed by combining primitive generators in sequence and in parallel (Section 2.1). But terms on their own are not enough, as there may be many terms that represent the same structure. To use the necessary 'structural' equations and identify these terms, we turn to *category theory*, in which mathematical structures are expressed using *categories*: collections of *objects* and composable *morphisms* between them (Section 2.4).

Category theory started out in the early 1940s, when Eilenberg and Mac Lane were working on problems in algebraic topology. The pair's original goal was to define what it meant for a homomorphism to be 'natural' [EM42], a concept which nowadays we would call a *natural transformation*. To properly define this required the notions of a *functor* and a *category*; the latter was first presented in 1945 [EM45].

It has since turned out that category theory can be used as something of a 'universal language' in mathematics, as many familiar mathematical structures can be expressed solely in terms of objects and morphisms. This makes category theory an incredibly powerful tool; rather than having to repeatedly prove things about concrete structures, abstract results can be instantiated to a plethora of settings.

**Example 2.1.** A common structure in set theory is the *Cartesian product*, in which two sets $A$ and $B$ can be combined to create a new set $A \times B$ containing pairs of elements $(a,b)$. A product is equipped with two functions $\mathsf{fst}\colon A \times B \to A$ and $\mathsf{snd}\colon A \times B \to B$ to extract the first and second element from a pair respectively.

Many properties of Cartesian products are not specific to sets, but arise from the presence of the two projections. Category theory has a notion of *categorical product* in which sets are generalised to objects and the projection functions to morphisms; this allows results about products to be applied to a wide range of structures in addition to the Cartesian product, such as set intersection, greatest lower bounds, greatest common divisors, and more complicated algebraic structures.

Category theory is especially relevant to computer scientists, and it has shed light upon topics such as semantics of programming languages [Ole82], databases [Spi12], probabilistic programming [CJ19; Fri20], and machine learning [FST19; CCG+20]. Category theory also underpins functional programming languages such as Haskell and OCaml, which use categorical concepts for computational purposes.

Using category theory, a circuit can be modelled as a morphism $A \to B$, but this is not expressive enough to work with complicated, two-dimensional circuits. By using functors (Section 2.5) and natural transformations (Section 2.6), we can define a class of categories known as *symmetric monoidal categories* [Mac63], in which the usual *sequential* composition is joined by *parallel* composition and morphisms for *swapping* inputs (Section 2.7). In digital circuits, we additionally create a *feedback loop* by attaching some outputs to inputs. This too can be viewed in a categorical perspective in the form of a *symmetric traced monoidal category* [JSV96] (Section 2.8).

To reason with circuits we must additionally make use of a *monoidal theory* (Section 2.9), a set of context-specific equations between the primitive components in a category. Monoidal theories have been successfully applied to study quantum protocols [CD08], dynamical systems [BE15; FSR16], signal flow diagrams [BSZ14; BSZ15; BHPS17; BSZ21], linear algebra [BSZ17; Zan15; BPSZ19; BP22], finite-state automata [PZ21; PZ22], electrical circuits [BS22], automatic differentiation [AGSZ23], synthetic chemistry [GLZ23], and first order logic [BDHS24], among many others.

To begin, we will take a meander through the various categorical definitions and notation that underpin the mathematical work of this thesis. A similar outline can be found in most category theory textbooks, but inspiration was taken in particular from the opening of [GZ23].

## 2.1 Terms

We are interested in using category theory as a tool to study *circuits* built from some primitive components, or *generators*.

**Definition 2.2** (Generators). A set of *generators* $\Sigma$ is a set equipped with two functions $\mathsf{dom}, \mathsf{cod} \colon \Sigma \to \mathbb{N}$.

Generators are the primitive building blocks of terms: their domains and codomains specify how many input and output wires they have. Terms are defined by combining these primitives together.

**Definition 2.3** (Term). Let $\Sigma$ be a set of generators. A $\Sigma$-term is written $f\colon m \to n$ where $m, n \in \mathbb{N}$. The set of $\Sigma$-terms, denoted $\Sigma_t$, is generated as follows:

$$\frac{\phi \in \Sigma}{\phi\colon \mathsf{dom}(\phi) \to \mathsf{cod}(\phi) \in \Sigma_t} \qquad \frac{}{\mathrm{id}_1\colon 1 \to 1 \in \Sigma_t}$$

$$\frac{}{\mathrm{id}_0\colon 0 \to 0 \in \Sigma_t} \qquad \frac{}{\sigma_{1,1}\colon 2 \to 2 \in \Sigma_t}$$

$$\frac{f\colon m \to n \in \Sigma_t \qquad g\colon n \to p \in \Sigma_t}{f \mathbin{;} g\colon m \to p \in \Sigma_t} \qquad \frac{f\colon m \to n \in \Sigma_t \qquad g\colon p \to q \in \Sigma_t}{f \otimes g\colon m + p \to n + q \in \Sigma_t}$$

$\Sigma$-terms are constructed recursively. There are four base cases: a generator from $\Sigma$ with appropriate inputs and outputs; an *identity* for single wires, empty space, and a *symmetry* for swapping over two wires. The two inductive cases are called *composition* and *tensor* respectively. Intuitively, these can be thought of as generating larger terms by composing subterms in sequence or parallel.

Although here identities and symmetries only operate on single wires, it is a simple exercise to define them for larger numbers of wires.

**Notation 2.4** (Composite identities). Composite identities $\mathrm{id}_n$ are defined inductively for $n \in \mathbb{N}$ as $\mathrm{id}_0 \coloneqq \mathrm{id}_0$ and $\mathrm{id}_{k+1} \coloneqq \mathrm{id}_k \otimes \mathrm{id}_1$.

**Notation 2.5** (Composite symmetries). Composite symmetries $\sigma_{m,n}$ for $m, n \in \mathbb{N}$ are defined inductively as

$$\sigma_{0,n} \coloneqq \mathrm{id}_n \quad \sigma_{m,0} \coloneqq \mathrm{id}_m \quad \sigma_{k+1,l+1} \coloneqq \mathrm{id}_k \otimes \sigma_{1,l} \otimes \mathrm{id}_1 \mathbin{;} \sigma_{k,l} \otimes \sigma_{1,1} \mathbin{;} \mathrm{id}_l \otimes \sigma_{k,1} \otimes \mathrm{id}_1$$

$\Sigma$-terms will be abbreviated to 'terms' when the signature is clear from context.

**Example 2.6.** Let $\Sigma_g \coloneqq \{\mathrm{AND}\colon 2 \to 1, \mathrm{OR}\colon 2 \to 1, \mathrm{NOT}\colon 1 \to 1\}$ be a set of *logic gate generators*. Examples of terms in $(\Sigma_g)_t$ include

$$(\mathrm{OR} \mathbin{;} \mathrm{id}_1) \otimes ((\mathrm{id}_1 \otimes \mathrm{NOT}) \mathbin{;} \mathrm{AND})\colon 4 \to 2$$
$$((\mathrm{id}_1 \otimes \mathrm{id}_1) \mathbin{;} \mathrm{OR}) \otimes ((\mathrm{id}_1 \otimes \mathrm{NOT}) \mathbin{;} \mathrm{AND})\colon 4 \to 2$$
$$((\mathrm{id}_1 \otimes \mathrm{id}_1) \otimes (\mathrm{id}_1 \otimes \mathrm{NOT})) \mathbin{;} (\mathrm{OR} \otimes \mathrm{AND})\colon 4 \to 2$$

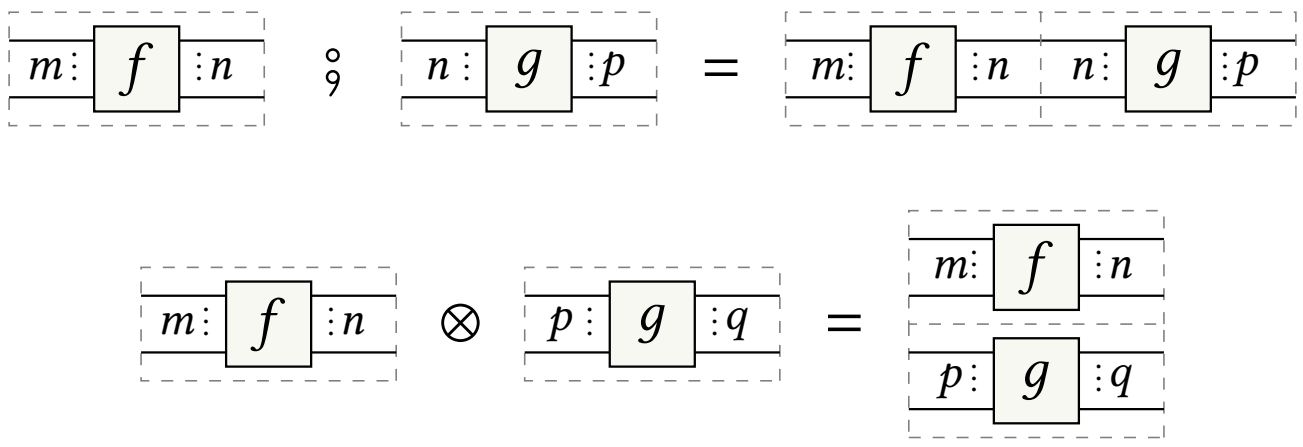

Figure 2.1: Sequential and parallel composition of string diagrams

## 2.2 String diagrams

Even simple terms described using one-dimensional text strings quickly become indecipherable. Fortunately, terms have a graphical syntax known as *string diagrams* [JS91] that makes reading terms far more intuitive. In a string diagram, a generator $\phi \colon m \to n$ is drawn as a box with $m$ inputs and $n$ outputs 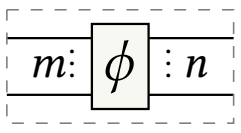, the identity $\mathrm{id}_1$ as a wire 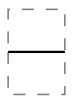, the empty identity $\mathrm{id}_0$ as empty space 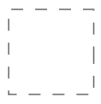, and the symmetry as two wires swapping over 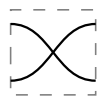. Composite terms are drawn as boxes 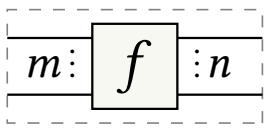; composition is then depicted as *horizontal juxtaposition* and tensor as *vertical juxtaposition*; both are illustrated in Figure 2.1.

**Remark 2.7.** The direction that the 'flow' of string diagrams travels from inputs to outputs is a hotly-debated topic; in this thesis we adopt the left-to-right approach. If you disagree, you could rotate the document by ninety degrees or use a mirror.

**Example 2.8.** Recall the set of generators from Example 2.6. We will draw each logic gate using the usual notation, i.e.

AND ≔ 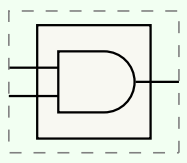 OR ≔ 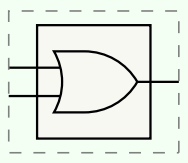 NOT ≔

Now the terms in Example 2.6 can be illustrated as:

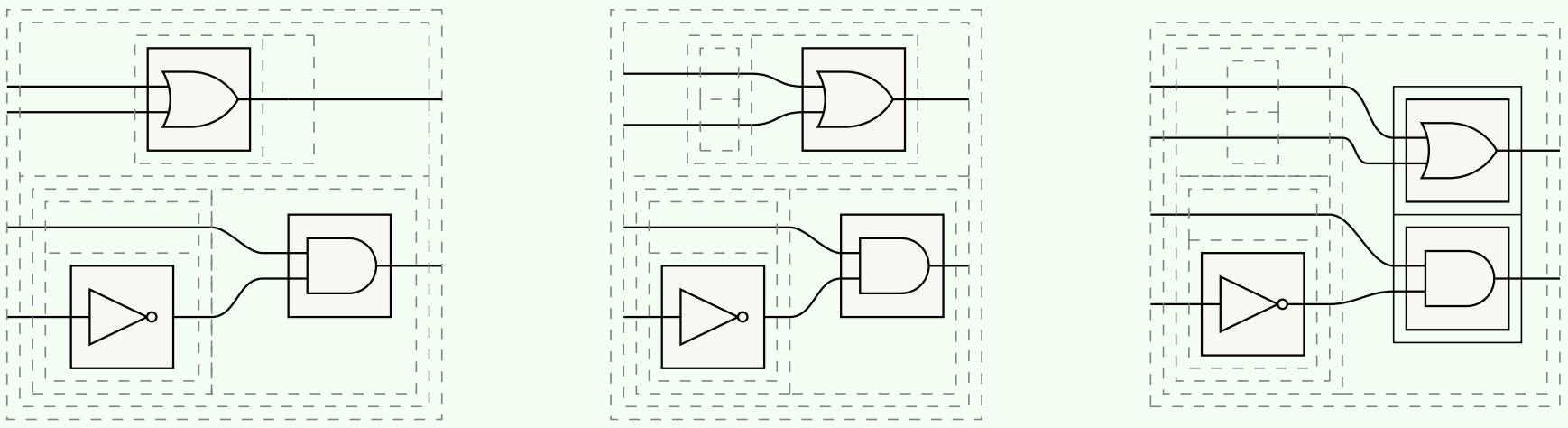

Note that while each term has a very different text string, and is tiled differently in each diagram, the connectivity of the thick wires is actually the same.

The above example illuminates the drawback of reasoning with term strings or even the 'tile' graphical syntax; there are terms which are 'morally' the same but because of variations in how they are constructed they are *not* the same term syntactically. We need some equations to identify terms appropriately; we will soon see how these equations can be derived using the mathematical structure of a *category*.

**Notation 2.9.** So far we have drawn multiple wires in parallel either explicitly or by using vertical dots. In the interests of clarity, we will now start to collapse multiple wires into a single wire, annotated with the appropriate number when not clear from context: so 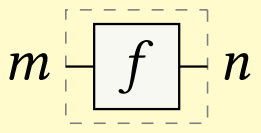 $:=$ 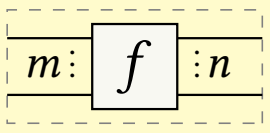 .

Note that these values $m$ or $n$ could be zero, and as such can just be drawn as empty space. It may be useful to explicitly draw wires of zero width for applying reasoning steps; we will do so with a 'faded' wire, like 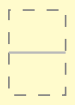 .

## 2.3 Coloured terms

In Σ-terms, the wires are *monochromatic*; there is no distinguishing between them. Sometimes it is advantageous to *annotate* wires with some information: in the realm of terms this is known as assigning the wires *colours* or *sorts*. When working with coloured terms, we need to fix the set of colours before specifying a set of generators.

**Notation 2.10.** We say that a set $C$ is *countable* if it is finite or countably infinite, i.e. there exists a set $X \subseteq \mathbb{N}$ such that there is a bijection $C \cong X$.

**Remark 2.11.** Usually the set of colours is finite, but we will see later in this thesis how having a colour for every single natural number might be useful.

In the monochromatic world the interface of a generator can be specified solely by two natural numbers $m$ and $n$, as there are $m$ input wires and $n$ output wires. When the wires are coloured, more information is needed: the inputs and outputs must be specified in terms of their colours and their ordering.

**Notation 2.12** (Words). Given a set $A$, the set of (finite) words of elements of $A$ is denoted $A^\star$. Words are written $x_0x_1x_2\cdots x_{n-1}$; arbitrary words are written with an overline $\overline{x}, \overline{y}, \overline{z} \ldots \in A^\star$. Given two words $\overline{x}, \overline{y}$, their concatenation is denoted $\overline{xy}$. For an element $a \in A$, the concatenation of this element $n$ times is written $a^n$, e.g. $a^3 \coloneqq aaa$. Given a word $\overline{x}$, its *length* is denoted $|\overline{x}|$; for $i < |\overline{x}|$ the *$i$-th element of* $\overline{x}$ is denoted $\overline{x}(i)$, i.e. in word $\overline{x} \coloneqq abc$, $\overline{x}(0) = a$, $\overline{x}(1) = b$ and $\overline{x}(2) = c$.

**Definition 2.13** (Coloured generators)**.** For a countable set $C$, a set of *$C$-coloured generators* $\Sigma$ is a set equipped with two functions $\mathrm{dom}, \mathrm{cod} \colon \Sigma \to C^\star$.

Coloured terms are generated in the same way as before, but terms can only be composed if the coloured wires in the interfaces match up.

**Definition 2.14** (Coloured terms)**.** For a countable set $C$ and a set of $C$-coloured generators, a $(C, \Sigma)$-term is written $f \colon \overline{m} \to \overline{n}$, where $\overline{m}, \overline{n} \in C^\star$. The set of $(C, \Sigma)$-terms, denoted $(C, \Sigma)_{\mathrm{t}}$, is generated as the monochromatic set of terms, but with an identity and symmetry for each $c \in C$, composition between terms whose output and input words agree on colours, and addition replaced by word concatenation.

**Remark 2.15.** When the set of colours is a singleton $C \coloneqq \{\bullet\}$, the $C$-coloured terms are just the monochromatic terms.

When drawing coloured terms string diagrammatically, the wires that connect the generators are coloured appropriately.

**Example 2.16.** Recall the logic gate generators from Example 2.6; let us now say that the OR gate is rated for a different voltage, so it is no longer compatible with the other components. We can model this with coloured terms; let $C_{\mathrm{g}} \coloneqq \{\bullet, {\color{red}\bullet}\}$ be a set of *voltage colours* and let $\Sigma_g^+ \coloneqq \{\mathrm{AND} \colon \bullet\bullet \to \bullet, \mathrm{OR} \colon {\color{red}\bullet\bullet} \to {\color{red}\bullet}, \mathrm{NOT} \colon \bullet \to \bullet\}$. We draw the 'red' OR appropriately as 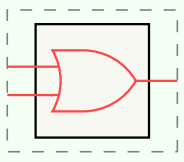. Now we can recreate the diagrams from Example 2.8 in the coloured setting. Note how all wires attached to the OR gate must also be red to ensure the colours all match up.

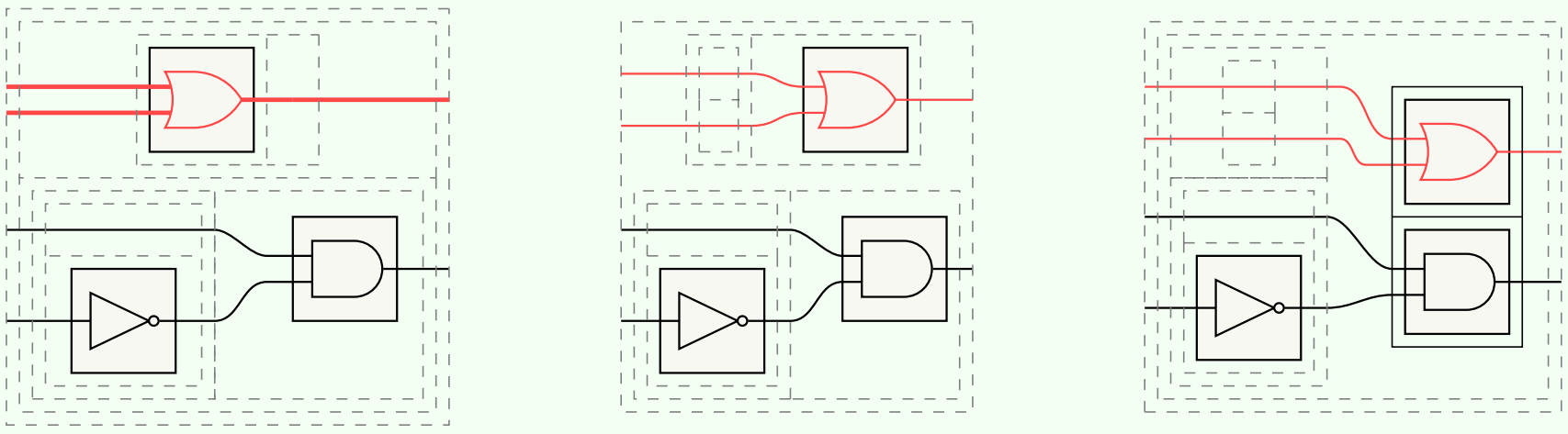

## 2.4 Categories

Terms are purely syntax, so two terms are only equal if they are constructed in precisely the same way. As we have already seen, this is far too strong a relation; there may be many terms that, while constructed in different ways, look the same when drawn out as a diagram modulo the tiling. We can identify terms that describe the same process using *equations*, but which ones?

**Example 2.17.** Consider the term $\mathrm{id}_m \mathbin{;} f$, read as 'do nothing and then run $f$', and drawn as 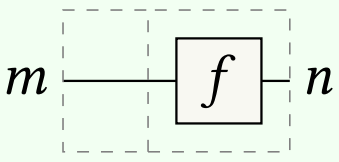 . Clearly this is the same as 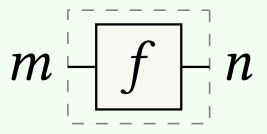 but with an elongated input wire. So one equation we need is 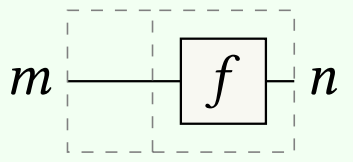 = 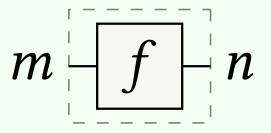 .

It turns out that the required equations are the equations of *symmetric monoidal categories*. To grasp how these equations are derived one requires quite a bit of technical knowledge, so we will build it up one step at a time. We start with a *category*.

**Definition 2.18** (Category)**.** A *category* $\mathcal{C}$ consists of a class of *objects* $\mathrm{ob}(\mathcal{C})$ ; a class of *morphisms* $\mathcal{C}(A, B)$ for every pair of objects $A, B \in \mathrm{ob}(\mathcal{C})$; and a *composition* operation $- \circ -\colon \mathcal{C}(B, C) \times \mathcal{C}(A, B) \to \mathcal{C}(A, C)$ such that

- for any object $A \in \mathrm{ob}(\mathcal{C})$ there exists a unique *identity* morphism $\mathrm{id}_A$;
- for any $f \in \mathcal{C}(A, B)$, it holds that $f \circ \mathrm{id}_A = f = \mathrm{id}_B \circ f$; and
- for any morphisms $f \in \mathcal{C}(A, B)$, $g \in \mathcal{C}(B, C)$ and $h \in \mathcal{C}(C, D)$, it holds that $(h \circ g) \circ f = h \circ (g \circ f)$.

A morphism $f \in \mathcal{C}(A, B)$ is also called an *arrow* , and will often be written $f\colon A \to B$ accordingly. When clear from context, we will use the notation $A \in \mathcal{C}$ or $f \in \mathcal{C}$ for objects and morphisms belonging to a particular category.

**Remark 2.19.** The definition of a category uses *classes* rather than sets as one might expect; this is due to size issues and Russell's paradox regarding the impossibility of the 'set of all sets'.

We interpret terms as morphisms; one difference here is that we previously used 'left-to-right' composition $\mathbin{;}$ rather than 'right-to-left' composition $\circ$.

**Notation 2.20.** *Diagrammatic order* composition is written as $f \mathbin{;} g \coloneqq g \circ f$.

The equations of categories are illustrated with string diagram notation in Figure 2.2.

$$\mathrm{id}_A \fatsemi f \quad = \quad f \qquad\qquad f \fatsemi \mathrm{id}_B \quad = \quad f$$

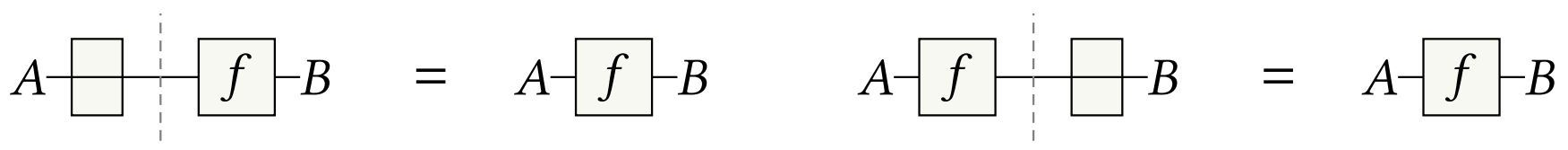

$$(f \fatsemi g) \fatsemi h \quad = \quad f \fatsemi (g \fatsemi h)$$

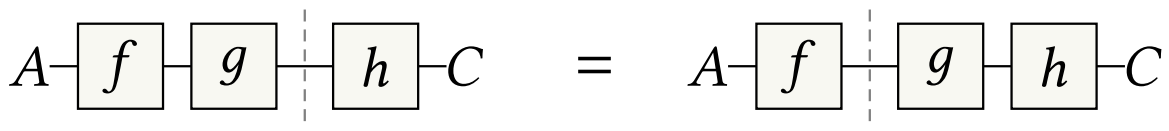

Figure 2.2: Equations of a category

### 2.4.1 Commutative diagrams

Equations in category theory can be expressed using *commutative diagrams*. For example, the unitality and associativity of composition can be illustrated as follows:

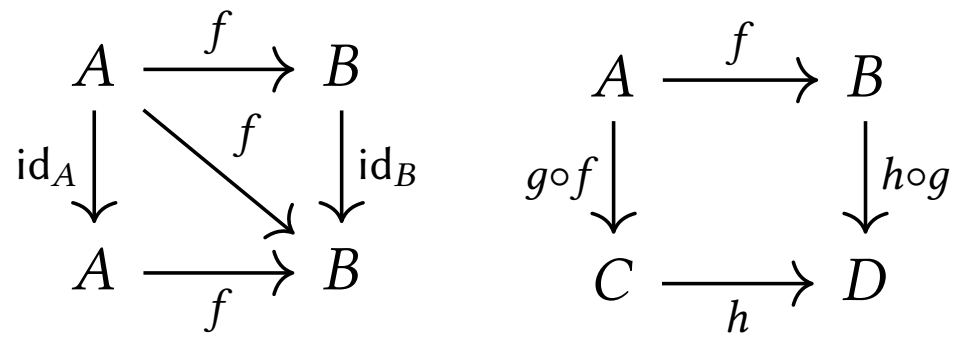

We say that the above two diagrams *commute* precisely because $\mathrm{id}_B \circ f = f = f \circ \mathrm{id}_A$ and $(h \circ g) \circ f = h \circ (g \circ f)$: no matter which path one takes, the results are equal.

### 2.4.2 Examples of categories

The definition of a category is quite abstract and might take some getting used to: it can be helpful to consider some concrete examples.

**Example 2.21** (Preorder)**.** A *preorder* is a reflexive, transitive binary relation $\lesssim$ on a set $X$. Any preorder generates a category $\mathcal{C}_{\leq}$: the objects are the elements of $X$ and $\mathcal{C}_{\leq}(x, y)$ contains exactly one morphism if $x \leq y$ and none otherwise.

**Example 2.22** (Sets)**.** The category **Set** has sets as objects and functions as morphisms, with composition of functions as composition. There are other categories with sets as objects, such as **Rel**, which has *relations* as morphisms and composition of relations as morphisms. There is also a category **FinSet** containing the *finite* sets and functions between them.

**Example 2.23** (Posets)**.** A *partial order* on a set $A$ is a reflexive, antisymmetric and transitive relation $\leq \subseteq A \times B$. A set equipped with a partial order is called a *partially ordered set*, or *poset* for short. For posets $(A, \leq_A)$ and $(B, \leq_B)$, a function $f \colon A \to B$ is called *monotone* if $a \leq_A a'$ implies that $f(a) \leq_B f(a')$.

Much like how sets form a category, posets form the category $\mathbf{Pos}$, where ob($\mathbf{Pos}$) are posets and $\mathbf{Pos}(X, Y)$ are the monotone functions $X \to Y$.

**Example 2.24** (Monoids)**.** A *monoid* is a tuple $(A, *, e)$ where $A$ is a set called the *carrier*, $* \colon A \times A \to A$ is a binary operation called the *multiplication*, and $e \in A$ is an element called the *unit*, such that $a * e = a = e * a$ for any $a \in A$. A *monoid homomorphism* between two monoids $(A, *, e_A)$ and $(B, +, e_B)$ is a map $h \colon A \to B$ such that $h(a * a') = h(a) + h(a')$ and $h(e_A) = e_B$. There is a category $\mathbf{Mon}$ with monoids as the objects and monoid homomorphisms as the morphisms.

**Example 2.25** (Product category)**.** Given two categories $\mathcal{C}$ and $\mathcal{D}$, their *product category* $\mathcal{C} \times \mathcal{D}$ is the category with objects defined as $\text{ob}(\mathcal{C} \times \mathcal{D}) \coloneqq \text{ob}(\mathcal{C}) \times \text{ob}(\mathcal{D})$ and the morphisms as defined as

$$(\mathcal{C} \times \mathcal{D})((A, A'), (B, B')) \coloneqq \{(f, f') \mid f \in \mathcal{C}(A, B), f' \in \mathcal{D}(A', B')\}$$

For morphisms $f \colon A \to B, g \colon B \to C \in \mathcal{C}$ and $f' \colon A' \to B', g' \colon B' \to C' \in \mathcal{D}$, the composition of $(f, g) \colon (A, A') \to (B, B')$ and $(g, g') \colon (B, B') \to (C, C')$ is defined as $(g, g') \circ (f, f') \coloneqq (g \circ f, g' \circ f')$.

### 2.4.3 Universal properties

Category theory is an appealing foundation because it can be used to *abstract* away from concrete constructions. Rather than proving results about particular objects and morphisms, we can show how they are an instantiation of some more abstract concept. One such way we can do this is by considering the properties a morphism might have.

**Definition 2.26** (Monomorphism)**.** A morphism $f \colon A \to B \in \mathcal{C}$ is called a *monomorphism* (or simply *mono* for short) if for any two morphisms $g_1, g_2 \colon C \to A$, if $f \circ g_1 = f \circ g_2$, then $g_1 = g_2$.

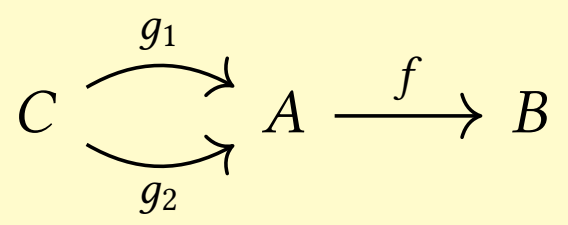

One can think of monomorphisms as morphisms which are *left-cancellative*. There is also a way to describe *invertible* morphisms.

**Definition 2.27** (Isomorphism)**.** A morphism $f\colon A \longrightarrow B \in \mathcal{C}$ is called an *isomorphism* (or simply *iso* for short) if there also exists a morphism $f^{-1}\colon B \longrightarrow A \in \mathcal{C}$ such that $f^{-1} \circ f = \mathrm{id}_A$ and $f \circ f^{-1} = \mathrm{id}_B$.

$$A \xrightarrow{f} B \xrightarrow{f^{-1}} A \qquad\qquad B \xrightarrow{f^{-1}} A \xrightarrow{f} B$$
$$A \xrightarrow{\mathrm{id}_A} A \qquad\qquad\qquad B \xrightarrow{\mathrm{id}_B} B$$

**Example 2.28.** In **Set**, the monomorphisms are the injective functions and the isomorphisms are the bijective functions.

Often we are concerned with particular *constructions* in a category; some interaction of objects and morphisms which has special properties. These are specified in terms of a *universal property*: a unique morphism that indicates the 'best' way to describe something. We will first consider universal properties concerning special objects.

**Definition 2.29** (Initial object)**.** An object $C$ in a category $\mathcal{C}$ is *initial* if, for any other object $X \in \mathcal{C}$ there exists a unique morphism $C \longrightarrow X$.

**Example 2.30.** In **Set**, the initial object is the empty set $\emptyset$, as there is a unique function from $\emptyset$ to any set $X$, the so-called 'absurd function'.

Most categorical notions also have a *dual*, in which all the arrows are flipped. This means that when defining constructions and proving results about them, we often also get results for free about the dual case.

**Definition 2.31** (Terminal object)**.** An object $C$ in a category $\mathcal{C}$ is *terminal* if, for any other object $X \in \mathcal{C}$ there exists a unique morphism $X \longrightarrow C$.

**Example 2.32.** In **Set**, the terminal object is the set containing a single object $\star$: from any set $X$ there is a unique function $X \longrightarrow \{\star\}$; namely the function $x \mapsto \star$.

If a category has an initial or terminal object, then it is *unique up to unique isomorphism*; this means that if we have two objects that satisfy the universal property, then these objects are isomorphic in a unique way.

We will now explore some universal properties that illustrate how common structures can be expressed in terms of the morphisms between them.

**Definition 2.33** (Product)**.** Given a category $\mathcal{C}$ and objects $A, B \in \mathcal{C}$, their *product* is an object $A \times B$ equipped with a pair of morphisms $p_0 : A \times B \longrightarrow A$ and $p_1 : A \times B \longrightarrow B$ called *projections* such that for every other object $Z$ with pair of morphisms $f : Z \longrightarrow A$ and $g : Z \longrightarrow B$, there exists a unique morphism $u : Z \longrightarrow A \times B$ such that the following diagram commutes:

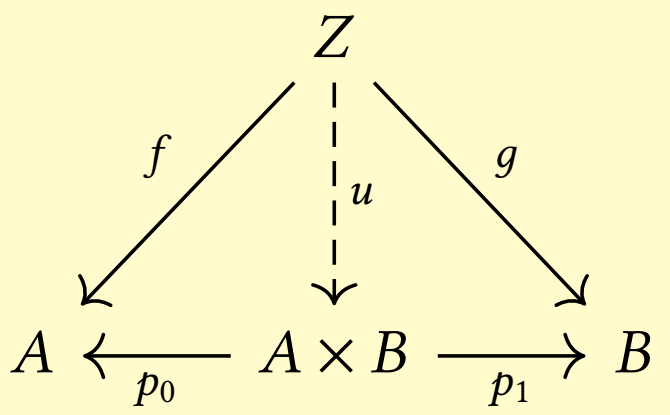

A category $\mathcal{C}$ is said to *have products* if the product exists for all objects $A, B \in \mathcal{C}$.

Again, if a category has a product $A \times B$ then it is unique up to unique isomorphism: any other construction $A \times' B$ also satisfying the universal property must be isomorphic to $A \times B$. This means that we are justified in referring to 'the' product if one exists.

**Example 2.34.** The product in **Set** is the Cartesian product.

The dual of a product is a construction with *in*jections rather than *pro*jections.

**Definition 2.35** (Coproduct)**.** Given a category $\mathcal{C}$ and objects $A, B \in \mathcal{C}$, their *coproduct* is an object $A + B$ equipped with a pair of morphisms $i_0 : A \longrightarrow A + B$ and $i_1 : B \longrightarrow A + B$ called *injections* such that for every other object $Z$ with pair of morphisms $f : A \longrightarrow B$ and $g : B \longrightarrow Z$, there exists a unique morphism $u : A + B \longrightarrow Z$ such that the following diagram commutes:

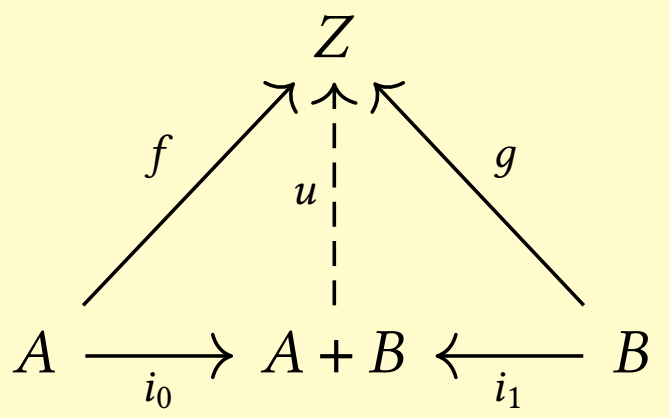

A category $\mathcal{C}$ is said to *have coproducts* if the coproduct exists for all pairs of objects $A, B \in \mathcal{C}$.

**Example 2.36.** The coproduct in **Set** is the disjoint union.

Finally, we will look at properties showing how pairs of morphisms with a common domain or codomain can be in some way 'unified'.

**Definition 2.37** (Pushout)**.** Given a category $\mathcal{C}$ and morphisms $f\colon A \to B$ and $g\colon A \to C$, a *pushout* is an object $D \in \mathcal{C}$ and a pair of morphisms $h\colon B \to D$ and $k\colon C \to D$ such that for any other pair of morphisms $h'\colon B \to Z$ and $k'\colon C \to Z$ there exists a unique morphism $u\colon D \to Z$, i.e. the following diagram commutes:

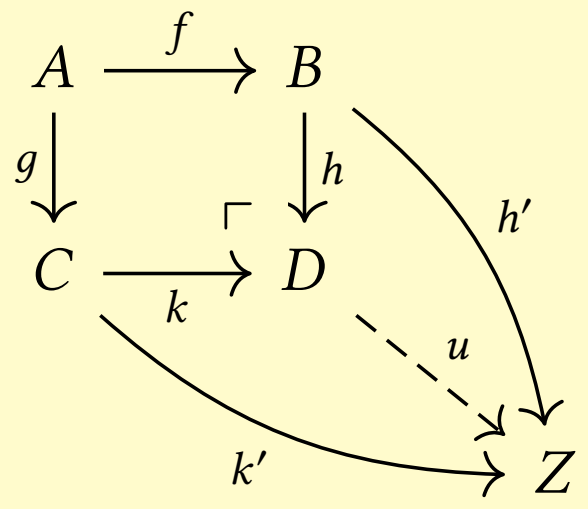

A category $\mathcal{C}$ is said to *have pushouts* if a pushout exists for any pair of morphisms.

**Example 2.38.** Set has pushouts: given morphisms $f\colon A \to B$ and $g\colon A \to C$, the pushout is the union of $B$ and $C$ identifying elements with a common preimage in $A$. Concretely, let $\sim \subseteq B \times C$ be a relation defined as $\{(b,c) \mid \exists a \in A.\, b = f(a) \wedge c = g(a)\}$. Then the pushout set $D$ is defined as $B \cup C/\sim$ with the morphisms $h\colon B \to D$ and $k\colon C \to D$ sending elements in $B$ and $C$ to the appropriate element in $D$.

A pushout square is normally indicated with a $\ulcorner$ symbol as shown above. The dual of the pushout is a *pullback*.

**Definition 2.39** (Pullback)**.** Given a category $\mathcal{C}$ and morphisms $f\colon B \to A$ and $g\colon C \to A$, a *pullback* is an object $D \in \mathcal{C}$ and a pair of morphisms $h\colon D \to B$ and $k\colon D \to C$ such that for any other pair of morphisms $h'\colon Z \to B$ and $k'\colon Z \to C$ there exists a unique morphism $u\colon Z \to D$, i.e. the following diagram commutes:

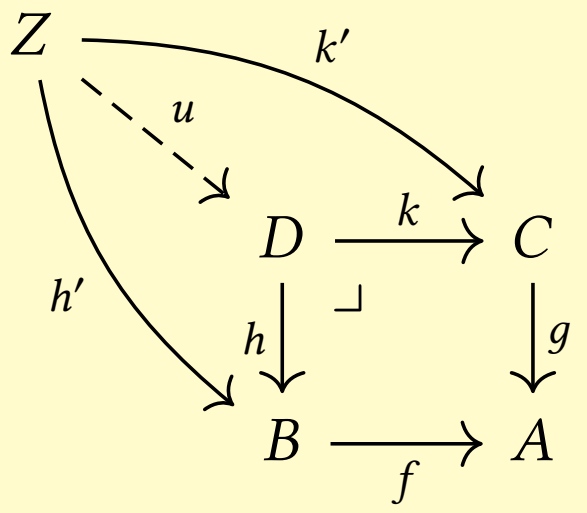

A category $\mathcal{C}$ is said to *have pullbacks* if a pullback exists for any pair of morphisms.

**Example 2.40.** Set also has pullbacks: given morphisms $f\colon B \to A$ and $g\colon C \to A$, the pullback is defined as $\{(b,c) \in B \times C \mid f(b) = g(c)\}$ and the maps $h\colon D \to B$ and $k\colon D \to C$ are defined as the first and second projection respectively.

Universal properties mean we do not need to restrict ourselves to a concrete category when using or proving results; we can instead say, for example, that the result holds in all categories with products.

## 2.5 Functors

It is actually quite rare that we are only making use of one category at a time. In order to compare categories, we need a notion of *mapping between* them. Such a map $F\colon \mathcal{C} \longrightarrow \mathcal{D}$ sends an object $A \in \mathcal{C}$ to an object $FA \in \mathcal{D}$, and a morphism $f\colon A \longrightarrow B \in \mathcal{C}$ to $Ff\colon FA \longrightarrow FB$. This is generally not a strong enough definition to be useful, so some additional restrictions must be added to define the notion of a *functor*.

**Definition 2.41** (Functor)**.** Given two categories $\mathcal{C}$ and $\mathcal{D}$, a *functor* $F\colon \mathcal{C} \longrightarrow \mathcal{D}$ maps objects and morphisms in $\mathcal{C}$ to objects and morphisms in $\mathcal{D}$ such that

- $F(\mathrm{id}_A) = \mathrm{id}_{FA}$ for every $A \in \mathrm{ob}(\mathcal{C})$; and
- $F(g \circ f) = F(g) \circ F(f)$ for every $f\colon A \longrightarrow B$ and $g\colon B \longrightarrow C$.

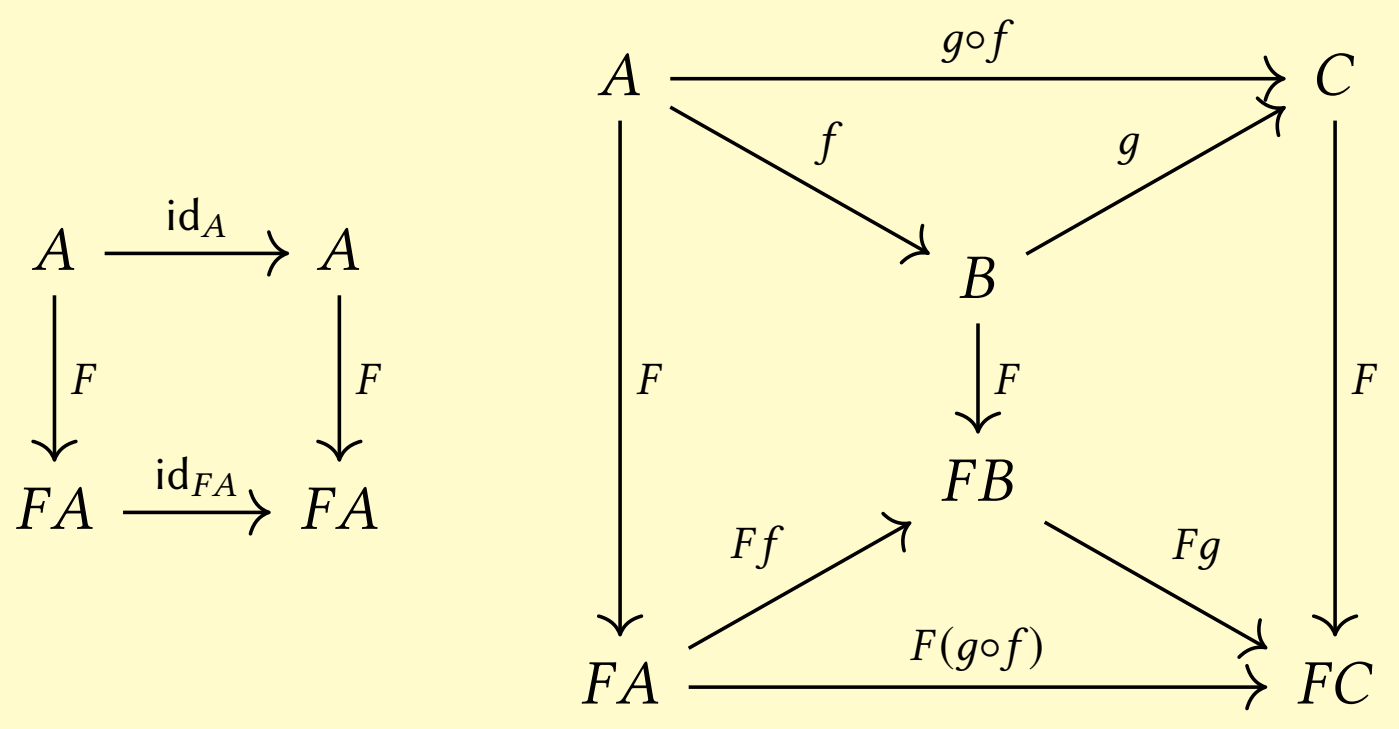

The two equations above are known as the *functoriality* equations; if a map satisfies these it is said to be *functorial*.

Functors have a graphical representation as 'functorial boxes' [Mel06]; the notation for applying a functor $F\colon \mathcal{C} \longrightarrow \mathcal{D}$ to a morphism $f\colon X \longrightarrow Y$ is shown in Figure 2.3, along with the depictions of the functoriality equations. As always, the wire labels are optional and will be omitted if unambiguous.

### 2.5.1 Examples of functors

Many notions in mathematics and computer science can be viewed as functors.

**Definition 2.42** (Endofunctor)**.** An *endofunctor* on category $\mathcal{C}$ is a functor $\mathcal{C} \longrightarrow \mathcal{C}$.

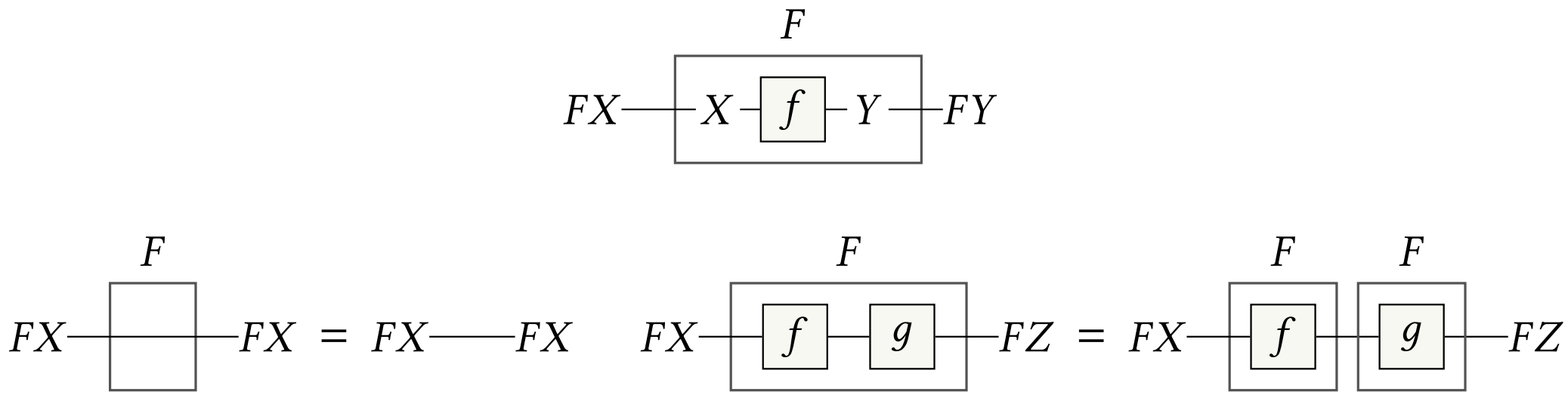

Figure 2.3: Diagrammatic depiction of 'functorial boxes', and graphical equations of functoriality

**Example 2.43** (Powerset functor). The notion of powerset can be interpreted as an endofunctor $\mathcal{P}\colon \mathbf{Set} \to \mathbf{Set}$, mapping a set $X$ to its powerset $\mathcal{P}(X)$ and a morphism $f\colon X \to Y$ to the function $\mathcal{P}(X) \to \mathcal{P}(Y)$ which applies $f$ pointwise.

**Example 2.44** (List functor). A functor that crops up frequently in computer science is the *list functor* List: $\mathbf{Set} \to \mathbf{Set}$, which sends a set $X$ to its set of lists $X^\star$, and sends a function $f\colon X \to Y$ to the function $f^\star\colon X^\star \to Y^\star$: which applies $f$ to each element of a list.

**Example 2.45** (Free monoid). The *free construction* of a mathematical structure can be viewed as its most 'bare-bones' version. Recall the definition of monoids from Example 2.24: the set of words $X^\star$ as defined in Notation 2.12 is the carrier of the *free monoid* on $X$. This means there is a functor $F\colon \mathbf{Set} \to \mathbf{Mon}$ (the *free functor*) that acts on objects as $X \mapsto (X^\star, (\overline{v}, \overline{w}) \mapsto \overline{vw}, \varepsilon)$ and sends morphisms $X \to Y$ to the corresponding monoid homomorphism $X^\star \to Y^\star$.

There is also a *forgetful* or *underlying* functor $U\colon \mathbf{Mon} \to \mathbf{Set}$ which sends a monoid $(X, *, e)$ to its carrier set $X$ and 'forgets' the monoid structure. The functors $F$ and $U$ form an *adjunction*, but this is beyond the scope of this thesis.

### 2.5.2 Functors as morphisms

It is possible to view functors purely as a standalone concept, but the core tenet of category theory is to view everything in terms of morphisms; functors are no different.

**Example 2.46** (Identity functor). A trivial endofunctor for any category $\mathcal{C}$ is the *identity functor* Id: $\mathcal{C} \to \mathcal{C}$ which acts as the identity on objects and morphisms.

**Example 2.47** (Functor composition)**.** If $F\colon \mathcal{C} \to \mathcal{D}$ and $G\colon \mathcal{D} \to \mathcal{E}$ are functors then their composite $G \circ F\colon \mathcal{C} \to \mathcal{E}$ is also a functor.

Identity and (associative) composition are all we need to define a category. Although we cannot define a category containing *all* categories, we can define a category containing all 'small' categories.

**Definition 2.48.** A category $\mathcal{C}$ is *small* if $\mathrm{ob}(\mathcal{C})$ and $\mathcal{C}(A, B)$ are sets for all $A, B \in \mathcal{C}$.

**Example 2.49** (Category of categories)**.** **Cat** is the category in which $\mathrm{ob}(\mathbf{Cat})$ are small categories and $\mathbf{Cat}(\mathcal{C}, \mathcal{D})$ contains functors $\mathcal{C} \to \mathcal{D}$. Identity is the identity functor, and composition is functor composition.

Using the category of categories means we can reason about categories using the same line of reasoning as for objects and morphisms.

### 2.5.3 Full and faithful functors

One important way we can class functors is based on how they act on the classes of morphisms between two objects.

**Notation 2.50.** Given a functor $F\colon \mathcal{C} \to \mathcal{D}$, let $F_{A,B}\colon \mathcal{C}(A, B) \to \mathcal{D}(FA, FB)$ be the induced map sending classes of morphisms $A \to B$ in $\mathcal{C}$ to the classes of morphisms $FA \to FB$ in $\mathcal{D}$.

**Definition 2.51** (Faithful functor [Mac78])**.** A functor $F\colon \mathcal{C} \to \mathcal{D}$ is *faithful* if $F_{A,B}$ is injective for all $A, B \in \mathcal{C}$.

A faithful functor $F\colon \mathcal{C} \to \mathcal{D}$ does not coalesce morphisms: every morphism $f \in \mathcal{C}(A, B)$ has a unique morphism $Ff \in \mathcal{D}(FA, FB)$.

**Definition 2.52** (Full functor [Mac78])**.** A functor $F\colon \mathcal{C} \to \mathcal{D}$ is *full* if $F_{A,B}$ is surjective for all $A, B \in \mathcal{C}$.

Every morphism $FA \to FB$ is in the image of a full functor.

**Definition 2.53** (Fully faithful functor [Mac78])**.** A functor $F\colon \mathcal{C} \to \mathcal{D}$ is *fully faithful* if $F_{A,B}$ is bijective for all $A, B \in \mathcal{C}$; i.e. the functor is full and faithful.

**Example 2.54.** Recall the categories **Set** and **Mon**, and the functor $U: \mathbf{Mon} \to \mathbf{Set}$ defined in Example 2.45. $U$ is faithful as monoid homomorphisms are just functions, but it is not full as not all functions are monoid homomorphisms. Note that even though $U$ is faithful, it is not injective on objects because there may be many monoids with the same carrier set.

Functors can be used to compare categories, and the notions of fullness and faithfulness show how exactly two categories are related. One way a category $\mathcal{C}$ can be related to another category $\mathcal{D}$ is that the former is the latter with 'some of the bits taken out'.

**Definition 2.55** (Subcategory)**.** Given a category $\mathcal{C}$, a *subcategory* $\mathcal{D}$ of $\mathcal{C}$ is a subclass of the objects in $\mathcal{C}$ and a subclass of the morphisms in $\mathcal{D}$ such that

- for every object $A \in \mathcal{D}$, the identity $\mathrm{id}_A$ is in $\mathcal{D}$;
- for every morphism $f: A \to B \in \mathcal{D}$, the source and targets $A$ and $B$ are in $\mathcal{D}$; and
- for morphisms $f: A \to B, g: B \to C \in \mathcal{D}$, the composition $g \circ f$ is in $\mathcal{D}$.

These conditions on a subcategory $\mathcal{D}$ enforce that $\mathcal{D}$ also has the structure of a category. This means there is an obvious induced functor $S: \mathcal{D} \to \mathcal{C}$ mapping objects and morphisms in $\mathcal{D}$ to the same objects in $\mathcal{C}$; this is called an *inclusion functor*, and is often written with a hooked arrow $\mathcal{D} \hookrightarrow \mathcal{C}$ . An inclusion functor is clearly faithful, since there cannot be two morphisms in the subcategory that map to the same morphism in the parent category. Inclusion functors that are also *full* are of particular interest.

**Definition 2.56** (Full subcategory)**.** A subcategory $\mathcal{D}$ is a *full subcategory* if its inclusion functor $\mathcal{D} \to \mathcal{C}$ is full and faithful; i.e. for all objects $A, B \in \mathcal{D}$, the morphisms $\mathcal{D}(A, B) = \mathcal{C}(A, B)$.

**Example 2.57.** **FinSet** is a full subcategory of **Set**, as every function between finite sets is a morphism in both **FinSet** and **Set**. **Set** is a subcategory of **Rel** as every function is a relation, but it is not a full subcategory because there are more relations $A \sim B$ than there are functions $A \to B$.

Sometimes a category is not merely a subcategory of another, but the two categories are actually (almost) the same.

**Definition 2.58.** Two categories $\mathcal{C}$ and $\mathcal{D}$ are *isomorphic* if there exist functors $F: \mathcal{C} \to \mathcal{D}$ and $G: \mathcal{D} \to \mathcal{C}$ such that $G \circ F = \mathrm{Id}_{\mathcal{C}}$ and $F \circ G = \mathrm{Id}_{\mathcal{D}}$.

It can be inconvenient to construct the functors in both directions; fortunately isomorphism can be shown by constructing just the functor in one direction, as long as it has the required properties.

**Lemma 2.59** ([Mac78])**.** Two categories $\mathcal{C}$ and $\mathcal{D}$ are isomorphic if and only if there exists a fully faithful functor $F\colon \mathcal{C} \to \mathcal{D}$ which is also bijective on objects.

**Remark 2.60.** Usually, isomorphism of categories is too restrictive; often the weaker notion *equivalence of categories* is used, in which the two functors $F$ and $G$ need only be *naturally isomorphic* to the identity functor rather than equal. However, in this thesis it turns out that all the results we need really are strong enough to be isomorphisms.

### 2.5.4 Universal properties through functors

Previously we defined some *universal properties* such as initial objects and pullbacks. Using functors, these universal constructions can be viewed as special cases of even more abstract notions: *limits* and *colimits*.

**Definition 2.61** (Cone)**.** Let $\mathcal{J}$ and $\mathcal{C}$ be categories, and let $F\colon \mathcal{J} \to \mathcal{C}$ be a functor. A *cone to* $F$ is an object $N \in \mathcal{C}$ equipped with a family of morphisms $\phi_X\colon N \to FX$ for each $X \in \mathcal{J}$, such that for each $f\colon X \to Y \in \mathcal{J}$, $Ff \circ \phi_X = \phi_Y$.

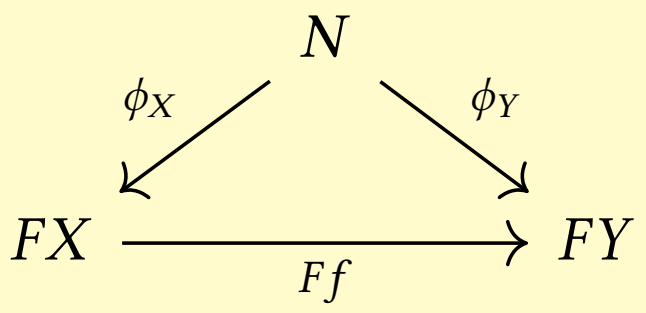

The limit is the 'best possible cone'.

**Definition 2.62** (Limit)**.** Let $\mathcal{J}$ and $\mathcal{C}$ be categories, and let $F\colon \mathcal{J} \to \mathcal{C}$ be a functor. The *limit of* $F$ is a cone to $F$ $(L, \phi)$ such that for every cone to $F$ $(N, \psi)$, there exists a unique morphism $u\colon N \to L$ such that $\phi_X \circ u = \psi_X$ for all $X \in \mathcal{J}$.

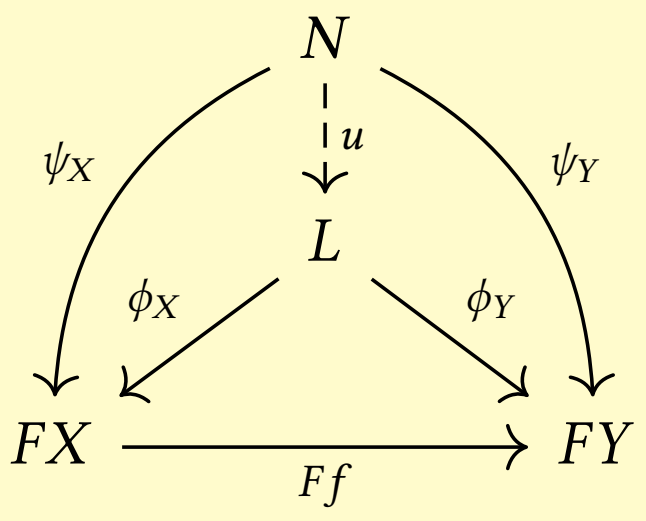

Limits generalise several of the universal constructions we have encountered so far.

**Example 2.63** (Terminal object)**.** When the category $\mathcal{J}$ is the empty category, the only functor $F\colon \mathcal{J} \to \mathcal{C}$ is the empty functor. Since there are no objects in $\mathcal{J}$, a cone of $F$ is just an object $N$, so the limit of $F$ is an object $L$ to which there is a unique morphism $N \to L$; the terminal object.

**Example 2.64** (Products)**.** When the category $\mathcal{J}$ has objects but no morphisms other than identities, a functor $F\colon \mathcal{J} \to \mathcal{C}$ is one that indexes objects of $\mathcal{C}$ by objects in $\mathcal{J}$. A cone of $F$ is an object $N$ with morphisms that pick each of these indexed elements, so the limit of $F$ is the product.

**Example 2.65** (Pullbacks)**.** Let $\mathcal{J}$ be a category containing objects $A$, $B$, $C$ with non-identity morphisms $B \to A$ and $C \to A$. A cone of $F\colon \mathcal{J} \to \mathcal{C}$ is an object $N$ and morphisms $N \to FA$, $N \to FB$ and $N \to FC$, so the limit of $F$ is the pullback.

Since limits define so many categorical structures, the ability to define arbitrary limits in a category makes it a much more appealing setting to work in.

**Definition 2.66.** For a small category $\mathcal{J}$, a category $\mathcal{C}$ *has limits of shape* $\mathcal{J}$ if every functor $F\colon \mathcal{J} \to \mathcal{C}$ has a limit in $\mathcal{C}$.

**Definition 2.67** (Complete category)**.** A category $\mathcal{C}$ is *complete* if it has limits for all functors $F\colon \mathcal{J} \to \mathcal{C}$, where $\mathcal{J}$ is a small category.

Limits do not generalise all the universal constructions we have seen so far; for the rest we must flip the arrows and consider the dual version.

**Definition 2.68** (Cocone)**.** Let $\mathcal{J}$ and $\mathcal{C}$ be categories, and let $F\colon \mathcal{J} \to \mathcal{C}$ be a functor. A *cocone to F* is an object $N \in \mathcal{C}$ equipped with a family of morphisms $\phi_X\colon FX \to N$ for each $X \in \mathcal{J}$, such that for each $f\colon X \to Y \in \mathcal{J}$, $\phi_Y \circ Ff = \phi_X$.

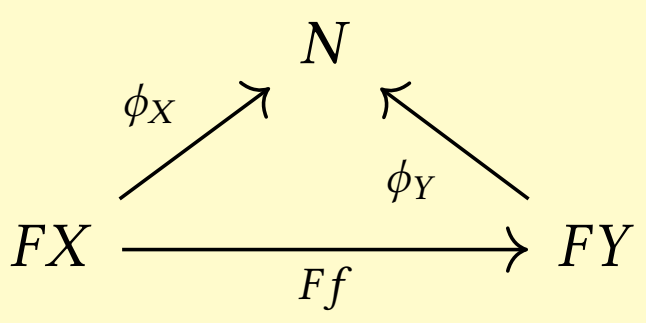

**Definition 2.69** (Colimit)**.** Let $\mathcal{J}$ and $\mathcal{C}$ be categories, and let $F\colon \mathcal{J} \to \mathcal{C}$ be a functor. The *colimit of F* is a cocone to $F$ $(L, \phi)$ such that for every other cocone to $F$ $(N, \psi)$, there exists a unique morphism $u\colon L \to N$ such that $u \circ \phi_X = \psi_X$ for all $X \in \mathcal{J}$.

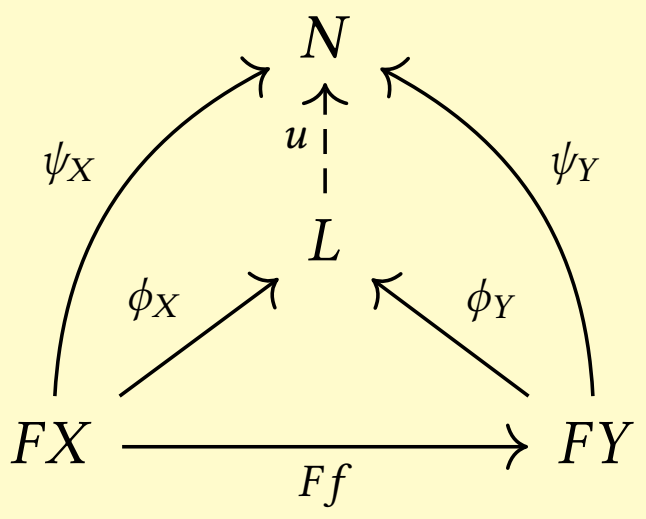

**Example 2.70.** By using the reasoning in Examples 2.63 to 2.65 and reversing the arrows, it is straightforward to see that initial objects, coproducts, and pushouts are examples of colimits.

Once again, we often want to work in a setting in which we can construct all colimits.

**Definition 2.71.** For a small category $\mathcal{J}$, a category $\mathcal{C}$ *has colimits of shape* $\mathcal{J}$ if every functor $F\colon \mathcal{J} \to \mathcal{C}$ has a colimit in $\mathcal{C}$.

**Definition 2.72** (Cocomplete category)**.** A category $\mathcal{C}$ is *cocomplete* if it has colimits for all functors $F\colon \mathcal{J} \to \mathcal{C}$, where $\mathcal{J}$ is a small category.

It is often the case that we are interested in functors that *preserve* structure, so we can exploit it in both categories. Since a lot of structure can be expressed in terms of limits or colimits, this can be expressed succinctly by saying that a functor merely preserves these limits or colimits.

**Definition 2.73.** A functor $F\colon \mathcal{C} \to \mathcal{D}$ *preserves all (co)limits* if $(FL, F\phi)$ is a (co)limit whenever $(L, \phi)$ is a (co)limit.

Sometimes a functor may not preserve *all* (co)limits but only some of them. For example, we may talk about $F$ being a coproduct-preserving functor . This is defined in the same way as above: $F(A + B)$ must itself be a coproduct $FA + FB$.

## 2.6 Natural transformations

As we have seen, functors are useful for mapping objects and morphisms from one category to another in a way that respects the underlying compositional structure. Of course, there may be many such functors $\mathcal{C} \to \mathcal{D}$; the logical next step is to consider maps between functors themselves. These maps are known as *natural transformations*.

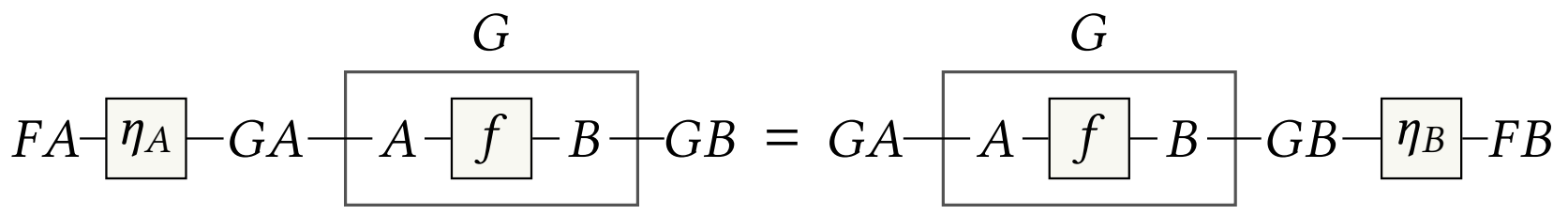

Figure 2.4: Naturality of transformations in string diagram notation

**Definition 2.74** (Natural transformation)**.** Given two functors $F, G\colon \mathcal{C} \to \mathcal{D}$, a *natural transformation* $\eta\colon F \to G$ is a family of morphisms $\eta_A \in \mathcal{D}(FA, GA)$ for each $A \in \mathrm{ob}(\mathcal{C})$, called the *components* of $\eta$, such that $\eta_B \circ Ff = Gf \circ \eta_A$, i.e. the following diagram commutes:

$$\begin{array}{ccc} A & \quad & FA \xrightarrow{\eta_A} GA \\ \downarrow{\scriptstyle f} & & {\scriptstyle Ff}\downarrow \qquad \downarrow{\scriptstyle Gf} \\ B & & FB \xrightarrow{\eta_B} GB \end{array}$$

For functors $F, G\colon \mathcal{C} \to \mathcal{D}$, a natural transformation $\eta\colon F \to G$ is a *family* of morphisms $\eta_A\colon FA \to GA \in \mathcal{D}$ (the components) for each object $A \in \mathcal{D}$. One can think of a natural transformation as a way of inducing morphisms of a certain structure across an *entire category*. Graphically, the naturality equation can be seen as how a natural transformation can be 'pushed through' morphisms, as shown in Figure 2.4.

We established that functors are morphisms between categories; in a similar vein natural transformations are morphisms between functors.

**Definition 2.75** (Functor category)**.** Given two categories $\mathcal{C}$ and $\mathcal{D}$, a *functor category* $[\mathcal{C}, \mathcal{D}]$ has as objects the functors $\mathcal{C} \to \mathcal{D}$ and as morphisms $F \to G$ the natural transformations between these functors.

Like with the category of categories, viewing functors in this way allows us to reason with them in the same way as regular objects and morphisms.

### 2.6.1 Examples of natural transformations

Natural transformations also often arise across mathematics and computer science.

**Example 2.76** (Singleton transformation)**.** Recall the List functor from Example 2.44. An example of a natural transformation is the *singleton transformation* $[-]\colon \mathrm{Id} \to \mathrm{List}$, which induces a function $[-]\colon X \to X^\star$ for each set $X$, defined as $x \mapsto [x]$.

**Example 2.77** (Reduce)**.** Recall the functors $F\colon \mathbf{Set} \to \mathbf{Mon}$ and $U\colon \mathbf{Mon} \to \mathbf{Set}$ from Example 2.45. Functors can be composed just like morphisms, so $F \circ U$ is a functor $\mathbf{Mon} \to \mathbf{Mon}$: such a functor has action $(X, *, e) \mapsto (X^\star, +\!\!+, [\,])$. The component of a natural transformation $F \circ U \to \mathrm{Id}$ at object $(X, *, e)$ is a monoid homomorphism $(X^\star, +\!\!+, [\,]) \to (X, *, e)$, i.e. a function $X^\star \to X$.

One example of such a natural transformation is the *reduce* or *fold* operation, which takes a list in $X^\star$ and reduces it to an element of $X$ by starting with the unit $e$ and multiplying it with each element of the list in turn.

As natural transformations are defined in terms of families of morphisms, they can inherit properties of the components.

**Definition 2.78** (Natural isomorphism)**.** A natural transformation is called a *natural isomorphism* if every component is an isomorphism.

## 2.7 Monoidal categories

The concepts of functors and natural transformations are used to interpret the *parallel* composition $\otimes$. To do this, a special kind of functor known as a *bifunctor* is used.

**Definition 2.79** (Bifunctor)**.** A *bifunctor* is a functor with a product category as its domain, i.e. a functor of the form $\mathcal{C} \times \mathcal{D} \to \mathcal{E}$.

The notation for functor boxes can be extended in order to show how bifunctors map from two categories into one.

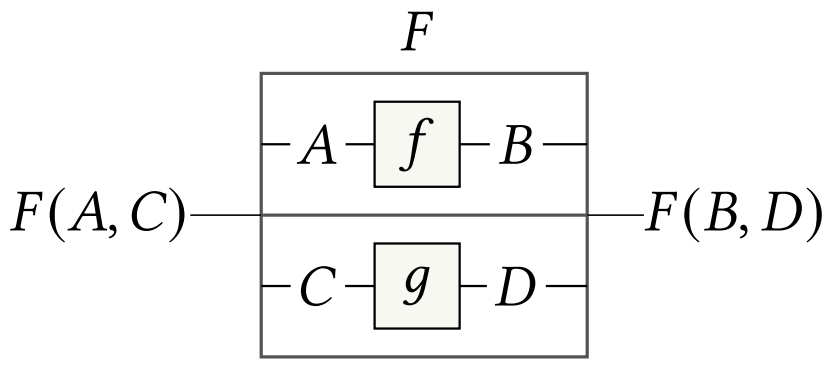

This suggests that a bifunctor is what we need to model parallel composition.

**Definition 2.80** (Monoidal category)**.** A *monoidal category* is a category $\mathcal{C}$ equipped with a bifunctor $-\otimes = \colon \mathcal{C} \times \mathcal{C} \to \mathcal{C}$ called the *tensor product* and an additional object $I$ called the *monoidal unit*, along with natural isomorphisms

- $\alpha_{A,B,C}\colon A \otimes (B \otimes C) \cong (A \otimes B) \otimes C$ called the *associator*;
- $\lambda_A\colon I \otimes A \cong A$ called the *left unitor*; and
- $\rho_A\colon A \otimes I \cong A$ called the *right unitor*

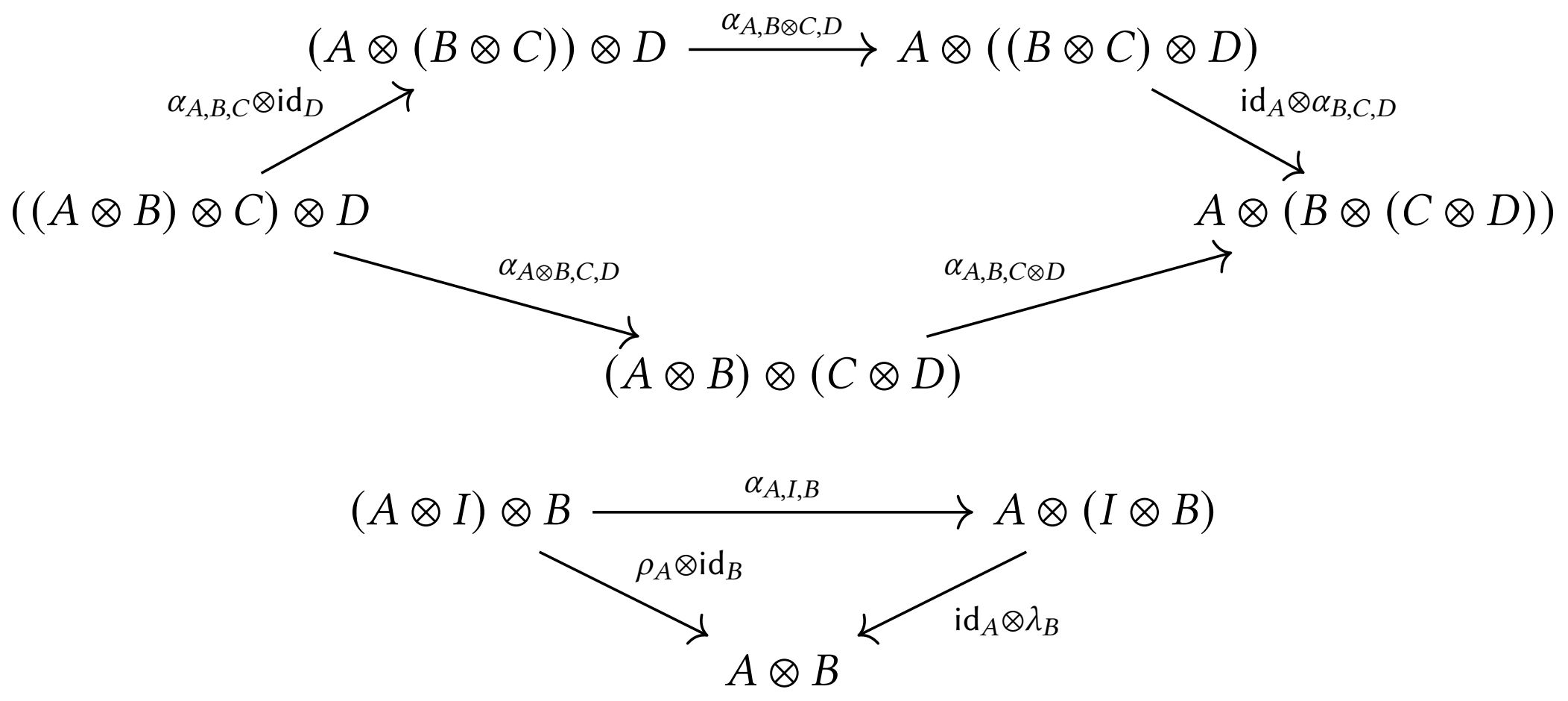

Figure 2.5: Commutative diagrams for monoidal categories

such that the *pentagon* and the *triangle* diagrams in Figure 2.5 commute.

**Example 2.81.** Set is a monoidal category, with the tensor product defined as the Cartesian product ($A \otimes B \coloneqq A \times B$) and the unit as the singleton set ($I \coloneqq \{\bullet\}$).

**Notation 2.82.** We adopt the convention that $\otimes$ takes precedence over $\fatsemi$, i.e. $f \otimes g \fatsemi h \otimes k$ should be bracketed as $(f \otimes g) \fatsemi (h \otimes k)$.

We will use the $\otimes$ bifunctor extensively through this thesis. For this reason, when drawing string diagrams for monoidal categories we will forego drawing the (bi)functor boxes and usually draw wires exclusively in their 'deconstructed' state: instead of a single wire $A \otimes B$ we will draw two wires $A$ and $B$.

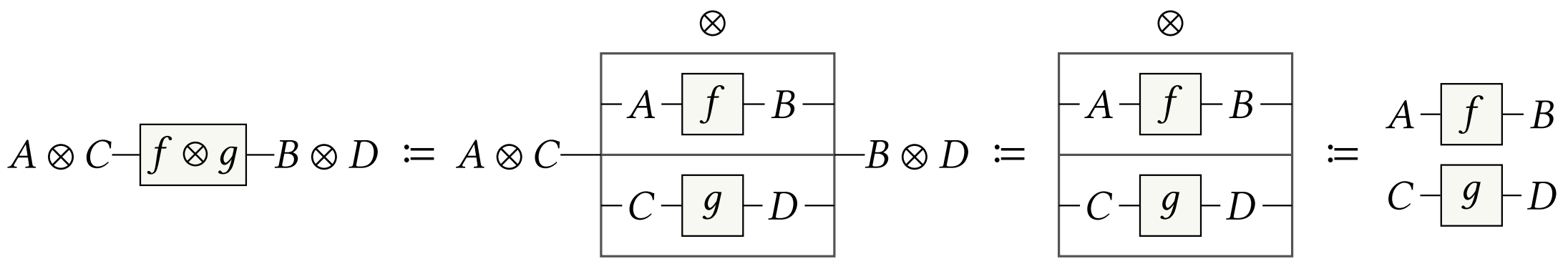

The definition of monoidal category we have presented is quite general, particularly with regards to the natural isomorphisms for unitors and associators. In our setting, it is normally sufficient for these isomorphisms to hold 'on the nose'.

**Definition 2.83** (Strict monoidal category)**.** A monoidal category is *strict* if $\lambda$, $\rho$ and $\alpha$ are identities.

In a strict monoidal category, the unitality and associativity of the tensor hold as equations, as they do for regular composition in a category. With this in mind, it can be instructive to view a strict monoidal category in terms of equations: these are illustrated in Figure 2.6.

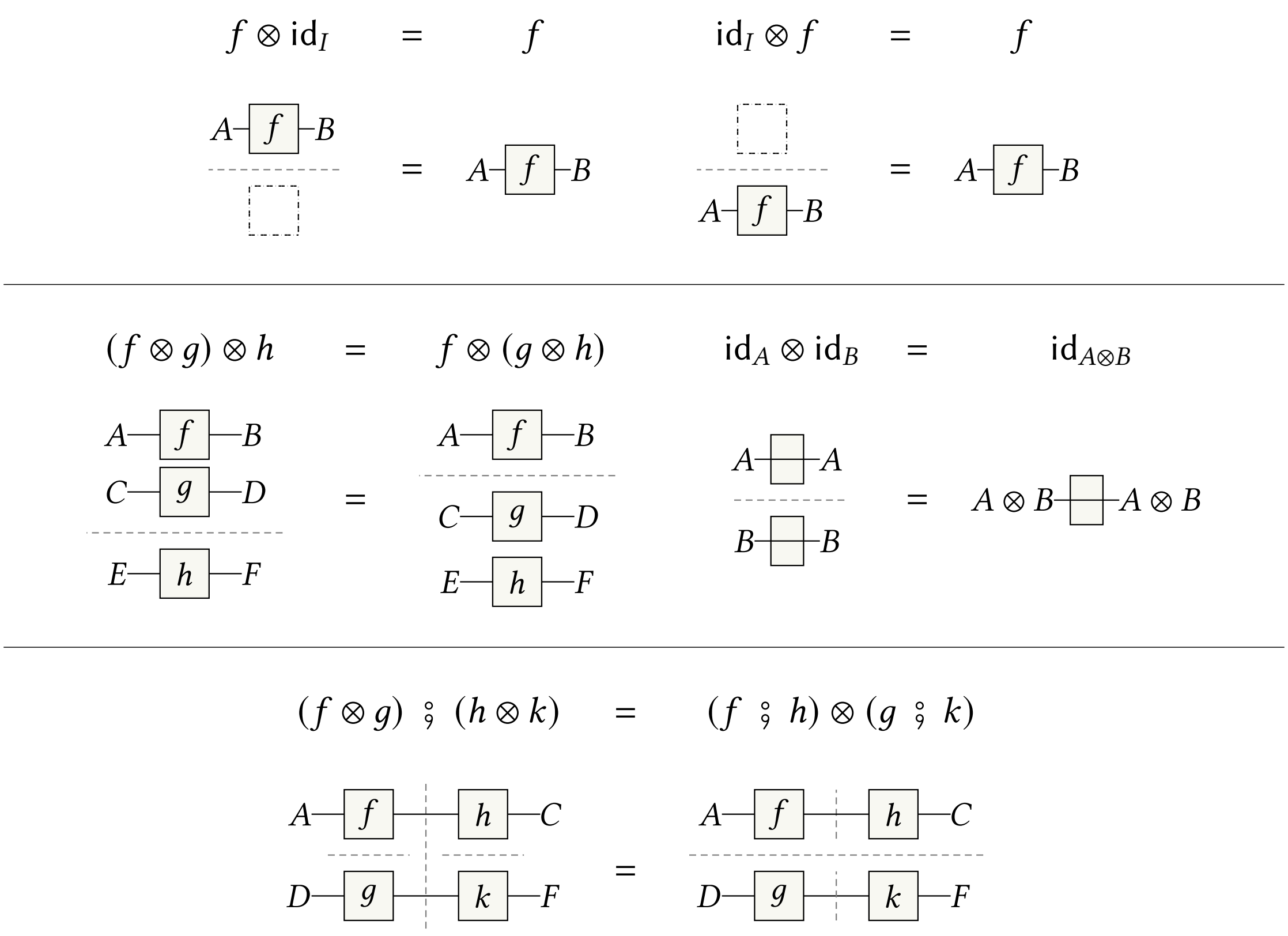

Figure 2.6: Equations of a strict monoidal category

## 2.7.1 Symmetric monoidal categories

We can now construct morphisms by composing them in sequence and in parallel, but there is no way to cross over the wires that connect boxes together. This can be achieved by equipping the categorical setting with another natural isormorphism.

**Definition 2.84** (Symmetric monoidal category)**.** A *symmetric monoidal category* (SMC) is a monoidal category $\mathcal{C}$ equipped with a natural isomorphism $\sigma_{A,B} \colon A \otimes B \cong B \otimes A$ such that the diagrams in Figure 2.7 commute.

As $\sigma$ is a natural isomorphism, it induces a *family* of morphisms $\sigma_{A,B} \colon A \otimes B \to B \otimes A$ for every pair of objects $A$ and $B$. In string diagrams, each such morphism $\sigma_{A,B}$ is drawn as 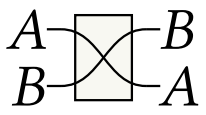: the swapping of wires $A$ and $B$. As with identities, the use of this morphism is so common that we usually drop the box around it.

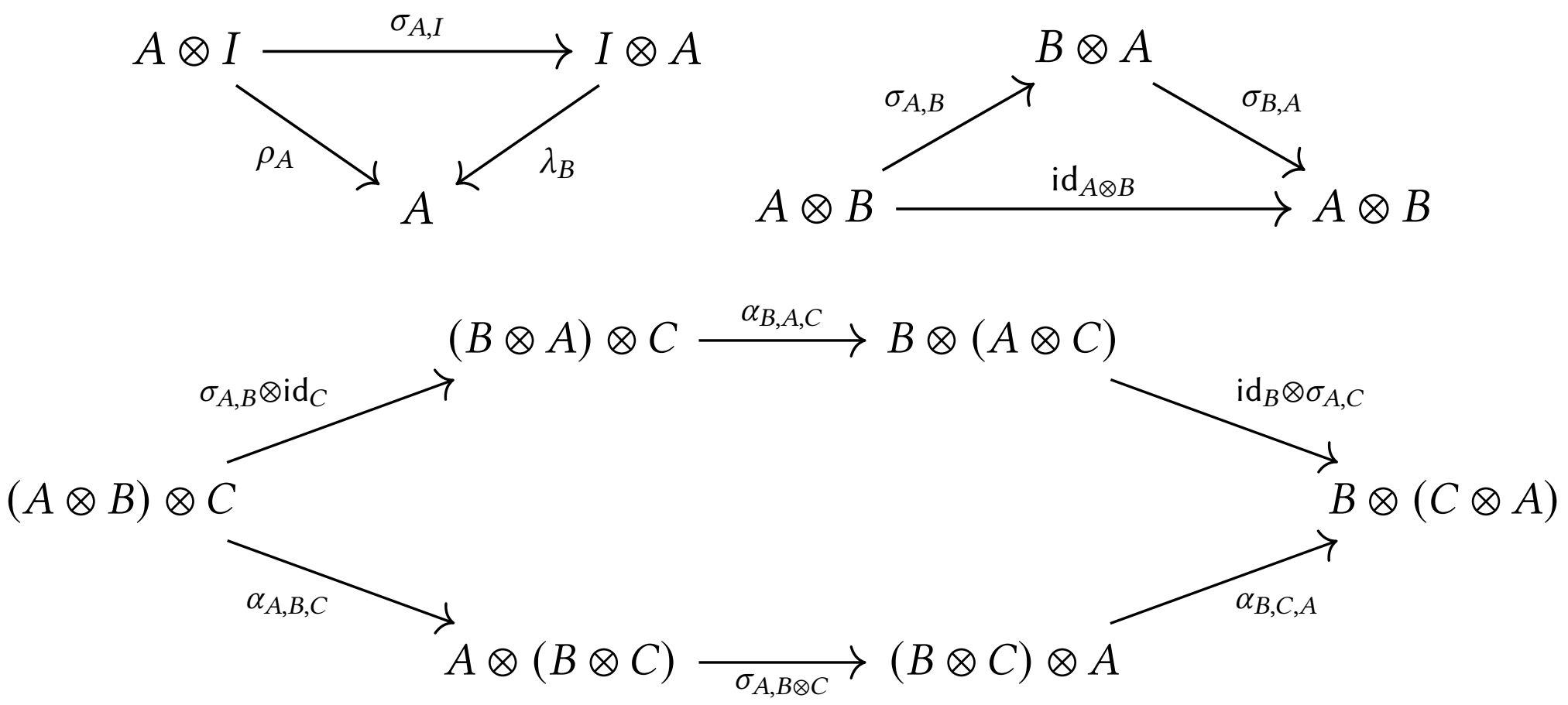

Figure 2.7: Commutative diagrams for symmetric monoidal categories

Once again, we are primarily concerned with *strict* symmetric monoidal categories. The equations of strict symmetric monoidal categories are listed in Figure 2.8.

**Example 2.85.** Set is a symmetric monoidal category, with $\sigma_{A,B} \colon A \times B \to B \times A$ defined as the function $\sigma_{A,B,(}a,b) = (b,a)$.

### 2.7.2 PROPs

Symmetric monoidal categories are an excellent setting for reasoning modulo 'structural equations'. We are especially interested in a subclass of SMCs called *PROP*s: categories of *PRO*ducts and *P*ermutations.

**Definition 2.86** (PROP [Mac65])**.** A *PROP* is a strict symmetric monoidal category with the natural numbers as objects and addition as tensor product on objects.

PROPs are a good fit for reasoning with string diagrams; as any object $n$ is equal to $\bigotimes_{i<n} 1$, a morphism $m \to n$ is a box with $m$ incoming wires and $n$ outgoing wires.

**Definition 2.87.** Given a set of generators $\Sigma$, let $\mathrm{S}_\Sigma$ be the PROP where $\mathrm{S}_\Sigma(m,n)$ is the set of $\Sigma$-terms of type $m \to n$ quotiented by the equations of SMCs.

This is known as a category *freely generated over* $\Sigma$, in that all of the morphisms in $\mathrm{S}_\Sigma$ have been 'generated' by combining elements of $\Sigma$ in various ways using composition and tensor. Crucially, many $\Sigma$-terms correspond to the *same* morphism in $\mathrm{S}_\Sigma$, as the latter are subject to the equations of SMCs.

In this thesis we will use multiple PROPs to represent different processes; often we will need to map between them. As PROPs are just special categories the most natural

$$f \otimes g \fatsemi \sigma_{B,D} = \sigma_{A,C} \fatsemi g \otimes f \qquad\qquad \sigma_{I,A} = \mathrm{id}_A$$

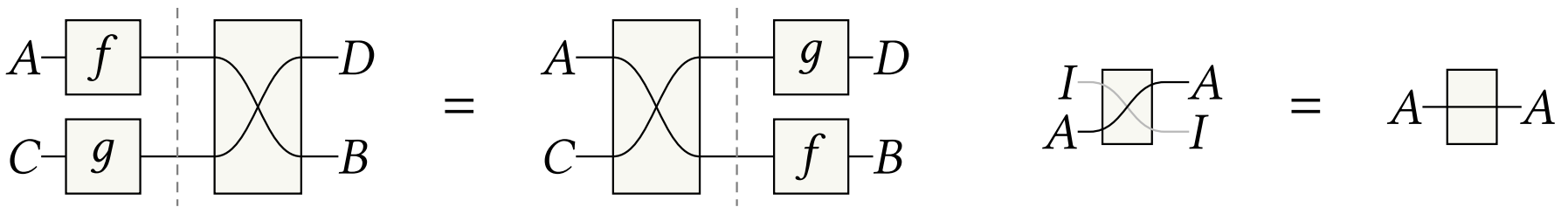

$$\sigma_{A,B} \otimes \mathrm{id}_C \fatsemi \mathrm{id}_B \otimes \sigma_{A,C} = \sigma_{A,B\otimes C} \qquad\qquad \sigma_{A,I} = \mathrm{id}_A$$

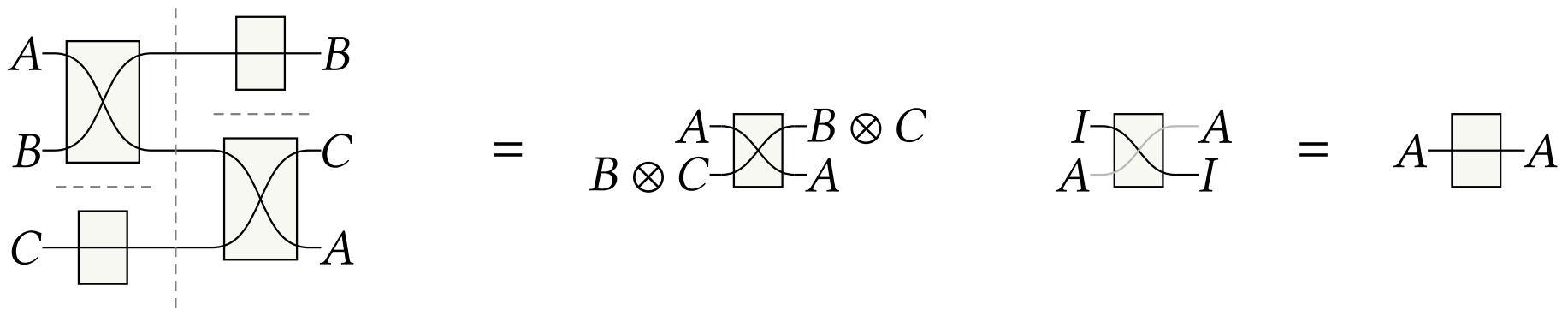

$$\sigma_{A,B} \fatsemi \sigma_{B,A} = \mathrm{id}_{A\otimes B}$$

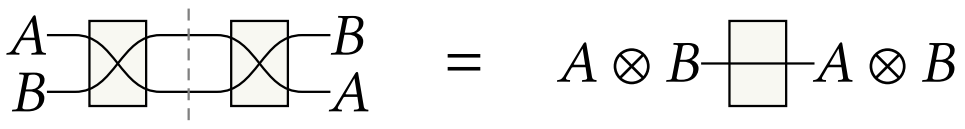

Figure 2.8: Equations of a strict symmetric monoidal category

way to do this is by using functors.

**Definition 2.88** (PROP morphism)**.** A *PROP morphism* is a strict symmetric monoidal functor between two PROPs i.e. a functor that preserves the strict symmetric monoidal structure.

As we saw earlier, functors can be composed and there is an identity functor, so PROPs themselves form a category.

**Definition 2.89.** Let **PROP** be the category with PROPs as objects and PROP morphisms as morphisms.

When we discussed terms we also mentioned the notion of *coloured* terms where the wires can be of different colours; appropriately, there are also coloured PROPs.

**Definition 2.90** (Coloured PROP)**.** Given a set of *colours* $C$, a $C$*-coloured PROP* is a strict symmetric monoidal category with the objects as lists in $C^\star$ and tensor product as word concatenation.

**Remark 2.91.** A 'regular' PROP as defined in Definition 2.86 is isomorphic to a coloured PROP with only one colour.

Note that this means the empty list $\varepsilon$ is the unit object in any $\mathcal{C}$-coloured PROP.

**Definition 2.92.** Given a countable set of colours $C$ and $C$-coloured generators $\Sigma$, let $\mathsf{S}_{C,\Sigma}$ be the $C$-coloured PROP where $\mathsf{S}_{C,\Sigma}(\overline{m}, \overline{n})$ is the set of $(C, \Sigma)$-terms of type $\overline{m} \to \overline{n}$ quotiented by the equations of SMCs.

Just like regular PROPs, there are morphisms of coloured PROPs and these form a category.

**Definition 2.93.** Let **CPROP** be the category with coloured PROPs as objects and coloured PROP morphisms as morphisms.

It can also be useful to consider the category of coloured PROPs over a fixed set of colours $C$.

**Definition 2.94.** For a countable set of colours $C$, let $\mathbf{CPROP}_C$ be the category with $C$-coloured PROPs as objects and $C$-coloured PROP morphisms as morphisms.

## 2.8 Reversing the wires

When considering string diagrams for symmetric monoidal categories, there is a strict notion of causality: it is not possible to create a cycle from the output of a box to its input, and outputs may only be joined to inputs. This enforces an implicit *left-to-right* directionality across the page.

This may not always be desirable: one may wish to model a feedback loop, or perhaps enforce some condition by unifying two outputs. To do this sort of thing, we must examine symmetric monoidal categories with some extra structure.

### 2.8.1 Symmetric traced monoidal categories

First we consider removing the acyclicity condition.

**Definition 2.95** (Symmetric traced monoidal category [JSV96])**.** A *symmetric traced monoidal category* (STMC) is a symmetric monoidal category $\mathcal{C}$ equipped with a family of functions $\mathrm{Tr}^X_{A,B}(-) \colon \mathcal{C}(X \otimes A, X \otimes B) \to \mathcal{C}(A, B)$ for any three objects $A$, $B$ and $X$ subject to the equations in Figure 2.9.

**Tightening**

$$\mathrm{Tr}^{X}_{A,D}\left(\mathrm{id}_X \otimes f \fatsemi g \fatsemi \mathrm{id}_X \otimes h\right) \quad = \quad f \fatsemi \mathrm{Tr}^{X}_{B,C}\left(g\right) \fatsemi h$$

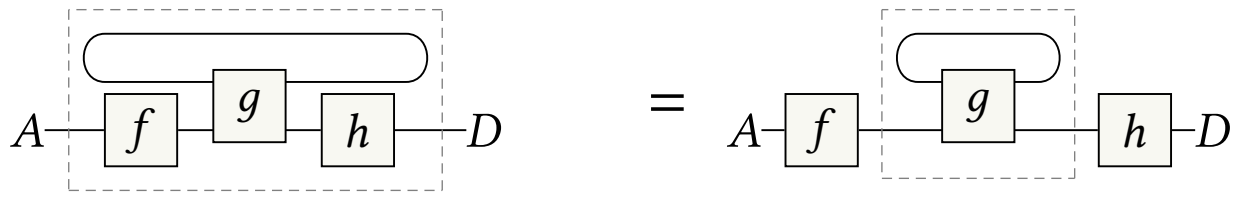

**Sliding**

$$\mathrm{Tr}^{X}_{A,B}\left(f \fatsemi g \otimes \mathrm{id}_B\right) = \mathrm{Tr}^{Y}_{A,B}\left(g \otimes \mathrm{id}_A \fatsemi f\right)$$

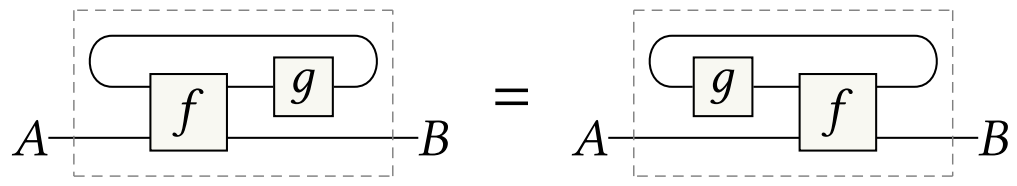

**Yanking**

$$\mathrm{Tr}^{X}_{X,X}\left(\sigma_{X,X}\right) \quad = \quad \mathrm{id}_X$$

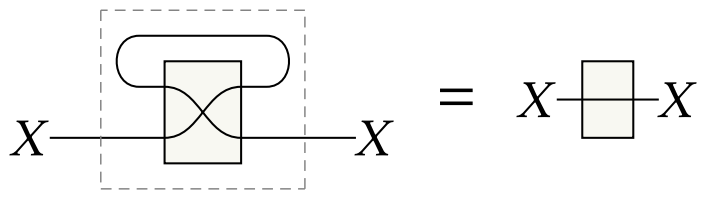

**Vanishing**

$$\mathrm{Tr}^{X}_{A,B}\left(\mathrm{Tr}^{Y}_{X\otimes A, X\otimes B}\left(f\right)\right) \quad = \quad \mathrm{Tr}^{X\otimes Y}_{A,B}\left(f\right)$$

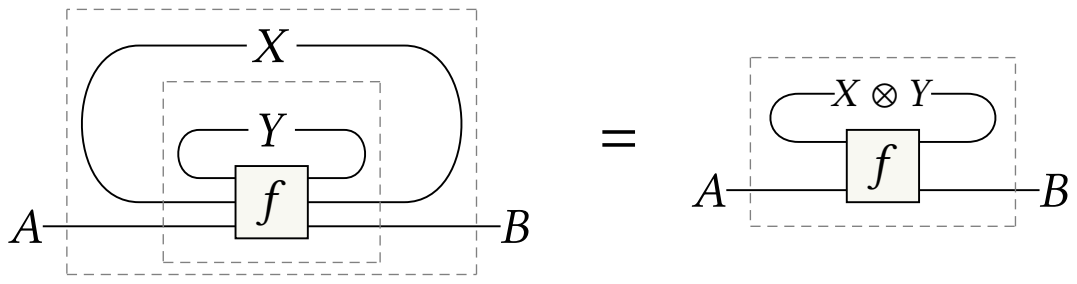

**Superposing**

$$\mathrm{Tr}^{X}_{A,B}\left(f\right) \otimes g \quad = \mathrm{Tr}^{X}_{A\otimes C, B\otimes D}\left(f \otimes g\right)$$

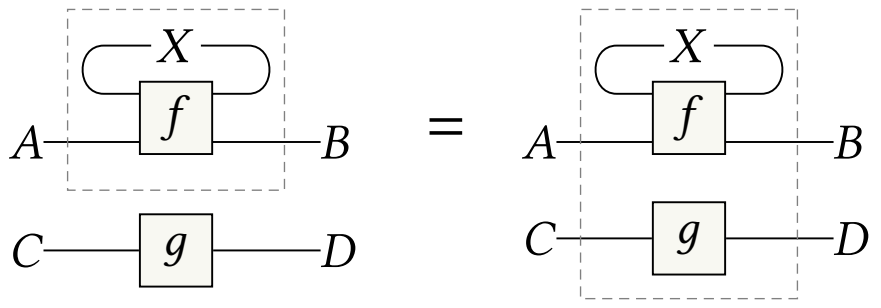

Figure 2.9: Equations of STMCs in string diagram notation

This means that if we have a morphism $f : X \otimes A \to X \otimes B$ in a STMC, we also have the morphism $\mathrm{Tr}^{X}_{A,B}\left(f\right) : A \to B$. Traced categories were given the string diagrammatic treatment in [JSV96], in which the trace is depicted as a loop.

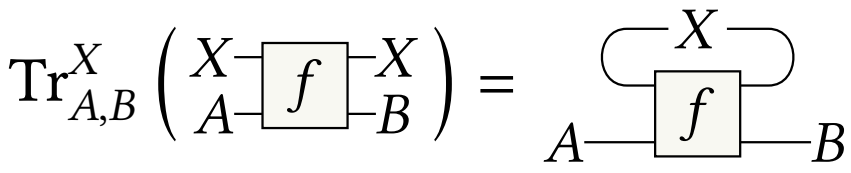

The equations of STMCs then have pleasant graphical interpretations, shown in Figure 2.9. Note that as with the equations of SMCs, these equations amount to deforming the diagram without altering connections between boxes, so do not need to be applied explicitly when performing equational reasoning.

Usually we will omit the subscripts from $\mathrm{Tr}^{X}_{A,B}\left(f\right)$ and write it simply as $\mathrm{Tr}^{X}\left(f\right)$.

**Example 2.96.** The classic example of a symmetric traced monoidal category is the category $\mathbf{FinVect}_k$, with finite dimensional vector spaces over a field $k$ as objects, and linear maps as morphisms. The monoidal product is the tensor product of vector spaces and the trace is an operation known as the 'partial trace'.

In the context of computer science, traces are often used to model *fixpoints* [Has97]; we will examine this in further detail in Section 10.1 as a possible application of our work on traced string diagram rewriting.

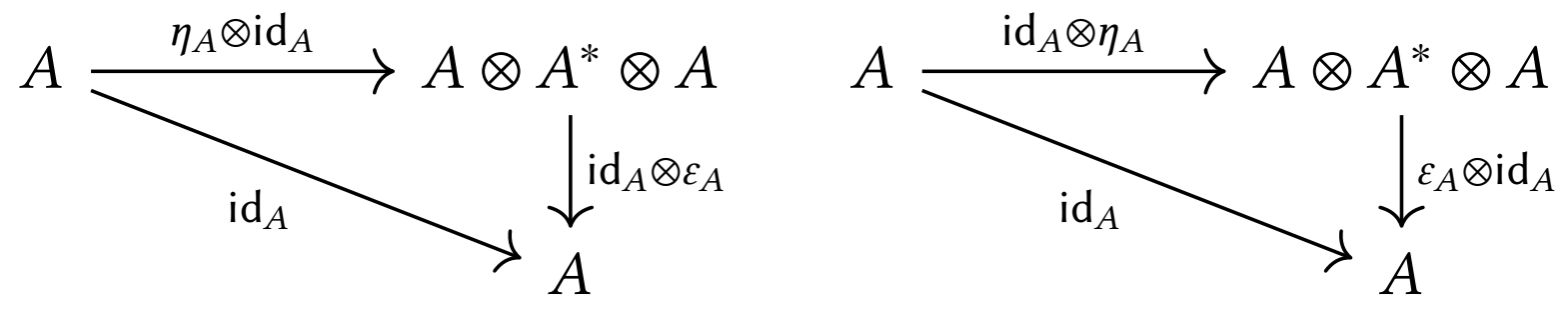

Figure 2.10: Commutative diagrams of a compact closed category

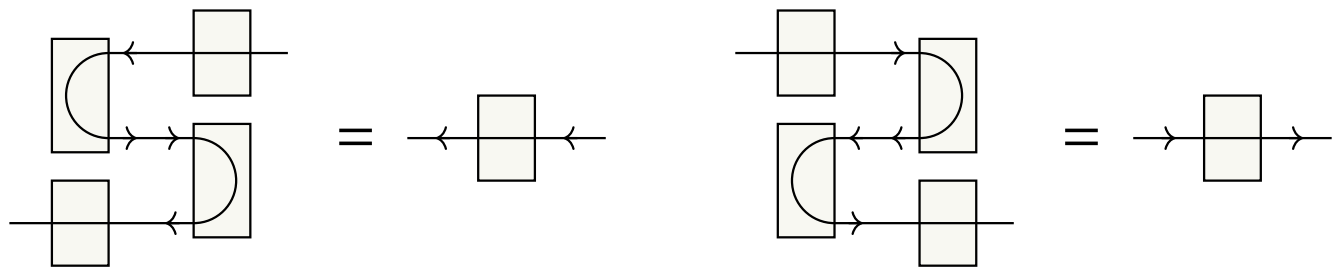

Figure 2.11: Equations of CCCs string diagram notation

### 2.8.2 Compact closed categories

While wires can flow backwards across the page in a traced category, regular left-to-right flow must still be in effect at a wire's *endpoints*, so outputs must be connected to inputs. We will now consider another setting in which this is not the case.

**Definition 2.97** (Compact closed category)**.** A *compact closed category* (CCC) is a symmetric monoidal category in which every object $X$ has a *dual* $X^*$ equipped with morphisms called the *unit* $\eta_A \colon I \to A^* \otimes A$ and the *counit* $\varepsilon_A \colon A \otimes A^* \to I$ such that the diagrams in Figure 2.10 commute.

In string diagrams, the dual is drawn as a wire flowing from right-to-left instead of left-to-right; when labelling wires with objects we will drop the notation for duals and recover the information solely from directionality of the wires. The unit and counit 'bend' wires: the unit is drawn as

and the counit as

. As a result of this units and counits are often referred to as *cups* and *caps* respectively. The two equations of compact closed categories are depicted as in Figure 2.11; this should provide some insight as to why they are often referred to as *snake equations*.

There are some cases where the actual directionality of wires is irrelevant; we only care about the ability to bend wires.

**Definition 2.98** (Self-dual compact closed category)**.** A compact closed category is *self-dual* if for every object $A$, $A^* \coloneqq A$.

In a self-dual compact closed category, we do not need to label wires with arrows.

### 2.8.3 Traced vs compact closed

The graphical notation is particularly suggestive of links between the trace, the cup and the cap. This is no coincidence, as there is a well-known result that allows one to construct a trace in a compact closed setting.

**Proposition 2.99** (Canonical trace [JSV96, Prop. 3.1])**.** Given a compact closed category $\mathcal{C}$, a trace on $\mathcal{C}$ can be constructed as follows:

$$\mathrm{Tr}^X(f) \coloneqq \eta_X \otimes \mathrm{id}_A \fatsemi \mathrm{id}_{X^*} \otimes f \fatsemi (\sigma_{X^*,X} \fatsemi \varepsilon_X) \otimes \mathrm{id}_B$$

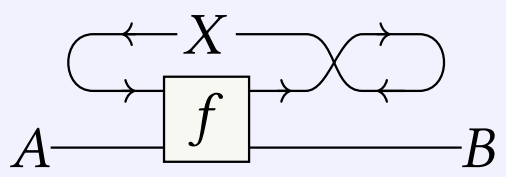

This is called the *canonical trace*.

In this thesis we are primarily concerned with traced categories, but a plethora of related work is based in the compact closed realm. The canonical trace allows us to adapt existing results for our setting as well.

**Remark 2.100.** It is also possible to consider the other direction: using the *Int-construction* [JSV96], given any STMC $\mathcal{C}$ one can construct a compact closed category $\mathrm{Int}(\mathcal{C})$. However, this will not be of relevance to us.

### 2.8.4 Adding more structure

While working in a traced or compact closed category allows one to model more processes than working in a vanilla symmetric monoidal category, it is by no means the only way in which categories can have structure. However, when stacking multiple structures on top of each other, we must be careful that we do not accidentally create something unwanted.

One common structure in a category is *finite products*, which were discussed in Definition 2.33.

**Theorem 2.101** ([Hou08, Thm. 1])**.** If a compact closed category has finite products then it also has finite coproducts.

It may not be the case that we want coproducts, so in this case a compact closed category is not suitable. On the other hand, adding structure to a category can cause it to become *trivial*: it can only have one object and one morphism. This can arise when enforcing more structure on the tensor product.

**Definition 2.102** (Cartesian category)**.** A monoidal category is *Cartesian* if its tensor product is given by the category-theoretic product.

**Lemma 2.103.** In a Cartesian category, the monoidal unit is the terminal object.

*Proof.* We must show that any arrow $f\colon A \to I$ is unique. Using the left unitor $\lambda_A\colon A \to I \times A$ and the projection maps $\pi_0\colon I \times A \to I$ and $\pi_1\colon I \times A \to A$, we can construct the following diagram:

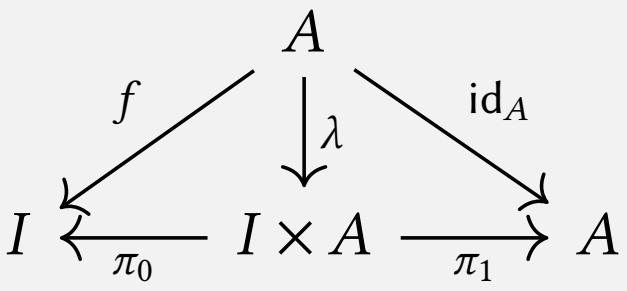

By the universal property of the product, any morphism $f\colon A \to I$ can be uniquely factored as $\lambda \fatsemi \pi_0$. □

Accordingly, the unique maps $!_A\colon A \to I$ are drawn string diagrammatically as 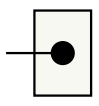 , which will help with reasoning.

**Proposition 2.104.** If a compact closed category is Cartesian, then it must be trivial i.e. it must have only one object and morphism.

*Proof.* We show that any object $A$ is isomorphic to the monoidal unit $I$ by proving that the composites $A \to I \to A$ and $I \to A \to I$ are both the identity. For the former, $!_A\colon A \to I$ is unique because $I$ is the terminal object, and $I \to A$ is uniquely determined to be $\eta_A \fatsemi \pi_1$ by the universal property of the product.

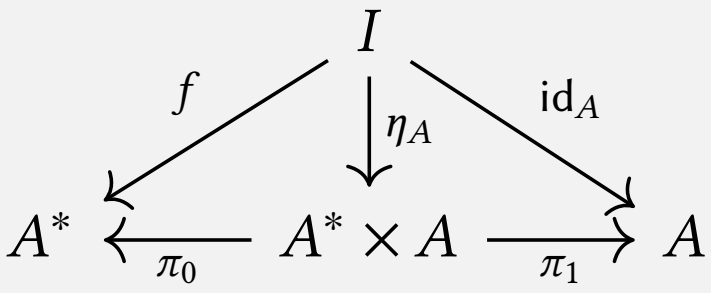

So we must show that $!_A \fatsemi \eta_A \fatsemi \pi_0 = \mathrm{id}_A$. This is clear when using string diagrams, noting that the cap is equal to the unique morphism $!_{A\times A^*}\colon A \times A^* \to I$.

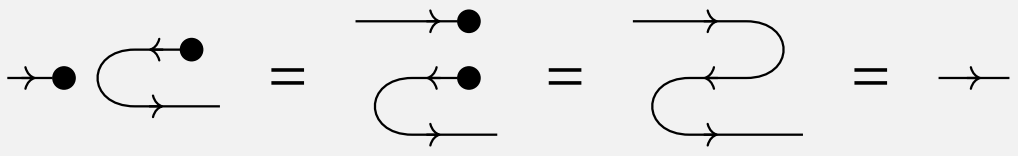

The latter equality holds because there is exactly one morphism $I \to I$. So any object $A$ is isomorphic to $I$. □

This means that if we want to reason about wires going backwards in Cartesian categories, we must do so in a traced setting.

## 2.9 Monoidal theories

So far we have only concerned ourselves with *structural* equations: equations that show how the same term can be constructed using different combinations of composition, tensor, the identity and symmetry. However, these only serve as a form of housekeeping: the true 'computational content' of processes comes from equations that show how the generators interact with *each other*. These equations are provided by a *monoidal theory*.

**Definition 2.105.** An *equation* is a pair of terms with the same domain and codomain.

**Definition 2.106** (Monoidal theory)**.** A *monoidal theory* is a tuple $(\Sigma, \mathcal{E})$ where $\Sigma$ is a set of generators and $\mathcal{E}$ is a set of equations.

An equation $f = g$ in a monoidal theory *identifies* the two morphisms $f$ and $g$, so that they are actually equal. When reasoning with a monoidal theory, we therefore need to work in a category subject to this identification of morphisms.

**Definition 2.107** (Quotient category)**.** Given a category $\mathcal{C}$ and a set of equations $\mathcal{E}$ between morphisms in $\mathcal{C}$ with the same source and target, the *quotient category* $\mathcal{C}/\mathcal{E}$ is the category in which $\mathrm{ob}(\mathcal{C}/\mathcal{E}) \coloneqq \mathrm{ob}(\mathcal{C})$ and $(\mathcal{C}/\mathcal{E})(X, Y) \coloneqq \mathcal{C}(X, Y)/\mathcal{E}$.

In a quotient category $\mathcal{C}/\mathcal{E}$ the morphisms are the *equivalence classes* of morphisms in $\mathcal{C}$ modulo the equations in $\mathcal{E}$.

**Definition 2.108.** Given a monoidal theory $(\Sigma, \mathcal{E})$, let $\mathrm{S}_{\Sigma,\mathcal{E}} \coloneqq \mathrm{S}_{\Sigma}/\mathcal{E}$.

When the set of equations $\mathcal{E}$ is empty, $\mathrm{S}_{\Sigma,\emptyset} = \mathrm{S}_{\Sigma}$, and we recover an ordinary PROP. We can also use the same procedure to define *coloured* monoidal theories.

**Definition 2.109** (Coloured monoidal theory)**.** A *coloured monoidal theory* is a tuple $(C, \Sigma, \mathcal{E})$ where $C$ is a countable set of colours, $\Sigma$ is a set of $C$-coloured generators, and $\mathcal{E}$ is a set of equations.

**Definition 2.110.** Given a coloured monoidal theory $(C, \Sigma, \mathcal{E})$, let $\mathrm{S}_{C,\Sigma,\mathcal{E}} \coloneqq \mathrm{S}_{C,\Sigma}/\mathcal{E}$.

### 2.9.1 Case study: commutative monoids

Monoidal theories for PROPs can be used to reason with many structures in mathematics. As we have seen, viewing terms in terms of diagrams rather than text strings is far more intuitive, so we will often forgo writing the terms at all and reason exclusively using diagrams. This extends to defining generators, which allows us to give them suggestive graphical representations rather than having to stick to a symbol with some domain and codomain.

As an example, we will explore the monoidal theory of *commutative monoids*; using the graphical notation the intended behaviour of the two generators is much clearer.

**Definition 2.111** (Commutative monoids). The monoidal theory of *commutative monoids* is $(\Sigma_{\mathbf{CMon}}, \mathcal{E}_{\mathbf{CMon}})$, where $\Sigma_{\mathbf{CMon}} \coloneqq \{$ 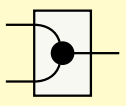 , 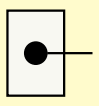 $\}$ called the *multiplication* and the *unit* respectively, and $\mathcal{E}_{\mathbf{CMon}}$ comprises the equations

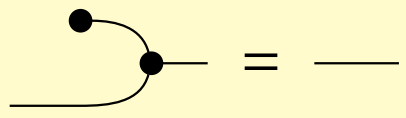 (left unitality) 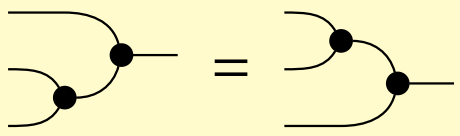 (associativity)

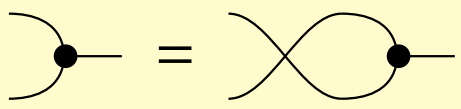 (commutativity)

We write $\mathbf{CMon} \coloneqq \mathsf{S}_{\Sigma_{\mathbf{CMon}}, \mathcal{E}_{\mathbf{CMon}}}$.

The equations describe the properties of the multiplication: it is unital with respect to the unit; it is associative; and it is commutative. These equations could be described textually, but the string diagrams provide intuitive visual interpretations; often it is insightful to reason *diagrammatically*.

**Example 2.112** (Right unitality). 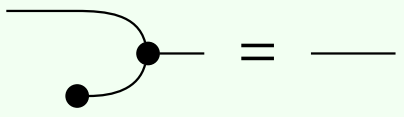 is a valid equation in **CMon**.

*Proof.* 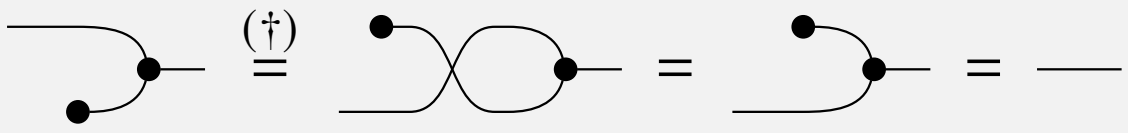
 □

Note that the first step (†) of the proof is performed solely by deforming the string diagram; this is permitted so long as connectivity is preserved. Deforming the string diagram corresponds to implicitly applying equations of symmetric monoidal categories. These explicit steps are shown below, with the unit wire in grey:

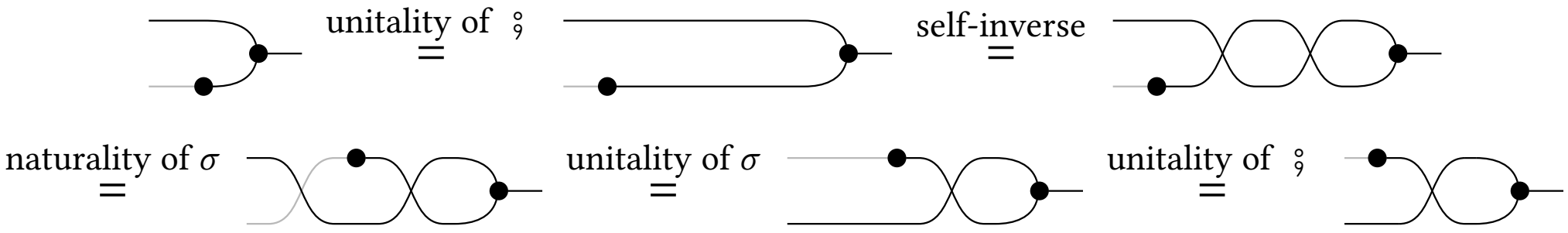

Already much more verbose than the simple deformation, this does not even take into account the repeated applications of associativity of both composition and tensor required if reasoning in the term language. If we write 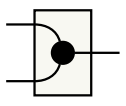 as $\mu$ and 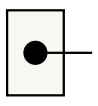 as $\eta$, then

the proof on terms becomes:

$$
\begin{aligned}
\mathsf{id}_1 \otimes \eta \fatsemi \mu &= \mathsf{id}_1 \otimes (\eta \fatsemi \mathsf{id}_1) \fatsemi \mu && \text{unitality of } \fatsemi \\
&= (\mathsf{id}_1 \fatsemi \mathsf{id}_1) \otimes (\eta \fatsemi \mathsf{id}_1) \fatsemi \mu && \text{unitality of } \fatsemi \\
&= ((\mathsf{id}_1 \otimes \eta) \fatsemi (\mathsf{id}_1 \otimes \mathsf{id}_1)) \fatsemi \mu && \text{functoriality of } \otimes \\
&= ((\mathsf{id}_1 \otimes \eta) \fatsemi (\sigma_{n,n} \fatsemi \sigma_{n,n})) \fatsemi \mu && \sigma \text{ is self inverse} \\
&= (((\mathsf{id}_1 \otimes \eta) \fatsemi \sigma_{n,n}) \fatsemi \sigma_{n,n}) \fatsemi \mu && \text{associativity of } \fatsemi \\
&= ((\sigma_{\varepsilon,n} \fatsemi (\eta \otimes \mathsf{id}_1)) \fatsemi \sigma_{n,n}) \fatsemi \mu && \text{naturality of } \sigma \\
&= ((\mathsf{id}_1 \fatsemi (\eta \otimes \mathsf{id}_1)) \fatsemi \sigma_{n,n}) \fatsemi \mu && \text{unitality of } \sigma \\
&= ((\eta \otimes \mathsf{id}_1) \fatsemi \sigma_{n,n}) \fatsemi \mu && \text{unitality of } \fatsemi \\
&= (\eta \otimes \mathsf{id}_1) \fatsemi (\sigma_{n,n} \fatsemi \mu) && \text{associativity of } \fatsemi \\
&= (\eta \otimes \mathsf{id}_1) \fatsemi \mu && \text{commutativity of } \mu \\
&= \mathsf{id}_1 && \text{left unitality of } \mu
\end{aligned}
$$

This is already far more verbose than the string diagram proof, but the term notation also blocks the insight required to make a proof step, which is often much easier to see in the string diagrammatic representation.

# Part I

# Semantics of Digital Sequential Circuits

# Chapter 3

# Syntax

Our *soup du jour* is that of *sequential synchronous digital circuits* constructed from primitive components such as logic gates or transistors. These circuits are *sequential* as they have a notion of *state*: outputs can be impacted by inputs in previous cycles rather than solely those in the current cycle, and *synchronous* because their state changes in time with some global clock.

**Remark 3.1.** Digital (electric) circuits are not to be confused with *electronic* circuits of switches and resistors. Essentially, the difference boils down to the difference between traced categories and compact closed categories: digital circuits have a clear notion of *causality* whereas electronic circuits are *relational* in nature. For a study of the latter, see [BS22].

**Remark 3.2.** The content of this section is a refinement of [GKS24, Section 2].

## 3.1 Circuit signatures

To construct circuits, we define a category in which a morphism $m \to n$ is a circuit with $m$ inputs and $n$ outputs. Rather than restricting to any particular gate set, we parameterise a given category of circuits over a *circuit signature* containing details about the available components.

**Definition 3.3** (Circuit signature, value, primitive symbol)**.** A *circuit signature* $\Sigma$ is a tuple $(\mathbf{V}, \bullet, \mathcal{P}, \mathrm{dom}, \mathrm{cod})$ where $\mathbf{V}$ is a finite set of *values*, $\bullet \in \mathbf{V}$ is a *disconnected* value, $\mathcal{P}$ is a (usually finite) set of *primitive symbols*, $\mathrm{dom}\colon \mathcal{P} \to \mathbb{N}$ is an *arity* function and $\mathrm{cod}\colon \mathcal{P} \to \mathbb{N}$ is a *coarity* function.

A particularly important signature, and one which we will turn to for the majority of examples in this thesis, is that of gate-level circuits.

**Example 3.4** (Gate-level circuits)**.** The circuit signature for *gate-level circuits* is $\Sigma_{\mathrm{B}} \coloneqq (\mathbf{V}_{\mathrm{B}}, \bot, \mathcal{P}_{\mathrm{B}}, \mathrm{dom}_{\mathrm{B}}, \mathrm{cod}_{\mathrm{B}})$, where $\mathbf{V}_{\mathrm{B}} \coloneqq \{\bot, \mathsf{f}, \mathsf{t}, \top\}$, respectively representing *no* signal, a *false* signal, a *true* signal and *both* signals at once, $\mathcal{P}_{\mathrm{B}} \coloneqq \{\mathrm{AND}, \mathrm{OR}, \mathrm{NOT}\}$, $\mathrm{dom}_{\mathrm{B}} \coloneqq \mathrm{AND} \mapsto 2, \mathrm{OR} \mapsto 2, \mathrm{NOT} \mapsto 1$ and $\mathrm{cod}_{\mathrm{B}} \coloneqq - \mapsto 1,$

**Remark 3.5.** Using four values may come as a surprise to those expecting the usual 'true' and 'false' logical values. This logical system is an old idea going back to Belnap [Bel77] who remarked that this is 'how a computer should think'. Rather than just thinking about how much *truth content* a value carries, the four value system adds a notion of *information content*: the $\bot$ value is no information at all (a disconnected wire), whereas the $\top$ value is both true and false information at once (a short circuit).

## 3.2 Combinational circuits

Before diving straight into sequential circuits, we will define a category of *combinational circuits*. These are circuits with no state; they compute *functions* of their inputs.

**Definition 3.6** (Combinational circuit generators)**.** Given a circuit signature $\Sigma = (\mathcal{V}, \bullet, \mathcal{P}, \mathrm{dom}, \mathrm{cod})$, let the set $\Sigma_{\mathrm{CCirc}}$ of *combinational circuit generators* be defined as the set containing $\mathrm{dom}(g)$–$\boxed{g}$–$\mathrm{cod}(g)$ for each $g \in \mathcal{P}$, 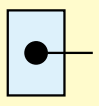 , 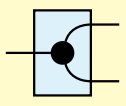 , 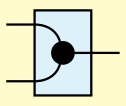 , and 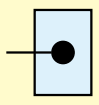 . We write $\mathbf{CCirc}_{\Sigma}$ for the freely generated PROP $\mathrm{S}_{\Sigma_{\mathrm{CCirc}}}$.

Each primitive symbol $p \in \mathcal{P}$ has a corresponding generator in $\mathbf{CCirc}_{\Sigma}$. The remaining generators are *structural* generators for manipulating wires: these are present regardless of the signature. In order, they are for *introducing* wires, *forking* wires, *joining* wires and *eliminating* wires.

**Example 3.7.** The gate generators of $\mathbf{CCirc}_{\Sigma_B}$ are 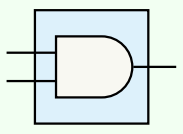, 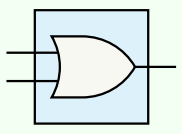, and 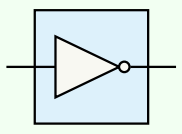.

When drawing circuits, the coloured backgrounds of generators will often be omitted in the interests of clarity. Since the category is freely generated, morphisms are defined by juxtaposing the generators in a given signature sequentially or in parallel with each other, the identity and the symmetry. Arbitrary combinational circuit morphisms defined in this way are drawn as light boxes 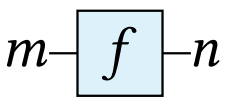.

**Notation 3.8.** The structural generators are only defined on single bits, but it is straightforward to define versions for arbitrary bit wires. Much like we often draw multiple wires as one (Notation 2.9), we can also draw these 'thicker' constructs in a similar way to the single-bit versions:

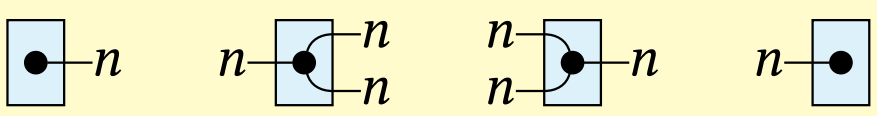

These composite constructs are defined inductively over the width of the wires.

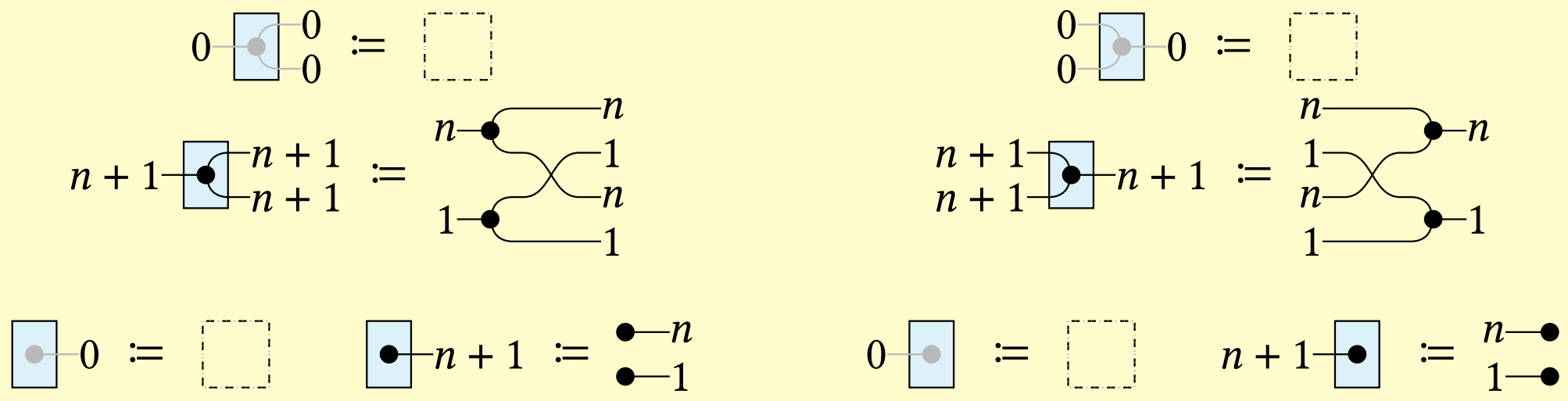

**Remark 3.9.** As mentioned during Notation 2.9, wires of zero width are usually drawn as empty space; in a similar fashion the zero width fork, join, and elimination constructs can be drawn as empty space or using 'faded' wires.

**Example 3.10** (More logic gates)**.** The AND, OR, and NOT gates are not the only logic gates used in circuit design. A NAND gate 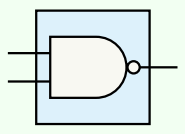 acts as the inverse of an AND gate: it outputs true if none of the inputs are true. Similarly, a NOR gate 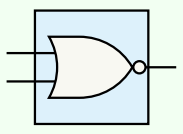 is the inverse of an OR gate. These two gates can be constructed in terms of the primitive gates in $\Sigma_B$:

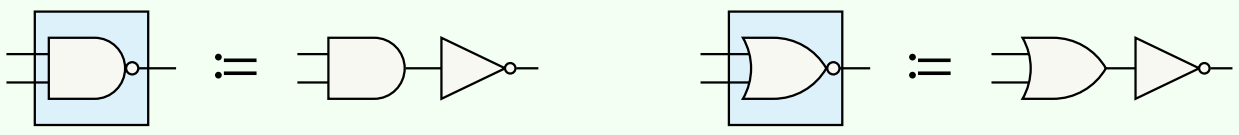

Another type of gate is the XOR gate 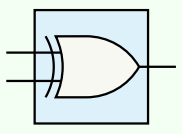, which outputs true if and only

if exactly one of the inputs is true. In $\mathbf{CCirc}_{\Sigma_B}$ this is constructed as

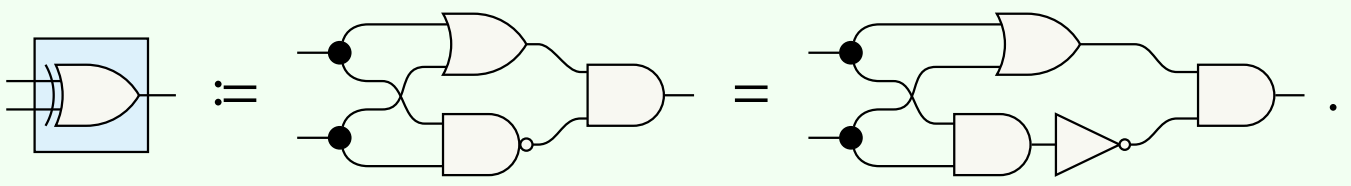

**Example 3.11** (Half adder)**.** The XOR gate is used in a classic combinational circuit known as a *half adder*, the basic building block of circuit arithmetic. A half adder takes two inputs and computes their *sum* modulo 2 and the resulting *carry*. That is to say, $0+0$ has sum $0$ and carry $0$, $1+0$ and $0+1$ have sum $1$ and carry $0$, and $1+1$ has sum $0$ and carry $1$.

The sum is computed using an XOR gate and the carry by an AND gate. The design of a half adder along with its construction in $\mathbf{CCirc}_{\Sigma_B}$ is shown below.

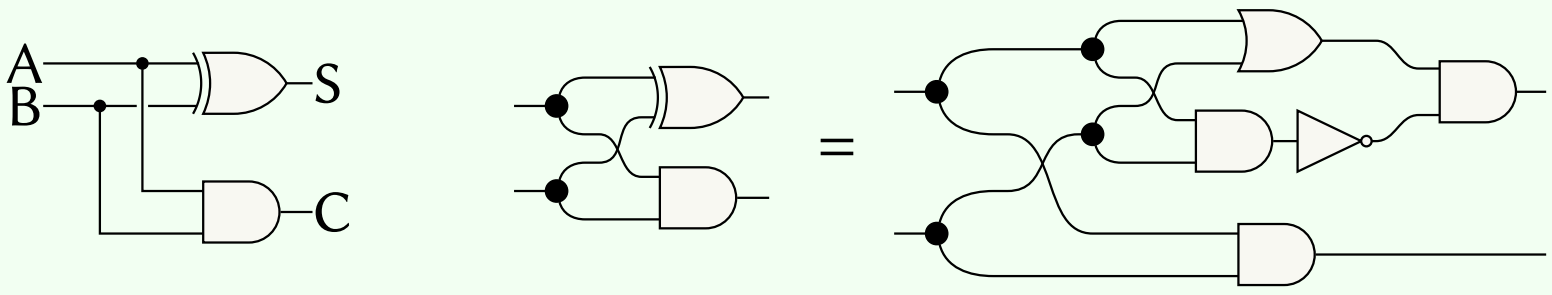

## 3.3 Sequential circuits

Combinational circuits compute functions of their inputs, but have no internal state. This is all very well for doing simple calculations, but for all but the most simple of circuits we need to be able to have *memory*. As we have mentioned earlier, such circuits are called *sequential circuits*.

Circuits gain state with *delay* and *feedback*. The latter means we need to move into a symmetric *traced* monoidal category.

**Definition 3.12** (Sequential circuits)**.** For a circuit signature $\Sigma$ with value set $\mathbf{V}$, let $\mathbf{SCirc}_\Sigma$ be the STMC freely generated over the generators of $\mathbf{CCirc}_\Sigma$ along with new generators 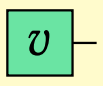 for each $v \in \mathbf{V} \setminus \bullet$, and a generator 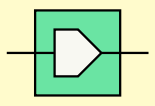 .

The first set of generators are *instantaneous values* for each value in $\mathbf{V} \setminus \bullet$. Value generators are intended to be interpreted as an *initial state*: in the first cycle of execution they will emit their value, and produce the disconnected $\bullet$ value after that. This is why there is no $\bullet$ value generator; instead it is a *combinational* generator 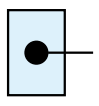 intended to always produce the $\bullet$ value.

**Notation 3.13.** Although $\bullet$— is itself not a sequential value, when we refer to an arbitrary value $v$— , $v$ can be any value in $\mathbf{V}$ including $\bullet$. For a word of values $\overline{v} \in \mathbf{V}^n$ (again possibly including $\bullet$), we may draw multiple value generators collapsed into one as $\overline{v}$—$n$ , defined inductively over $\overline{v}$ as

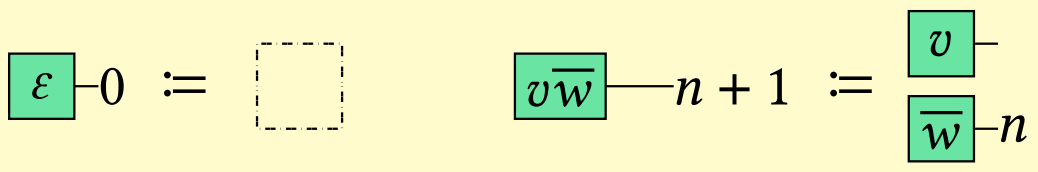

**Example 3.14.** The 'values' of $\mathbf{SCirc}_{\Sigma_B}$ are $\bullet$— , f— , t— , ⊤— ; the first is a combinational generator and the latter three are sequential.

The delay component is the opposite of a value; in the first cycle of execution it is intended to produce the $\bullet$ value, but in future cycles it outputs the signal it received in the previous cycle.

**Remark 3.15.** While the mathematical interpretation of a delay is straightforward, the physical aspect of a digital circuit it models is less clear. An obvious interpretation could be that the delay models a D flipflop in a clocked circuit, so the delay is one clock cycle. A more subtle interpretation is the 'minimum obervable duration'; in this case the delay models inertial delay on wires up to some fixed precision.

**Notation 3.16.** Like values, we can derive delay components for arbitrary-bit wires, drawn like $n$—▷—$n$ .

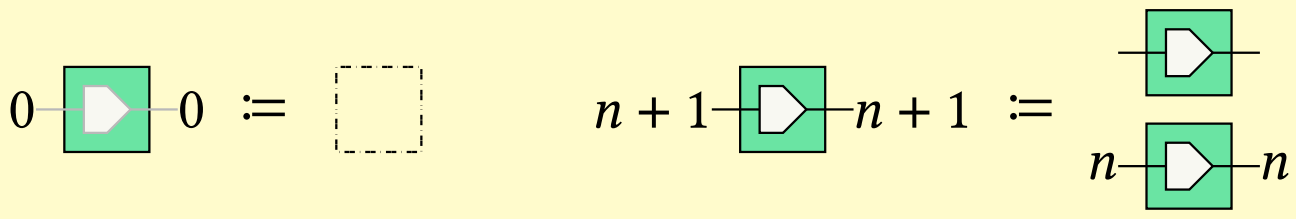

Often one may also want to think about delays with some explicit 'initial value', like a sort of register. This is so common that we introduce special notation for it.

**Notation 3.17** (Register)**.** For a word $\overline{v} \in \mathbf{V}^m$, let

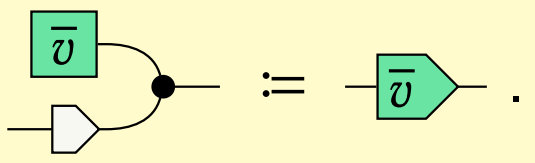

To distinguish them from combinational circuits, arbitrary sequential circuit morphisms are drawn as green boxes $m$—$f$—$n$ .

**Example 3.18** (SR latch)**.** A sequential circuit one might come across early on in an electronics textbook is the *SR NOR latch*, one of the simplest registers. A possible design and interpretation in $\mathbf{SCirc}_{\Sigma_B}$ are illustrated below.

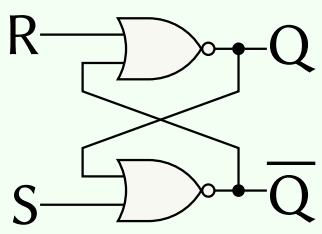

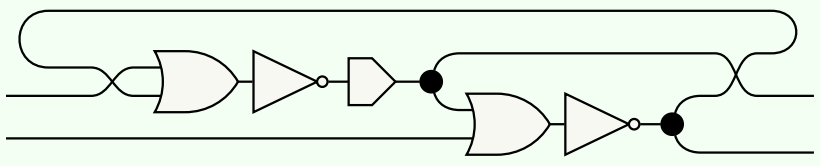

SR NOR latches are used to hold state. They have two inputs: Reset (R) and Set (S), and two outputs Q and $\overline{Q}$ which are always negations of each other. When Set receives a true signal, the Q output is forced true, and will remain as such even if the Set input stops being pulsed true. It is only when the Reset input is pulsed true that the Q output will return to false. (It is illegal for both Set and Reset to be pulsed high simultaneously; this issue is fixed in more complicated latches).

SR latches work because of delays in how gates and wires transmit signals; one of the feedback loops between the two NOR gates will 'win'. We can model this in $\mathbf{SCirc}_{\Sigma}$ by using a different number of delay generators between the top and the bottom of the latch. We have opted for just the one because otherwise the upcoming examples become excessively complicated, but any number would do, so long as the top and bottom differ.

## 3.4 Generalised circuit signatures

In a circuit signature, gates are assigned a number of input and output wires. This serves us well when we want to model lower level circuits in which we really are dealing with single-bit wires. However, when designing circuits it is often advantageous to work at a higher level of abstraction with 'thicker' wires carrying words of information. For example, the values in the circuits could be used to represent binary numbers.

This can still be modelled in $\mathbf{SCirc}_{\Sigma}$ 'as is' by using lots of parallel wires to connect to the various primitives, but this can get messy with wires all over the place. Instead, we will introduce a generalisation of circuit signatures in which these thicker buses of wires are treated as first-class objects.

**Definition 3.19** (Generalised circuit signature)**.** A *generalised circuit signature* $\Sigma$ is a tuple $(\mathbf{V}, \bullet, \mathcal{P}, \mathrm{dom}, \mathrm{cod})$ where $\mathbf{V}$ is a finite set of values, $\bullet \in \mathbf{V}$ is a *disconnected* value, $\mathcal{P}$ is a (usually finite) set of *gate symbols*, $\mathrm{dom} \colon \mathcal{P} \to \mathbb{N}_+^\star$ is an *arity* function and $\mathrm{cod} \colon \mathcal{P} \to \mathbb{N}_+^\star$ is a *coarity* function.

In a generalised circuit signature, primitives are typed with input and output *words* rather than just natural numbers.

**Example 3.20.** The generalised circuit signature for *simple arithmetic circuits* is $\Sigma^+_{\mathrm{B}} := (\mathbf{V}_{\mathrm{B}}, \bot, \mathcal{P}^+_{\mathrm{B}}, \mathrm{dom}^+_{\mathrm{B}}, \mathrm{cod}^+_{\mathrm{B}})$, where

$$\mathcal{P}_{\mathrm{B}} := \{\mathrm{AND}_{k,n}, \mathrm{OR}_{k,n}, \mathrm{NOT}_n, \mathrm{add}_n \mid n \in \mathbb{N}_+\}$$

$$\mathrm{dom}^+_{\mathrm{B}}(\mathrm{AND}_{k,n}) := [n \mid i < k] \quad \mathrm{dom}^+_{\mathrm{B}}(\mathrm{OR}_{k,n}) := [n \mid i < k]$$

$$\mathrm{dom}^+_{\mathrm{B}}(\mathrm{NOT}_n) := [n] \quad \mathrm{dom}^+_{\mathrm{B}}(\mathrm{add}_n) := [n, n]$$

$$\mathrm{cod}^+_{\mathrm{B}} := \mathrm{AND}_{k,n} \mapsto [n], \mathrm{OR}_{k,n} \mapsto [n], \mathrm{NOT}_n \mapsto [n], \mathrm{add}_n \mapsto [n]$$

The gates $\mathrm{AND}_{k,n}$ and $\mathrm{OR}_{k,n}$ are gates that operate on $k$ input wires of width $n$; similarly the $\mathrm{NOT}_n$ gate operates on input wires of width $n$. The $\mathrm{add}_n$ component represents an adder that takes as input two $n$-bit wires and outputs their $n$-bit sum.

Just like a monochromatic circuit signature generates monochromatic PROPs, a generalised circuit signature generates $\mathbb{N}_+$-coloured PROPs.

**Definition 3.21.** For a generalised circuit signature $\Sigma$, let the set $\Sigma_{\mathbf{CCirc}^+}$ of *generalised combinational circuit generators* be defined as the set containing

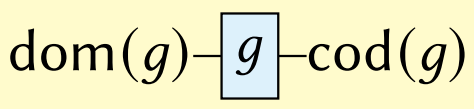

for each $g \in \mathcal{P}$

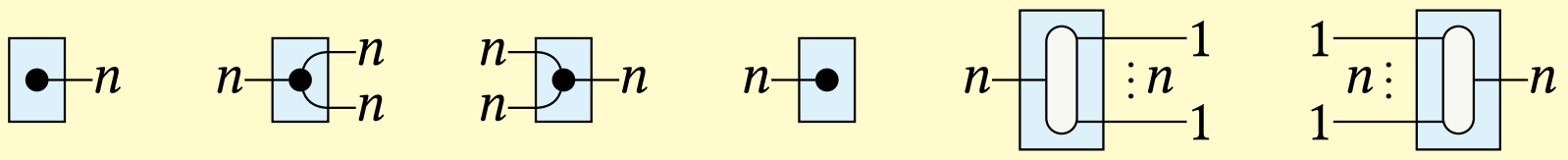

for each $n \in \mathbb{N}_+$

We write $\mathbf{CCirc}^+_\Sigma$ for the freely generated $\mathbb{N}_+$-coloured PROP $\mathrm{S}_{\Sigma_{\mathbf{CCirc}^+}}$.

Most of the generators in $\mathbf{CCirc}^+_\Sigma$ are fairly straightforward generalisations of the primitives in $\mathbf{CCirc}_\Sigma$ to act on each colour (width) of wires. The only new generators are the *bundlers* at the end of the bottom row; their intended meaning is that they can be used to *split* and *combine* bundles of wire buses into bundles with different widths. These constructs were first proposed by Wilson et al in [WGZ23] as a notation for *non-strict categories*. We take inspiration from their observation that a similar idea could also be applied to strict categories.

**Example 3.22** (ALU)**.** The computation of a CPU is performed by an *arithmetic logic unit*, or ALU for short. An ALU takes some input wires of a fixed width and performs an operation on them given some control signal. While ALUs can often perform many different operations, we will look at an example operating on four-bit wires that performs a bitwise AND operation when the control is false, and an addition when the control is true. This ALU will also produce an output indicating if the sum is zero, and the sign of the sum; these auxiliary outputs only produce useful output when the addition operation is selected.

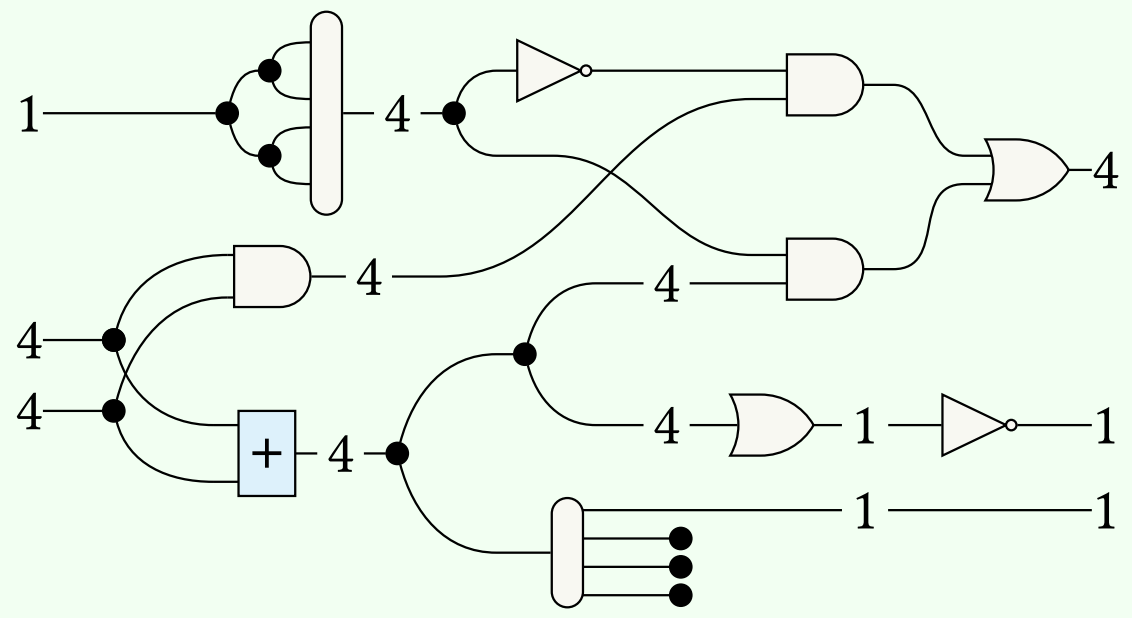

To apply the single-bit control to the four-bit AND gates, the top bundler and forks are used to create a wire containing only the original bit.

The sum is zero if all of the bits are false. To determine this, the $\mathrm{OR}_{1,4}$ gate folds the four-bit sum into a one-bit value which is true if at least one of the bits is true. The $\mathrm{NOT}_{1,1}$ inverts the output to produce true if there are no true bits.

In two's complement representation, the most significant bit indicates if the sum is negative. To model this, the lower bundler splits the four-bit sum into its constituent bits, discarding the least significant three.

Sequential circuits are generalised in the same way.

**Definition 3.23.** For a generalised circuit signature $\Sigma$, let the set $\Sigma_{\mathbf{SCirc}^+}$ of *generalised sequential circuit generators* be the set of generalised combinational circuit generators along with $\boxed{\bar{v}}\!-n$ for each $n \in \mathbb{N}_+$ and $\bar{v} \in \mathbf{V}^n$, and $n\!-\!\boxed{\triangleright}\!-n$ for each $n \in \mathbb{N}_+$. We write $\mathbf{SCirc}_\Sigma^+$ for the freely generated PROP $\mathbf{T}_{\Sigma_{\mathbf{SCirc}^+}}$.

Most of the upcoming results will be shown for the monochromatic case, as the proofs are more elegant. However, most of the results also generalise to the coloured case, and this will be remarked upon throughout.

# Chapter 4

# Denotational semantics

Circuits in $\mathbf{SCirc}_\Sigma$ are just *syntax*; they have no *behaviour* (or *semantics*), assigned to them. A semantics for digital circuits relates circuits which have 'the same behaviour given some interpretation'. But there is not just one way to construct such a relation. In this thesis we will examine three such ways: a *denotational* semantics, an *operational* semantics and an *algebraic* semantics. Each approach comes with advantages and drawbacks; skillful use of all three will lead to a powerful, fully compositional, perspective on sequential circuits.

Although each semantics relation is constructed differently, they must relate the *same* circuits; the behaviour of a circuit should not be different depending on which lens we are viewing it through. Formally, this means that each semantics is *sound and complete* with respect to the others; two circuits $m\text{–}\boxed{f}\text{–}n$ and $m\text{–}\boxed{g}\text{–}n$ are related by one semantics if and only if they are related by the others.

First of all we will define the *denotational semantics* of digital circuits, which will act as the gold standard against which the other semantics will be compared. Denotational semantics is the notion of assigning meaning to structures using values in some *semantic domain*: a partially ordered set with some extra structure. The idea is an old one in computer science, going back to the work of Scott and Strachey [Sco70; SS71].

**Example 4.1.** Consider a language of mathematical expressions defined as follows:

$$n \in N ::= \overline{0} \mid \overline{1} \mid \overline{2} \mid \cdots \qquad e \in E ::= add\, e\, e \mid mul\, e\, e \mid n$$

To define a denotational semantics for terms $E$ in this language, we need to pick a *semantic domain* for the denotations of terms to belong to. An obvious one here is the natural numbers $\mathbb{N}$; given a term $e \in E$, we write $[\![e]\!]$ for its denotation in $\mathbb{N}$. For each $\overline{n} \in N$, $[\![\overline{n}]\!]$ is the corresponding natural number, and the operations are interpreted as $[\![add\, e_1\, e_2]\!] = [\![e_1]\!] + [\![e_2]\!]$ and $[\![mul\, e_1\, e_2]\!] = [\![e_1]\!] \cdot [\![e_2]\!]$ respectively.

The above example illustrates quite nicely how a denotational semantics should be *compositional*; the denotations of a composite term should be constructed by combining the denotations of its components.

**Remark 4.2.** The content of this section is a refinement of [GKS24, Section 3].

## 4.1 Denotational semantics of digital circuits

Now that we comprehend what exactly denotational semantics is, we turn to our goal of defining a denotational semantics for digital circuits. We will interpret digital circuits as a certain class of *stream functions*, functions that operate on infinite sequences of values. This represents how the output of a circuit may not just operate on the current input, but all of the previous ones as well.

**Remark 4.3.** In [MSB12], the semantics of digital circuits with delays and cycles are presented using *timed ternary simulation*, an algorithm to compute how a sequence of circuit outputs stabilises over time given the inputs and value of the current state. This differs from our approach as we assign each circuit a concrete stream function describing its behaviour, rather than having to solve a system of equations in terms of the gates inside a circuit in order to determine its behaviour.

### 4.1.1 Interpreting circuit components

Before assigning stream functions to a given circuit in $\mathbf{SCirc}_\Sigma$, we will first decide how to interpret the individual *components* of a given circuit signature. This is crucial because a denotational semantics is defined *compositionally*; eventually we will need to refer to the interpretations of particular components.

First we consider the interpretation of the *values* that flow through the wires in our circuits. In the denotational semantics the set of values will need to have a bit more structure, as it must be ordered by how much *information* each value carries.

**Definition 4.4** (Partially ordered set)**.** A *partially ordered set*, or *poset* for short, is a set $A$ equipped with a reflexive, antisymmetric, and transitive relation $\leq$, i.e.

- for all $a \in A$, $a \leq a$;
- for all $a, b \in A$, if $a \leq b$ and $b \leq a$ then $a = b$; and
- for all $a, b, c \in A$, if $a \leq b$ and $b \leq c$ then $a \leq c$.

A poset $(A, leq)$ is called a *finite poset* if $A$ is finite.

**Definition 4.5** (Least and greatest elements)**.** In a poset $(A, \leq)$, a *least element* is an element $\bot \in A$ such that for all elements $y \in A$, $\bot \leq y$. Similarly, a *greatest element* is an element $\top \in A$ such that for all elements $x \in A$, $x \leq \top$.

**Example 4.6.** The natural numbers $\mathbb{N}$ are a poset ordered in the usual way; they have a least element $0$ but not a greatest element. However, any finite subset of the natural numbers *does* have a maximal element.

In our context of digital circuits the least and greatest elements respectively represent signals with a complete *lack* of information and *every* piece of information at once. However, we need to add more structure than just this; given two signals we want a way to be able to combine their information into one signal.

**Definition 4.7** (Lower and upper bounds)**.** Given a poset $(A, \leq)$, a *lower bound* of a subset $B \subseteq A$ is an element $x \in B$ such that for all $b \in B$, $x \leq b$. Similarly, an *upper bound* of $C \subseteq A$ is an element $y \in C$ such that for all $b \in B$, $b \leq y$.

We are interested in the 'closest' lower and upper bounds.

**Definition 4.8** (Meet and join)**.** Given a poset $(A, \leq)$, a lower bound $x$ of $B \subseteq A$ is called an *meet* (or infimum, or greatest lower bound) if for all lower bounds $b \in B$, $b \leq x$. Similarly, an upper bound $y$ of $B$ is called a *join* (or supremum, or least upper bound) if for all upper bounds $c \in B$, $y \leq c$.

We usually draw the meet and the join operations as rotated versions of the order operation. For example, in the above definitions we have used $\wedge$ and $\vee$ for the order $\leq$; in subsequent sections we will use $\sqcap$ and $\sqcup$ for the order $\sqsubseteq$. In general, the join and meet of a pair of elements in a poset need not exist. We are interested in the structures in which they *always* exist, which are known as *lattices*.

**Definition 4.9** (Lattice)**.** A *lattice* is a poset $(A, \leq)$ in which each pair of elements $a, b \in A$ has a meet and join. A lattice is called a *finite lattice* if $A$ is finite, and *bounded* if it has a least element and a greatest element.

Much like how every non-empty finite poset has at least one least and greatest element, every non-empty finite lattice has a join and meet.

**Notation 4.10.** For a poset $(A, \leq)$, we write $\bigwedge A$ for the meet of all the elements in $A$ (if it exists) and $\bigvee A$ for the join of all the elements in $A$ (again, if it exists).

**Lemma 4.11.** A non-empty finite lattice $(A, \leq)$ is bounded.

*Proof.* As each pair of elements in $A$ has a meet, and as $A$ is finite, we can define the least element $\bot$ as $\bigwedge A$. This element is the least element of $A$ by definition of the meet: $(a_0 \wedge a_1) \leq a_0$ and $(a_0 \wedge a_1) \leq a_1$, $((a_0 \wedge a_1) \wedge a_2) \leq (a_0 \wedge a_1)$ and $((a_0 \wedge a_1) \wedge a_2) \leq a_2$, and so on. The proof holds in reverse for the greatest element $\top$ by using the join $\bigvee A$. □

**Example 4.12.** Let $A = \{0, 1, 2\}$, and let $(\mathcal{P}A, \subseteq)$ be the poset defined as the powerset of $A$ ordered by subset inclusion. This poset can be illustrated by the following *Hasse diagram*, in which a line going up from $a$ to $b$ indicates that $a \leq b$.

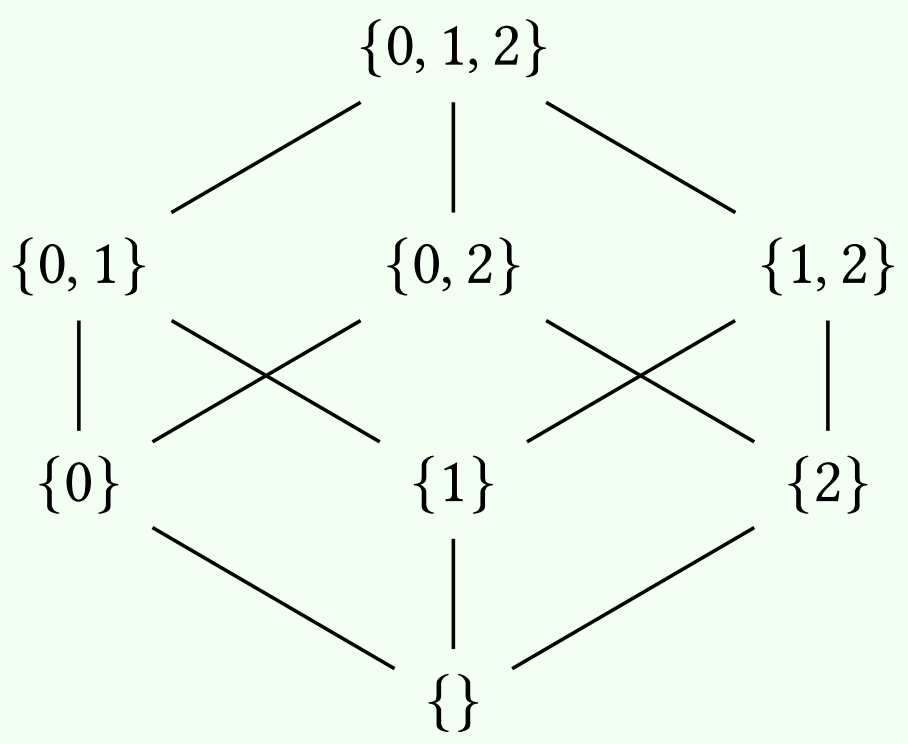

The diagram clearly illustrates the lattice structure on this poset: the join is union and the meet is intersection. Subsequently the greatest element $\top$ is the set $A$ and the least element $\bot$ is the empty set $\{\}$.

**Remark 4.13.** If $A$ is a lattice, then for any $n \in \mathbb{N}$, $A^n$ is also a lattice by comparing elements pointwise. The $\bot$ is then the word containing $n$ copies of the $\bot$ element in $A$, and similarly for the $\top$ element.

Recall the set $A \coloneqq \{0,1,2\}$ from Example 4.12 and the lattice structure on its powerset $\mathcal{P}A$. This induces an ordering on $(\mathcal{P}A)^2$: $\{0,1\}\{1\} \leq \{0,1,2\}\{1,2\}$ because $\{0,1\} \leq \{0,1,2\}$ and $\{1\} \leq \{1,2\}$, and conversely $\{0,1\}\{1\} \nleq \{0\}\{1,2\}$ because $\{0,1\} \nleq \{0\}$. The join and meet are computed by taking the join and meet of each component: $\{0,1\}\{1\} \vee \{0,1,2\}\{1,2\} = \{0,1,2\}\{1,2\}$ and $\{0,1\}\{1\} \vee \{0\}\{1,2\} = \{0,1\}\{1,2\}$.

Values in a circuit signature are interpreted as elements of a finite lattice, so that combining two signals into one wire can be modelled using the join. Now the primitives in the signature must be interpreted. Clearly they should be interpreted as functions between the values, but these functions must respect the order on the value lattice; one should not be able to lose information by performing a computation.

**Definition 4.14.** Let $(A, \leq_A)$ and $(B, \leq_B)$ be partial orders. A function $f\colon A \to B$ is *monotone* if, for every $x, y \in A$, $x \leq_A y$ if and only if $f(x) \leq_B f(y)$.

Another condition on the primitives is that when all the inputs to a primitive are $\bot$, then it should produce $\bot$; we cannot produce information from nothing.

**Definition 4.15.** Let $A, B$ be finite lattices, where $\bot_A$ is the least element of $A$ and $\bot_B$ is the least element of $B$. A function $f\colon A \to B$ is *$\bot$-preserving* if $f(\bot_A) = \bot_B$.

Assigning interpretations to the combinational components of a circuit sets the stage for the entire denotational semantics.

**Definition 4.16** (Interpretation). For a signature $\Sigma = (\mathbf{V}, \bullet, \mathcal{P}, \mathsf{dom}, \mathsf{cod})$, an *interpretation* of $\Sigma$ is a tuple $(\sqsubseteq, [\![-]\!])$ where $(\mathbf{V}, \sqsubseteq)$ is a lattice with $\bullet$ as the least element, and $[\![-]\!]$ maps each $p \in \mathcal{P}$ to a $\bot$-preserving monotone function $\mathbf{V}^{\mathsf{dom}(p)} \to \mathbf{V}^{\mathsf{cod}(p)}$.

**Example 4.17.** Recall the Belnap signature $\Sigma_\mathrm{B} = (\mathbf{V}_\mathrm{B}, \bot, \mathcal{P}_\mathrm{B}, \mathsf{dom}_\mathrm{B}, \mathsf{cod}_\mathrm{B})$ from Example 3.4. We assign a partial order $\leq_\mathrm{B}$ to values in $\mathbf{V}_\mathrm{B}$ as follows:

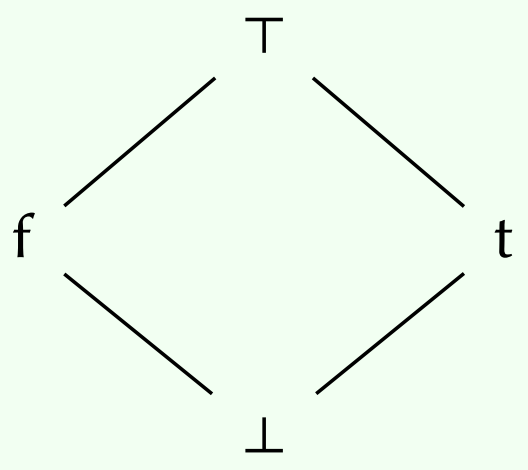

The gate interpretation $[\![-]\!]_\mathrm{B}$ has action $\mathrm{AND} \mapsto \wedge, \mathrm{OR} \mapsto \vee, \mathrm{NOT} \mapsto \neg$ where $\wedge$, $\vee$ and $\neg$ are defined by the following truth tables [Bel77]:

| $\wedge$ | $\bot$ | f | t | $\top$ |
|---|---|---|---|---|
| $\bot$ | $\bot$ | f | $\bot$ | f |
| f | f | f | f | f |
| t | $\bot$ | f | t | $\top$ |
| $\top$ | f | f | $\top$ | $\top$ |

| $\vee$ | $\bot$ | f | t | $\top$ |
|---|---|---|---|---|
| $\bot$ | $\bot$ | $\bot$ | t | t |
| f | $\bot$ | f | t | $\top$ |
| t | t | t | t | t |
| $\top$ | t | $\top$ | t | $\top$ |

| $\neg$ | |
|---|---|
| $\bot$ | $\bot$ |
| t | f |
| f | t |
| $\top$ | $\top$ |

The Belnap interpretation is then $(\leq_B, [\![-]\!]_B)$. An online tool for experimenting with the Belnap interpretation can be found at https://belnap.georgejkaye.com.

### 4.1.2 Denotational semantics of combinational circuits

The semantic domain for *combinational* circuits is straightforward: each circuit maps to a monotone function.

**Definition 4.18.** Let $\mathbf{Func}_{\mathcal{I}}$ be the PROP in which the morphisms $m \to n$ are the $\bot$-preserving monotone functions $\mathbf{V}^m \to \mathbf{V}^n$.

To map between PROPs we must use a PROP morphism.

**Definition 4.19.** Let $[\![-]\!]^{C}_{\mathcal{I}} \colon \mathbf{CCirc}_{\Sigma} \to \mathbf{Func}_{\mathcal{I}}$ be the PROP morphism with action defined as

$[\![$ 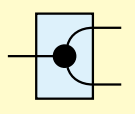 $]\!]^{C}_{\mathcal{I}} \coloneqq (v) \mapsto (v, v)$

$[\![$ 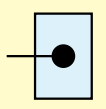 $]\!]^{C}_{\mathcal{I}} \coloneqq (v) \mapsto ()$

$[\![$ 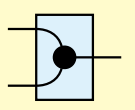 $]\!]^{C}_{\mathcal{I}} \coloneqq (v, w) \mapsto v \sqcup w$

$[\![$ 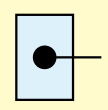 $]\!]^{C}_{\mathcal{I}} \coloneqq () \mapsto \bot$

$$[\![ m \text{—} \boxed{p} \text{—} n ]\!]^{C}_{\mathcal{I}} \coloneqq [\![p]\!]$$

**Remark 4.20.** One might wonder why the fork and the join have different semantics, as they would be physically realised by the same wiring. This is because digital circuits have a notion of *static causality*: outputs can only connect to inputs. This is why the semantics of combinational circuits is *functions* and not *relations*.

In real life one could force together two digital devices, but this might lead to undefined behaviour in the digital realm. This is reflected in the semantics by the use of the join; for example, in the Belnap interpretation if one tries to join together t and f then the overspecified $\top$ value is produced.

### 4.1.3 Denotational semantics of sequential circuits

As one might expect, sequential circuits are slightly trickier to deal with. In a combinational circuit, the output only depends on the inputs at the current cycle, but for sequential circuits inputs can affect outputs many cycles after they occur.

We therefore have to reason with *infinite sequences* of inputs rather than individual values; these are known as *streams*.

**Notation 4.21.** Given a set $A$, we denote the set of streams (infinite sequences) of $A$ by $A^\omega$. As a stream can equivalently be viewed as a function $\mathbb{N} \to A$, we write $\sigma(i) \in A$ for the $i$-th element of stream $\sigma \in A^\omega$. Individual streams are written as $\sigma \in A^\omega \coloneqq \sigma(0) :: \sigma(1) :: \sigma(2) :: \cdots$.

Streams can be viewed a bit like lists, in that they have a head component and an (infinite) tail component.

**Definition 4.22** (Operations on streams)**.** Given a stream $\sigma \in A^\omega$, its *initial value* $\mathsf{hd}(\sigma) \in A$ is defined as $\sigma \mapsto \sigma(0)$ and its *stream derivative* $\mathsf{tl}(\sigma) \in A^\omega$ is defined as $\sigma \mapsto (i \mapsto \sigma(i+1))$.

**Notation 4.23.** For a stream $\sigma$ with initial value $a$ and stream derivative $\tau$ we write it as $\sigma \coloneqq a :: \tau$.

Streams will serve as the inputs and outputs to circuits, so the denotations of sequential circuits will be *stream functions*, which consume and produce streams. Just like with functions, we cannot claim that all streams are the denotations of sequential circuits.

**Definition 4.24** (Causal stream function [Rut06])**.** A stream function $f : A^\omega \to B^\omega$ is *causal* if, for all $i \in \mathbb{N}$ and $\sigma, \tau \in A^\omega$ we have that $\sigma(j) = \tau(j)$ for all $j \leq i$, then $f(\sigma)(i) = f(\tau)(i)$.

Causality is a form of continuity; a causal stream function is a stream function in which the $i$-th element of its output stream only depends on elements 0 through $i$ inclusive of the input stream; it cannot look into the future. A neat consequence of causality is that it enables us to lift the stream operations of initial value and stream derivative to stream *functions*.

**Definition 4.25** (Initial output [Rut06])**.** For a causal stream function $f : A^\omega \to B^\omega$ and $a \in A$, the *initial output of $f$ on input $a$* is an element $f[a] \in A$ defined as $f[a] \coloneqq \mathsf{hd}(f(a :: \sigma))$ for an arbitrary $\sigma \in A^\omega$.

Since $f$ is causal, the stream $\sigma$ in the definition of initial output truly is arbitrary; the hd function only depends on the first element of the stream.

**Definition 4.26** (Functional stream derivative [Rut06])**.** For a stream function $f\colon A^\omega \to B^\omega$ and $a \in A$, the *functional stream derivative of $f$ on input $a$* is a stream function $f_a$ defined as $f_a \coloneqq \sigma \mapsto \mathsf{tl}(f(a :: \sigma))$.

The functional stream derivative $f_a$ is a new stream function which acts as $f$ would 'had it seen $a$ first'.

**Remark 4.27.** One intuitive way to view stream functions is to think of them as the states of some Mealy machine; the initial output is the output given some input, and the functional stream derivative is the transition to a new state. As with most things in mathematics, this is no coincidence; there is a homomorphism from any Mealy machine to a stream function. We will exploit this fact in the next section.

This leads us to the next property of denotations of sequential circuits. Although they may have infinitely many inputs and outputs, circuits themselves are built from a finite number of components. This means they cannot specify infinite *behaviour*.

**Notation 4.28.** Given a finite word $\overline{a} \in A^\star$, we abuse notation and write $f_{\overline{a}}$ for the repeated application of the functional stream derivative for the elements of $\overline{a}$, i.e. $f_\varepsilon \coloneqq f$ and $f_{a::\overline{b}} \coloneqq (f_a)_{\overline{b}}$.

**Definition 4.29.** Given a stream function $f\colon A^\omega \to B^\omega$, we say it is *finitely specified* if the set $\{f_{\overline{a}} \mid \overline{a} \in A^\star\}$ is finite.

As the components of our circuits are monotone and $\bot$-preserving, a denotational semantics for circuits must also be monotone and $\bot$-preserving. This means we need to lift the order on values to an order on streams.

**Notation 4.30.** For a poset $(A, \leq_A)$ and streams $\sigma, \tau \in A^\omega$, we say that $\sigma \leq_{A^\omega} \tau$ if $\sigma(i) \leq_A \tau(i)$ for all $i \in \mathbb{N}$.

For these properties to be suitable as a denotational semantics for sequential circuits, we must show that stream functions with these properties form a category we can map into from $\mathbf{SCirc}_\Sigma$. We will first show that these categories form a symmetric monoidal category, so we need a suitable candidate for composition and tensor. There are fairly obvious choices here: for the former we use regular function composition and for the latter we use the Cartesian product.

**Lemma 4.31.** Causality is preserved by composition and Cartesian product.

*Proof.* If the $i$-th element of two stream functions $f$ and $g$ only depends on the first $i+1$ elements of the input, then so will their composition and product. □

**Lemma 4.32.** Finite specification is preserved by composition and Cartesian product.

*Proof.* For both the composition and product of two stream functions $f$ and $g$, the largest the set of stream derivatives could be is the product of stream derivatives of $f$ and $g$, so this will also be finite. □

**Lemma 4.33.** $\perp$-preserving monotonicity is preserved by composition and Cartesian product.

*Proof.* The composition and product of any monotone function is monotone, and must preserve the $\perp$ element. □

As the categorical operations preserve the desired properties, these stream functions form a PROP.

**Proposition 4.34.** There is a PROP $\mathbf{Stream}_{\mathcal{I}}$ in which the morphisms $m \to n$ are the causal, finitely specified and $\perp$-preserving monotone stream functions $(\mathbf{V}^m)^\omega \to (\mathbf{V}^n)^\omega$.

*Proof.* Identity is the identity function, the symmetry swaps streams, composition is composition of functions, and tensor product on morphisms $f\colon (\mathbf{V}^m)^\omega \to (\mathbf{V}^n)^\omega$ and $g\colon (\mathbf{V}^p)^\omega \to (\mathbf{V}^q)^\omega$ is the Cartesian product of functions composed with the components of the isomorphism $(\mathbf{V}^m)^\omega \times (\mathbf{V}^n)^\omega \cong (\mathbf{V}^{m+n})^\omega$.

As these constructs satisfy the categorical axioms, and as function composition and Cartesian product preserve causality (Lemma 4.31), finite specification (Lemma 4.32), and monotonicity (Lemma 4.33), this data defines a symmetric monoidal category. □

Modelling sequential circuits as stream functions deals with temporal aspects, but what about feedback? As the assignment of denotations needs to be compositional, we need to map the trace on $\mathbf{SCirc}_\Sigma$ to a trace on $\mathbf{Stream}_{\mathcal{I}}$. A usual candidate for the trace when considering partially ordered settings is the *least fixed point*.

**Definition 4.35** (Least fixed point)**.** For a poset $(A, \leq)$ and function $f : A \to A$, the least fixed point of $f$ is a value $\mu_f$ such that $f(\mu_f) = f$ and, for all values $v$ such that $f(v) = v$, $\mu_f \leq v$.

Least fixed points are ubiquitous in program semantics, where they are often used to model *recursion*; since feedback is a form of recursion it seems apt that we should also follow this route. As fixed points are so important, they are the subject of many theorems; one that will come in very useful for us is the *Kleene fixed-point theorem*, which is concerned with a special class of monotone functions.

**Definition 4.36** (Directed subset)**.** For a poset $(A, \leq)$, a non-empty subset $B \subseteq A$ is called a *directed subset* if every pair of elements in $B$ has an upper bound in $B$. If this set has a join $\bigvee B$ then this element is called a *directed join*.

**Notation 4.37** (Image)**.** For a function $f : A \to B$ and subset $C \subseteq A$, we write $f[C] \subseteq B$ for the *image* of $C$ under $f$.

**Definition 4.38** (Scott continuity)**.** Given two posets $(A, \leq_A)$ and $(B, \leq_B)$, a function $f : A \to B$ is *Scott-continuous* if for every directed subset $C \subseteq A$ it holds that $f(\bigvee_A C) = \bigvee_B (f[C])$ i.e. $f$ preserves directed joins.

**Theorem 4.39** (Kleene fixed-point theorem [Tar55])**.** Let $(A, \leq)$ be a poset in which each of its directed subsets has a join, and let $f : L \to L$ be a Scott-continuous function. Then $f$ has a least fixed point, defined as $\bot \vee f(\bot) \vee f(f(\bot)) \vee \cdots$.

**Remark 4.40.** The Kleene fixed-point theorem was not actually proved by Kleene, but is only named after him! The result is often instead attributed to Tarski.

As we have so far only considered monotone functions, it is useful to get some intuition for what Scott-continuity brings to the table.

**Example 4.41.** An example of a directed subset of $\mathbf{V}_{\mathrm{B}}^{\omega}$ is the set $T$ defined as $\{\mathrm{t}^n \bot \mid n \in \mathbb{N}\}$; the join of this set is $\mathrm{t}^{\omega}$. One Scott-continuous function $\mathbf{V}_{\mathrm{B}}^{\omega} \to \mathbf{V}_{\mathrm{B}}^{\omega}$ is defined as $f(\sigma)(0) = \bot$ and $f(\sigma)(i+1) = f(\sigma)(i)$; this is Scott-continuous because finding the join of a set and then prepending it with $\bot$ is the same as prepending $\bot$ to each stream in the set and then finding their join.

An example of a stream function that is monotone but *not* Scott-continuous is the function defined as $g(\mathrm{t}^n \bot^{\omega}) \coloneqq \mathrm{f}^{\omega}(n)$ and $g(\mathrm{t}^{\omega}) \coloneqq \top^{\omega}$ (the other inputs do not matter for this example). We can show this is not Scott-continuous by considering the set $T$ above, as $f(\bigsqcup T) = f(\mathsf{t}^\omega) = \top^\omega \neq \mathsf{f}^\omega = \bigsqcup f[T]$. However, note that this function is *not* causal: as $\mathsf{t}^\omega$ has the same prefix as every element in $T$, $f(\mathsf{t}^\omega)$ must also share output prefixes.

So far we have not explicitly enforced Scott-continuity on stream functions; it turns out that it is implied by causality and monotonicity.

**Proposition 4.42.** Let $(A, \leq_A)$ and $(B, \leq_B)$ be finite lattices, and let $(A^\omega, \leq_{A^\omega})$ and $(B^\omega, \leq_{B^\omega})$ be the induced lattices on streams. Then a causal and monotone function $A^\omega \longrightarrow B^\omega$ must also be Scott-continuous.

*Proof.* Consider a directed subset $D \subseteq A^\omega$; we need to show that for an arbitrary causal, monotone and finitely specified function $f$ we have that $f(\bigvee D) = \bigvee f[D]$.

First consider the case when there is a greatest element in $D$. In this case, $\bigvee D$ must be the greatest element, and as such $\bigvee D \in D$. As $f$ is monotone then $f(\bigvee D)$ must be the greatest element in $f[D]$; subsequently, $f(\bigvee D) = \bigvee f[D]$.

Now consider the case where there is no greatest element in $D$ and subsequently $\bigvee D \notin D$; if there is no greatest element, $D$ must be infinite. Even though it is infinite, $D$ is a directed subset so each pair of elements must have an upper bound, and as $\leq_{A^\omega}$ is computed pointwise by using $\leq_A$ we can consider what the upper bounds are pointwise too. Because $A$ is finite, there cannot be an infinite chain of upper bounds for each element $i$; there must exist an element $a_i \in A$ such that $D$ contains infinitely many streams $\sigma$ such that $\sigma(i) = a_i$. This means that $(\bigvee D)(i) = a$, so every prefix of $\bigvee D$ must exist as a prefix of a stream in $D$. As $f$ is causal, for each prefix $f(\bigvee D)$ there must also exist a $d \in D$ such that $f(d)$ has that prefix, and as such $\bigvee f[D] = f(\bigvee D)$. □

This means we can use the Kleene fixed point theorem as a tool to show that the least fixed point is a trace on $\mathbf{Stream}_{\mathcal{I}}$.

**Notation 4.43** (Concatenation)**.** For a set $A$, $\overline{a} \in A^m$, and $\overline{b} \in A^n$, we write $\overline{a} +\!\!+ \overline{b}$ for the *concatenation* of $\overline{a}$ and $\overline{b}$: the tuple of length $m+n$ containing the elements of $\overline{a}$ followed by the elements of $\overline{b}$.

**Notation 4.44** (Projection)**.** For a set $A$, let $\overline{a} \in A^m$ and $\overline{b} \in A^n$. Then for their concatenation $\overline{c} \coloneqq \overline{a} +\!\!+ \overline{b} \in A^{m+n}$, we write $\pi_0(\overline{c}) = \overline{a}$ and $\pi_1(\overline{c}) = \overline{b}$ for the *projections*.

**Lemma 4.45.** Given a morphism $f\colon (\mathbf{V}^{x+m})^\omega \to (\mathbf{V}^{x+n})^\omega \in \mathbf{Stream}_{\mathcal{I}}$, and stream $\sigma \in (\mathbf{V}^m)^\omega$, the function $\tau \mapsto \pi_0(f(\tau,\sigma))$ has a least fixed point.

*Proof.* The function $\tau \mapsto \pi_0(f(\tau,\sigma))$ is causal and monotone because $f$ and the projection function are causal and monotone, so it is Scott-continuous by Proposition 4.42. By the Kleene fixed point theorem, this function has a least fixed point, defined as $\pi_0(f(\bot^\omega,\sigma)) \sqcup \pi_0(f(\pi_0(f(\bot^\omega,\sigma)),\sigma)) \sqcup \cdots$. □

We must show that this notion of least fixed point is a trace on $\mathbf{Stream}_{\mathcal{I}}$. The first step is to show that taking the least fixed point of a stream function in $\mathbf{Stream}_{\mathcal{I}}$ produces another causal, finitely specified, $\bot$-preserving, and monotone stream function. The original proof idea for this is due to David Sprunger, and relies on an ordering on stream functions themselves.

**Definition 4.46.** Let $A$ and $B$ be posets and let $f, g\colon A^\omega \to B^\omega$ be stream functions. We say $f \preceq g$ if $f(\sigma) \leq_{B^\omega} g(\sigma)$ for all $\sigma \in A^\omega$.

**Theorem 4.47.** For a function $f\colon (\mathbf{V}^{x+m})^\omega \to (\mathbf{V}^{x+n})^\omega$, let $\mu_f(\sigma)$ be the least fixed point of the function $\tau \mapsto \pi_0(f(\tau,\sigma))$. Then, the stream function $\sigma \mapsto \pi_1\big(f(\mu_f(\sigma),\sigma)\big)$ is in $\mathbf{Stream}_{\mathcal{I}}$.

*Proof.* To show this, we need to prove that $\sigma \mapsto \pi_1\big(f(\mu_f(\sigma)\sigma)\big)$ is in $\mathbf{Stream}_{\mathcal{I}}$: it is causal, finitely specified, and $\bot$-preserving monotone.

Since $f\colon (\mathbf{V}^{x+m})^\omega \to (\mathbf{V}^{x+n})^\omega$ is a morphism of $\mathbf{Stream}_{\mathcal{I}}$, it has finitely many stream derivatives. For each stream derivative $f_{\overline{w}}$, let the function $\widehat{f_{\overline{w}}}\colon (\mathbf{V}^{x+m})^\omega \to (\mathbf{V}^x)^\omega$ be defined as $\tau\sigma \mapsto \pi_0(f_{\overline{w}}(\tau\sigma))$. Note that each of these functions are causal, $\bot$-preserving, and monotone, because they are constructed from pieces that are causal $\bot$-preserving and monotone.

In particular, $\mu_f(\sigma)$ is the least fixed point of $\widehat{f_\varepsilon}((-)\sigma)$. Using the Kleene fixed point theorem, the least fixed point of $\widehat{f}((-)\sigma)$ can be obtained by composing $\widehat{f}((-)\sigma)$ repeatedly with itself. This means that $\mu_f(\sigma) = \bigsqcup_{k\in\mathbb{N}} \widehat{f}^k(\bot^\omega,\sigma)$ where $\widehat{f}^k$ is the $k$-fold composition of $f(-,\sigma)$ with itself, i.e. $\widehat{f}^0(\tau\sigma) = \tau$ and $\widehat{f^{k+1}}(\tau\sigma) = \widehat{f}\left(\left(\widehat{f}^k(\sigma,\tau)\right)\sigma\right)$. That the mapping $\mu_f$ is causal and monotone is straightforward: each of the functions in the join is causal and monotone, and join preserves these properties. It remains to show that this mapping has finitely many stream derivatives.

When equipped with $\preceq$, the set of functions $(\mathbf{V}^{x+m})^\omega \to (\mathbf{V}^x)^\omega$ is a poset, of which $\{\widehat{f_w} \mid w \in (\mathbf{V}^{x+m})^\star\}$ is a finite subset. Restricting the ordering $\preceq$ to this set yields a finite poset. Since this poset is finite, the set of strictly increasing sequences in this poset is also finite. We will now demonstrate a relationship between these sequences and stream derivatives of $\mu_f$.

Suppose $S = \widehat{f_{\overline{w_0}}} \prec \widehat{f_{\overline{w_1}}} \prec \cdots \prec \widehat{f_{\overline{w_{\ell-1}}}}$ is a strictly increasing sequence of length $\ell$ in the set of stream functions $\{\widehat{f_w} \mid w \in (\mathbf{V}^{x+m})^\star\}$. We define a function $g_S \colon (\mathbf{V}^m)^\omega \to (\mathbf{V}^x)^\omega$ as $(\sigma) \mapsto \bigsqcup_{k\in\mathbb{N}} g_k(\sigma)$ where

$$g_k(\sigma) = \begin{cases} \bot^\omega & \text{if } k = 0 \\ \widehat{f_{\overline{w_k}}}((g_{k-1}(\sigma), \sigma)) & \text{if } 1 \leq k \leq \ell \\ \widehat{f_{\overline{w_{\ell-1}}}}((g_{k-1}(\sigma), \sigma)) & \text{if } \ell < k \end{cases}.$$

Let the set $G \coloneqq \{g_S \mid S \text{ is a strictly increasing sequence}\}$. When $S$ is set to the one-item sequence $\widehat{f}$, $g_S$ is $\mu_f$, so $\mu_f \in G$. As $G$ is finite, this means that if $G$ is closed under stream derivative, $\mu_f$ has finitely many stream derivatives. Any element of $G$ is either $\bot^\omega$ or has the form $\sigma \mapsto \widehat{f_{\overline{w}}}(g_k(\sigma), \sigma)$ for some $\sigma \in (\mathbf{V}^m)^\omega$ and $k > 0$. As $\bot^\omega$ is its own stream derivative, we need to show that applying stream derivative to the latter produces another element of $G$.

$$\begin{aligned}
\sigma \mapsto \left(\widehat{f_{\overline{w}}}\,(g_{k-1}(\sigma), \sigma)\right)_{(a,b)} &= \sigma \mapsto \mathsf{tl}\left(\widehat{f_{\overline{w}}}((a,b) :: (g_{k-1}(\sigma), \sigma))\right) \\
&= \sigma \mapsto \mathsf{tl}\,(\pi_0(f_{\overline{w}}((a,b) :: (g_{k-1}(\sigma), \sigma)))) \\
&= \sigma \mapsto \pi_0(\mathsf{tl}\,(f_{\overline{w}}((a,b) :: (g_{k-1}(\sigma), \sigma)))) \\
&= \sigma \mapsto \pi_0\Big((f_{\overline{w}}(g_{k-1}(\sigma), \sigma))_{(a,b)}\Big) \\
&= \sigma \mapsto \pi_0\big(f_{(a,b)::\overline{w}}(g_{k-1}(\sigma), \sigma)\big) \\
&= \sigma \mapsto \widehat{f_{(a,b)::\overline{w}}}(g_{k-1}(\sigma), \sigma)
\end{aligned}$$

As $\pi_0(f_{(a,b)::\overline{w}})$ is in $G$, the latter is closed under stream derivative. Subsequently, $\mu_f$ has finitely many stream derivatives.

This means that all the components of $\sigma \mapsto \pi_1(f(\mu_f(\sigma), \sigma))$ are causal, monotone and finitely specified, and as these properties are preserved by composition, the composite must also have them, so $\sigma \mapsto \pi_1(f(\mu_f(\sigma), \sigma))$ is in $\mathbf{Stream}_\mathcal{I}$. □

Even if $\mathbf{Stream}_\mathcal{I}$ is closed under least fixed point, this does not mean that it is a valid trace. To verify this we must establish that the categorical axioms of the trace hold.

**Theorem 4.48.** A trace on $\mathbf{Stream}_{\mathcal{I}}$ is given for a function $f\colon (\mathbf{V}^{x+m})^\omega \to (\mathbf{V}^{x+n})^\omega$ by the stream function $\sigma \mapsto \pi_1(f(\mu_f(\sigma), \sigma))$, where $\mu_f(\sigma)$ is the least fixed point of the function $\tau \mapsto \pi_0(f(\tau, \sigma))$ for fixed $\sigma$.

*Proof.* By Theorem 4.47, $\mathbf{Stream}_{\mathcal{I}}$ is closed under taking the least fixed point, so we just need to show that the axioms of STMCs hold. Most of these follow in a straightforward way; the only interesting one is the sliding axiom. We need to show that for stream functions $f\colon (\mathbf{V}^{x+m})^\omega \to (\mathbf{V}^{y+n})^\omega$ and $g\colon (\mathbf{V}^{y})^\omega \to (\mathbf{V}^{x})^\omega$, we have that $\mathrm{Tr}^y\left((\tau, \sigma) \mapsto f(g(\tau), \sigma)\right) = \mathrm{Tr}^x\left((\tau, \sigma) \mapsto g(\pi_0(f(\tau, \sigma)), \pi_1(f(\tau, \sigma)))\right)$.

Let $l \coloneqq (\tau, \sigma) \mapsto f(g(\tau), \sigma)$ and $r \coloneqq (\tau, \sigma) \mapsto g(\pi_0(f(\tau, \sigma)), \pi_1(f(\tau, \sigma)))$; we must apply the candidate trace construction to both of these and check they are equal. For $l$, the least fixed point of $\tau \mapsto \pi_0(f(g(\tau), \sigma))$ is

$$\mu_l(\sigma) = \pi_0(f(g(\bot^\omega), \sigma)) \sqcup \pi_0(f(g(\pi_0(f(g(\bot^\omega), \sigma))), \sigma)) \sqcup \cdots .$$

Plugging this into the candidate trace construction we have that

$$\begin{aligned} &\sigma \mapsto \pi_1\Big(f(g(\mu_l^l(\sigma)), \sigma)\Big) \\ &\quad = \sigma \mapsto \pi_1(f(g(\pi_0(f(g(\dots f(g(\pi_0(f(g(\bot^\omega), \sigma))), \sigma))))), \sigma)) \end{aligned}$$

For the right-hand side, the least fixed point of $\tau \mapsto g(\pi_0(f(\tau, \sigma)))$ is

$$\mu_r(\sigma) = g(\pi_0(f(\bot^\omega, \sigma))) \sqcup g(\pi_0(f(g(\pi_0(f(g(\bot^\omega), \sigma))), \sigma))) \sqcup \cdots$$

When plugged into the candidate trace construction this produces

$$\begin{aligned} &\sigma \mapsto \pi_1(g(\pi_0(f(\mu_r(\sigma), \sigma))), \pi_1(f(\mu_r(\sigma), \sigma))) \\ &\quad = \sigma \mapsto \pi_1(f(\mu_\sigma^r, \sigma)) \\ &\quad = \sigma \mapsto \pi_1(f(g(\pi_0(f(g(\dots f(g(\pi_0(f(g(\bot^\omega), \sigma))), \sigma))), \sigma)))) \\ &\quad = \sigma \mapsto \pi_1(f(g(\pi_0(f(g(\dots f(g(\pi_0(f(\bot^\omega, \sigma))), \sigma))), \sigma)))) \end{aligned}$$

Both the left-hand and the right-hand sides of the sliding equation are equal, so the construction is indeed a trace. □

We now have two traced PROPs: a *syntactic* PROP of sequential circuit terms $\mathbf{SCirc}_\Sigma$ and a *semantic* PROP of causal, finitely specified, monotone stream functions $\mathbf{Stream}_{\mathcal{I}}$. It would be straightforward to now define a map from circuits to these stream functions; indeed, this is the strategy used in [GKS24]. Instead, we will first examine another structure with close links to both circuits and stream functions; that of *Mealy machines*. The structure of Mealy machines will come in useful when considering the *completeness* of the denotational semantics.

## 4.2 Monotone Mealy machines

It is not immediately obvious how to translate back from stream functions in $\mathbf{Stream}_{\mathcal{I}}$ to circuits in $\mathbf{SCirc}_{\Sigma}$. Even though these stream functions have finitely many stream derivatives, how does one encapsulate this behaviour into a circuit? Fortunately, we have a secret weapon: the *Mealy machine* [Mea55].

Mealy machines are used in circuit design to specify the behaviour of a circuit without having to use concrete components. They also have a very useful *coalgebraic* viewpoint which we will wield in order to build a bridge from circuits into stream functions. In particular, there is a unique homomorphism from a Mealy machine to a causal, finitely specified stream function. Our strategy is to assemble a special class of Mealy machines which we dub *monotone Mealy machines* into another traced PROP.

**Definition 4.49** (Mealy machine [Mea55])**.** Let $A$ and $B$ be finite sets. A (finite) $(A, B)$-*Mealy machine* is a tuple $(S, f)$ where $S$ is a finite set called the *state space*, $f \colon S \to (S \times B)^A$ is the *Mealy function*.

An $(A, B)$-Mealy machine is parameterised over a set $A$ of *inputs* and a set $B$ of *outputs*, and is comprised of a set $S$ of *states* and a function transforming a pair $(s, a)$ of a current state and an input into a pair $\langle t, b \rangle$ of a transition state and an output.

**Notation 4.50.** We will use the shorthand $f_0 \coloneqq (s, a) \mapsto \pi_0(f(s)(a))$ and $f_1 \coloneqq (s, a) \mapsto \pi_1(f(s)(a))$ for the transition and output component of the Mealy function respectively.

**Example 4.51.** Let the set of Booleans be defined as $\mathbf{B} \coloneqq \{\mathsf{f}, \mathsf{t}\}$. We define a $(\mathbf{B}, \mathbf{B})$-Mealy machine $(S, f)$ as follows:

$$S \coloneqq \{s_0, s_1\} \qquad \begin{array}{ll} f(s_0, \mathsf{f}) \mapsto \langle s_0, \mathsf{f} \rangle & f(s_0, \mathsf{t}) \mapsto \langle s_1, \mathsf{t} \rangle \\ f(s_1, \mathsf{f}) \mapsto \langle s_1, \mathsf{t} \rangle & f(s_1, \mathsf{t}) \mapsto \langle s_0, \mathsf{f} \rangle \end{array}$$

This is a Mealy machine with two states; at state $s_0$ the output is the input, and at state $s_1$ the output is the negation. If the input is true then the state switches. To illustrate Mealy machines we draw states as circles; an arrow between states labelled $v|w$ represents a transition on input $v$ producing output $w$.

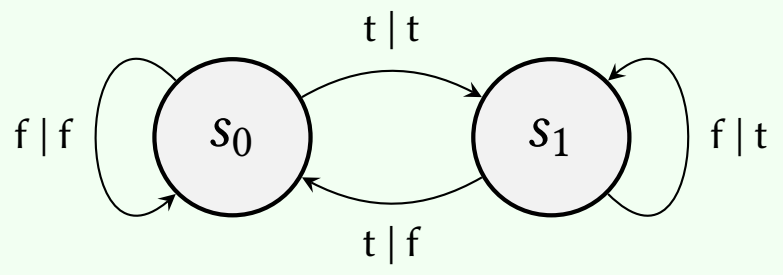

## 4.2.1 The coalgebraic perspective

The definition of Mealy machine above is timeless and forms the basis for most of modern electronics. The natural question for the categorist to ask is *can we make it more categorical?* And as is often the case, we can, using the notion of a *coalgebra*.

**Definition 4.52** (Coalgebra)**.** For a category $\mathcal{C}$, let $F\colon \mathcal{C} \to \mathcal{C}$ be an endofunctor. A *coalgebra* for $F$, or $F$-coalgebra, is an object $A \in \mathcal{C}$ along with a morphism $\alpha\colon A \to FA \in \mathcal{C}$, usually written $(A, \alpha)$.

A Mealy machine is a pair of a set and a function, so this is a coalgebra in **Set**.

**Definition 4.53.** For sets $A$ and $B$, an $(A, B)$-*Mealy coalgebra* is a coalgebra of the functor $Y\colon \mathbf{Set} \to \mathbf{Set}$, defined as $S \mapsto (S \times B)^A$.

**Example 4.54** ([BRS08])**.** Given sets $A$ and $B$, let $\Gamma$ be the set of causal stream functions $A^\omega \to B^\omega$, and let $v\colon \Gamma \to (\Gamma \times B)^A$ be the function defined as $f \mapsto a \mapsto \langle f_a, f[a] \rangle$. Then $(\Gamma, v)$ is a $(A, B)$-Mealy coalgebra.

The above example lays the groundwork to establish connections between circuits, stream functions and Mealy machines. If we inspect it a little closer, we find that stream functions are even more special than just being 'an' $(A, B)$-Mealy coalgebra.

**Definition 4.55** (Mealy homomorphism)**.** For sets $A$ and $B$, a *Mealy homomorphism* between two $(A, B)$-Mealy coalgebra $(S, f)$ and $(T, g)$ is a function $h\colon S \to T$ preserving transitions and outputs, i.e. if $f(s, a) = (r, b)$, then $g(h(s), a) = (h(r), b)$.

The *final* $(A, B)$-Mealy coalgebra is a $(A, B)$-Mealy coalgebra to which every other $(A, B)$-Mealy coalgebra has a unique homomorphism.

**Proposition 4.56** ([Rut06], Prop. 2.2)**.** For every $(A, B)$-Mealy coalgebra $(S, f)$, there exists a unique $(A, B)$-Mealy homomorphism $!\colon (S, f) \to (\Gamma, v)$.

*Proof.* An $(A, B)$-Mealy homomorphism $g\colon (S, f) \to (\Gamma, v)$ is a function $S \to \Gamma$, so for a state $s_0 \in S$, $g(s)$ will be a stream function $A^\omega \to B^\omega$. Let $\sigma \in A^\omega$ be an input stream; there is a (unique) series of transitions

$$s_0 \xrightarrow{\sigma(0)\,|\,b_0} s_1 \xrightarrow{\sigma(1)\,|\,b_1} s_2 \xrightarrow{\sigma(2)\,|\,b_2} s_3 \xrightarrow{\sigma(3)\,|\,b_3} \cdots$$

Then $!(s)$ is defined for input $\sigma$ and index $i \in \mathbb{N}$ as $!(s)(\sigma)(i) \coloneqq b_i$. □

For a Mealy coalgebra $(S, f)$ and a start state $s_0$, $!(s_0)(\sigma)$ maps to the stream of outputs that $(S, f)$ would produce by applying $f$ to each element of $\sigma$, starting from $s_0$.

### 4.2.2 Mealy machines on posets

To use Mealy machines as a bridge between $\mathbf{SCirc}_\Sigma$ and $\mathbf{Stream}_\mathcal{I}$ they must be assembled into another traced PROP. Not all Mealy machines defined so far correspond to circuits in $\mathbf{SCirc}_\Sigma$; we must refine our notion of Mealy machine in order to find those that do: those that map to stream functions in $\mathbf{Stream}_\mathcal{I}$.

**Lemma 4.57.** For a Mealy machine $(S, f)$ and state $s_0 \in S$, $!(s_0)$ is finitely specified.

*Proof.* $S$ is finite, and $!(-)$ must preserve transitions. □

We must also impose a monotonicity condition.

**Definition 4.58** (Monotone Mealy machine)**.** Let $A$, $B$ be posets; an $(A, B)$-Mealy machine $(S, f)$ is called a *monotone* Mealy machine if $S$ is also a poset and $f$ is $\bot$-preserving monotone with respect to the ordering on $A$, $B$, and $S$.

To map to Mealy machines from circuits we need to assemble the former into another PROP, in which the morphisms $m \to n$ are $(\mathbf{V}^m, \mathbf{V}^n)$-Mealy machines; we must also take into account the 'initial state' of circuits in $\mathbf{SCirc}_\Sigma$.

**Definition 4.59** (Initialised Mealy machine)**.** An *initialised* Mealy machine is a tuple $(S, f, s_0)$, where $(S, f)$ is a Mealy machine, and $s_0 \in S$ is an *initial state*.

**Example 4.60.** We can initialise the $(\mathbf{B}, \mathbf{B})$-Mealy machine $(\{s_0, s_1\}, f)$ from Example 4.51 in two ways; here we will choose to initialise it as $(S, f, s_0)$. In the diagrams, we label the initial state with an arrow.

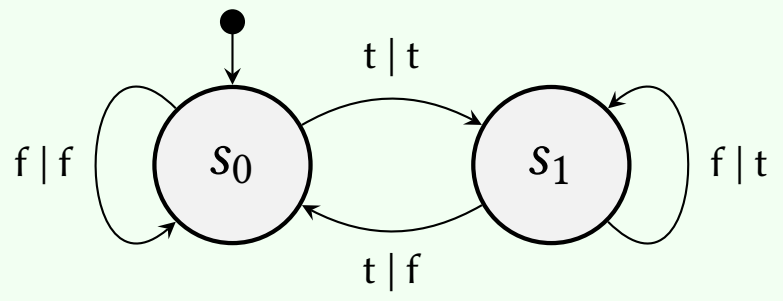

All that remains to define is the composition of Mealy machines, which is standard.

**Definition 4.61** (Cascade product of Mealy machines [Gin14]). Given an initialised $(A, B)$-Mealy machine $(S, f, s_0)$ and an initialised $(B, C)$-Mealy machine $(T, g, t_0)$, their *cascade product* is an initialised $(A, C)$-Mealy machine defined as

$$(S \times T, ((s, t), a) \mapsto \langle (f_0(s, a), g_0(t, f_1(s, a))), g_1(t, f_1(s, a)) \rangle, (s_0, t_0)) .$$

The cascade product of two Mealy machines effectively executes the first on the inputs, then executes the second on the outputs of the first.

**Example 4.62.** Recall the initialised $(\mathbb{B}, \mathbb{B})$-Mealy machine $(S, f, s_0)$ from Example 4.60; we will now compose this with $(\{t_0, t_1\}, g, t_0)$ where $g$ is defined as follows:

$$g(t_0, \mathsf{f}) \coloneqq \langle t_0, \mathsf{f} \rangle \qquad g(t_0, \mathsf{t}) \coloneqq \langle t_1, \mathsf{t} \rangle \qquad g(t_1, \mathsf{f}) \coloneqq \langle t_1, \mathsf{t} \rangle \qquad g(t_1, \mathsf{t}) \coloneqq \langle t_1, \mathsf{t} \rangle$$

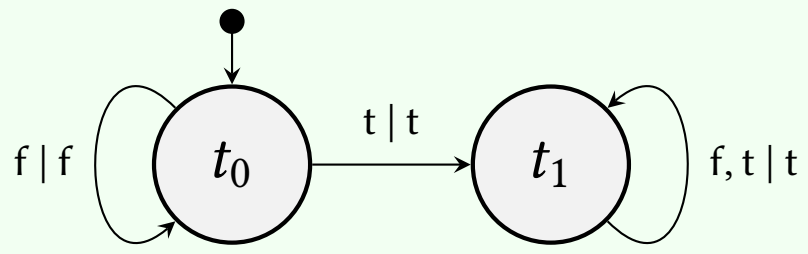

The cascade product $(R, h, r_0)$ of these two machines defined as follows:

$$\begin{aligned}
R &\coloneqq \{(s_0, t_0), (s_1, t_0), (s_0, t_1), (s_1, t_1)\} & r_0 &\coloneqq (s_0, t_0) \\
h((s_0, t_0), \mathsf{f}) &\coloneqq \langle (s_0, t_0), \mathsf{f} \rangle & h((s_0, t_0), \mathsf{t}) &\coloneqq \langle (s_0, t_1), \mathsf{t} \rangle \\
h((s_1, t_0), \mathsf{f}) &\coloneqq \langle (s_1, t_1), \mathsf{t} \rangle & h((s_1, t_0), \mathsf{t}) &\coloneqq \langle (s_0, t_0), \mathsf{f} \rangle \\
h((s_0, t_1), \mathsf{f}) &\coloneqq \langle (s_0, t_1), \mathsf{t} \rangle & h((s_0, t_1), \mathsf{t}) &\coloneqq \langle (s_1, t_1), \mathsf{t} \rangle \\
h((s_1, t_1), \mathsf{f}) &\coloneqq \langle (s_1, t_1), \mathsf{t} \rangle & h((s_1, t_1), \mathsf{t}) &\coloneqq \langle (s_0, t_1), \mathsf{t} \rangle
\end{aligned}$$

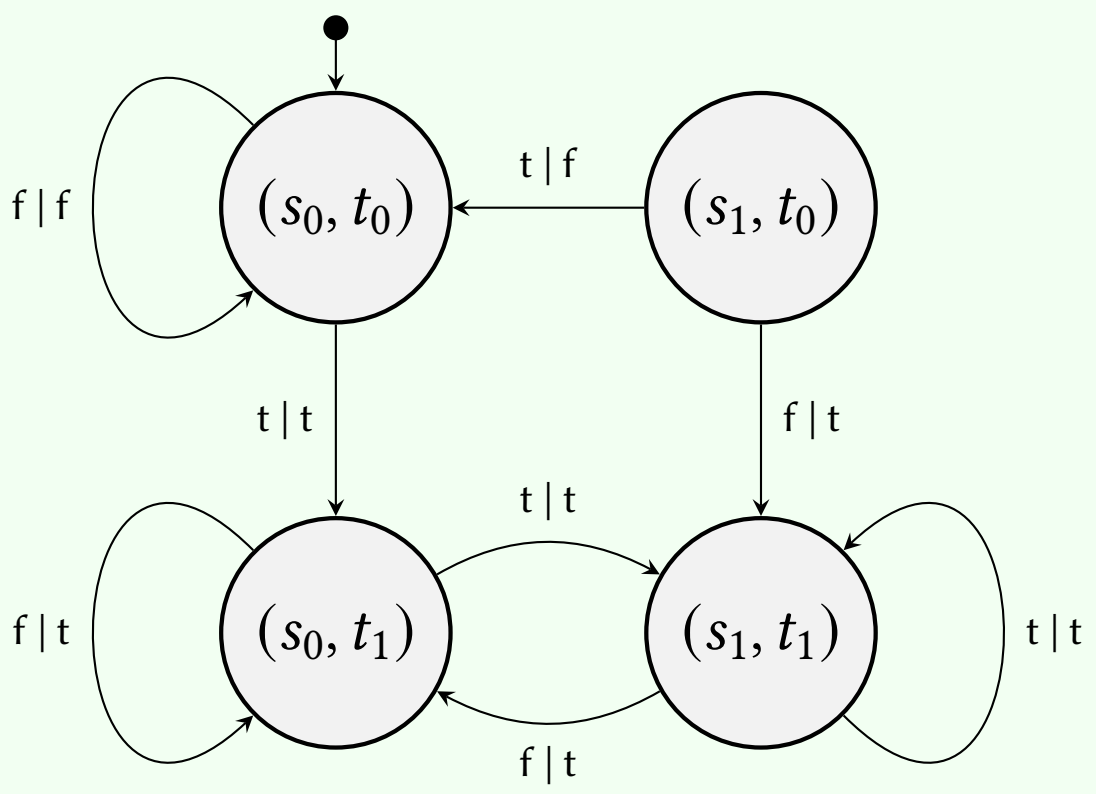

Tensor product is far more straightforward.

**Definition 4.63** (Direct product of Mealy machines)**.** Given an initialised $(A, B)$-Mealy machine $(S, f, s_0)$ and an initialised $(C, D)$-Mealy machine $(T, g, t_0)$, their *direct product* is an initialised $(A \times C, B \times D)$-Mealy machine defined as

$$(S \times T, ((s,t),(a,c)) \mapsto \langle (f_0(s,a), g_0(s,a)), (f_1(s,a), g_1(s,a)) \rangle, (s_0, t_0)).$$

**Example 4.64.** The direct product of the two initialised $(\mathbf{B}, \mathbf{B})$-Mealy machines introduced in Example 4.60 and Example 4.62 is a $(\mathbf{B}^2, \mathbf{B}^2)$-Mealy machine $(Q, k, q_0)$ defined as follows:

$$
\begin{aligned}
Q &\coloneqq \{(s_0,t_0),(s_1,t_0),(s_0,t_1),(s_1,t_1)\} & q_0 &\coloneqq (s_0,t_0)\\
h((s_0,t_0),\mathrm{ff}) &\coloneqq \langle (s_0,t_0),\mathrm{ff}\rangle & h((s_0,t_0),\mathrm{tf}) &\coloneqq \langle (s_1,t_0),\mathrm{tf}\rangle\\
h((s_0,t_0),\mathrm{ft}) &\coloneqq \langle (s_0,t_1),\mathrm{ft}\rangle & h((s_0,t_0),\mathrm{tt}) &\coloneqq \langle (s_1,t_1),\mathrm{tt}\rangle\\
h((s_1,t_0),\mathrm{ff}) &\coloneqq \langle (s_1,t_0),\mathrm{tf}\rangle & h((s_1,t_0),\mathrm{tf}) &\coloneqq \langle (s_0,t_0),\mathrm{ff}\rangle\\
h((s_1,t_0),\mathrm{ft}) &\coloneqq \langle (s_1,t_1),\mathrm{tt}\rangle & h((s_1,t_0),\mathrm{tt}) &\coloneqq \langle (s_0,t_1),\mathrm{ft}\rangle\\
h((s_0,t_1),\mathrm{ff}) &\coloneqq \langle (s_0,t_1),\mathrm{ft}\rangle & h((s_0,t_1),\mathrm{tf}) &\coloneqq \langle (s_1,t_1),\mathrm{tt}\rangle\\
h((s_1,t_1),\mathrm{ft}) &\coloneqq \langle (s_0,t_1),\mathrm{ft}\rangle & h((s_0,t_1),\mathrm{tt}) &\coloneqq \langle (s_1,t_1),\mathrm{tt}\rangle\\
h((s_1,t_1),\mathrm{ff}) &\coloneqq \langle (s_1,t_1),\mathrm{tt}\rangle & h((s_1,t_1),\mathrm{tf}) &\coloneqq \langle (s_0,t_1),\mathrm{ft}\rangle\\
h((s_1,t_1),\mathrm{ft}) &\coloneqq \langle (s_1,t_1),\mathrm{tt}\rangle & h((s_1,t_1),\mathrm{tt}) &\coloneqq \langle (s_0,t_1),\mathrm{ft}\rangle
\end{aligned}
$$

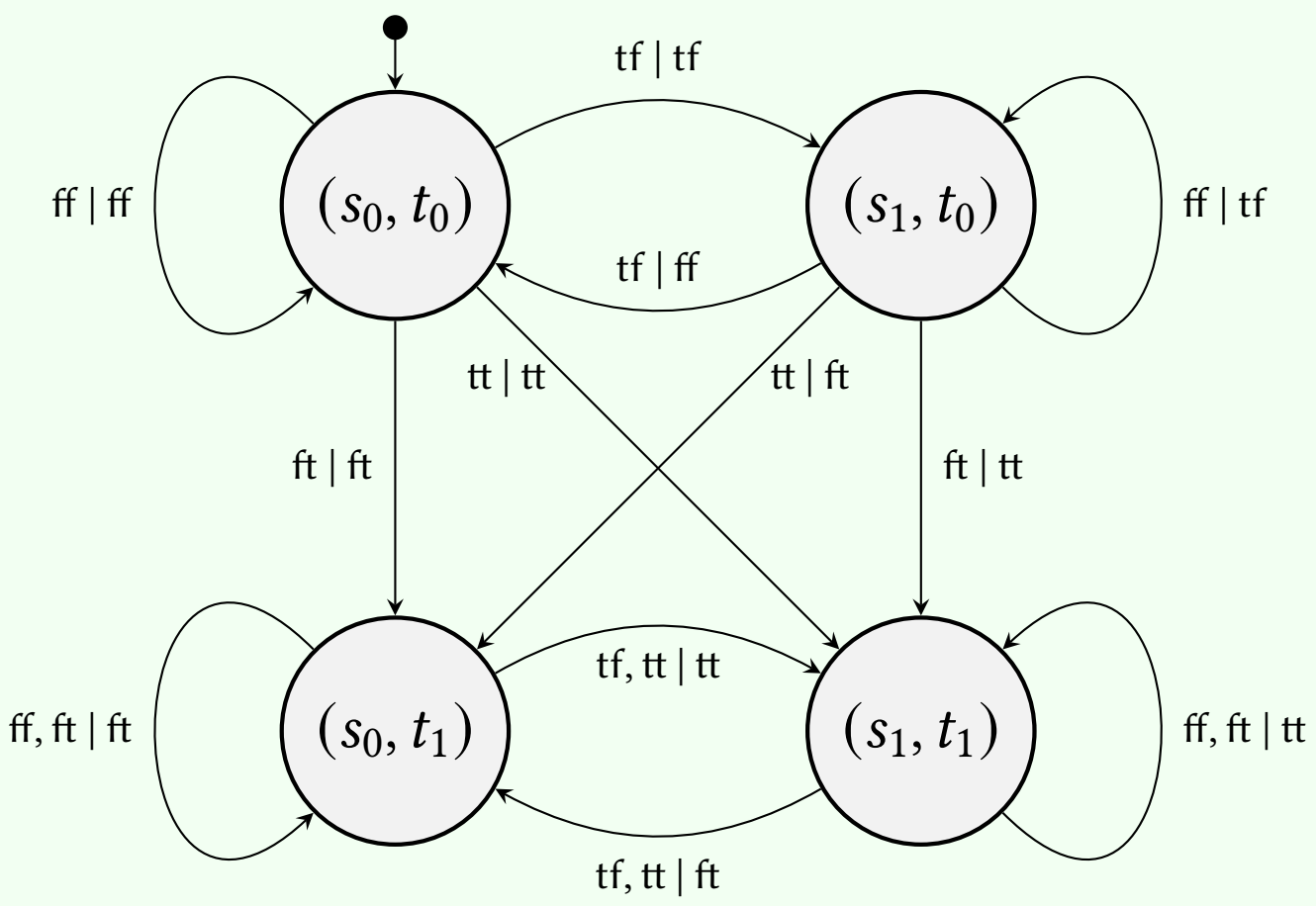

With cascade product as composition and direct product as tensor, initialised monotone Mealy machines form a PROP.

**Definition 4.65.** Let $\mathbf{Mealy}_{\mathcal{I}}$ be the PROP in which the morphisms $m \to n$ are the initialised monotone $(\mathbf{V}^m, \mathbf{V}^n)$-Mealy machines. Composition is by cascade product and tensor on morphisms is by direct product. The identity and the symmetry are the single-state machines that output the input and swap the inputs respectively.

Once again, we must show that this category has a trace. This can be computed in much the same way as it was for stream functions.

**Definition 4.66.** Let $(S, f)$ be a monotone $(\mathbf{V}^{x+m}, \mathbf{V}^{x+n})$-Mealy machine. For a state $s \in S$ and input $\overline{a} \in \mathbf{V}^m$, let $\mu_s(\overline{a})$ be the least fixed point of $\overline{r} \mapsto \pi_0(f(s, \overline{r} \mathbin{+\!\!+} \overline{a}))$. The *least fixed point* of an initialised Mealy machine $(S, f, s_0)$ is a $(\mathbf{V}^m, \mathbf{V}^n)$-Mealy machine $(S, (s, \overline{a}) \mapsto f(\mu_s \overline{a} \mathbin{+\!\!+} \overline{a}), s_0)$

**Example 4.67.** Consider the monotone $(\mathbf{V}^3, \mathbf{V}^3)$-Mealy machine with state set $\mathbf{V}_{\mathrm{B}}$, initial state $\bot$, and Mealy function

$$g \coloneqq (s, (x, y, z)) \mapsto \langle \neg y \wedge \neg x, (\neg s \wedge \neg z, s, \neg s \wedge \neg z) \rangle).$$

To take the trace of this machine, we must first compute the least fixed point of $v \mapsto \neg s \wedge \neg z$, which is clearly just $\neg s \wedge \neg z$. Therefore the Mealy function of the traced $(\mathbf{V}^2, \mathbf{V}^2)$ machine is $(s, (y, z)) \mapsto g(s, (\neg s \wedge \neg z, y, \neg s \wedge \neg z))$.

We must show that this is the trace on $\mathbf{Mealy}_{\mathcal{I}}$.

**Proposition 4.68.** The least fixed point is a trace on $\mathbf{Mealy}_{\mathcal{I}}$.

*Proof.* Let $(S, f)$ be a monotone $(\mathbf{V}^{x+m}, \mathbf{V}^{x+n})$-Mealy machine; this means that $S$ must be a poset. The Mealy function $f : S \times \mathbf{V}^{x+m} \to S \times \mathbf{V}^{x+n}$ is monotone with regards to the orders on $S$ and $\mathbf{V}^{x+m}$ and $S \times (x + n)$ is finite, so $f$ has a least fixed point. The function $f' \coloneqq (s, \overline{a}) \mapsto \pi_1(f(\mu_s(\overline{a}) +\!\!+ \overline{a}))$ is a composition of $\bot$-preserving monotone functions, so it is itself $\bot$-preserving monotone; this makes $(S, f')$ a monotone $(\mathbf{V}^m, \mathbf{V}^n)$-Mealy machine. This construction is a trace for the same reason as the trace of $\mathbf{Stream}_{\mathcal{I}}$ is. □

With monotone Mealy machines in a PROP, we can now represent the unique homomorphism from a Mealy machine to a set of state functions as a PROP morphism.

**Proposition 4.69.** There is a traced PROP morphism $!_{\mathcal{I}}(-) : \mathbf{Mealy}_{\mathcal{I}} \to \mathbf{Stream}_{\mathcal{I}}$ sending a monotone Mealy machine $(S, f, s_0) : m \to n$ to $!(s_0)$, where ! is the unique homomorphism $(S, f) \to (\Gamma, v)$.

*Proof.* Since every stream function also has a Mealy coalgebra structure and Mealy homomorphisms preserve transitions and outputs, composition of Mealy machines also coincides with composition of stream functions. □

### 4.2.3 Streams to Mealy machines

So far we have seen how a causal, finitely specified, and $\bot$-preserving monotone stream function can be retrieved from a monotone Mealy machine. It is also possible to retrieve a Mealy machine for a given stream function in $\mathbf{Stream}_{\mathcal{I}}$ by repeatedly taking stream derivatives; since we know there are finitely many we will be able to compute a finite set of states in a Mealy machine.

**Example 4.70.** Let $f : \mathbf{V}_{\mathrm{B}} \to \mathbf{V}_{\mathrm{B}}$ be a stream function defined as $f(\sigma)(0) = \sigma(0)$ and $f(\sigma)(k+1) = f(\sigma)(k) \wedge \sigma(k+1)$. We will derive a Mealy machine in $\mathbf{Mealy}_{\mathcal{I}}$ from this stream function. The complete set of states is $\{f, f_\bot, f_{\mathsf{f}}, f_\top\}$:

- $f_{\mathsf{t}} = f$;
- $(f_\bot)_\bot = (f_\bot)_{\mathsf{t}} = f_\bot$;
- $(f_\top)_\top = (f_\top)_{\mathsf{t}} = f_\top$; and
- $(f_\bot)_{\mathsf{f}} = (f_\bot)_\top = (f_\top)_\bot = (f_\top)_{\mathsf{f}} = f_{\mathsf{f}}$.

The Mealy function is defined for each state as the initial value and stream derivative of the original stream function. The initial state of the Mealy machine is $f$.

In fact, for a function $f$ this procedure produces a *minimal* Mealy machine.

**Corollary 4.71** (Corollary 2.3, [Rut06])**.** For a causal, finitely specified, stream function $f\colon M^\omega \to N^\omega$, let $S$ be the least set of causal stream functions including $f$ and closed under stream derivatives: i.e. for all $h \in S$ and $a \in M$, $h_a \in S$. Then the initialised Mealy machine $\langle\!\langle f \rangle\!\rangle_{\mathcal{I}} = (S, g, f)$, where $g(h)(a) = \langle h[a], h_a \rangle$, has the smallest state space of Mealy machines with the property $!_{\mathcal{I}} \langle\!\langle f \rangle\!\rangle_{\mathcal{I}} = f$.

*Proof.* Since $S$ is generated from the function $f$ and is the *smallest* possible set, there are no unreachable states in $S$. Since the output and transition of states in $\langle\!\langle f \rangle\!\rangle_{\mathcal{I}}$ are the initial output and stream derivative, two states can only have the same 'behaviour' if they are derived from the same original stream function. □

We will encode this data into a PROP morphism from $\mathbf{Stream}_{\mathcal{I}}$ to $\mathbf{Mealy}_{\mathcal{I}}$; in order to do this we must verify that the produced Mealy machine is monotone.

**Lemma 4.72.** The functions $\sigma \mapsto \mathsf{hd}(\sigma)$ and $\sigma \mapsto \mathsf{tl}(\sigma)$ are monotone.

*Proof.* Let $\sigma \coloneqq a :: \sigma'$ and $\tau \coloneqq b :: \tau'$ be streams in $A^\omega$ such that $\sigma \leq_{A^\omega} \tau$; subsequently $a \leq b$ and $\sigma' \leq_{A^\omega} \tau'$. So $\mathsf{hd}(\sigma) \coloneqq \mathsf{hd}(a :: \sigma') = a \leq_A b = \mathsf{hd}(b :: \tau') \coloneqq \mathsf{hd}(\tau)$ and $\mathsf{tl}(\sigma) \coloneqq \mathsf{tl}(a :: \sigma') = \sigma' \leq_{A^\omega} \tau' = \mathsf{tl}(b :: \tau') = \mathsf{tl}(\tau)$. □

**Lemma 4.73.** For posets $A$ and $B$ and a monotone causal stream function $f\colon A^\omega \to B^\omega$, the functions $a \mapsto f[a]$ and $a \mapsto f_a$ are monotone.

*Proof.* Let $a, b \in A$ such that $a \leq_A b$; then by monotonicity $f(a :: \sigma) \leq_{B^\omega} f(b :: \sigma)$ for all $\sigma \in A^\omega$. By Lemma 4.72, $\mathsf{hd} \circ f$ and $\mathsf{tl} \circ f$ are monotone. First we show that the initial output is monotone: $f[a] \coloneqq \mathsf{hd}(f(a :: \sigma)) \leq_A \mathsf{hd}(f(b :: \sigma)) = f[b]$. For the stream derivative, $f_a(\sigma) \coloneqq \mathsf{tl}(f(a :: \sigma)) \leq_{B^\omega} \mathsf{tl}(f(b :: \sigma)) \coloneqq f_a(\sigma)$. □

**Lemma 4.74.** Given $f \in \mathbf{Stream}_{\mathcal{I}}$, $\langle\!\langle f \rangle\!\rangle_{\mathcal{I}}$ is a monotone Mealy machine.

*Proof.* Each state in the derived Mealy machine is a monotone stream function, so this is a poset ordered by $\preceq$ as defined in Definition 4.46. and The Mealy function is the pairing of the initial output and stream derivative; by Lemma 4.73 these are monotone so the Mealy function must also be monotone. □

**Corollary 4.75.** The procedure $\langle\!\langle - \rangle\!\rangle_{\mathcal{I}}$ is a PROP morphism $\mathbf{Stream}_{\mathcal{I}} \to \mathbf{Mealy}_{\mathcal{I}}$.

This means we can map between monotone Mealy machines and causal, finitely specified, monotone stream functions in either direction. Mealy machines are perhaps more intuitive to work with, but stream functions are the 'purest' specification of the behaviour in that they have the smallest possible state set. Ideally we would be able to work in whichever setting is most beneficial at a given time, so we need to show that the translations are *behaviour-preserving*.

**Proposition 4.76.** $!_{\mathcal{I}}(-) \circ \langle\!\langle - \rangle\!\rangle_{\mathcal{I}} = \mathrm{id}_{\mathbf{Stream}_{\mathcal{I}}}$.

*Proof.* Stream functions are equal if they have the same initial output and stream derivative. $\langle\!\langle - \rangle\!\rangle_{\mathcal{I}}$ preserves outputs and derivatives by definition, and $!_{\mathcal{I}}(-)$ preserves transitions and outputs because it is a Mealy homomorphism. □

The reverse does not hold as many Mealy machines may map to the same stream function, but the result of $!_{\mathcal{I}}(-) \circ \langle\!\langle - \rangle\!\rangle_{\mathcal{I}} \circ !_{\mathcal{I}}(-)$ is of course equal to $!_{\mathcal{I}}(-)$.

## 4.3 Between circuits and Mealy machines

The close links between $\mathbf{Stream}_{\mathcal{I}}$ and $\mathbf{Mealy}_{\mathcal{I}}$ are nice to have but hardly groundbreaking; the main contribution of this chapter is to introduce $\mathbf{SCirc}_{\Sigma}$ to the mix by defining maps $\mathbf{SCirc}_{\Sigma} \to \mathbf{Mealy}_{\mathcal{I}}$ and $\mathbf{Mealy}_{\mathcal{I}} \to \mathbf{SCirc}_{\Sigma}$. This allows us to use monotone Mealy machines as a stepping stone in the correspondence between sequential circuits and stream functions. By exploiting the coalgebraic properties shared between Mealy machines and stream functions, this can be used to show that $\mathbf{Stream}_{\mathcal{I}}$ is both a *sound* and *complete* denotational semantics: there is a stream function in $\mathbf{Stream}_{\mathcal{I}}$ for every circuit in $\mathbf{SCirc}_{\Sigma}$, and there is a circuit in $\mathbf{SCirc}_{\Sigma}$ for every stream function in $\mathbf{Stream}_{\mathcal{I}}$.

### 4.3.1 Circuits to monotone Mealy machines

Circuits have a very natural interpretation as Mealy machines, so the action of a PROP morphism from $\mathbf{SCirc}_{\Sigma}$ to $\mathbf{Mealy}_{\mathcal{I}}$ is fairly intuitive.

**Definition 4.77.** Let $[-]_{\mathcal{I}} \colon \mathbf{SCirc}_{\Sigma} \to \mathbf{Mealy}_{\mathcal{I}}$ be the traced PROP morphism de-

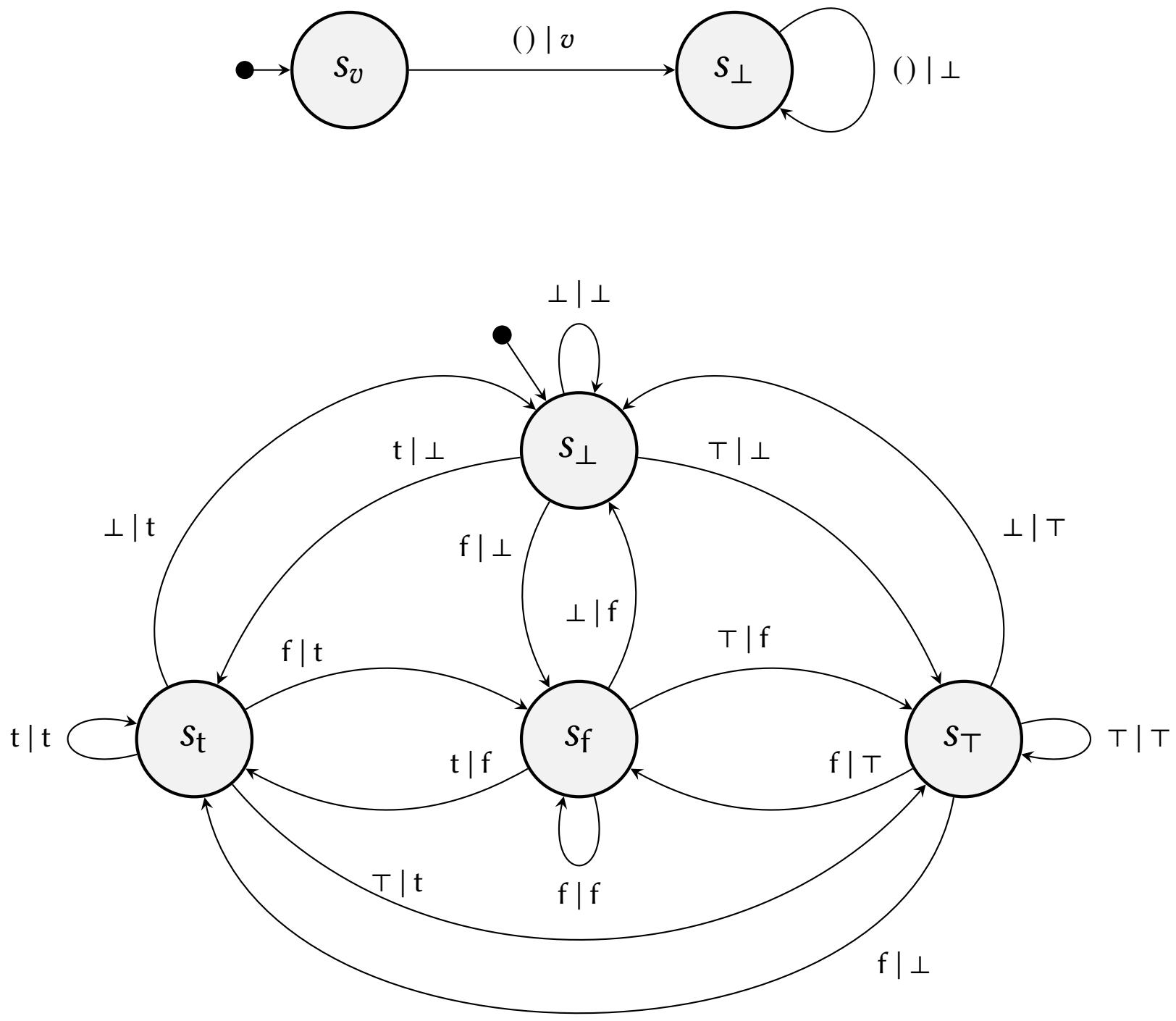

Figure 4.1: Mealy machines for Belnap values and delays

fined on generators as

$$\left[\,\boxed{g}\,\right]_{\mathcal{I}} \coloneqq (\{s\}, \quad (s, \overline{v}) \mapsto \langle s, [\![g]\!]\,(\overline{v})\rangle\,, \quad s)$$

$\left[\right.$ 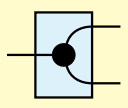 $\left.\right]_{\mathcal{I}} \coloneqq (\{s\}, \quad (s, v) \mapsto \langle s, (v, v)\rangle\,, \quad s)$

$\left[\right.$ 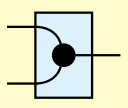 $\left.\right]_{\mathcal{I}} \coloneqq (\{s\}, \quad (s, (v, w)) \mapsto \langle s, (v \sqcup w)\rangle\,, \quad s)$

$\left[\right.$ 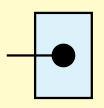 $\left.\right]_{\mathcal{I}} \coloneqq (\{s\}, \quad (s, v) \mapsto \langle s, ()\rangle\,, \quad s)$

$$\left[\,\boxed{v}\,\right]_{\mathcal{I}} \coloneqq (\{s_v, s_\bot\}, \quad \{s_v \mapsto \langle s_\bot, v\rangle\,, s_\bot \mapsto \langle s_\bot, \bot\rangle\}, \quad s_v)$$

$\left[\right.$ 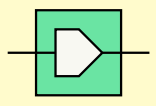 $\left.\right]_{\mathcal{I}} \coloneqq (\{s_v \mid v \in \mathbf{V}\}, \quad (s_v, a) \mapsto \langle v, s_a\rangle\,, \quad s_\bot)$

**Example 4.78.** The action of $[-]_{\mathcal{I}_B}$ on values and delays in $\mathbf{SCirc}_{\Sigma_B}$ is illustrated in Figure 4.1.

**Example 4.79.** Applying $[-]_{\mathcal{I}_B}$ to the SR NOR latch from Example 3.18 produces the monotone Mealy machine in Example 4.67, which is illustrated in Figure 4.2.

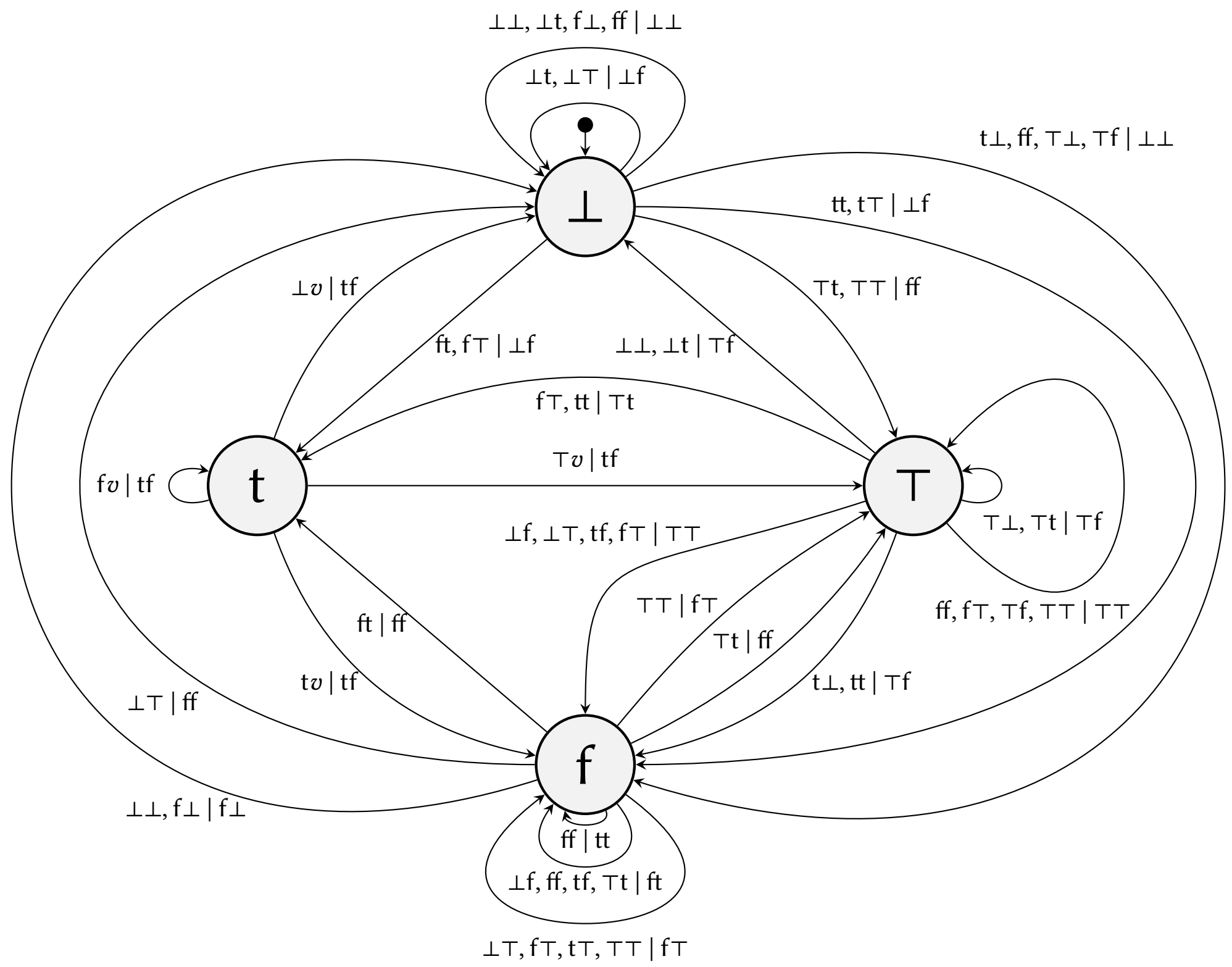

Figure 4.2: Mealy machine for the SR NOR latch

Mealy machines are a reasonable semantics for sequential circuits, but the image of $[-]_{\mathcal{I}}$ does not always lead to minimal Mealy machines, and there are many Mealy machines that may correspond to the same behaviour. The 'purest' semantics of a sequential circuit is a stream function in $\mathbf{Stream}_{\mathcal{I}}$.

**Definition 4.80.** Let $[\![-]\!]^{\mathsf{S}}_{\mathcal{I}} \colon \mathbf{SCirc}_{\Sigma} \to \mathbf{Stream}_{\mathcal{I}}$ be defined as $!_{\mathcal{I}}(-) \circ [-]_{\mathcal{I}}$.

We have now finally established the *denotation* of a sequential circuit $m$–$\boxed{f}$–$n$: it is the stream function $[\![ -\boxed{f}- ]\!]^{\mathsf{S}}_{\mathcal{I}} \colon (\mathbf{V}^m)^{\omega} \to (\mathbf{V}^n)^{\omega}$. The existence of the PROP morphism $[\![-]\!]^{\mathsf{S}}_{\mathcal{I}}$ confirms that causal, finitely specified and $\bot$-preserving monotone stream functions are a *sound* denotational semantics for sequential circuits, as every circuit in $\mathbf{SCirc}_{\Sigma}$ has a corresponding stream function in $\mathbf{Stream}_{\mathcal{I}}$.

It is useful to verify that this denotational semantics of sequential circuits agrees with the denotational semantics we defined earlier for *combinational* circuits in Section 4.1.1.

**Lemma 4.81.** Let $m$–$\boxed{f}$–$n$ be a combinational circuit; for $\sigma \in (\mathbf{V}^m)^{\omega}$ and $i \in \mathbb{N}$, $[\![ -\boxed{f}- ]\!]^{\mathsf{S}}_{\mathcal{I}}(\sigma)(i) = [\![ -\boxed{f}- ]\!]^{\mathsf{C}}_{\mathcal{I}}(\sigma(i))$.

*Proof.* Since $m\text{-}\boxed{f}\text{-}n$ is combinational, $\left[ m\text{-}\boxed{f}\text{-}n \right]_{\mathcal{I}}$ is a Mealy machine with a single state $s$, i.e. there is a function $g\colon \mathbf{V}^m \to \mathbf{V}^n$ such that $\left[ m\text{-}\boxed{f}\text{-}n \right]_{\mathcal{I}} = (\{s\}, (s,\bar{v}) \mapsto \langle s, g(\bar{v})\rangle, s)$. By definition of $!_{\mathcal{I}}(-)$, we have that $\left\| \left[ m\text{-}\boxed{f}\text{-}n \right]_{\mathcal{I}} \right\|^{\leq}_{\mathcal{I}}(\sigma)(i) = g(\sigma(i))$. To complete the proof, we need to show that $g(\sigma)(i) = \left[\!\left[ \text{-}\boxed{f}\text{-} \right]\!\right]^{\mathsf{C}}_{\mathcal{I}}(\sigma(i))$; this holds because $[-]_{\mathcal{I}}$ and $[\![-]\!]^{\mathsf{C}}_{\mathcal{I}}$ freely build functions using the same constructs. □

Using this idea, it will be convenient to have a mapping from functions to these constant stream functions.

**Definition 4.82.** Let $\ulcorner - \urcorner_{\mathcal{I}}\colon \mathbf{Func}_{\mathcal{I}} \to \mathbf{Stream}_{\mathcal{I}}$ be defined as the PROP morphism with action $\ulcorner f \urcorner_{\mathcal{I}} \coloneqq \sigma \mapsto i \mapsto f(\sigma)(i)$

### 4.3.2 Monotone Mealy machines to circuits

We now need a way to retrieve a circuit morphism in $\mathbf{SCirc}_{\Sigma}$ from a stream function $f \in \mathbf{Stream}_{\mathcal{I}}$. To prevent us from picking an arbitrary circuit, the denotation of the circuit must also be $f$.

We already know by Corollary 4.71 that given a stream function $f$ we can retrieve a monotone Mealy machine $\langle\!\langle f \rangle\!\rangle_{\mathcal{I}}$. All that remains is to translate this into a circuit morphism. For regular Mealy machines, there is a standard procedure in circuit design [KJ09] in which each state of a Mealy machine is *encoded* as a power of values, and the Mealy function is interpreted as a circuit using combinational logic.

**Example 4.83.** Consider the following Mealy machine operating on Boolean values.

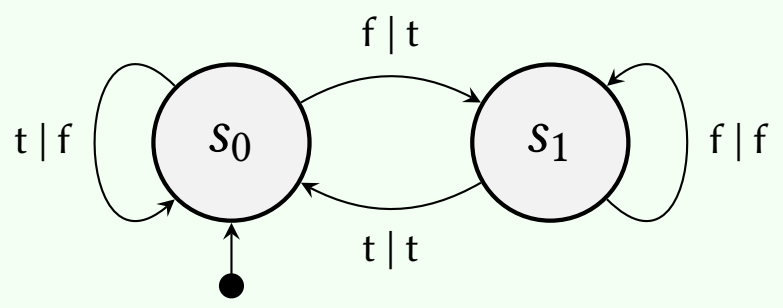

To convert this machine to a circuit, we assign each state a boolean value: in this case $s_0 \mapsto \mathsf{f}, s_1 \mapsto \mathsf{t}$. We can now construct a truth table to show how a state and an input map to a transition and an output:

| | | | |
|---|---|---|---|
| f | f | t | t |
| f | t | f | f |
| t | f | t | f |
| t | t | f | t |

It is possible to describe these truth tables as logical expressions: in this case the expression for the next state is $(v_0, v_1) \mapsto (\neg v_0 \wedge \neg v_1) \vee (v_0 \wedge \neg v_1)$ and the expression for the output is $(v_0, v_1) \mapsto (\neg v_0 \wedge \neg v_1) \vee (v_0 \wedge v_1)$. These expressions can clearly be constructed as combinational circuits using AND, OR and NOT gates; the entire circuit corresponding to the Mealy machine is constructed by combining the combinational logic with registers to hold the state.

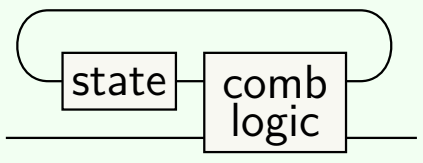

We will use a variation of this procedure to map from $\mathbf{Mealy}_{\mathcal{I}}$ to $\mathbf{SCirc}_{\Sigma}$. However, when considering *monotone* Mealy machines, this procedure must additionally respect monotonicity as the combinational logic is constructed using monotone components. This means that an arbitrary encoding cannot be used; we will now show how to select something suitable.

**Definition 4.84** (Encoding)**.** Let $S$ be a set equipped with a partial order $\preceq$ and a total order $\leq$ such that $S$ can be represented as $s_0 \leq s_1 \leq \ldots s_{k-1}$. The $\leq$-*encoding* for this assignment is a function $\gamma_{\leq} \colon S \to \mathbf{V}^k$ defined as $\gamma_{\leq}(s)(i) \coloneqq \top$ if $s_i \preceq s$ and $\gamma_{\leq}(s)(i) \coloneqq \bot$ otherwise.

**Example 4.85.** Recall the monotone Mealy machine from Example 4.79, which has state set $\mathbf{V}_{\mathrm{B}} \coloneqq \{\bot, \mathrm{f}, \mathrm{t}, \top\}$. We choose the total order on $\mathbf{V}_{\mathrm{B}}$ as $\bot \leq \mathrm{f} \leq \mathrm{t} \leq \top$; subsequently, the $\leq$-encoding is defined as $\bot \mapsto \top\bot\bot\bot, \mathrm{f} \mapsto \top\top\bot\bot, \mathrm{t} \mapsto \top\bot\top\bot, \top \mapsto \top\top\top\top$.

It is essential that a $\leq$-encoding respects the original ordering of the states.

**Lemma 4.86.** For an ordered state space $(S, \preceq)$ and a $\leq$-encoding $\gamma_{\leq}$, $s \preceq s'$ if and only if $\gamma_{\leq}(s) \sqsubseteq \gamma_{\leq}(s')$.

*Proof.* First the $(\Rightarrow)$ direction. Let $s_i, s_j \in S$ such that $s_i \preceq s_j$; we need to show that for every $l < |S|$, $\gamma_{\leq}(s_i)(l) \sqsubseteq \gamma_{\leq}(s_j)(l)$. The only way this can be violated is if $s_i(l) = \top$ and $s_j(l) = \bot$; this can only occur if $s_l \preceq s_i$ and $s_l \npreceq s_j$. But since $s_i \preceq s_j$, this is a contradiction due to transitivity so $\gamma_{\leq}(s_l) \sqsubseteq \gamma_{\leq}(s_j)$ also holds.

Now the $(\Leftarrow)$ direction. Assume that $\gamma_{\leq}(s_i) \sqsubseteq \gamma_{\leq}(s_j)$; we need to show that $s_i \preceq s_j$; i.e. that $\gamma_{\leq}(s_j)(i) = \top$ If $\gamma_{\leq}(s_i) \sqsubseteq \gamma_{\leq}(s_j)$, then for each $l < k$ then $\gamma_{\leq}(s_i)(l) \sqsubseteq \gamma_{\leq}(s_j)(l)$; in particular $\gamma_{\leq}(s_i)(i) \sqsubseteq \gamma_{\leq}(s_j)(i)$ By definition of $\gamma_{\leq}$, $\gamma_{\leq}(s_i)(i) = \top$, so if $\gamma_{\leq}(s_i) \sqsubseteq \gamma_{\leq}(s_j)$ then $\gamma_{\leq}(s_j)(i)$ is also $\top$. □

Using this encoding, we will construct a circuit morphism that, when interpreted as a function, implements the output and transition function of the Mealy machine. There is no reason for such a morphism to exist for an arbitrary interpretation: why should we expect some collection of gates to be able to model every function? The useful interpretations are those that *can* model every function.

**Definition 4.87** (Functional completeness). A *complete interpretation* is a tuple $(\mathcal{I}, ||-||)$ in which $\mathcal{I}$ is an interpretation of a signature $\Sigma$ and $||-|| \colon \mathbf{Func}_{\mathcal{I}} \to \mathbf{SCirc}_{\Sigma}$ is a map that sends functions $f \colon \mathbf{V}^m \to \mathbf{V}^n$, to circuits of the form 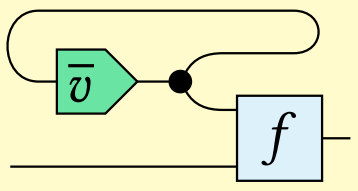

for some word $\overline{v} \in \mathbf{V}^\star$ such that $[\![||f||]\!]^{\mathsf{S}}_{\mathcal{I}}(\sigma)(i) = f(\sigma(i))$.

For a given complete interpretation $(\mathcal{I}, ||-||)$, we refer to a circuit $||f||$ as the *normalised circuit for* $f$.

**Remark 4.88.** Even though $||-||$ maps combinational functions, its codomain is the category of *sequential* circuits $\mathbf{SCirc}_{\Sigma}$. This is because instantaneous values may be required to create the normalised circuit. Despite the use of sequential components, the loop enforces that the state is *constant*: it will always produce the word $\overline{v}$, so the the circuit still has combinational behaviour.

Sometimes this is the only way to ensure every function can be modelled. For example, consider the Boolean function $\mathbf{B} \to \mathbf{B}$ that always produces $\mathsf{f}$. Using the strategy from Example 4.83, no lines of the truth table are true, so the expression can only be defined using the unit of the disjunction, the constant false.

Note also that this sequential component is by no means mandatory: the functional completeness map may actually map only to combinational circuits, in which case the width of the sequential component would be $0$.

**Example 4.89.** The Belnap interpretation from Example 4.17 is functionally complete; for interests of space we postpone the proof to Section 4.5. This is due to a variation of the standard functional completeness method for Boolean values.

With the knowledge that any monotone function has a corresponding circuit in $\mathbf{SCirc}_{\Sigma}$, we set about encoding an arbitrary Mealy function $S \times \mathbf{V}^m \to S \times \mathbf{V}^n$ into a function $\mathbf{V}^k \times \mathbf{V}^n \to \mathbf{V}^k \times \mathbf{V}^n$. One point to note here is that there may be more values in $\mathbf{V}^k$ than there are states in $S$, so we may need to 'fill in the gaps' in a way that is compatible with monotonicity.

**Definition 4.90** (Monotone completion)**.** Let $A$ be a finite poset and let $B$ be a finite lattice such that $A \subseteq B$. Then for another finite lattice $C$ and a monotone function $f : A \to C$, let the *monotone $B$-completion* of $f$ be the function $f_{\mathsf{m}} : B \to C$ recursively defined as

$$f_{\mathsf{m}}(v) = \begin{cases} f(v) & \text{if } v \in A \\ \bot_C & \text{if } v = \bot_B, \bot \notin A \\ \bigvee\{f_{\mathsf{m}}(w) \mid w \leq_B v, w \neq v\} & \text{otherwise} \end{cases}$$

**Example 4.91.** For $n \in \mathbb{N}$, let $\mathbb{N}_n$ be the subset of the natural numbers containing the numbers $0, 1, \ldots, n-1$ with the usual order. Let $f : \{2, 4\} \to \mathbb{N}$ be defined as $2 \mapsto 6$ and $4 \mapsto 7$. The monotone $\mathbb{N}_5$-completion of $f$ is a function $f_{\mathsf{m}} : \mathbb{N}_5 \to \mathbb{N}$, defined as follows: $f_{\mathsf{m}}(0) = 0$ as $0$ is the least element of $\mathbb{N}_5$; $f_{\mathsf{m}}(1) = 0$ as $0 \leq 1$ and $g_{\mathsf{m}}(1) = 0$; $f_{\mathsf{m}}(2) = 6$ as $2 \in \{2, 4\}$ and $g(2) = 6$; $f_{\mathsf{m}}(3) = 6$ because $f_{\mathsf{m}}(3) = f_{\mathsf{m}}(0) \vee f_{\mathsf{m}}(1) \vee f_{\mathsf{m}}(2) = 0 \vee 0 \vee 6 = 6$; and $f_{\mathsf{m}}(4) = 7$ because $f(4) = 7$.

A Mealy function can now be encoded over powers of $\mathbf{V}$ by using the monotone completion of some encoding function.

**Definition 4.92** (Monotone Mealy encoding)**.** For a monotone Mealy machine $(S, f, s_0)$ with $k$ states and an encoding $\gamma_\leq : S \to \mathbf{V}^k$, let $\gamma_\leq^p : \gamma_\leq[S] \times \mathbf{V}^m \to \mathbf{V}^k \times \mathbf{V}^n$ be defined as the function $(\gamma_\leq(s), \overline{x}) \mapsto (\gamma_\leq(f_0(s, \overline{x})), f_1(s, \overline{x}))$. The *monotone Mealy encoding* of $(S, f, s_0)$ is a function $\gamma_\leq(f) : \mathbf{V}^k \times \mathbf{V}^m \to \mathbf{V}^k \times \mathbf{V}^n$ defined as the $(\mathbf{V}^k \times \mathbf{V}^m)$-completion of $\gamma_\leq^p$.

To obtain the syntactic circuit for a monotone Mealy function encoded in this way, it needs to be a morphism in $\mathbf{Func}_{\mathcal{I}}$. It is monotone by definition, but we need to make sure it is also $\bot$-preserving.

**Lemma 4.93.** A monotone Mealy encoding is in $\mathbf{Func}_{\mathcal{I}}$.

*Proof.* A Mealy encoding is monotone as it is a monotone completion. There cannot be a state $\bot^k$ since at least one bit must be $\top$; this means the monotone completion will send the input $(\bot^k, \bot^m)$ to $(\bot^k, \bot^n)$: it is $\bot$-preserving. □

The foundations are now set for establishing the image of a PROP morphism from Mealy machines to circuit terms. There is one more thing to consider: Definition 4.84 depends on some arbitrary total ordering on the states in a given monotone Mealy machine. While this may not seem much of an issue, when defining a PROP morphism this must be *fixed*, otherwise the mapping of Mealy machines to circuits might be nondeterministic.

**Definition 4.94** (Chosen state order)**.** Let $(S, f, s_0)$ be a monotone Mealy machine with input space $\mathbf{V}^m$, and let $\leq$ be a total order on $\mathbf{V}$; $\leq$ can be extended to a total order $\leq_\star$ on $(\mathbf{V}^m)^\star$ using the lexicographic order. Let $S'$ be the set of accessible states of $S$. For each state $s \in S'$, let $t_{s,\leq} \in (\mathbf{V}^m)^\star$ be the minimal element of the subset of words that transition from $s_0$ to $s$, ordered by $\leq$. Then the *chosen state order* $\leq_{S'}$ is a total order on $S'$ defined as $s \leq_{S'} s'$ if $t_{s,\leq} \leq_\star t_{s',\leq}$.

The PROP morphism from monotone Mealy machines to circuits can then be parameterised by some ordering on the set of values $\mathbf{V}$, ensuring that there is a canonical term in $\mathbf{SCirc}_\Sigma$ for each monotone Mealy machine.

**Definition 4.95.** For a functionally complete interpretation $\mathcal{I}$ and total order $\leq$ on $\mathbf{V}$, let $||-||^{\leq}_{\mathcal{I}} \colon \mathbf{Mealy}_\mathcal{I} \to \mathbf{SCirc}_\Sigma$ be the traced PROP morphism with action defined for a monotone Mealy machine $(S, f, s)$ as producing 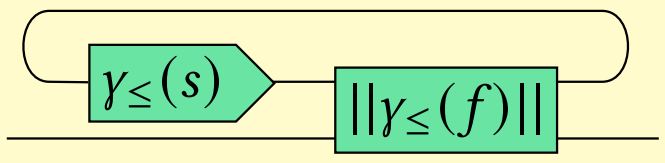

.

Before proceeding to the result that this PROP morphism is behaviour-preserving, we must show a lemma linking the behaviour of circuits in the image of $||-||^{\leq}_{\mathcal{I}}$ to initial outputs and stream derivatives.

**Proposition 4.96.** For a combinational circuit 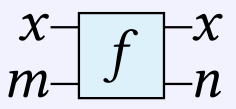

, let $\mathsf{mf}(f)$ be the map with action $(\bar{s}) \mapsto$ 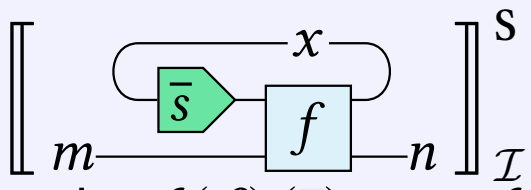

and let $g :=$ 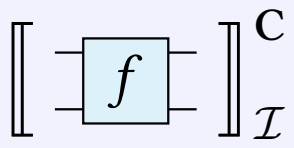

. Then, $\mathsf{mf}(f)(\bar{s})[\bar{a}] = \pi_1(g(\bar{s}, \bar{a}))$ and $\mathsf{mf}(f)(\bar{s})_{\bar{a}} = \mathsf{mf}(f)(\pi_0(g(\bar{s}, \bar{a})))$.

*Proof.* The machine 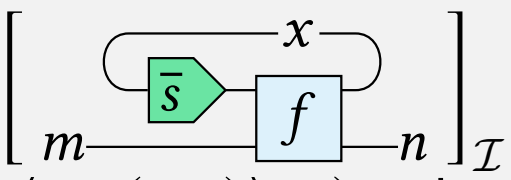

is computed as the fixed point of the machine $(\mathbf{V}^x, (\bar{r}, \bar{a}) \mapsto \langle \bar{r}, g(\bar{s}, \bar{a}) \rangle, \bar{s})$, which is $(\mathbf{V}^x, \bar{v} \mapsto \langle \pi_0(g(\bar{s}, \bar{a})), \pi_1(g(\bar{s}, \bar{a})) \rangle, \bar{s})$. The output and derivative of 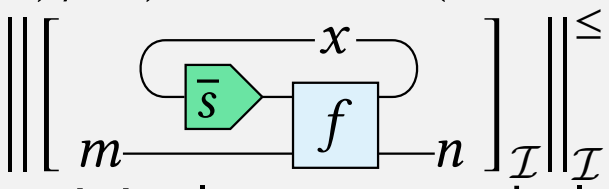

are the output and transition of the Mealy machine, so the original statement holds by Lemma 4.81. □

The goal of this section is to show that the translation from Mealy machines to circuits and back again using $[-]_\mathcal{I} \circ ||-||^{\leq}_{\mathcal{I}}$ is *behaviour-preserving*: while the mapping may not be the identity in $\mathbf{Mealy}_\mathcal{I}$, the stream functions obtained using $\langle\!\langle - \rangle\!\rangle_\mathcal{I}$ should be equal. This is an important new result, as it means that rather than showing results about the denotational semantics of circuits in $\mathbf{SCirc}_\Sigma$ by interpreting them in $\mathbf{Stream}_\mathcal{I}$, we can view morphisms of the former as Mealy machines instead.

**Theorem 4.97.** $!(-) = [\![-]\!]^{\mathsf{S}}_{\mathcal{I}} \circ ||-||^{\leq}_{\mathcal{I}}$.

*Proof.* Let $(S, f)$ be a monotone Mealy machine and let $s \in S$ be an arbitrary state. By Proposition 4.69, the initial output of $!_\mathcal{I}(S, f, s)$ is $\overline{a} \mapsto f_1(s, \overline{a})$ and the stream derivative of $!_\mathcal{I}(S, f, s)$ is $\overline{a} \mapsto !_\mathcal{I}(f_0(s, \overline{a}))$.

Now we consider the composite $[\![||(S, f, s)||^{\leq}_{\mathcal{I}}]\!]^{\mathsf{S}}_{\mathcal{I}}$. By Definition 4.95 we have that $||(S, f, s_0)||^{\leq}_{\mathcal{I}} =$ 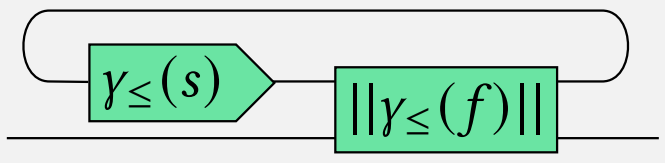 ; by applying $||-||^{\leq}_{\mathcal{I}}$ and $||-||$, there exists a combinational circuit 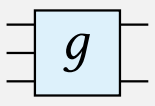 such that

$$||(S, f, s_0)||^{\leq}_{\mathcal{I}} =$$

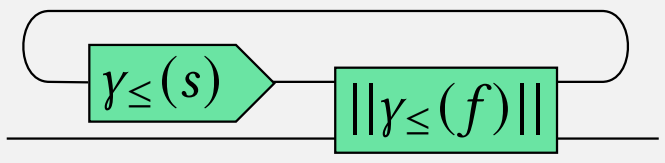

$$=$$

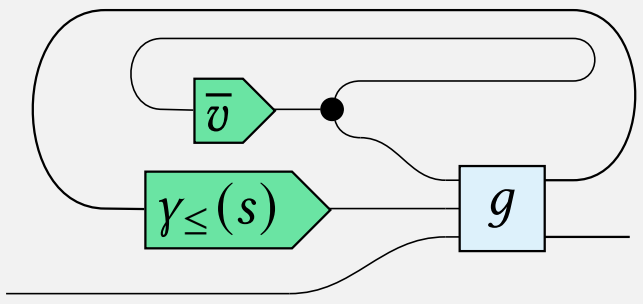

.

Let $g' = [\![$ 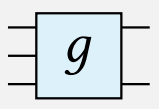 $]\!]^{\mathsf{C}}_{\mathcal{I}}$; note that for all $\overline{r} \in \mathbf{V}^x$ and $\overline{a} \in \mathbf{V}^m$, $g'(\overline{v}, \overline{r}, \overline{a}) = \gamma_\leq(f)(\overline{r}, \overline{a})$.

We can now use Proposition 4.96 to compute the initial output and stream derivative of $[\![||(S, f, s)||^{\leq}_{\mathcal{I}}]\!]^{\mathsf{S}}_{\mathcal{I}}$. To show that $!_\mathcal{I}(-) = [\![-]\!]^{\mathsf{S}}_{\mathcal{I}} \circ !_\mathcal{I}(-)$, we need to show that these 'agree' with those of $!_\mathcal{I}(S, f, s)$. For the initial output, this means we just need to show they are equal:

$$\begin{aligned}
[\![||(S, f, s)||^{\leq}_{\mathcal{I}}]\!]^{\mathsf{S}}_{\mathcal{I}}\,[\overline{a}] &= \pi_1\left([\![g]\!]^{\mathsf{C}}_{\mathcal{I}}\,(\overline{v}, \gamma_\leq(s), \overline{a})\right) \\
&= \pi_1\left([\![||\gamma_\leq(f)||]\!]^{\mathsf{C}}_{\mathcal{I}}\,(\gamma_\leq(s), \overline{a})\right) \\
&= \pi_1\left(\gamma_\leq(f)(\gamma_\leq(s), \overline{(a)})\right) \\
&= \pi_1\,(\gamma_\leq(f_0(s, \overline{a})), f_1(s, \overline{a})) \\
&= f_1(s, \overline{a})
\end{aligned}$$

For the stream derivative, we need to show that as states vary over $s \in S$, the stream derivative of $[\![ \,||(S, f, s)||^{\leq}_{\mathcal{I}}\, ]\!]^{\mathsf{S}}_{\mathcal{I}}$ is the $\gamma_{\leq}$-encoding of $!_{\mathcal{I}}(S, f, s)$.

$$\begin{aligned}
\left([\![ \,||(S, f, s)||^{\leq}_{\mathcal{I}}\, ]\!]^{\mathsf{S}}_{\mathcal{I}}\right)_{\overline{a}} &= \pi_0 \left([\![ g ]\!]^{\mathsf{C}}_{\mathcal{I}} \left(\overline{v}, \gamma_{\leq}(s), \overline{a}\right)\right) \\
&= \pi_0\left([\![ \,||\gamma_{\leq}(f)||\, ]\!]^{\mathsf{C}}_{\mathcal{I}} \left(\gamma_{\leq}(s), \overline{a}\right)\right) \\
&= \pi_0\left(\gamma_{\leq}(f)(\gamma_{\leq}(s), \overline{(a)})\right) \\
&= \pi_0 \left(\gamma_{\leq}(f_0(s, \overline{a})), f_1(s, \overline{a})\right) \\
&= \gamma_{\leq}(f_0(s, \overline{a}))
\end{aligned}$$

The initial outputs and stream derivatives agree, so $!(-) = [\![ - ]\!]^{\mathsf{S}}_{\mathcal{I}} \circ ||-||^{\leq}_{\mathcal{I}}$. □

On its own, this is a nice result to have; if we only know the specification of a circuit in terms of a (monotone) Mealy machine, we can use the PROP morphism $||-||^{\leq}_{\mathcal{I}}$ to generate a circuit in $\mathbf{SCirc}_{\Sigma}$ which has the same behaviour as a stream function. However, this is but one ingredient in our ultimate goal: the completeness of the denotational semantics.

## 4.4 Completeness of the denotational semantics

We want $\mathbf{Stream}_{\mathcal{I}}$ to be a *complete* denotational semantics for digital circuits. This means that for every stream function $f \in \mathbf{Stream}_{\mathcal{I}}$, there must be at least one one circuit in $\mathbf{SCirc}_{\Sigma}$ such that its behaviour under $\mathcal{I}$ is $f$.

**Corollary 4.98.** $[\![ - ]\!]^{\mathsf{S}}_{\mathcal{I}} \circ ||-||^{\leq}_{\mathcal{I}} \circ \langle\!\langle - \rangle\!\rangle_{\mathcal{I}} = \mathrm{id}_{\mathbf{Stream}_{\mathcal{I}}}$.

*Proof.* This follows immediately from Theorem 4.97 and Proposition 4.76, as we have that $[\![ - ]\!]^{\mathsf{S}}_{\mathcal{I}} \circ ||-||^{\leq}_{\mathcal{I}} \circ \langle\!\langle - \rangle\!\rangle_{\mathcal{I}} = !_{\mathcal{I}}(-) \circ \langle\!\langle - \rangle\!\rangle_{\mathcal{I}} = \mathrm{id}_{\mathbf{Stream}_{\mathcal{I}}}$. □

There is no isomorphism between $\mathbf{SCirc}_{\Sigma}$ and $\mathbf{Stream}_{\mathcal{I}}$ as many circuits may have the same semantics but different syntax. Instead, we can work in *equivalence classes* of syntactic circuits based on their behaviour in $\mathbf{Stream}_{\mathcal{I}}$.

**Definition 4.99** (Denotational equivalence)**.** Two sequential circuits are *denotationally equivalent* under $\mathcal{I}$, written $m$–$\boxed{f}$–$n$ $\approx_{\mathcal{I}}$ $m$–$\boxed{g}$–$n$ if $[\![ -\boxed{f}- ]\!]^{\mathsf{S}}_{\mathcal{I}} = [\![ -\boxed{g}- ]\!]^{\mathsf{S}}_{\mathcal{I}}$. Let $\mathbf{SCirc}_{\Sigma/\approx_{\mathcal{I}}}$ be the result of quotienting $\mathbf{SCirc}_{\Sigma}$ by $\approx_{\mathcal{I}}$.

Every morphism in $\mathbf{SCirc}_{\Sigma/\approx_{\mathcal{I}}}$ is a class of circuits that share the same behaviour under $\mathcal{I}$. As we have a PROP morphism $||-||^{\leq}_{\mathcal{I}} \circ \langle\langle - \rangle\rangle_{\mathcal{I}}$, we know that for every such behaviour there must be at least one such syntactic circuit, and subsequently exactly one equivalence class $\mathbf{SCirc}_{\Sigma/\approx_{\mathcal{I}}}$. Using Corollary 4.98, we know that all the circuits in this equivalence class have the same behaviour as the original stream function, so we can derive an isomorphism between $\mathbf{SCirc}_{\Sigma/\approx_{\mathcal{I}}}$ and $\mathbf{Stream}_{\mathcal{I}}$.

**Corollary 4.100.** $\mathbf{SCirc}_{\Sigma/\approx_{\mathcal{I}}} \cong \mathbf{Stream}_{\mathcal{I}}$.

This gives us, for the first time, a fully compositional, sound and complete, denotational semantics for sequential circuits with delay and (possibly non-delay-guarded) feedback. This formal model will serve as a backdrop against the operational and algebraic semantics presented in the upcoming chapters.

## 4.5 Denotational semantics for Belnap logic

For an interpretation to admit a sound and complete denotational semantics it needs to be *functionally complete*. One may wonder if this is a reasonable assumption to make, as if it is not then the denotational semantics is not particularly useful.

To this end, we will demonstrate how the functional completeness condition holds for the Belnap interpretation introduced in Example 4.17. This will make use of the well-known functional completeness of Boolean logic.

**Definition 4.101.** Let $\mathbf{B} \coloneqq \{0, 1\}$ be the set of *Boolean* values, and let $\wedge_{\mathbf{B}}, \vee_{\mathbf{B}}, \neg_{\mathbf{B}}$ be the usual operations on Booleans.

**Lemma 4.102.** All functions $\mathbf{B}^m \to \mathbf{B}$ can be expressed using the set of operations $\{0, \wedge_{\mathbf{B}}, \vee_{\mathbf{B}}, \neg_{\mathbf{B}}\}$.

*Proof.* Let $f : \mathbf{B}^m \to \mathbf{B}$ be a Boolean function: we need to create a Boolean expression using variables $v_0, v_1, \ldots, v_{m-1}$. For each $\overline{v} \in \mathbf{B}^m$, we construct a conjunction of all $m$ variables, in which $v_i$ is negated if $\overline{v}(i) = 0$. We can then define a disjunction of the conjunctions for words $\overline{v}$ such that $f(\overline{v}) = 1$. If there are no such words, then the expression is $0$. It is simple to check that this expression is equivalent to the original function. □

We use of this result by 'simulating' the Boolean operations in the Belnap realm.

**Lemma 4.103.** There is an isomorphism $\mathbf{V} \cong \mathbf{B}^2$.

*Proof.* There are several mappings one could choose, but for this section we will use $\phi \coloneqq \bot \mapsto 00, \mathsf{f} \mapsto 10, \mathsf{t} \mapsto 01, \top \mapsto 11$. □

The Belnap values f and $\top$ are *falsy* whereas the t and $\top$ are *truthy*. The value $\bot$ is neither falsy nor truthy. This is reflected in the mapping shown above; $\phi(v)(0)$ is 1 if and only if $v$ is falsy, and $\phi(v)(1)$ is 1 if and only if $v$ is truthy. We write $\phi_0(v) \coloneqq \phi(v)(0)$ and $\phi_1(v) \coloneqq \phi(v)(1)$.

This means that rather than trying to divine an expression directly from a Belnap function, we can instead define *two* functions; one for how falsy the output is, and one for how truthy is.

**Definition 4.104.** Let $\mathbf{V}_0 \coloneqq \{\bot, \mathsf{f}\}$ and let $\mathbf{V}_1 \coloneqq \{\bot, \mathsf{t}\}$.

For $v \in \mathbf{V}_0$, $\phi(v)(1) = 0$ (all the information is contained within the falsy bit), and for $v' \in \mathbf{V}_1$, $\phi(v)(0) = 0$ (all the information is contained in the truthy bit). These sets are of particular interest when comparing to Boolean operations.

**Lemma 4.105.** $\mathbf{V}_0$ and $\mathbf{V}_1$ are closed under $\wedge$ and $\vee$.

*Proof.* This can be verified by inspecting the truth tables:

| $\wedge$ | $\bot$ | f |
|---|---|---|
| $\bot$ | $\bot$ | f |
| f | f | f |

| $\vee$ | $\bot$ | f |
|---|---|---|
| $\bot$ | $\bot$ | $\bot$ |
| f | $\bot$ | f |

| $\wedge$ | $\bot$ | t |
|---|---|---|
| $\bot$ | $\bot$ | $\bot$ |
| t | $\bot$ | t |

| $\vee$ | $\bot$ | t |
|---|---|---|
| $\bot$ | $\bot$ | t |
| t | t | t |

□

If one looks closer, these are just the truth tables for $\vee_{\mathbf{B}}$ and $\wedge_{\mathbf{B}}$ but with different symbols. This means that any expression we make using $\wedge_{\mathbf{B}}$ and $\vee_{\mathbf{B}}$ in the Boolean realm can be 'simulated' in the falsy and truthy Belnap subsets. Formally, we have the following.

**Lemma 4.106.** The following diagrams commute:

$$\begin{CD} (\mathbf{V}_0)^2 @>{\wedge}>> \mathbf{V}_0 \\ @V{(\phi_0,\phi_0)}VV @VV{\phi}V \\ \mathbf{B}^2 @>>{\vee_{\mathbf{B}}}> \mathbf{B} \end{CD} \qquad \begin{CD} (\mathbf{V}_0)^2 @>{\vee}>> \mathbf{V}_0 \\ @V{(\phi_0,\phi_0)}VV @VV{\phi}V \\ \mathbf{B}^2 @>>{\wedge_{\mathbf{B}}}> \mathbf{B} \end{CD}$$

$$\begin{array}{ccc} (\mathbf{V}_1)^2 & \xrightarrow{\wedge} & \mathbf{V}_1 \\ {\scriptstyle (\phi_1,\phi_1)}\downarrow & & \downarrow{\scriptstyle \phi_0} \\ \mathbf{B}^2 & \xrightarrow[\wedge_{\mathbf{B}}]{} & \mathbf{B} \end{array} \qquad \begin{array}{ccc} (\mathbf{V}_1)^2 & \xrightarrow{\vee} & \mathbf{V}_1 \\ {\scriptstyle (\phi_1,\phi_1)}\downarrow & & \downarrow{\scriptstyle \phi_1} \\ \mathbf{B}^2 & \xrightarrow[\vee_{\mathbf{B}}]{} & \mathbf{B} \end{array}$$

*Proof.* By testing the four values in each case. □

We have not discussed how the Boolean $\neg_{\mathbf{B}}$ can be simulated using Belnap operations; this is because it is not possible to do this while remaining in the two Belnap subsets. We must make use of a certain subset of Boolean functions that can be constructed *without* using $\neg_{\mathbf{B}}$.

**Definition 4.107.** Let the total order $\leq_{\mathbf{B}}$ be defined as $0 \leq 1$.

As with $\mathbf{V}$, $\mathbf{B}^m$ inherits the order on $\mathbf{B}$ pointwise. Subsequently, a Boolean function $f : \mathbf{B}^m \to \mathbf{B}$ is *monotone* if $f(\overline{v}) \leq_{\mathbf{B}} f(\overline{w})$ whenever $\overline{v} \leq_{\mathbf{B}} \overline{w}$. Intuitively, flipping an input bit from 0 to 1 can never flip an output bit from 1 to 0.

**Lemma 4.108.** All monotone functions $\mathbf{B}^m \to \mathbf{B}$ can be expressed with the set of operations $\{\wedge, \vee, 1\}$.

*Proof.* This progresses as with Lemma 4.102, but if the element of a word $\overline{v}(i) = 0$, it is omitted from the conjunction rather than the variable being negated.

To show that this expresses the same truth table as the original function, consider an omitted variable $v_i$; there exists an assignment of the other variables such that if $v_i = 0$ then $f(\ldots, v_i, \ldots) = 1$. By monotonicity, it must be the case that if $v_i = 1$ then $f(\ldots, v_i, \ldots) = 1$, so no information is lost by omitting the negation.

If $f(0, 0, \ldots, 0) = 1$, then the inner conjunction is empty and must be represented by the constant 1, (the unit of $\wedge_{\mathbf{B}}$). This is valid due to monotonicity, as if $f$ produces 1 for the least element, then it must produce 1 for all inputs. □

**Corollary 4.109.** All monotone functions $(\mathbf{V}_0)^m \to \mathbf{V}_0$ can be expressed with the operations $\{\wedge, \vee, \mathsf{f}\}$, and all monotone functions $(\mathbf{V}_1)^m \to \mathbf{V}_1$ can be expressed with the operations $\{\wedge, \vee, \mathsf{t}\}$.

*Proof.* As there is an order isomorphism $\mathbf{V}_0 \cong \mathbf{V}_1 \cong \mathbf{B}$, any monotone function in the Belnap subsets can be viewed as a monotone Boolean function. This means the strategy of Lemma 4.108 can be applied using Lemma 4.106 to substitute the appropriate Belnap operation. □

All the pieces are now in place to express the final functional completeness result; we just need to 'explode' a Belnap value into its falsy and truthy components, and then unify the two at the end.

**Definition 4.110.** Let the functions $\psi_0^0, \psi_0^1 \colon \mathbf{V} \to \mathbf{V}_0$ and $\psi_1^0, \psi_1^1 \colon \mathbf{V} \to \mathbf{V}_1$ be defined according to the table below.

| | $\psi_0^0$ | $\psi_0^1$ | $\psi_1^0$ | $\psi_1^1$ |
|---|---|---|---|---|
| $\bot$ | $\bot$ | $\bot$ | $\bot$ | $\bot$ |
| t | $\bot$ | f | $\bot$ | t |
| f | f | $\bot$ | t | $\bot$ |
| $\top$ | f | f | t | t |

The functions $\psi_0^0$ and $\psi_1^0$ send a value $v$ to f or t respectively if $v$ is falsy; $\psi_0^1$ and $\psi_1^1$ send a value $v$ to f or t if $v$ is truthy. Otherwise, they produce $\bot$.

**Lemma 4.111.** The functions in Definition 4.110 can be expressed using the operations $\{\wedge, \vee, \neg, \bot\}$.

*Proof.* From left to right, the columns in the table above are the functions $v \mapsto - \wedge \bot$, $v \mapsto \neg(- \vee \bot)$, $v \mapsto \neg(- \wedge \bot)$ and $v \mapsto - \vee \bot$. □

**Definition 4.112.** For a monotone function $f \colon \mathbf{V}^m \to \mathbf{V}$, let $f_0 \colon ((\mathbf{V}_0)^m)^2 \to \mathbf{V}_0$ be defined as $f_0(\psi_0^0(\overline{v}), \psi_0^1(\overline{v})) \coloneqq \phi_0(f(\overline{v}))$. Similarly, let $f_1 \colon ((\mathbf{V}_1)^m)^2 \to \mathbf{V}_1$ be defined as $f_1(\psi_1^0(\overline{v}), \psi_1^1(\overline{v})) \coloneqq \phi_1(f(\overline{v}))$.

By putting these pieces all together we can express all monotone Belnap functions.

**Theorem 4.113.** All monotone functions $\mathbf{V}^m \to \mathbf{V}$ can be expressed using the operations $\{\wedge, \vee, \neg, \sqcup, \bot, \mathsf{t}, \mathsf{f}\}$.

*Proof.* This follows by defining a function with the same behaviour as the original, but made up of components known to be expressible using the operations specified.

Let $f' \colon \mathbf{V}^m \to \mathbf{V}^2$ be defined as $f'(\overline{v}) \coloneqq \left(f_0(\psi_0^0(\overline{v}), \psi_0^1(\overline{v})), f_1(\psi_1^0(\overline{v}), \psi_1^1(\overline{v}))\right)$ .. By Corollary 4.109, $f_0$ and $f_1$ can be defined using $\{\wedge, \vee, \mathsf{t}, \mathsf{f}\}$, and by Lemma 4.111, $\psi_0^0, \psi_0^1, \psi_1^0$ and $\psi_1^1$ can be defined using $\{\wedge, \vee, \bot\}$.

The output of $f'(\overline{v})$ is $(\phi_0(f(\overline{v})), \phi_1(f(\overline{v})))$ by definition; the falsiness and the truthiness of $f(\overline{v})$. To combine the two outputs into a single output we want to implement the following truth table:

$$\begin{array}{cc|c} \bot & \bot & \bot \\ \bot & \mathsf{t} & \mathsf{t} \\ \mathsf{f} & \bot & \mathsf{f} \\ \mathsf{f} & \mathsf{t} & \top \end{array}$$

This is clearly just the truth table for $\sqcup$, so the entire expression can be defined using the operations $\{\wedge, \vee, \neg, \sqcup, \bot, \mathsf{t}, \mathsf{f}\}$. □

**Example 4.114.** Consider the following truth table (in fact just the table for $\neg$).

$$\begin{array}{c|c} \bot & \bot \\ \mathsf{t} & \mathsf{f} \\ \mathsf{f} & \mathsf{t} \\ \top & \top \end{array}$$

We translate these into the falsy and truthy tables as follows:

$$\begin{array}{c|c} \bot\bot & \bot \\ \bot\mathsf{f} & \mathsf{f} \\ \mathsf{f}\bot & \bot \\ \mathsf{ff} & \mathsf{f} \end{array} \qquad \begin{array}{c|c} \bot\bot & \bot \\ \bot\mathsf{t} & \bot \\ \mathsf{t}\bot & \mathsf{t} \\ \mathsf{tt} & \mathsf{t} \end{array}$$

Using Corollary 4.109, the corresponding Belnap expressions are

$$(v_0, v_1) \mapsto v_1 \wedge (v_0 \vee v_1) \qquad (v_0, v_1) \mapsto v_0 \vee (v_0 \wedge v_1)$$

To combine these expressions on two inputs into a single expression on one input, we need to add the appropriate translators. We obtain the expression

$$v_0 \mapsto \neg(\bot \vee v_0) \wedge ((\bot \wedge v_0) \vee \neg(\bot \vee v_0)) \sqcup \neg(\bot \wedge v_0) \vee (\neg(\bot \wedge v_0) \wedge (\bot \vee v_0))$$

which can be verified to act the same as the original table.

Although this result only applies to functions with a single output, it is easily generalised to arbitrary-output functions.

**Corollary 4.115.** All monotone functions $\mathbf{V}^{m+1} \to \mathbf{V}^n$ can be expressed using the operations $\{\wedge, \vee, \neg, \bot, \sqcup\}$.

*Proof.* By repeating the process in Theorem 4.113 for each output. □

Since an expression can be using only operations with counterparts in the syntactic realm, the map $||-||_B : \mathbf{Func}_{\mathcal{I}_B} \to \mathbf{SCirc}_{\Sigma_B}$ sends functions $f$ to circuits of the form

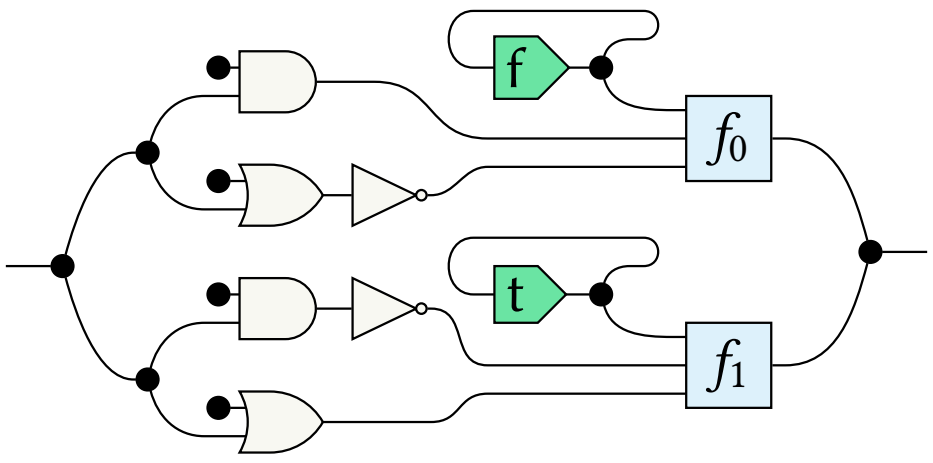

in which $f_0$ and $f_1$ are 'syntactic' falsy and truthy disjunctive normal forms respectively. While the truthy circuit is just a regular disjunctive normal form, because the falsy operations are simulated by the opposite gate, it looks a bit different.

**Definition 4.116** (Conjunction)**.** A Belnap circuit is a *truthy conjunction* if it is the infinite register 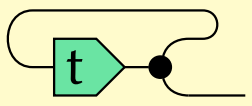 or of the form 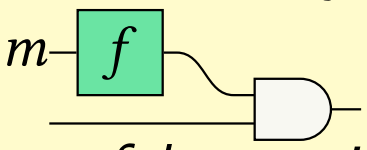 , where $f$ is another truthy conjunction. A Belnap circuit is a *falsy conjunction* if it is the infinite register 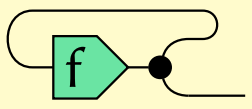 or of the form 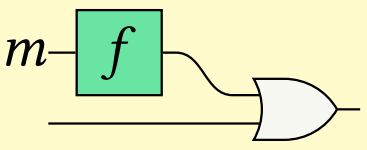 , where $f$ is another falsy conjunction.

**Definition 4.117** (Disjunctive normal form)**.** A Belnap circuit is in *truthy disjunctive normal form* if it is the eliminator 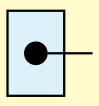 or of the form 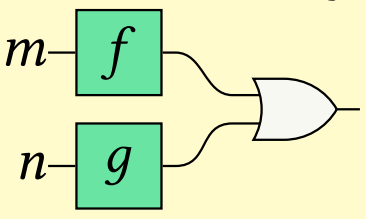 , where $f$ is in truthy disjunctive normal form and $g$ is a truthy conjunction. A Belnap circuit is in *falsy disjunctive normal form* if it is the eliminator 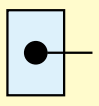 or of the form 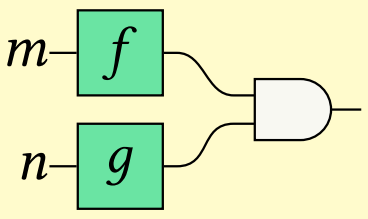 , where $f$ is in falsy disjunctive normal form and $g$ is a falsy conjunction.

The two subcircuits are falsy and truthy disjunctive normal forms, and can be defined syntactically by using a 'composite fork' to copy the inputs for each clause in the normal form.

**Definition 4.118.** For $n \in \mathbb{N}$, an *m, k-fork* $m$–$\Delta_{m,k}$–$km$ is defined inductively with $m$–$\Delta_{m,0}$–$km$ ≔ $m$——$m$ and –$\Delta_{m,k+1}$– ≔ 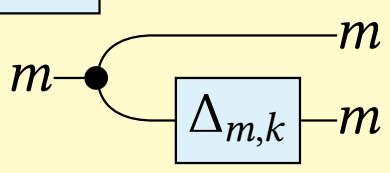 .

**Definition 4.119.** Let $||-||_{\mathrm{B}} \colon \mathbf{Func}_{\mathcal{I}_{\mathrm{B}}} \to \mathbf{CCirc}_{\Sigma_{\mathrm{B}}}$ be defined as the map sending a function $f \colon \mathbf{V}^m \to \mathbf{V}^n$ to a circuit

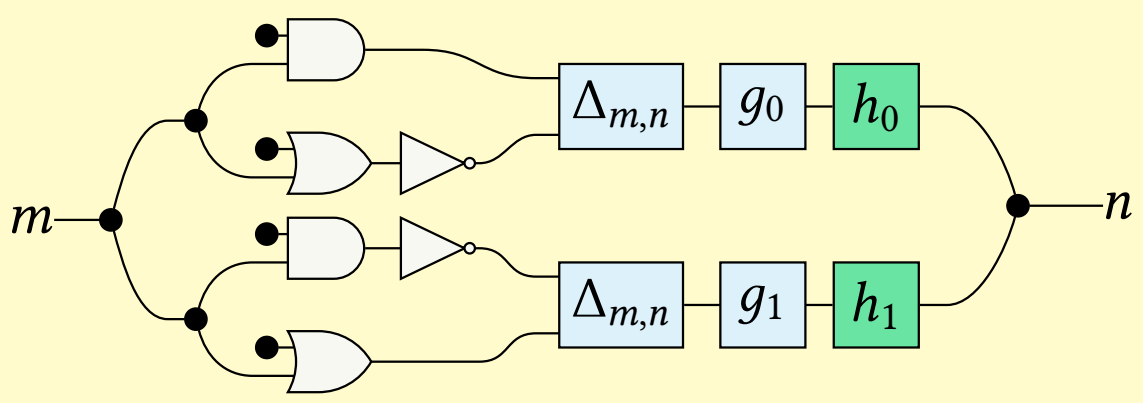

where $g_0$ and $g_1$ only contain identity and elimination constructs, and $h_0$ and $h_1$ and in falsy and truthy conjunctive normal form respectively, defined in the obvious way derived from the procedure detailed in this section.

This means that the denotational semantics for sequential circuits can definitely be used for the Belnap interpretation $\mathcal{I}_{\mathrm{B}}$. In particular, this means we can translate any Mealy machine in $\mathbf{Mealy}_{\mathcal{I}_{\mathrm{B}}}$ (and subsequently, any stream function in $\mathcal{I}_{\mathrm{B}}{}^{\omega}$) into a syntactic circuit in $\mathbf{SCirc}_{\Sigma_{\mathrm{B}}}$.

Even when applying the above techniques to small concrete examples, the results quickly balloon in size; a tool has been developed to generate Belnap expressions from functions and truth tables, and it can be found at https://belnap.georgejkaye.com.

## 4.6 Denotational semantics for generalised circuits

Although we have discussed the denotational semantics in terms of monochromatic circuit signatures, it is straightforward to extend the results to categories generated over *generalised* circuit signatures

In the semantic categories $\mathbf{Func}_{\mathcal{I}}$, $\mathbf{Stream}_{\mathcal{I}}$, and $\mathbf{Mealy}_{\mathcal{I}}$, the morphisms are all variants on functions of the form $\mathbf{V}^m \to \mathbf{V}^n$ that operate on powers of elements in $\mathbf{V}$: one element for each (single-bit) input or output wire. In the generalised setting, these input and output wires may not all be the same width, so the input and output sets must be *powers of powers* of values.

**Notation 4.120.** Given a set $A$ and a word $\overline{v} \in \mathbb{N}_+^\star$ of length $n$, we write $A^{\overline{v}} \coloneqq A^{\overline{v}(0)} \times A^{\overline{v}(1)} \times \cdots \times A^{\overline{v}(n-1)}$.

Note that for a word $\overline{m} \coloneqq 11\ldots1$ of length $k$, the set $A^{\overline{m}} = A^1 \times A^1 \times \ldots A^1$ is isomorphic to $A^k$, much like how setting the set of colours in a coloured PROP to the singleton recovers a monochromatic PROP.

The semantic categories can now be extended to these *coloured* interfaces.

**Definition 4.121.** Let $\mathbf{Func}^+_{\mathcal{I}}$ be the $\mathbb{N}_+$-coloured PROP in which the morphisms $\overline{m} \to \overline{n}$ are the monotone functions $\mathbf{V}^{\overline{m}} \to \mathbf{V}^{\overline{n}}$. Let $\mathbf{Stream}^+_{\mathcal{I}}$ be the $\mathbb{N}_+$-coloured PROP in which the morphisms $\overline{m} \to \overline{n}$ are the causal, monotone, and finitely specified stream functions $(\mathbf{V}^{\overline{m}})^{\omega} \to (\mathbf{V}^{\overline{n}})^{\omega}$. Let $\mathbf{Mealy}^+_{\mathcal{I}}$ be the $\mathbb{N}_+$-coloured PROP in which the moprhisms $\overline{m} \to \overline{n}$ are the monotone $(\mathbf{V}^{\overline{m}}, \mathbf{V}^{\overline{n}})$-Mealy machines.

The various PROP morphisms between these categories are defined in a similar way to the monochromatic versions, but now we have to account for the structural generators for each $n \in \mathbb{N}_+$, as well as the bundlers.

**Definition 4.122.** Let $\llbracket - \rrbracket^{\mathrm{C}+}_{\mathcal{I}} \colon \mathbf{CCirc}^+_{\Sigma} \to \mathbf{Func}^+_{\mathcal{I}}$ be the coloured PROP morphism with action defined as

$\llbracket$ 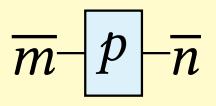 $\rrbracket^{\mathrm{C}+}_{\mathcal{I}} \coloneqq \llbracket p \rrbracket$

$\llbracket$ 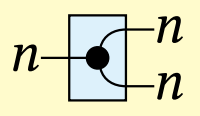 $\rrbracket^{\mathrm{C}+}_{\mathcal{I}} \coloneqq (\overline{v}) \mapsto (\overline{v}, \overline{v})$

$\llbracket$ 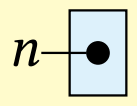 $\rrbracket^{\mathrm{C}+}_{\mathcal{I}} \coloneqq (\overline{v}) \mapsto ()$

$\llbracket$ 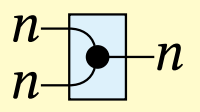 $\rrbracket^{\mathrm{C}+}_{\mathcal{I}} \coloneqq (\overline{v}, \overline{w}) \mapsto \overline{v} \sqcup \overline{w}$

$\llbracket$ 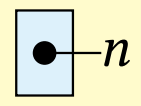 $\rrbracket^{\mathrm{C}+}_{\mathcal{I}} \coloneqq () \mapsto \bot^n$

$\llbracket$ 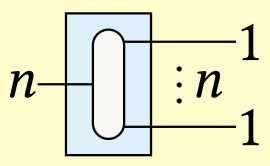 $\rrbracket^{\mathrm{C}+}_{\mathcal{I}} \coloneqq (\overline{v}) \mapsto (\overline{v}(0), \overline{v}(1), \dots, \overline{v}(n-1))$

$\llbracket$ 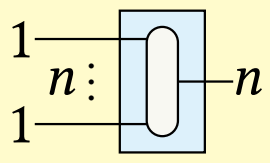 $\rrbracket^{\mathrm{C}+}_{\mathcal{I}} \coloneqq (v_0, v_1, \dots, v_{n-1}) \mapsto (v_0 v_1 \dots v_{n-1})$

The map from coloured circuits to Mealy machines proceeds in a similar manner.

**Definition 4.123.** Let $[-]^+_{\mathcal{I}} \colon \mathbf{SCirc}^+_{\Sigma} \to \mathbf{Mealy}^+_{\mathcal{I}}$ be the traced PROP morphism

defined on generators as

$\left[\right.$ 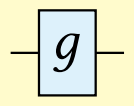 $\left.\right]^{+}_{\mathcal{I}} \coloneqq (\{s\}, \quad (\overline{v}_0, \ldots, \overline{v}_{m-1}) \mapsto \langle s, [\![g]\!] (\overline{v}_0, \overline{v}_1, \ldots, \overline{v}_{m-1}) \rangle, \quad s)$

$\left[\right.$ 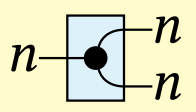 $\left.\right]^{+}_{\mathcal{I}} \coloneqq (\{s\}, \quad \overline{v} \mapsto \langle s, (\overline{v}, \overline{v}) \rangle, \quad s)$

$\left[\right.$ 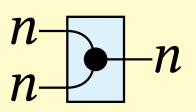 $\left.\right]^{+}_{\mathcal{I}} \coloneqq (\{s\}, \quad (\overline{v}, \overline{w}) \mapsto \langle s, \overline{v} \sqcup \overline{w} \rangle, \quad s)$

$\left[\right.$ 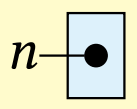 $\left.\right]^{+}_{\mathcal{I}} \coloneqq (\{s\}, \quad \overline{v} \mapsto \langle s, s \rangle, \quad s)$

$\left[\right.$ 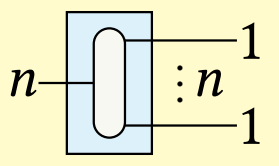 $\left.\right]^{+}_{\mathcal{I}} \coloneqq (\{s\}, \quad (v_0, v_1, \ldots, v_{n-1}) \mapsto ((v_0), (v_1), \ldots, (v_{n-1})), \quad s)$

$\left[\right.$ 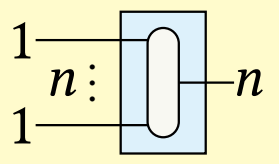 $\left.\right]^{+}_{\mathcal{I}} \coloneqq (\{s\}, \quad ((v_0), (v_1), \ldots, (v_{n-1})) \mapsto (v_0, v_1, \ldots, v_{n-1}), \quad s)$

$\left[\right.$ 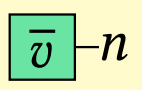 $\left.\right]^{+}_{\mathcal{I}} \coloneqq (\{s_{\overline{v}}, s_{\bot}\}, \quad \{s_{\overline{v}} \mapsto \langle s_{\bot}, \overline{v} \rangle, s_{\bot} \mapsto \langle s_{\bot}, \bot \rangle\}, \quad s_{\overline{v}})$

$\left[\right.$ 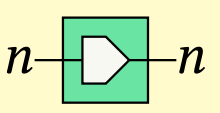 $\left.\right]^{+}_{\mathcal{I}} \coloneqq (\{s_{\overline{v}} \mid \overline{v} \in \mathbf{V}^n\}, \quad (s_{\overline{v}}, \overline{w}) \mapsto \langle s_{\overline{w}}, \overline{v} \rangle, \quad s_{\bot^n})$

Although morphisms in $\mathbf{Mealy}^{+}_{\mathcal{I}}$ have different interfaces to those in $\mathbf{Stream}^{+}_{\mathcal{I}}$, they are still monotone Mealy machines so it is simple to translate them into stream functions or coloured circuits.

**Definition 4.124.** Let the coloured PROP morphisms $!^{+}_{\mathcal{I}}(-) \colon \mathbf{Mealy}^{+}_{\mathcal{I}} \to \mathbf{Stream}^{+}_{\mathcal{I}}$, $\langle\!\langle - \rangle\!\rangle^{+}_{\mathcal{I}} \colon \mathbf{Stream}^{+}_{\mathcal{I}} \to \mathbf{Mealy}^{+}_{\mathcal{I}}$ and $||-||^{\leq +}_{\mathcal{I}} \colon \mathbf{Mealy}^{+}_{\mathcal{I}} \to \mathbf{SCirc}^{+}_{\Sigma}$ be defined as before, and let $[\![-]\!]^{\mathrm{S}+}_{\mathcal{I}} \colon \mathbf{SCirc}^{+}_{\Sigma} \to \mathbf{Stream}^{+}_{\mathcal{I}}$ be defined as $!^{+}_{\mathcal{I}}(-) \circ [-]^{+}_{\mathcal{I}}$.

By putting all these coloured PROP morphisms together, we can show the same results as we did in the previous section.

**Theorem 4.125.** $!^{+}_{\mathcal{I}}(-) = [\![-]\!]^{\mathrm{S}+}_{\mathcal{I}} \circ !^{+}_{\mathcal{I}}(-)$ and $[\![-]\!]^{\mathrm{S}+}_{\mathcal{I}} \circ ||-||^{\leq +}_{\mathcal{I}} \circ \langle\!\langle - \rangle\!\rangle^{+}_{\mathcal{I}} = \mathrm{id}_{\mathbf{Stream}^{+}_{\mathcal{I}}}$.

As before, we derive a notion of denotational equivalence for generalised circuits.

**Definition 4.126.** Two generalised sequential circuits are *denotationally equivalent* under $\mathcal{I}$, written $\overline{m}-\boxed{f}-\overline{n} \approx^{+}_{\mathcal{I}} \overline{m}-\boxed{g}-\overline{n}$ if $\left[\!\left[ -\boxed{f}- \right]\!\right]^{\mathrm{S}+}_{\mathcal{I}} = \left[\!\left[ -\boxed{g}- \right]\!\right]^{\mathrm{S}+}_{\mathcal{I}}$. Let $\mathbf{SCirc}^{+}_{\Sigma/\approx^{+}_{\mathcal{I}}}$ be the result of quotienting $\mathbf{SCirc}_{\Sigma}$ by $\approx^{+}_{\mathcal{I}}$.

**Corollary 4.127.** $\mathbf{SCirc}^{+}_{\Sigma/\approx^{+}_{\mathcal{I}}} \cong \mathbf{Stream}^{+}_{\mathcal{I}}$.

# Chapter 5

# Operational semantics

With the sound and complete denotational semantics, the behaviour of circuits is determined by observing their behaviour as stream functions. This already gives us a perspective on digital circuits closer to that of programming languages. To compare the behaviour of two circuits in $\mathbf{SCirc}_\Sigma$, we examine their corresponding stream functions.

Denotational semantics is not the be-all and end-all of circuit semantics. Crucially, it obscures the *structure* of a circuit by compressing all the behaviour into one function: we don't know *why* the circuit is behaving the way it does, just that something has caused it to do so. When it comes to circuit design, the structure of the circuit is important, as that is what is going to be printed onto silicon. Space is at a premium, so knowing how each component contributes to the output behaviour is of critical importance.

We now turn our attention to the next course in our menu of semantics: *operational semantics*. This is quite a different beast to denotational semantics: rather than assigning a mathematical structure to each circuit, semantics are derived from how something is *executed*. One can think of an operational semantics as stepping through a program using a debugger, with rules applied in order to derive the next state.

Operational semantics is another classic concept in computer science; 'steps' of execution were used to define the semantics of ALGOL 68 [VMP+76]. The name itself, as with many topics of the time, is attributed to Dana Scott [Sco70], who acknowledged that even with the abstraction of denotational semantics, 'the operational aspects cannot be completely ignored'.

**Example 5.1.** Recall the language of expressions from Example 4.1. We can define an observational semantics on this language with the following set of rules:

$$\frac{}{\overline{n} \Rightarrow \overline{n}}\ \text{(Value)} \qquad \frac{e_0 \Rightarrow \overline{n_0} \qquad e_1 \Rightarrow \overline{n_1}}{add\ e_0\ e_1 \Rightarrow \overline{n_0 + n_1}}\ \text{(Add)} \qquad \frac{e_0 \Rightarrow \overline{n_0} \qquad e_1 \Rightarrow \overline{n_1}}{mul\ e_0\ e_1 \Rightarrow \overline{n_0 \cdot n_1}}\ \text{(Mul)}$$

We can use these rules to *reduce* expressions to values. Two expressions have the same semantics if they reduce to the same expression. Formally this can be written as a proof tree:

$$\cfrac{\cfrac{\cfrac{}{\overline{4} \Rightarrow \overline{4}} \qquad \cfrac{}{\overline{2} \Rightarrow \overline{2}}}{mul\ \overline{4}\ \overline{2} \Rightarrow \overline{8}} \qquad \cfrac{\cfrac{}{\overline{2} \Rightarrow \overline{2}} \qquad \cfrac{}{\overline{3} \Rightarrow \overline{3}}}{add\ \overline{2}\ \overline{3} \Rightarrow \overline{5}}}{add\ (mul\ \overline{4}\ \overline{2})\ (add\ \overline{2}\ \overline{3}) \Rightarrow \overline{13}}$$

The operational semantics described thus far is commonly known as *structural operational semantics* [Plo81]; it shows how the behaviour of the whole is represented by the behaviour of its parts. However, deriving a proof tree as above can be clunky: to find the meaning of a large term one has to 'drill down' into the contexts until a value is reached before propagating these values back up the tree.

A more intuitive way to view this style of operational semantics is using a *reduction semantics*. First applied by Plotkin [Plo75] before being properly coined and generalised in the subsequent decade [FF87; Fel87], a reduction semantics specifes a set of rules which can be successively applied to individual components of some larger context. These reduction rules can be derived from the rules in the main operational semantics by replacing subexpressions in derivations by the primitives in the language. These rules can then be applied to the 'smallest' terms in the expression (those higher up in the proof tree), reducing these to primitives themselves, such that reductions may be applied to their parent expressions and so on.

**Example 5.2.** Returning to Example 5.1, the two rules of the corresponding reductions semantics are $add\ \overline{n_0}\ \overline{n_1} \overset{\text{(Add)}}{\rightsquigarrow} \overline{n_0 + n_1}$ and $mul\ \overline{n_0}\ \overline{n_1} \overset{\text{(Mul)}}{\rightsquigarrow} \overline{n_0 \cdot n_1}$. Note that as the numbers $\overline{n}$ are the 'primitives' of the language, there is no corresponding reduction for the (Value) rule. Using these reductions an expression can be reduced to a value.

$$add\ (mul\ \overline{4}\ \overline{2})\ (add\ \overline{2}\ \overline{3}) \overset{\text{(Mul)}}{\rightsquigarrow} add\ \overline{8}\ (add\ \overline{2}\ \overline{3}) \overset{\text{(Add)}}{\rightsquigarrow} add\ \overline{8}\ \overline{5} \overset{\text{(Add)}}{\rightsquigarrow} \overline{13}$$

We write $\overset{*}{\rightsquigarrow}$ for a sequence of reduction steps. Such a reduction sequence to a value is not necessarily canonical; while there is one canonical proof tree for a given term there may be many reduction sequences. In an ideal reduction system this should not be an issue.

**Definition 5.3.** A reduction system is *confluent* if, for any term $e$, if there exists distinct reductions $e \rightsquigarrow e_1$ and $e \rightsquigarrow e_2$, then there exists term $e_3$ along with reduction sequences $e_1 \overset{*}{\rightsquigarrow} e_3$ and $e_2 \overset{*}{\rightsquigarrow} e_3$.

While it is important that different reduction sequences should always converge, it is equally important that we are not stuck performing reductions forever.

**Definition 5.4.** A reduction system is *terminating* if, for every term $e$, there is no infinite reduction sequence starting from $e$.

If a reduction system is terminating, then repeatedly applying reductions will eventually lead to a term with no opportunity to apply any more.

**Definition 5.5.** A term is in *normal form* if no reductions apply to it.

If a terminating reduction system is also confluent, every term must have a *unique* such normal form. In the setting of an operational semantics, this normal form is the behaviour of the term.

**Example 5.6.** The reduction rules in Example 5.2 are clearly terminating since they both reduce the number of operations in a term. The rules are also confluent: we could have chosen a different order of reductions but the result is the same.

$$add\,(mul\,\overline{4}\,\overline{2})\,(add\,\overline{2}\,\overline{3}) \overset{\text{(Mul)}}{\rightsquigarrow} add\,\overline{8}\,(add\,\overline{2}\,\overline{3}) \overset{\text{(Add)}}{\rightsquigarrow} add\,\overline{8}\,\overline{5} \overset{\text{(Add)}}{\rightsquigarrow} \overline{13}$$

This means that $\overline{13}$ is the normal form of $add\,(mul\,\overline{4}\,\overline{2})\,(add\,\overline{2}\,\overline{3})$ and subsequently the behaviour of the term.

When defining an operational semantics for digital circuits we prefer to use the reduction semantics style of presentation, as one of our motivations is for a computer to evaluate circuits step-by-step.

As we will come to see, it is not feasible to create a small-step reduction semantics for digital circuits while remaining terminating and confluent. However, what we *can* do is specify some larger transformations to apply to a circuit followed by some more traditional exhaustive reductions.

Our goal for this chapter is to develop a sound and complete notion of *observational equivalence* i.e. circuits are related if and only if 'executing' them using the operational semantics produces the same values. To determine which transformations and reduc-tions are sound, we turn to the denotational semantics; a reduction between circuits $m - \boxed{f} - n \rightsquigarrow m - \boxed{g} - n$ is sound with respect to some interpretation $\mathcal{I}$ if and only if $\left[\!\left[ m - \boxed{f} - n \right]\!\right]^{\mathrm{S}}_{\mathcal{I}} = \left[\!\left[ m - \boxed{g} - n \right]\!\right]^{\mathrm{S}}_{\mathcal{I}}$.

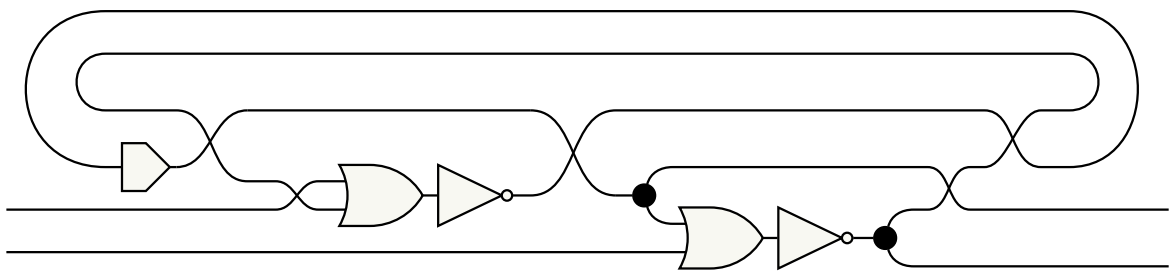

Figure 5.1: The SR NOR latch from Example 3.18 in global trace-delay form

**Remark 5.7.** The content of this chapter is a refined version of [GKS24, Sec. 4].

## 5.1 Feedback

One of the major issues that comes with trying to reduce circuits in $\mathbf{SCirc}_\Sigma$ is the presence of feedback. Without proper attention, one could end up infinitely unfolding and we never produce any output values. The first portion of our operational semantics revolves around some *global transformations* to make a circuit suitable for reduction.

The first observation we make does not even need anything new to be defined as it follows immediately from axioms of STMCs.

**Lemma 5.8** (Global trace-delay form)**.** For a sequential circuit $m - \boxed{f} - n$ there exists a combinational circuit

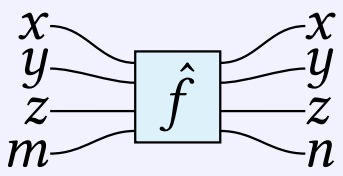

and values $\overline{v} \in \mathbf{V}^z$ such that $m - \boxed{f} - n$ =

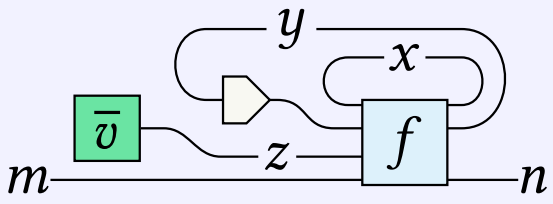

by axioms of STMCs.

*Proof.* By applying the axioms of traced categories; any trace can be 'pulled' to the outside of a term by superposing and tightening. For the delays, a trace can be introduced using yanking and then the same procedure as above followed. ☐

**Example 5.9.** The SR NOR latch circuit from Example 3.18 is assembled into global trace-delay form in Figure 5.1.

This form is evocative of what we saw when mapping from Mealy machines to circuits in the previous section, but rather than the state being determined by one word, the instantaneous values and the delays are kept separate.

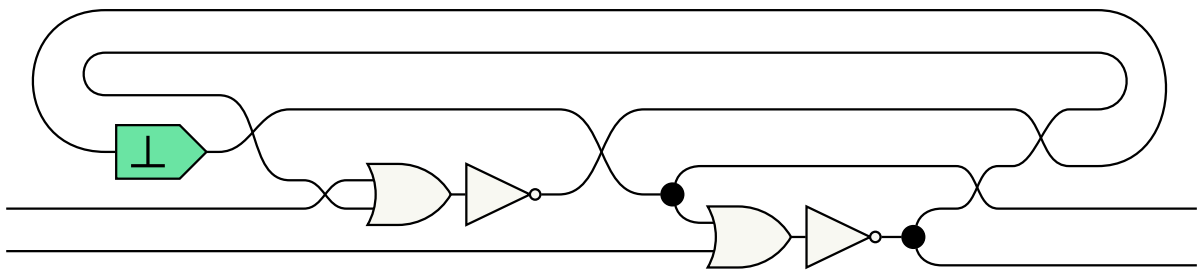

Figure 5.2: Applying the (Mealy) rule to the circuit in Figure 5.1

**Definition 5.10** (Pre-Mealy form)**.** A sequential circuit is in *pre-Mealy form* if it is in the form 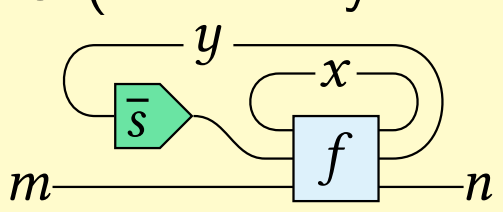

.

Our first reduction transforms a circuit from global trace-delay form to pre-Mealy form.

**Lemma 5.11.** The following rule is sound:

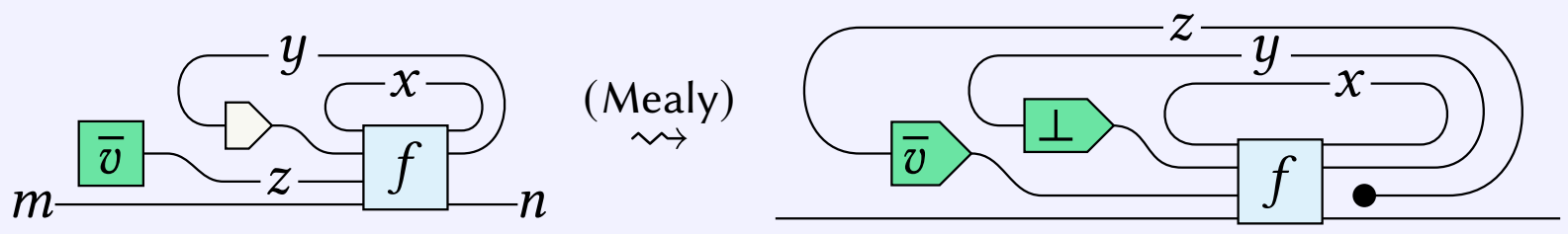

*Proof.* It is a simple exercise to check the corresponding stream functions. □

By assembling a circuit into global trace-delay form and applying the (Mealy) rule, we can construct a word $\bar{s}$ from the juxtaposition of $\bot$ elements for each register combined with the instantaneous values $\bar{v}$ i.e. using the notation of the above lemma $\bar{s} \coloneqq \bot^y\bar{v}$. This word represents the initial state of the circuit, but it is by no means unique: it depends on how the circuit is put into global trace-delay form. What matters most is that we *can* do it.

**Corollary 5.12.** For any sequential circuit 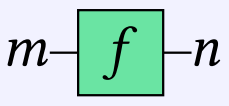

, there exists at least one valid application of the Mealy rule.

**Example 5.13.** The (Mealy) rule is applied to the global trace-delay form SR NOR latch in Figure 5.1. Here the initial state word is just $\bot$.

The result of applying the (Mealy) reduction still differs from the image of $||-||_{\mathcal{I}}^{\leq}$ as it may have a trace with no delay on it: an instance of *non-delay-guarded feedback*.

The mere mention of non-delay-guarded feedback may trigger alarm bells in the minds of those well-acquainted with circuit design. It is often common in industry to enforce that circuits have no non-delay-guarded feedback; one might ask if we should also enforce this tenet in order to stick to 'well-behaved' circuits.

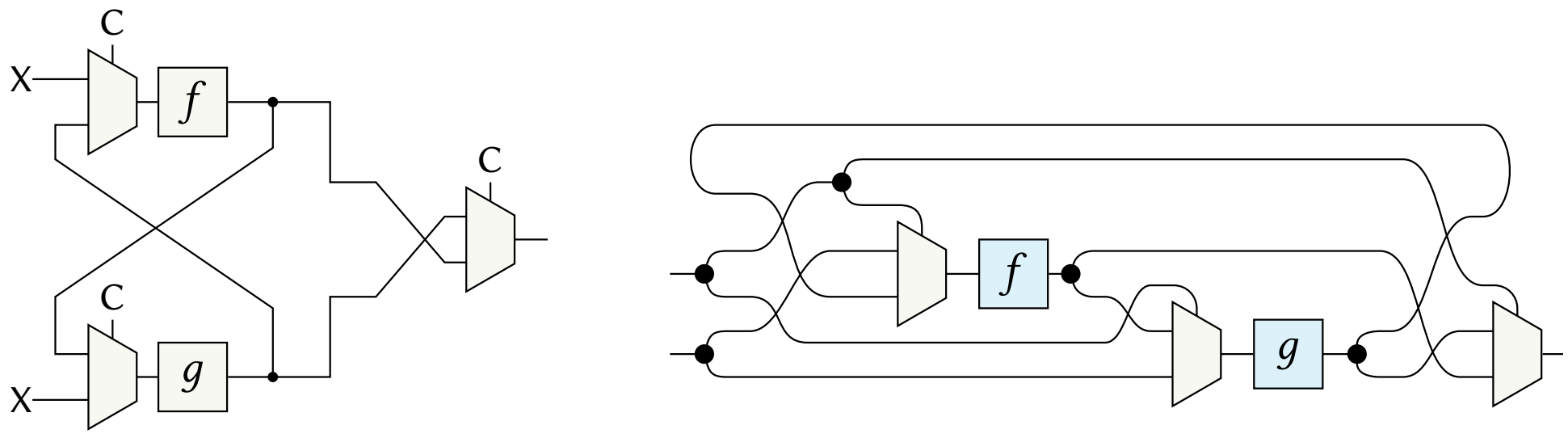

Figure 5.3: A useful cyclic combinational circuit [MSB12, Fig. 1], and a possible interpretation in $\mathbf{SCirc}_\Sigma$.

**Remark 5.14.** *Categories with feedback* [KSW02] are a weakening of traced categories that remove the yanking axiom: this effectively makes all traces delay-guarded. *Categories with delayed trace* [SK19] weaken this further by removing the sliding axiom, so no components can be 'pushed round' into the next tick of execution. Neither of these are suitable for us as we actually *want* to allow non-delay-guarded feedback.

In fact, careful use of non-delay-guarded feedback can still result in useful circuits as a clever way of sharing resources [Mal94; Rie04; MSB12]. The minimal circuit to implement a function often *must* be constructed using cycles [Riv77; RB03].

**Example 5.15.** A particularly famous circuit [Mal94] which is useful despite the presence of non-delay-guarded feedback is shown in Figure 5.3, where $f$ and $g$ are arbitrary combinational circuits. The trapezoidal gate is a *multiplexer*; it has a vertical *control* input and two horizontal *data* inputs. The multiplexer is defined as 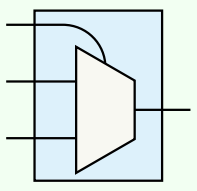 $:=$ 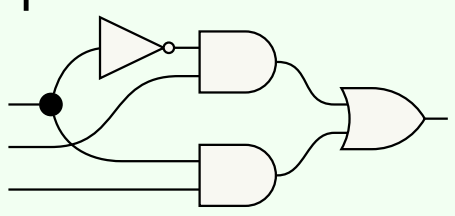 . The multiplexer is effectively an if statement: when the control is f the output is the first data input and when it is t the output is the second data input.

The circuit in Figure 5.3 has no state and its trace is global so it is already in pre-Mealy form, and has non-delay-guarded feedback. Despite this, it produces useful output when the control signal is true or false: when the control signal is f then the behaviour of the circuit is $\llbracket\, f \; g \,\rrbracket^{\mathsf{C}}_{\mathcal{I}_\star}$ and when the control is t then the behaviour is $\llbracket\, g \; f \,\rrbracket^{\mathsf{C}}_{\mathcal{I}_\star}$. The feedback is just used as a clever way to share circuit components.

A combinational circuit surrounded by non-delay-guarded feedback still implements a function, as there are no delay components. Nevertheless, non-delay-guarded feedback does still block our path to future transformations, so it must be eliminated. Using a methodology also employed by [RB12], we turn to the Kleene fixed-point theorem.

**Lemma 5.16.** For a monotone function $f\colon \mathbf{V}^{n+m} \to \mathbf{V}^n$ and $i \in \mathbb{N}$, let $f^i\colon \mathbf{V}^m \to \mathbf{V}^n$ be defined as $f^0(x) = f(\bot^n, x)$ and $f^{k+1}(x) = f(f^k(x), x)$. Let $c$ be the length of the longest chain in the lattice $\mathbf{V}^n$. Then, for $j > c$, $f^c(x) = f^j(x)$.

*Proof.* Since $f$ is monotone and $\mathbf{V}^n$ is finite, the former has a least fixed point by the Kleene fixed-point theorem. This will either be some value $v$ or the $\top$ element. The most iterations of $f$ it would take to obtain this fixed point is $c$, i.e. the function produces a value one step up the lattice each time. □

**Definition 5.17** (Iteration)**.** Given a combinational circuit $\begin{smallmatrix}x\\m\end{smallmatrix}\!\!-\!\boxed{f}\!-\!\!\begin{smallmatrix}x\\n\end{smallmatrix}$, its *$n$-th iteration* $m\!-\!\boxed{f^n}\!-\!\!\begin{smallmatrix}x\\n\end{smallmatrix}$ is defined inductively over $n$ in the following way:

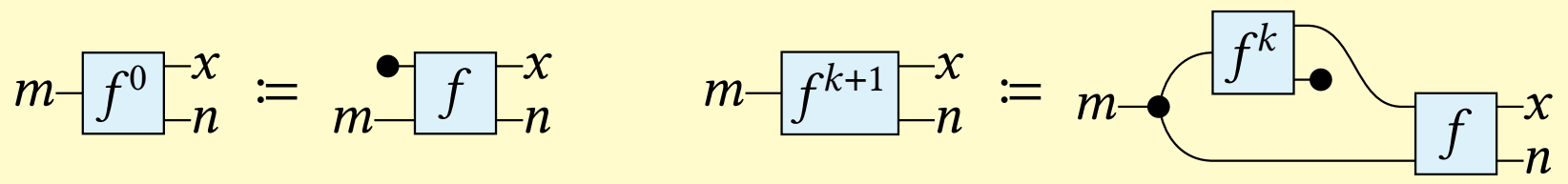

The trace in $\mathbf{Stream}_{\mathcal{I}}$ is by the least fixed point, computed by repeatedly applying $f$ to itself starting from $\bot$. The above lemma gives a fixed upper bound for the number of times we need to apply $f$ to reach this fixed point, based on the size of the lattice. We can use this in the syntactic setting.

**Definition 5.18** (Unrolling)**.** For an interpretation with values $\mathbf{V}$, the *unrolling* of a combinational circuit $\begin{smallmatrix}x\\m\end{smallmatrix}\!\!-\!\boxed{f}\!-\!\!\begin{smallmatrix}x\\n\end{smallmatrix}$, written $m\!-\!\boxed{f^\dagger}\!-\!\!\begin{smallmatrix}x\\n\end{smallmatrix}$, is defined as $m\!-\!\boxed{f^{c+1}}\!-\!\!\begin{smallmatrix}x\\n\end{smallmatrix}$ where $c$ is the length of the longest chain in $\mathbf{V}^x$.

Using these constructs we can eliminate non-delay-guarded feedback around a combinational circuit.

**Proposition 5.19.** The instant feedback rule 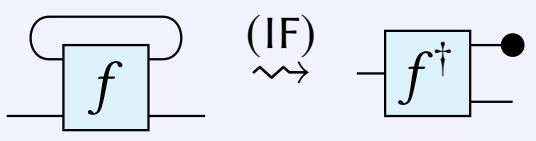 is sound.

*Proof.* By Lemma 5.16, applying the function $(\overline{x}) \mapsto \pi_x\left(\left[\!\left[\,\boxed{f}\,\right]\!\right]^{\mathrm{C}}_{\mathcal{I}}\right)(\overline{x}, \overline{v})$ to itself $c$ times reaches a fixed point. The circuit is combinational so each element of the output $\left[\!\left[\,\boxed{f}\,\right]\!\right]^{\mathrm{S}}_{\mathcal{I}}(\sigma)(i)$ is a function; this means that Lemma 5.16 can be applied to each element. □

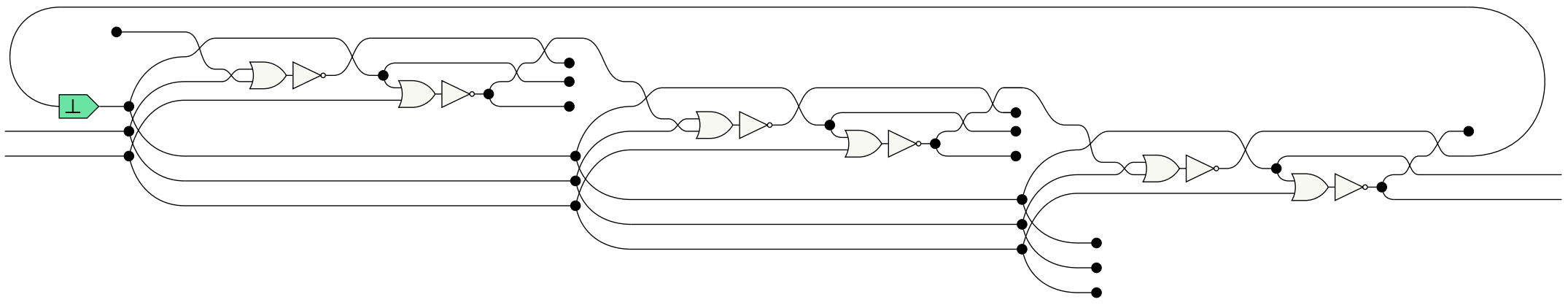

Figure 5.4: Applying the (IF) rule to the circuit in Figure 5.2

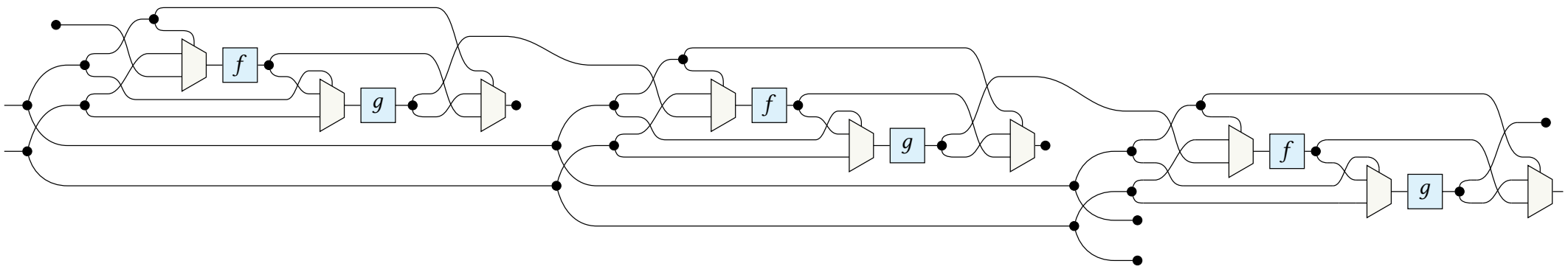

Figure 5.5: Applying the (IF) rule to the circuit in Figure 5.3

**Example 5.20.** In Figure 5.4, the IF is applied to the SR latch circuit in pre-Mealy form from Figure 5.1.

**Example 5.21.** In Figure 5.5, the (IF) rule is applied to the cyclic combinational circuit from Figure 5.3.

If applied locally for every feedback loop, the (IF) rule would cause an exponential blowup, but if a circuit is in global trace-delay form, the rule need only be applied once to the global loop. Although the value of $c$ increases as the number of feedback wires increases, it only does so linearly in the height of the lattice.

**Remark 5.22.** [MSB12] uses a ternary set of values and monotone functions to present *constructive* circuits: the circuits that stabilise to unique Boolean values for all Boolean inputs. This definition excludes circuits that oscillate between two values, as these are not considered to be monotone circuits. Conversely, in our model such circuits *can* be monotone. For example a Belnap circuit may alternate between t and f because these occupy the same level of the lattice.

With a method to eliminate non-delay-guarded feedback, we can establish the class of circuits which will act as the keystone of both the operational semantics in this section and the algebraic semantics of the next.

**Definition 5.23** (Mealy form)**.** A sequential circuit 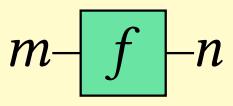

is in *Mealy form* if it is in the form 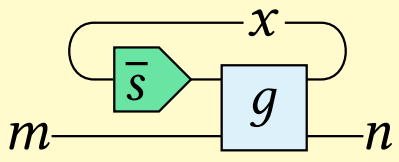

.

**Theorem 5.24.** For a sequential circuit $m$–[$f$]–$n$, there exists at least one combinational circuit $x$, $m$–[$g$]–$x$, $n$ and values $\overline{s} \in \mathbf{V}^x$ such that $m$–[$f$]–$n$ $\overset{*}{\rightsquigarrow}$ 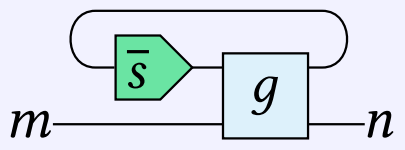 by applying (Mealy) followed by (IF).

*Proof.* Any circuit can be assembled into global trace-delay form by Lemma 5.8 and furthermore transformed into pre-Mealy form by using (Mealy). Since the core of a circuit in pre-Mealy form is combinational and has a non-delay-guarded trace, (IF) can be applied to it to produce a circuit with only delay-guarded feedback: a circuit in Mealy form. □

Non-delay-guarded feedback can be exhaustively unrolled because the circuit essentially models a function despite the presence of the trace: this means that we can transform the circuit without having to 'look into the future'. This is not the case for delay-guarded feedback as the internal state of the circuit may depend on future inputs. Indeed, a circuit with delay-guarded feedback may never 'settle' on one internal configuration but rather oscillate between multiple states. This is simply a facet of sequential circuits and there is nothing we can do about that. What we *can* do is show how to *process* inputs at a given tick of the clock.

## 5.2 Productivity

It is not possible to reduce an open circuit to some output values, as there will be open wires awaiting the next inputs. Nevertheless, if we precompose a circuit with some inputs we can provide some rules for propagating them across the circuit.

Formally, for sequential circuit $m$–[$f$]–$n$ and values $\overline{v} \in (\mathbf{V}^m)^\omega$, this corresponds to finding reductions such that $m$–[$\overline{v}$▷]–[$f$]–$n$ $\overset{*}{\rightsquigarrow}$ $m$–[$g$]–[$\overline{w}$▷]–$n$. We first consider the combinational case, with our final global transformation.

**Lemma 5.25** (Streaming)**.** The following *streaming rule* is sound:

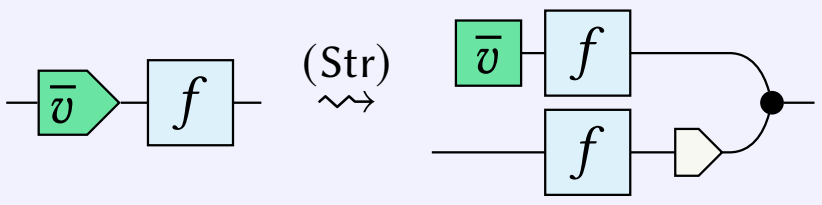

*Proof.* Once again this can be shown by considering the stream semantics. First note that by unfolding the notation, –[$\overline{v}$▷]–[$f$]– $:=$ 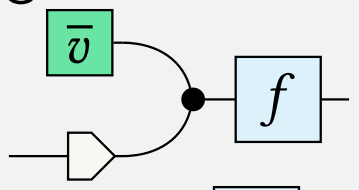. The streaming rule is then effectively 'pushing' the combinational circuit –[$f$]– across the join.

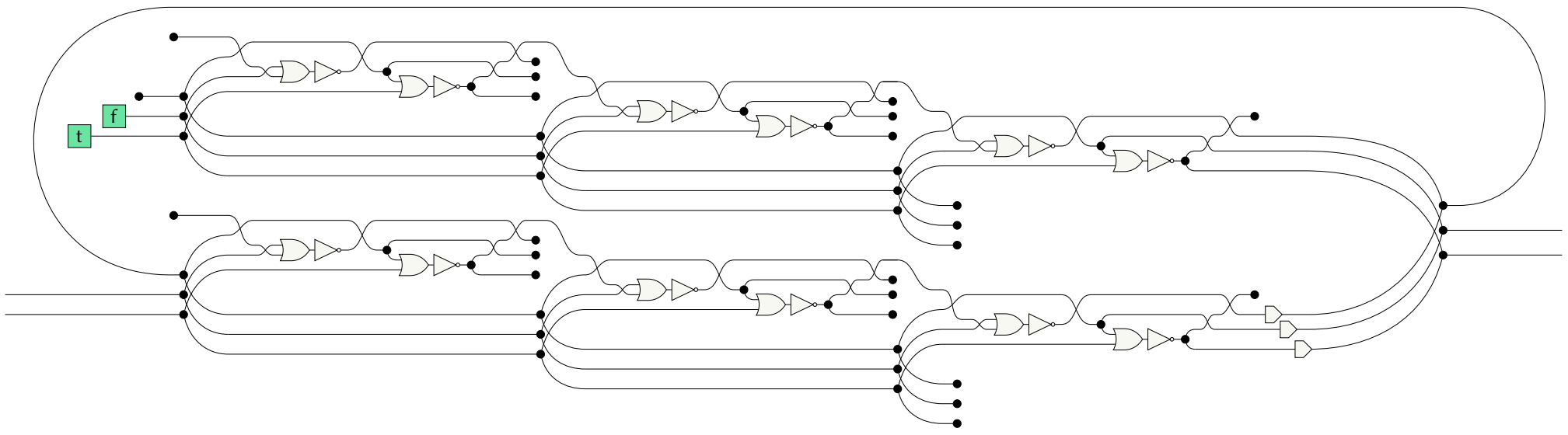

Figure 5.6: Applying Str with inputs tf to the circuit from Figure 5.4

The join is *not* a natural transformation so this does not hold in general, but because one argument is an instantaneous value and the other is a delay, at least one of the inputs to the join will be $\bot$ for a given circuit. As the interpretations of combinational circuits must be $\bot$-preserving, the circuit can safely be pushed across the join and delay. □

The streaming rule shows that when a combinational circuit is applied to an input with an instantaneous and a delayed component, the circuit can be copied so that one copy handles what is happening 'now' and the other handles what is happening 'later'.

**Example 5.26.** Pulsing the signals ft to the inputs of an SR NOR latch starts the procedure for 'setting' the latch, causing it to output tf. Figure 5.6 shows how the Str rule is applied to the unrolled SR NOR circuit from Figure 5.4 with these inputs to create a copy for what is happening 'now' and another for what is happening 'later'.

As there is a delay on the bottom argument of the join, the output of a streamed circuit at the current tick is now contained entirely in the top argument of the join. The final rules we present will reduce this copy to values, as desired.

**Lemma 5.27** (Value rules)**.** The following *value rules* are sound:

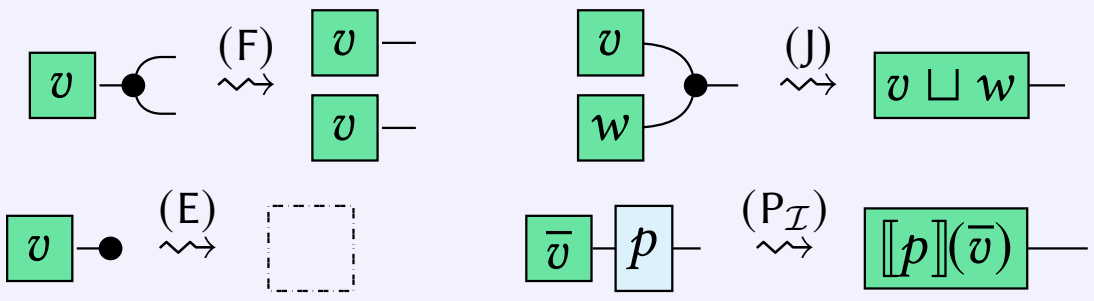

*Proof.* Straightforward by considering the interpretations of values as stream functions. □

Reducing the 'now' core is the only time in which exhaustive application is required, as more is involved than just copying circuit components.

**Lemma 5.28.** Applying the value rules is confluent.

*Proof.* There are no overlaps between the rules. □

**Lemma 5.29.** For a combinational circuit $m\!-\!\boxed{f}\!-\!n$ and $\overline{v} \in \mathbf{V}^m$, there exists $\overline{w} \in \mathbf{V}^n$ such that applying the value rules to $\boxed{\overline{v}}\!-\!\boxed{f}\!-$ terminates at $\boxed{\overline{w}}\!-$ .

*Proof.* By induction on the structure of $m\!-\!\boxed{f}\!-\!n$ . □

These rules are all we need to propagate input values across a circuit.

**Corollary 5.30.** For a circuit 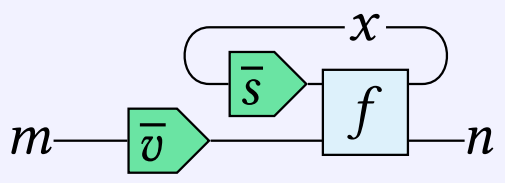 there exist $\overline{t} \in \mathbf{V}^x$ and $\overline{w} \in \mathbf{V}^n$ such that 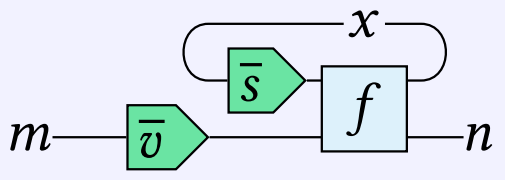 $\overset{*}{\rightsquigarrow}$ 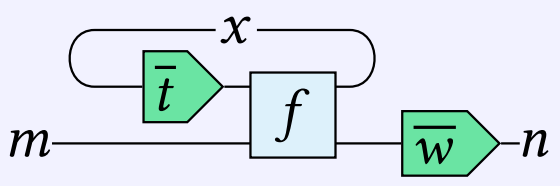 by applying Str once followed by the value rules exhaustively.

**Example 5.31.** Figure 5.7 shows how the value rules are applied to the streamed circuit from Figure 5.6. After performing all the reductions exhaustively on the 'now' circuit, the next state is t, the first output is ⊥ and the second is f. While the next state and second output make sense (if we apply Set, the state of the latch should turn true and the negated output false), the first output may raise eyebrows. This arises due to the presence of the delay; it will take another cycle to produce the expected output tf.

By now putting together all the components in this section and the previous, we have a productive strategy for processing inputs to *any* sequential circuit.

**Corollary 5.32** (Productivity)**.** For sequential circuit $m\!-\!\boxed{f}\!-\!n$ and inputs $\overline{v} \in \mathbf{V}^m$, there exists $\overline{w} \in \mathbf{V}^n$ such that $m\!-\!\boxed{\overline{v}}\!\!\rhd\!-\!\boxed{f}\!-\!n \overset{*}{\rightsquigarrow} m\!-\!\boxed{g}\!-\!\boxed{\overline{w}}\!\!\rhd\!-\!n$ by applying (Mealy), (IF) and (Str) once successively followed by the value rules exhaustively.

**Remark 5.33.** As we saw in Corollary 5.30, applying (Str) followed by the value rules to a circuit in Mealy form produces another circuit in Mealy form. This means that for one circuit and a long waveform stream of inputs, (Mealy) and (IF) need only be applied *once* at the very start before processing values.

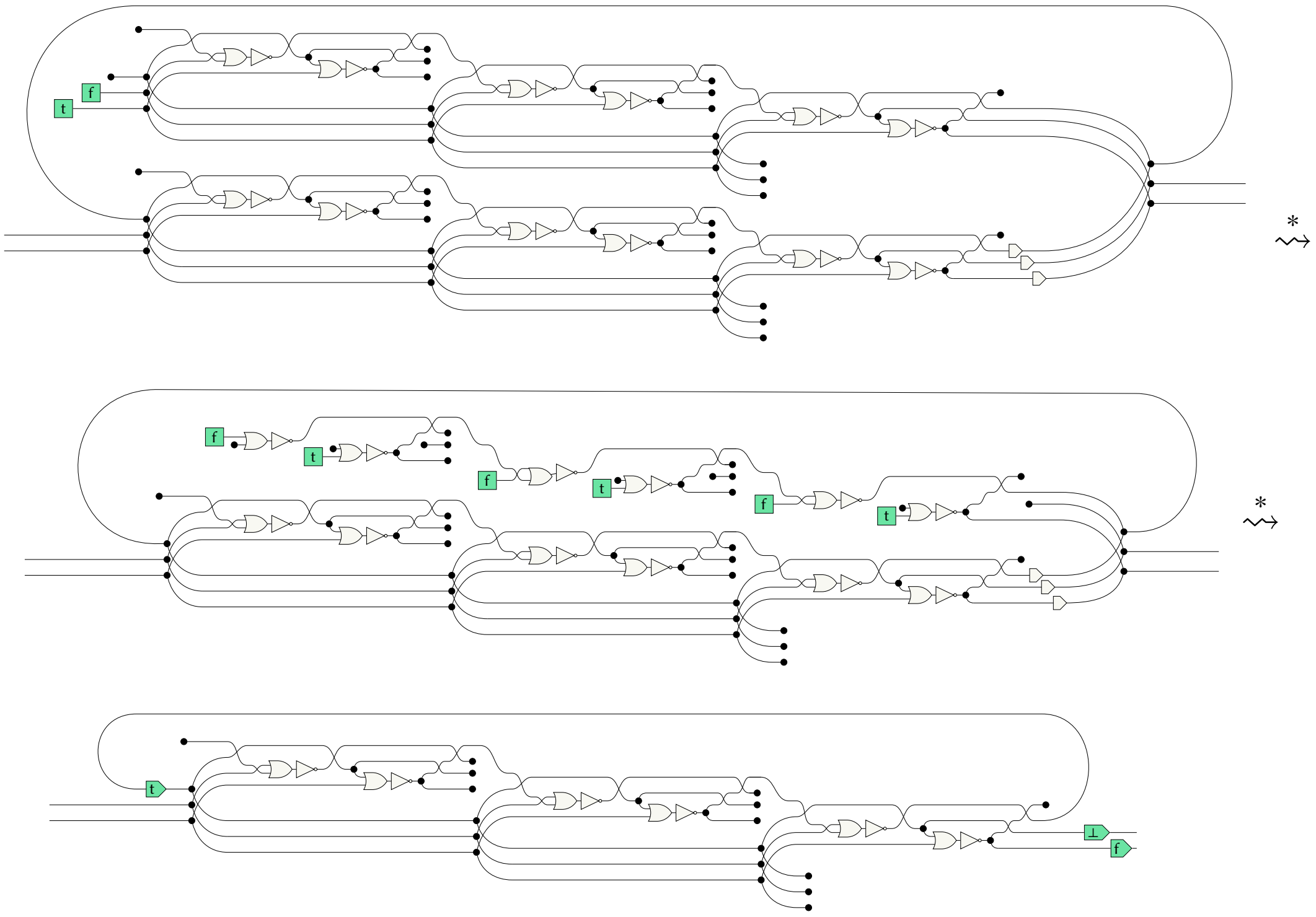

Figure 5.7: Using the value rules to reduce the streamed SR NOR latch circuit from Figure 5.6.

**Remark 5.34.** This style of operational semantics differs from some other approaches in the area, such as the work on signal flow graphs [BSZ21]. In these works, the operational semantics is specified in terms of the state transitions that take place in a circuit over time. For example, the rule that applies to the fork in signal flow graphs is

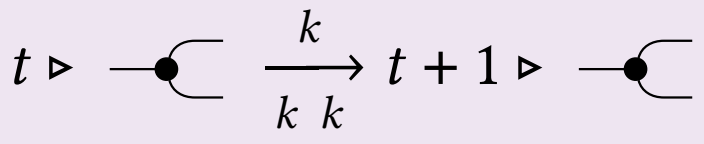

where $t$ is the current timestep, $k$ is the input signal and $k\ k$ is the (forked) output signal. Note that the fork itself does not change; the 'computation' occurring is contained entirely within the inputs and outputs.

In our world of digital circuits we are more interested in propagating values to see how this affects the internal structure of a circuit; this is another instance of how we are working with a *causal* rather than *relational* semantics. This means we specify inputs as explicit components, and the reductions actually change the structure of the circuit.

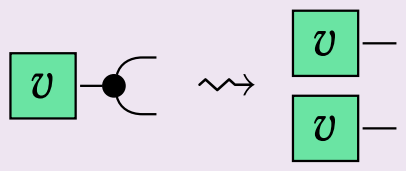

## 5.3 Observational equivalence

In the denotational semantics, we defined the relation of *denotational equivalence*, in which circuits are related if their denotations as streams are equal. For operational semantics we have another notion of relation on circuits: that of *observational equivalence*. This is due to Morris [Mor69], who named it 'extensional equivalence': essentially, two processes are observationally equivalent if they cannot be distinguished by their input-output behaviour.

Testing for observational equivalence is traditionally performed by checking that a program behaves the same in all *contexts*. For digital circuits, this means that for all possible streams of inputs, the circuit produces the same outputs. Of course, there are infinitely many streams of inputs, despite the set of values being finite. Fortunately, since circuits are constructed from a finite number of *components*, we need not check them all.

**Lemma 5.35.** Let

be a sequential circuit with $c$ delay components. Then applying Corollary 5.32 successively to a Mealy form of this circuit will produce at most $|\mathbf{V}|^c$ unique states.

*Proof.* The only varying elements of the state word are contributed by the $c$ delay components, as the values transition to $\perp$. □

To test all of the possible internal states of a circuit, we must use sequences of inputs long enough in time to guarantee that every possible state of a circuit is triggered.

**Notation 5.36** (Waveform)**.** The empty waveform is defined as 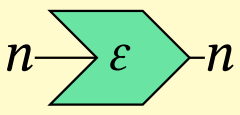

$\coloneqq$ 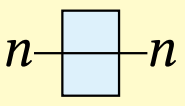

. Given values $\overline{v} \in \mathbf{V}^n$ and sequence $\underline{w} \in (\mathbf{V}^n)^\star$, the waveform for sequence $\overline{v} :: \underline{w}$ is drawn as 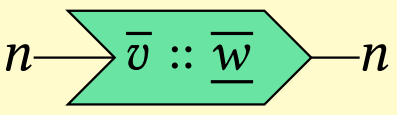

$\coloneqq$ 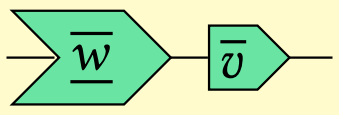

.

As a circuit with $c$ delay components has at most $|\mathbf{V}|^c$ states, to fully evaluate the behaviour of a circuit it suffices to check every waveform of length $|\mathbf{V}|^c + 1$. This is because even if such a waveform causes all $|\mathbf{V}|^c$ states to occur, the final element must produce a previously visited state, as there are no other states that could arise.

**Corollary 5.37.** Given a circuit in Mealy form 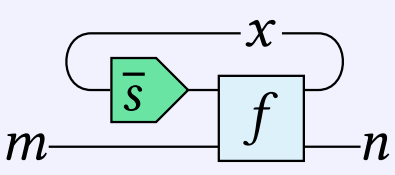

and input sequence $\underline{v} \in (\mathbf{V}^m)^\star$ of length $|\mathbf{V}|^c + 1$, there exists a state $\overline{r} \in \mathbf{V}^x$, an input sequence $\underline{u} \in (\mathbf{V}^m)^\star$ and output sequences $\underline{w}, \underline{z} \in (\mathbf{V}^n)^\star$ such that applying Corollary 5.32 yields the following reduction pattern:

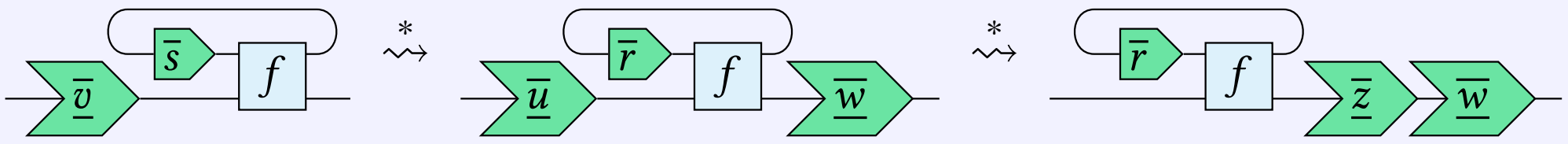

This means that every possible behaviour of a circuit can be evaluated using a finite number of sequences. This can be used to define our notion of observational equivalence for digital circuits.

**Definition 5.38** (Observational equivalence of circuits)**.** We say that two sequential circuits $m\text{-}\boxed{f}\text{-}n$ and $m\text{-}\boxed{g}\text{-}n$ with no more than $c$ delays are said to be *observationally equivalent under* $\mathcal{I}$, written $\text{-}\boxed{f}\text{-} \sim_{\mathcal{I}} \text{-}\boxed{g}\text{-}$ if applying productivity produces the same output waveforms for all input waveforms $\underline{v} \in (\mathbf{V}^m)^\star$ of length $|\mathbf{V}^c| + 1$.

Observational equivalence is our semantic relation for operational semantics, which relates two circuits based on their *execution*. To ensure it is suitable, it must be sound and complete with respect to the denotational semantics.

**Theorem 5.39.** Two sequential circuits $m\text{-}\boxed{f}\text{-}n$ and $m\text{-}\boxed{g}\text{-}n$ are observationally equivalent $m\text{-}\boxed{f}\text{-}n \sim_{\mathcal{I}} m\text{-}\boxed{g}\text{-}n$ if and only if $\left[\!\left[ m\text{-}\boxed{f}\text{-}n \right]\!\right]^{\mathrm{S}}_{\mathcal{I}} = \left[\!\left[ m\text{-}\boxed{g}\text{-}n \right]\!\right]^{\mathrm{S}}_{\mathcal{I}}$.

*Proof.* The ($\Rightarrow$) direction follows by Corollary 5.37, as every possible internal configuration of the circuit will be tested. For ($\Leftarrow$), if $\left[\!\left[ m\text{-}\boxed{f}\text{-}n \right]\!\right]^{\mathrm{S}}_{\mathcal{I}} = \left[\!\left[ m\text{-}\boxed{g}\text{-}n \right]\!\right]^{\mathrm{S}}_{\mathcal{I}}$, then this means $\left[\!\left[ m\text{-}\boxed{f}\text{-}n \right]\!\right]^{\mathrm{S}}_{\mathcal{I}}(\underline{v} :: \sigma) = \left[\!\left[ m\text{-}\boxed{g}\text{-}n \right]\!\right]^{\mathrm{S}}_{\mathcal{I}}(\underline{v} :: \sigma)$ for any $\sigma, \tau \in (\mathbf{V}^m)^\omega$. By definition of $[\![-]\!]^{\mathrm{S}}_{\mathcal{I}}$, we then have that $\left[\!\left[ m \rhd \underline{v} \rhd \boxed{f}\text{-}n \right]\!\right]^{\mathrm{S}}_{\mathcal{I}}(\sigma) = \left[\!\left[ m \rhd \underline{v} \rhd \boxed{g}\text{-}n \right]\!\right]^{\mathrm{S}}_{\mathcal{I}}(\sigma)$. Since this holds for *all* sequences $\underline{v}$, it must hold for those of length $|\mathbf{V}|^c + 1$, so the condition for observational equivalence is met. □

To verify that this is the 'best' equivalence relation, we turn to a definition of observational equivalence in terms of universal properties [Gor98]. Gordon states that a relation is an *adequate* observational semantics if it only relates circuits that have the same denotational semantics; observational equivalence is defined as the largest adequate congruence.

**Corollary 5.40.** $\sim_{\mathcal{I}}$ is the largest adequate congruence on $\mathbf{SCirc}_\Sigma$.

*Proof.* For $\sim_{\mathcal{I}}$ to be a congruence it must be preserved by composition, tensor and trace, and for it to be the largest there must be no denotationally equal circuit it does not relate. These, along with adequacy, all follow by Theorem 5.39. □

This makes $\sim_{\mathcal{I}}$ a suitable notion of observational equivalence for sequential circuits.

**Definition 5.41.** Let $\mathbf{SCirc}_{\Sigma/\sim_{\mathcal{I}}}$ be defined as $\mathbf{SCirc}_{\Sigma}/\sim_{\mathcal{I}}$.

**Corollary 5.42.** There is an isomorphism $\mathbf{SCirc}_{\Sigma/\approx_{\mathcal{I}}} \cong \mathbf{SCirc}_{\Sigma/\sim_{\mathcal{I}}}$.

The results of the previous section give us an upper bound on the length of waveforms required to establish observational equivalence, so we have a terminating strategy for comparing digital circuits. Unfortunately, this is still an *exponential* upper bound, so it is infeasible to check for the equivalence of circuits with more than a few delay components. Nevertheless, the operational semantics gives us a straightforward way to *evaluate* circuits while respecting their internal structure, unlocking more insight as to *why* circuits are behaving the way they are.

Moreover, while it may be infeasible to check *every single possible input* to a circuit, it is often the case that one knows a particular input is fixed. By precomposing the circuit with appropriate infinite waveforms to represent the fixed inputs, insight and potential optimisations may be gleaned; this is known as *partial evaluation*, which will be examined more in Chapter 7.

## 5.4 Operational semantics for generalised circuits

When dealing with arbitrary-width wires, the only part of the operational semantics that does not completely generalise in the obvious way are the value rules.

**Lemma 5.43** (Generalised value rules)**.** The following *generalised value rules* are sound:

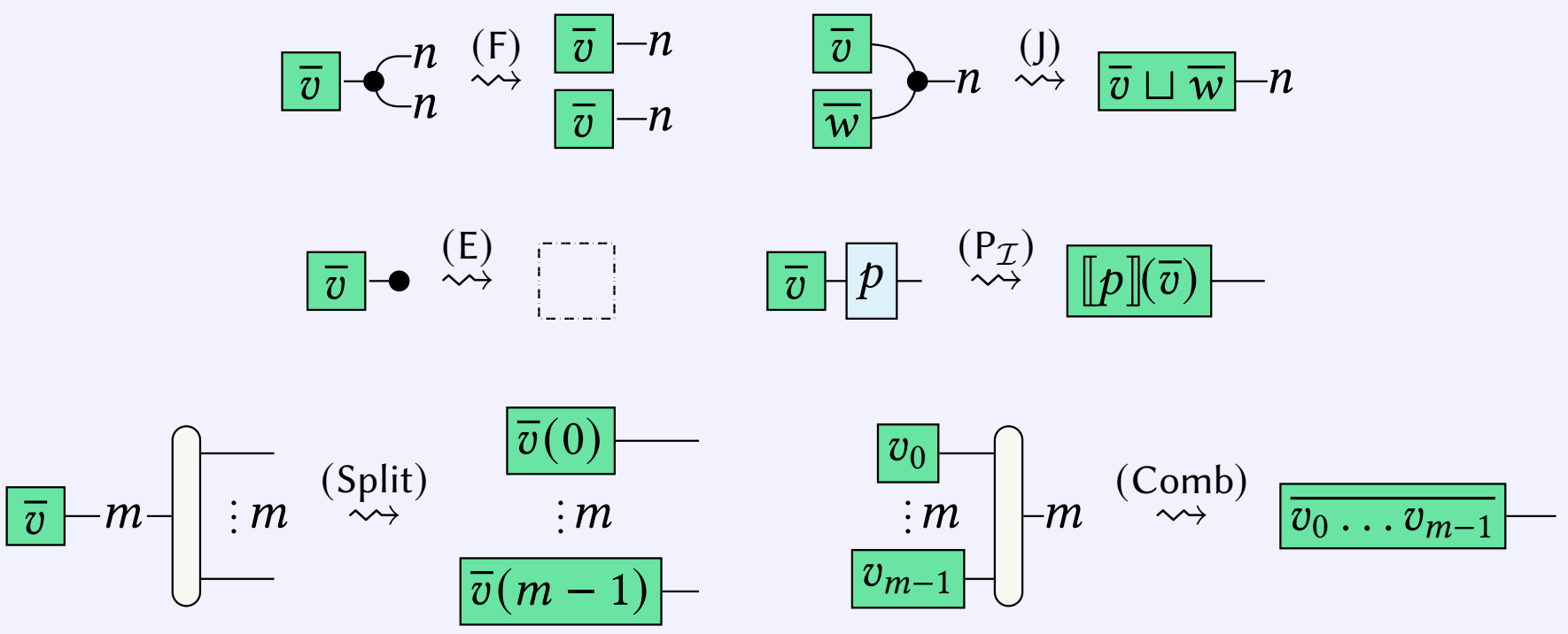

**Lemma 5.44.** Applying the generalised value rules is confluent.

*Proof.* There are no overlaps between the rules. □

**Lemma 5.45.** | For a generalised combinational circuit $\overline{m}\!-\!\boxed{f}\!-\!\overline{n}$ and $\overline{v} \in \mathbf{V}^{\overline{m}}$, there exists a word $\overline{w} \in \mathbf{V}^{\overline{n}}$ such that applying the value rules exhaustively to $\boxed{\overline{v}}\!-\!\boxed{f}\!-$ terminates at $\boxed{\overline{w}}\!-$ .

With these rules, the inputs to a generalised circuit can be processed.

**Corollary 5.46.** For generalised circuit 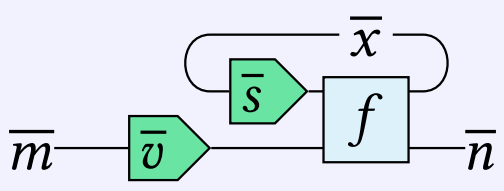 there exist $\overline{t} \in \mathbf{V}^{\overline{x}}$ and $\overline{w} \in \mathbf{V}^{\overline{n}}$ such that 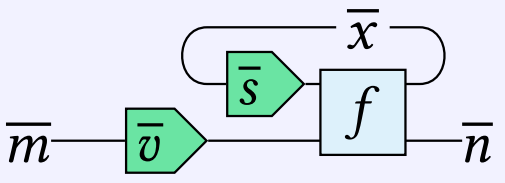 $\overset{*}{\rightsquigarrow}$ 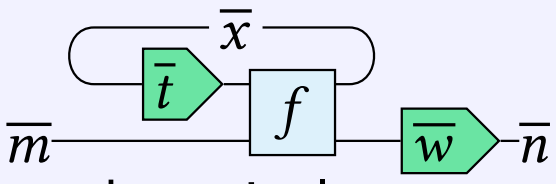 by applying (Str) once followed by the generalised value rules exhaustively.

**Corollary 5.47** (Generalised productivity)**.** For sequential circuit $\overline{m}\!-\!\boxed{f}\!-\!\overline{n}$ and inputs $\overline{v} \in \mathbf{V}^{\overline{m}}$, there exists $\overline{w} \in \mathbf{V}^{\overline{n}}$ such that $\overline{m}\!-\!\boxed{\overline{v}\,\triangleright}\!-\!\boxed{f}\!-\!\overline{n} \overset{*}{\rightsquigarrow} \overline{m}\!-\!\boxed{g}\!-\!\boxed{\overline{w}\,\triangleright}\!-\!\overline{n}$ by applying Mealy, IF and Str once successively followed by the value rules exhaustively.

Since register components can now hold words rather than just values, for observational equivalence we must consider longer input waveforms.

**Definition 5.48** (Register width)**.** Given a generalised sequential circuit $\overline{m}\!-\!\boxed{f}\!-\!\overline{n}$ , let $c_n$ be the number of $n$-width delay components $n\!-\!\boxed{\triangleright}\!-\!n$ ; the *register width* of $\overline{m}\!-\!\boxed{f}\!-\!\overline{n}$ is computed as $\Sigma_{n \in \mathbb{N}}\, c_n \cdot n$.

**Definition 5.49.** We say that two generalised sequential circuits $\overline{m}\!-\!\boxed{f}\!-\!\overline{n}$ and $\overline{m}\!-\!\boxed{g}\!-\!\overline{n}$ with register width at most $c$ are said to be *observationally equivalent under $\mathcal{I}$*, written $-\!\boxed{f}\!- \sim^{+}_{\mathcal{I}} -\!\boxed{g}\!-$ if applying productivity produces the same output waveforms for all input waveforms $\underline{\overline{v}} \in (\mathbf{V}^{\overline{m}})^{\star}$ of length $|\mathbf{V}^{c}| + 1$.

The observational equivalence results from the previous section then generalise nicely to the multicoloured case.

**Theorem 5.50.** Given two sequential circuits $\overline{m}\!-\!\boxed{f}\!-\!\overline{n}$ and $\overline{m}\!-\!\boxed{g}\!-\!\overline{n}$ , we have that $\overline{m}-\boxed{f}-\overline{n} \sim^{+}_{\mathcal{I}} \overline{m}-\boxed{g}-\overline{n}$ if and only if $\left[\!\left[ \overline{m}-\boxed{f}-\overline{n} \right]\!\right]^{\mathrm{S}}_{\mathcal{I}} = \left[\!\left[ \overline{m}-\boxed{g}-\overline{n} \right]\!\right]^{\mathrm{S}}_{\mathcal{I}}$.

**Corollary 5.51.** $\sim^{+}_{\mathcal{I}}$ is the largest adequate congruence on $\mathsf{SCirc}^{+}_{\Sigma}$.

**Definition 5.52.** Let $\mathsf{SCirc}^{+}_{\Sigma/\sim^{+}_{\mathcal{I}}}$ be defined as $\mathsf{SCirc}^{+}_{\Sigma}/\sim^{+}_{\mathcal{I}}$.

**Corollary 5.53.** There is an isomorphism $\mathsf{SCirc}^{+}_{\Sigma/\approx^{+}_{\mathcal{I}}} \cong \mathsf{SCirc}^{+}_{\Sigma/\sim^{+}_{\mathcal{I}}}$.

Chapter 6

# Algebraic semantics

The operational semantics and notion of observational equivalence means that the behaviour of two circuits can be compared by checking whether every input produces the same output for both circuits. But this is somewhat of a nuclear option; rather than using what we know about the structure of a circuit's components, we just blast away exhaustively trying all the inputs to find a contradiction.

A more elegant method of reasoning is by defining a set of *equations* between subcircuits and *quotienting* $\mathbf{SCirc}_\Sigma$ by these equations. A proof of equivalence between two circuits is then presented using algebraic reasoning: applying equations to translate one circuit into the other. This is often far more efficient than having test every input, and is known formally as an *algebraic* semantics.

**Example 6.1.** For the last time we return to the language of arithmetical expressions from Example 4.1. An algebraic semantics for this language can be defined using a set of equations: the familiar equations of associativity, commutativity, unitality, annihiliation and distributivity, along with equations for performing arithmetic.

$$\mathit{add}\ (\mathit{add}\ e_1\ e_2)\ e_3 = \mathit{add}\ e_1\ (\mathit{add}\ e_2\ e_3) \qquad \mathit{mul}\ (\mathit{mul}\ e_1\ e_2)\ e_3 = \mathit{mul}\ e_1\ (\mathit{mul}\ e_2\ e_3)$$
$$\mathit{add}\ e_1\ e_2 = \mathit{add}\ e_2\ e_1 \qquad \mathit{mul}\ e_1\ e_2 = \mathit{mul}\ e_2\ e_1$$
$$\mathit{add}\ e_1\ \overline{0} = e_1 \qquad \mathit{mul}\ e_1\ \overline{1} = e_1 \qquad \mathit{mul}\ e_1\ \overline{0} = \overline{0}$$
$$\mathit{mul}\ e_1\ (\mathit{add}\ e_2\ e_3) = \mathit{add}\ (\mathit{mul}\ e_1\ e_2)\ (\mathit{mul}\ e_1\ e_3)$$
$$\mathit{add}\ \overline{n_1}\ \overline{n_2} = \overline{n_1 + n_2} \qquad \mathit{add}\ \overline{n_1}\ \overline{n_2} = \overline{n_1 \cdot n_2}$$

If everything is specified concretely as values then one could easily just use the last two equations to compare two expressions by reducing two expressions to values as in the operational semantics. The power of the algebraic semantics comes from the fact we can reason abstractly with expressions containing *blackboxes*. Take the following example, containing some arbitrary component $e$.

$$
\begin{aligned}
mul\ (add\ e\ (mul\ e\ \overline{3}))\ \overline{2} &= mul\ (add\ (mul\ e\ 1)\ (mul\ e\ \overline{3}))\ \overline{2} \\
&= mul\ (mul\ e\ add(\overline{1}\ \overline{3}))\ \overline{2} \\
&= mul\ (mul\ e\ \overline{4})\ \overline{2} \\
&= mul\ e\ (mul\ \overline{4}\ \overline{2}) \\
&= mul\ e\ \overline{8} \\
&= mul\ \overline{8}\ e
\end{aligned}
$$

Despite not specifying the structure of $e$, we have shown how the expression is equal to a slightly simpler one. This process creates new general equations which can be used as 'shortcuts' in future reasoning, potentially saving many steps.

As with the operational semantics, we are especially interested in defining a *sound and complete* algebraic semantics with respect to the denotational semantics. That is to say, for each equation $-\boxed{f}- = -\boxed{g}-$ then $\left[\!\left[ -\boxed{f}- \right]\!\right]^{\mathsf{S}}_{\mathcal{I}} = \left[\!\left[ -\boxed{g}- \right]\!\right]^{\mathsf{S}}_{\mathcal{I}}$, and there must be enough equations such that if $\left[\!\left[ -\boxed{f}- \right]\!\right]^{\mathsf{S}}_{\mathcal{I}} = \left[\!\left[ -\boxed{g}- \right]\!\right]^{\mathsf{S}}_{\mathcal{I}}$ then there exists a series of equations identifying $-\boxed{f}-$ and $-\boxed{g}-$.

**Remark 6.2.** An 'equational theory' for sequential circuits was one of the first things presented in the previous work [GJ16; GJL17a]. In that paper, equations that 'seemed right' were used to quotient the syntax, with the ultimate aim of showing that the resulting category was *Cartesian*. This was done quite informally, and was made more confusing as the categories of circuits were subsequently quotiented by some notion of 'extensional equivalence', an attempt to rectify the fact that the equations only dealt with closed circuits. Soundness and completeness of the equational theory was not considered because there was nothing to compare it against.

In essence, the previous work was almost 'the wrong way round': equations were defined and semantics drawn from them. In the more recent version of the work [GKS24, Sec. 5], which forms the basis for this chapter, the equations are derived from the denotational semantics. Not only does this give us a formal way of verifying that these equations are sound, it sets the backdrop against which we can test if the algebraic semantics are sufficient: are any two denotationally equal circuits identified by equations?

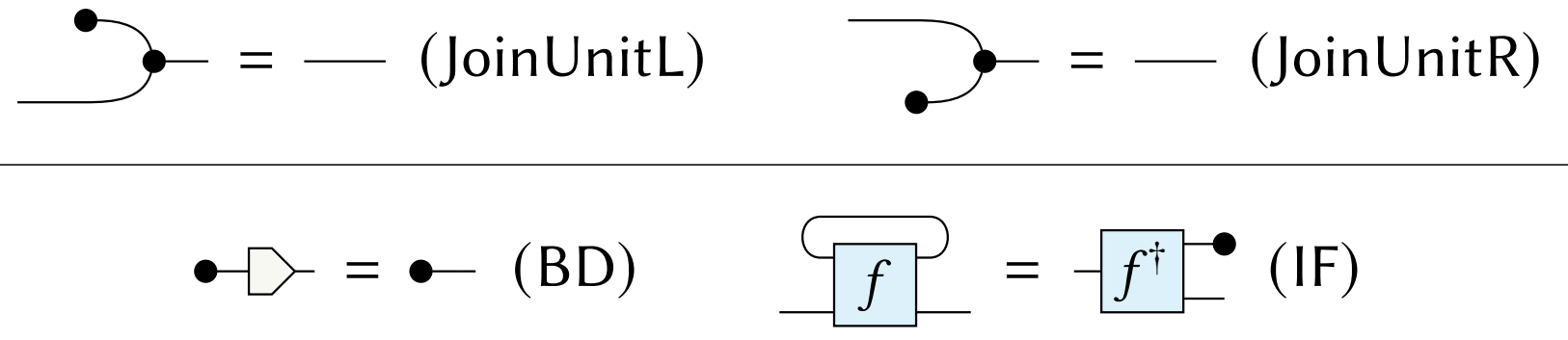

Figure 6.1: Set of Mealy equations $\mathcal{M}$.

When defining such an equational theory, there may be several different sound and complete formulations. Ideally, we want to stick to simple *local* equations that concern the interactions of concrete generators as much as possible, but as we will see we will sometimes have no choice but to define *families* of equations parameterised over some arbitrary subcircuit.

## 6.1 Normalising circuits

How does one start when trying to define a complete set of equations for some framework? Usually the strategy is to define enough equations to bring any term to some sort of (pseudo-)*normal form*; the theory is then complete if terms with the same semantics have the same normal form.

We have already seen something that looks a bit like a normal form: the *Mealy form* from the previous section. This is by no means a true normal form, as there are many different Mealy forms that represent the same behaviour. Nevertheless, it is a useful starting point so we will need equations to bring circuits to Mealy form in our theory.

Instead of just turning the Mealy reduction rules into equations, we will show how Mealy form can be derived using smaller equations.

**Definition 6.3.** The set $\mathcal{M}$ of *Mealy equations* in Figure 6.1 are sound.

*Proof.* The first two rules hold as the join is a monoid in the stream semantics. The (BD) holds because the semantics of the delay component are to output a $\perp$ value first and then the (delayed) inputs: as the semantics of the 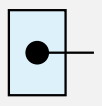 component are to *always* produce $\perp$, then it does not make a difference how delayed it is. The final equation is the instant feedback rule, which is sound by Proposition 5.19. □

**Proposition 6.4.** Given a sequential circuit $m$–$\boxed{f}$–$n$, there exists a combinational circuit $\begin{smallmatrix} x \\ m \end{smallmatrix}$–$\boxed{g}$–$\begin{smallmatrix} x \\ n \end{smallmatrix}$ and values $\bar{s} \in \mathbf{V}^x$ such that 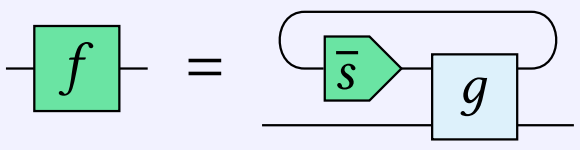 in $\mathbf{SCirc}_\Sigma/\mathcal{M}$.

*Proof.* Any circuit can be assembled into global trace-delay form solely using the axioms of STMCs. From this, a circuit in pre-Mealy form can be obtained by translating delays and values into registers using the following equations:

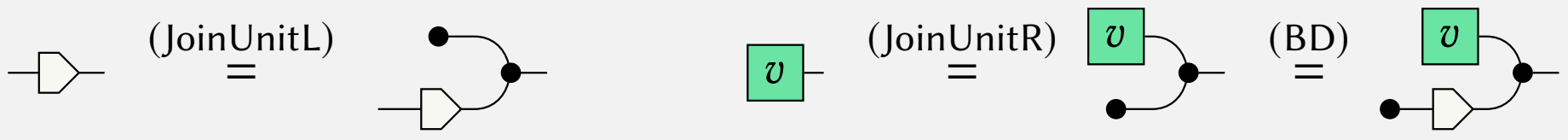

Subsequently a circuit in Mealy form can be obtained by applying the (IF) rule. □

$\mathbf{SCirc}_\Sigma/\mathcal{M}$ is a category in which all circuits are equal to at least one circuit in Mealy form. In general, there will be many Mealy forms depending on the ordering one picks for the delays and value; our task is to provide equations to map any two denotationally equivalent circuits to the *same* Mealy form.

Even if the combinational cores of two Mealy forms have the same behaviour, they may not have the same structure. To reduce the number of cores we have to consider, we will first establish equations for translating any combinational circuit into some canonical circuit. We already met a method for determining what this canonical circuit is: the functional completeness map $||-||$ from $\mathbf{Func}_\mathcal{I}$ to $\mathbf{SCirc}_\Sigma$.

**Definition 6.5** (Normalised circuit)**.** A sequential circuit $m$–$\boxed{f}$–$n$ is *normalised* if it is in the image of $||-||$.

As a shorthand, we will often abuse notation and write $||f|| :=$ –$\boxed{||f||}$–. Recall that even though $||-||$ maps into $\mathbf{SCirc}_\Sigma$, every circuit in its image has combinational behaviour. This is quite an important distinction to make, so we will give it a proper name.

**Definition 6.6** (Essentially combinational)**.** A sequential circuit is *essentially combinational* if it is in the form 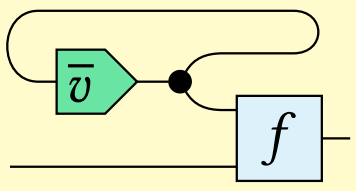 .

Such circuits are sequential circuits that exhibit combinational behaviour: any value components are only used to introduce constants which do not alter over time.

As the normalised version of a given circuit is interpretation-dependent, there is no standard set of equations for normalising a circuit. Instead, these must be specified on an interpretation-by-interpretation basis.

**Definition 6.7** (Normalising equations)**.** For a complete interpretation $(\mathcal{I}, ||-||)$, a set of equations $\mathcal{N}_{\mathcal{I}}$ is *normalising* if any essentially combinational circuit $m$–$\boxed{f}$–$n$ is equal to a circuit in the image of $||-||$ by equations in $\mathcal{N}_{\mathcal{I}}$.

**Definition 6.8** (Normalisable interpretation)**.** A complete interpretation $\mathcal{I}$ is called *normalisable* if there exists a set of normalising equations $\mathcal{N}_{\mathcal{I}}$.

The normalising equations for a given interpretation can be used to translate a combinational core into a circuit from which it is easy to read off a truth table.

**Theorem 6.9.** For every sequential circuit $m$–$\boxed{f}$–$n$ in a normalisable complete interpretation $(\mathcal{I}, ||-||)$ over $\Sigma$, there exists essentially combinational 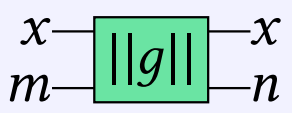 and $\bar{s} \in \mathbf{V}^x$ such that $m$–$\boxed{f}$–$n$ = 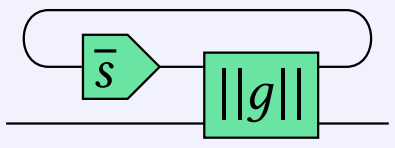 in $\mathbf{SCirc}_\Sigma/\mathcal{M} + \mathcal{N}_{\mathcal{I}}$.

*Proof.* By Proposition 6.4, $m$–$\boxed{f}$–$n$ = 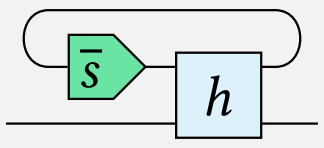 and by equations in $\mathcal{N}_{\mathcal{I}}$, $\boxed{h}$ = $\boxed{||g||}$ . ☐

## 6.2 Encoding equations

A circuit in Mealy form is a syntactic representation of a Mealy machine: the combinational core is the Mealy function, and the registers are the initial state. It is important to determine the states that the circuit might assume, as these determine whether or not an equation is valid.

**Definition 6.10** (States)**.** Let $f : \mathbf{V}^{x+m} \to \mathbf{V}^{x+n}$ be a monotone function and let $\bar{s} \in \mathbf{V}^x$ be a state. Then the *states of $f$ from $\bar{s}$*, denoted $S_{f,\bar{s}}$, is the smallest set containing $\bar{s}$ and closed under $\bar{r} \mapsto \pi_0(f(\bar{r}, \bar{v}))$ for any $\bar{v} \in \mathbf{V}^m$.

**Example 6.11.** Consider the circuit 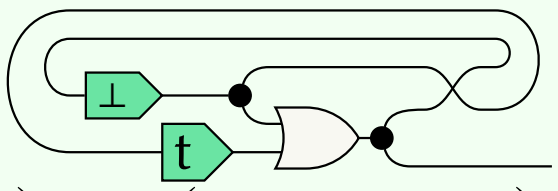 . The semantics of the combinational core are clearly $(s, r) \mapsto (s \vee r, s, s \vee r)$, where the first two characters are the next state and the third is the output. The initial state is $\bot\mathsf{t}$, so the subsequent states are $(\bot \vee \mathsf{t}, \bot) = (\mathsf{t}, \bot)$ and $(\mathsf{t} \vee \bot, \mathsf{t}) = (\mathsf{t}, \mathsf{t})$. As $(\mathsf{t} \vee \mathsf{t}, \mathsf{t}) = (\mathsf{t}, \mathsf{t})$, there are no more circuit states and the complete set is $\{(\bot, \mathsf{t}), (\mathsf{t}, \bot), (\mathsf{t}, \mathsf{t})\}$.

Note that as the output of the circuit is computed as $s \vee r$, for each circuit

state the output is t. This means that the circuit is denotationally equivalent to 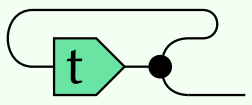 , but this circuit only has a single state t.

We need to *encode* the states of one circuit as another; we have already encountered this notion using *Mealy homomorphisms* (Definition 4.55); functions between the state sets that preserve transitions and outputs. While two 'inverse' homomorphisms may not be isomorphisms, the round trip will always map to a state with the same behaviour.

**Lemma 6.12.** Given two Mealy homomorphisms $h\colon (S, f) \to (T, g)$ and $h'\colon (T, g) \to (S, f)$, any state $s \in S$ and input $a \in A$, $f_0(s, a) = f_0(h'(h(s)), a)$.

*Proof.* Immediate as Mealy homomorphisms preserve outputs. □

We will use Mealy homomorphisms as circuits to encode state sets; this means we need to ensure the encoders and decoders are monotone.

**Lemma 6.13.** For partial orders $S$ and $T$ and monotone Mealy coalgebra $(S, f)$ and $(T, g)$, any Mealy homomorphism $h\colon (S, f) \to (T, g)$ is monotone.

*Proof.* In a monotone Mealy coalgebra, the functions $f$ and $g$ are monotone, and for $h$ to be a Mealy homomorphism, $f_1(s) = g_1(h(s))$. For states $s, r \in S$ we have $g_1(h(s), a) = f_1(s, a) \leq f_1(r, a) = g_1(h(r), a)$. This means that the function $s \mapsto g_1(h(s), a)$ is monotone; as $g_1$ is monotone, $h$ must also be monotone. □

For two circuits 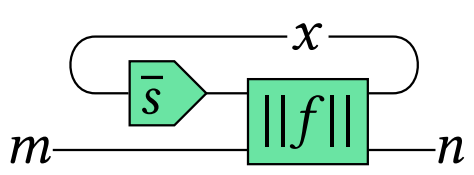 and 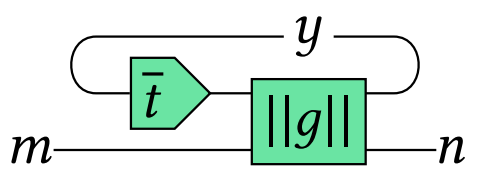 , the encoders and decoders will be homomorphisms $(S_{f,\bar{s}}, f) \to (S_{g,\bar{t}}, g)$ and $(S_{g,\bar{t}}, g) \to (S_{f,\bar{s}}, f)$. These are homomorphisms on the subset of states that a circuit can assume, *not* the entire set of words that can fit into the state. This means that encoding and decoding circuits cannot be inserted arbitrarily but only in certain contexts.

**Proposition 6.14** (Encoding equation)**.** For a normalised circuit 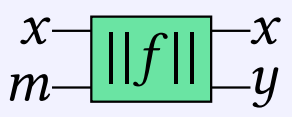 and $\bar{s} \in \mathbf{V}^x$, let $\mathsf{enc}\colon S_{f,\bar{s}} \to \mathbf{V}^y$ and $\mathsf{dec}\colon \mathbf{V}^y \to S_{f,\bar{s}}$ be functions such that $\mathsf{dec} \circ \mathsf{enc}\colon S_{f,\bar{s}} \to S_{f,\bar{s}}$ is a Mealy homomorphism. Then the *encoding equation* (Enc) in Figure 6.2 is sound, where $\mathsf{enc}_\mathsf{m}$ and $\mathsf{dec}_\mathsf{m}$ are monotone completions as defined in Definition 4.90.

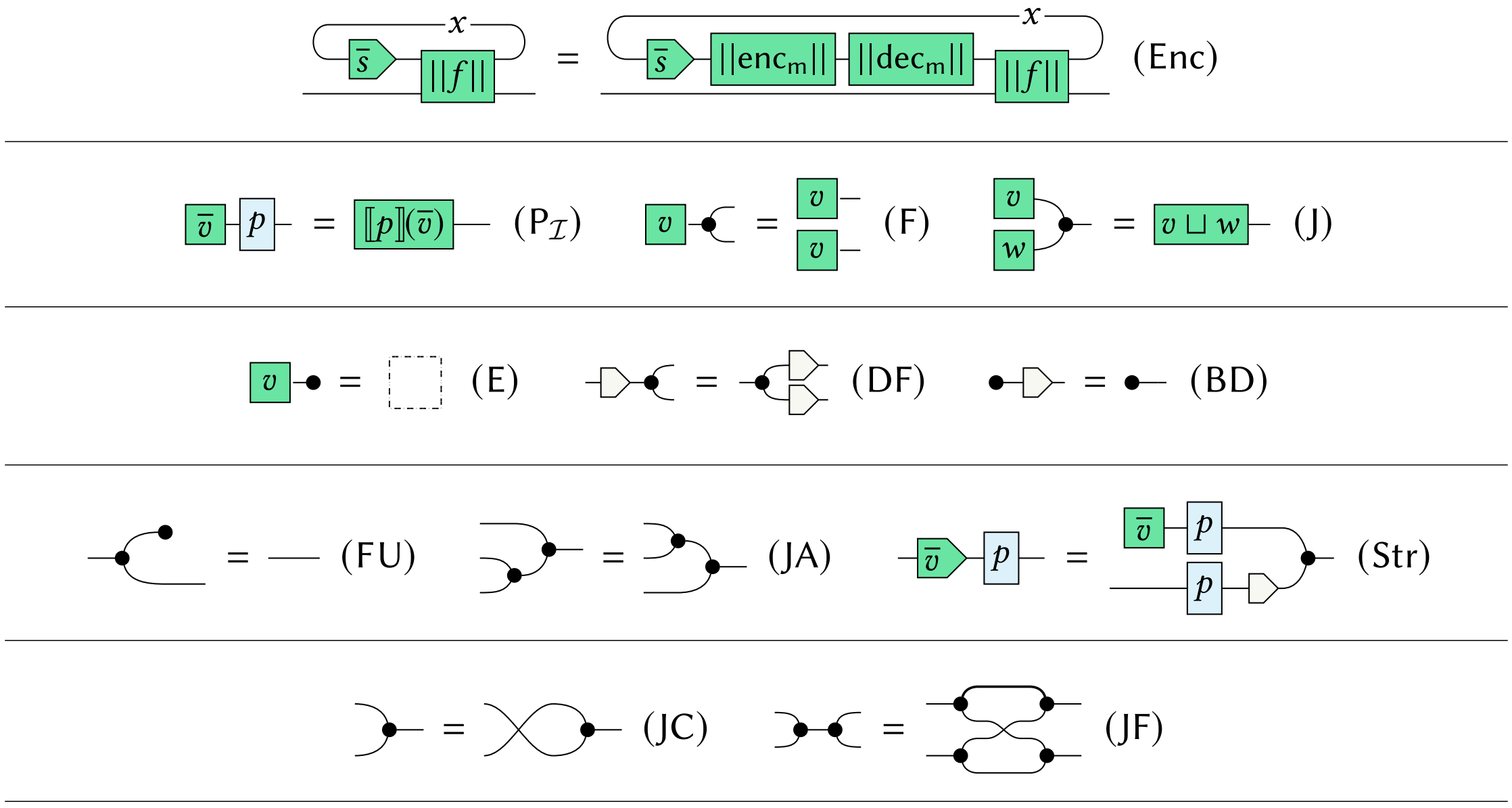

Figure 6.2: Set $\mathcal{H}$ of equations for encoding circuit states

*Proof.* Let $g$ be defined as the map $\bar{r} \mapsto$ $\left[\!\left[$ 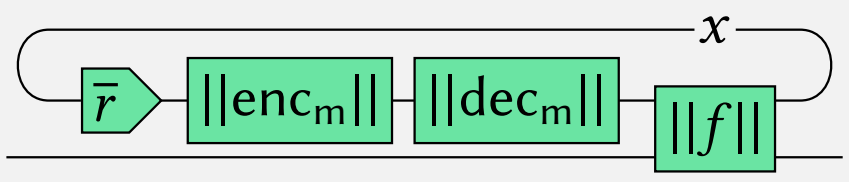 $\right]\!\right]^{\mathsf{S}}_{\mathcal{I}}$; by Proposition 4.96 we know that $g(\bar{t})[\bar{v}] = \pi_1(f(\mathrm{dec}(\mathrm{enc}(\bar{t})), \bar{v}))$ and $g(\bar{t})_{\bar{v}} = g(\pi_0(f(\mathrm{dec}(\mathrm{enc}(\bar{t})), \bar{v})))$. As dec $\circ$ enc is a Mealy homomorphism, for $\bar{t} \in S_{f,\bar{s}}$ we have that $g(\bar{t})[\bar{v}] = \pi_1(f(\bar{t}), \bar{v})$ and that $g(\bar{t})_{\bar{v}}$ shares outputs and transitions with $g(\pi_0(f(\bar{t})), \bar{v})$. As 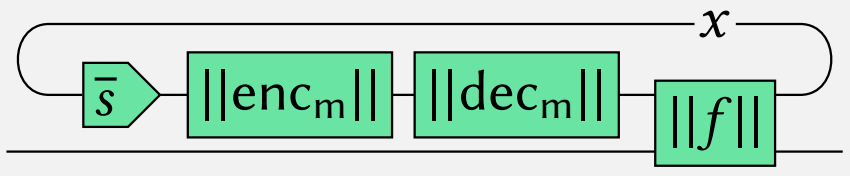 $\coloneqq g(\bar{s})$ and $\bar{s} \in S_{f,\bar{s}}$, every subsequent stream derivative will also be of the form $g(\bar{t})$ where $\bar{t} \in S_{f,\bar{s}}$, so the equation is sound. □

**Remark 6.15.** The encoding equation is an equation *schema*: this is required because the width of a circuit state can be arbitrarily large, and each extra bit adds a whole new set of Mealy homomorphisms to consider.

The encoding equation only inserts encoder circuits; to actually change the state we need some more equations.

**Lemma 6.16.** The equations on the bottom four rows of Figure 6.2 are sound.

*Proof.* It is a straightforward exercise to compare the stream functions. □

To show the final result we must prove some lemmas; first we show how we can 'pump' a value out of an infinite waveform.

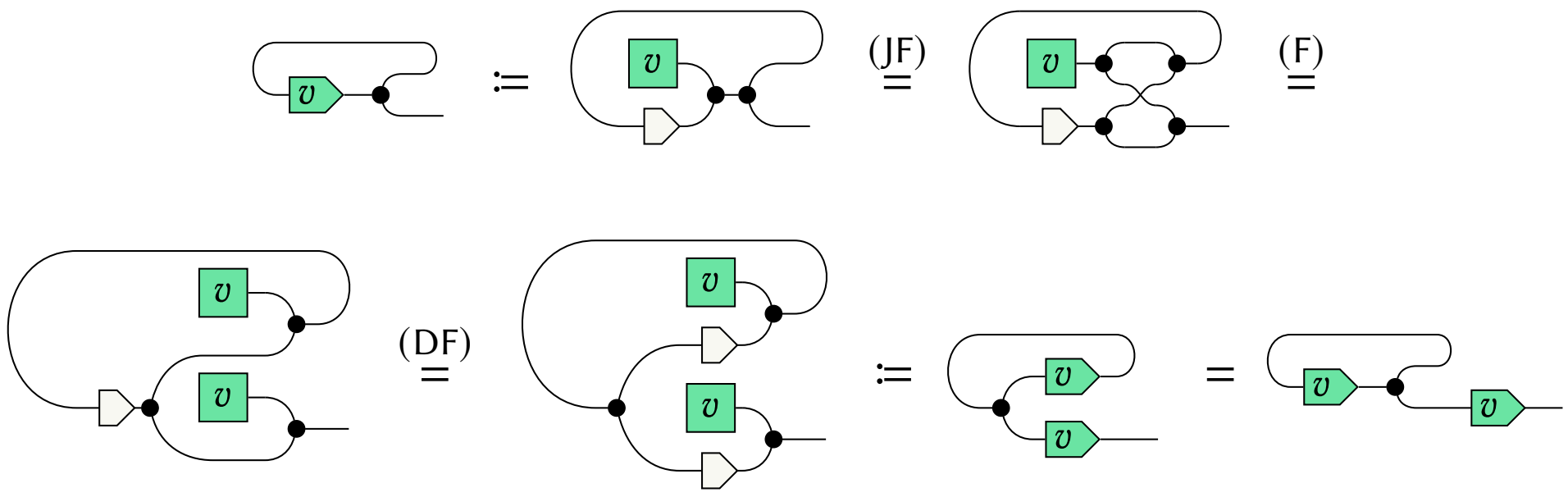

Figure 6.3: Proof of Lemma 6.17

**Lemma 6.17.** For $v \in \mathbf{V}$, 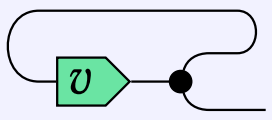 $=$ 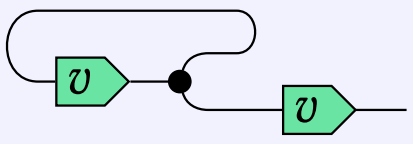 using the encoding equations.

*Proof.* The proof is straightforward and is illustrated in Figure 6.3. □

The next lemma shows how the familiar 'streaming' rule from the operational semantics can be derived equationally.

**Lemma 6.18.** For a combinational circuit 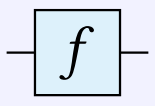, 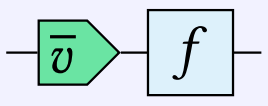 $=$ 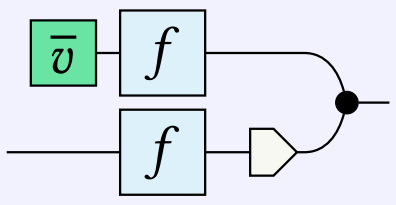 by the encoding equations.

*Proof.* This is by induction on the structure of 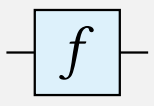. The case for the primitive is immediate by (Str). For 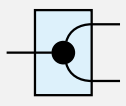 we have that

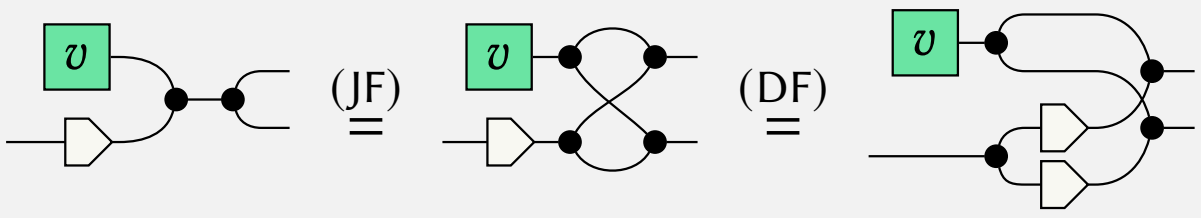

The proof for 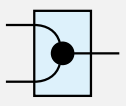 is illustrated in Figure 6.4. The case for 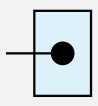 is trivial, and the case for 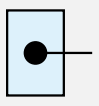 follows by (CU) and (BD). The cases for 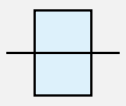 and 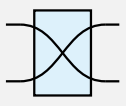 follow by axioms of STMCs. Since the underlying circuit is combinational, for the inductive cases we just need to check composition and tensor, which are trivial. □

We next show how the encoding equations can be used to translate combinational circuits with inputs into values.

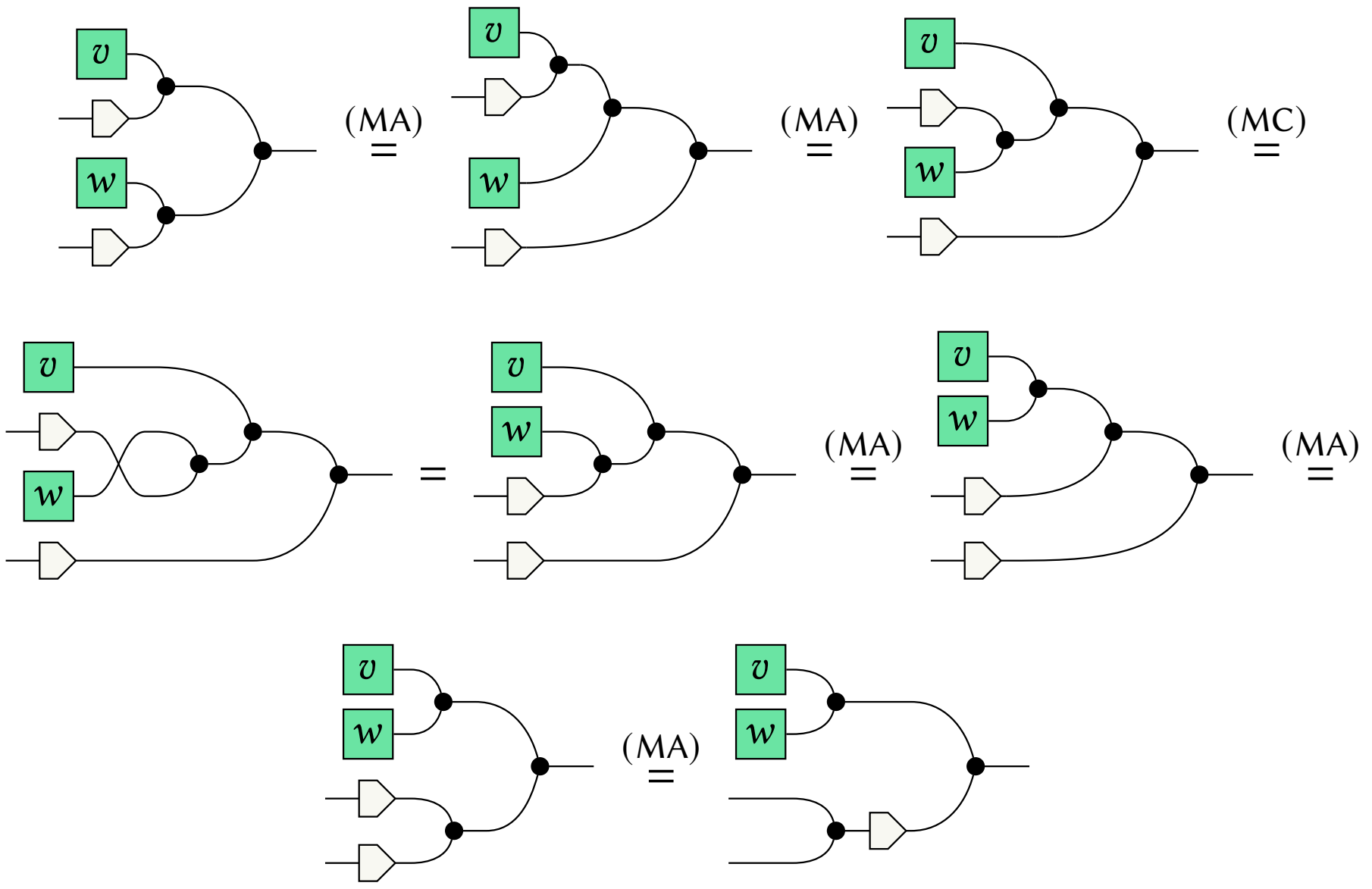

Figure 6.4: Proof of Lemma 6.18 for the join case

**Lemma 6.19.** Let $m\!-\!\boxed{f}\!-\!n$ be a combinational circuit such that $\left[\!\!\left[ -\boxed{f}- \right]\!\!\right]^{\mathrm{C}}_{\mathcal{I}} = g$. Then $\boxed{\overline{v}}\!-\!\boxed{f}\!- = \boxed{g(\overline{v})}\!-$ by the encoding equations.

*Proof.* For the same reasoning as Lemma 5.29, the $(\mathrm{P}_{\mathcal{I}})$, (F), (J) and (E) equations can be used to show that there exists $\overline{w} \in \mathbf{V}^n$ such that $m\!-\!\boxed{f}\!-\!n = \boxed{\overline{w}}\!-$.

Now we need to show that $\left[\!\!\left[ \boxed{\overline{v}}\!-\!\boxed{f}\!- \right]\!\!\right]^{\mathrm{S}}_{\mathcal{I}} = \left[\!\!\left[ \boxed{g(\overline{v})}\!- \right]\!\!\right]^{\mathrm{S}}_{\mathcal{I}}$. By functoriality of $[\![-]\!]^{\mathrm{S}}_{\mathcal{I}}$, $\left[\!\!\left[ \boxed{\overline{v}}\!-\!\boxed{f}\!- \right]\!\!\right]^{\mathrm{S}}_{\mathcal{I}} = \left[\!\!\left[ \boxed{\overline{v}}\!- \right]\!\!\right]^{\mathrm{S}}_{\mathcal{I}} \mathbin{;} \left[\!\!\left[ -\boxed{f}- \right]\!\!\right]^{\mathrm{S}}_{\mathcal{I}}$. By Lemma 4.81 we know that $\left[\!\!\left[ -\boxed{f}- \right]\!\!\right]^{\mathrm{S}}_{\mathcal{I}}(\sigma)(i) = \left[\!\!\left[ -\boxed{f}- \right]\!\!\right]^{\mathrm{C}}_{\mathcal{I}} = g(\sigma)(i)$ for all $\sigma \in (\mathbf{V}^m)^\omega$ and $i \in \mathbb{N}$. Since $\left[\!\!\left[ \boxed{\overline{v}}\!- \right]\!\!\right]^{\mathrm{S}}_{\mathcal{I}} = \overline{v} :: \bot :: \bot :: \ldots$, we have that $\left[\!\!\left[ \boxed{\overline{v}}\!-\!\boxed{f}\!- \right]\!\!\right]^{\mathrm{S}}_{\mathcal{I}} = g(\overline{v}) :: \bot :: \bot :: \ldots$, which is the interpretation of $\boxed{g(\overline{v})}\!-$. As the equations are sound they must preserve the stream semantics, so $\overline{w} = g(\overline{v})$. □

Finally, we use the above lemma to show how values can be applied to *essentially* combinational circuits.

**Lemma 6.20.** Let $f \colon \mathbf{V}^m \to \mathbf{V}^n$ be a monotone function such that $-\boxed{\|f\|}- :=$

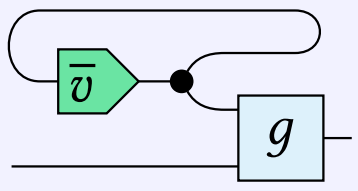

. Then

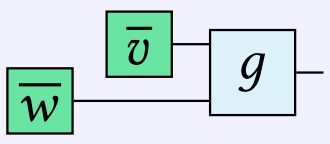

$= \boxed{f(\overline{w})}\!-$ .

*Proof.* Let $h \coloneqq \llbracket$ 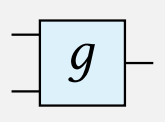 $\rrbracket^{\mathrm{C}}_{\mathcal{I}}$; using Lemma 6.19, we have that 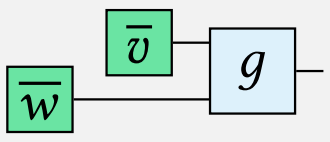 = 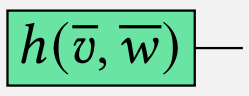 . So we must show that $f(\overline{w}) = h(\overline{v}, \overline{w})$.

$f(\overline{w}) = \llbracket$ 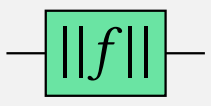 $\rrbracket^{\mathrm{S}}_{\mathcal{I}} (\overline{w} :: \bot^{\omega})(0)$ — Definition 4.87

$\coloneqq \llbracket$ 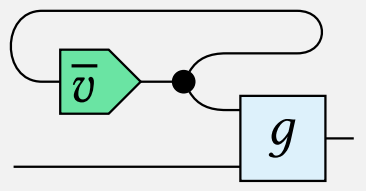 $\rrbracket^{\mathrm{S}}_{\mathcal{I}} (\overline{w} :: \bot^{\omega})(0)$

$= \llbracket$ 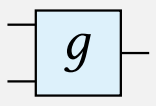 $\rrbracket^{\mathrm{S}}_{\mathcal{I}} (\overline{v}^{\omega}, \overline{w} :: \bot^{\omega})(0)$

$= \llbracket$ 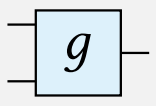 $\rrbracket^{\mathrm{C}}_{\mathcal{I}} (\overline{v}, \overline{w})$ — Lemma 4.81

$= h(\overline{v}, \overline{w})$

This completes the proof. □

With these lemmas in our toolkit, we can now show that the encoding equations allow us to translate a circuit into one with an encoded state, and therefore translate between the state sets of any two denotationally equivalent circuits.

**Theorem 6.21.** For a circuit 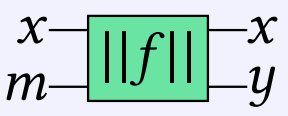 and initial state $\overline{s} \in \mathbf{V}^{x}$, the equation 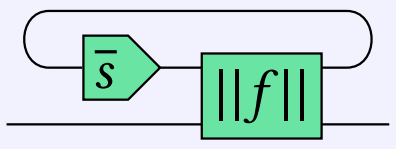 = 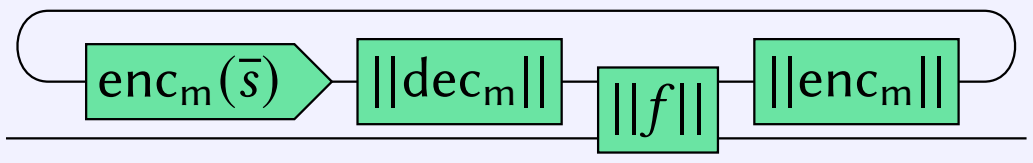 is derivable by the equations in $\mathcal{H}$.

*Proof.* We have that 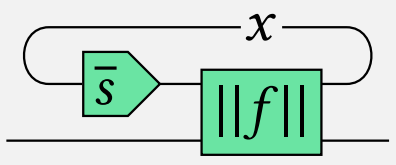 = 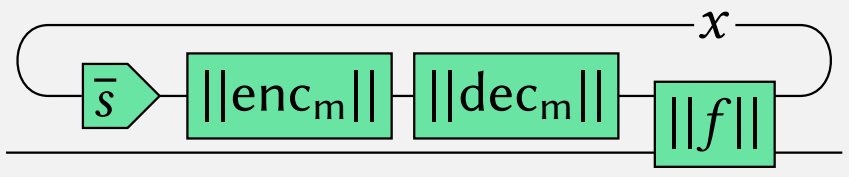 by the (Enc) equation; we need to 'push' the encoder 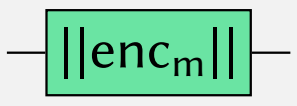 through the state. Although the encoder is sequential, by the definition of $||-||$, it must be of the form

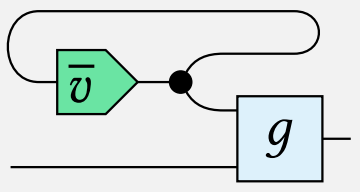 by definition of complete interpretations. This means we have

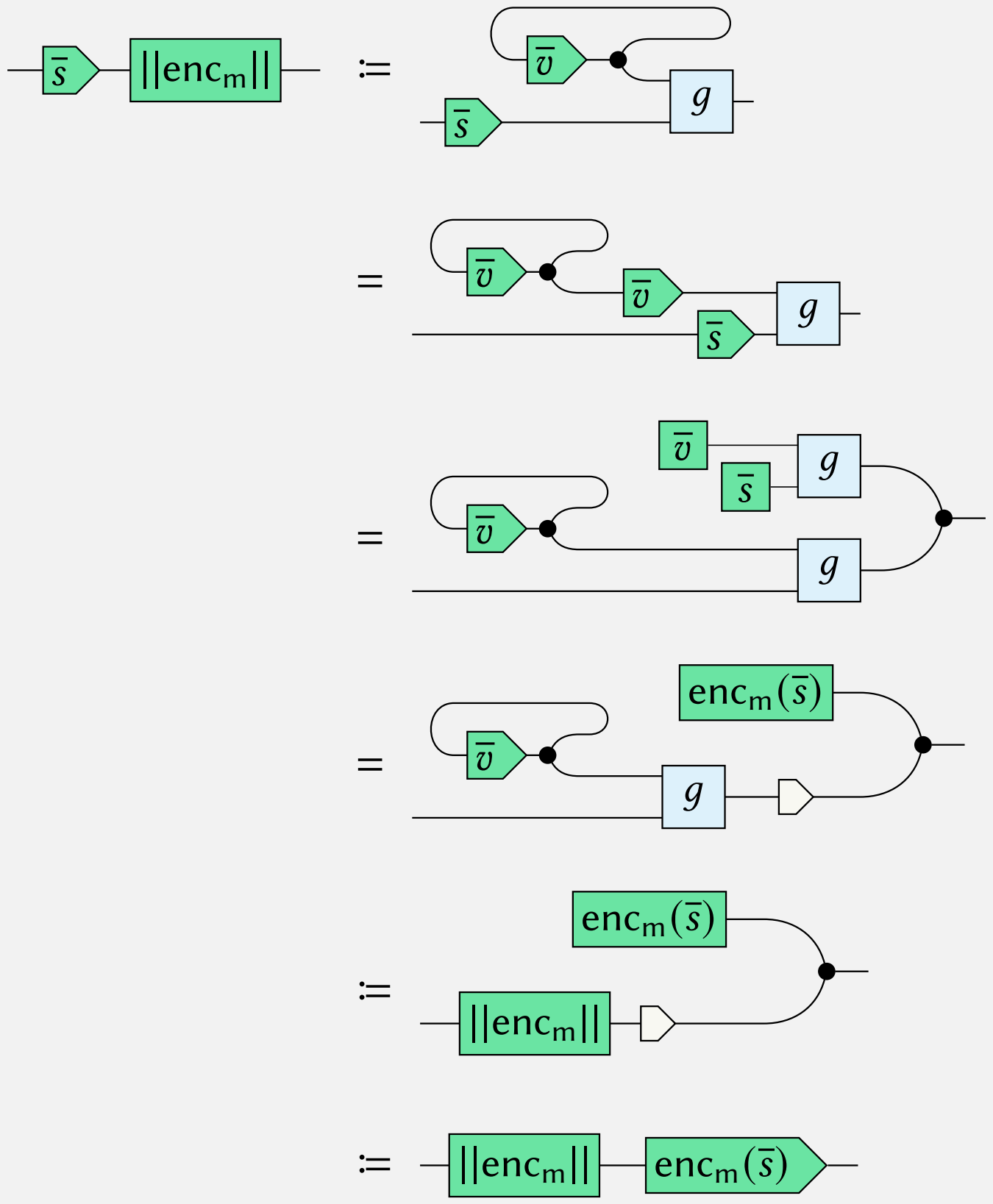

Lemma 6.17

Lemma 6.18

Lemma 6.20

The proof is completed by sliding the encoder around the trace. □

With the right encoders, the initial state of a circuit can be translated into a different word, giving us a new circuit in Mealy form. As all the involved components are essentially combinational, the circuit can be normalised again to produce a circuit in normalised Mealy form.

## 6.3 Restriction equations

We can now map the state set of one circuit to another using encodings. Does this mean that the two circuits will now be structurally equal? Unfortunately not: all it means is that the circuits agree on the set of circuit states.

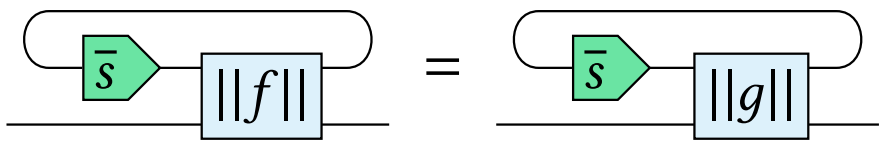

(Res) where $f|(S_{f,\bar{s}} \times \mathbf{V}^{\overline{m}}) = g|(S_{g,\bar{s}} \times \mathbf{V}^{\overline{m}})$

Figure 6.5: The schema of *restriction* equations

**Example 6.22.** Consider the following two circuits in $\mathbf{SCirc}_{\Sigma_{\mathrm{B}}}$:

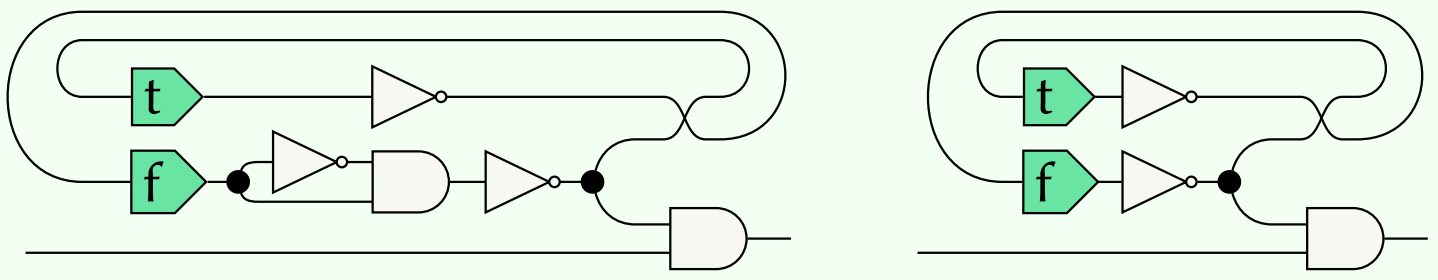

Both circuits have circuit states $\{\mathrm{tf}\}$, but their combinational cores do *not* have the same semantics. They only act the same because they receive certain inputs.

The final family of equations required is one for mapping between combinational circuits that agree on the subset of possible inputs they actually receive.

**Notation 6.23.** Given sets $A$, $B$ and $C$ where $C \subseteq A$ and a function $f\colon A \to B$, the *restriction of $f$ to $C$* is a function $f|C\colon C \to B$, defined as $f|C(c) \coloneqq f(c)$.

**Definition 6.24** (Restriction equations)**.** Let the schema of *restriction equations* be defined as in Figure 6.5.

**Example 6.25.** By a restriction equation, the circuits in Example 6.22 are now equal, as the cores produce equal outputs for inputs where the state is tf.

## 6.4 Completeness of the algebraic semantics

It is now possible to collect all the equations together and define a sound and complete algebraic theory of sequential digital circuits.

**Definition 6.26.** For a complete interpretation $\mathcal{I}$, let $\mathcal{E}_{\mathcal{I}}$ be $\mathcal{M} + \mathcal{N}_{\mathcal{I}} + \mathcal{H} + (\mathrm{Res})$, and let $\mathbf{SCirc}_{\Sigma/\mathcal{E}_{\mathcal{I}}}$ be defined as $\mathbf{SCirc}_{\Sigma}/\mathcal{E}_{\mathcal{I}}$.

For this to be a *complete* set, we must be able to translate a circuit $m$–$\boxed{f}$–$n$ into another circuit $m$–$\boxed{g}$–$n$ with the same behaviour by only using these equations.

**Theorem 6.27.** For a complete interpretation $\mathcal{I}$, $m\text{-}\boxed{f}\text{-}n = m\text{-}\boxed{g}\text{-}n$ in $\mathbf{SCirc}_{\Sigma/\mathcal{E}_{\mathcal{I}}}$ if and only if $[\![ \text{-}\boxed{f}\text{-} ]\!]^{\mathsf{S}}_{\mathcal{I}} = [\![ \text{-}\boxed{g}\text{-} ]\!]^{\mathsf{S}}_{\mathcal{I}}$.

*Proof.* All the equations are sound, so we only need to consider the ($\Leftarrow$) direction. Using Theorem 6.9, the circuits $\text{-}\boxed{f}\text{-}$ and $\text{-}\boxed{g}\text{-}$ can be brought to Mealy form, so we have that

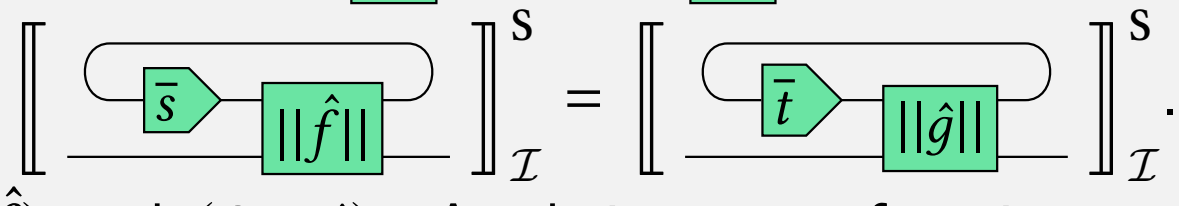

. This induces Mealy machines $(S_{\hat{f},\bar{s}}, \hat{f})$ and $(S_{\hat{g},\bar{t}}, \hat{g})$. As their stream functions are equal, there are Mealy homomorphisms $\phi \colon S_{\hat{f},\bar{s}} \to S_{\hat{g},\bar{t}}$ and $\psi \colon S_{\hat{g},\bar{t}} \to S_{\hat{f},\bar{s}}$ and subsequently the composite of these homomorphisms is also a Mealy homomorphism; these will act as a decoder-encoder pair.

Using the encoding equation, we have by Theorem 6.21 that

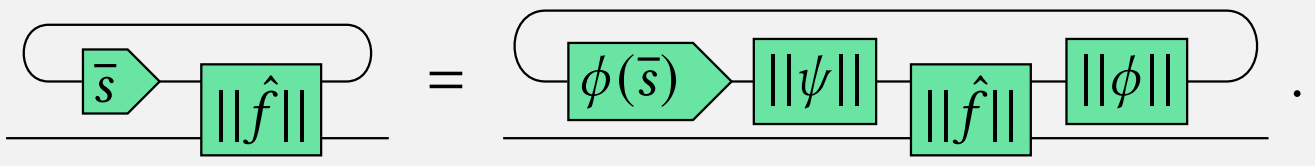

The circuit

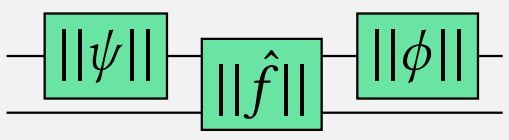

is a composition of normalised circuits, so it is essentially combinational; when restricted to the set $S_{\hat{g},\bar{t}}$ its truth table is the same as that of

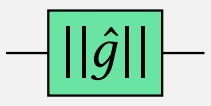

, as the encoder-decoder pair were defined precisely as the Mealy homomorphisms that translate between the two Mealy machines. Using the normalisation equations again, the encoded circuit can be brought into normalised Mealy form. Finally, the restriction equations can be used to translate from

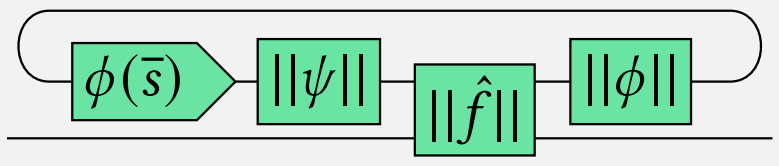

into

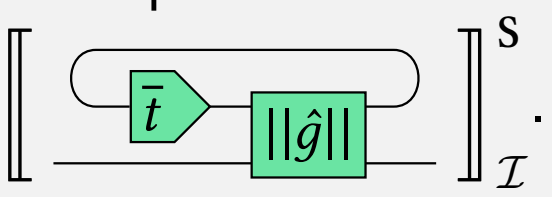

. □

As always, the soundness and completeness of the algebraic semantics means we can establish another isomorphism of PROPs.

**Corollary 6.28.** $\mathbf{SCirc}_{\Sigma/\approx_{\mathcal{I}}} \cong \mathbf{SCirc}_{\Sigma/\mathcal{E}_{\mathcal{I}}}$.

One might wonder how this improves on the operational approach, as the normal form is quite complicated. The beauty of the *algebraic* semantics is that equations can be proven as lemmas and used in the future as shortcuts; in time, the algebraicist will build up a repertoire of equations and use them to bend circuits to their will.

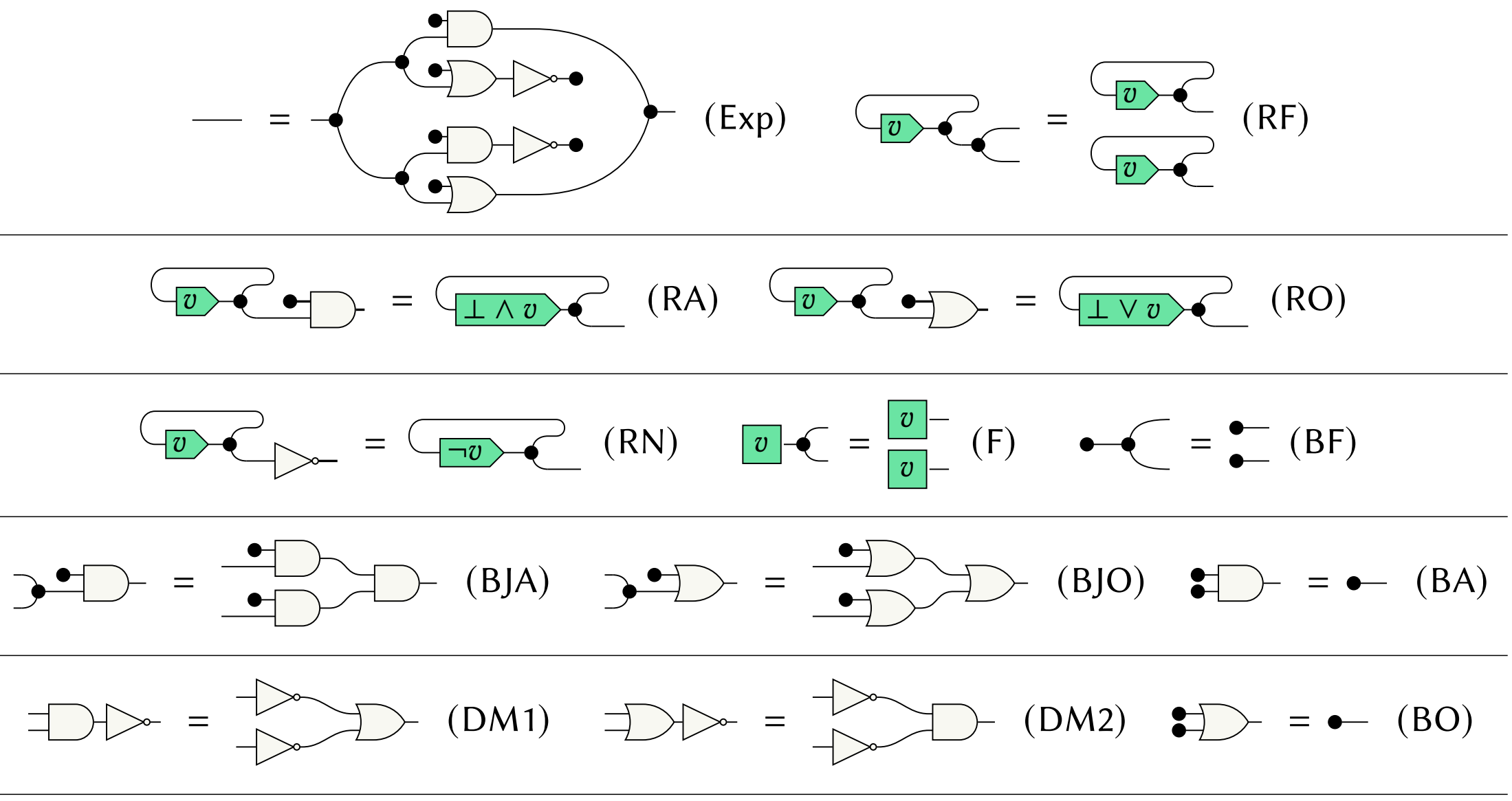

Figure 6.6a: First part of the set of *explosion equations* $\mathcal{X}$

## 6.5 Algebraic semantics for Belnap logic

For a sound and complete equational theory, equations are required to bring any essentially combinational circuit into a canonical form. We will now demonstrate this for the Belnap interpretation; recall from Section 4.5 that the canonical form for Belnap circuits is a circuit that ‘explodes’ its inputs into circuits for the ‘falsy’ and ‘truthy’ components of the output, before joining these together. The first equations we will define translate any essentially combinational circuit into such an exploded circuit.

**Definition 6.29** (Explosion equations)**.** Let the set of *explosion equations* $\mathcal{X}$ be defined as the equations listed in Figures 6.6a and 6.6b.

**Lemma 6.30.** The explosion equations are sound.

*Proof.* By checking all the inputs. □

Most of the equations in $\mathcal{X}$ are well-known; the only interesting one is (Exp), which says that we can always ‘explode’ a wire and join it back together. To translate a circuit into exploded form, we use this equation to introduce an ‘empty explosion’, then propagate the original components across with the other equations. The first obstacle to this is the forks at the left of the explosion.

**Lemma 6.31.** For any essentially combinational Belnap circuit $m$–$\boxed{f}$–$n$, the equation 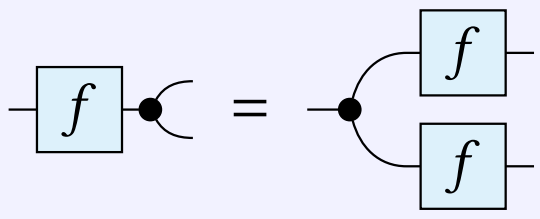 in $\mathbf{SCirc}_{\Sigma_B}/\mathcal{X}$.

*Proof.* This follows for the combinational generators by applying (JF), (BF), (AF), (OF), (NF), and is immediate for the fork. The infinite register 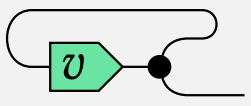 is also a base case and is covered by (RF). The inductive cases are trivial. □

Once the circuit is past the opening forks, the remaining equations are used to push them across the translators.

**Proposition 6.32.** Given an essentially combinational Belnap circuit $m$–$\boxed{f}$–$1$, there exist combinational Belnap circuits $1, m, m$–$\boxed{f_0}$–$1$ and $1, m, m$–$\boxed{f_1}$–$1$ containing no 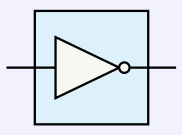 or 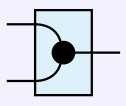 generators, such that

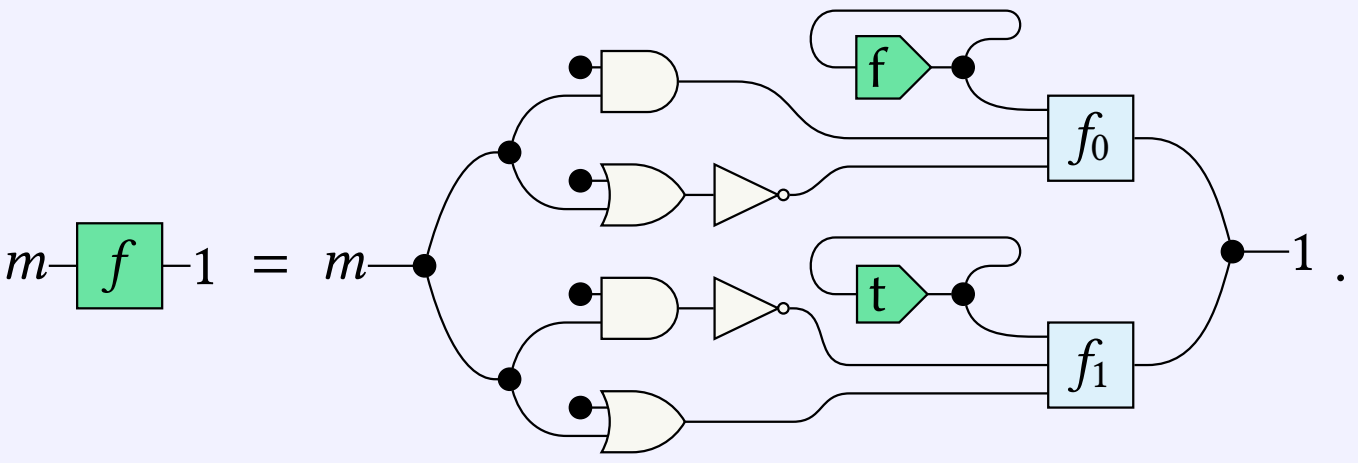

*Proof.* First we consider the base cases. If –$\boxed{f}$– is the identity, then it can be transformed into the desired form with (Exp). Since –$\boxed{f}$– has codomain 1 it cannot be a symmetry. For the other generators and the infinite register, (Exp) can be applied to the output wire to create the exploded 'skeleton', followed by using Lemma 6.31 to copy the components into four. The four copies can be pushed through the translators using (JF), (BJA), (BJO), (BF), (BO) (AC), (OC), (AOD), (OAD), (RA), (RO), (RN), (DM1), (DM2), (BN), (BA), and (BO). As propagating the 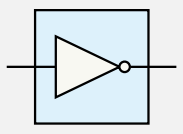 flips the translators by using (DM1) and (DM2), (FC) must be used to restore the correct order, and (DNE) is used to eliminate additional 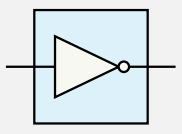 gates. Any infinite registers containing $\perp$ can be converted to 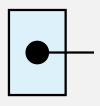 components using (Bot), and other registers can be combined using (RF).

For the composition inductive case, we have two exploded circuits and we need to push the first inside the second. Using Lemma 6.31, the first circuit can be propagated across the forks at the start of the second, so each of the four translators has as input a copy of the first circuit. Using the same strategy as for the base case the components of the circuit can then be propagated across the translators. For tensor, the circuits can be interleaved using axioms of STMCs. □

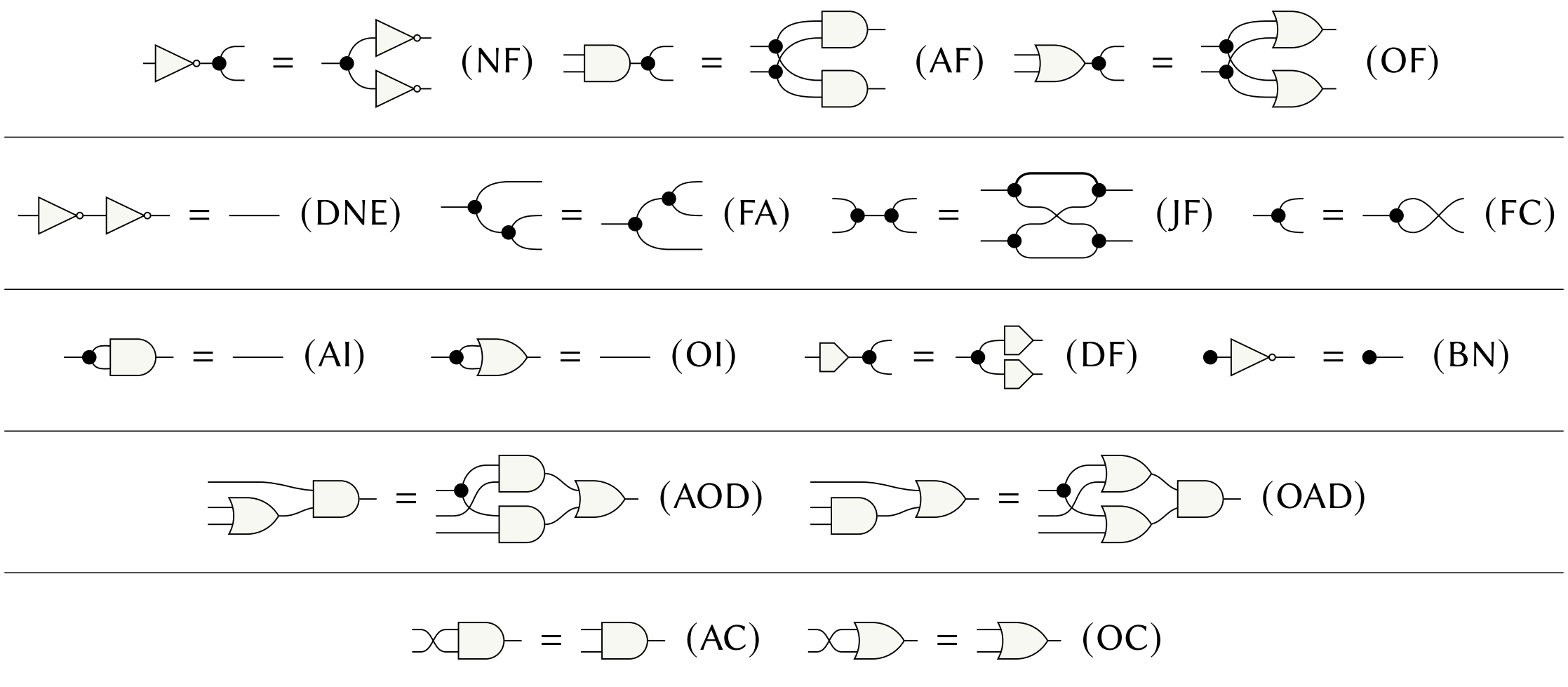

Figure 6.6b: Second part of the set of *explosion equations* $\mathcal{X}$

This already looks similar to a circuit in the image of $||-||_{\mathrm{B}}$, but the two subcircuits must also be translated into truthy or falsy disjunctive normal form.

**Definition 6.33** (Normal form equations)**.** Let the set of *normal form equations* $\mathcal{F}$ be defined as the equations listed in Figure 6.7.

**Lemma 6.34.** The equations in $\mathcal{F}$ are sound.

*Proof.* By checking all the inputs. □

We will now show that these equations suffice to translate the subcircuits in the exploded circuit into falsy or truthy disjunctive normal form.

**Lemma 6.35.** Given a Belnap circuit 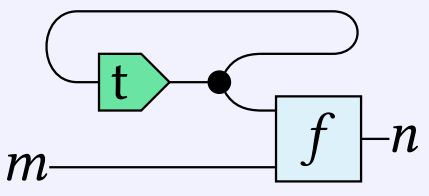 containing no 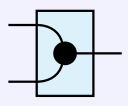, 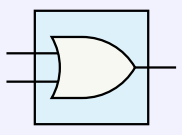 or 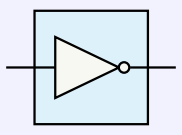 components, there exists a circuit 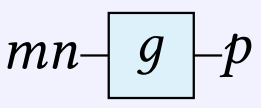 containing only identity and elimination constructs, and a circuit 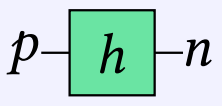 defined as the tensor of $n$ truthy conjunctions, such that 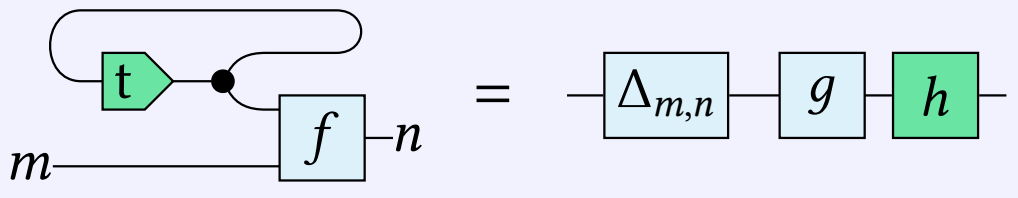 in $\mathbf{SCirc}_{\Sigma_{\mathrm{B}}}/\mathcal{F}$.

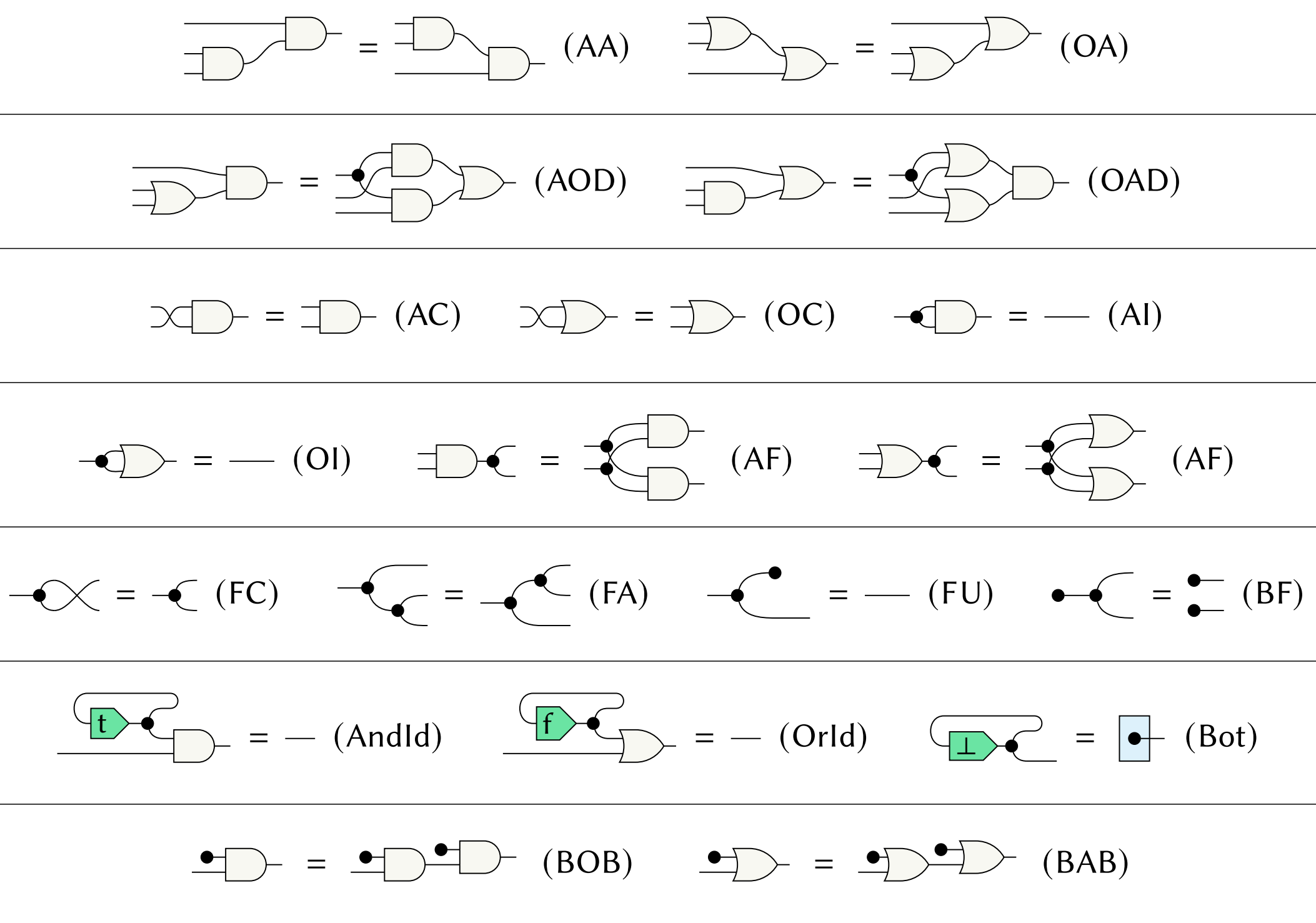

Figure 6.7: Set of *normal form equations* $\mathcal{F}$.

*Proof.* Repeatedly applying (AF) to $f$ propagates the 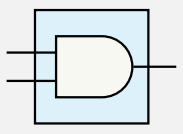 components in the circuit as far to the right as possible, so all the fork and eliminate constructs are in the left half of the term. Using (FC), (FA) to rearrange the forks, and (FU) to introduce forks where necessary, we can manipulate these forks such that each of the $m$ inputs has a connection to each of the $n$ outputs. Similarly, we can use (AndId) to introduce infinite registers where appropriate, so we have a circuit of the form below.

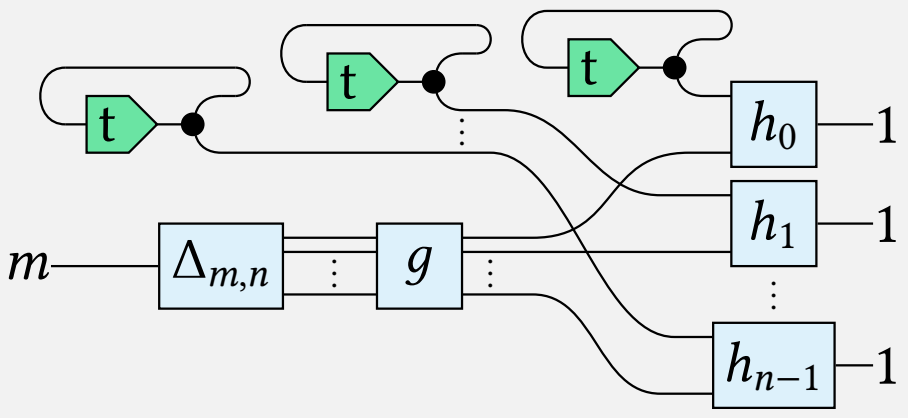

The subcircuits may not be truthy conjunctions yet, as the input wires may be used more than once. Using the (AI), (FC), (FA), (AA) and (AC), each subcircuit can be translated into a truthy conjunction. □

The proof for the falsy circuit is almost exactly the same.

**Lemma 6.36.** Given a Belnap circuit 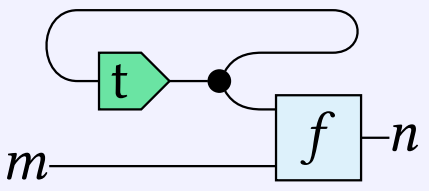 containing no 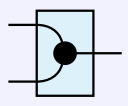, 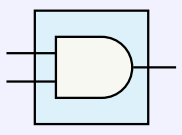 or 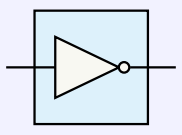 components, there exists a circuit $mn$–$g$–$p$ containing only identity and elimination constructs, and a circuit $p$–$h$–$n$ defined as the tensor of $n$ falsy conjunctions, such that $-f- = -\Delta_{m,n}-g-h-$ in $\mathbf{SCirc}_{\Sigma_B}/\mathcal{F}$.

*Proof.* As Lemma 6.35, but with the equations on 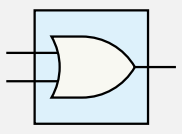 components. □

Truthy and falsy conjunctions can then be used to create truthy and falsy conjunctive normal forms.

**Proposition 6.37.** Given a Belnap circuit

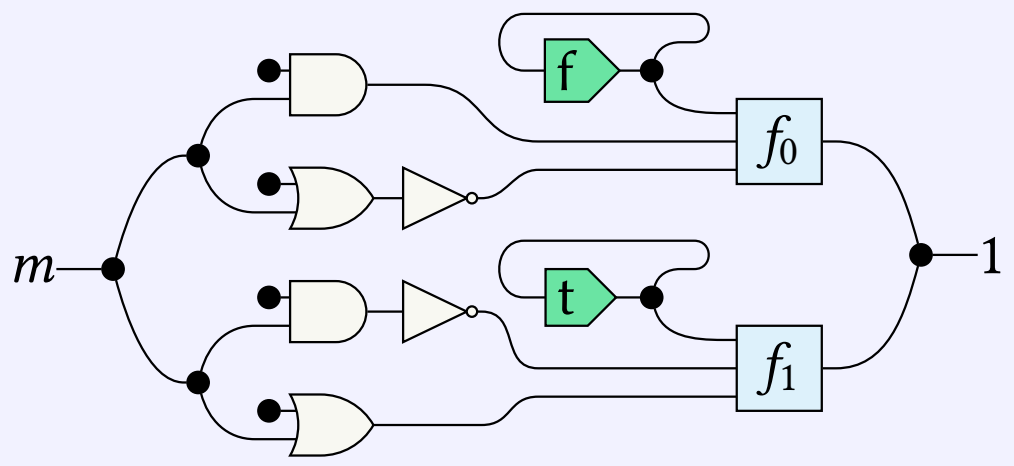

in which $f_0$ and $f_1$ contain no 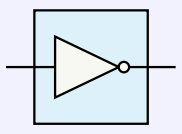 or 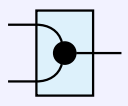 generators, there exists

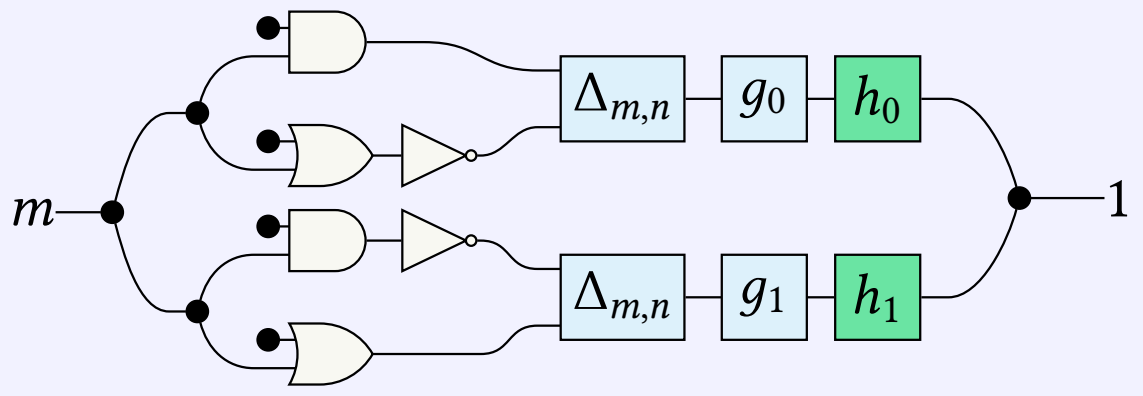

in which $g_0$ and $g_1$ contain only identity and elimination components, $h_0$ is in falsy conjunctive normal form and $h_1$ is in truthy conjunctive normal form, such that in $\mathbf{SCirc}_{\Sigma_B}/\mathcal{F}$.

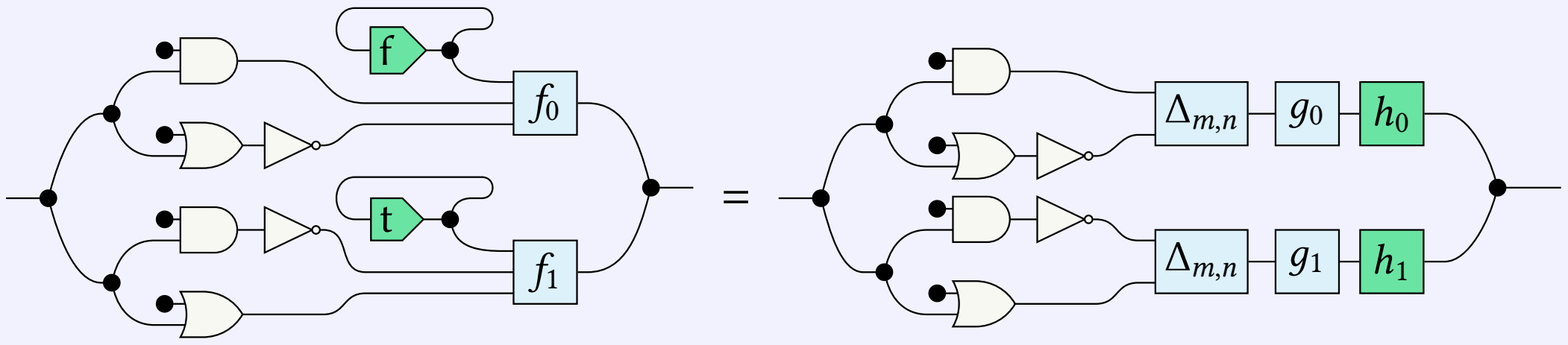

*Proof.* First, all the 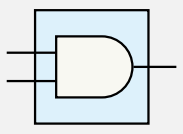 components need to be propagated to the far right of the $f_0$ circuit using (OAD), and all the 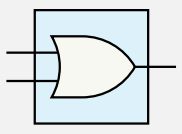 components need to be propagated to the far right of the $f_1$ circuit using (AOD). This means that the circuits are split into two halves, each containing one type of gate.

For these circuits to be in truthy or falsy disjunctive normal form, they need to contain exactly one 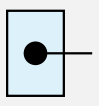 component. If there is not already such a component inside the $f_0$ or $f_1$ subcircuits, one can be inserted using the (BOB) equation for the former and the (BAB) equation for the latter. If there are multiple unit components, these can be propagated through the circuit using (OA), (OAD), (AA), and (AOD), and combined into one by using (BA) or (BO).

Now we have circuits that have the 'root' of disjunctive normal forms, but the 'leaves' are not conjunctions. This is remedied by applying Lemma 6.35 and Lemma 6.36 to the left half of each circuit. □

Putting this all together gives us the desired canonical form theorem.

**Theorem 6.38.** Given an essentially combinational Belnap circuit $f$, there exists a circuit $g$ in the image of $||-||_B$ such that $f = g$ in $\mathbf{CCirc}_{\Sigma_B}/\mathcal{X} + \mathcal{F}$.

*Proof.* This follows by applying Proposition 6.32 followed by Proposition 6.37. □

This shows that the equations in this section can translate any essentially combinational Belnap circuit into a circuit in the image of the functional completeness map.

## 6.6 Algebraic semantics for generalised circuits

When it comes to lifting the algebraic semantics to the generalised case, all we need to do is to extend some of the equations to arbitrary width wires, and to add equations to handle the bundlers.

**Definition 6.39.** Let the set $\mathcal{M}^+$ be defined as $\mathcal{M}$ but with equations (MUL), (MUR), and (BD) for wires of each with $n \in \mathbb{N}_+$.

Since the set of normalising equations $\mathcal{N}_\mathcal{I}$ is determined by the interpretation, we do not need to do anything there. The encoding equations need to be extended to act on all wire widths, and we need to be able to handle encoders that contain bundlers.

**Definition 6.40.** Let the set $\mathcal{H}^{+}$ be defined as $\mathcal{H}$ but with all equations adjusted to operate on wires of all widths $n \in \mathbb{N}_{+}$, and with the addition of equations

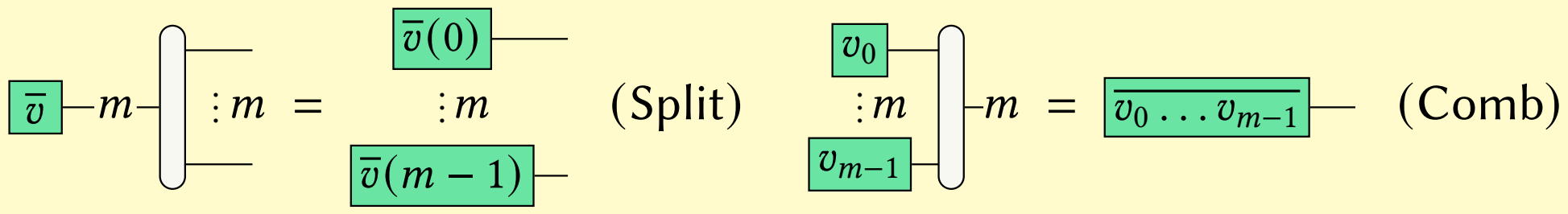

on bundlers.

Finally the restriction schema also needs to operate on arbitrary-width wires.

**Definition 6.41.** Let ($\text{Res}^{+}$) be defined as (Res) but extended to operate on wires of widths $n \in \mathbb{N}_{+}$.

Putting this all together gives us a set of equations for generalised circuits.

**Definition 6.42.** For a generalised interpretation $\mathcal{I}^{+}$, let $\mathcal{E}^{+}_{\mathcal{I}} \coloneqq \mathcal{M}^{+} + \mathcal{N}^{+}_{\mathcal{I}} + \mathcal{H}^{+} +$ ($\text{Res}^{+}$), and let $\mathbf{SCirc}^{+}_{\Sigma/\mathcal{E}^{+}_{\mathcal{I}}}$ be defined as $\mathbf{SCirc}^{+}_{\Sigma}/\mathcal{E}^{+}_{\mathcal{I}}$.

This set of equations is sound and complete.

**Theorem 6.43.** For a functionally complete generalised interpretation $\mathcal{I}$, $\overline{m}\!-\!\boxed{f}\!-\!\overline{n} = \overline{m}\!-\!\boxed{g}\!-\!\overline{n}$ in $\mathbf{SCirc}^{+}_{\Sigma/\mathcal{E}^{+}_{\mathcal{I}}}$ if and only if $[\![ -\boxed{f}- ]\!]^{\mathrm{S+}}_{\mathcal{I}} = [\![ -\boxed{g}- ]\!]^{\mathrm{S+}}_{\mathcal{I}}$.

Subsequently we obtain another isomorphism of categories.

**Corollary 6.44.** $\mathbf{SCirc}^{+}_{\Sigma/\approx^{+}_{\mathcal{I}}} \cong \mathbf{SCirc}^{+}_{\Sigma/\mathcal{E}^{+}_{\mathcal{I}}}$.

Chapter 7

# Potential applications

So far, we have been concerned largely with *theoretical* concepts; we have shown how the categorical framework of sequential digital circuits is rigorous enough to handle composing circuits in sequence, parallel or with the trace without causing any of the three semantic models to become degenerate. Unfortunately, this is not enough for people in industry, who are more concerned with the *practical* benefits of the framework: what can we do with it that cannot already be done?

Circuit design is already a very well-studied area and existing technologies are incredibly successful. Our framework is therefore not meant to *replace* the existing technologies, but to *complement* them by highlighting different perspectives on reasoning with sequential digital circuits. While the ideas provided in this section are certainly not industry-grade applications, they are intended to demonstrate the potential of what the compositional theory can bring to the table.

## 7.1 Tidying up

When building a circuit, it is often desirable to reduce the number of wires and components used; this reduces both the physical size of the circuit and its power consumption. We can use partial evaluation to transform a circuit into a more minimal form.

**Definition 7.1** (Tidying rules)**.** Let the set of *tidying rules* be defined as the rules in Figure 7.1.

Most of the tidying rules are self-explanatory; the final rule is necessary in order to deal with traced circuits with no outputs. Since all circuits with no outputs have the same behaviour, we are permitted to cut the trace to obtain a circuit we can apply more tidying rules to. As non-delay-guarded feedback is already handled by the (IF) rule, we only need to consider the delay-guarded case.

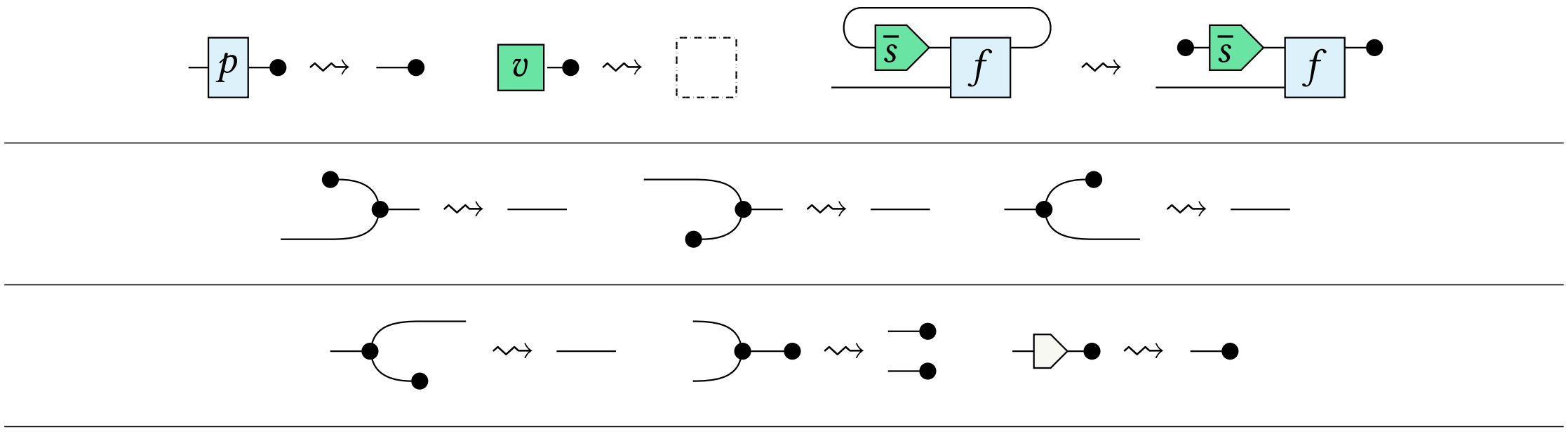

Figure 7.1: Rules for tidying up circuits in Mealy form

**Proposition 7.2.** Applying the tidying rules to a circuit in Mealy form is confluent and terminating.

*Proof.* The tidying rules always decrease the size of the circuit. The only choice is raised when there is a trace around a combinational circuit, but this does not change the internal structure of the subcircuit, so rule applications are prevented. Moreover, since all this rule does is cut a trace, it does not matter if this is performed all in one go, or each feedback loop is cut one by one. □

## 7.2 Partial evaluation

Partial evaluation [Jon96] is a paradigm used in software optimisation in which programs are 'evaluated as much as possible' while only some of the inputs are specified. For example, it may be the case that a particular input to a program is fixed for long periods of time; using partial evaluation, we can define a program specialised for this input. This program might run significantly faster than the original.

There has been work into partial evaluation for hardware, such as constant propagation [SHM96; SM99] and unfolding [TM06]. However, this has been relatively informal, and can be made rigorous using the categorical framework. In this section we will focus on how we could extend the reduction-based operational semantics to define automatic procedures for applying partial evaluation to circuits.

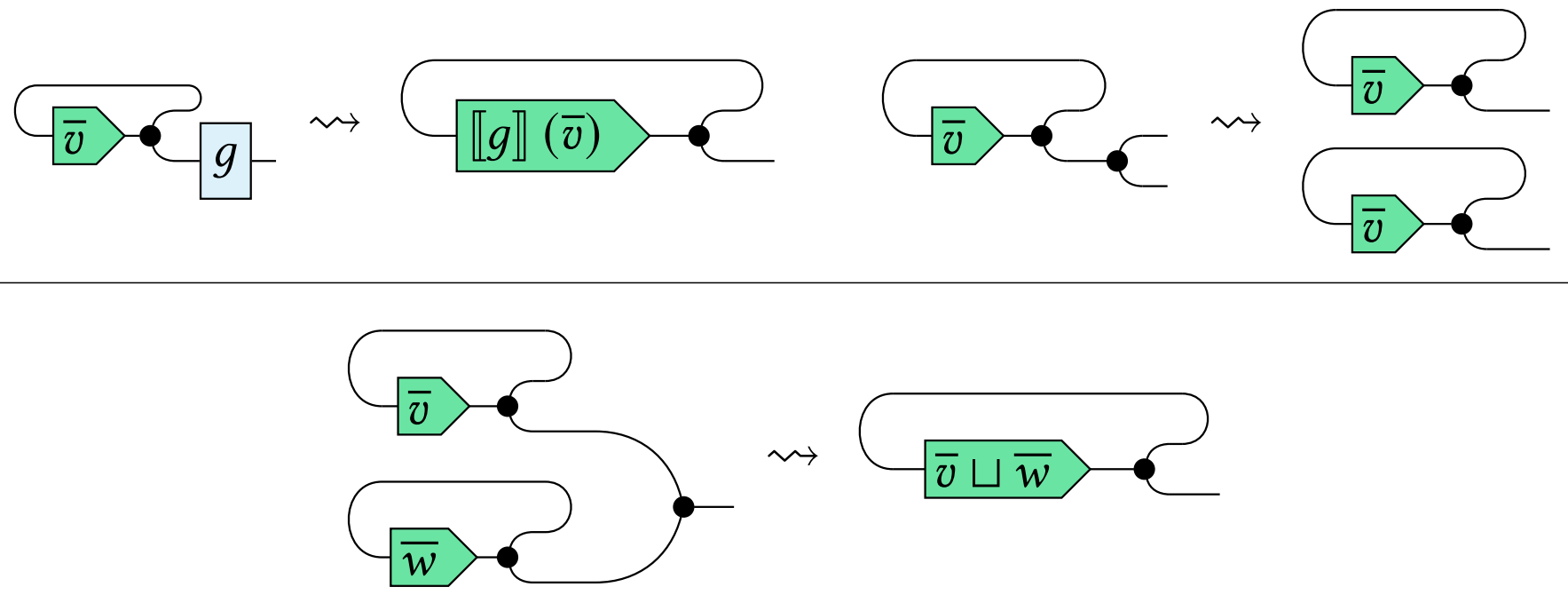

Figure 7.2: Rules for infinite waveforms

### 7.2.1 Shortcut rules

It is often the case that we know that some of the inputs to a circuit are fixed. This can be modelled by precomposing the relevant input with an *infinite waveform* 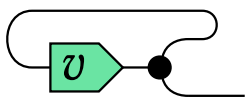 . We can propagate these waveforms across a circuit to see if we can reduce it to a circuit *specialised* for these inputs.

To propagate waveforms across circuits we need to derive a version of the ($\mathrm{P}_{\mathcal{I}}$) rule for applying waveforms to primitives. These rules are illustrated in Figure 7.2.

This is not the only way we can partially evaluate with some inputs. In some interpretations, it may be that we learn something about the output of a primitive with only some of the inputs specified.

**Example 7.3** (Belnap shortcuts)**.** In the Belnap interpretation $\mathcal{I}_{\mathrm{B}}$, if one applies a false value to an AND gate then it will output false regardless of the other input. Similarly, if one applies a true value to an OR gate it will output true. Conversely, if one applies a true value to an AND gate or a false value to an OR gate, it will act as the identity on the other input.

These 'shortcuts' can also be implemented as rules, as illustrated in Figure 7.3. Note that here the value that 'triggers' the shortcut must be contained within an infinite waveform; if we applied the rule with just an instantaneous value, this value would produce $\perp$ on ticks after the first and the rule would be unsound.

**Example 7.4** (Control switches)**.** Recall that a *multiplexer* is a circuit component constructed as 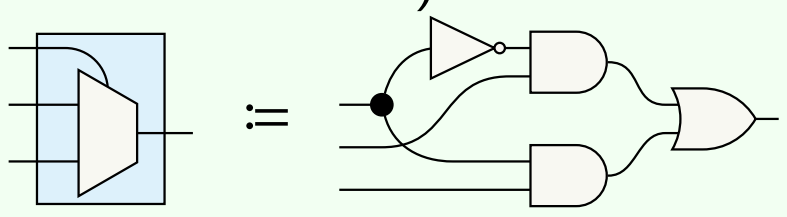 . The first input is a *control* which specifies which of the two other input signals is produced as the output signal. It is often the case that these control signals will be fixed for long periods of time;

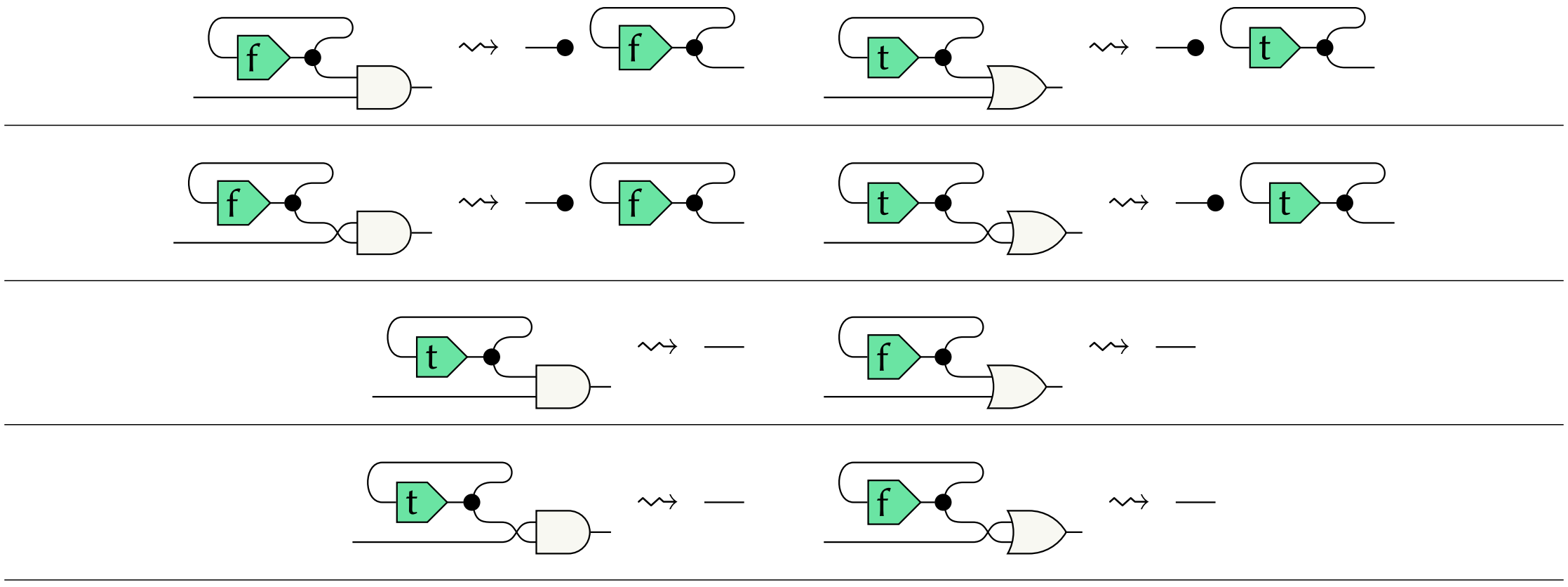

Figure 7.3: Belnap shortcut rules for waveforms

perhaps they specify some sort of global circuit configuration.

Consider the circuit 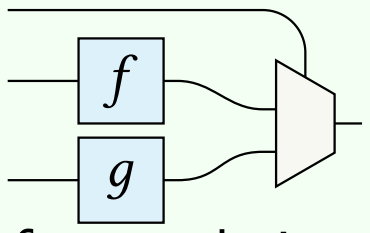 , in which the control signal to the multiplexer determines which of two subcircuits will become the output. We will assume that the control signal is held at false, and reduce accordingly by instantiating the rule in Figure 7.2 detailing the interaction of gates and waveforms to the NOT case.

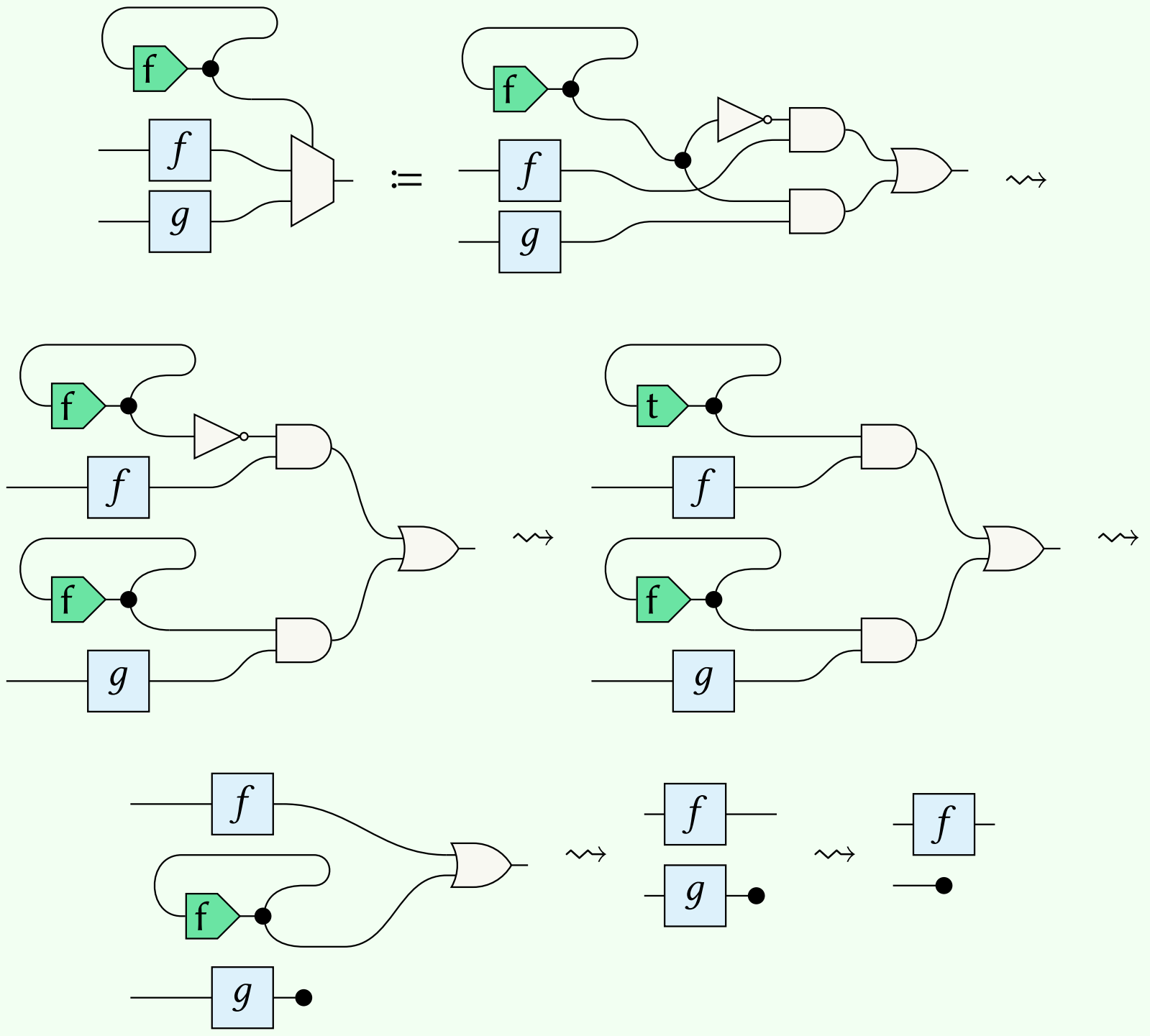

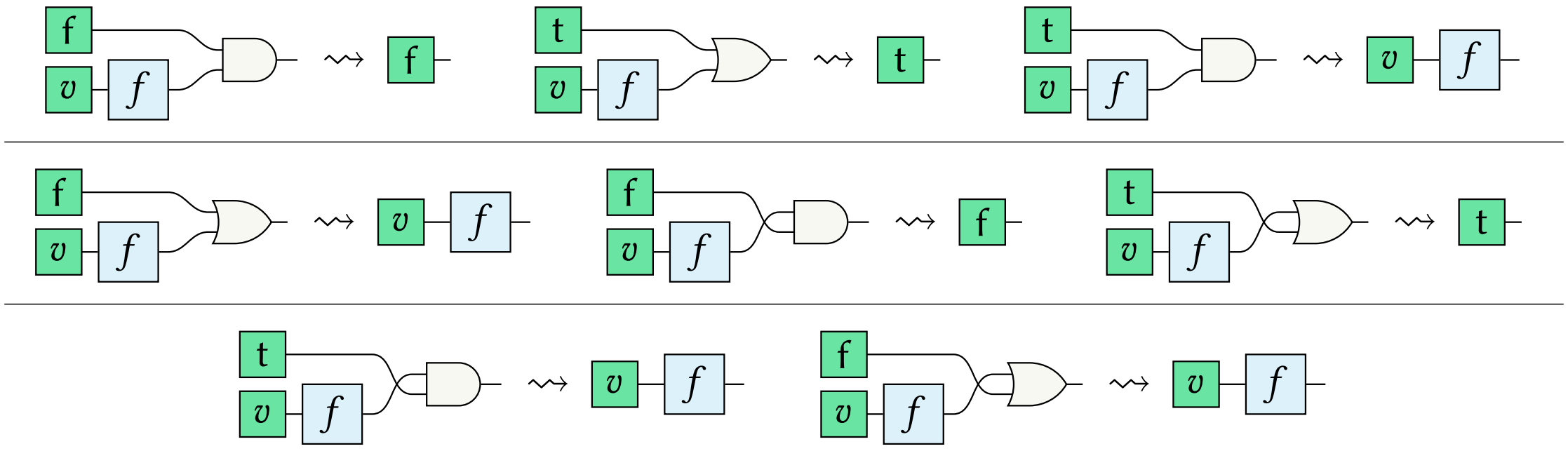

Figure 7.4: Examples of 'instantaneous' shortcut rules

### 7.2.2 Shortcuts after streaming

The rules in the previous sections are intended for use on circuits before we even apply values to them. However, there is still potential for partial evaluation when we consider the outputs of a circuit one step at a time. To do this, we can apply variants of the shortcut rules *after* performing streaming for some inputs. These variants are illustrated in Figure 7.4.

**Example 7.5** (Blocking boxes)**.** Consider the circuit 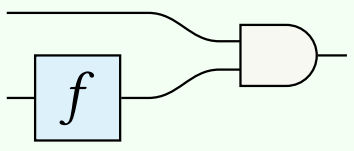 , which contains a 'blackbox' combinational circuit 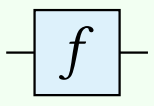 with unknown behaviour.

Even though we cannot directly reduce the blackbox, if we set the first input to false and use the shortcut rule above, we can still produce an output value.

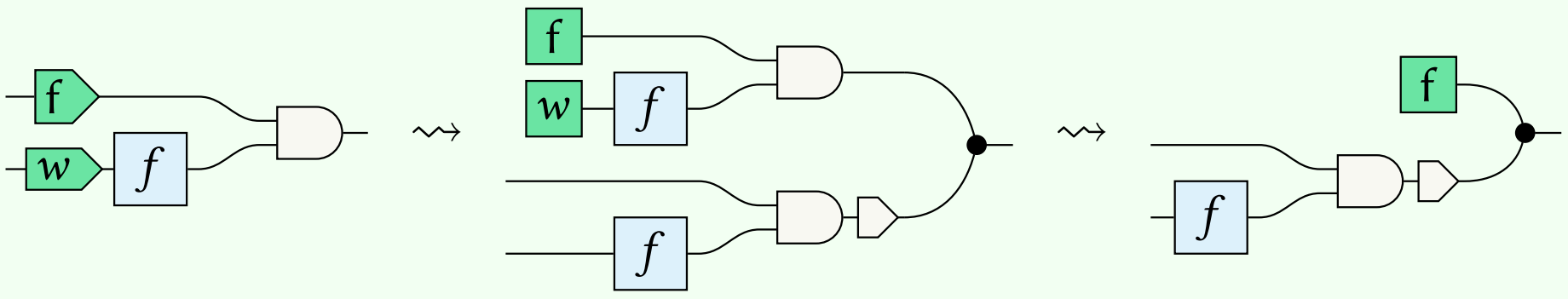

As well as removing redundant blackboxes, judicious use of shortcut reductions can dramatically reduce the reductions needed to get the outputs of a circuit.

### 7.2.3 Protocols

Sometimes we may not know the exact inputs to a circuit, but know that they make up a fixed subset of all possible inputs, or they follow some sort of protocol. We can implement this in our reduction framework with *uncertain values* which we either know nothing about or know can only take some specified values.

**Definition 7.6.** Let $\mathbf{SCirc}^{\mathrm{P}}_{\Sigma}$ be the result of extending $\mathbf{SCirc}_{\Sigma}$ with value generators for each word $\overline{v?} \in \mathbf{V}^{\star}$.

The additional value generators indicate that they could produce one of multiple possible values. When a circuit contains uncertain values $\boxed{\overline{v_0?}}\!-$ , $\boxed{\overline{v_1?}}\!-$ , . . . $\boxed{\overline{v_{n-1}?}}\!-$ , where the maximum length of a given $v_i$ is $k$, there are $k$ possible value assignments. For a given assignment $i < k$, each value will produce a concrete value defined as $v_i(j)$ if $|v_i| > j$ or $\bot$ otherwise.

To avoid confusion with our syntax sugar for arbitrary-width values, we will always end uncertain value lists with ?. When writing out specific uncertain words, we delimit the elements with vertical bars like f|t to allude to the fact that this value is either the first *or* the second element.

**Example 7.7.** If a circuit contains uncertain values $\boxed{\mathsf{f|t}}\!-$ and $\boxed{\mathsf{t|f}}\!-$ in a circuit, then there are two universes to consider, one where the values output ft and one where they output tf. If we add in another uncertain value with three possible values, $\boxed{\mathsf{t|f|\top}}\!-$ , we now have three possible universes, in which the values output ftt, tff, and $\bot\bot\top$ respectively.

To reason with uncertain values in the reductional framework we need to add rules for processing them. Once again it is useful to have versions for both waveforms and values, for reasoning before and during execution.

**Definition 7.8** (Uncertain rules)**.** The *uncertain rules* are listed in Figure 7.5.

After applying uncertain values to a primitive, it may turn out that all the possibilities are in fact the same. This removes any uncertainty, and means the value can be treated as an ordinary value in future reductions and outputs.

**Example 7.9** (Protocols)**.** One sticking point that arises when using the categorical framework is the presence of the $\bot$ and $\top$ values, which would not normally be explicitly provided to a circuit. These values mean that some well-known Boolean identities do not always hold. By using uncertain values, we can specify the values that *will* be applied to a circuit and apply reductions that are not valid in general but are in this context.

In the following example, setting the first two inputs to true/false inverses

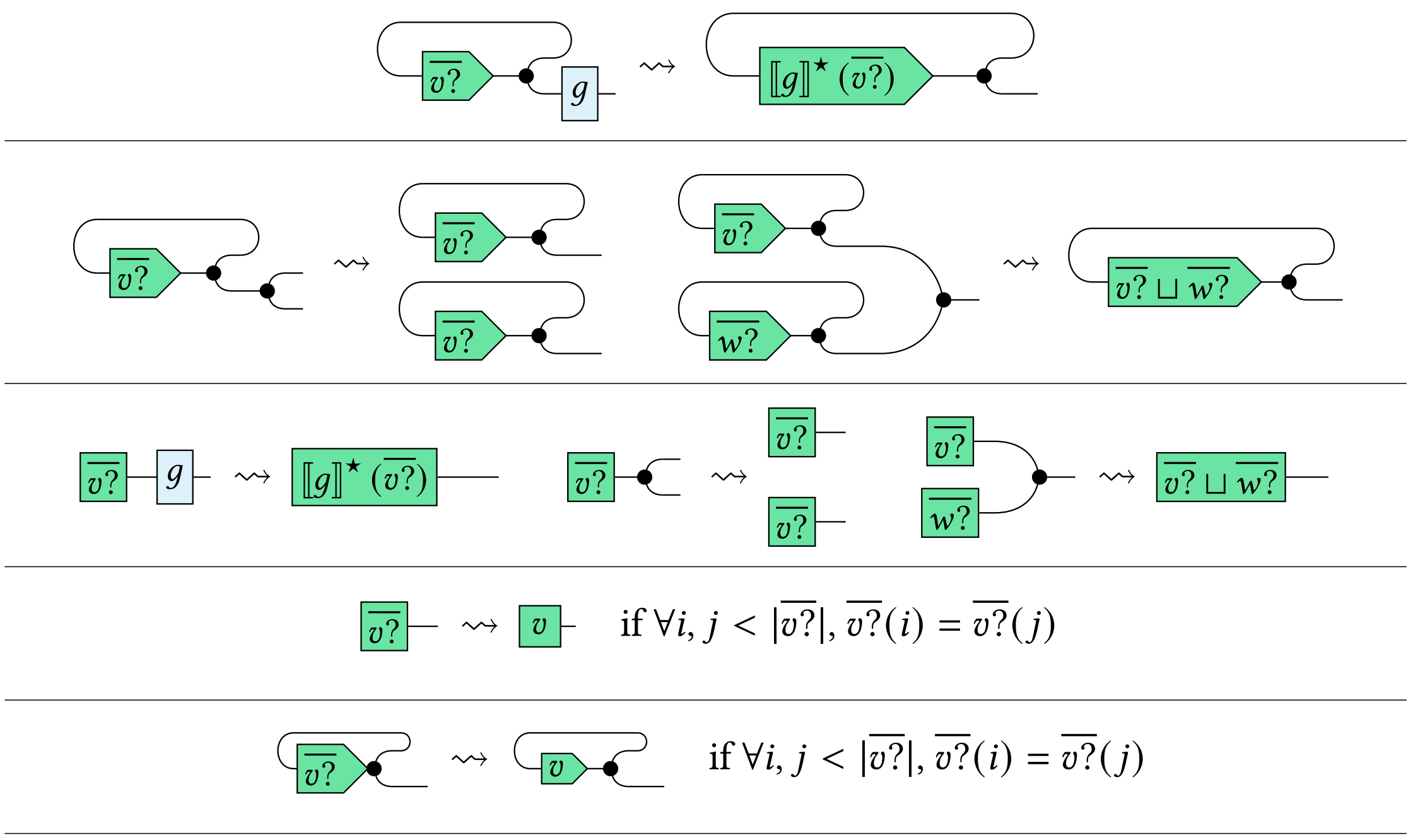

Figure 7.5: Rules for uncertain values

reduces the circuit to one with combinational behaviour.

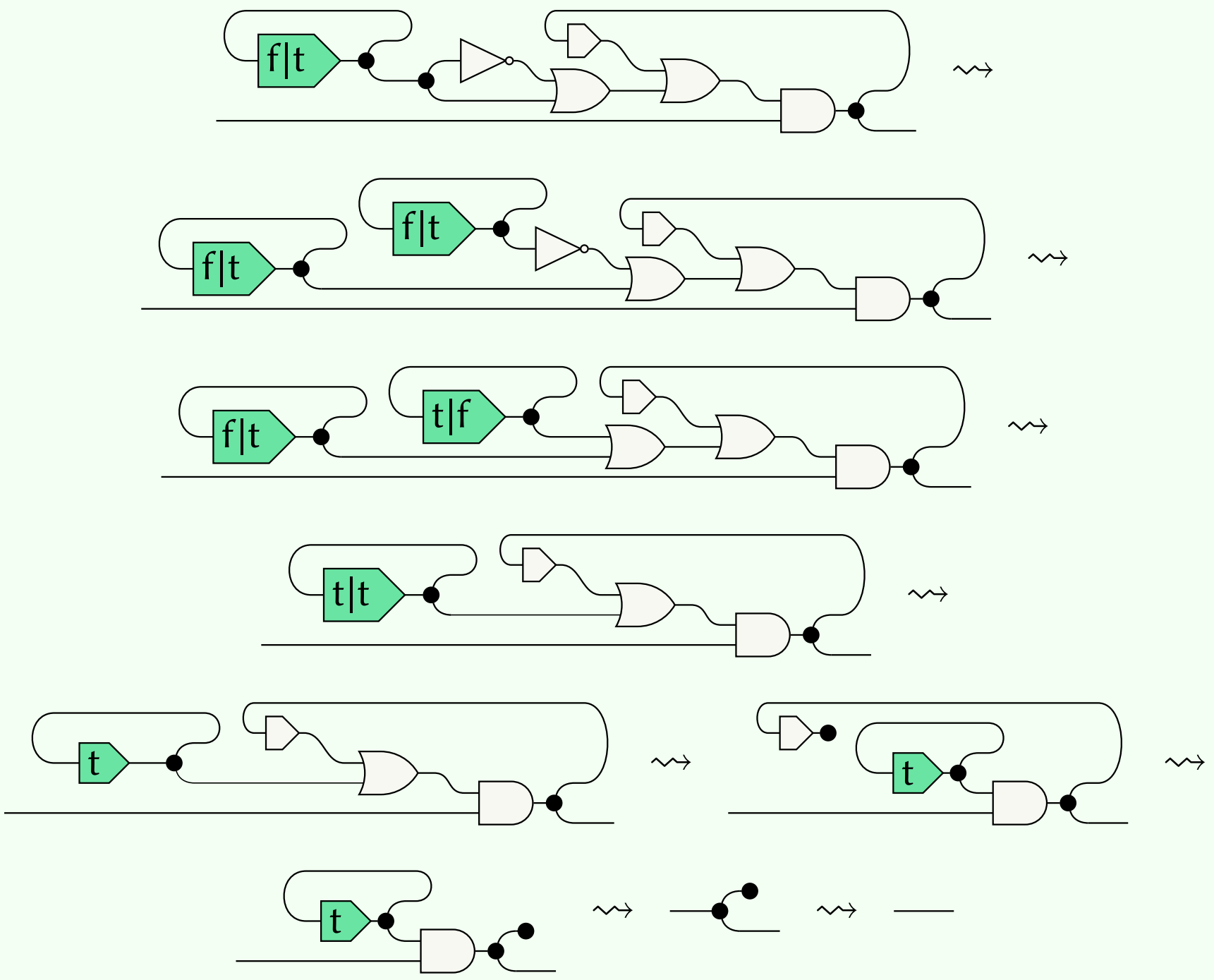

## 7.3 Layers of abstraction

Circuits can be viewed at multiple levels of abstraction. One could drop down to the level of transistors, as illustrated in [GJL17a, Sec. 4.1]. Alternatively, one could become more abstract, setting the generators to be *subcircuits*, such as arithmetic operations.

The levels of abstraction need not remain isolated. Using *layered explanations* [LZ22], multiple signatures can be mixed in one diagram, with the subcircuits acting as 'windows' into different levels of abstraction, and drawn using 'functorial boxes' [Mel06].

**Example 7.10** (Implementation)**.** Suppose one is working in a high-level signature $\Sigma_+$ containing a generator $-\boxed{?}-$ , representing an *IP core*: a circuit that has a known behaviour but with an unknown implementation. This component can be left as a blackbox and evaluated as demonstrated above.

The designer then attempts to design their own implementation using the gate-level signature $\Sigma_B$. To synthesise the final circuit, a map is defined from generators in $\Sigma_+$ to morphisms in $\mathbf{SCirc}_{\Sigma_B}$, which induces a functor $\mathbf{SCirc}_{\Sigma_+} \to \mathbf{SCirc}_{\Sigma_B}$. Different implementations can be defined as different maps, and hence different functors. These circuits can then be tested to see if they act as intended.

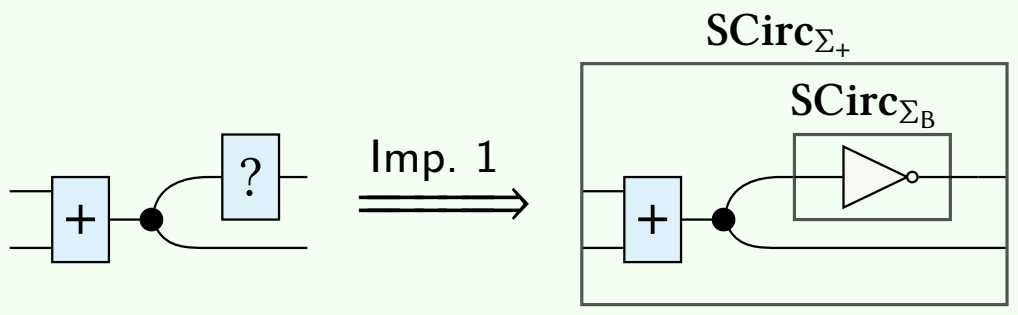

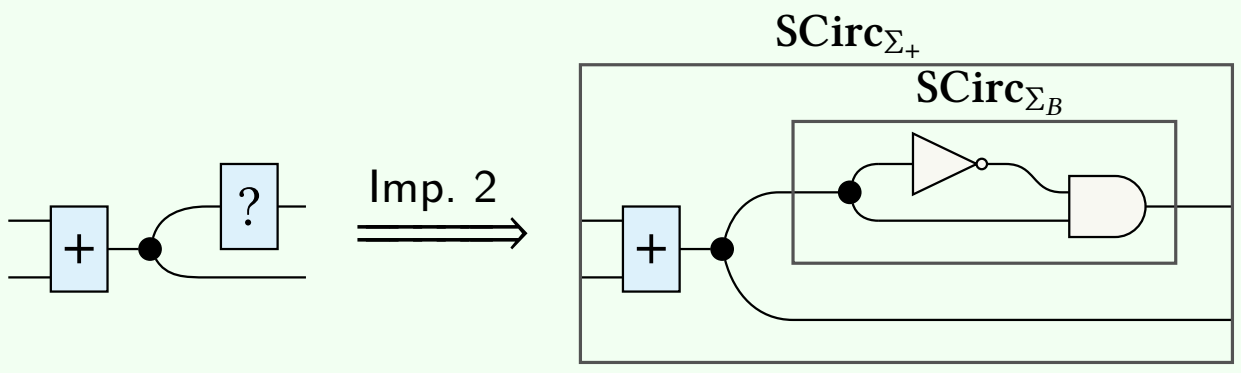

## 7.4 Refining circuits

A key part of circuit design comes in *optimising circuits*: making them run as fast as possible and reduce the *clock cycle*.

**Example 7.11** (Retiming)**.** The clock cycle of a circuit is determined by the longest paths between registers. Altering the paths between registers can be achieved using *retiming* [LS91]: moving registers across gates. This is modelled by the streaming rule (Lemma 5.25); forward retiming (streaming left to right) is always possible but for backward retiming (streaming right to left), the value in the register must be in the image of the gates.

The streaming rule permits retiming using the composite register construct, but can also be used to retime raw delay components.

**Lemma 7.12** (Timelessness)**.** For any primitive 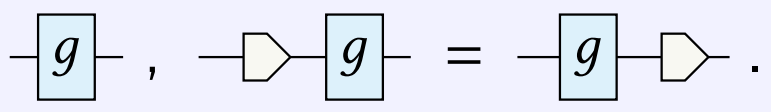

*Proof.* 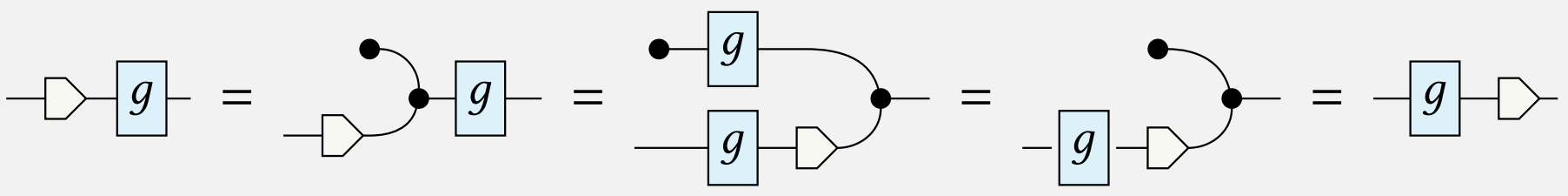 □

When reasoning equationally, the behaviour of the circuits on either side of the equation must have exactly the same behaviour. However, when reasoning with circuits it is sometimes the case that this is too strict an assertion; we are looking for circuits that output the same outputs but over a shorter period of time. This means we may wish to use transformations that only 'morally' preserve the behaviour of a circuit.

**Definition 7.13.** For two finite sequences $\underline{v}, \underline{w} \in (\mathbf{V}^m)^k$, we say that $\underline{w}$ is a *stretching* of $\underline{v}$, written $\underline{v} \ll \underline{w}$, if $\underline{w}$ contains the characters of $\underline{v}$ but possibly repeated or with additional $\bot$ characters e.g. tf $\ll$ $\bot\bot$tt$\bot$f.

**Definition 7.14.** For two sequential circuits $m$–[$f$]–$n$ and $m$–[$g$]–$n$ with $c$ and $c'$ delay components respectively, we say that $m$–[$f$]–$n$ is *logically equivalent* to $m$–[$g$]–$n$, written $m$–[$f$]–$n$ $\ll$ $m$–[$g$]–$n$, if for all sequences $\underline{v}, \underline{w}$ produced by the productive operational semantics for inputs of length $\max(c, c')$, $\underline{v} \ll \underline{w}$

Including this notion of equivalence in algebraic reasoning allows us to reason with *inequalities* as well as equalities, so more efficient circuits can be identified. The simplest form of reasoning with logical equivalence is where we have the same circuit but guarded by different numbers of delays.

**Notation 7.15.** We write 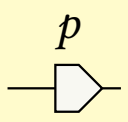 for the composition of $p$ delay components, i.e. 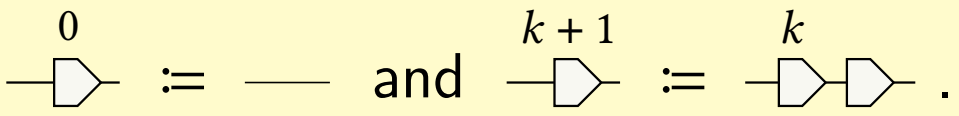

**Lemma 7.16.** For a combinational circuit $m$–[$f$]–$n$ and $p, q \in \mathbb{N}$ such that $p < q$,

then 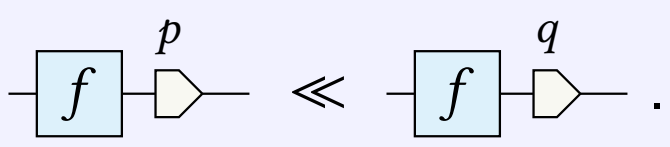

This means that to reason with delays we can use streaming and timelessness to propagate them across combinational components, and then use logical equivalence to

**Example 7.17.** One source of delay in circuits is the time gates take to process input signals. We can model this by inserting delay components after each gate, such as in the following circuit:

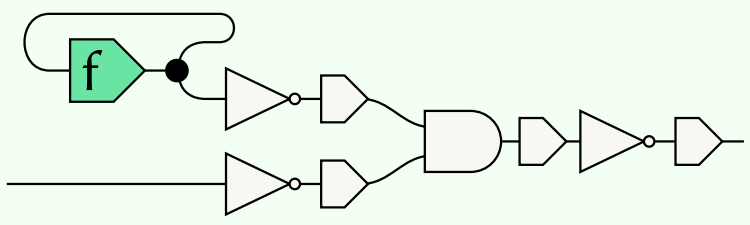

During reasoning we can permit these delays to be moved around, so long as when we finish any gates are still guarded by delays.

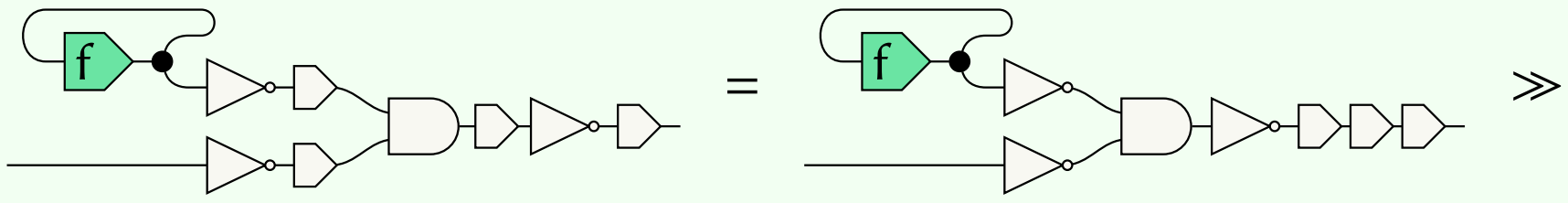

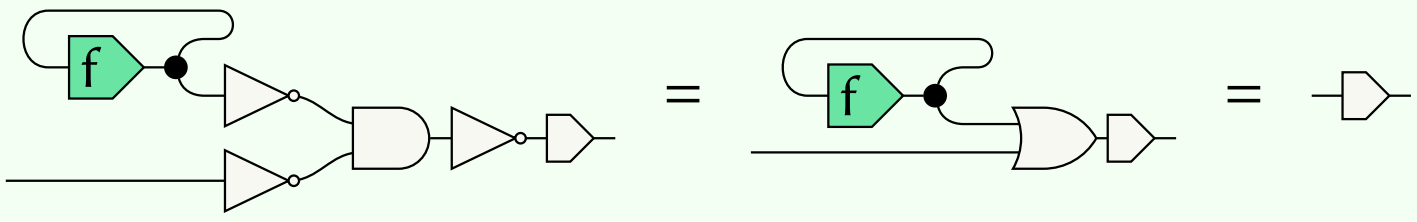

While this is a somewhat contrived toy example, it is possible that this technique could be applied to actual circuit optimisation procedures.

**Example 7.18** (Pipelining)**.** *Pipelining* [Par99] is a technique in which more registers are inserted into a circuit to increase throughput. This can be emulated in the compositional framework by applying transformations locally to registers. Ordinarily, such transformations can obfuscate a circuit's behaviour since the state space dramatically changes. In the compositional model, the structure of the circuit is left relatively untouched so this is less of an issue.

Not all circuit transformations are for the purpose of improving performance. Sometimes additional components must be bolted onto a circuit for *testing* purposes.

**Example 7.19** (Scan chains)**.** A common way of testing circuits is by using a *scan chain* [MZ00], a way of forcing the inputs to flipflops to test how specific states affect the outputs of the circuit. Adding a flipflop to a scan chain requires some extra inputs: the $\mathsf{scan}_{\mathsf{en}}$ wire toggles if the flipflop operates in normal mode or if it takes $\text{scan}_{\text{in}}$ as its value.

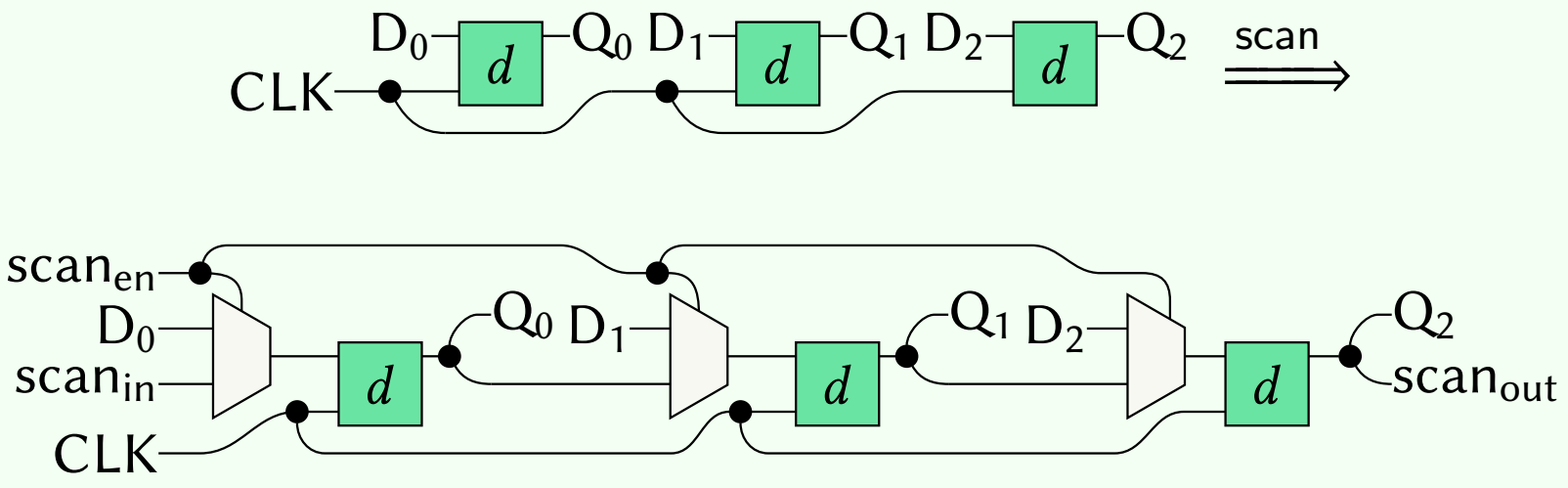

One could factor in these transformations when designing the circuit, but this can obfuscate the design of the actual logic. Additionally, applying these transformations where the remaining part of the circuit is *not* combinational can be quite complex. With the compositional approach the two tasks can be kept isolated by using blackboxes, layered explanations, and graphical reasoning.

## 7.5 Implementation

Throughout this section we have discussed some potential applications for the compositional theory for digital circuits. However, the examples have been kept to relatively small toy examples for ease of presentation and explanation. For readers less convinced by theoretical results, this might not be enough; how can the framework be adapted for real-life examples? Because examples can quickly balloon in size, it becomes impractical to develop and work through them by hand. Instead, it is necessary to pass the work along to a computer to generate and test things *automatically*. But how do we even communicate such things with a computer? This will be answered in great detail in the next part of this thesis.

# Part II

# Graph Rewriting for Sequential Digital Circuits

# Chapter 8

# String diagrams as hypergraphs

String diagrams are an appealing way of reasoning with pen and paper: they bring intuition to one-dimensional text strings and can often shed light on the next step of a proof. Unfortunately, they do have drawbacks: they take up a lot of time and space, and if not drawn with care can end up being messy, removing any benefit of using them in the first place.

Instead it is desirable to perform reasoning with string diagrams *computationally*. This presents some questions: how can we encode circuits constructed categorically in such a way that a computer can understand them? Perhaps we could input terms using traditional one-dimensional text representations? Text is something that computers are very good at processing, but as we have already established it is verbose, unintuitive, and, most importantly, means we have to apply the axioms of STMCs explicitly. Sticking to string diagrams would be ideal, but the representation needs some thought. Although computers do not deal well with topological objects like string diagrams, they are very well acquainted with combinatorial objects; for a computer to reason effectively with string diagrams, they must first be interpreted as *graphs*.

## 8.1 String diagram rewriting

Graph rewriting specialised for rewriting sting diagrams is a relatively new field. One of the first approaches was developed at the turn of the 2010s using *string graphs* [DDK10; DK13; Kis12], an example of which is illustrated in Figure 8.1. String graphs have two classes of vertices for *boxes* and *wires*; the former vertices represent generators in string diagrams and the latter vertices represent the wires between them. One nuance of string graphs is that a wire in a string diagram can be represented by arbitrarily many wire vertices connected together; all of these different depictions are identified by a notion of *wire homeomorphism* , in which adjacent wire vertices can be collapsed into one.

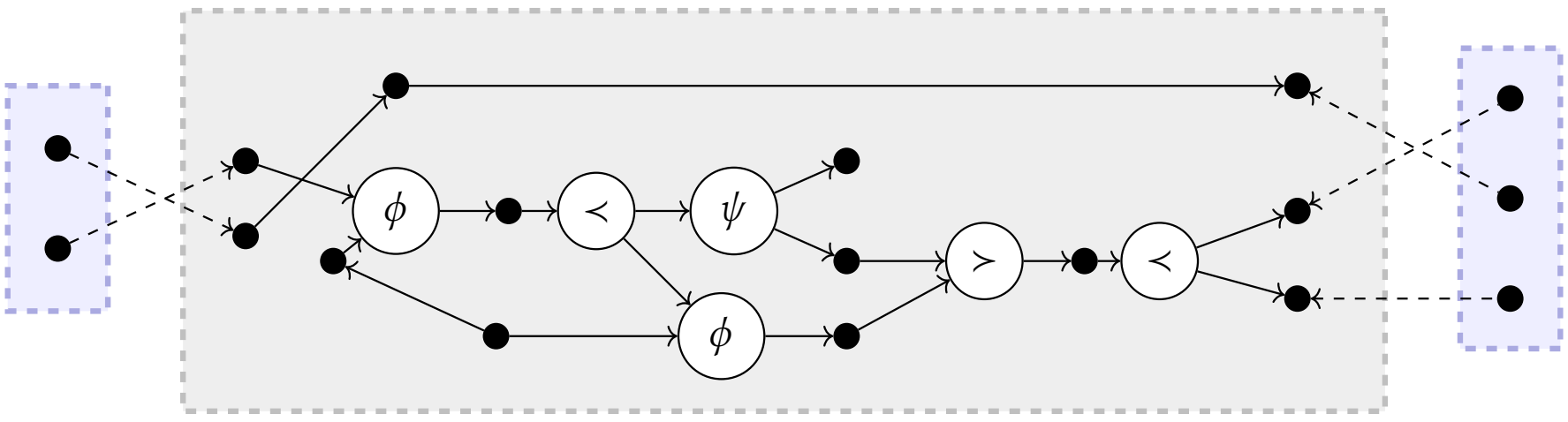

Figure 8.1: Example of an interfaced string graph

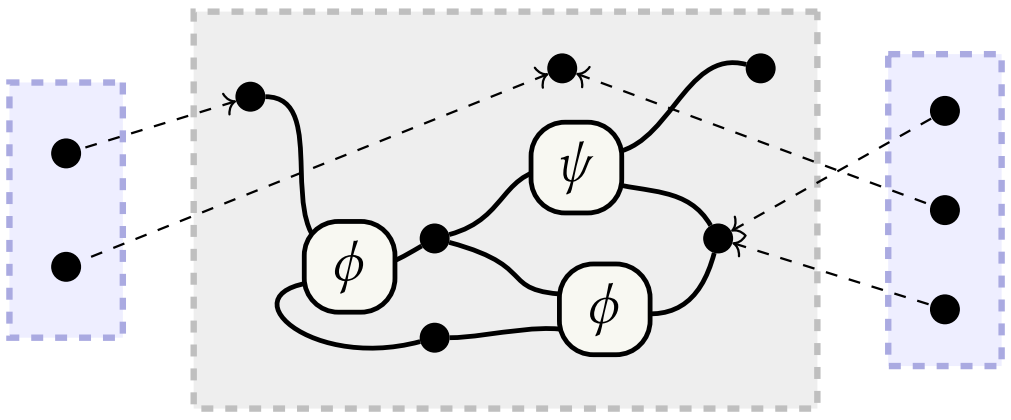

Figure 8.2: Example of an interfaced hypergraph

String graphs modulo wire homeomorphism are a suitable setting for modelling traced or compact closed categories, but their main drawback is that a given term may correspond to many different graphs thanks to wire homeomorphism.

More recently, there has been a flurry of work on string diagram rewriting modulo *Frobenius structure* using *hypergraphs* [BGK+16; Zan17; BGK+17; BGK+18; BGK+22a; BGK+22b; BGK+22c], such as that in Figure 8.2. Hypergraphs are a generalisation of graphs in which edges can have arbitrarily many sources and targets, rather than just one each. When interpreting string diagrams as hypergraphs, generators are represented as hyperedges and connections between generators indicated by shared source or target vertices. As there is no restriction for vertices to only be incident on a single source and target, one can model structures such as monoids or comonoids.

While string diagrams 'absorb' the equations of SMCs, hypergraphs go one further and absorb the equations of a *special commutative Frobenius algebra*: string diagrams equal by Frobenius equations are interpreted as isomorphic hypergraphs. This means rewriting using hypergraphs can be even more advantageous than using string diagrams.

Naturally, there have also been variations on this work where the complete Frobenius structure is not present. Suitable restrictions on hypergraphs and the graph rewriting process are also identified in [BGK+16] for rewriting *symmetric monoidal structure*. Research followed on rewriting modulo *(co)monoid structure* [MPZ23] ('half a Frobenius') and our work [GK23] on rewriting modulo *traced comonoid structure*. The latter is the basis for this part of the thesis.

## 8.2 Hypergraphs

We will begin by defining categories of hypergraphs, following the pattern outlined in [BGK+22a]. Hypergraphs are formally defined as a functor category.

**Definition 8.1** (Hypergraph [BGK+16]). Let $\mathbf{X}$ be the category with object set $(\mathbb{N} \times \mathbb{N}) + \star$ and morphisms $\mathsf{s}_i \colon (k, l) \to \star$ for each $i < k$ and $\mathsf{t}_j \colon (k, l) \to \star$ for each $j < l$. The category of hypergraphs $\mathbf{Hyp}$ is the functor category $[\mathbf{X}, \mathbf{Set}]$.

One can think of the category $\mathbf{X}$ as a 'template' for the structure of a hypergraph: the object $\star$ represents the vertices and each object $(k, l)$ represents hyperedges with $k$ sources and $l$ targets; each such edge must pick $k$ sources and $l$ targets from $\star$. Objects in $\mathbf{Hyp}$ are functors that instantiate each object in $\mathbf{X}$ to a concrete set. For a hypergraph $F \in \mathbf{Hyp}$ we write $F_\star$ for its vertices and $F_{k,l}$ for its edges with $k$ sources and $l$ targets.

**Example 8.2.** Let a hypergraph $F$ be defined as follows:

$$F_\star \coloneqq \{v_0, v_1, v_2, v_3, v_4, v_5\} \quad F_{2,1} \coloneqq \{e_0\} \quad F_{1,2} \coloneqq \{e_1\}$$
$$\mathsf{s}_0(e_0) \coloneqq v_1 \quad \mathsf{s}_1(e_0) \coloneqq v_0 \quad \mathsf{s}_0(e_1) \coloneqq v_1$$
$$\mathsf{t}_0(e_0) \coloneqq v_3 \quad \mathsf{t}_0(e_1) \coloneqq v_4 \quad \mathsf{t}_1(e_1) \coloneqq v_5$$

Much like with regular graphs, it is more intuitive to draw out hypergraphs rather than look at their combinatorial representation. We draw vertices as black dots and hyperedges as 'bubbles' with ordered tentacles on the left and right that connect to source and target vertices respectively, as illustrated in Figure 8.3. Note that the vertices do not have any notion of ordering or directionality.

Since it is a functor category, the morphisms in $\mathbf{Hyp}$ are natural transformations: structure-preserving maps between hypergraphs.

**Definition 8.3** (Hypergraph homomorphism). For two hypergraphs $F, G \in \mathbf{Hyp}$, a *hypergraph homomorphism* $f \colon F \to G$ is a pair of functions $f_\star \colon F_\star \to G_\star$ and $f_{k,l} \colon F_{k,l} \to G_{k,l}$ such that the following diagrams commute:

$$\begin{array}{ccc} F_{k,l} & \xrightarrow{f_{k,l}} & G_{k,l} \\ \downarrow{\scriptstyle \mathsf{s}_i} & & \downarrow{\scriptstyle \mathsf{s}_i} \\ F_\star & \xrightarrow{f_\star} & G_\star \end{array} \qquad \begin{array}{ccc} F_{k,l} & \xrightarrow{f_{k,l}} & G_{k,l} \\ \downarrow{\scriptstyle \mathsf{t}_i} & & \downarrow{\scriptstyle \mathsf{t}_i} \\ F_\star & \xrightarrow{f_\star} & G_\star \end{array}$$

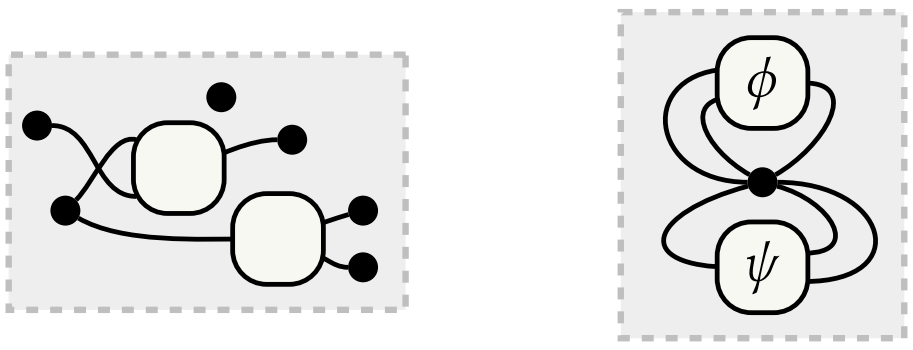

Figure 8.3: Illustration of the hypergraph from Example 8.2, the hypergraph signature from Example 8.6, and the labelling of the former with the latter from Example 8.9

**Example 8.4.** Consider the following hypergraph $G$:

$$G_\star \coloneqq \{v_6, v_7, v_8\} \quad G_{2,1} \coloneqq \{e_2\}$$
$$\mathsf{s}_0(e_2) \coloneqq v_6 \quad \mathsf{s}_1(e_2) \coloneqq v_7 \quad \mathsf{t}_0(e_2) \coloneqq v_8$$

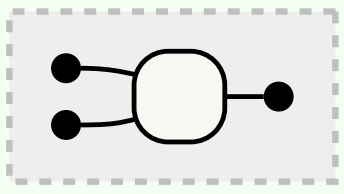

Recall the hypergraph $F$ from Example 8.2. A homomorphism $h\colon G \to F$ is a map from the vertices and edges of the former to those of the the latter preserving sources and targets; one possible homomorphism could be

$$h_\star(v_6) \coloneqq v_1 \quad h_\star(v_7) \coloneqq v_0 \quad h_\star(v_8) \coloneqq v_3$$
$$h_{2,1}(e_2) \coloneqq e_0$$

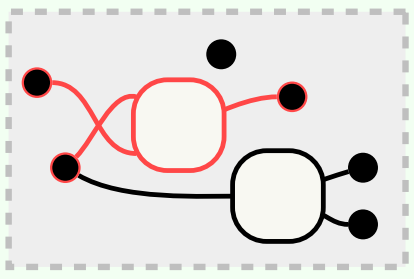

Injective hypergraph homomorphisms are often known as *embeddings*. However, there is no requirement for hypergraph homomorphisms to be injective. Consider another hypergraph $H$ defined as

$$H_\star \coloneqq \{v_9\} \quad H_{2,1} \coloneqq \{e_3\}$$
$$\mathsf{s}_0(e_3) = v_9 \quad \mathsf{s}_1(e_3) = v_9 \quad \mathsf{t}_0(e_3) = v_9$$

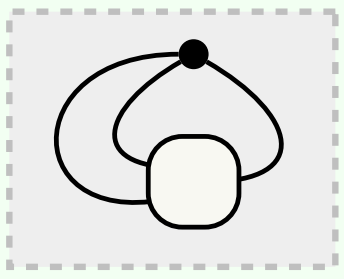

There is a non-injective homomorphism $k\colon G \to H$ defined as follows:

$$h_\star(v_6) \coloneqq v_9 \quad h_\star(v_7) \coloneqq v_9 \quad h_\star(v_8) \coloneqq v_9 \quad h_{2,1}(e_2) \coloneqq e_3$$

Although the vertices of $G$ are merged by $h$, the sources and targets are preserved.

### 8.2.1 Labelled hypergraphs

The graphical notation for hypergraphs is particularly evocative of string diagrams: generators correspond to hyperedges and wires to the vertices between them. The only thing missing is that the hyperedges are not *labelled* with generator symbols.

**Definition 8.5** (Hypergraph signature [BGK+16]). For a set of generators $\Sigma$ with arities and coarities as defined in Definition 2.2, the corresponding *hypergraph signature* $[\![\Sigma]\!]$ is an object of **Hyp** defined as follows:

$$[\![\Sigma]\!]_\star \coloneqq \{v\} \quad [\![\Sigma]\!]_{k,l} \coloneqq \{e_g \mid g \in \Sigma\} \quad \mathsf{s}_i(e_g) \coloneqq v \quad \mathsf{t}_j(e_g) \coloneqq v$$

**Example 8.6.** Let $\Sigma_m \coloneqq \{\phi \colon 2 \to 1, \psi \colon 1 \to 2\}$ be a monoidal signature. The corresponding hypergraph signature $[\![\Sigma]\!]$ is

$$[\![\Sigma]\!]_\star \coloneqq \{v\} \quad [\![\Sigma]\!]_{2,1} \coloneqq \{e_\phi\} \quad [\![\Sigma]\!]_{1,2} \coloneqq \{e_\psi\}$$
$$\mathsf{s}_0(e_\phi) \coloneqq v \quad \mathsf{s}_1(e_\phi) \coloneqq v \quad \mathsf{s}_0(e_\psi) \coloneqq v \quad \mathsf{t}_0(e_\phi) \coloneqq v \quad \mathsf{t}_0(e_\psi) \coloneqq v \quad \mathsf{t}_1(e_\psi) \coloneqq v$$

and is drawn as in the middle of Figure 8.3, where the edges are annotated with the appropriate label for clarity.

A hypergraph $F$ labelled over $\Sigma$ is a hypergraph homomorphism $\Gamma \colon F \to [\![\Sigma]\!]$; an edge $e \in F_{k,l}$ is labelled with generator $\phi$ if $\Gamma_{k,l}(e) = e_\phi$. This means a category of *labelled* hypergraphs is a category in which the objects are hypergraph homomorphisms to $[\![\Sigma]\!]$. This is a well-studied categorical template, and it has a special name.

**Definition 8.7** (Slice category [Law63]). For a category $\mathcal{C}$ and an object $C \in \mathcal{C}$, the *slice category* $\mathcal{C} \downarrow C$ has as objects the morphisms of $\mathcal{C}$ with target $C$ and as morphisms $(f \colon X \to C) \to (g \colon X' \to C)$ the morphisms $g \colon X \to X' \in \mathcal{C}$ such that $f' \circ g = f$.

Each object in a slice category $\mathcal{C} \downarrow C$ is a morphism with $C$ as its target. When we set $C$ to some hypergraph signature, this is a perfect setting for labelled hypergraphs.

**Definition 8.8** (Labelled hypergraphs [BGK+16]). For a set of generators $\Sigma$, the category of $\Sigma$-*labelled hypergraphs* is the slice category $\mathbf{Hyp}_\Sigma \coloneqq \mathbf{Hyp} \downarrow [\![\Sigma]\!]$.

**Example 8.9.** One labelling of the hypergraph $F$ from Example 8.2 could be defined using the homomorphism $\Gamma \colon F \to [\![\Sigma_m]\!]$ with action

$$\Gamma_\star(-) \coloneqq v \quad \Gamma_{2,1}(e_0) \coloneqq e_\phi \quad \Gamma_{1,2}(e_1) \coloneqq e_\psi$$

We draw a labelled hypergraph as a regular hypergraph but with labelled edges, as shown in Figure 8.3. If there are multiple generators with the same arity and coarity

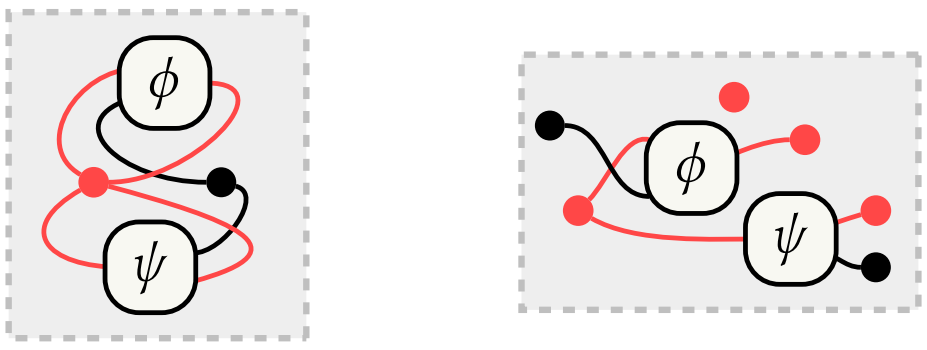

Figure 8.4: Illustration of the coloured hypergraph signature from Example 8.11 and the labelling of the hypergraph in Example 8.2 from Example 8.13

in a signature, there may well be multiple valid labellings of a hypergraph.

### 8.2.2 Coloured hypergraphs

We may additionally work over a countable set of *colours*. Accordingly, hypergraph signatures can be generalised to *coloured* hypergraph signatures.

**Definition 8.10** (Coloured hypergraph signature [BGK+16])**.** For a countable set set $C$ and a set of $C$-coloured generators $\Sigma$ as defined in Definition 2.2, the *coloured hypergraph signature* $[\![(C,\Sigma)]\!]$ is an object of **Hyp** defined as follows:

$$[\![(C,\Sigma)]\!]_\star \coloneqq \{v_c \mid c \in C\} \quad [\![(C,\Sigma)]\!]_{k,l} \coloneqq \{e_g \mid g \in \Sigma\}$$
$$\mathsf{s}_i(e_g) \coloneqq v_{\mathsf{dom}(e_g)(i)} \quad \mathsf{t}_j(e_g) \coloneqq v_{\mathsf{cod}(e_g)(j)}$$

**Example 8.11.** Let $C \coloneqq \{\bullet, {\color{red}\bullet}\}$ be let $\Sigma_c \coloneqq \{\phi \colon {\color{red}\bullet}\bullet \to {\color{red}\bullet}, \psi \colon {\color{red}\bullet} \to \bullet{\color{red}\bullet}\}$ be a monoidal signature; the coloured hypergraph signature $[\![(C,\Sigma_c)]\!]$ is

$$[\![(C,\Sigma_c)]\!]_\star \coloneqq \{v_\bullet, v_{\color{red}\bullet}\} \quad [\![(C,\Sigma_c)]\!]_{2,1} \coloneqq \{e_\phi\} \quad [\![(C,\Sigma_c)]\!]_{1,2} \coloneqq \{e_\psi\}$$
$$\mathsf{s}_0(e_\phi) \coloneqq v_{\color{red}\bullet} \quad \mathsf{s}_1(e_\phi) \coloneqq v_\bullet \quad \mathsf{s}_0(e_\psi) \coloneqq v_{\color{red}\bullet} \quad \mathsf{t}_0(e_\phi) \coloneqq v_{\color{red}\bullet} \quad \mathsf{t}_0(e_\psi) \coloneqq v_\bullet \quad \mathsf{t}_1(e_\psi) \coloneqq v_{\color{red}\bullet}$$

and is drawn by labelling edges appropriately as in Figure 8.4.

**Definition 8.12** (Coloured hypergraphs [BGK+16])**.** For a countable set $C$ and set of $C$-coloured generators $\Sigma$, let $\mathbf{Hyp}_{C,\Sigma}$ be the category of $(C,\Sigma)$-*labelled hypergraphs*, defined as the slice category $\mathbf{Hyp} \downarrow [\![(C,\Sigma)]\!]$.

**Example 8.13.** Returning again to the hypergraph $F$ in Example 8.2, we can label

it with colours and generators from $(C, \Sigma_c)$ with the hypergraph homomorphism

$$\Gamma_\star(v_0) \coloneqq v_{\color{red}\bullet} \quad \Gamma_\star(v_1) \coloneqq v_{\bullet} \quad \Gamma_\star(v_2) \coloneqq v_{\color{red}\bullet}$$
$$\Gamma_\star(v_3) \coloneqq v_{\color{red}\bullet} \quad \Gamma_\star(v_4) \coloneqq v_{\color{red}\bullet} \quad \Gamma_\star(v_5) \coloneqq v_{\bullet}$$
$$\Gamma_{2,1}(e_0) \coloneqq e_\phi \quad \Gamma_{1,2}(e_1) \coloneqq e_\psi$$

Coloured hypergraphs are drawn as labelled hypergraphs, but their vertices are additionally coloured, as shown in Figure 8.4.

### 8.2.3 Cospans of hypergraphs

String diagrams also have *input* and *output* interfaces. (Labelled) hypergraphs may have suggestively dangling vertices in the pictures, but this is not actually encoded in the definition; moreover we may wish to set a non-dangling vertex as an input or output. To set the interfaces of a hypergraph, hypergraph homomorphisms are used to 'pick' the appropriate vertices.

**Definition 8.14** (Cospan)**.** A *cospan* is a pair of morphisms $X \to A$ and $Y \to A$, written $X \to A \leftarrow Y$. A *cospan morphism* $(X \xrightarrow{f} A \xleftarrow{g} Y) \to (X \xrightarrow{h} B \xleftarrow{k} Y)$ is a morphism $\alpha\colon A \to B$ such that the following diagram commutes:

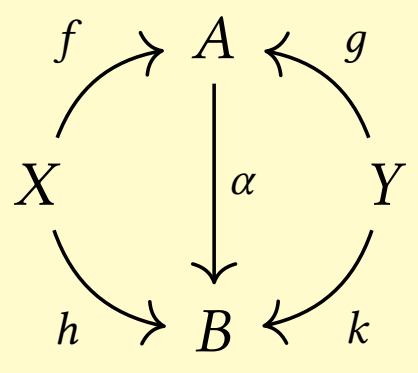

Two cospans $X \to A \leftarrow Y$ and $X \to B \leftarrow Y$ are *isomorphic* if there exists a morphism of cospans as above where $\alpha$ is an isomorphism.

Cospans will be used to model interfaced hypergraphs; for a cospan $X \to A \leftarrow Y$ the 'apex' $A$ will be a hypergraph and the 'legs' $X \to A$ and $Y \to A$ are input and output maps respectively. Since we will be comparing interfaced hypergraphs with string diagram terms, we require a categorical setting in which the objects are *cospans* of hypergraphs, so they must be composable.

**Definition 8.15** (Composition of cospans)**.** In a category $\mathcal{C}$ with pushouts, the composition of cospans $X \xrightarrow{f} A \xleftarrow{g} Y$ and $Y \xrightarrow{h} B \xleftarrow{k} Z$ is a cospan $X \to D \leftarrow Y$ where $D$ is computed by pushout:

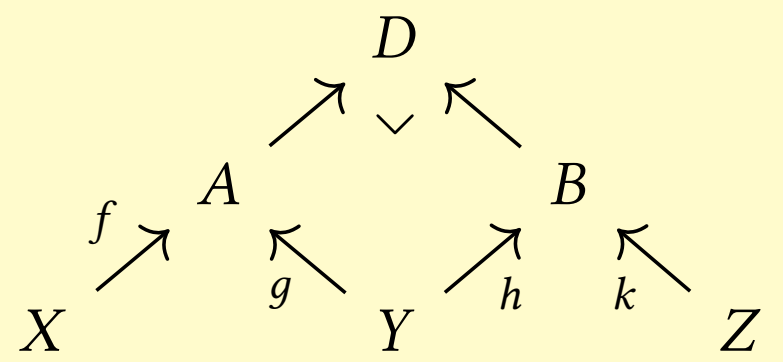

**Definition 8.16** (Categories of cospans)**.** Let $\mathcal{C}$ be a category with finite colimits. The category of cospans over $\mathcal{C}$, denoted $\mathsf{Csp}(\mathcal{C})$, has as objects the objects of $\mathcal{C}$ and as morphisms $A \to B$ the isomorphism classes of cospans $A \to X \leftarrow B$ for some $X \in \mathcal{C}$. Composition is by pushout as detailed in Definition 8.15 and the identity is $X \xrightarrow{\mathrm{id}_X} X \xleftarrow{\mathrm{id}_X} X$.

**Lemma 8.17.** For a category $\mathcal{C}$ with finite colimits, $\mathsf{Csp}(\mathcal{C})$ is symmetric monoidal with the tensor given by the coproduct in $\mathcal{C}$, the monoidal unit given by the initial object $0 \in \mathcal{C}$, and the symmetry defined as the cospan $A + B \to A + B \leftarrow B + A$.

*Proof.* It is a simple exercise to check that the equations of SMCs hold. □

As mentioned above, the legs of the cospan will act as the interfaces of the hypergraph: the hypergraph homomorphisms from the legs to the apex will pick out the input and output. But this means that not every hypergraph can act as an interface to a hypergraph, as any edges in the hypergraphs would also need to be mapped somewhere. We must restrict the interface hypergraphs to those that contain only vertices.

**Definition 8.18** (Discrete hypergraph)**.** A *discrete hypergraph* is a hypergraph in which each edge set is empty.

The legs of cospans representing interfaced hypergraphs will be discrete hypergraphs; if a hypergraph $F$ has $m$ inputs then the input hypergraph $A$ will contain $m$ vertices. This is not enough to fully specify the interfaces, as there must also be a notion of *order*. We need a way of specifying which vertex corresponds to which numbered input or output; this is formally performed by another functor.

**Theorem 8.19** ([BGK+22a], Thm. 3.6)**.** Let $\mathbb{X}$ be a PROP in which the monoidal product is a coproduct, let $\mathcal{C}$ be a category with finite colimits, and let $F \colon \mathbb{X} \to \mathcal{C}$ be a coproduct-preserving functor. Then there exists a PROP $\mathsf{Csp}_F(\mathcal{C})$ whose arrows $\overline{m} \to \overline{n}$ are isomorphism classes of $\mathcal{C}$ cospans $Fn \to C \leftarrow Fn$.

*Proof.* Composition is given by pushout. $F$ preserves coproducts so $F(m+p) \cong Fm + Fp$ and $F(n+q) \cong Fn + Fq$; subsequently the coproduct of $Fm \to C \leftarrow Fn$ and $Fp \to C \leftarrow Fq$ is given by $Fm + Fp \to C + D \leftarrow Fp + Fq$. Symmetries in $\mathbb{X}$ are determined by the universal property of the coproduct; they are inherited by $\mathsf{Csp}_F(\mathcal{C})$ because $F$ preserves coproducts. □

$F$ is the functor that will be used to map the objects in the cospan legs with some notion of ordering. For our purpose, the domain of this functor will be the PROP of finite sets $\{0, 1, 2, \ldots, m-1\}$.

**Definition 8.20.** Let $\mathbb{F}$ be the PROP with morphisms $m \to n$ the functions between finite sets $[m] \to [n]$.

We can now state the functor used to assemble interfaces of hypergraphs into words.

**Definition 8.21.** Let $D \colon \mathbb{F} \to \mathbf{Hyp}_\Sigma$ be defined as the functor sending an object $n$ to the discrete hypergraphs with $n$ vertices, and sending a function $m \to n$ to the induced homomorphism of discrete hypergraphs.

Instantiating Theorem 8.19 with $D$ produces the category of interfaced hypergraphs.

**Corollary 8.22.** There is a PROP $\mathsf{Csp}_D(\mathbf{Hyp}_\Sigma)$ where the morphisms are isomorphism classes of interfaced hypergraphs.

**Example 8.23.** Recall the labelled hypergraph $F$ from Example 8.9. We assign interfaces to it as the cospan $3 \xrightarrow{f} F \xleftarrow{g} 3$, where

$$f(0) = v_0 \quad f(1) = v_1$$
$$g(0) = v_5 \quad g(1) = v_4 \quad g(2) = v_2$$

Interfaces are drawn on the left and right of a main graph, with numbers illustrating the action of the cospan maps.

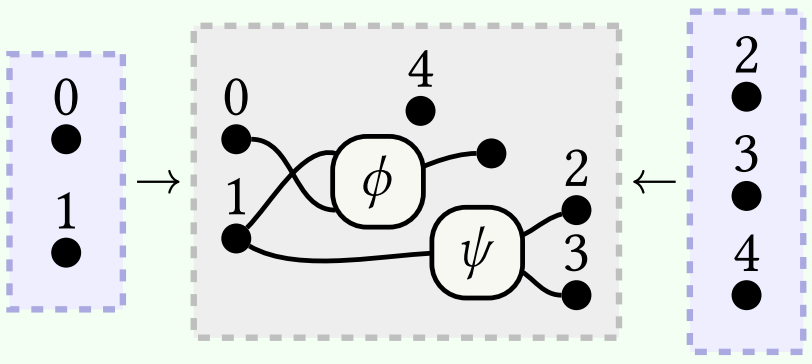

For clarity, we number the outputs after the inputs, but this does not reflect the mapping performed by $D$. Composition in $\mathsf{Csp}_D(\mathbf{Hyp}_\Sigma)$ is by pushout; effectively the vertices in the output of the first cospan are 'glued together' with the inputs of the second. Note that although we write different numbers for the glued cospan legs, they are actually mapped from the *same* finite set of three elements.

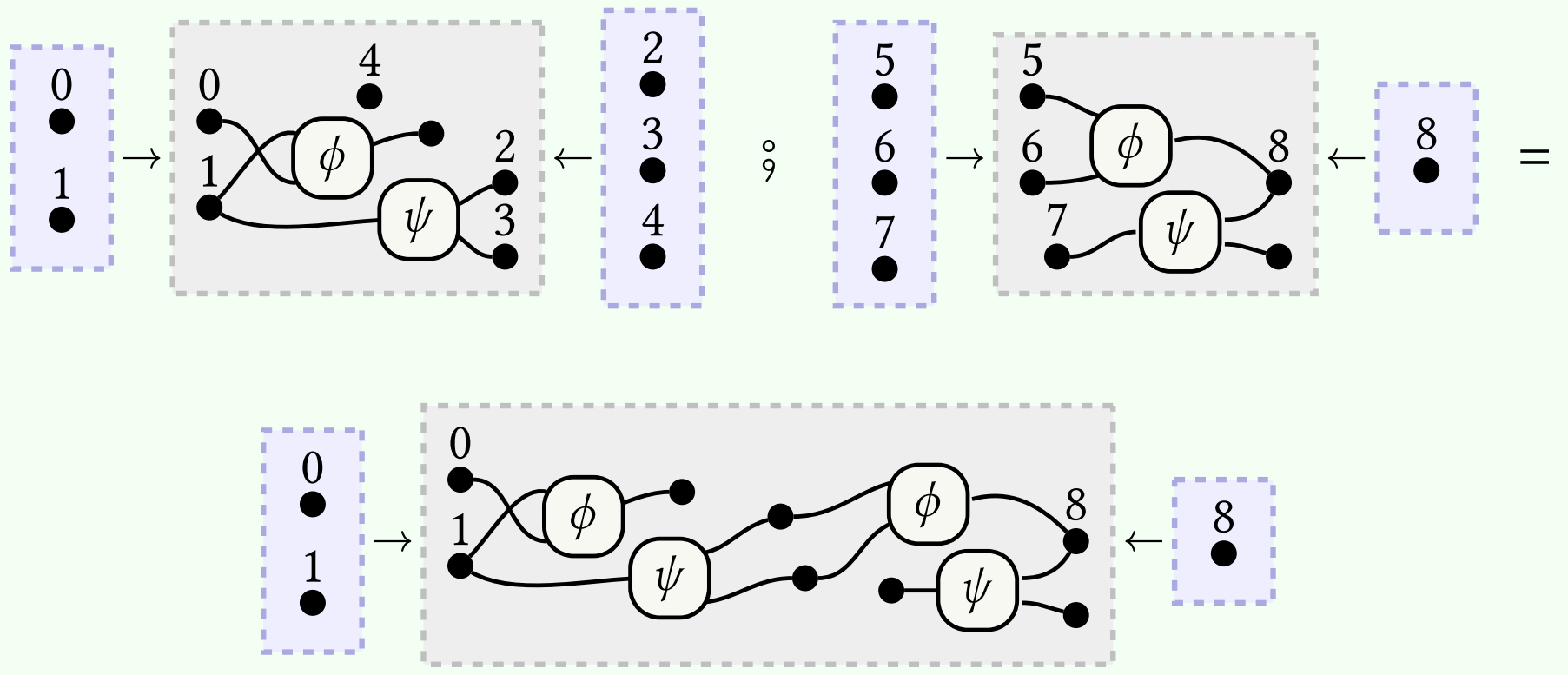

Tensor in $\mathsf{Csp}_D(\mathbf{Hyp}_\Sigma)$ is by direct product; putting cospans on top of each other.

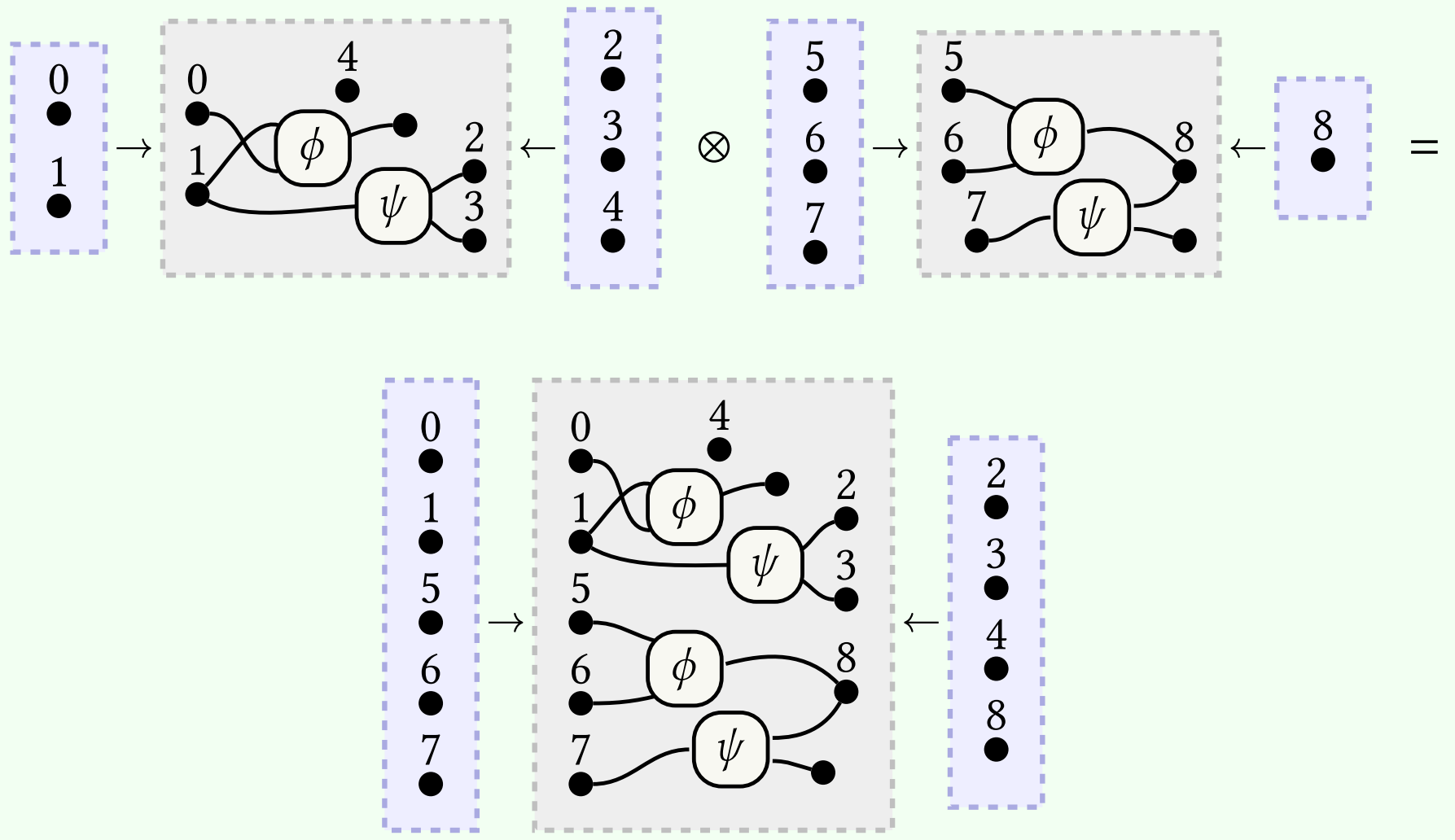

## 8.2.4 Coloured cospans

In the coloured setting there is slightly more nuance as the legs of the cospans are not just numbers but *words* of colours in some countable set $\mathcal{C}$.

**Theorem 8.24.** Let $\mathbb{X}$ be a coloured PROP whose monoidal product is a coproduct, $\mathcal{C}$ a category with finite colimits, and $F\colon \mathbb{X} \to \mathcal{C}$ a coproduct-preserving functor. Then there exists a coloured PROP $\mathsf{Csp}_F(\mathcal{C})$ whose arrows $\overline{m} \to \overline{n}$ are isomorphism classes of $\mathcal{C}$ cospans $F\overline{m} \to C \leftarrow F\overline{n}$.

*Proof.* As Theorem 8.19 but with word concatenation rather than addition. □

It only remains to determine exactly what the functor $F$ should be. In [BGK+22a], objects $[m] \in \mathbb{F}$ are be coloured over some finite set of colours $\mathcal{C}$ using a morphism $[m] \to \mathcal{C}$. Since we are working with potentially countably infinite sets of colours, the definition of $\mathbb{F}$ must first be tweaked.

**Definition 8.25.** Let $\hat{\mathbb{F}}$ be the category $\mathbb{F}$ augmented with the set of natural numbers and the functions $[m] \to \mathbb{N}$ for each finite set $[m]$.

Adding in morphisms $[m] \to \mathbb{N}$ allows for colourings with countably infinite sets of colours; the PROP of finite sets *coloured* over some countable set $C$ is the slice $\hat{\mathbb{F}} \downarrow C$. Objects of this category are pairs $([m], w\colon [m] \to C)$; this pair can be viewed as a word in $C^\star$ of length $m$, with the $i$th letter as $w(i)$.

**Remark 8.26.** Note that we do not include the morphisms $\mathbb{N} \to [m]$ in $\hat{\mathbb{F}}$; this is because when we view objects of $\hat{\mathbb{F}} \downarrow C$ as words in $C^\star$, we still only want to consider finite words despite there being potentially countably infinite colours.

All that remains is to verify that $\hat{\mathbb{F}} \downarrow C$ is indeed a coloured PROP. To assist in this endeavour, we recall a property of slice categories.

**Lemma 8.27.** For a category $\mathcal{C}$ with coproducts, $\mathcal{C} \downarrow X$ has coproducts.

*Proof.* Let $A, B, X$ be objects in $\mathcal{C}$; as $\mathcal{C}$ has coproducts $A + B$ is also an object in $\mathcal{C}$. Then the coproduct of $(A, A \to X)$ and $(B, B \to X)$ in $\mathcal{C} \downarrow X$ is $A + B \to X$; the universal morphism is $[f, g]$. □

Following the strategy of [BGK+22a, Prop. 2.23], we now show that $\hat{\mathbb{F}} \downarrow C$ is a coloured PROP.

**Proposition 8.28.** For a countable set $C$, $\hat{\mathbb{F}} \downarrow C$ is a coloured PROP.

*Proof.* As established, the objects of $\hat{\mathbb{F}} \downarrow C$ can be viewed as words in $C^\star$. As slice categories preserve coproducts by Lemma 8.27, $\hat{\mathbb{F}} \downarrow C$ is strict symmetric monoidal, and the coproduct acts as concatenation of words. □

The category of interfaced coloured hypergraphs is then constructed in the same way as the monochromatic version.

**Definition 8.29** ([BGK+22a], Rem. 3.12)**.** Let $D_C : \hat{\mathbb{F}} \downarrow C \to \mathbf{Hyp}_{C,\Sigma}$ be defined as the functor sending a word $\overline{n}$ to the corresponding discrete coloured hypergraph containing vertices coloured as in $\overline{n}$, and sending a function $\overline{m} \to \overline{n}$ to the induced homomorphism of discrete hypergraphs.

As with the monochromatic case, applying Theorem 8.24 to this functor gives us a category of *coloured* interfaced hypergraphs $\mathrm{Csp}_{D_C}(\mathbf{Hyp}_{C,\Sigma})$.

**Example 8.30.** Composition in $\mathrm{Csp}_{D_C}(\mathbf{Hyp}_{C,\Sigma})$ is as in $\mathrm{Csp}_D(\mathbf{Hyp}_\Sigma)$ but now the glueing must also preserve colours.

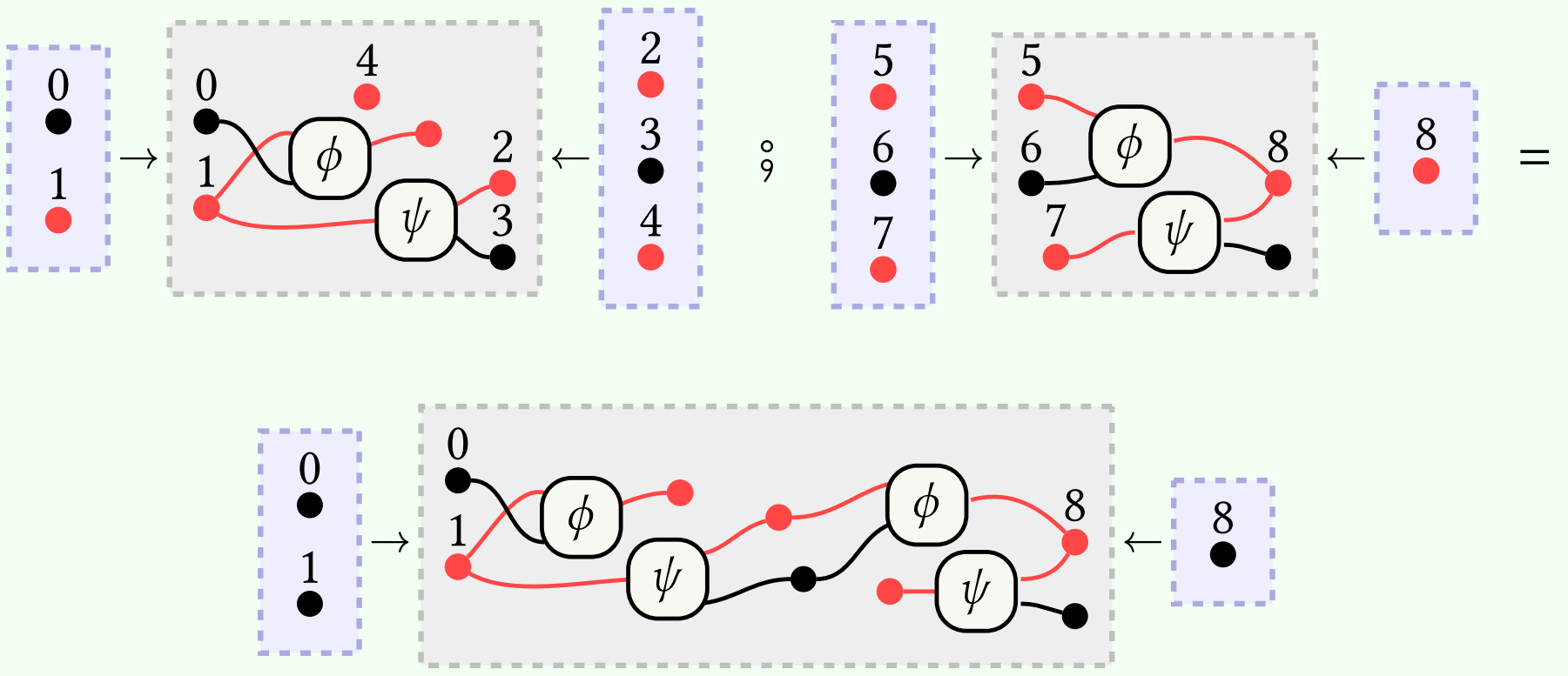

The tensor is exactly the same as in $\mathrm{Csp}_D(\mathbf{Hyp}_\Sigma)$.

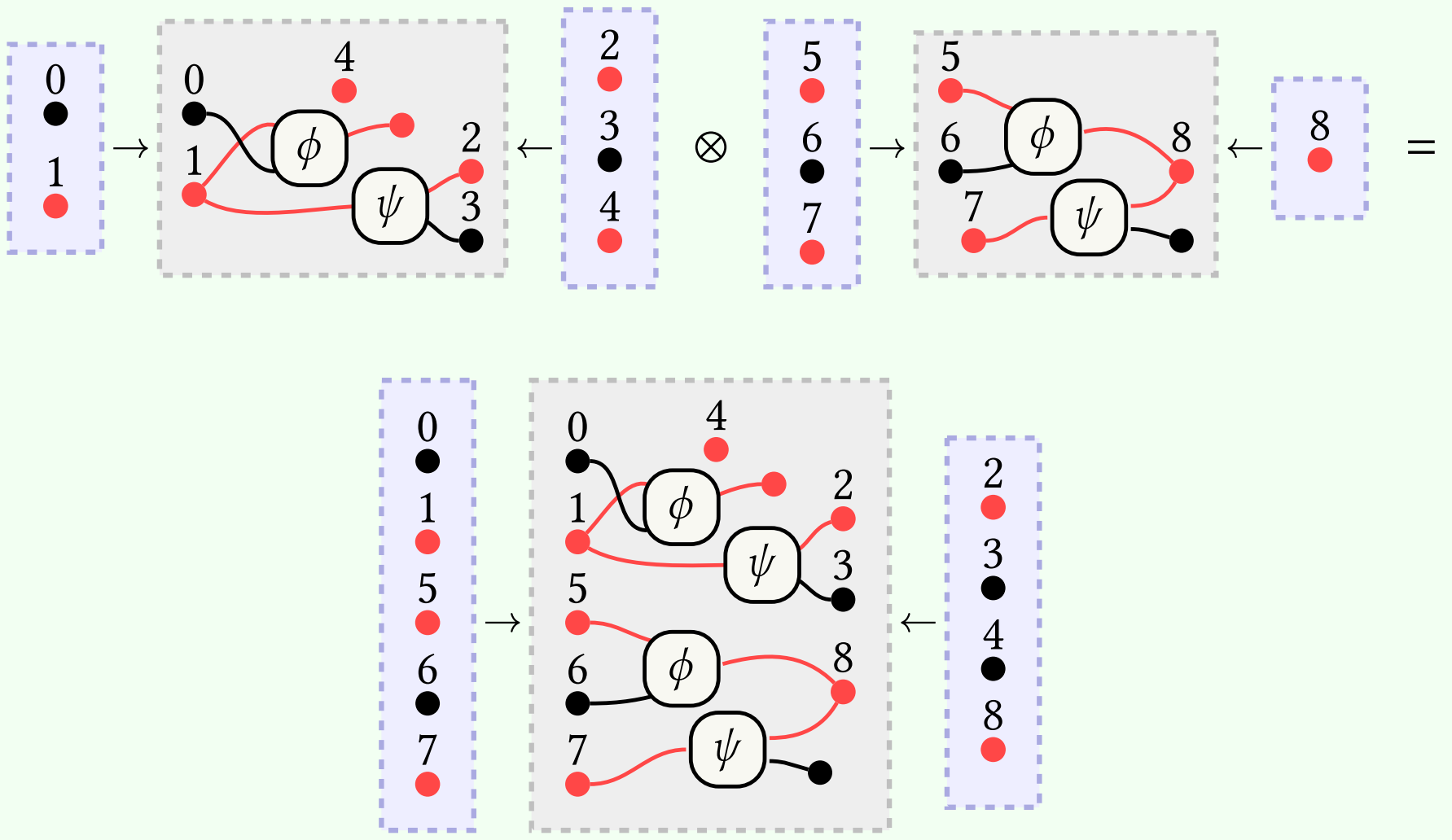

## 8.3 Frobenius terms as hypergraphs

In order to perform graph rewriting on string diagrams, we will interpret the latter as cospans of hypergraphs. We will first recount the constructions used by Bonchi et al in [BGK+22a] for a broader class of terms before showing their recipe can be adapted for a *traced* setting either with or without a comonoid structure.

### 8.3.1 Frobenius structure

When reasoning with monoidal theories and string diagrams, two structures that often appear are a *commutative monoid* (for joining and introducing wires) and a *cocommutative comonoid* (for forking and eliminating wires).

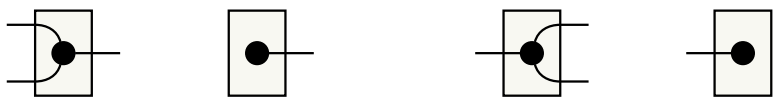

The monoid and comonoid are subject to the usual equations of (co)unitality, (co)associativity and (co)commutativity.

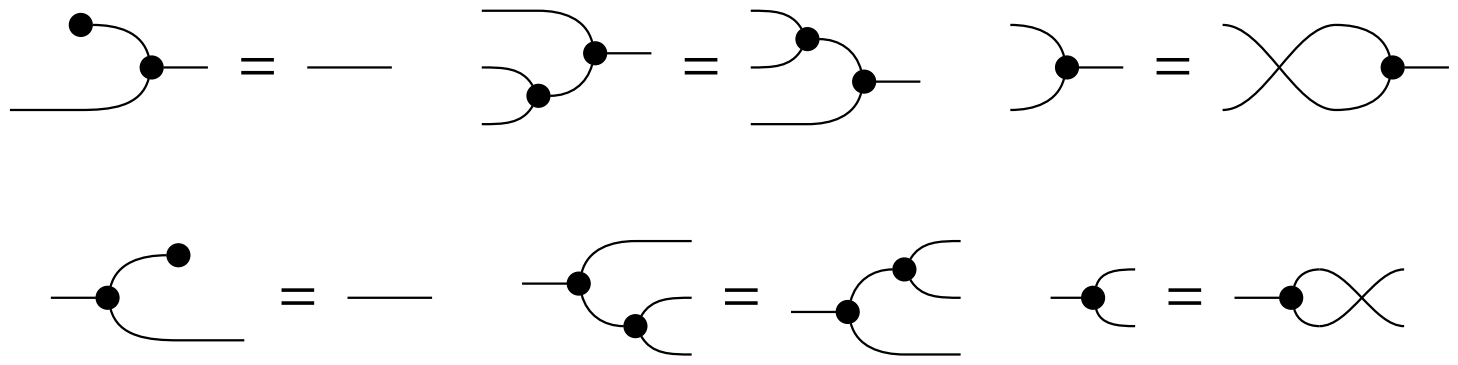

When monoids and comonoids appear together, there are multiple ways they can interact. One way is by using the equations of a *Frobenius algebra*; this is particularly relevant to us because symmetric monoidal terms equipped with a Frobenius structure correspond precisely to the cospans of hypergraphs defined in the previous section.

**Definition 8.31.** The monoidal theory of *special commutative Frobenius algebras* is defined as $(\Sigma_{\mathbf{Frob}}, \mathcal{E}_{\mathbf{Frob}})$, where $\Sigma_{\mathbf{Frob}} \coloneqq \{$ 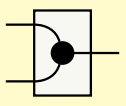 , 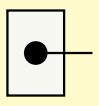 , 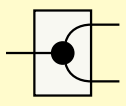 , 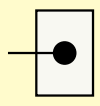 $\}$ and the equations of $\mathcal{E}_{\mathbf{Frob}}$ are listed in Figure 8.5. We write $\mathbf{Frob} \coloneqq \mathsf{S}_{\Sigma_{\mathbf{Frob}}, \mathcal{E}_{\mathbf{Frob}}}$.

The equations of special Frobenius algebras are those of commutative monoids and cocommutative comonoids along with the 'Frobenius' and 'special' equations.

**Example 8.32.** The following are all terms in **Frob**:

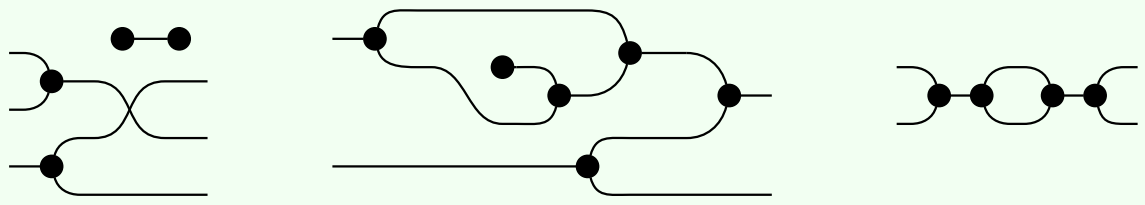

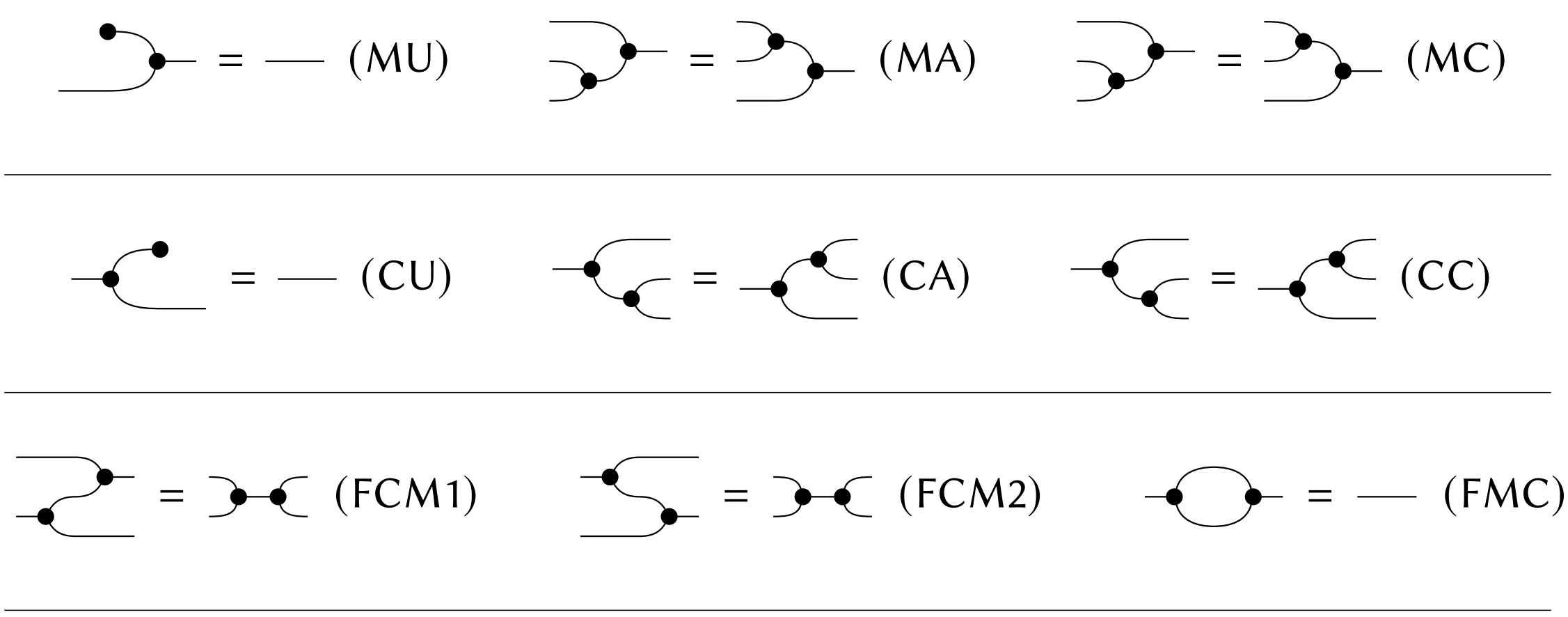

Figure 8.5: Equations $\mathcal{E}_{\mathbf{Frob}}$ of a *special commutative Frobenius algebra.*

Using the equations of $\mathcal{E}_{\mathbf{Frob}}$, it can be shown that the latter two terms are equal:

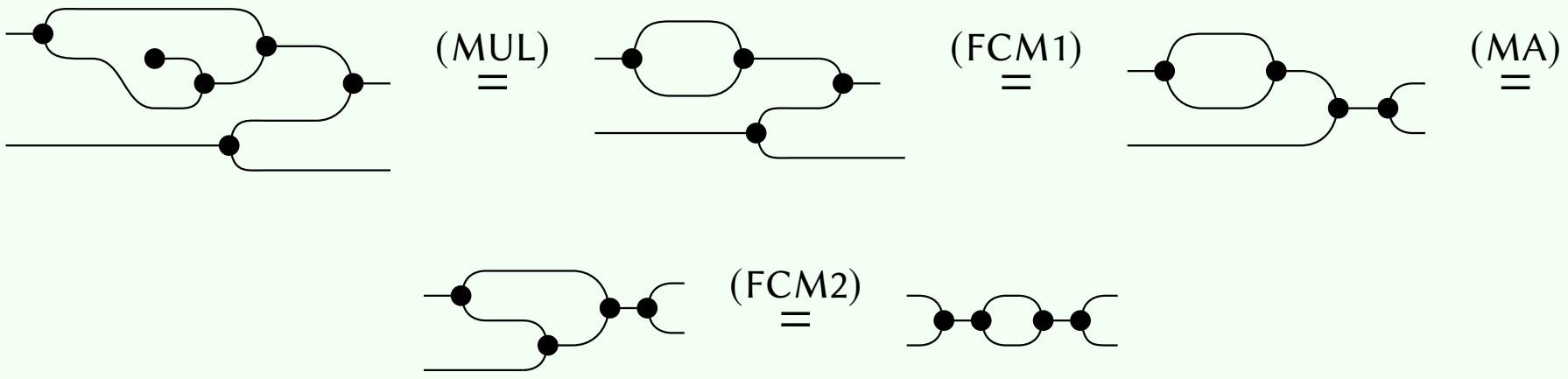

Effectively, any terms in **Frob** with the same input-output connectivity are equal.

### 8.3.2 Coloured Frobenius

**Frob** is a monochromatic PROP. To define a *coloured* version of **Frob** we simply use a different copy of **Frob** to represent each colour, using a fact about **PROP** and **CPROP**.

**Theorem 8.33** ([BCR18], Corollary 5.3)**.** **PROP** has coproducts.

This generalises to **CPROP** by replacing natural numbers with words. This means that given coloured PROPs $\mathcal{C}$ and $\mathcal{D}$ with objects the words in $C^\star$ and $D^\star$ respectively, there is also a coloured PROP $\mathcal{C} + \mathcal{D}$ with objects the words in $(C + D)^\star$ and morphisms defined in the obvious way. We can use this to define a multi-coloured version of **Frob** as a coproduct of copies of **Frob**.

**Definition 8.34** ([BGK+22a])**.** For a countable set $C$, let $\mathbf{Frob}_C \in \mathbf{CPROP}$ be defined as $\mathbf{Frob}_C \coloneqq \sum_{c \in C} \mathbf{Frob}$.

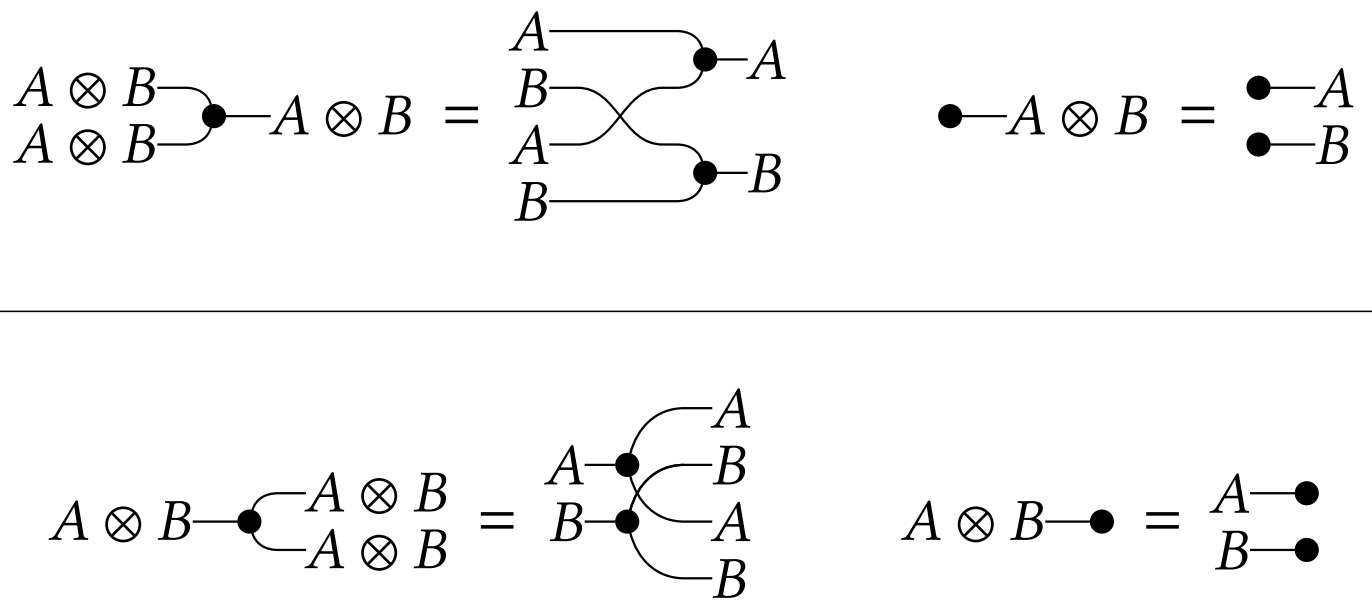

Figure 8.6: Equations of a hypergraph category

**Example 8.35.** In $\mathbf{Frob}_C$, there is a copy of the Frobenius structure for each colour in $C$. For example, when $\mathcal{C} \coloneqq \{\bullet, {\color{red}\bullet}\}$, the following are terms in $\mathbf{Frob}_C$.

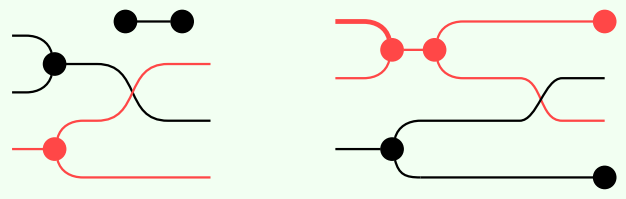

Although there are two different colours of wires, these wires cannot interact without the addition of other generators to map between them.

### 8.3.3 Hypergraph categories

Frobenius structures have turned out to be very useful in studying compositional processes such as quantum processes [CD08] and signal flow graphs [BSZ14; BSZ15]. It is useful to talk about the setting in which *every* object has such a structure.

**Definition 8.36** (Hypergraph category [FS19]). A *hypergraph category* is a category in which every object is equipped with a special commutative Frobenius algebra subject to the *coherence equations* in Figure 8.6.

**Remark 8.37.** The notion of a hypergraph category has been rediscovered numerous times over the years. They were originally called *well-supported compact closed categories* by Carboni and Walters [CW87], and have subsequently appeared as *dgs-monoidal categories* [KSW97; GH98; GHL99; BGM02] and *dungeon categories* [Mor14]. The term *hypergraph categories* was coined more recently but has become the standard in the compositional processes community [Kis15; Fon15; BFP16; BF18].

As the Frobenius structure is their entire *raison d'être*, it is unsurprising that the categories of Frobenius terms we encountered earlier are hypergraph categories.

**Lemma 8.38.** $\mathbf{Frob}$ is a hypergraph category.

*Proof.* The generators in $\mathbf{Frob}$ provide the Frobenius structure for the object $1$; for the other objects the structure is derived by following the recipes given by the coherence equations in Figure 8.6. □

It is now possible to formally define what we mean when we say 'Frobenius terms'.

**Definition 8.39.** For a set of generators $\Sigma$, let $\mathbf{H}_\Sigma$ be the PROP freely generated over $\Sigma + \Sigma_{\mathbf{Frob}}$.

While we can view $\mathbf{H}_\Sigma$ as just 'the category containing all the Frobenius terms', it can be advantageous to view it as a coproduct.

**Lemma 8.40.** $\mathbf{H}_\Sigma \cong \mathbf{S}_\Sigma + \mathbf{Frob}$.

*Proof.* Every term in $\mathbf{H}_\Sigma$ can be expressed as a combination of generators either in $\mathbf{S}_\Sigma$ or $\mathbf{Frob}$. □

We can also proceed similarly for the multi-coloured case.

**Lemma 8.41.** $\mathbf{Frob}_C$ is a hypergraph category.

*Proof.* As Lemma 8.38, but there is now a 'base' Frobenius structure for each colour $c \in C$. □

**Definition 8.42.** For a set of $C$-coloured generators over $\Sigma$, let $\mathbf{H}_{C,\Sigma}$ be the $C$-coloured PROP freely generated over $\Sigma + \Sigma_{\mathbf{Frob}_C}$.

**Lemma 8.43.** $\mathbf{H}_{C,\Sigma} \cong \mathbf{S}_{C,\Sigma} + \mathbf{Frob}_C$.

Viewing $\mathbf{H}_\Sigma$ and $\mathbf{H}_{C,\Sigma}$ as coproducts will prove to be beneficial when establishing a correspondence between terms and graphs in the the next section, as it allows us to consider the symmetric monoidal and the Frobenius components separately.

### 8.3.4 Hypergraph categories and hypergraphs

Perhaps confusingly, the category of *hypergraphs* $\mathbf{Hyp}_\Sigma$ is *not* a hypergraph category, but the category of *cospans* of hypergraphs is. This can be shown by exploiting a correspondence between **Frob** and the PROPs of finite sets we encountered earlier.

**Proposition 8.44** ([Lac04], Ex. 5.4)**.** $\mathbf{Frob} \cong \mathrm{Csp}(\mathbb{F})$.

We omit the formal proof and sketch the correspondence. Terms in **Frob** are formed of all the ways of combining 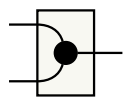, 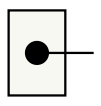, 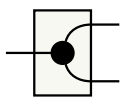, 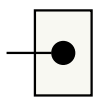, 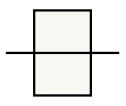, and 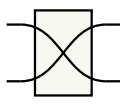 in sequence and parallel, so a string diagram for a term $f : m \to n$ is depicted as $x$ connected components drawing paths from $m$ inputs to $n$ outputs, such as in the example below.

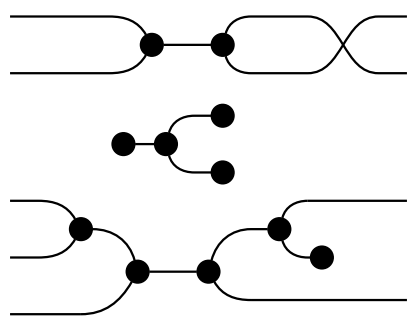

Note there is no requirement for each component to connect to one or both interfaces as the 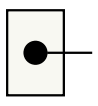 and 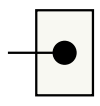 generators can introduce and stub wires. A term $f : m \to n$ with $x$ connected components corresponds to a cospan of finite sets $[m] \xrightarrow{i} [x] \xleftarrow{j} [n]$, where the functions $i$ and $j$ map the inputs and outputs to the components they connect to.

**Example 8.45.** Consider the term $f : 5 \to 4$ drawn on the left below. This corresponds to a cospan $[5] \to [3] \leftarrow [4]$ as shown on the right below.

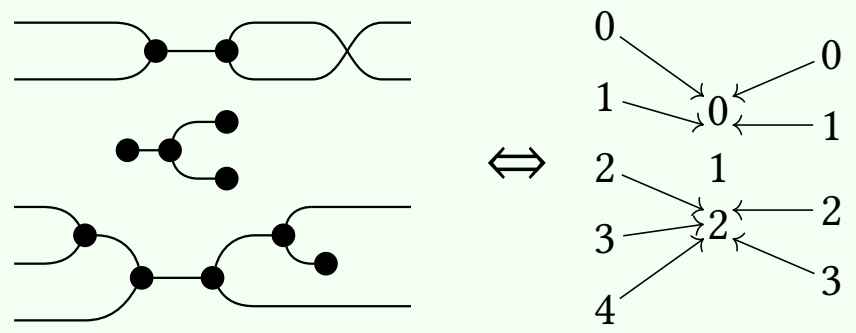

The cospan representation shows how all connected Frobenius components can be 'squished' into a single blob.

We have now ascertained the relationship between **Frob** and $\mathrm{Csp}(\mathbb{F})$. The missing link is the relationship between the latter and $\mathrm{Csp}_D(\mathbf{Hyp}_\Sigma)$; this arises as a special case of the following theorem.

**Theorem 8.46** ([BGK+22a], Thm. 3.8)**.** Let $\mathbb{X}$ be a PROP in which the monoidal product is a coproduct, let $\mathcal{C}$ be a category such that $\mathbb{X}$ and $\mathcal{C}$ have finite limits, and let $F : \mathbb{X} \to \mathcal{C}$ be a colimit-preserving functor. Then there is a homomorphism of PROPs $\tilde{F} : \mathrm{Csp}(\mathbb{X}) \to \mathrm{Csp}_F(\mathcal{C})$ that sends $m \xrightarrow{f} X \xleftarrow{g} n$ to $Fm \xrightarrow{Ff} FX \xleftarrow{Fg} Fn$. If $F$ is full and faithful, then $\tilde{F}$ is faithful.

*Proof.* Since $F$ preserves finite colimits, it preserves composition (pushout) and monoidal product (coproduct); symmetries are clearly preserved. To show that $\tilde{F}$ is faithful when $F$ is full and faithful, suppose that $\tilde{F}(m \xrightarrow{f} X \xleftarrow{g} n) = \tilde{F}(m \xrightarrow{f'} X \xleftarrow{g'} n)$. This gives us the following commutative diagram in $\mathcal{C}$:

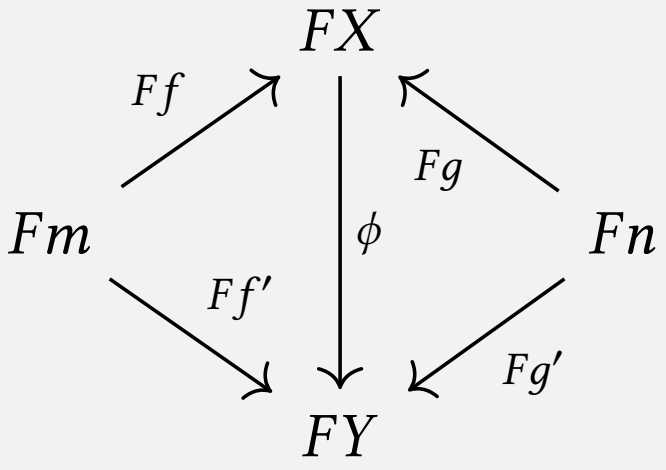

where $\phi$ is an isomorphism because morphisms in $\mathrm{Csp}_F(\mathbf{C})$ are isomorphism classes of cospans. As $F$ is full, there exists $\psi\colon X \to Y$ such that $F\psi = \phi$. As $F$ is faithful, $\psi$ is an isomorphism; this means $m \xrightarrow{f} X \xleftarrow{g} n$ and $m \xrightarrow{f'} X \xleftarrow{g'} n$ are equal in $\mathrm{Csp}(\mathbb{X})$, so $\tilde{F}$ is faithful. □

**Corollary 8.47** ([BGK+22a], Cor. 3.9)**.** There is a faithful PROP morphism $\tilde{D}\colon \mathrm{Csp}(\mathbb{F}) \to \mathrm{Csp}_D(\mathbf{Hyp}_\Sigma)$

With this we can derive a map from Frobenius terms to cospans of hypergraphs.

**Definition 8.48.** Let $[-]_\Sigma\colon \mathbf{Frob} \to \mathrm{Csp}_D(\mathbf{Hyp}_\Sigma)$ be the PROP morphism defined by using Proposition 8.44 followed by Corollary 8.47.

**Example 8.49.** The action of $[-]_\Sigma$ on the Frobenius generators is as follows:

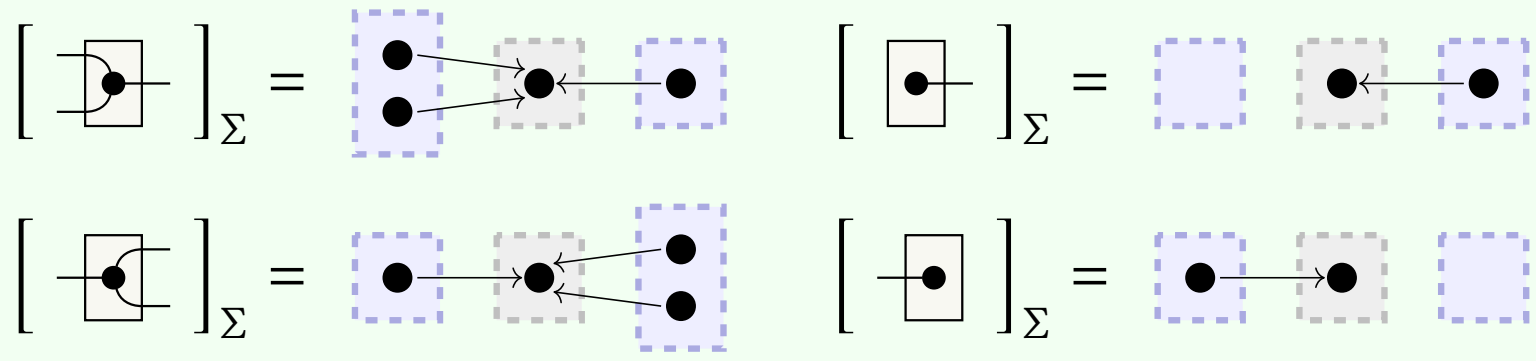

As there is a faithful embedding of $\mathbf{Frob}$ into $\mathrm{Csp}_D(\mathbf{Hyp}_\Sigma)$ and both categories share the same objects, we also get the result alluded to at the start of this section.

**Corollary 8.50.** $\mathrm{Csp}_D(\mathbf{Hyp}_\Sigma)$ is a hypergraph category.

### 8.3.5 From coloured terms to coloured graphs

This result shows how the correspondence works for the monochromatic case; what about for coloured terms? Here, we replace $\mathbb{F}$ with the coloured version $\hat{\mathbb{F}} \downarrow C$ seen in the previous section. A coloured version of Proposition 8.44 was shown for a *finite* set of colours in [BGK+22a]; we recall its proof before extending this to the *countable* setting we work in.

**Lemma 8.51.** In a category $\mathcal{C}$ with a terminal object 1, $\mathcal{C} \cong \mathcal{C} \downarrow 1$.

*Proof.* Since 1 is terminal, there is a unique morphism $A \to 1$ for each object $A$ in $\mathcal{C}$, so there is an object $(A, !_A : A \to 1)$ in $\mathcal{C} \downarrow 1$ for each object $A \in \mathcal{C}$. In $\mathcal{C} \downarrow 1$ there is a morphism $(A, !_A : A \to 1) \to (B, !_B : B \to 1)$ in for every morphism $f : A \to B \in \mathcal{C}$ such that $f \mathbin{;} !_B = !_A$; since both $f \mathbin{;} !_B$ and $!_A$ are morphisms $A \to 1$ they must be the same unique morphism. Therefore $\mathcal{C} \cong \mathcal{C} \downarrow 1$. □

**Theorem 8.52** ([BGK+22a], Theorem 2.24)**.** For a finite set of colours $C \in \mathbb{F}$, there is an isomorphism of coloured PROPs $\mathbf{Frob}_C \cong \mathsf{Csp}(\mathbb{F} \downarrow C)$.

*Proof.* By definition of $\mathbf{Frob}_C$, Definition 8.34, Proposition 8.44, and Lemma 8.51 we have that

$$\mathbf{Frob}_C \coloneqq \sum_{c \in C} \mathbf{Frob} \cong \sum_{c \in C} \mathsf{Csp}(\mathbb{F}) \cong \sum_{c \in C} \mathsf{Csp}(\mathbb{F} \downarrow 1)$$

In the other direction we have that $\mathsf{Csp}(\mathbb{F} \downarrow C) \cong \mathsf{Csp}(\mathbb{F} \downarrow \sum_{c \in C} 1)$ as $C$ is countable. So we need to show that $\sum_{c \in C} \mathsf{Csp}(\mathbb{F} \downarrow 1) \cong \mathsf{Csp}(\mathbb{F} \downarrow \sum_{c \in C} 1)$. The objects of the former are coproducts of objects in $\mathbb{F} \downarrow C$; as this is a coloured prop the coproduct is concatenation and subsequently the objects can be viewed as words in $C^\star$. Similarly, the objects of the latter are objects of $\mathbb{F} \downarrow \sum_{c \in C} 1$, which can clearly also be seen as words in $C^\star$.

The morphisms of the former are coproducts of cospans, which can equivalently be viewed as a single cospan with coproducts in the legs and apex; using the reasoning above this means it is a cospan of words in $C^\star$; it is easy to see that this is also the case for morphisms in the latter. □

We need to show a version of this for the case where $C$ may be *countably infinite*. The strategy is much the same as, but relies on one small observation.

**Lemma 8.53.** Let $C \in \mathbb{F}$ be a finite cardinal. Then $\hat{\mathbb{F}} \downarrow C \cong \mathbb{F} \downarrow C$.

*Proof.* The morphisms in $\hat{\mathbb{F}} \downarrow C$ are the morphisms $[m] \to C$ for finite $C$, which are precisely the morphisms of $\mathbb{F} \downarrow C$. □

This slips in to the proof above to extend it to *countable* sums.

**Theorem 8.54.** For a countable set $C$, there is an isomorphism of coloured PROPs $\mathbf{Frob}_C \cong \mathsf{Csp}(\hat{\mathbb{F}} \downarrow C)$.

*Proof.* The proof is almost the same as Theorem 8.52 but with the addition of Lemma 8.53. We have that

$$\mathbf{Frob}_C \coloneqq \sum_{c \in C} \mathbf{Frob} \cong \sum_{c \in C} \mathsf{Csp}(\mathbb{F}) \cong \sum_{c \in C} \mathsf{Csp}(\mathbb{F} \downarrow 1) \cong \sum_{c \in C} \mathsf{Csp}(\hat{\mathbb{F}} \downarrow 1).$$

In the other direction we still have that $\mathsf{Csp}(\hat{\mathbb{F}} \downarrow C) \cong \mathsf{Csp}(\hat{\mathbb{F}} \downarrow \sum_{c \in C} 1)$ as $C$ is still countable. As before we need to show that $\sum_{c \in C} \mathsf{Csp}(\hat{\mathbb{F}} \downarrow 1) \cong \mathsf{Csp}(\hat{\mathbb{F}} \downarrow \sum_{c \in C} 1)$, which follows by the same reasoning as in the prequel. □

As with the monochromatic case, we must now define a map from finite sets to discrete hypergraphs.

**Definition 8.55.** For a countable set $C$, let $D_C \colon \hat{\mathbb{F}} \downarrow C \to \mathbf{H}_{C,\Sigma}$ be defined as the functor that maps a coloured word $\overline{w}$ to the discrete coloured hypergraph containing an appropriately coloured vertex for each element of $\overline{w}$.

**Corollary 8.56** ([BGK+22a], Rem. 3.12)**.** There is a faithful PROP morphism $\widetilde{D_C} \colon \mathsf{Csp}(\hat{\mathbb{F}} \downarrow C) \to \mathsf{Csp}_{D_C}(\mathbf{Hyp}_{C,\Sigma})$.

It is now possible to map from coloured Frobenius terms to cospans of hypergraphs.

**Definition 8.57.** Let $[-]_{C,\Sigma} \colon \mathbf{Frob}_C \to \mathsf{Csp}_{D_C}(\mathbf{Hyp}_{C,\Sigma})$ be the homomorphism obtained by composing the isomorphism of Theorem 8.54 with Corollary 8.56.

**Corollary 8.58.** $\mathsf{Csp}_{D_C}(\mathbf{Hyp}_{C,\Sigma})$ is a hypergraph category.

### 8.3.6 From terms to graphs

Our goal is to map from terms in $\mathbf{H}_\Sigma$ into cospans in $\mathsf{Csp}_D(\mathbf{Hyp}_\Sigma)$. As we know that $\mathbf{H}_\Sigma$ can be viewed as the coproduct $\mathbf{S}_\Sigma + \mathbf{Frob}$, it suffices to define this map in terms of a map from $\mathbf{S}_\Sigma$ and a map from $\mathbf{Frob}$. The results of the previous section gives us the latter, so all that remains is the former.

**Definition 8.59** ([BGK+22a, Sec. 4.1])**.** Let $[\![-]\!]_\Sigma \colon \mathsf{S}_\Sigma \to \mathsf{Csp}_D(\mathbf{Hyp}_\Sigma)$ be a PROP morphism with the action on generators defined as

$$[\![\; m \!-\! \boxed{\phi} \!-\! n \;]\!]_\Sigma := m \longrightarrow$$

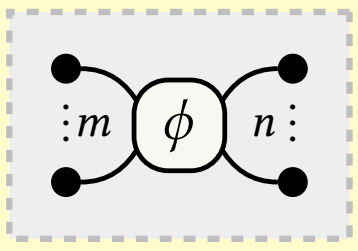

$$\longleftarrow n$$

To map from terms in a hypergraph category to cospans of hypergraphs, we simply put the two maps together.

**Definition 8.60.** Let $\langle\!\langle - \rangle\!\rangle_\Sigma \colon \mathsf{S}_\Sigma + \mathbf{Frob} \to \mathsf{Csp}_D(\mathbf{Hyp}_\Sigma)$ be the PROP morphism defined as the copairing of $[\![-]\!]_\Sigma$ and $[-]_\Sigma$.

Already we have all we need to state one of the key results of [BGK+22a]: the correspondence between terms with a Frobenius structure and cospans of hypergraphs. We will state one corollary concerning cospans of discrete hypergraphs before proceeding to the main result.

**Corollary 8.61.** Given a discrete hypergraph $k \in \mathbf{Hyp}_\Sigma$, any cospan $m \to k \leftarrow n$ in $\mathsf{Csp}_D(\mathbf{Hyp}_\Sigma)$ is in the image of $[-]_\Sigma$.

*Proof.* By Proposition 8.44. □

Cospans of this form will play a part in the main theorem, in which we show the isomorphism between Frobenius terms and cospans of hypergraphs by decomposing a given cospan into a particular form.

**Theorem 8.62** ([BGK+22a], Theorem 4.1)**.** There is an isomorphism of PROPs $\mathsf{S}_\Sigma + \mathbf{Frob} \cong \mathsf{Csp}_D(\mathbf{Hyp}_\Sigma)$.

*Proof.* Since $\mathsf{S}_\Sigma + \mathbf{Frob}$ is a coproduct in $\mathbf{PROP}$, this can be shown by proving that $\mathsf{Csp}_D(\mathbf{Hyp}_\Sigma)$ satisfies the universal property of the coproduct: given a coloured PROP $\mathbb{A}$ and PROP morphisms $\mathsf{S}_\Sigma \to \mathbb{A}$ and $\mathbf{Frob} \to \mathbb{A}$, there exists a unique morphism $u \colon \mathsf{Csp}_D(\mathbf{Hyp}_\Sigma) \to \mathbb{A}$ as below:

$$\begin{array}{ccccc} \mathsf{S}_{C,\Sigma} & \xrightarrow{[\![-]\!]_\Sigma} & \mathsf{Csp}_D(\mathbf{Hyp}_\Sigma) & \xleftarrow{[-]_\Sigma} & \mathbf{Frob} \\ & {\scriptstyle f}\searrow & \downarrow{\scriptstyle u} & \swarrow{\scriptstyle g} & \\ & & \mathbb{A} & & \end{array}$$

All the PROP morphisms involved are identity-on-objects, so all that is required to show the existence of $u$ is to show that any morphism in $\mathsf{Csp}_{D_C}(\mathbf{Hyp}_{C,\Sigma})$ can be expressed as a composition of components either in the image of $[\![-]\!]_\Sigma$ or $[-]_\Sigma$.

Consider a cospan $m \xrightarrow{f} G \xleftarrow{g} n$ in $\mathsf{Csp}_{D_C}(\mathbf{Hyp}_{C,\Sigma})$; let $N$ be the set of vertices, let $E$ be the set of hyperedges, and let $\chi\colon E \to \Sigma$ be the induced labelling function. Pick an order $e_0, e_1, e_{j-1}$ on the edges; then define $\tilde{m} \xrightarrow{s} \tilde{E} \xleftarrow{t} \tilde{n}$ as the cospan $\bigotimes_{0 \leq i < j} [\![\chi(e_i)]\!]_\Sigma$. This cospan 'stacks up' the edges in $G$ without connecting them together; the legs of the cospan are the sources and targets of these edges concatenated in the order specified. It is easy to define functions $f'\colon m \to N$, $g'\colon n \to N$, $h\colon \tilde{m} \to N$ and $k\colon \tilde{n} \to N$ that send vertices to the corresponding vertex in the set of all vertices in the graph.

With this data, the original cospan $m \xrightarrow{f} G \xleftarrow{g} n$ can be viewed as the following composition of cospans:

$$(m \xrightarrow{f'} N \xleftarrow{\mathrm{id},h} N \otimes \tilde{n}) \fatsemi (N \otimes \tilde{n} \xrightarrow{\mathrm{id}\otimes s} N \otimes \tilde{E} \xleftarrow{\mathrm{id}\otimes t} N \otimes \tilde{m}) \fatsemi (N \otimes \tilde{m} \xrightarrow{\mathrm{id},k} N \xleftarrow{g'} n)$$

This is well-defined because $\otimes$ is the coproduct in $\mathbf{Hyp}_{C,\Sigma}$. By computing the composition by pushout, it can be shown that the composite above is isomorphic to the original cospan $m \xrightarrow{f} G \xleftarrow{g} n$.

Now it must be verified that each cospan in the composite is in the image of either $[\![-]\!]_\Sigma$ or $[-]_\Sigma$. The outer cospans are discrete so they are in the image of $[-]_\Sigma$ by Corollary 8.61. The centre cospan is constructed from the identity and cospans in the image of $[\![-]\!]_\Sigma$, so the entire cospan is in the image of $[\![-]\!]_\Sigma$.

The morphism $u$ can therefore be defined based on the actions of $[\![-]\!]_\Sigma$ and $[-]_\Sigma$. This morphism is unique: although different orders can be assigned on the edges, all the categories are symmetric monoidal so this is not an issue. □

At first glance, the composite cospan described above might look confusing. As mentioned, the central cospan $\tilde{m} \to \tilde{E} \leftarrow \tilde{n}$ serves to 'stack up' the edges in some order, all detached from each other. To make the entire cospan isomorphic to the original, the connections of the sources and targets must be the same: the job of the two outer cospans is to 'join them up' appropriately by connecting targets on the right to sources on the left by going 'over the top' of the edges via the identity cospan $N \to N \leftarrow N$.

**Example 8.63.** Consider the following term and its cospan interpretation:

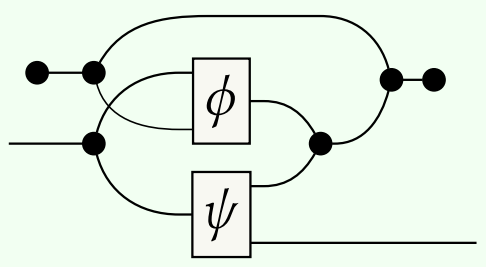

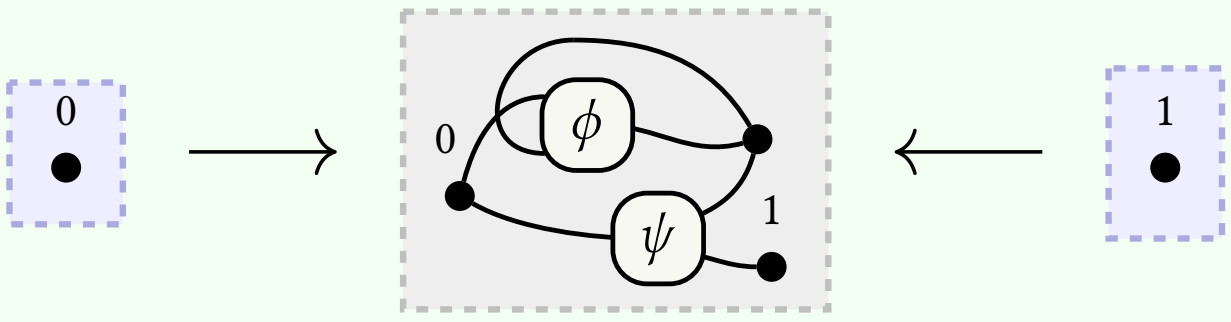

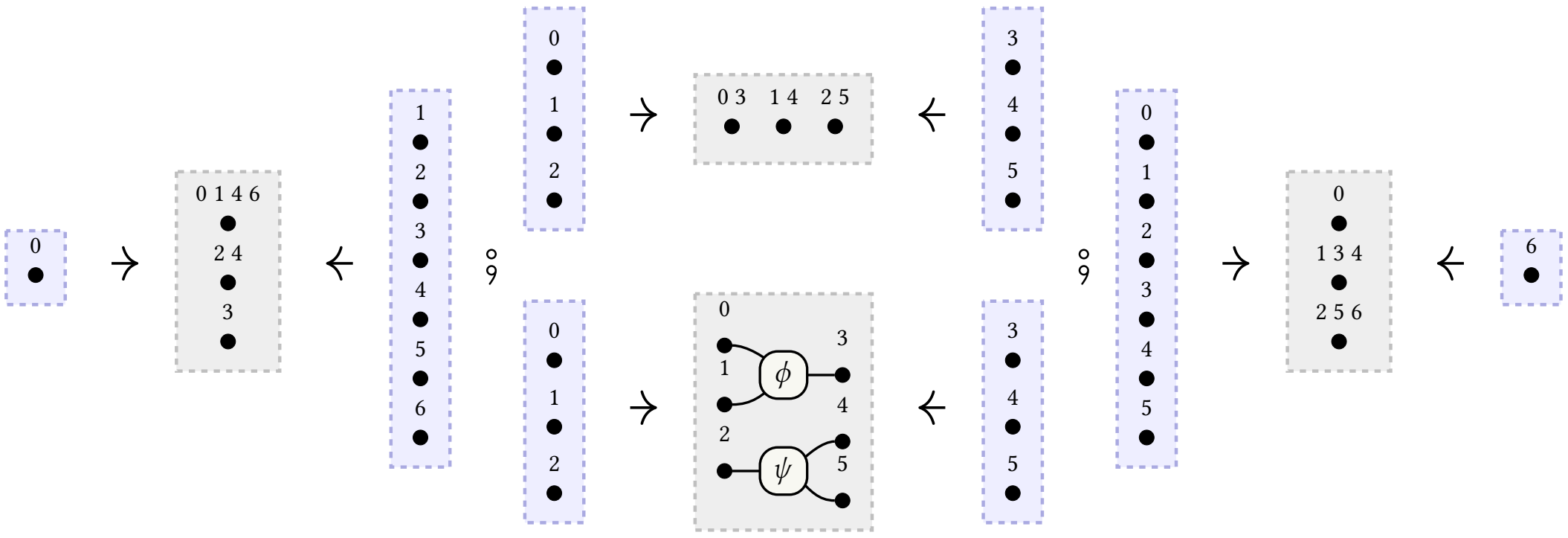

Figure 8.7: The cospan of Example 8.63 in the form of Theorem 8.62

This cospan can be assembled into the form detailed in the above proof as shown in Figure 8.7. By following the vertex maps, one can verify that this is indeed isomorphic to the original cospan. The outermost components correspond to terms in **Frob** and the innermost to a term in $\mathsf{S}_\Sigma$.

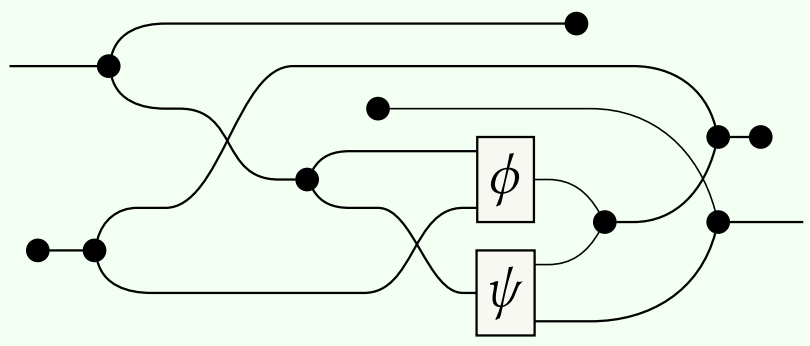

This term is equal to the original term by the Frobenius equations.

This result means that any two terms in $\mathsf{S}_\Sigma + \mathbf{Frob}$ which are equal by the Frobenius equations can be mapped to isomorphic cospans of hypergraphs.

**Example 8.64.** Recall the following terms in **Frob** from Example 8.32, which we showed were equal by the Frobenius equations.

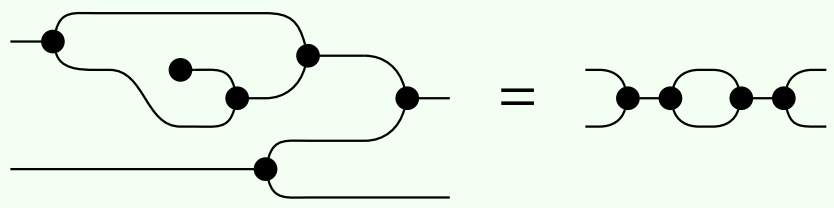

By the isomorphism of Theorem 8.62, these two terms should map to the same cospan of hypergraphs and, indeed, they both map to the following:

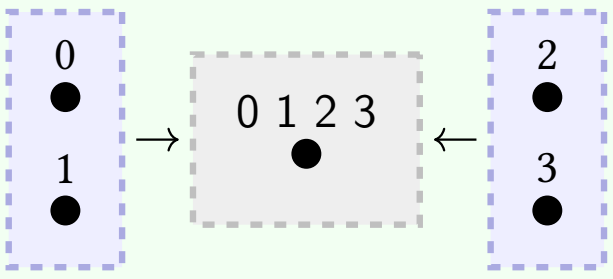

All of the Frobenius structure collapses into one vertex, much like when we considered the correspondence between Frobenius terms and finite sets.

### 8.3.7 The coloured correspondence

The results for the monochromatic case also follow for the coloured case, so we can restate the correspondence for coloured terms and hypergraphs.

**Definition 8.65.** Let $[\![-]\!]_{C,\Sigma} \colon \mathbf{S}_{C,\Sigma} \to \mathsf{Csp}_{D_C}(\mathbf{Hyp}_{C,\Sigma})$ be defined as $\langle\!\langle - \rangle\!\rangle_{\Sigma}$ but assigning appropriate colours to the danging vertices.

**Definition 8.66.** Let $\langle\!\langle - \rangle\!\rangle_{C,\Sigma} \colon \mathbf{S}_{C,\Sigma} + \mathbf{Frob}_C \to \mathsf{Csp}_{D_C}(\mathbf{Hyp}_{C,\Sigma})$ be the copairing of $[\![-]\!]_{C,\Sigma}$ and $[-]_{C,\Sigma}$.

**Theorem 8.67** ([BGK+22a], Prop. 4.4)**.** There is an isomorphism of $C$-coloured PROPs $\mathbf{H}_{C,\Sigma} \cong \mathsf{Csp}_{D_C}(\mathbf{Hyp}_{C,\Sigma})$.

*Proof.* In the same manner as Theorem 8.62, but the components of the composite cospan now have appropriately coloured vertices. □

## 8.4 Symmetric monoidal terms

We have now seen that that cospans of hypergraphs are an excellent fit for reasoning about terms in a freely generated hypergraph category. However, there are times we might not have so much structure in our terms; indeed for our case of digital circuits we only operate in a setting with a trace. This means that not every cospan of hypergraphs will correspond to a valid term. Fortunately, Bonchi et al also characterised the cospans of hypergraphs that correspond to *symmetric monoidal* terms without any additional structure. We will use some of this machinery when it comes to tackling the traced case.

### 8.4.1 Monogamous acyclic cospans

There are two features that distinguish vanilla symmetric monoidal terms from Frobenius terms; wires cannot arbitrarily fork or join, and cycles may not be created. The former is tackled by a condition on the connectivity of vertices.

**Definition 8.68** (Degree [BGK+22b, Def. 12])**.** For a hypergraph $F \in \mathbf{Hyp}$, the *degree* of a vertex $v \in F_\star$ is a tuple $(i, o)$ where $i$ is the number of hyperedges with with $v$ as a target, and $o$ is the number of hyperedges with $v$ as a source.

**Definition 8.69** (Monogamy [BGK+22b, Def. 13])**.** For a cospan $m \xrightarrow{f} F \xleftarrow{g} n$ in $\mathsf{Csp}_D(\mathbf{Hyp}_\Sigma)$, let $\mathsf{in}(F)$ and $\mathsf{out}(F)$ be the image of $f$ and $g$ respectively. We call the cospan $m \xrightarrow{f} F \xleftarrow{g} n$ *monogamous* if $f$ and $g$ are mono and, for all vertices $v$, the degree of $v$ is

$$\begin{array}{llll} (0,0) & \text{if } v \in \mathsf{in}(F) \wedge v \in \mathsf{out}(F) & (0,1) & \text{if } v \in \mathsf{in}(F) \\ (1,0) & \text{if } v \in \mathsf{out}(F) & (1,1) & \text{otherwise} \end{array}$$

**Example 8.70.** The following cospans of hypergraphs are monogamous:

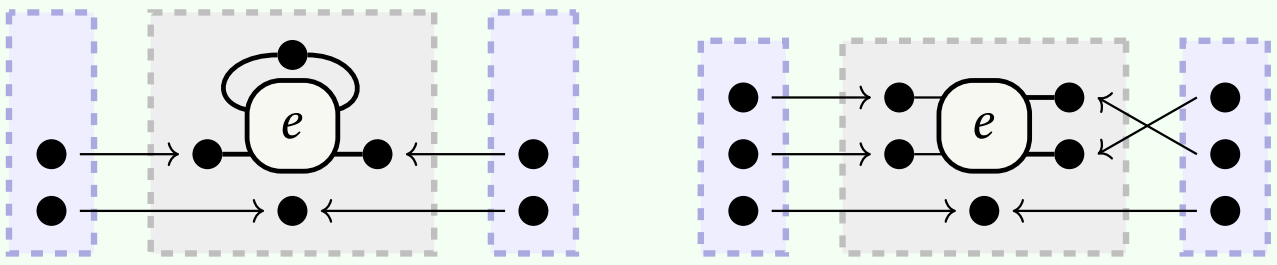

The following cospans of hypergraphs are not monogamous:

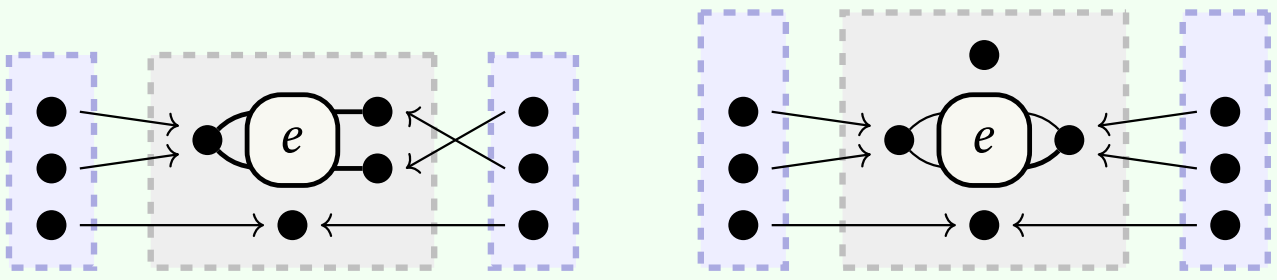

Since our goal is to assemble monogamous cospans into a category, it is necessary to check that the property is preserved by the various categorical operations.

**Lemma 8.71** ([BGK+22b], Lem. 15)**.** Identities and symmetries are monogamous.

*Proof.* The cospans involved are discrete and all vertices are in both interfaces, so the cospans are monogamous. □

**Lemma 8.72** ([BGK+22b], Lem. 16)**.** Monogamicity is preserved by composition.

*Proof.* Assume we compose two monogamous acyclic cospans $m \xrightarrow{f} F \xleftarrow{g} n$ and $n \xrightarrow{h} G \xleftarrow{k} p$. The interfaces remain mono as pushouts along monos are monos in $\mathbf{Hyp}_{C,\Sigma}$. The only altered vertices are those in the image of $g$ and $h$, which are merged pointwise; vertices in the image of $g$ have out-degree $0$ and those in the image of $h$ have in-degree $0$ so the merged vertices have at most degree $(1,1)$. □

**Lemma 8.73** ([BGK+22b], Lem. 17)**.** Monogamicity is preserved by tensor.

*Proof.* The degrees of vertices are unaffected as tensor is by coproduct and only vertices in the original interfaces will be in the new interfaces. □

As seen in Example 8.70, monogamy makes no guarantees about cycles. Since symmetric monoidal terms cannot have cycles, a notion of *acyclicity* must also be enforced.

**Definition 8.74** (Predecessor [BGK+22b, Def. 18])**.** A hyperedge $e$ is a *predecessor* of hyperedge $e'$ if there exists a vertex $v$ in the sources of $e$ and the targets of $e'$.

**Definition 8.75** (Path [BGK+22b, Def. 19])**.** A *path* between two hyperedges $e$ and $e'$ is a sequence of hyperedges $e_0, \dots, e_{n-1}$ such that $e = e_0$, $e' = e_{n-1}$, and for each $i < n - 1$, $e_i$ is a predecessor of $e_{i+1}$. A subgraph $H$ is the *start* or *end* of a path if it contains a vertex in the sources of $e$ or the targets of $e'$ respectively.

Since vertices are single-element subgraphs, one can also talk about paths from vertices.

**Definition 8.76** (Acyclicity [BGK+22b, Def. 20])**.** A hypergraph $F$ is acyclic if there is no path from a vertex to itself. A cospan $m \to F \leftarrow n$ is acyclic if $F$ is.

**Example 8.77.** The following cospans of hypergraphs are acyclic:

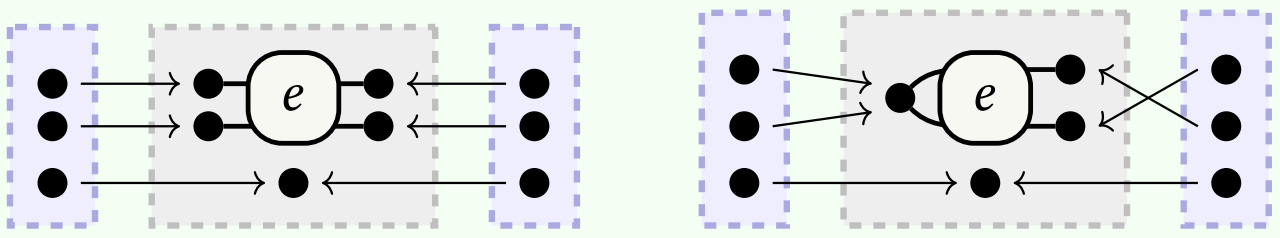

The following cospans of hypergraphs are not acyclic:

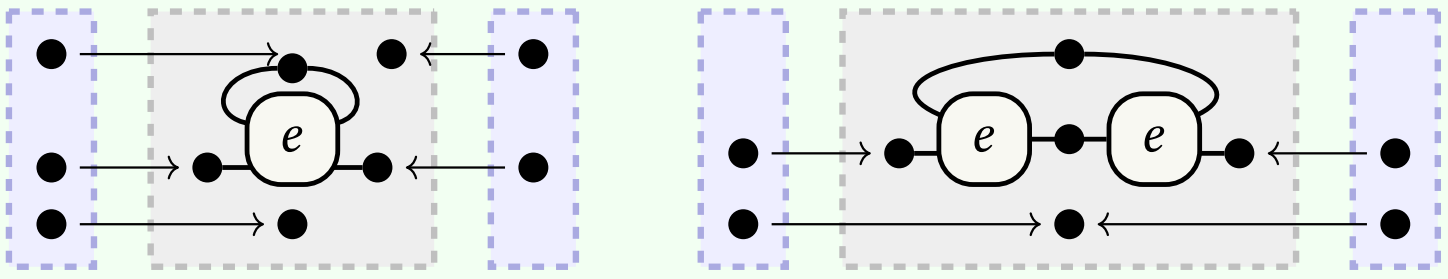

Once again, for acyclicity to be a suitable condition on a category of cospans, it needs to be preserved by categorical operations.

**Lemma 8.78** ([BGK+22b, Prop. 21])**.** Identities and symmetries are acyclic.

*Proof.* By Lemma 8.71 as discrete hypergraphs cannot contain cycles. □

**Lemma 8.79** ([BGK+22b, Prop. 21])**.** Acyclicity is preserved by tensor.

*Proof.* The original graphs are not altered. □

When turning to composition, we run into a problem; composition of arbitrary cospans may not preserve acyclicity. It is only when acyclicity is combined with monogamy that composition can be safely performed.

**Lemma 8.80** ([BGK+22b, Prop. 21])**.** Monogamous acyclicity is preserved by composition.

*Proof.* Assume we compose two monogamous acyclic cospans $m \xrightarrow{f} F \xleftarrow{g} n$ and $n \xrightarrow{h} G \xleftarrow{k} p$. A cycle cannot be created by composition because there cannot be a path in $F$ that starts in the image of $g$ or a path in $G$ that ends in the image of $h$, because these vertices have out-degree and in-degree $0$ respectively. □

This shows that monogamous acyclic cospans of hypergraphs form a category.

**Definition 8.81** ([BGK+22b])**.** Let $\mathsf{MACsp}_D(\mathbf{Hyp}_\Sigma)$ be defined as the sub-PROP of $\mathsf{Csp}_D(\mathbf{Hyp}_\Sigma)$ containing the monogamous acyclic cospans of hypergraphs.

**Example 8.82.** The following cospans of hypergraphs are monogamous acyclic:

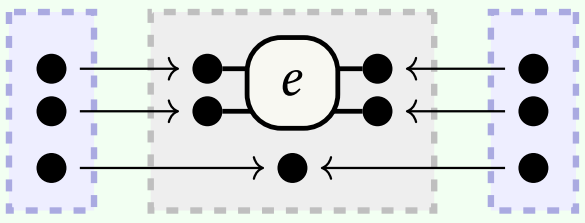

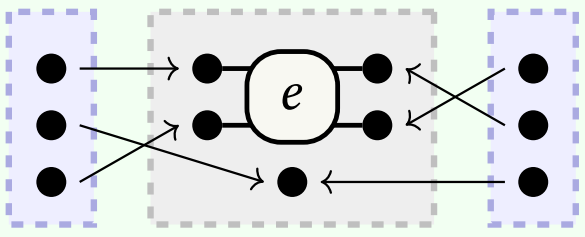

The following cospans of hypergraphs are not monogamous acyclic:

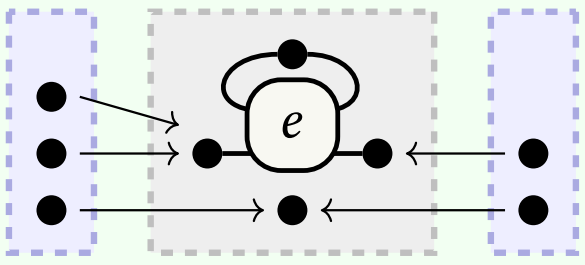

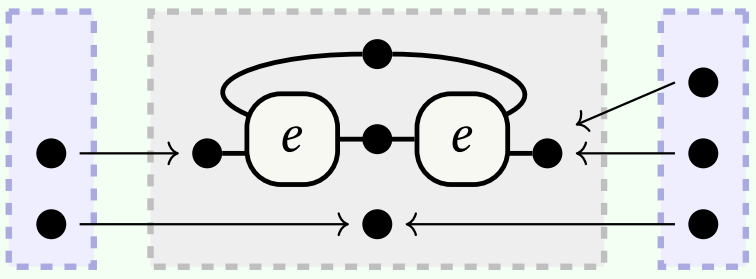

Just like how cospans of hypergraphs correspond to string diagrams of Frobenius terms, monogamous acyclic terms correspond to string diagrams of symmetric monoidal terms. Bonchi et al showed this by proving that $\mathsf{S}_\Sigma$ is isomorphic to $\mathsf{MACsp}_D(\mathbf{Hyp}_\Sigma)$; to do this they needed a few more ingredients. The first is a lemma showing that a special class of subgraphs can always be 'extracted' from a parent graph.

**Definition 8.83** (Convex subgraph [BGK+22b, Def. 23]). A subgraph $G \subseteq F$ is convex if for any vertices $v, v'$ in $G$ and any path $p$ from $v$ to $v'$, every edge $e$ in $p$ is also in $G$.

**Lemma 8.84** (Decomposition [BGK+22b, Lem. 24]). For a monogamous acyclic cospan $m \to F \leftarrow n$ and and convex subgraph $L$ of $G$, there exist $k \in \mathbb{N}$ and a unique cospan $i \to L \leftarrow j$ such that $G$ can be factored as the following composite of monogamous acyclic cospans:

$$(m \to C_1 \leftarrow k+i) \ ; \ (k+i \to k+L \leftarrow k+i) \ ; \ (k+j \to C_2 \leftarrow n)$$

Essentially, we can always 'pull out' a convex subgraph of a monogamous acyclic cospan in such a way that the remaining cospans are still monogamous acyclic. This is an important part of characterising the image of $[\![-]\!]_\Sigma$.

**Theorem 8.85** ([BGK+22b], Thm. 25). A cospan $m \to F \leftarrow n$ is in the image of $[\![-]\!]_\Sigma$ if and only if $m \to F \leftarrow n$ is monogamous acyclic.

*Proof.* The ($\Rightarrow$) direction is by induction on the structure of terms in $\mathsf{S}_\Sigma$: the interpretation of generators is monogamous acyclic and the inductive cases are by Lemmas 8.71, 8.73 and 8.78 to 8.80.

The ($\Leftarrow$) direction is by induction on the number of edges in $F$. If there are none, then $m \to F$ and $n \to F$ are bijections by monogamy so the term is in the image of identities or symmetries in $\mathsf{S}_\Sigma$. For the inductive step, pick a single edge $e$. This is a convex subgraph of $F$, so $m \to F \leftarrow n$ can be factored as in Lemma 8.84. The edge $e$ has a label $\chi(e) \in \Sigma$, so the subgraph $i \to e \leftarrow j$ is the result of $[\![\chi(e)]\!]_\Sigma$. Since the remaining cospans are monogamous acyclic by Lemma 8.84, they are in the image of $[\![-]\!]_\Sigma$ by inductive hypothesis, so the original cospan $m \to F \leftarrow n$ is also in the image of $[\![-]\!]_\Sigma$. □

This shows that $[\![-]\!]_\Sigma$ is full; to conclude the isomorphism we need to show that it is also faithful. We know that the copairing $[\![-]\!]_\Sigma + [-]_\Sigma$ is faithful by Theorem 8.62, so we just need to show the same is true for its components, using a result about pushouts in **PROP**.

**Definition 8.86** ([MS09], Defs. 3.1, 3.2). A functor $F \colon \mathcal{C} \to \mathcal{D}$ satisfies the *3-for-2 property* if, for each triple of morphisms $f, g, h \in \mathcal{D}$ such that $h = g \circ f$, if any two of $f$, $g$ and $h$ are in the image of $F$, then the third is also in the image of $F$.

**Theorem 8.87** ([MS09], Thm. 3.3)**.** Let $F_A : \mathcal{C} \to \mathcal{A}$ and $F_B : \mathcal{C} \to \mathcal{B}$ be faithful functors such that the following diagram is a pushout.

$$\begin{array}{ccc} \mathcal{C} & \xrightarrow{F_\mathcal{D}} & \mathcal{D} \\ {\scriptstyle F_\mathcal{E}}\downarrow & & \downarrow{\scriptstyle G_\mathcal{D}} \\ \mathcal{E} & \xrightarrow[G_\mathcal{E}]{} & \mathcal{B} \end{array}$$

Then, if $F_A$ and $F_B$ both satisfy the 3-for-2 property, then the functors $G_A$ and $G_B$ are also faithful.

To apply this result, we need to show that $\mathbf{Hyp}_\Sigma$ is a pushout.

**Definition 8.88.** Let $\mathbb{P}$ be the sub-PROP of $\mathbb{F}$ containing the bijective functions.

A morphism in $\mathbb{P}$ is a permutation of wires. As all the functions are bijections, there can only be morphisms $m \to m$.

**Lemma 8.89.** $\mathbb{P}$ is the initial object in $\mathbf{PROP}$.

*Proof.* All the morphisms in $\mathbb{P}$ are identities and symmetries; the coloured PROP morphism to any other PROP maps these to the corresponding constructs. □

Subsequently, $\mathrm{S}_\Sigma + \mathbf{Frob}$ can be expressed as a pushout and the '3-for-2' condition applied to show the faithfulness of $[\![-]\!]_\Sigma$, using another well-known categorical lemma.

**Lemma 8.90** ([Bor94], Prop. 2.8.2)**.** If a category $\mathcal{C}$ has pushouts and an initial object, then $\mathcal{C}$ also has coproducts.

*Proof.* Given objects $A, B \in \mathcal{C}$, the coproduct $A + B$ is constructed as follows:

$$\begin{array}{ccc} 0 & \longrightarrow & A \\ \downarrow & & \downarrow \\ B & \longrightarrow & A + B \end{array}$$

This is a coproduct due to the universal property of pushouts. □

**Proposition 8.91.** $[\![-]\!]_\Sigma : \mathrm{S}_\Sigma \to \mathrm{Csp}_D(\mathbf{Hyp}_\Sigma)$ is faithful.

*Proof.* From Theorem 8.62, we know that $\mathsf{Csp}_D(\mathbf{Hyp}_\Sigma) \cong \mathsf{S}_\Sigma + \mathbf{Frob}$. Both $\mathsf{S}_\Sigma$ and $\mathbf{Frob}$ are objects of $\mathbf{PROP}$, which has $\mathbb{P}$ as its initial object by Lemma 8.89. As coproducts are pushouts from the initial object (Lemma 8.90), we can construct the following diagram in $\mathbf{PROP}$:

$$\begin{array}{ccc} \mathbb{P} & \xrightarrow{!_2} & \mathbf{Frob} \\ {\scriptstyle !_1}\downarrow & & \downarrow{\scriptstyle [-]_\Sigma} \\ \mathsf{S}_\Sigma & \xrightarrow[\llbracket - \rrbracket_\Sigma]{} & \mathsf{S}_\Sigma + \mathbf{Frob} \end{array}$$

where $!_1$ and $!_2$ are the unique morphisms from $\mathbb{P}$ induced by initiality: these are both faithful. $!_1$ and $!_2$ clearly satisfy the 3-for-2 condition as every morphism in $\mathbb{P}$ is an isomorphism, so $\llbracket - \rrbracket_\Sigma$ must also be faithful by Theorem 8.87. □

Since $\langle\!\langle - \rangle\!\rangle_\Sigma$ is full and faithful, we have reached our final destination.

**Corollary 8.92** ([BGK+22b], Cor. 26)**.** There is an isomorphism of PROPs $\mathsf{S}_\Sigma \cong \mathsf{MACsp}_D(\mathbf{Hyp}_\Sigma)$.

### 8.4.2 Coloured symmetric monoidal terms

To generalise the above results to the countably coloured case, the only modification is to apply the 3-for-2 condition in the category of $C$-coloured PROPs.

**Lemma 8.93.** Let $\hat{\mathbb{P}}$ be the sub-PROP of $\hat{\mathbb{F}}$ containing only the bijective functions.

**Lemma 8.94.** For a countable set of colours $C$, $\hat{\mathbb{P}}$ is the initial object in $\mathbf{CPROP}_C$.

This means that we obtain correspondence results in the coloured case.

**Proposition 8.95.** $\langle\!\langle - \rangle\!\rangle_{C,\Sigma} \colon \mathsf{S}_{C,\Sigma} \to \mathsf{Csp}_{D_C}(\mathbf{Hyp}_{C,\Sigma})$ is faithful.

**Corollary 8.96.** For a countable set of colours $C$, there is an isomorphism of $C$-coloured PROPs $\mathsf{S}_{C,\Sigma} \cong \mathsf{MACsp}_{D_C}(\mathbf{Hyp}_{C,\Sigma})$.

## 8.5 Traced terms

We have now seen the classes of cospans of hypergraphs that correspond to terms in a hypergraph category and terms in a symmetric monoidal category. Terms in a symmetric traced monoidal category sit somewhere in the middle of these two: cycles are permitted but wires cannot fork or join arbitrarily. Our contribution is to characterise the cospans of hypergraphs that correspond to traced terms by weakening the conditions of monogamy and acyclicity described in the previous section.

**Definition 8.97.** For a set of generators $\Sigma$, let $\mathbf{T}_\Sigma$ be the traced PROP freely generated over $\Sigma$.

First we will establish the morphism $\mathbf{T}_\Sigma \to \mathsf{Csp}_D(\mathbf{Hyp}_\Sigma)$ we will use to map traced terms to cospans of hypergraphs. We could do this manually fairly easily, but this would mean we would have to redo all the proofs in the previous section from scratch. Instead we will reuse the previous results, by exploiting the correspondence between traced and compact closed categories.

**Lemma 8.98** ([RSW05], Prop. 2.8)**.** Every hypergraph category is self-dual compact closed.

*Proof.* In a hypergraph category, the cup on a given object is defined as 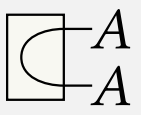  $:=$ 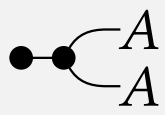  and the cap as 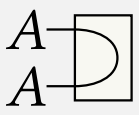  $:=$ 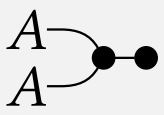 . The snake equations follow by applying the Frobenius equation and unitality:

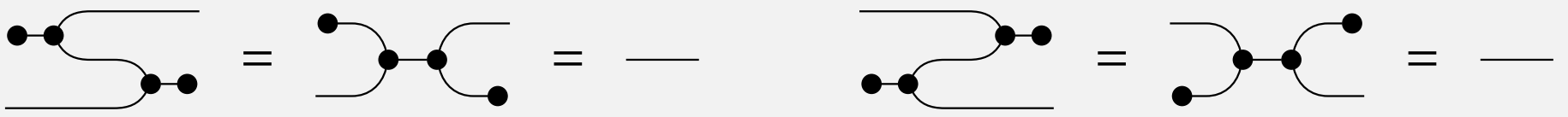

□

**Lemma 8.99.** $\mathbf{T}_\Sigma$ is a subcategory of $\mathbf{H}_\Sigma$.

*Proof.* Since $\mathbf{Hyp}_\Sigma$ is compact closed, it has a (canonical) trace. For $\mathbf{T}_\Sigma$ to be a subcategory of $\mathbf{H}_\Sigma$, every morphism of the former must also be a morphism on the latter. Since the two categories are freely generated (with the trace constructed through the Frobenius generators in the latter), all that remains is to check that every morphism in $\mathbf{T}_\Sigma$ is a unique morphism in $\mathbf{H}_\Sigma$, i.e. the equations of $\mathbf{Frob}$ do not merge any together. This is trivial since the equations do not apply to the construction of the canonical trace. □

**Definition 8.100.** Let $\lfloor - \rfloor_\Sigma^{\mathrm{T}} \colon \mathbf{T}_\Sigma \to \mathbf{H}_\Sigma$ be defined as the inclusion functor induced by Lemma 8.99.

**Corollary 8.101.** $\lfloor - \rfloor_\Sigma^{\mathrm{T}}$ is faithful.

To translate a term in $\mathbf{T}_\Sigma$ into a cospan of hypergraphs, one uses the inclusion functor $\lfloor - \rfloor_\Sigma^{\mathbf{T}}$ to elevate to the Frobenius realm, before applying $\langle\!\langle - \rangle\!\rangle_\Sigma$ from the previous section to obtain a cospan of hypergraphs.

**Corollary 8.102.** $\langle\!\langle - \rangle\!\rangle_\Sigma \circ \lfloor - \rfloor_\Sigma^{\mathbf{T}}$ is faithful.

### 8.5.1 Partial monogamy

Since $\langle\!\langle - \rangle\!\rangle_\Sigma \circ \lfloor - \rfloor_\Sigma^{\mathbf{T}}$ is faithful, every traced term in $\mathbf{T}_\Sigma$ has a unique corresponding cospan of hypergraphs. This functor is not clearly not full; there are more terms in $\mathbf{H}_\Sigma$ than there are in $\mathbf{T}_\Sigma$. The next step is to characterise the image of $\langle\!\langle - \rangle\!\rangle_\Sigma \circ \lfloor - \rfloor_\Sigma^{\mathbf{T}}$.

Since monogamous acyclic cospans correspond exactly to symmetric monoidal terms, this property is too restrictive to be used as a setting for modelling traced terms. Clearly, we can drop the acyclicity condition, as the trace can introduce cycles. However, there is also a foible that arises with the monogamicity condition: although wires are also not permitted to arbitrarily fork or join in a traced category, it is possible to have a case where wires do not connect to any generators while also remaining disconnected from the interfaces. This special case is the trace of the identity, which in string diagrams is depicted as a closed loop $\mathrm{Tr}^1$ ( 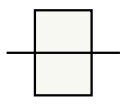 ) = 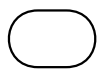 .

**Remark 8.103.** One might think a closed loop can be discarded, i.e. 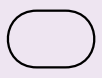 = 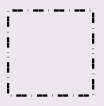 , but this is *not* always the case, such as in $\mathbf{FinVect}_k$ [Has97, Sec. 6.1].

These closed loops can be modelled as vertices disconnected from either interface.

**Definition 8.104** (Partial monogamy)**.** For a cospan $m \xrightarrow{f} F \xleftarrow{g} n$ in $\mathsf{Csp}_D(\mathbf{Hyp}_\Sigma)$, let $\mathsf{in}(F)$ be defined as the image of $f$ and let $\mathsf{out}(F)$ be defined as the image of $g$. A cospan $m \xrightarrow{f} F \xleftarrow{g} n \in \mathsf{Csp}_D(\mathbf{Hyp}_\Sigma)$ is *partial monogamous* if $f$ and $g$ are mono and, for all vertices $v$, the degree of $v$ is

$$\begin{array}{llrl} (0,0) & \text{if } v \in \mathsf{in}(F) \wedge v \in \mathsf{out}(F) & (0,1) & \text{if } v \in \mathsf{in}(F) \\ (1,0) & \text{if } v \in \mathsf{out}(F) & (0,0) \text{ or } (1,1) & \text{otherwise} \end{array}$$

**Example 8.105.** The following cospans of hypergraphs are partial monogamous:

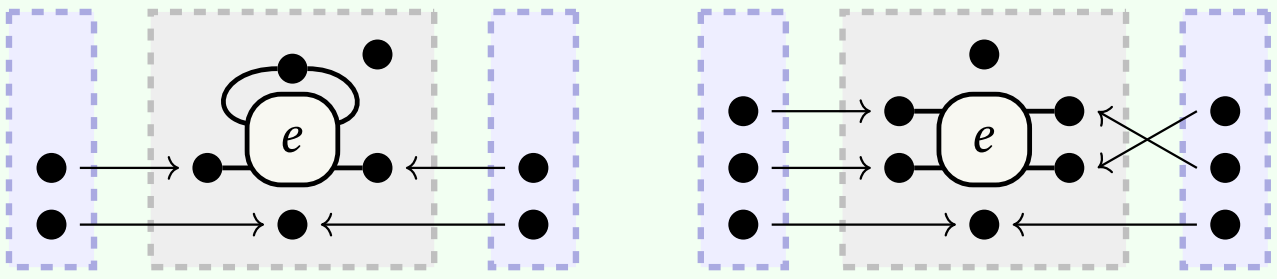

The following cospans of hypergraphs are not partial monogamous:

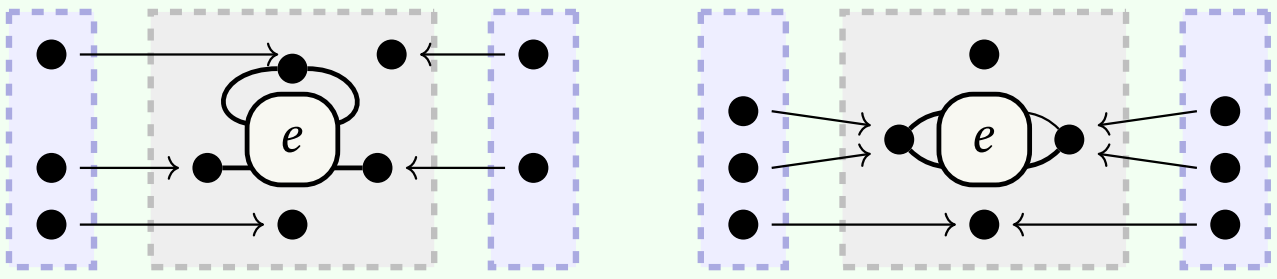

To use partial monogamous cospans as a setting for interpreting traced terms, they must be assembled into a sub-PROP of $\mathrm{Csp}_D(\mathbf{Hyp}_\Sigma)$.

**Lemma 8.106.** Identities and symmetries are partial monogamous.

*Proof.* Identities and symmetries are monogamous by Lemma 8.71 so they must also be partial monogamous. □

**Lemma 8.107.** Partial monogamy is preserved by composition.

*Proof.* By Lemma 8.72, composition preserves monogamicity. The only difference between partial monogamous cospans and monogamous ones is that the former may have cycles and vertices of degree $(0, 0)$ not in the interfaces. However, since neither of these can be interfaces they cannot be altered by composition, so partial monogamy must also be preserved. □

**Lemma 8.108.** Partial monogamy is preserved by tensor.

*Proof.* Tensor preserves monogamicity by Lemma 8.73, and as tensor does not affect the degree of vertices then it preserves partial monogamy as well. □

As partial monogamicity is preserved by both forms of composition, the partial monogamous cospans themselves form a PROP.

**Definition 8.109.** Let $\mathrm{PMCsp}_D(\mathbf{Hyp}_\Sigma)$ be the sub-PROP of $\mathrm{Csp}_D(\mathbf{Hyp}_\Sigma)$ containing only the partial monogamous cospans of hypergraphs.

We must show that $\mathrm{PMCsp}_D(\mathbf{Hyp}_\Sigma)$ is also traced, by making sure the canonical trace does not create cospans that are not partial monogamous.

**Theorem 8.110.** The canonical trace is a trace on $\mathrm{PMCsp}_D(\mathbf{Hyp}_\Sigma)$.

*Proof.* Consider a partial monogamous cospan $x + m \xrightarrow{f+h} F \xleftarrow{g+k} x + n$; we must show that its trace $m \xrightarrow{h} F' \xleftarrow{k} n$ is partial monogamous. For each vertex $a \in x$, $f(a)$ and $g(a)$ are merged together in the traced graph, summing their degrees. If a vertex is in the image of $h$ or $k$, this is also the case in the traced cospan. We consider the various cases:

- if $f(a) = g(a)$, then this vertex must have degree $(0,0)$; the traced vertex will still have degree $(0,0)$ and will no longer be in the interface;
- if $f(a)$ is also in the image of $n \to F$ and $g(i)$ is also in the image of $m \to F$, then both $f(a)$ and $g(a)$ have degree $(0,0)$; the traced vertex will still have degree $(0,0)$ and be in both interfaces of the traced cospan;
- if $f(a)$ is in the image of $n \to F$, then $f(i)$ has $(0,0)$ and $g(a)$ has degree $(1,0)$, so the traced vertex has degree $(1,0)$ and is in the image of $n \to F'$;
- if $g(i)$ is in the image of $m \to F$, then the above applies in reverse; and
- if neither vertex is in the image of $m \to F$ and $n \to F$, then the traced vertex will have degree $(1,1)$ and be in the image of no interface.

In all these cases, partial monogamy is preserved. □

Crucially, while we leave $\mathsf{PMCsp}_D(\mathbf{Hyp}_\Sigma)$ in order to construct the trace using the cup and cap, the resulting cospan *is* in $\mathsf{PMCsp}_D(\mathbf{Hyp}_\Sigma)$.

### 8.5.2 The traced correspondence

Now that we have a traced sub-PROP of cospans of hypergraphs, it is time to show that this particular sub-PROP is the one that corresponds to traced terms.

**Theorem 8.111.** A cospan $m \to F \leftarrow n$ is in the image of $\langle\!\langle - \rangle\!\rangle_\Sigma \circ \lfloor - \rfloor_\Sigma^{\mathbf{T}}$ if and only if it is partial monogamous.

*Proof.* For the ($\Rightarrow$) direction, the generators of $\mathbf{T}_\Sigma$ are mapped to monogamous cospans by $\langle\!\langle - \rangle\!\rangle_\Sigma \circ \lfloor - \rfloor_\Sigma^{\mathbf{T}}$, and partial monogamy is preserved by composition (Lemma 8.107), tensor (Lemma 8.108), and trace (Theorem 8.110).

For the ($\Leftarrow$) direction, we show that any partial monogamous cospan $m \xrightarrow{f} F \xleftarrow{g} n$ is in the image of $\langle\!\langle - \rangle\!\rangle_\Sigma \circ \lfloor - \rfloor_\Sigma^{\mathbf{T}}$ by constructing an isomorphic trace of cospans, in which each component under the trace is in the image of $[\![ - ]\!]_\Sigma$. The components of the new cospan are as follows:

- let $L$ be the discrete hypergraph containing vertices with degree $(0,0)$ that are not in the image of $f$ or $g$;

- let $E$ be the hypergraph containing hyperedges of $F$ disconnected from each other along with their source and target vertices;
- let $V$ be the discrete hypergraph containing all the vertices of $F$; and
- let $S$ and $T$ be the discrete hypergraphs containing the source and target vertices of hyperedges in $F$ respectively, with the ordering determined by some order $e_1, e_2, \cdots, e_n$ on the edges in $F$.

These parts can be composed to form the following composite:

$$L+T+m \xrightarrow{\mathrm{id}+\mathrm{id}+f} L+V \xleftarrow{\mathrm{id}+\mathrm{id}+g} L+S+n \ \fatsemi\ L+S+n \xrightarrow{\mathrm{id}} L+E+n \xleftarrow{\mathrm{id}} L+T+n$$

We take the trace of $L+T$ over this composite to obtain a cospan isomorphic to the original. The components of the composite under the trace are all monogamous acyclic so are in the image of $[\![-]\!]_\Sigma$ by Theorem 8.85; this means there is a term $f \in \mathbf{S}_\Sigma$ such that $[\![f]\!]_\Sigma$ is isomorphic to the original composite. The trace of $f$ is in $\mathbf{T}_\Sigma$, so the trace of the composite is in the image of $\langle\!\langle - \rangle\!\rangle_\Sigma \circ \lfloor - \rfloor^{\mathbf{T}}_\Sigma$. □

As with the Frobenius case (Theorem 8.62), the large composite cospan may seem confusing. We stack up the edges in the cospan $\tilde{m} \to \tilde{E} \leftarrow \tilde{n}$, but now join up the connections by tracing the targets of the edges around, and shuffling them to the correct source. The graph $L$ contains any identity loops.

**Example 8.112.** Consider the following term and its cospan interpretation:

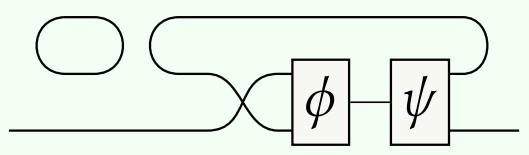

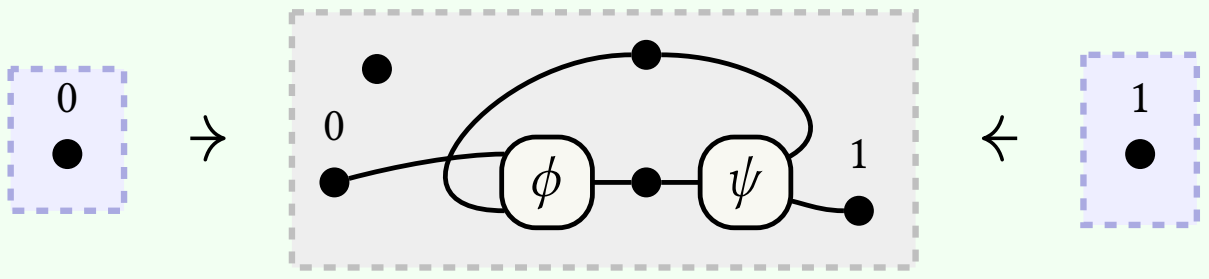

We assemble the latter into the composite cospan of Theorem 8.111 as shown in Figure 8.8. Both of the components under the trace correspond to terms in $\mathbf{S}_\Sigma$, so applying the trace to this produces a term in $\mathbf{T}_\Sigma$ which is equal to the original by string diagrammatic deformations.

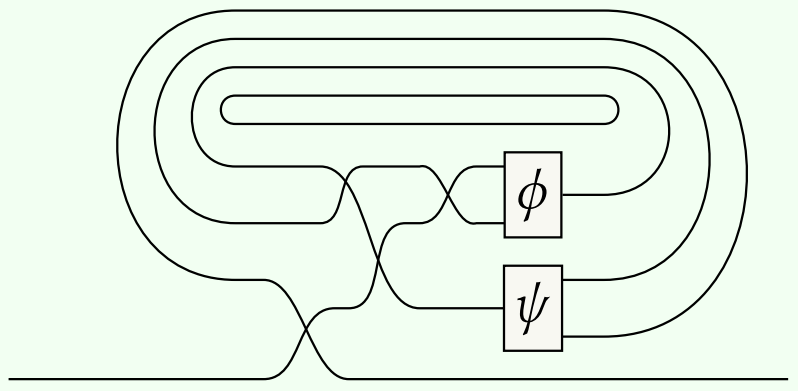

This shows that $\langle\!\langle - \rangle\!\rangle_\Sigma \circ \lfloor - \rfloor^{\mathbf{T}}_\Sigma$ is a *full* mapping from $\mathbf{T}_\Sigma$ to $\mathrm{PMCsp}_{D_C}(\mathbf{Hyp}_{C,\Sigma})$. As it is both full and faithful, we have that there is an isomorphism between the category of terms and the category of cospans of partial monogamous hypergraphs.

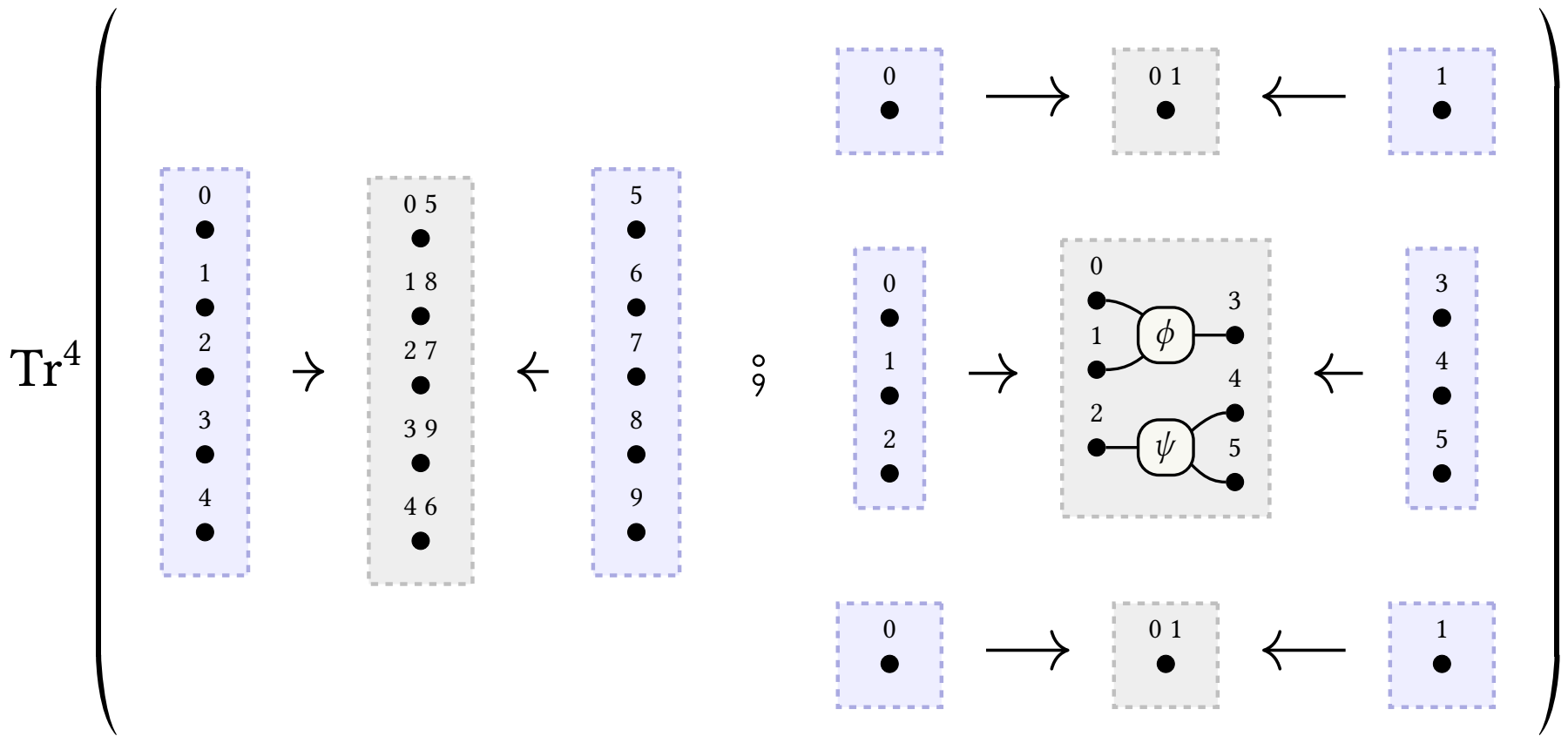

Figure 8.8: The cospan of Example 8.112 in the form of Theorem 8.111

**Corollary 8.113.** $\mathbf{T}_\Sigma \cong \mathrm{PMCsp}_D(\mathbf{Hyp}_\Sigma)$.

This means that $\mathrm{PMCsp}_{D_C}(\mathbf{Hyp}_{C,\Sigma})$ is a suitable setting for interpreting terms in $\mathbf{T}_\Sigma$: every term has a corresponding cospan of hypergraphs, and every cospan has a corresponding term.

**Example 8.114.** The partial monogamous cospans from Example 8.105 are shown below with their corresponding terms in $\mathbf{T}_\Sigma$.

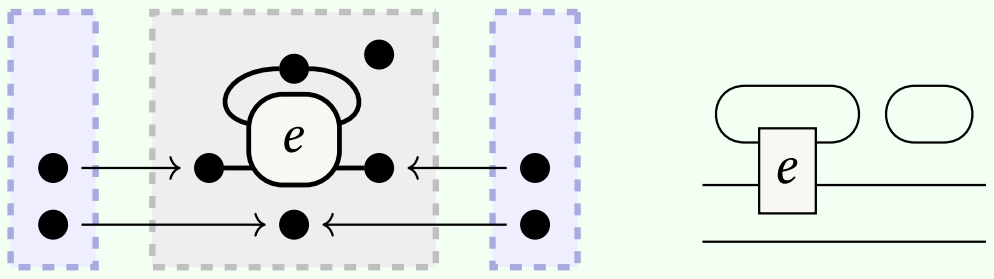

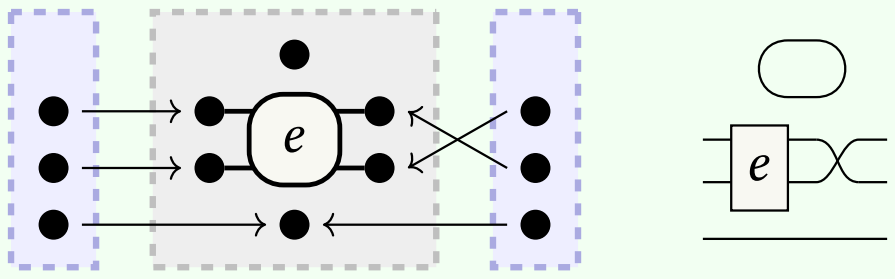

### 8.5.3 Traced coloured terms

The results above all also hold for the coloured case by following the same strategy but with the appropriate coloured PROPs.

**Definition 8.115.** For a countable set of colours $C$ and a set of generators $\Sigma$, let $\mathbf{T}_{C,\Sigma}$ be the traced $C$-coloured PROP freely generated over $\Sigma$.

**Lemma 8.116.** $\mathbf{T}_{C,\Sigma}$ is a subcategory of $\mathbf{H}_{C,\Sigma}$.

**Definition 8.117.** Let $\lfloor - \rfloor^{\mathbf{T}}_{C,\Sigma} \colon \mathbf{T}_{C,\Sigma} \to \mathbf{H}_{C,\Sigma}$ be the inclusion functor induced by Lemma 8.116.

**Corollary 8.118.** $\lfloor - \rfloor^{\mathrm{T}}_{C,\Sigma}$ is faithful.

The partial monogamy condition works the same for coloured cospans, so the partial monogamous cospans of coloured hypergraphs form a traced coloured category.

**Definition 8.119.** Let $\mathsf{PMCsp}_{D_C}(\mathbf{Hyp}_{C,\Sigma})$ be the sub-PROP of $\mathsf{Csp}_{D_C}(\mathbf{Hyp}_{C,\Sigma})$ containing only the partial monogamous cospans of hypergraphs.

**Theorem 8.120.** The canonical trace is a trace on $\mathsf{PMCsp}_{D_C}(\mathbf{Hyp}_{C,\Sigma})$.

The fullness proof then proceeds as before.

**Theorem 8.121.** A cospan is in the image of $\langle\!\langle - \rangle\!\rangle_{C,\Sigma} \circ \lfloor - \rfloor^{\mathrm{T}}_{C,\Sigma}$ if and only if it is partial monogamous.

**Corollary 8.122.** $\mathrm{T}_{C,\Sigma} \cong \mathsf{PMCsp}_{D_C}(\mathbf{Hyp}_{C,\Sigma})$.

## 8.6 Hypergraphs for traced comonoid terms

By characterising the cospans of hypergraphs that correspond to *traced* terms, we already have a setting in which we can model sequential circuit morphisms combinatorially. But we can go further. When modelling *Frobenius* terms, we were modelling them modulo the Frobenius equations; when interpreted as cospans of hypergraphs the comonoid and monoid structures merged together into single vertices so we did not need to consider the equations of associativity, commutativity or unitality.

In the realm of sequential circuits we also have monoid and comonoid structures, but instead of forming a Frobenius structure they only form a *bialgebra*. The equations of a bialgebra are different to those of a Frobenius algebra in how the monoid and comonoid interact. Compare the two Frobenius equations with the bialgebra equation shown below:

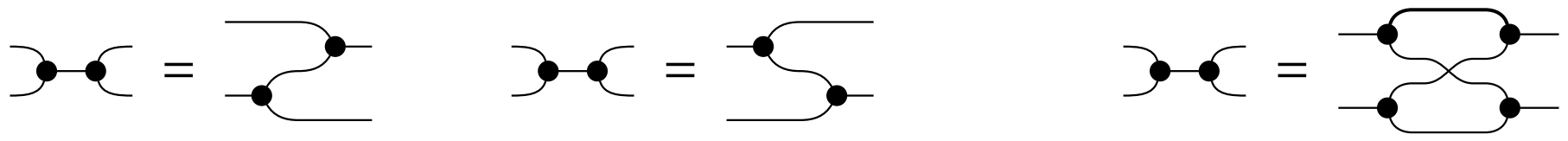

In a Frobenius setting, it is possible to derive the bialgebra equation BCM from the Frobenius equations combined with the equations of monoids and comonoids.

**Lemma 8.123.** In $\mathbf{H}_\Sigma$, 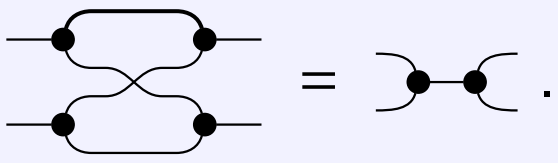 .

*Proof.*

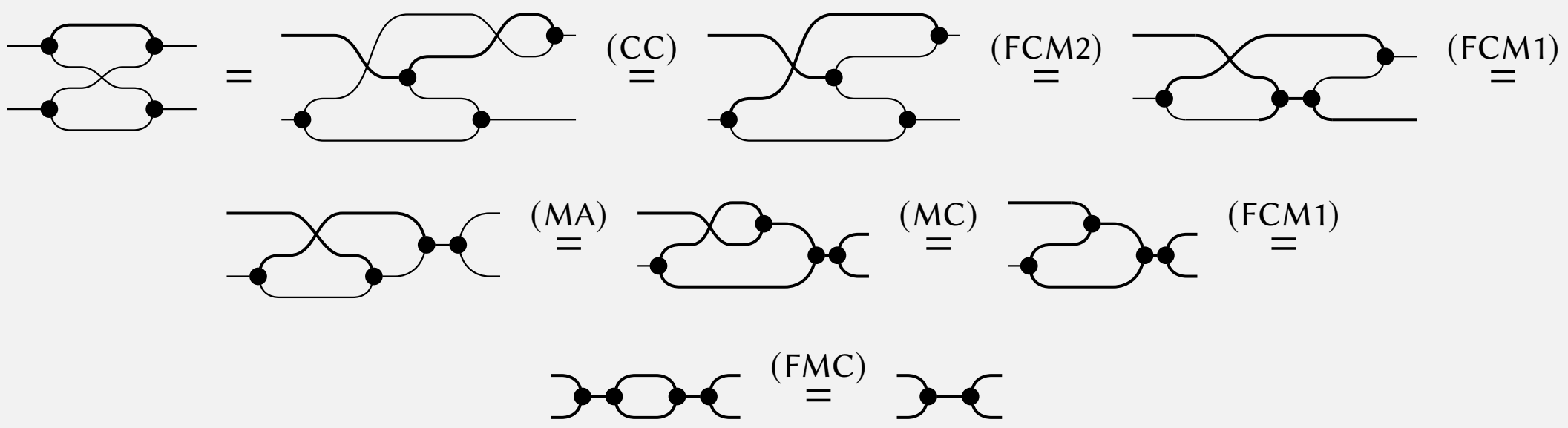

□

This is clear from the hypergraph interpretation, as all four terms involved map to the same (discrete) cospan of hypergraphs.

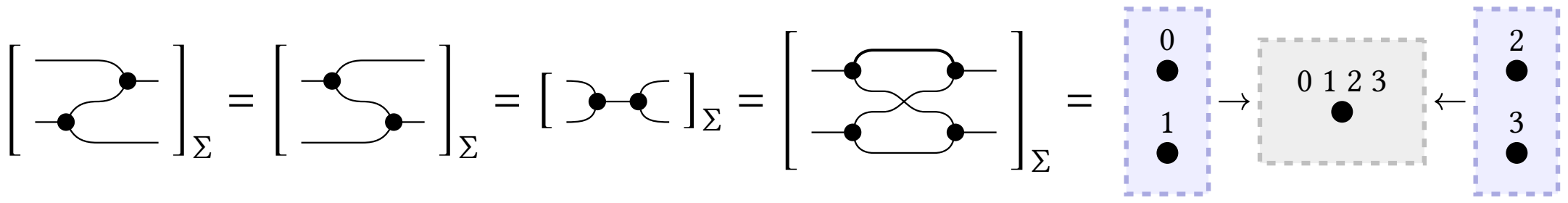

However, the converse does not hold: it is not possible to derive the Frobenius equations from the bialgebra equation without having one of the Frobenius equations to begin with. This poses a problem: we want to use $\mathrm{Csp}_{D_C}(\mathbf{Hyp}_{C,\Sigma})$ as a setting for rewriting digital circuits, but as it by default contains a Frobenius structure, *too many* equations would hold. Since the issue only arises with the interactions between the monoid and the comonoid, we can use cospans of hypergraphs to reason modulo the equations of just one of the two structures.

**Remark 8.124.** Alas, we cannot claim to have pioneered the idea of interpreting terms with just a monoid or comonoid structure as cospans of hypergraphs [FL23; MPZ23]. What we bring to the table is studying how such terms interact with the *trace*: does removing acyclicity lead to any degeneracies?

In the case of sequential circuits, it makes sense to focus on the comonoid structure, as forking wires is far more of a natural concept than joining them. To characterise categories of terms with a comonoid structure, we must first define the monoidal theory of cocommutative comonoids.

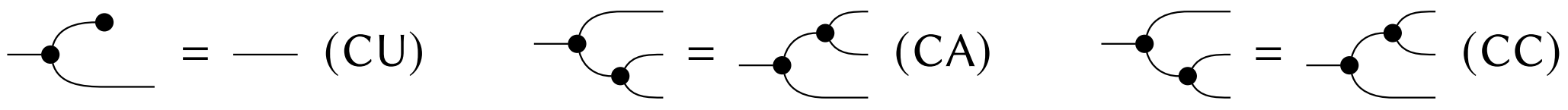

Figure 8.9: Equations $\mathcal{E}_{\mathbf{CComon}}$ of a *commutative comonoid*

**Definition 8.125.** Let $(\Sigma_{\mathbf{CComon}}, \mathcal{E}_{\mathbf{CComon}})$ be the symmetric monoidal theory of *cocommutative comonoids*, with $\Sigma_{\mathbf{CComon}} := \{$ 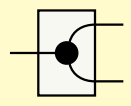 , 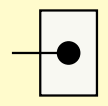 $\}$ and $\mathcal{E}_{\mathbf{CComon}}$ defined as in Figure 8.9. We write $\mathbf{CComon} := \mathbf{S}_{\Sigma_{\mathbf{CComon}}, \mathcal{E}_{\mathbf{CComon}}}$.

From now on, we write 'comonoid' to mean 'cocommutative comonoid'.

When identifying the cospans of hypergraphs that correspond to terms with traced comonoid structure, the notion of monogamy will once again need to be modified. Partial monogamy is now too strong, as this means wires cannot fork. Weakening to no monogamy at all is too much, as we do not want wires to join as well as fork. Effectively, vertices need to be 'monogamous on one side'.

**Definition 8.126** (Partial left-monogamy)**.** For a cospan $m \xrightarrow{f} H \xleftarrow{g} n$, we say it is *partial left-monogamous* if $f$ is mono and, for all vertices $v \in H_\star$, the degree of $v$ is $(0, m)$ if $v \in f_\star$ and $(0, m)$ or $(1, m)$ otherwise, for some $m \in \mathbb{N}$.

Partial left-monogamy is a weakening of partial monogamy that allows vertices to have multiple 'out' connections, which represent the use of the comonoid structure to fork wires; vertices must still only have one 'in' connection.

**Example 8.127.** The following cospans are partial left-monogamous:

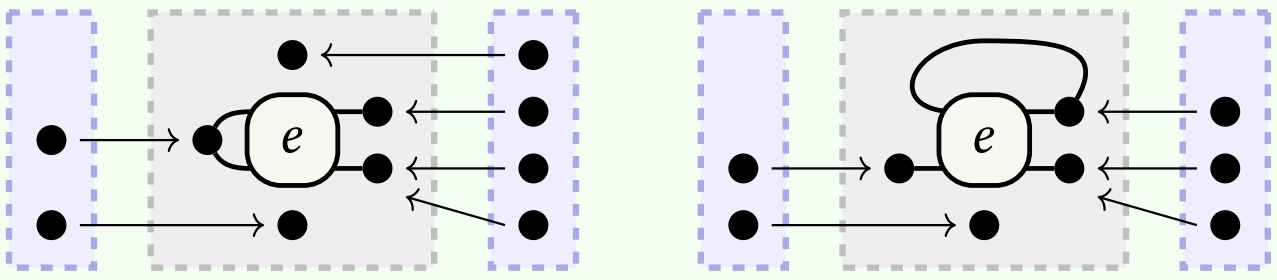

The following cospans are not partial left-monogamous:

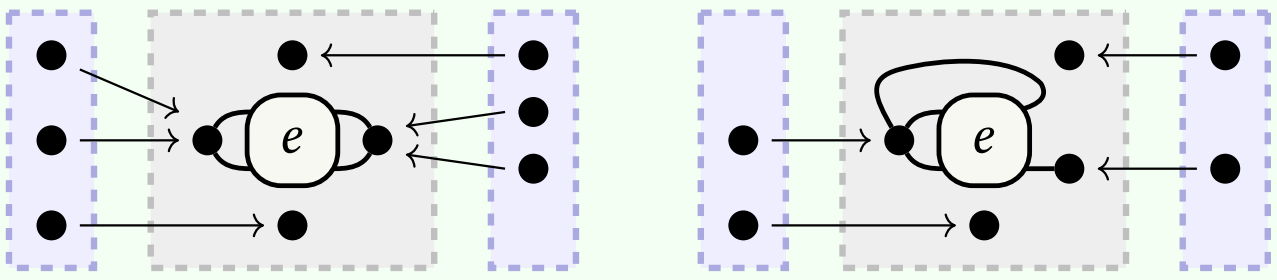

**Remark 8.128.** As with the vertices not in the interfaces with degree $(0, 0)$ in the vanilla traced case, the vertices not in the interface with degree $(0, m)$ allow for the

interpretation of terms such as Tr ( 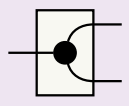 ).

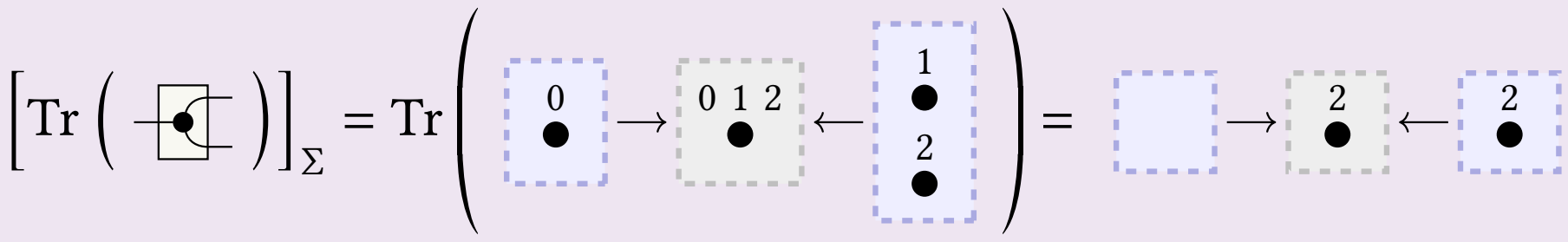

We must ensure that partial left-monogamy is preserved by the categorical operations, so that partial left-monogamous cospans form another PROP.

**Lemma 8.129.** Identities and symmetries are partial left-monogamous.

*Proof.* Again by Lemma 8.71, identities and symmetries are monogamous so they are also partial left-monogamous. □

**Lemma 8.130.** Given two partial left-monogamous cospans $m \longrightarrow F \longleftarrow n$ and $n \longrightarrow G \longleftarrow p$, the composition $(m \longrightarrow F \longleftarrow n) \fatsemi (n \longrightarrow G \longleftarrow p)$ is partial left-monogamous.

*Proof.* Only the vertices in the image of $n \longrightarrow G$ have their in-degree modified; they gain the in-tentacles of the corresponding vertices in the image of $n \longrightarrow F$. Initially the vertices in $n \longrightarrow G$ have in-degree $0$ by partial monogamy; they will gain at most one in-tentacle from vertices in $n \longrightarrow F$ as each of these vertices has in-degree $0$ or $1$ and $n \longrightarrow G$ is mono. So the composite graph is partial left-monogamous. □

**Lemma 8.131.** Given two partial left-monogamous cospans $m \longrightarrow F \longleftarrow n$ and $p \longrightarrow G \longleftarrow q$, the tensor $(m \longrightarrow F \longleftarrow n) \otimes (n \longrightarrow G \longleftarrow p)$ is partial left-monogamous.

*Proof.* The elements of the original graphs are unaffected. □

This means we can assemble the partial left-monogamous cospans of hypergraphs into the desired sub-PROP.

**Definition 8.132.** Let $\mathrm{PLMCsp}_D(\mathbf{Hyp}_\Sigma)$ be the sub-PROP of $\mathrm{Csp}_D(\mathbf{Hyp}_\Sigma)$ containing only the partial left-monogamous cospans of hypergraphs.

As this PROP is not restricted to acyclic cospans like those used for just terms with a (co)monoid structure, it has the additional structure of a trace.

**Proposition 8.133.** The canonical trace is a trace on $\mathsf{PLMCsp}_D(\mathbf{Hyp}_\Sigma)$.

*Proof.* We must show that for any vertices in the image of $x + n \to K$ merged by the canonical trace, at most one of them can have in-degree 1. This follows because anything in the image of $x+m \to K$ must have in-degree 0, and $x+m \to K$ is mono so it cannot merge vertices in the image of $x + n \to K$. □

We now have the setting in which we will model terms with a comonoid structure. To actually define the mapping from $\mathbf{T}_\Sigma + \mathbf{CComon}$ we will reuse some ingredients from the previous sections.

**Definition 8.134.** Let $\lfloor - \rfloor^{\mathbf{C}} \colon \mathbf{CComon} \to \mathbf{Frob}$ be the embedding of $\mathbf{CComon}$ into $\mathbf{Frob}$, and let $\lfloor - \rfloor_\Sigma \colon \mathbf{T}_\Sigma + \mathbf{CComon} \to \mathbf{S}_\Sigma + \mathbf{Frob}$ be the copairing $\lfloor - \rfloor_\Sigma^{\mathbf{T}} + \lfloor - \rfloor^{\mathbf{C}}$.

**Corollary 8.135.** $\lfloor - \rfloor^{\mathbf{C}}$ and $\lfloor - \rfloor_\Sigma^{\mathbf{T}}$ are faithful.

After translating from $\mathbf{T}_\Sigma + \mathbf{CComon}$ to $\mathbf{S}_\Sigma + \mathbf{Frob}$, we can then use the previously defined PROP morphism $\langle\!\langle - \rangle\!\rangle_\Sigma$ to obtain a cospan of hypergraphs; as before. To show that partial left-monogamy is the correct notion to characterise terms in a traced comonoid setting, it is necessary to ensure that the image of these PROP morphisms actually lands in $\mathsf{PLMCsp}_D(\mathbf{Hyp}_\Sigma)$. First we verify that this

**Lemma 8.136.** The image of $[-]_\Sigma \circ \lfloor - \rfloor^{\mathbf{C}}$ is in $\mathsf{PLMCsp}_D(\mathbf{Hyp}_\Sigma)$.

*Proof.* This is straightforward by inspecting the cases. □

To show the correspondence between $\mathbf{T}_\Sigma + \mathbf{CComon}$ and $\mathsf{PLMCsp}_D(\mathbf{Hyp}_\Sigma)$, we use a similar strategy to Theorem 8.111.

**Lemma 8.137.** Given a discrete hypergraph $X \in \mathbf{Hyp}_\Sigma$, any cospan $m \xrightarrow{f} X \leftarrow n$ with $f$ mono is in the image of $[-]_\Sigma \circ \lfloor - \rfloor^{\mathbf{C}}$.

*Proof.* By definition of $[-]_\Sigma \circ \lfloor - \rfloor^{\mathbf{C}}$. □

**Theorem 8.138.** A cospan of hypergraphs is in the image of $\mathbf{T}_\Sigma + \mathbf{CComon} \cong \mathsf{PLMCsp}_D(\mathbf{Hyp}_\Sigma)$ if and only if it is partial left-monogamous.

*Proof.* It suffices to show that a cospan $m \to F \leftarrow n$ in $\mathsf{PLMCsp}_D(\mathbf{Hyp}_\Sigma)$ can be decomposed into a traced cospan in which every component under the trace is in the image of either $\langle\!\langle - \rangle\!\rangle_\Sigma$ or $[-]_\Sigma \circ \lfloor - \rfloor^{\mathsf{C}}$. This is achieved by taking the construction of Theorem 8.111 and allowing the first component to be partial left-monogamous; by Lemma 8.137 this is in the image of $[-]_\Sigma \circ \lfloor - \rfloor^{\mathsf{C}}$. The remaining components remain in the image of $[\![-]\!]_\Sigma$. Subsequently, the entire traced cospan must be in the image of $\langle\!\langle - \rangle\!\rangle_\Sigma \circ \lfloor - \rfloor_\Sigma$. □

The composite cospan for the comonoid correspondence is broadly the same as that of the traced correspondence, but now the term derived from the discrete component may additionally contain the comonoid and the counit.

As $\langle\!\langle - \rangle\!\rangle_\Sigma$ and $\lfloor - \rfloor_\Sigma$ are faithful, we immediately find the following.

**Corollary 8.139.** There is an isomorphism of coloured PROPs $\mathbf{T}_\Sigma + \mathbf{CComon} \cong \mathsf{PLMCsp}_D(\mathbf{Hyp}_\Sigma)$.

This means the PROP $\mathsf{PLMCsp}_D(\mathbf{Hyp}_\Sigma)$ of partial left-monogamous cospans of hypergraphs is suitable for modelling terms in $\mathbf{T}_\Sigma + \mathbf{CComon}$: traced terms with a cocommutative comonoid structure.

**Example 8.140.** The partial monogamous cospans from Example 8.127 are shown below with their corresponding terms in $\mathbf{T}_\Sigma + \mathbf{CComon}$.

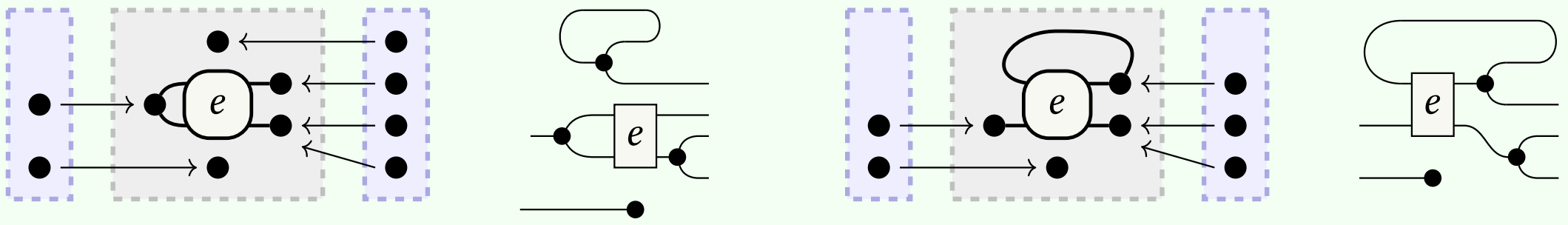

## 8.6.1 Traced coloured comonoids

As usual, the results of the monochromatic setting generalise in a straightforward manner to the multicoloured setting. As with **Frob**, a multicoloured version of **CComon** is defined as a coproduct in **CPROP**.

**Definition 8.141.** For a countable set $C$, let $\mathbf{CComon}_C \coloneqq \Sigma_{c \in C}\mathbf{CComon}$.

Partial left-monogamy follows as before, so we have a traced coloured PROP.

**Definition 8.142.** Let $\mathsf{PLMCsp}_{D_C}(\mathbf{Hyp}_{C,\Sigma})$ be the sub-PROP of $\mathsf{Csp}_{D_C}(\mathbf{Hyp}_{C,\Sigma})$ containing only the partial left-monogamous cospans of hypergraphs.

**Proposition 8.143.** The canonical trace is a trace on $\mathsf{PLMCsp}_{D_C}(\mathbf{Hyp}_{C,\Sigma})$.

As we embedded **CComon** into **Frob**, we embed $\mathbf{CComon}_C$ into $\mathbf{Frob}_C$.

**Definition 8.144.** Let $\lfloor - \rfloor^{\mathbf{C}}_C \colon \mathbf{CComon}_C \to \mathbf{Frob}_C$ be the embedding of $\mathbf{CComon}_C$ into $\mathbf{Frob}_C$, and let $\lfloor - \rfloor_{C,\Sigma} \colon \mathbf{T}_{C,\Sigma} + \mathbf{CComon}_C \to \mathbf{S}_{C,\Sigma} + \mathbf{Frob}_C$ be the copairing of $\lfloor - \rfloor^{\mathbf{T}}_{C,\Sigma}$ and $\lfloor - \rfloor^{\mathbf{C}}_C$.

**Corollary 8.145.** $\lfloor - \rfloor^{\mathbf{C}}_C$ and $\lfloor - \rfloor^{\mathbf{T}}_{C,\Sigma}$ are faithful.

The monochromatic results can then be lifted to the coloured case in the same way.

**Theorem 8.146.** A cospan of hypergraphs is in the image of $\mathbf{T}_{C,\Sigma} + \mathbf{CComon}_C \cong \mathsf{PLMCsp}_{D_C}(\mathbf{Hyp}_{C,\Sigma})$ if and only if it is partial left-monogamous.

**Corollary 8.147.** There is an isomorphism of coloured PROPs $\mathbf{T}_{C,\Sigma} + \mathbf{CComon}_C \cong \mathsf{PLMCsp}_{D_C}(\mathbf{Hyp}_{C,\Sigma})$.

# Chapter 9

# Graph rewriting

String diagrams equal by topological deformations or by the equations of commutative comonoids can be translated into isomorphic cospans of hypergraphs. This already gives us a straightforward way to interpret these structures computationally. However, it is very rare that we will only be working modulo the equations of traced comonoids; the equations of some *monoidal theory* will also be in play. Since an equation $-\boxed{f}- = -\boxed{g}-$ may actually change the components of the term, there is no reason for the interpretations $\langle\!\langle \lfloor -\boxed{f}- \rfloor_{\Sigma} \rangle\!\rangle_{\Sigma}$ and $\langle\!\langle \lfloor -\boxed{g}- \rfloor_{\Sigma} \rangle\!\rangle_{\Sigma}$ to be isomorphic. Instead, to reason about cospans of hypergraphs modulo the equations of some monoidal theory the graphs must be *rewritten.*

Traditional term rewriting is inherently one-dimensional, which can be somewhat restrictive as it enforces that each generator must have a coarity of 1. On the other hand, string diagram rewriting is an example of *higher dimensional* rewriting, which has its roots in Burroni's work on polygraphs [Bur93]. This higher-dimensional approach briefly entered the world of digital circuits in Lafont's work on Boolean circuits [Laf03], which used a rudimentary form of graphical syntactic rewriting. However, in this approach even the axioms of SMCs needed to be applied explicitly, greatly hampering the tractability of the system.

It is only more recently that rewriting string diagrams modulo the axioms of SMCs using *graph rewriting* has been studied, starting with the aforementioned *string graphs* of Dixon and Kissinger [DDK10; Kis12; DK13]. This only considered rewriting in the presence of a trace rather than considering any additional structure, but nevertheless was successfully implemented in a proof assistant called Quantomatic [KZ15]. Our work continues to follow that of Bonchi et al [BGK+22a; BGK+22b] on rewriting with *hypergraphs*, which has several advantages. Unlike string graphs, the category of hypergraphs is *adhesive* [LS04] which affords it some nice rewriting properties. Moreover, rewriting modulo Frobenius structure (or even just monoid or comonoid structure) can reveal 'hidden' rewrites that may not be seen with the naked eye. In this chapter, we will explore how to extend this to the traced and traced comonoid case.

**Remark 9.1.** The content of this section is a revised version of [GK23, Sec. 5].

## 9.1 Double pushout rewriting

Let us first consider how equations in a monoidal theory are applied to terms without any sort of special graph interpretation.

**Definition 9.2** (Rewriting system)**.** A *rewriting system* $\mathcal{R}$ for a PROP $\mathcal{C}$ is a set of *rewrite rules* $\left\langle i\text{-}\boxed{l}\text{-}j\,,\, i\text{-}\boxed{r}\text{-}j \right\rangle$. Given terms $m\text{-}\boxed{g}\text{-}n$ and $m\text{-}\boxed{h}\text{-}n$ in $\mathbf{T}_\Sigma$ we write $\text{-}\boxed{g}\text{-} \Rightarrow_{\mathcal{R}} \text{-}\boxed{h}\text{-}$ if there exists a rewrite rule $\left\langle i\text{-}\boxed{l}\text{-}j\,,\, i\text{-}\boxed{r}\text{-}j \right\rangle$ in $\mathcal{R}$ along with terms $m\text{-}\boxed{c_1}\text{-}{}^{i}_{k}$ and ${}^{j}_{k}\text{-}\boxed{c_2}\text{-}n$ in $\mathbf{S}_\Sigma$ such that by axioms of STMCs. We write $m\text{-}\boxed{g}\text{-}n \Rightarrow^{\star}_{\mathcal{R}} m\text{-}\boxed{h}\text{-}n$ for a sequence of such rules.

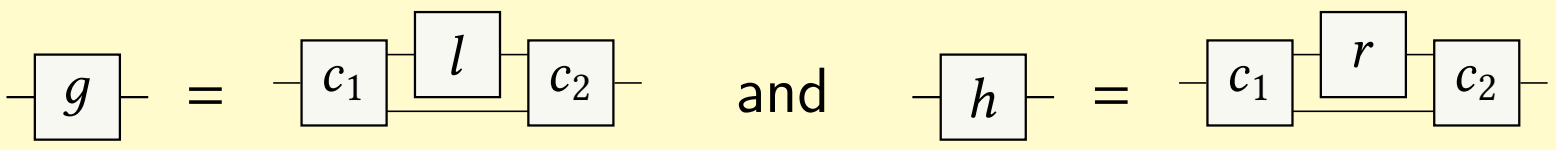

Since Frobenius terms are symmetric monoidal terms equipped with additional generators, this definition is also suitable for rewriting modulo Frobenius structure. For *traced* terms some tweaking will be required; this will be detailed in the next section.

Rules in a rewrite system are *directed*, whereas equations are not. Of course, it is straightforward to derive a rewriting system from an equational theory by adding the reductions for both directions of each equation.

**Definition 9.3** ([BGK+22b, Sec. 2.4])**.** For a monoidal theory $(\Sigma, \mathcal{E})$, let $\mathcal{R}_\mathcal{E}$ be the rewriting system containing rules $\left\langle i\text{-}\boxed{l}\text{-}j\,,\, i\text{-}\boxed{r}\text{-}j \right\rangle$ and $\left\langle i\text{-}\boxed{r}\text{-}j\,,\, i\text{-}\boxed{l}\text{-}j \right\rangle$ for each $i\text{-}\boxed{l}\text{-}j = i\text{-}\boxed{r}\text{-}j \in \mathcal{E}$.

**Proposition 9.4** ([BGK+22b, Prop. 2.18])**.** For two terms $\text{-}\boxed{g}\text{-}\,,\, \text{-}\boxed{h}\text{-} \in \mathbf{T}_{\Sigma,\mathcal{E}}$, $\text{-}\boxed{g}\text{-} = \text{-}\boxed{h}\text{-}$ if and only if $\text{-}\boxed{g}\text{-} \Rightarrow^{\star}_{\mathcal{R}_\mathcal{E}} \text{-}\boxed{h}\text{-}$.

The equivalent for graphs is *graph rewriting*; we use *double pushout (DPO)* rewriting, which was introduced in the early 70s by Ehrig, Pfender, and Schneider [EPS73] as one of the first *algebraic* approaches to graph rewriting. First defined for graphs, it has since been generalised for a variety of combinatorial structures.

A double pushout rewrite rule is defined by mapping to the left and right hand side of a rule from their shared interface.

**Definition 9.5** (Span). A *span* is a pair of morphisms $A \to B$ and $A \to C$, usually written $B \leftarrow A \to C$.

**Definition 9.6** (DPO rule). Given two interfaced hypergraphs $i \xrightarrow{a_1} L \xleftarrow{a_2} j$ and $i \xrightarrow{b_1} R \xleftarrow{b_2} j$ in $\mathsf{Csp}_D(\mathbf{Hyp}_\Sigma)$, their corresponding *DPO rule* is a span in $\mathbf{Hyp}_\Sigma$ defined as $L \xleftarrow{[a_1,a_2]} i + j \xrightarrow{[b_1,b_2]} R$.

A DPO rule is a span in the category of hypergraphs $\mathbf{Hyp}_\Sigma$, as we will identify occurrences of rules with hypergraph homomorphisms.

We will use an extension of DPO rewriting, known as *double pushout rewriting with interfaces* (DPOI rewriting) [BGK+17]. This framework enjoys the *Knuth-Bendix property* [KB70]; graph rewriting is confluent when all *critical pairs are joinable*. This means that a rewriting system is confluent if, whenever there is an overlap of rules in a graph $G$ such that $G$ could rewrite to $H$ or $H'$, there exists another graph $K$ and rewrites such that $H$ and $H'$ rewrite to $K$.

**Definition 9.7** (DPOI rewriting). Let $\mathcal{R}$ be a set of DPO rules. Then, for morphisms $G \leftarrow m + n$ and $H \leftarrow m + n$ in $\mathbf{Hyp}_{C,\Sigma}$, there is a rewrite $G \rightsquigarrow_{\mathcal{R}} H$ if there exist a span $L \leftarrow i + j \to R \in \mathcal{R}$ and cospan $i + j \to C \leftarrow m + n \in \mathbf{Hyp}_\Sigma$ such that the following diagram commutes.

$$\begin{array}{ccccc} L & \longleftarrow & i+j & \longrightarrow & R \\ \downarrow & \lrcorner & \downarrow & \llcorner & \downarrow \\ G & \longleftarrow & C & \longrightarrow & H \\ & \nwarrow & \uparrow & \nearrow & \\ & & m+n & & \end{array}$$

The first thing to note is that the graphs all have a *single* interface $G \leftarrow m + n$; to perform graph rewriting on graphs in $\mathbf{Hyp}_\Sigma$, interfaces of terms in $\mathsf{S}_\Sigma + \mathbf{Frob}$ must be 'folded' into one using the compact closed structure.

**Definition 9.8** ([BGK+22b]). Let $\ulcorner - \urcorner \colon \mathbf{S}_\Sigma + \mathbf{Frob} \to \mathbf{S}_\Sigma + \mathbf{Frob}$ be defined as having action $m\!-\!\boxed{f}\!-\!n \mapsto \supset\!\boxed{f}\!-\!n,\ m$ (with input $m$ bent around by a cup to become an output).

The image of $\ulcorner - \urcorner$ is not in the image of $\lfloor - \rfloor^{\mathsf{T}}_\Sigma$ or $\lfloor - \rfloor_\Sigma$ any more, as inputs of generators may now connect to outputs of the term. This is not an issue, as long as we 'unfold' the interfaces once rewriting is completed.

**Proposition 9.9** ([BGK+22a, Prop. 4.8]). For a term $m\!-\!\boxed{f}\!-\!n \in \mathbf{S}_\Sigma + \mathbf{Frob}$, if $\langle\!\langle -\boxed{f}- \rangle\!\rangle_\Sigma = m \xrightarrow{i} F \xleftarrow{o} n$ then $\langle\!\langle \ulcorner -\boxed{f}- \urcorner \rangle\!\rangle_\Sigma$ is isomorphic to $0 \to F \xleftarrow{i+o} m+n$.

*Proof.* Straightforward by definition of the cup using the Frobenius structure. □

We are now ready to begin rewriting. Say we have a DPO rule $L \leftarrow i+j \to R$ and a larger cospan of hypergraphs $m \to G \leftarrow n$. We suggestively assemble them as follows, using the transformation above to 'convert' the latter cospan to one with a single interface.

$$\begin{array}{ccccc} L & \longleftarrow & i+j & \longrightarrow & R \\ & & & & \\ G & & & & \\ & \nwarrow & & & \\ & & m+n & & \end{array}$$

To identify an occurrence of $L$ in $G$, we use a hypergraph homomorphism $L \to G$ to identify the components that will be rewritten.

$$\begin{array}{ccccc} L & \longleftarrow & i+j & \longrightarrow & R \\ \downarrow & & & & \\ G & & & & \\ & \nwarrow & & & \\ & & m+n & & \end{array}$$

We now need to identify the *context* in which the rewrite will occur. Essentially, the context is the 'graph $G$ with $L$ cut out', which can be formally defined with what is known as a *pushout complement*. This can be thought of as a 'reverse pushout'.

**Definition 9.10** (Pushout complement). Let $i+j \to L \to G \leftarrow m+n$ be morphisms in $\mathbf{Hyp}_{C,\Sigma}$; their *pushout complement* is an object $C$ with morphisms $i+j \to C \to G$ such that $L \to G \leftarrow C$ is a pushout and the diagram below commutes.

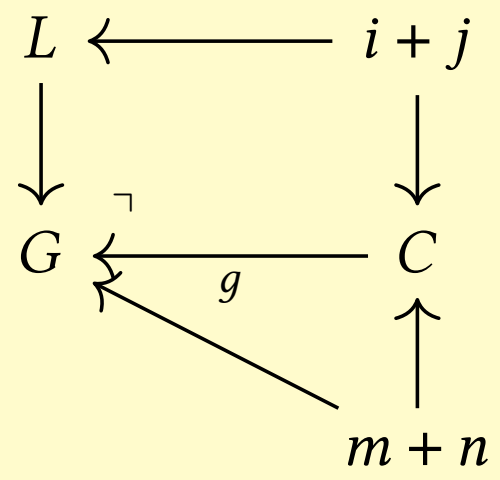

Is a pushout complement always guaranteed to exist for any morphism $L \to G$? The answer is no; this will be discussed at length in the next section. If a pushout complement does exist, it specifies the rewriting context. This leaves us a hole in which the other side of the rewrite rule can be glued in.

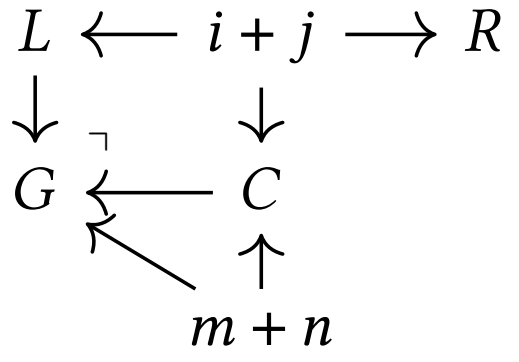

To actually compute the rewritten graph, we perform another pushout to retrieve the complete DPO diagram. Note that the interface $m + n$ of the original graph is preserved throughout the process; we can use $\ulcorner - \urcorner$ to return to a cospan of the form $m \to H \leftarrow n$.

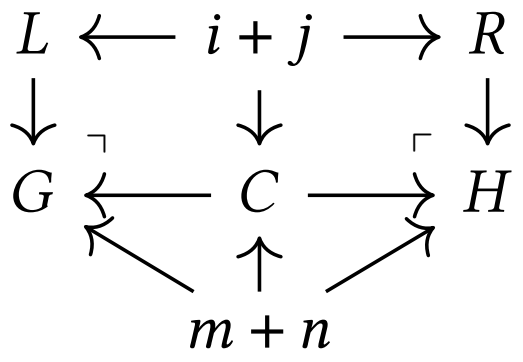

**Example 9.11.** Consider the following term rewrite rule and its interpretation as a DPO rule.

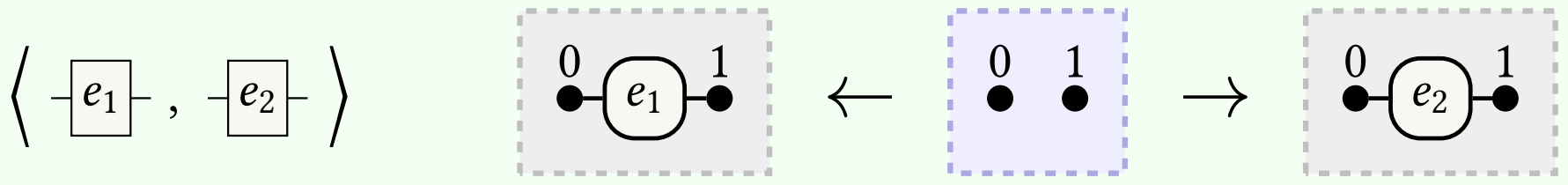

Consider the term 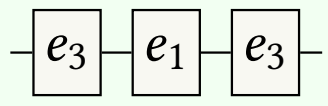 ; a DPO rewrite is performed as follows:

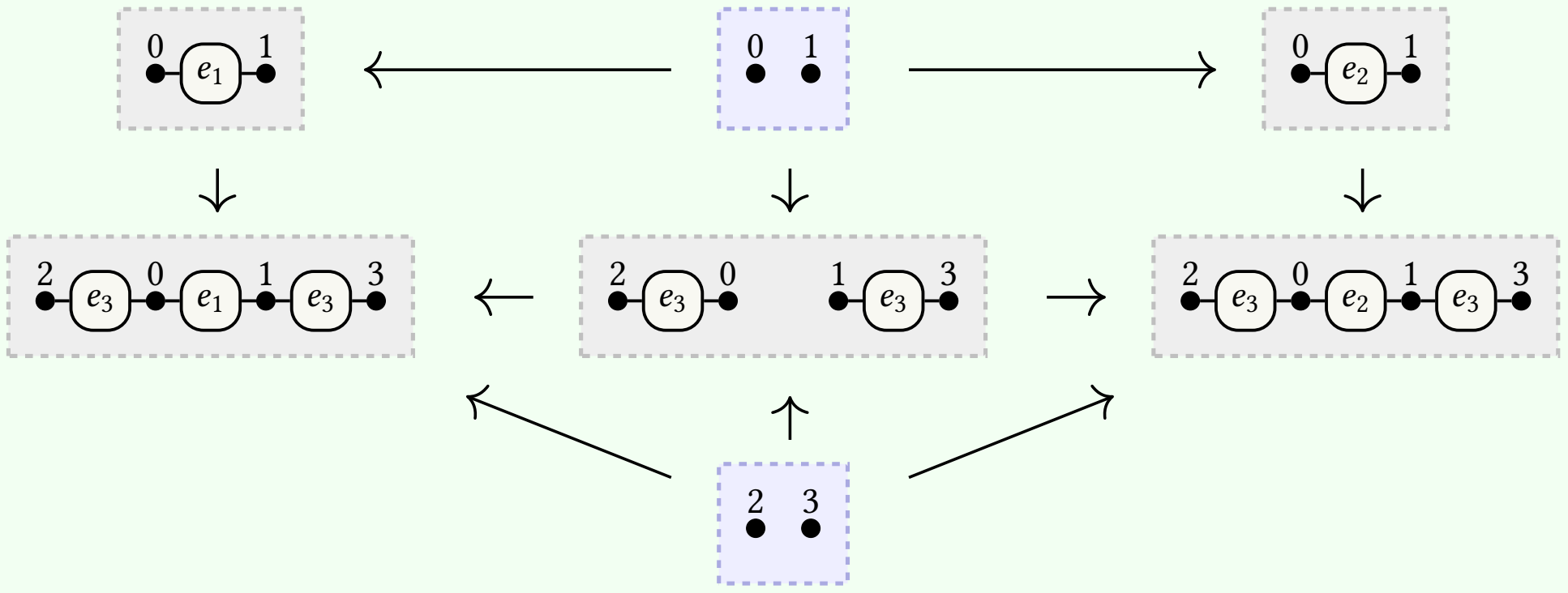

As expected, the result is the term 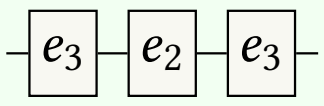 .

## 9.2 Pushout complements

While computing the rewritten graph from a context is a deterministic procedure, finding the pushout complement specifying said context is a little more subtle.

**Definition 9.12.** A morphism $i+j \longrightarrow L \longrightarrow G$ is called a *matching* if it has at least one pushout complement.

If there is no pushout complement, there is no possible rewrite, so it is important to know when one exists. Fortunately, there are two well-known conditions for existence of pushout complements when rewriting with hypergraphs. The first ensures that all the sources and targets of a hyperedge are present in a candidate context.

**Definition 9.13** (No-dangling-hyperedges condition [CMR+97], Prop. 3.3.4)**.** Given morphisms $i + j \xrightarrow{a} L \xrightarrow{f} G$ in $\mathbf{Hyp}_{\Sigma}$, they satisfy the *no-dangling* condition if, for every hyperedge not in the image of $f$, each of its source and target vertices is either not in the image of $f$ or are in the image of $f \circ a$.

**Example 9.14.** The following pair of morphisms does not satisfy the no-dangling-hyperedges condition.

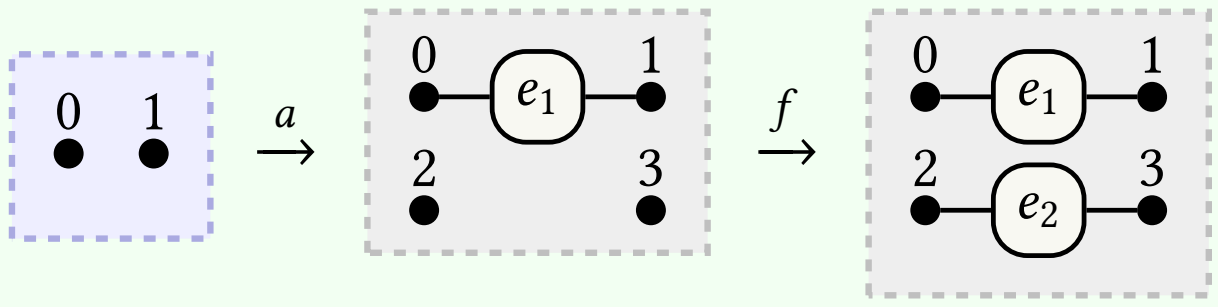

To obtain the pushout complement we ‘cut out’ any vertices in the rightmost graph which are in the image of $f$ but not the image of $f \circ a$, as the latter are the interfaces of the rule. However, if we cut out the vertices labelled 2 and 3, the edge $e_2$ will be left with 'dangling' tentacles connected to no vertices, a malformed hypergraph.

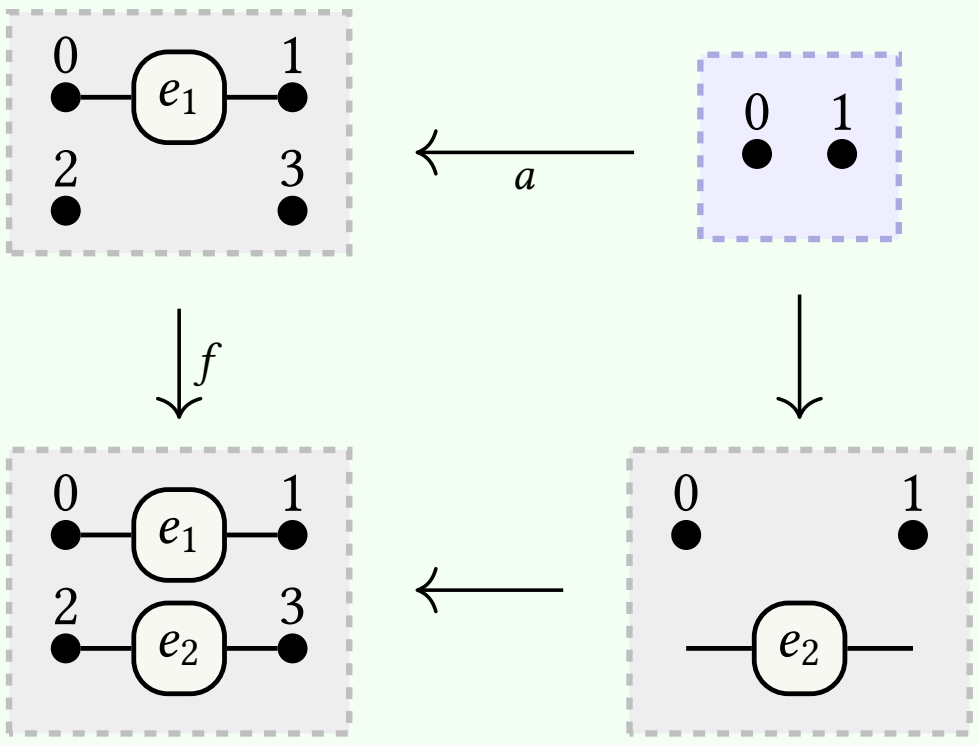

The second condition enforces that merging of vertices is well-defined.

**Definition 9.15** (No-identification condition [CMR+97], Prop. 3.3.4)**.** Given morphisms $i + j \xrightarrow{a} L \xrightarrow{f} G$ in $\mathbf{Hyp}_\Sigma$, they satisfy the *no-identification* condition if any two distinct elements merged by $f$ are also in the image of $f \circ a$.

**Example 9.16.** The following does not satisfy the no-identification condition.

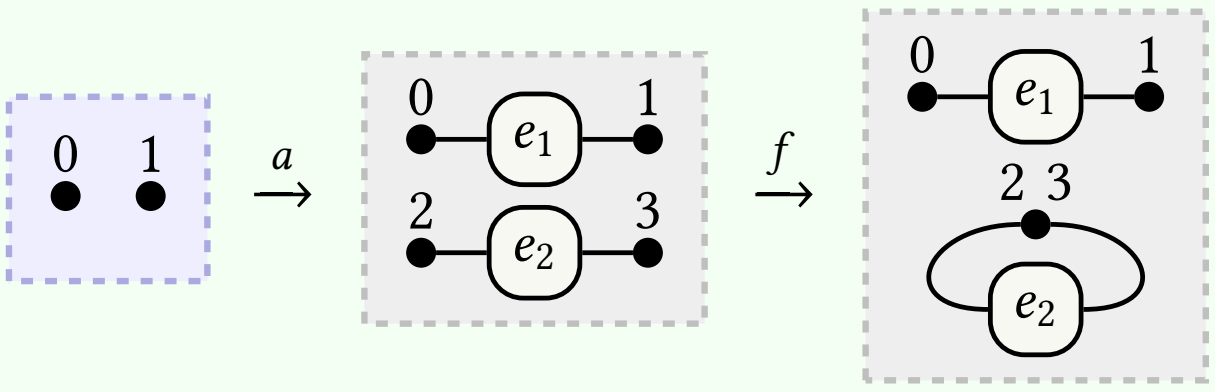

When trying to construct a pushout complement, the edge $e_2$ will be removed. However, since vertices 2 and 3 are not mapped from the rule interfaces, there is no reason that a pushout would glue them together so that they are merged in the final graph. Therefore no pushout complement can exist.

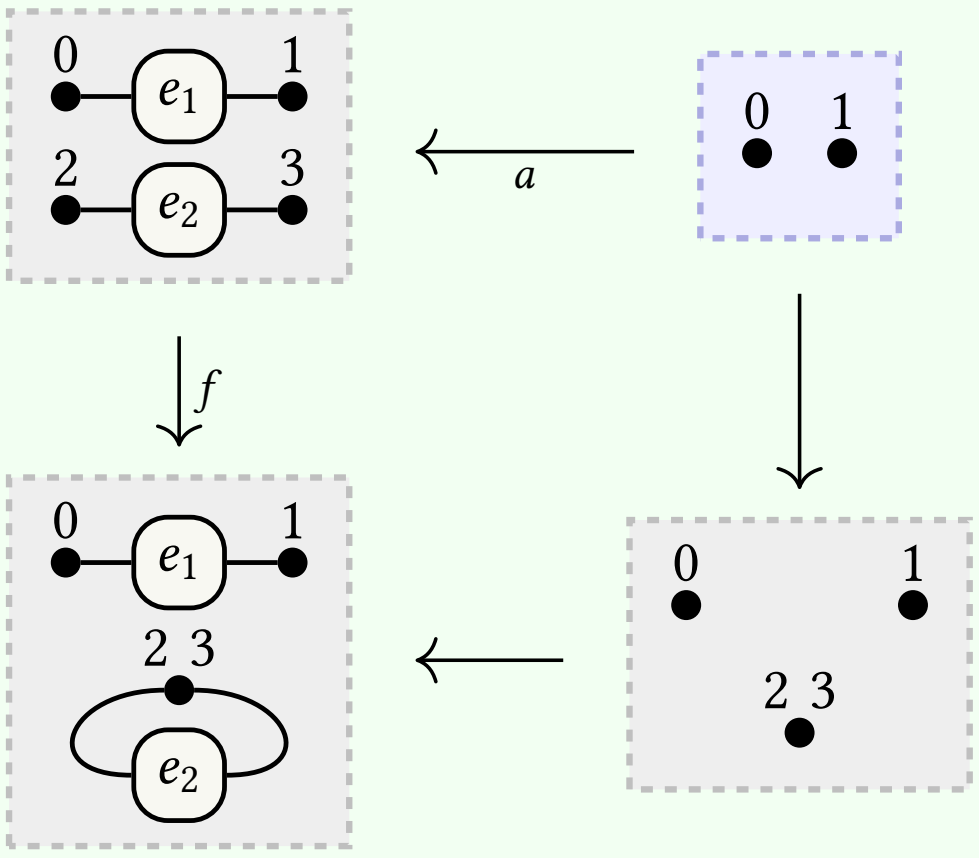

With these two conditions, we can establish when pushout complements exist for a pair of hypergraph homomorphisms. If there is a pushout complement, then there is an opportunity for a rewrite.

**Proposition 9.17** ([CMR+97], Prop. 3.3.4)**.** The morphisms $i + j \to L \to G \in \mathbf{Hyp}_\Sigma$ have at least one pushout complement if and only if they satisfy the no-dangling and no-identification conditions.

It is all very well knowing if there is at least one pushout complement, but what about when there is *exactly one* pushout complement? When this is the case, the rewrite is uniquely specified for a given rule and matching. To answer this question, we must examine a class of categories to which $\mathbf{Hyp}_\Sigma$ belongs, known as *adhesive* categories. One can think of these as categories in which graph rewriting 'plays nicely'.

To define what an adhesive category is, we must first define a special kind of pushout that interacts in a particular way with other pushouts and pullbacks.

**Definition 9.18** (van Kampen square [LS05, Def. 2.1])**.** Let there be a commutative cube as drawn below.

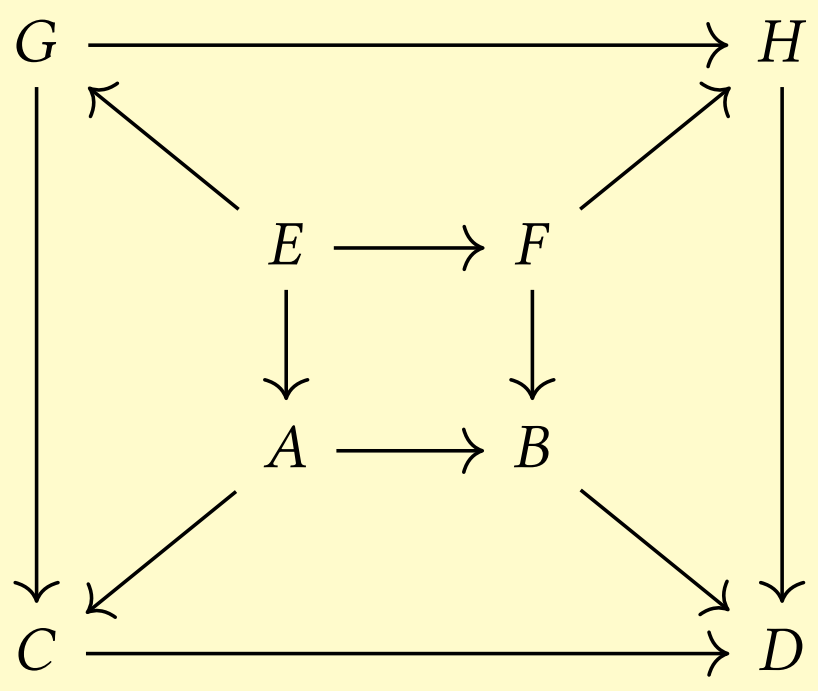

The bottom face of the cube ($ABCD$) is called a *van Kampen (VK) square* if it is a pushout and, when the back and left faces ($EFAB$ and $GECA$) are pullbacks, the front and right faces ($GHCD$ and $FHBD$) are pullbacks if and only if the top face ($GHEF$) is a pushout.

In an adhesive category, VK squares arise when performing pushouts of *monomorphisms*. This is important for graph rewriting because monomorphisms in categories of graphs generally correspond to *embeddings*; graph homomorphisms where subgraphs can be 'cut out' of a graph without causing the rest of the graph to become degenerate.

**Definition 9.19.** Given a span $A \overset{f}{\leftarrow} B \overset{g}{\rightarrow} C$ with a pushout $B \rightarrow D \leftarrow C$, the pushout is called a *pushout along a monomorphism* if $f$ or $g$ is a monomorphism.

**Definition 9.20** (Adhesive category [LS05, Def. 3.1])**.** A category is *adhesive* if

- it has pushouts along monomorphisms;
- it has pullbacks; and
- pushouts along monomorphisms are VK squares.

The definition of van Kampen square and subsequently adhesive categories may look a bit confusing to the uninitiated. Succinctly, objects in an adhesive category can be 'split apart' and 'glued together' by using pushouts and pullbacks.

**Example 9.21.** A natural example of an adhesive category is **Set**, which has pushouts and pullbacks. To get some idea how the van Kampen condition holds, consider the following commutative cube in **Set**, adapted from [Kis12, Sec. 4.3].

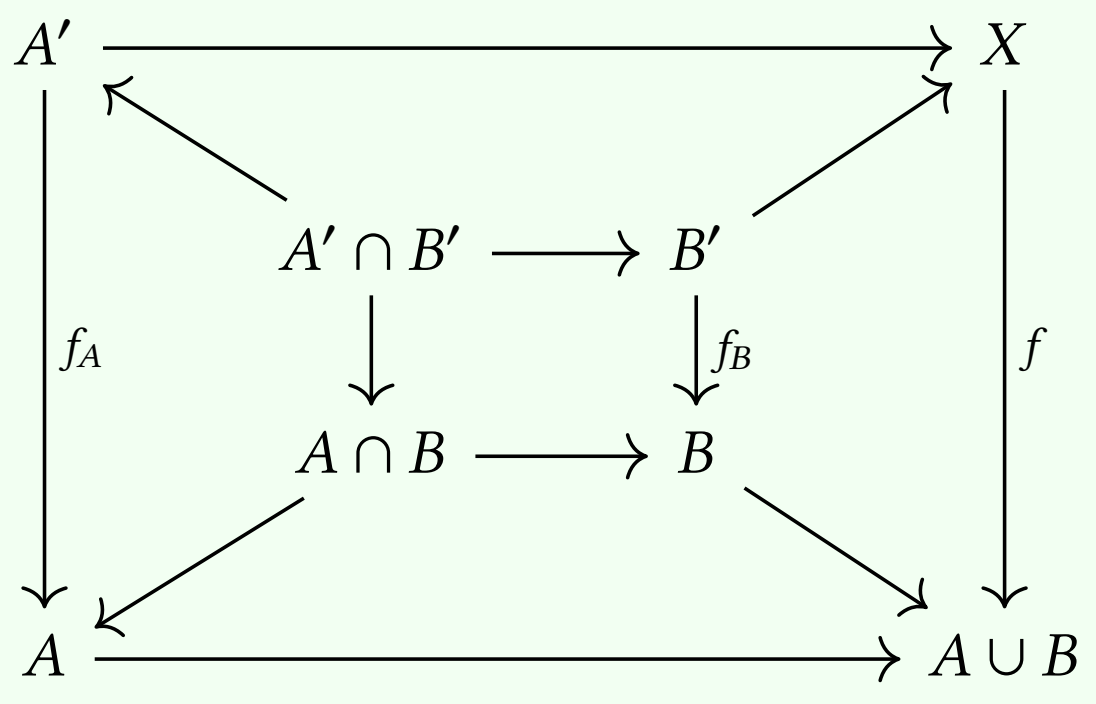

The bottom face is a pushout, and since $A \cup B \rightarrow A$ and $A \cup B \rightarrow B$ are monomorphisms, this is a pushout along a monomorphism. Furthermore, the left and back faces are pullbacks because $f_A$ and $f_B$ agree on the intersection of $A'$ and $B'$.

Now we must show that the front and right faces are pullbacks if and only if the top face is a pushout. For the front and right faces to be pullbacks, $f$ must restrict to $f_A$ and $f_B$ along $A$ and $B$. This can only be the case if and only if $X = A' \cap B'$, in which case the top square must be a pushout.

Proving the necessary van Kampen condition might be tricky. Fortunately, adhesivity is preserved by several categorical constructions, so by using **Set** as a base it is straightforward to show that more complicated categories are also adhesive.

**Proposition 9.22** ([LS05], Prop. 3.5)**.** For an adhesive category $\mathcal{C}$ and object $C$ of $\mathcal{C}$, $\mathcal{C} \downarrow C$ is adhesive. Given another category $\mathcal{X}$, $[\mathcal{X}, \mathcal{C}]$ is also adhesive.

**Corollary 9.23.** $\mathbf{Hyp}_\Sigma$ and $\mathbf{Hyp}_{C,\Sigma}$ are adhesive.

*Proof.* $\mathbf{Hyp}_\Sigma$ and $\mathbf{Hyp}_{C,\Sigma}$ are defined as the slice of a functor category over **Set**, so they are adhesive. □

The key property adhesive categories enjoy is that, for certain DPO rules, a pushout complement is *uniquely* defined for a given matching.

**Definition 9.24** (Left-linear rules)**.** A DPO rule $L \xleftarrow{f} i + j \rightarrow R$ is called *left-linear* if $f$ is mono.

**Theorem 9.25** ([LS05], Lem. 4.5)**.** In an adhesive category, if a pushout complement exists for morphisms $I \xrightarrow{m} L \rightarrow G$ and $m$ is a monomorphism, then it is unique up to isomorphism.

*Proof.* The proof relies on several non-trivial lemmas that hold in adhesive categories in addition to some other results about pushouts. We refer the interested reader to [Kis12, Lems. 4.3.6 - 4.3.9] for the grisly details. □

However, there may be useful rewrite rules which are *not* left-linear.

**Example 9.26.** Consider the following (reasonable) rule.

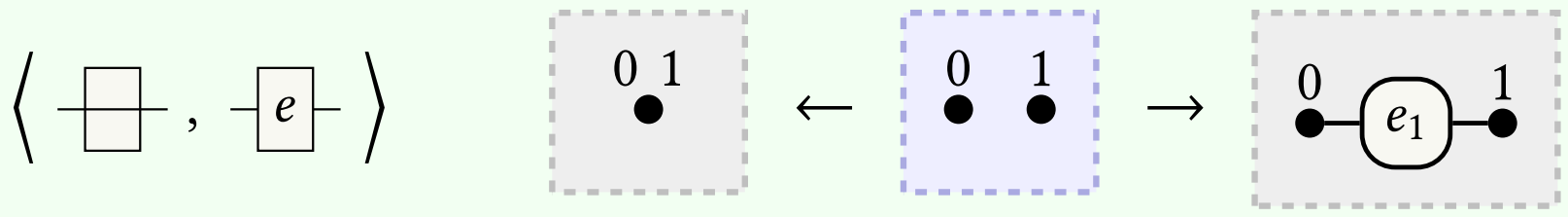

. Now consider applying the above rule to the term $e_2$—$e_3$ using the following

pair of morphisms:

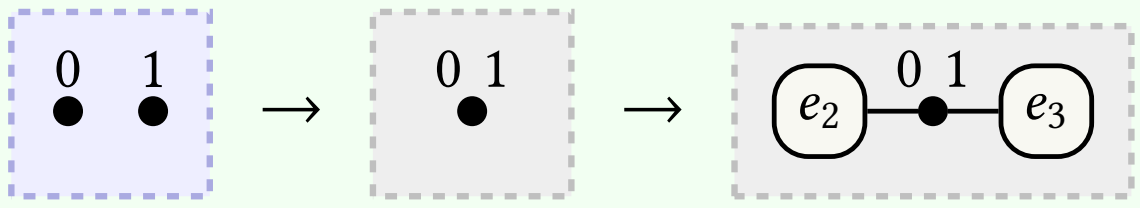

This matching yields the following pushout complements and rewrites:

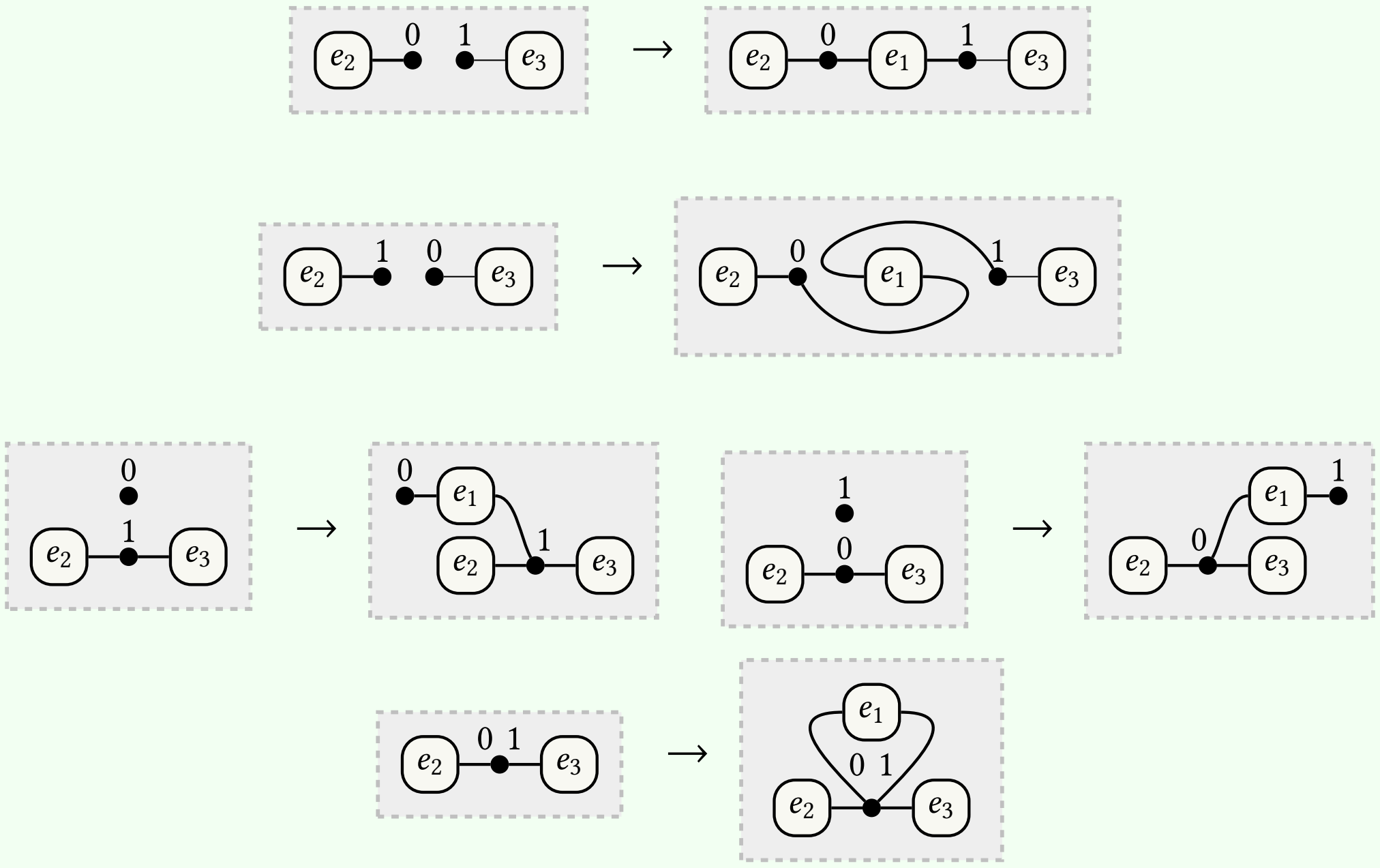

One might think this is undesirable, but these multiple rewrites actually arise due to the presence of the Frobenius algebra.

**Example 9.27.** Each of the five complements and rewrites in Example 9.26 corresponds to a valid application of equations on terms, perhaps modulo Frobenius equations. The first complement 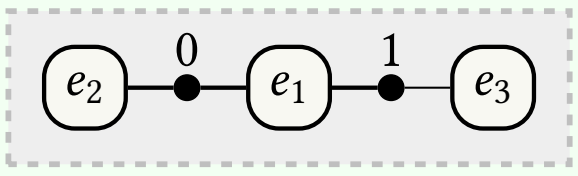 is the 'obvious' one:

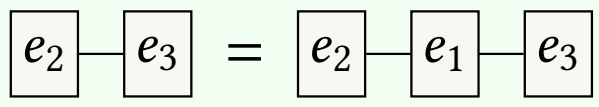

The complement 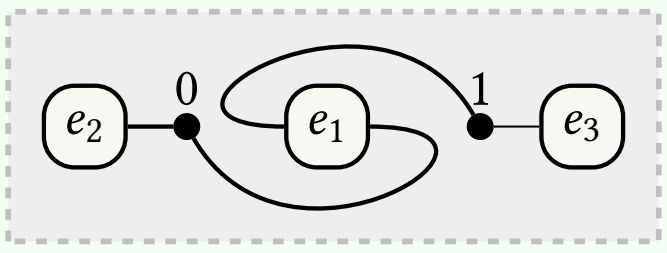 uses the compact closed structure:

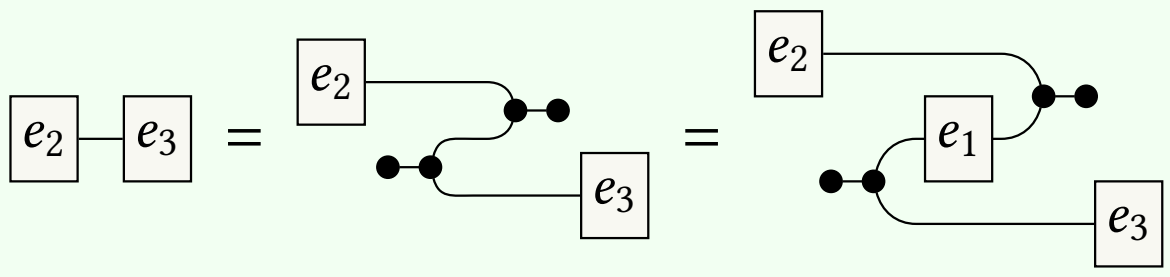

The complement 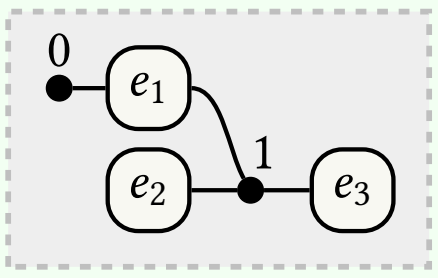 uses the unitality of the monoid:

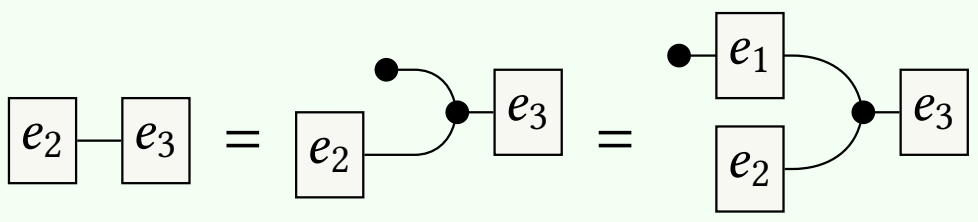

The complement 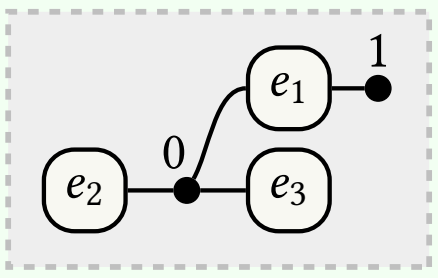 uses the counitality of the comonoid:

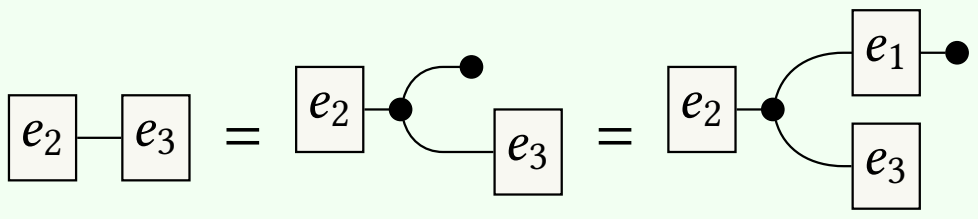

The complement 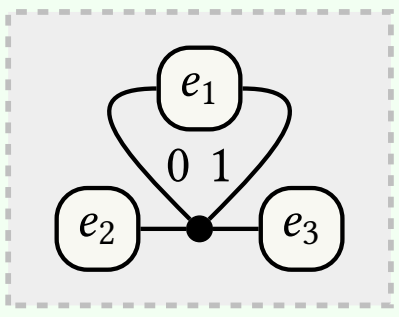 uses the Frobenius equations:

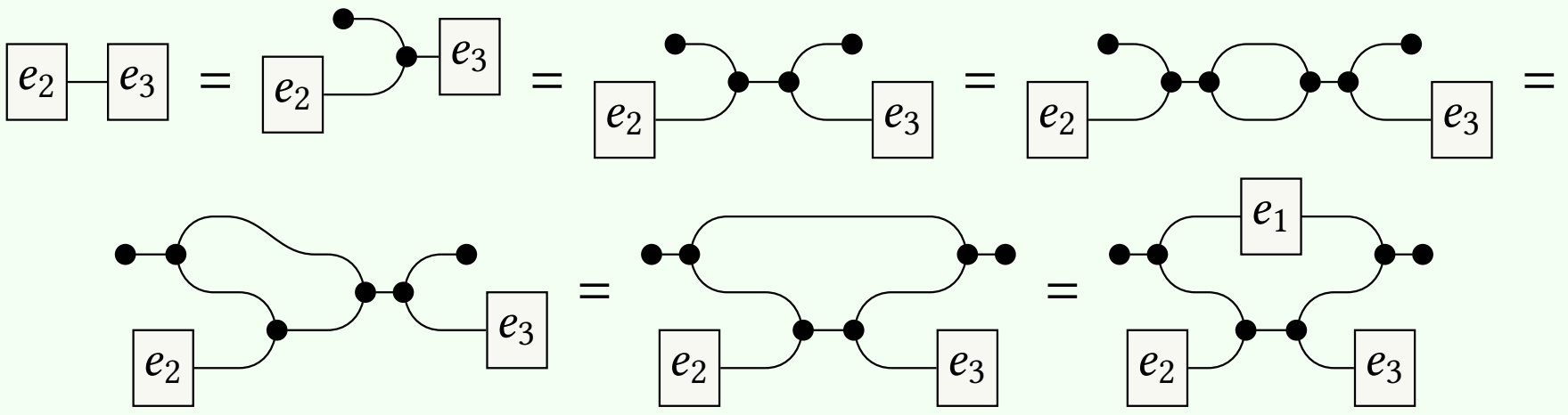

The problem of finding each possible pushout complement has already been tackled for hypergraphs [HJKS11]; they can be enumerated as quotients of an 'exploded' context.

**Definition 9.28** ([HJKS11, Const. 1]). For morphisms $i + j \to L \xrightarrow{f} G$ in $\mathbf{Hyp}_\Sigma$, their *exploded context* is the graph $i + j + \tilde{G}$ where $\tilde{G}$ constructed as follows:

1. for each vertex $v \in G$ not in the image of $f$, add one vertex to $\tilde{G}$;
2. for each hyperedge $e \in G$ not in the image of $f$, add one hyperedge to $\tilde{G}$;
3. for each hyperedge $e \in \tilde{G}$, let the $i$-th source $\tilde{s}_i(e)$ be defined as $s_i(h)$ if $s_i(h) \in \tilde{G}$ or a new, fresh vertex otherwise;
4. repeat the above for the targets.

Pushout complements can then be computed as quotients of this exploded context.

**Proposition 9.29** ([HJKS11] (Props. 3-4), [BGK+22a]). For a pair of morphisms $i + j \to L \to G$ in $\mathbf{Hyp}_\Sigma$, let $i + j + \tilde{G}$ be its exploded context. Define a map $q\colon i + j + \tilde{G} \to G$ sending elements in $\tilde{G}$ from $G$ to themselves, and sending vertices from $i + j$ to their image under $i + j \to L \to G$. Then a pushout complement $i + j \to C \to G$ is valid if and only the context $C$ is a quotient on the exploded context that only identifies vertices in the image of $q^{-1}(v)$ for each vertex $v \in G$.

Given a DPO rule and matching, we can enumerate all pushout complements; each of these corresponds to a valid rewrite in a Frobenius setting.

**Notation 9.30.** For a rule $\left\langle i\text{-}\boxed{l}\text{-}j\,,\; i\text{-}\boxed{r}\text{-}j \right\rangle \in \mathbf{S}_\Sigma + \mathbf{Frob}$, its DPO rule is defined as $\left\langle\!\left\langle \left\langle -\boxed{l}-\,,\; -\boxed{r}- \right\rangle \right\rangle\!\right\rangle_\Sigma \coloneqq \langle\!\langle \ulcorner -\boxed{l}- \urcorner \rangle\!\rangle_\Sigma \leftarrow i + j \rightarrow \langle\!\langle \ulcorner -\boxed{r}- \urcorner \rangle\!\rangle_\Sigma$.

**Theorem 9.31** ([BGK+22a, Thm. 4.9]). For rule $r \in \mathbf{S}_\Sigma + \mathbf{Frob}$, we have that $-\boxed{g}- \Rightarrow_r -\boxed{h}-$ if and only if $\langle\!\langle \ulcorner -\boxed{g}- \urcorner \rangle\!\rangle_\Sigma \rightsquigarrow_{\langle\!\langle r \rangle\!\rangle_\Sigma} \langle\!\langle \ulcorner -\boxed{g}- \urcorner \rangle\!\rangle_\Sigma$.

### 9.2.1 Multicoloured rewriting

The results generalise in the obvious way to the coloured setting.

**Notation 9.32.** For a term rewrite rule $\left\langle \bar{i}\text{-}\boxed{l}\text{-}\bar{j}\,,\; \bar{i}\text{-}\boxed{r}\text{-}\bar{j} \right\rangle$ in $\mathbf{S}_{C,\Sigma} + \mathbf{Frob}_C$, its interpretation as a DPO rule is defined as

$$\left\langle\!\left\langle \left\langle -\boxed{l}-\,,\; -\boxed{r}- \right\rangle \right\rangle\!\right\rangle_{C,\Sigma} \coloneqq \langle\!\langle \ulcorner -\boxed{l}- \urcorner_C \rangle\!\rangle_{C,\Sigma} \leftarrow \overline{ij} \rightarrow \langle\!\langle \ulcorner -\boxed{r}- \urcorner_C \rangle\!\rangle_{C,\Sigma}.$$

**Definition 9.33** ([BGK+22a]). Let $\ulcorner - \urcorner_C\colon \mathbf{S}_{C,\Sigma} + \mathbf{Frob}_C \to \mathbf{S}_{C,\Sigma} + \mathbf{Frob}_C$ be defined as having action $\bar{m}\text{-}\boxed{f}\text{-}\bar{n} \mapsto$ (the diagram with input $\bar{m}$ bent around to an output, $\boxed{f}$ with outputs $\bar{m}$ and $\bar{n}$).

**Theorem 9.34** ([BGK+22a, Prop. 4.10]). For rewrite rule $r \in \mathbf{S}_{C,\Sigma} + \mathbf{Frob}_C$, we have that $-\boxed{g}- \Rightarrow_r -\boxed{h}-$ if and only if $\langle\!\langle \ulcorner -\boxed{g}- \urcorner_C \rangle\!\rangle_{C,\Sigma} \rightsquigarrow_{\langle\!\langle r \rangle\!\rangle_{C,\Sigma}} \langle\!\langle \ulcorner -\boxed{g}- \urcorner_C \rangle\!\rangle_{C,\Sigma}$.

## 9.3 Rewriting with traced structure

Say we have a rewrite rule and a term as illustrated below:

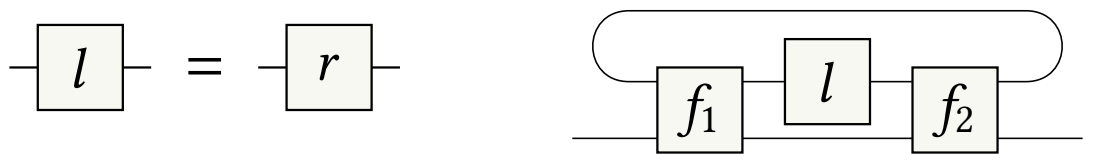

Clearly the rule should be valid in a traced setting, but when assembling the term into the form of Definition 9.35 not all of the pieces are traced terms.

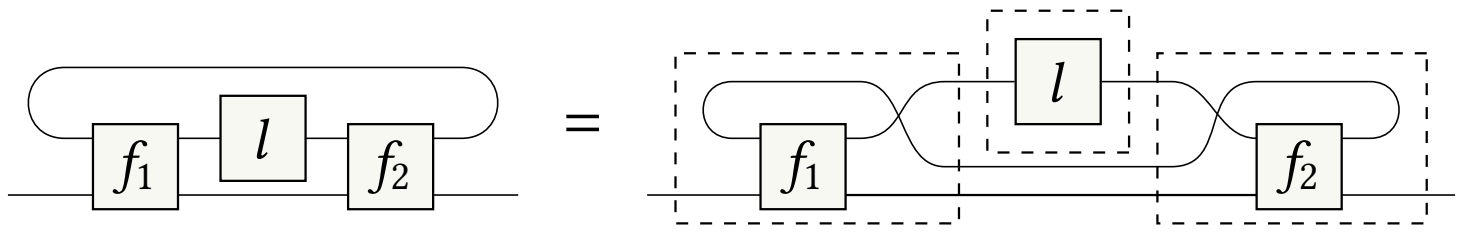

Fortunately, by tweaking the layout of the terms this term can be put into a form in which we can isolate the instance of a rewrite rule and the remaining context such that all of the pieces are valid traced monoidal terms.

**Definition 9.35** (Traced rewriting system)**.** A *rewriting system* $\mathcal{R}$ for a traced PROP $\mathcal{C}$ consists is a set of *rewrite rules* $\left\langle i\!-\!\boxed{l}\!-\!j\,,\; i\!-\!\boxed{r}\!-\!j \right\rangle$. Given terms $m\!-\!\boxed{g}\!-\!n$ and $m\!-\!\boxed{h}\!-\!n$ in $\mathbf{T}_\Sigma$ we write $-\boxed{g}- \Rightarrow_{\mathcal{R}} -\boxed{h}-$ if there exists a rewrite rule $\left\langle i\!-\!\boxed{l}\!-\!j\,,\; i\!-\!\boxed{r}\!-\!j \right\rangle$ in $\mathcal{R}$ and $\begin{smallmatrix} j- & & -i \\ m- & \boxed{c} & -n \end{smallmatrix}$ in $\mathbf{T}_\Sigma$ such that

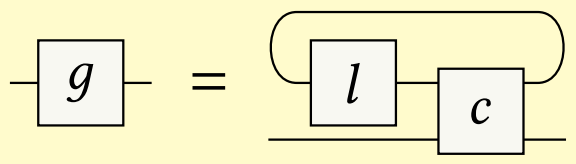

and

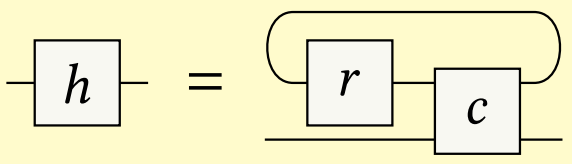

by axioms of STMCs. We write $-\boxed{l}- \Rightarrow^\star_{\mathcal{R}} -\boxed{r}-$ for a sequence of such rules.

In the Frobenius setting, every pushout complement is a valid rewrite, but there is no reason for the same to be the case for traced or traced comonoid rewriting. Bonchi et al showed in [BGK+22b] that *exactly one* pushout complement corresponds to a valid rewrite in the symmetric monoidal case by characterising it as a *boundary complement*.

**Definition 9.36** (Boundary complement [BGK+22b, Def. 30])**.** For monogamous cospans $i \xrightarrow{a_1} L \xleftarrow{a_2} j$ and $m \xrightarrow{b_1} G \xleftarrow{b_2} n$ and a monomorphism $f\colon L \to G$, a pushout complement as below

$$\begin{array}{ccc} L & \xleftarrow{a:=[a_1,a_2]} & i+j \\ {\scriptstyle f}\downarrow & & \downarrow{\scriptstyle c:=[c_1,c_2]} \\ G & \xleftarrow{g} & C \\ & \nwarrow{\scriptstyle [b_1,b_2]} & \uparrow{\scriptstyle d:=[d_1,d_2]} \\ & & m+n \end{array}$$

is called a *boundary complement* if the morphisms $c_1$ and $c_2$ are mono and $j + m \xrightarrow{[c_2,d_1]} C \xleftarrow{[d_2,c_1]} n + i$ is a monogamous cospan.

**Proposition 9.37** ([BGK+22b], Prop. 31)**.** When boundary complements exist in $\mathbf{Hyp}_\Sigma$, they are unique.

For rewriting in a traced setting we weaken boundary complements, replacing references to monogamy with partial monogamy.

**Definition 9.38** (Traced boundary complement)**.** For partial monogamous cospans $i \xrightarrow{a_1} L \xleftarrow{a_2} j$ and $m \xrightarrow{b_1} G \xleftarrow{b_2} n$, a pushout complement as below

$$\begin{array}{ccc} L & \xleftarrow{a:=[a_1,a_2]} & i+j \\ {\scriptstyle f}\downarrow & & \downarrow{\scriptstyle c:=[c_1,c_2]} \\ G & \xleftarrow{g} & C \\ & \nwarrow{\scriptstyle [b_1,b_2]} & \uparrow{\scriptstyle d:=[d_1,d_2]} \\ & & m+n \end{array}$$

is called a *traced boundary complement* if the morphisms $c_1$ and $c_2$ are mono and $j + m \xrightarrow{[c_2,d_1]} C \xleftarrow{[d_2,c_1]} n + i$ is a partial monogamous cospan.

By restricting to traced boundary complements, DPO rewriting can be formulated for terms in a traced setting.

**Definition 9.39** (Traced DPO)**.** For morphisms $G \leftarrow m + n$ and $H \leftarrow m + n$ in $\mathbf{Hyp}_\Sigma$, there is a traced rewrite $G \rightsquigarrow_{\mathcal{R}} H$ if there exists a rule $L \leftarrow i + j \rightarrow R \in \mathcal{R}$ and cospan $i + j \rightarrow C \leftarrow m + n \in \mathbf{Hyp}_\Sigma$ such that the diagram in Definition 9.7 commutes and $i + j \rightarrow C$ is a traced boundary complement.

In a traced boundary complement, the matching $L \rightarrow G$ is not required to be mono; permitting the matching to merge vertices means that incidences of a rewrite rule can be found inside a trace.

**Example 9.40.** Consider the rule 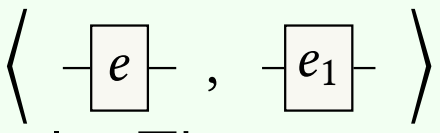 and the term 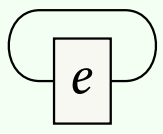, in which there is clearly an instance of the rule. The interpretation of this as a DPO derivation with a valid traced boundary complement is illustrated below.

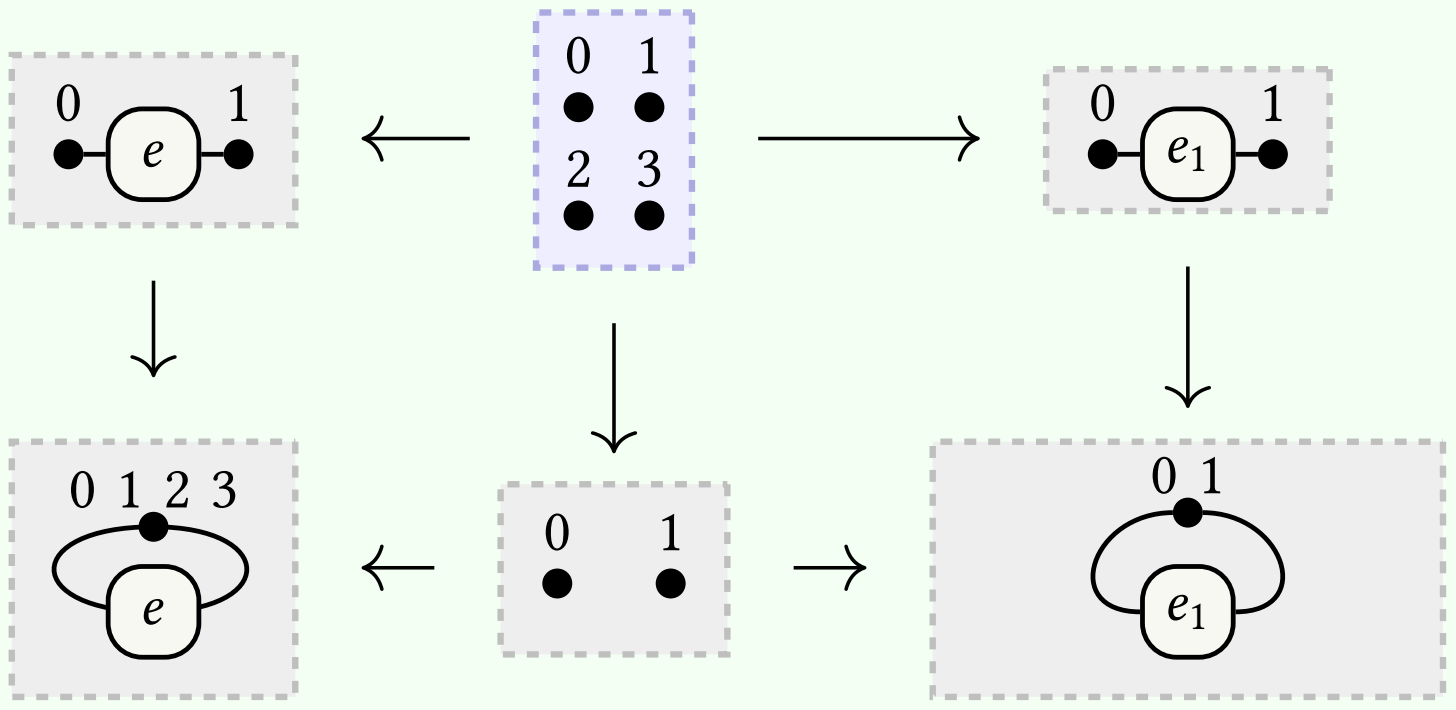

A key feature of rewriting modulo traced structure is the *yanking* axiom, which can lead to some non-obvious rewrites.

**Example 9.41.** Consider the rule 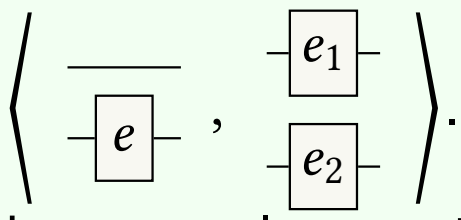. The interpretation of this as a DPO rule in a valid traced boundary complement is illustrated below.

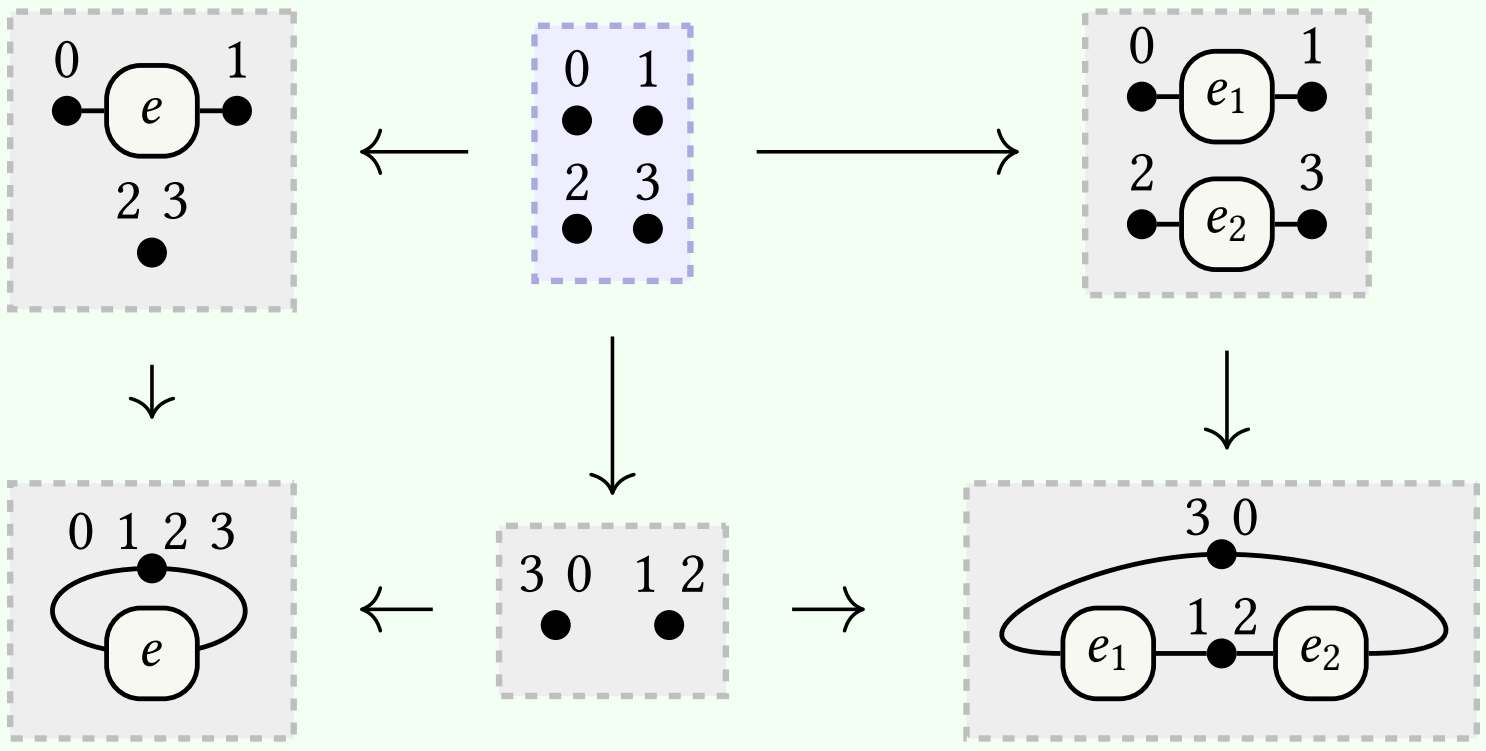

This corresponds to a valid term rewrite:

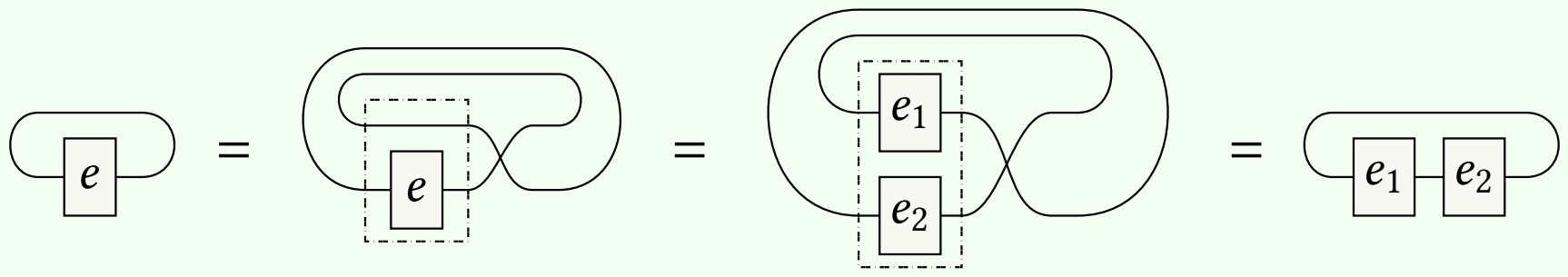

Use of yanking is also what can lead to multiple boundary complements, and hence a choice in rewrites.

**Example 9.42.** Consider the rule 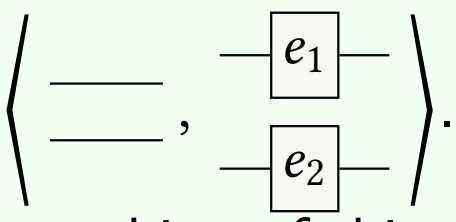. Below are two valid traced boundary complements involving a matching of this rule.

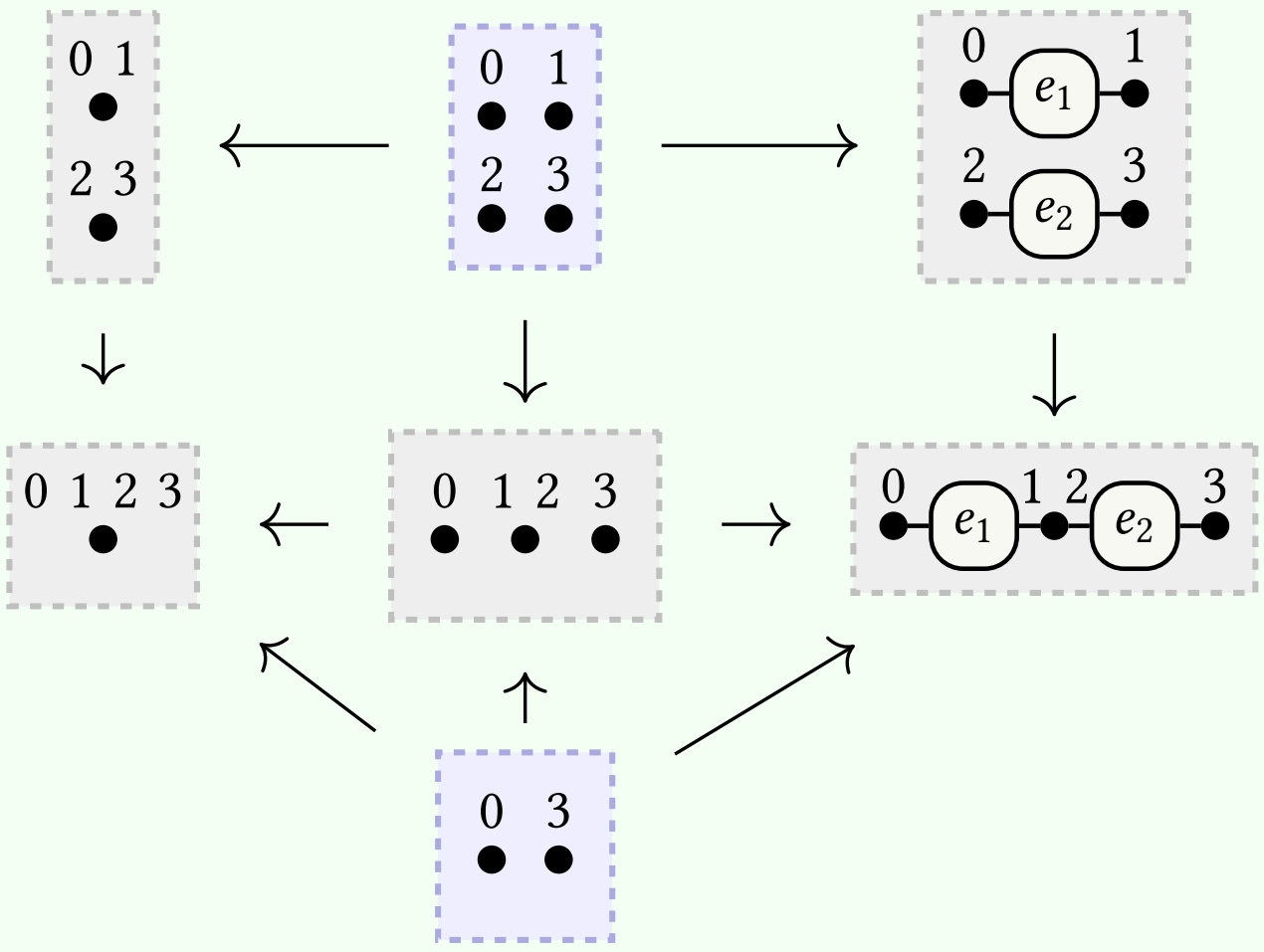

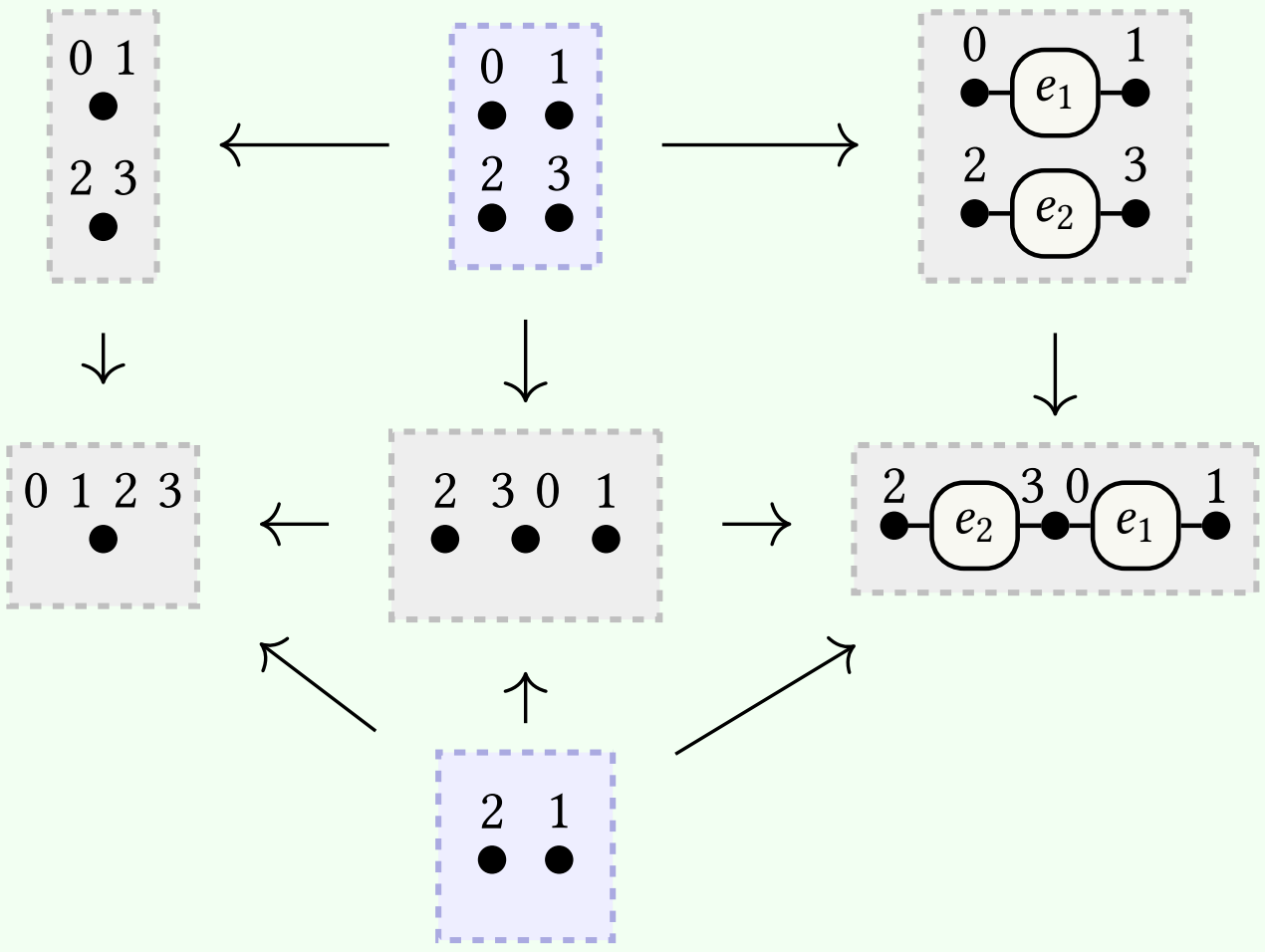

These two derivations arise through yanking:

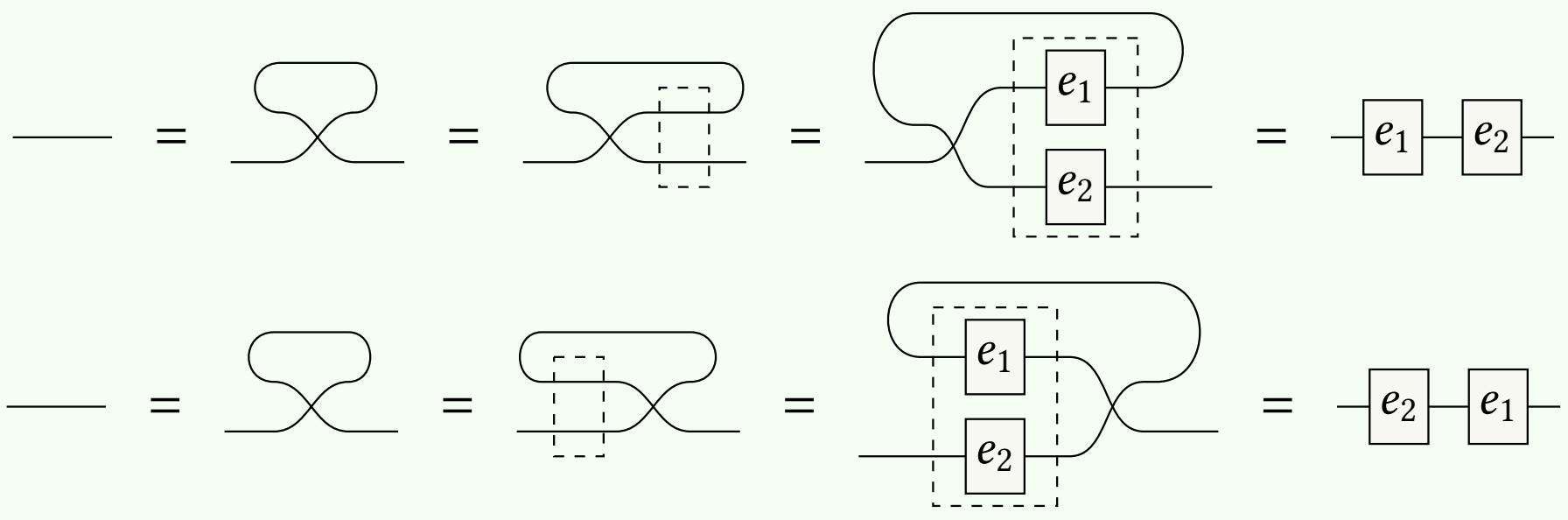

Another condition on symmetric monoidal graph rewriting is that the matching must be *convex*: any path between vertices must also be captured. Again thanks to yanking, this is not necessary in the traced case.

**Example 9.43.** Consider the rule $\left\langle\right.$ 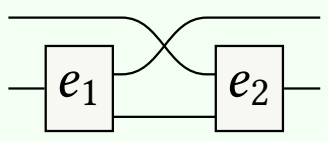 , 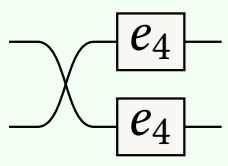 $\left.\right\rangle$ and the term 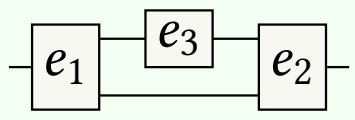 . Although it is not immediately obvious, there is in fact a matching of the former in the latter. Performing the DPO procedure yields the following:

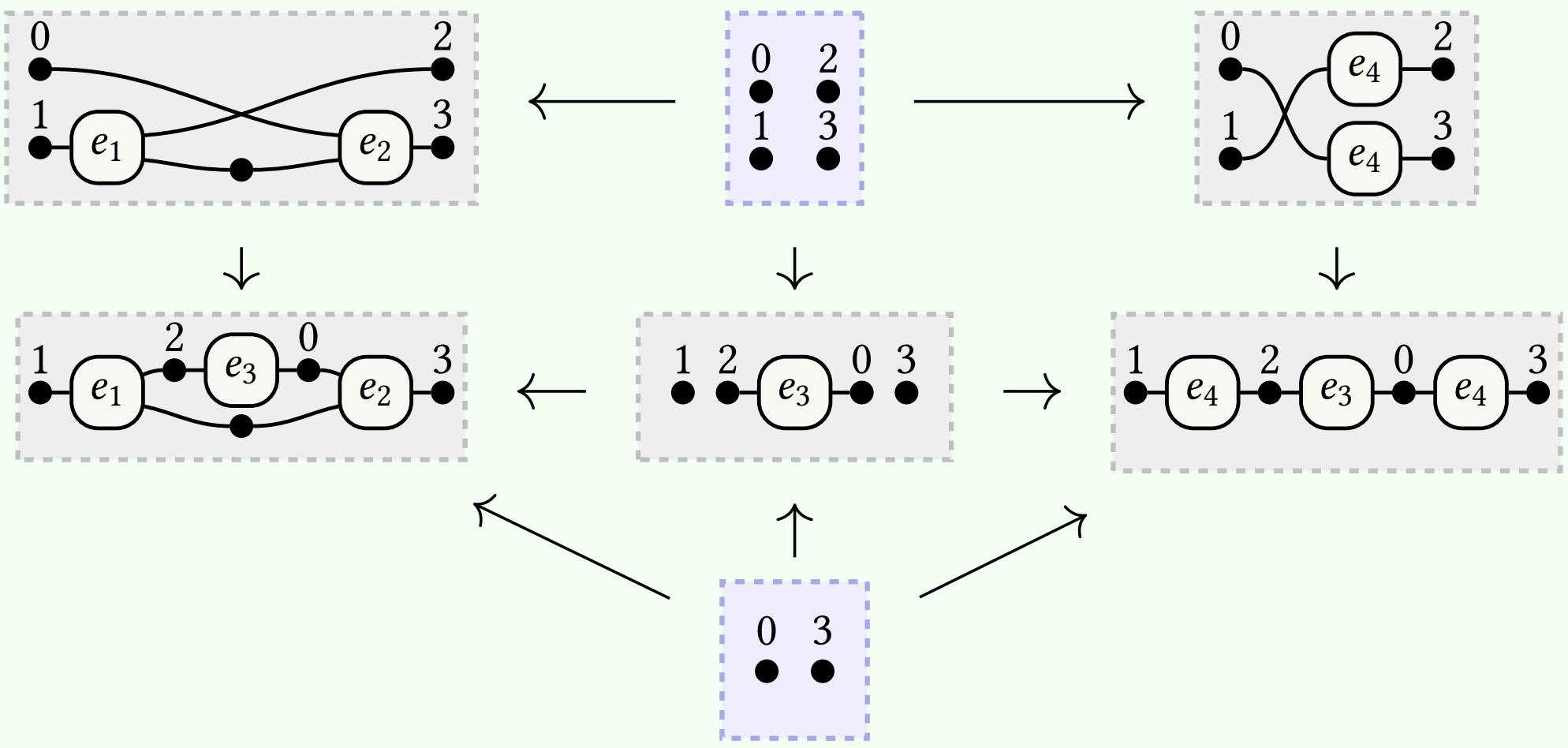

In a non-traced setting this is an invalid rule, but it is possible with yanking.

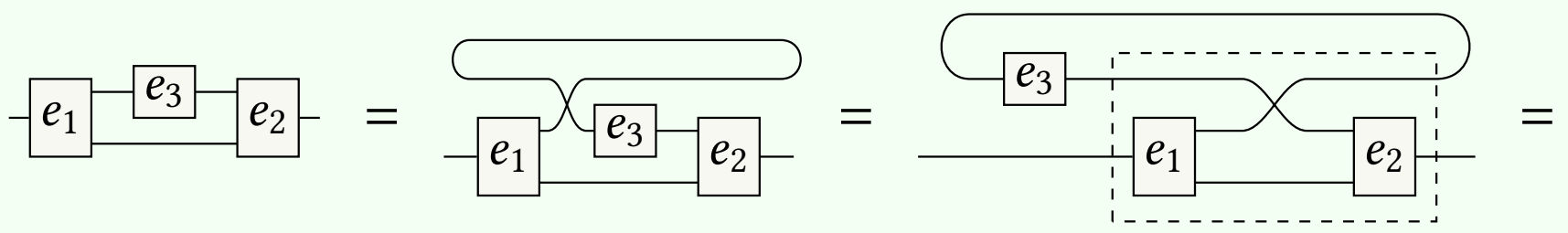

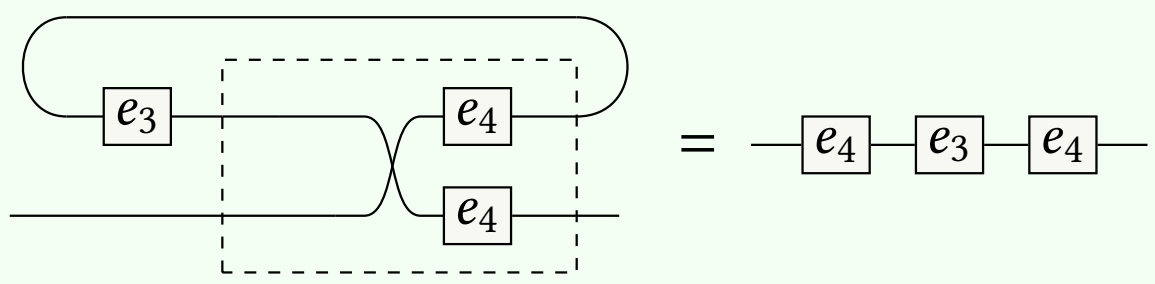

For traced DPO to be sound, the rewritten graph must correspond to a traced term. First we prove a lemma to show how using the compact closed structure of $\mathbf{S}_\Sigma + \mathbf{Frob}$ to reorganise interfaces corresponds to switching the cospan maps in $\mathsf{Csp}_D(\mathbf{Hyp}_\Sigma)$.

**Lemma 9.44.** Let 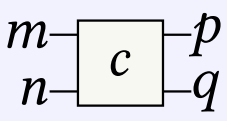 be a term in $\mathbf{S}_\Sigma + \mathbf{Frob}$. Then if $\langle\!\langle$ 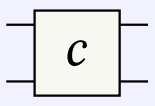 $\rangle\!\rangle_{C,\Sigma} = m + n \xrightarrow{[f_1, f_2]} F \xleftarrow{[g_1, g_2]} p + q$ then $\langle\!\langle$ 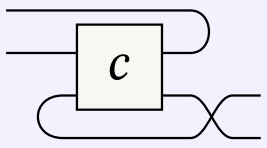 $\rangle\!\rangle_{C,\Sigma} = p + m \xrightarrow{[g_1, f_1]} F \xleftarrow{[f_2, g_2]} n + q$.

*Proof.* By definition of cups and caps in $\mathsf{Csp}_D(\mathbf{Hyp}_\Sigma)$. □

We need to show that rewriting a term with a rule $\left\langle\right.$ 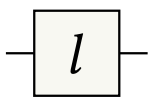 , 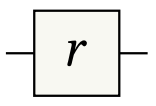 $\left.\right\rangle$ coincides with graph rewriting on the hypergraph interpretations of this rule.

**Notation 9.45.** For a traced rule $\left\langle\, i\!-\!\boxed{l}\!-\!j\,,\; i\!-\!\boxed{r}\!-\!j \,\right\rangle \in \mathbf{T}_\Sigma$, its DPO rule is defined as $\langle\!\langle \left\lfloor \left\langle -\boxed{l}-\,,\,-\boxed{r}- \right\rangle \right\rfloor_\Sigma^{\mathbf{T}} \rangle\!\rangle_\Sigma \coloneqq \langle\!\langle \ulcorner \left\lfloor -\boxed{l}- \right\rfloor_\Sigma^{\mathbf{T}} \urcorner \rangle\!\rangle_\Sigma \leftarrow i+j \rightarrow \langle\!\langle \ulcorner \left\lfloor -\boxed{r}- \right\rfloor_\Sigma^{\mathbf{T}} \urcorner \rangle\!\rangle_\Sigma$.

**Theorem 9.46.** For a rule $r \in \mathbf{T}_\Sigma$, we have that $-\boxed{g}- \Rightarrow_r -\boxed{h}-$ if and only if $\langle\!\langle \ulcorner \left\lfloor -\boxed{g}- \right\rfloor_\Sigma^{\mathbf{T}} \urcorner \rangle\!\rangle_\Sigma \rightsquigarrow_{\langle\!\langle \lfloor r \rfloor_\Sigma^{\mathbf{T}} \rangle\!\rangle_\Sigma} \langle\!\langle \ulcorner \left\lfloor -\boxed{h}- \right\rfloor_\Sigma^{\mathbf{T}} \urcorner \rangle\!\rangle_\Sigma$.

*Proof.* First the ($\Rightarrow$) direction. If $-\boxed{g}- \Rightarrow_{\mathcal{R}} -\boxed{h}-$ then we have $-\boxed{g}- =$ 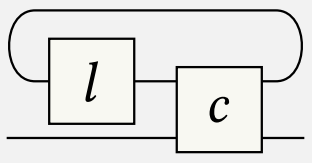 and 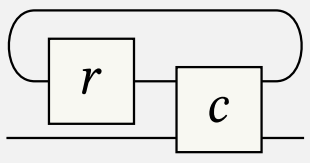 $= -\boxed{h}-$ ; we must derive the DPO diagram in $\mathbf{Hyp}_\Sigma$. First we give names to the following cospans:

$$0 \rightarrow L \leftarrow i+j \coloneqq \langle\!\langle \ulcorner \left\lfloor -\boxed{l}- \right\rfloor_\Sigma^{\mathbf{T}} \urcorner \rangle\!\rangle_\Sigma = \langle\!\langle$$ 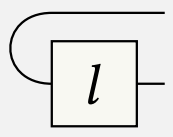 $$\rangle\!\rangle_\Sigma$$

$$0 \rightarrow R \leftarrow i+j \coloneqq \langle\!\langle \ulcorner \left\lfloor -\boxed{r}- \right\rfloor_\Sigma^{\mathbf{T}} \urcorner \rangle\!\rangle_\Sigma = \langle\!\langle$$ 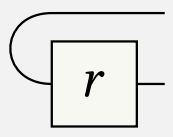 $$\rangle\!\rangle_\Sigma$$

$$0 \rightarrow G \leftarrow m+n \coloneqq \langle\!\langle \ulcorner \left\lfloor -\boxed{g}- \right\rfloor_\Sigma^{\mathbf{T}} \urcorner \rangle\!\rangle_\Sigma = \langle\!\langle$$ 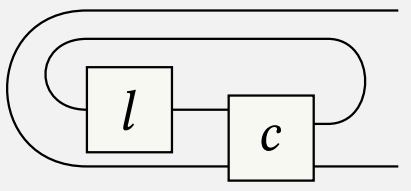 $$\rangle\!\rangle_\Sigma$$

$$0 \rightarrow H \leftarrow m+n \coloneqq \langle\!\langle \ulcorner \left\lfloor -\boxed{h}- \right\rfloor_\Sigma^{\mathbf{T}} \urcorner \rangle\!\rangle_\Sigma = \langle\!\langle$$ 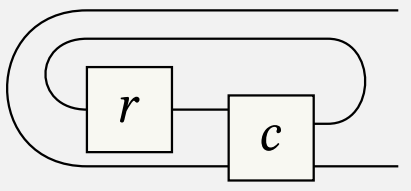 $$\rangle\!\rangle_\Sigma$$

Moving into $\mathbf{S}_\Sigma + \mathbf{Frob}$, we have that 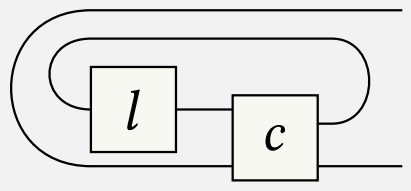 $=$ 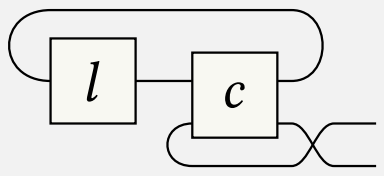 ; so by functoriality $\langle\!\langle \ulcorner \left\lfloor -\boxed{g}- \right\rfloor_\Sigma^{\mathbf{T}} \urcorner \rangle\!\rangle_\Sigma = \langle\!\langle$ 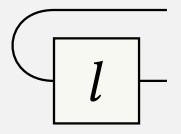 $\rangle\!\rangle_\Sigma \,;\, \langle\!\langle$ 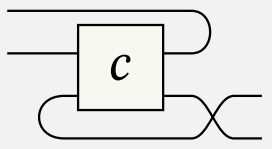 $\rangle\!\rangle_\Sigma$, i.e. $0 \rightarrow G \leftarrow m+n = 0 \rightarrow L \leftarrow i+j \,;\, i+j \rightarrow C \leftarrow m+n$. Cospan composition is by pushout, so $L \rightarrow G \leftarrow C$ is a pushout. Using the same reasoning, $R \rightarrow G \leftarrow C$ is also a pushout; this gives us the DPO diagram. All that remains is to check that the aforementioned pushouts are traced boundary complements; this follows by Lemma 9.44 as $\langle\!\langle \left\lfloor \boxed{c} \right\rfloor_\Sigma \rangle\!\rangle_\Sigma$ is partial monogamous.

Now for the ($\Leftarrow$) direction: we assume we have a traced DPO rewrite, so there exist cospans $0 \rightarrow L \leftarrow i+j, 0 \rightarrow R \leftarrow i+j, i+j \rightarrow C \leftarrow m+n$ as above such that the DPO diagram commutes and $i+j \rightarrow C \rightarrow G$ is a traced boundary complement. We must show that $-\boxed{g}- =$ 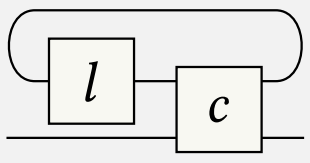 and $-\boxed{h}- =$ 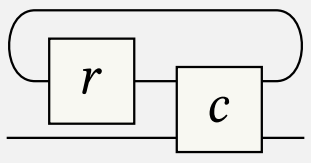 .

We have that $0 \rightarrow G \leftarrow m+n = 0 \rightarrow L \leftarrow i+j \,;\, i+j \xrightarrow{[c_1,c_2]} C \xleftarrow{[d_1,d_2]} m+n$ as cospan composition is by pushout. Let $\begin{smallmatrix} i \\ j \end{smallmatrix}\!\boxed{c'}\!\begin{smallmatrix} m \\ n \end{smallmatrix}$ be the term in $\mathbf{S}_\Sigma + \mathbf{Frob}$ such that $\langle\!\langle \boxed{c'} \rangle\!\rangle_\Sigma = i + j \xrightarrow{[c_1,c_2]} C \xleftarrow{[d_1,d_2]} m + n$, which exists as $\langle\!\langle - \rangle\!\rangle_\Sigma$ is full.

The cospan $j + m \xrightarrow{[c_2,d_1]} C \xleftarrow{[c_1,d_2]} i + n$ is partial monogamous because $i + j \to C \to G$ is a traced boundary complement. Let ${}^{j}_{m}\boxed{c}^{i}_{n}$ be the term in $\mathbf{S}_\Sigma + \mathbf{Frob}$ such that $\langle\!\langle \boxed{c} \rangle\!\rangle_\Sigma = j + m \xrightarrow{[c_2,d_1]} C \xleftarrow{[c_1,d_2]} i + n$. Using Lemma 9.44, we have that $\langle\!\langle$ 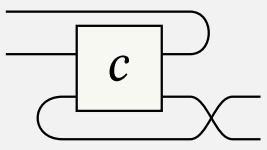 $\rangle\!\rangle_\Sigma = i + j \xrightarrow{[c_1,c_2]} C \xleftarrow{[d_1,d_2]} m + n$.

So we have that $\langle\!\langle \ulcorner \boxed{g} \urcorner \rangle\!\rangle_\Sigma = \langle\!\langle \ulcorner \boxed{l} \urcorner \rangle\!\rangle_\Sigma \,;\, \langle\!\langle \boxed{c'} \rangle\!\rangle_\Sigma$; by fullness we derive that 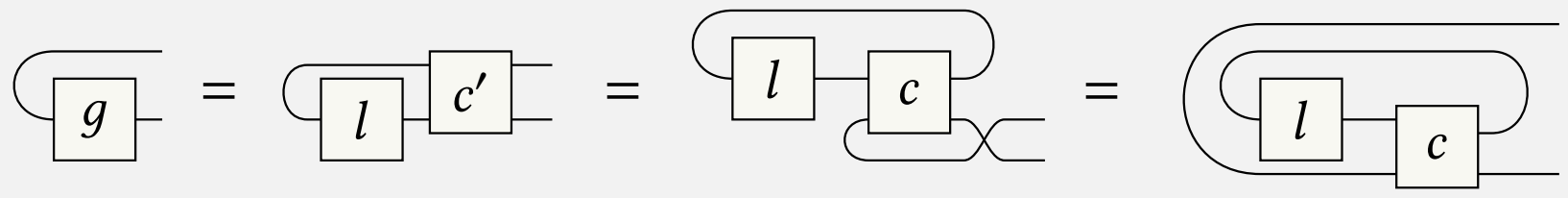. This means that $\ulcorner \boxed{g} \urcorner =$ 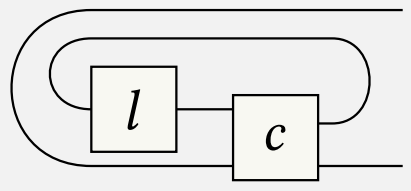 so 'unfolding' the interface gives us $\boxed{g} =$ 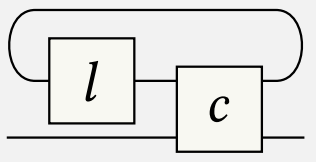. Since $\langle\!\langle \boxed{c} \rangle\!\rangle_\Sigma$ is partial monogamous, $\boxed{c}$ is in $\mathbf{T}_\Sigma$. As the trace in $\mathbf{T}_\Sigma$ is the canonical trace, the entire term is in $\mathbf{T}_\Sigma$, completing the proof. The same procedure holds for rewriting from the other direction. □

This gives us a sound and complete graph rewriting system for terms in $\mathbf{T}_\Sigma$, and can be generalised to the coloured setting as well.

**Notation 9.47.** For a rewrite rule $\left\langle \bar{i}\boxed{l}\bar{j}, \bar{i}\boxed{r}\bar{j} \right\rangle \in \mathbf{T}_{C,\Sigma}$, its DPO rule is $\langle\!\langle \left\lfloor \left\langle \boxed{l}, \boxed{r} \right\rangle \right\rfloor^{\mathbf{T}}_{C,\Sigma} \rangle\!\rangle_{C,\Sigma}$, defined as

$$\langle\!\langle \ulcorner \left\lfloor \boxed{l} \right\rfloor^{\mathbf{T}}_{C,\Sigma} \urcorner_C \rangle\!\rangle_{C,\Sigma} \leftarrow \overline{ij} \rightarrow \langle\!\langle \ulcorner \left\lfloor \boxed{r} \right\rfloor^{\mathbf{T}}_{C,\Sigma} \urcorner_C \rangle\!\rangle_{C,\Sigma}.$$

**Theorem 9.48.** For a rule $r \in \mathbf{T}_\Sigma$, we have that $\boxed{g} \Rightarrow_r \boxed{h}$ if and only if $\langle\!\langle \ulcorner \left\lfloor \boxed{g} \right\rfloor^{\mathbf{T}}_{C,\Sigma} \urcorner_C \rangle\!\rangle_{C,\Sigma} \rightsquigarrow_{\langle\!\langle \lfloor r \rfloor^{\mathbf{T}}_{C,\Sigma} \rangle\!\rangle_{C,\Sigma}} \langle\!\langle \ulcorner \left\lfloor \boxed{g} \right\rfloor^{\mathbf{T}}_{C,\Sigma} \urcorner_C \rangle\!\rangle_{C,\Sigma}$.

## 9.4 Rewriting with commutative comonoid structure

Hypergraphs are a good fit for rewriting terms in $\mathbf{S}_\Sigma + \mathbf{Frob}$ because they allow rewriting modulo *Frobenius structure*. We can take advantage of this to rewrite modulo *cocommutative comonoid* structure on top of the trace.

**Definition 9.49** (Traced left-boundary complement)**.** For partial left-monogamous cospans $i \xrightarrow{a_1} L \xleftarrow{a_2} j$ and $n \xrightarrow{b_1} G \xleftarrow{b_2} m \in \mathbf{Hyp}_\Sigma$, a pushout complement as in Definition 9.38 is called a *traced left-boundary complement* if $c_2$ is mono and $j + m \xrightarrow{[c_2,d_1]} C \xleftarrow{[c_1,d_2]} i + n$ is a partial left-monogamous cospan.

**Definition 9.50** (Traced comonoid DPO)**.** For morphisms $G \leftarrow m + n$ and $H \leftarrow m + n$ in $\mathbf{Hyp}_\Sigma$, there is a traced comonoid rewrite $G \rightsquigarrow_{\mathcal{R}} H$ if there exists a rule $L \leftarrow i + j \rightarrow G \in \mathcal{R}$ and cospan $i + j \rightarrow C \leftarrow m + n \in \mathbf{Hyp}_\Sigma$ such that diagram in Definition 9.7 commutes and $i + j \rightarrow C \rightarrow G$ is a traced left-boundary complement.

Just like traced rewriting, there may be multiple traced left-boundary complements, which arise because we are now absorbing the equations of a cocommutative comonoid.

**Example 9.51.** Consider the following rule and its interpretation.

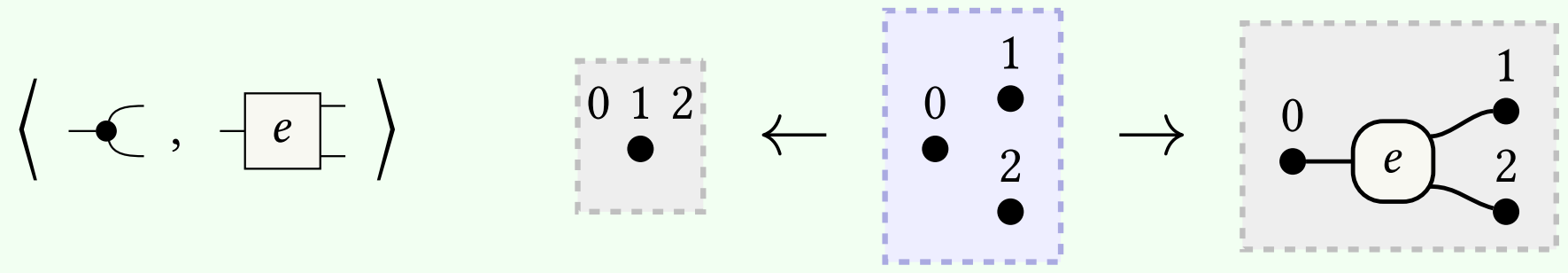

Two valid rewrites of this rule in 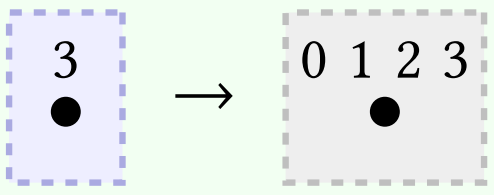 are:

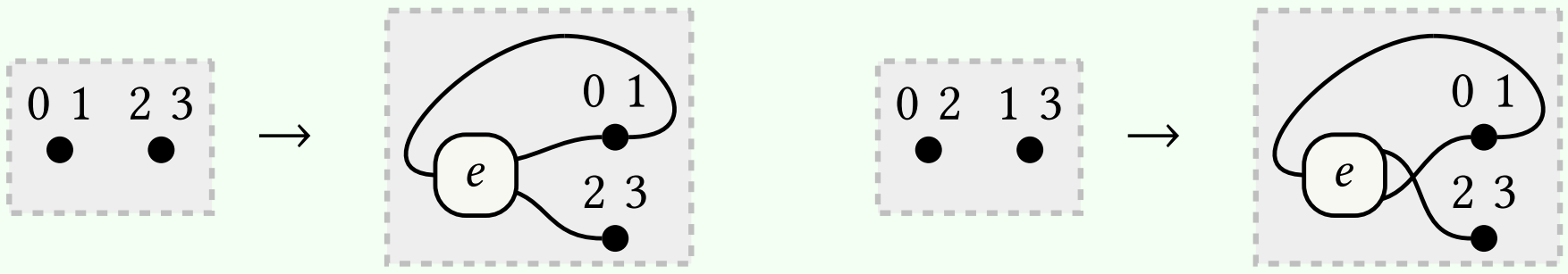

The first rewrite is the 'obvious' one, but the second also holds by cocommutativity:

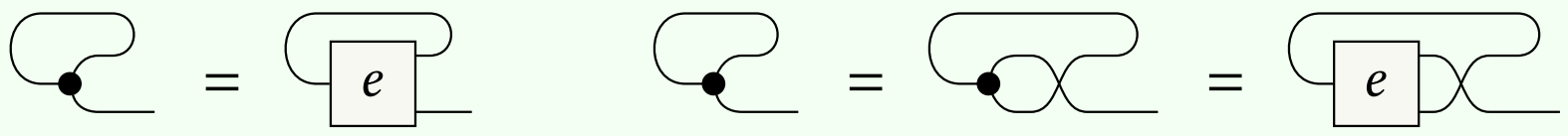

To show the soundness and completeness of traced comonoid rewriting, we follow the same procedure as in the traced setting.

**Notation 9.52.** For a rule $\left\langle i\text{-}\boxed{l}\text{-}j,\ i\text{-}\boxed{r}\text{-}j \right\rangle \in \mathbf{T}_\Sigma + \mathbf{CComon}$, its traced comonoid DPO rule is $\langle\!\langle \left\lfloor \left\langle -\boxed{l}-,\ -\boxed{r}- \right\rangle \right\rfloor_\Sigma \rangle\!\rangle_\Sigma$, defined as

$$\langle\!\langle \ulcorner \left\lfloor -\boxed{l}- \right\rfloor_\Sigma \urcorner \rangle\!\rangle_\Sigma \leftarrow i + j \rightarrow \langle\!\langle \ulcorner \left\lfloor -\boxed{r}- \right\rfloor_\Sigma \urcorner \rangle\!\rangle_\Sigma.$$

**Theorem 9.53.** For a rewrite rule $r \in \mathbf{T}_\Sigma + \mathbf{CComon}$, we have that $-\boxed{g}- \Rightarrow_r -\boxed{h}-$ if and only if $\langle\!\langle \ulcorner \left\lfloor -\boxed{g}- \right\rfloor_\Sigma \urcorner \rangle\!\rangle_\Sigma \rightsquigarrow_{\langle\!\langle \lfloor r \rfloor_\Sigma \rangle\!\rangle_\Sigma} \langle\!\langle \ulcorner \left\lfloor -\boxed{g}- \right\rfloor_\Sigma \urcorner \rangle\!\rangle_\Sigma$.

*Proof.* As Theorem 9.46, but with partial left-monogamy and traced left-boundary complements. □

This means that we can also safely perform rewriting modulo traced comonoid structure by building on the machinery used for the traced case. Predictably, the same also holds for the coloured setting.

**Notation 9.54.** For a rule $\left\langle \bar{i}-\boxed{l}-\bar{j}, \bar{i}-\boxed{r}-\bar{j} \right\rangle \in \mathbf{T}_{C,\Sigma} + \mathbf{CComon}_C$, its DPO rule is $\langle\!\langle \left\lfloor \left\langle -\boxed{l}-, -\boxed{r}- \right\rangle \right\rfloor_{C,\Sigma} \rangle\!\rangle_{C,\Sigma}$, defined as

$$\langle\!\langle \ulcorner \left\lfloor -\boxed{l}- \right\rfloor_{C,\Sigma} \urcorner_C \rangle\!\rangle_{C,\Sigma} \leftarrow \overline{ij} \rightarrow \langle\!\langle \ulcorner \left\lfloor -\boxed{r}- \right\rfloor_{C,\Sigma} \urcorner_C \rangle\!\rangle_{C,\Sigma}.$$

**Theorem 9.55.** For a rewrite rule $r \in \mathbf{T}_{C,\Sigma} + \mathbf{CComon}_C$, we have $-\boxed{g}- \Rightarrow_r -\boxed{h}-$ if and only if $\langle\!\langle \ulcorner \left\lfloor -\boxed{g}- \right\rfloor_{C,\Sigma} \urcorner_C \rangle\!\rangle_{C,\Sigma} \rightsquigarrow_{\langle\!\langle \lfloor r \rfloor_{C,\Sigma} \rangle\!\rangle_{C,\Sigma}} \langle\!\langle \ulcorner -\boxed{g}- \urcorner_C \rangle\!\rangle_{C,\Sigma}$.

CHAPTER 10

# Applications of graph rewriting

We can now rewrite traced string diagrams modulo *traced comonoid* structure, which can be applied to several areas across computed science. In this chapter, we will examine how the rewriting framework can be applied to settings with a *traced Cartesian* (dataflow) structure, and to our intended application of sequential circuits.

## 10.1 Cartesian structure

One important class of categories with a traced comonoid structure are *traced Cartesian*, or *dataflow*, categories [CŞ90]. Recall from Definition 2.102 that the tensor in a Cartesian category is given by the category-theoretic product. We can equivalently view this as a category in which each object is equipped with a commutative comonoid structure in which the comonoid and counit are *natural*: morphisms can be 'pushed' through them.

**Theorem 10.1** ([Fox76])**.** A category $\mathcal{C}$ equipped with a commutative comonoid structure is Cartesian if and only if the equations in Figure 10.1 hold for all $m$–$\boxed{f}$–$n$ .

Sequential circuits have a natural notion of copying and discarding data, so it makes sense that the semantic categories of circuits should be Cartesian. In [GJ16], the equational theory is used to show that this is the case, but with the stream semantics we have a much more elementary proof.

**Theorem 10.2.** $\mathbf{Stream}_{\mathcal{I}}$ is Cartesian.

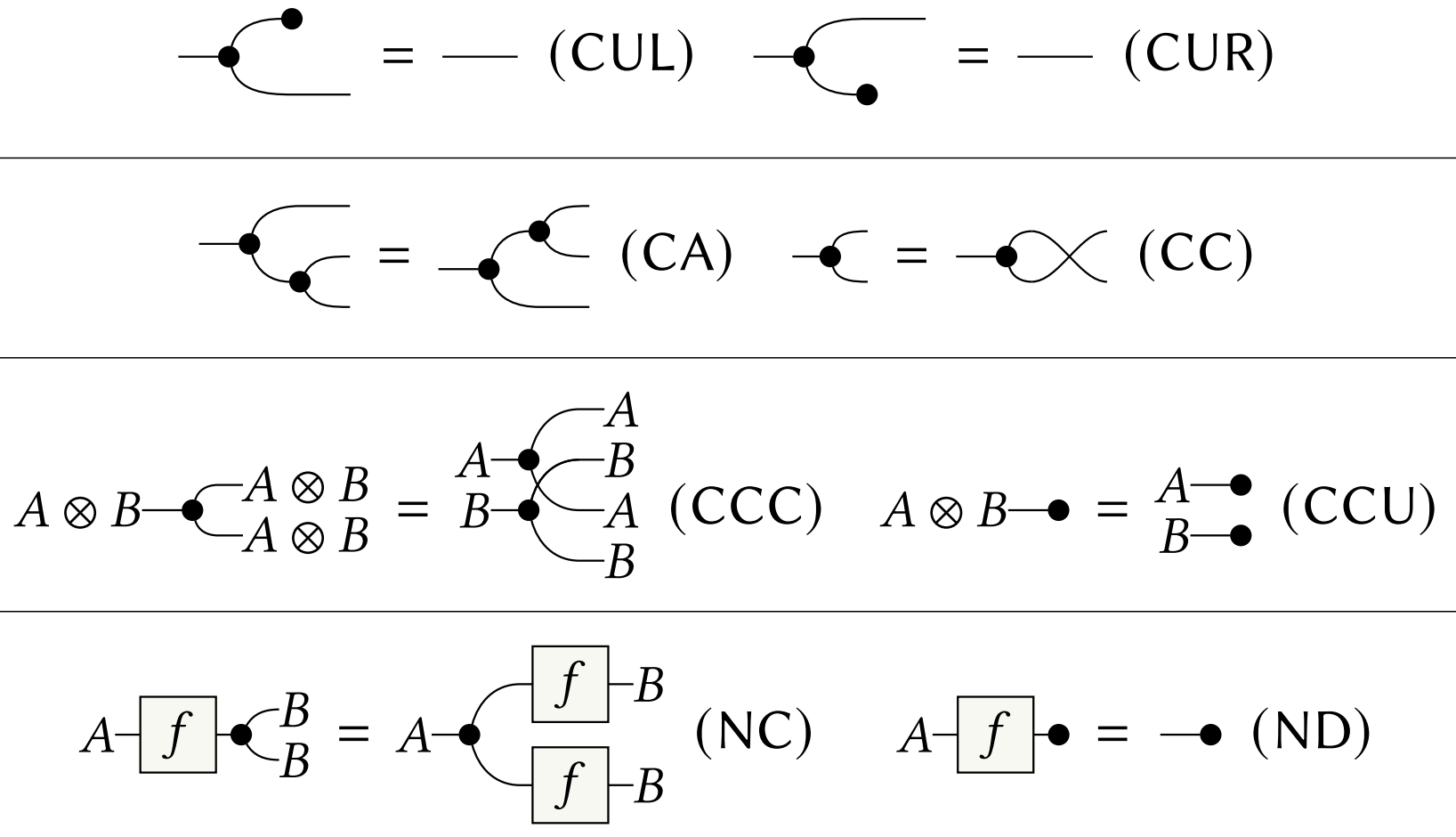

Figure 10.1: Equations that hold in any Cartesian category

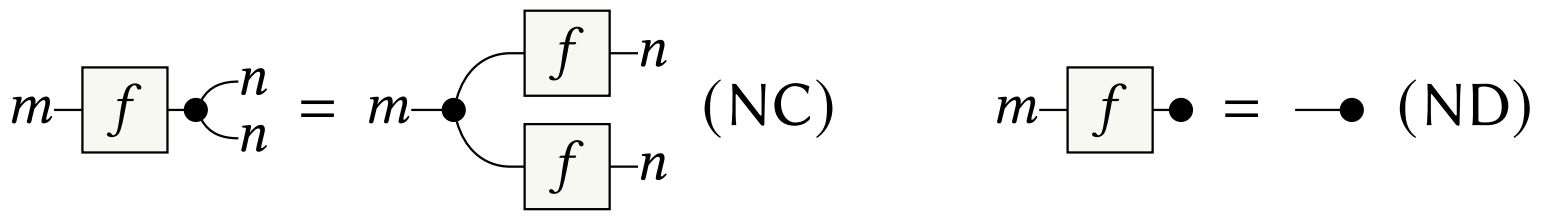

Figure 10.2: Equations of the monoidal theory **Cart**, for generator $f$

*Proof.* The tensor in $\mathbf{Stream}_{\mathcal{I}}$ is defined to be the Cartesian product. □

As the three semantic categories are isomorphic to $\mathbf{Stream}_{\mathcal{I}}$ they are also Cartesian.

**Corollary 10.3.** $\mathbf{SCirc}_{\Sigma/\approx_{\mathcal{I}}}$, $\mathbf{SCirc}_{\Sigma/\sim_{\mathcal{I}}}$ and $\mathbf{SCirc}_{\Sigma/\mathcal{E}_{\mathcal{I}}}$ are Cartesian.

We can express the data of a Cartesian category as an extension of the monoidal theory of commutative comonoids.

**Definition 10.4.** The Cartesian monoidal theory $(\Sigma_{\mathbf{Cart}}, \mathcal{E}_{\mathbf{Cart}})$ is defined as $\Sigma_{\mathbf{Cart}} \coloneqq \Sigma_{\mathbf{CComon}}$ and $\mathcal{E}_{\mathbf{Cart}} \coloneqq \Sigma_{\mathbf{CComon}} + (\mathrm{NC}) + (\mathrm{ND})$ where (NC) and (ND) are defined as in Figure 10.2.

**Remark 10.5.** Note that we do not need the two 'coherence' Cartesian equations when considering monoidal theories, because they follow immediately from the construction of multiple-bit structures (Notation 3.8).

The hypergraph interpretations of these rules are shown in Figure 10.3.

**Remark 10.6.** The combination of Cartesian equations with the underlying compact closed structure of $\mathrm{Csp}_D(\mathbf{Hyp}_\Sigma)$ may prompt alarm bells, as a compact closed category in which the tensor is the Cartesian product is trivial. However, $\mathrm{Csp}_D(\mathbf{Hyp}_\Sigma)$ is *not* subject to these equations: it is only a setting for performing graph rewrites.

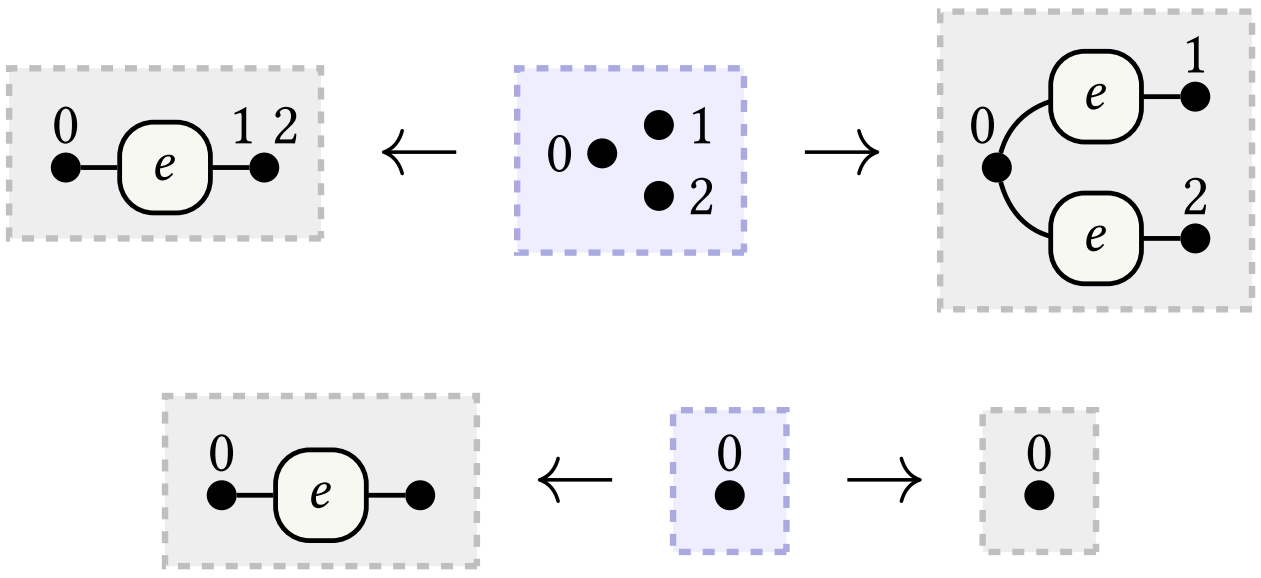

Figure 10.3: Interpretations of equations in **Cart** for generator $e$.

Using hypergraphs to reason about Cartesian categories is appealing because one can focus on applying the two naturality equations (NC) and (ND). As a case study, we will consider *fixed point operators*; several equivalent axiomatisations exist [Has97; SP00], but we use the equations presented in [Has09].

**Definition 10.7** ([Has09])**.** Let $\mathcal{C}$ be a Cartesian category. A *Conway fixed point operator* in $\mathcal{C}$ is a family of functions $(-)^{\dagger_{A,X}} : \mathcal{C}(A \times X, X) \to \mathcal{C}(A, X)$ drawn as

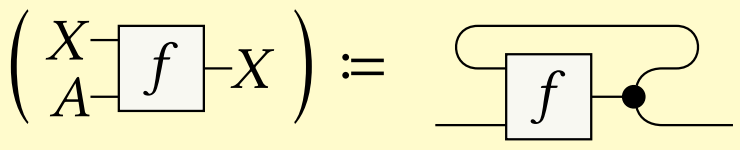

subject to the following equations:

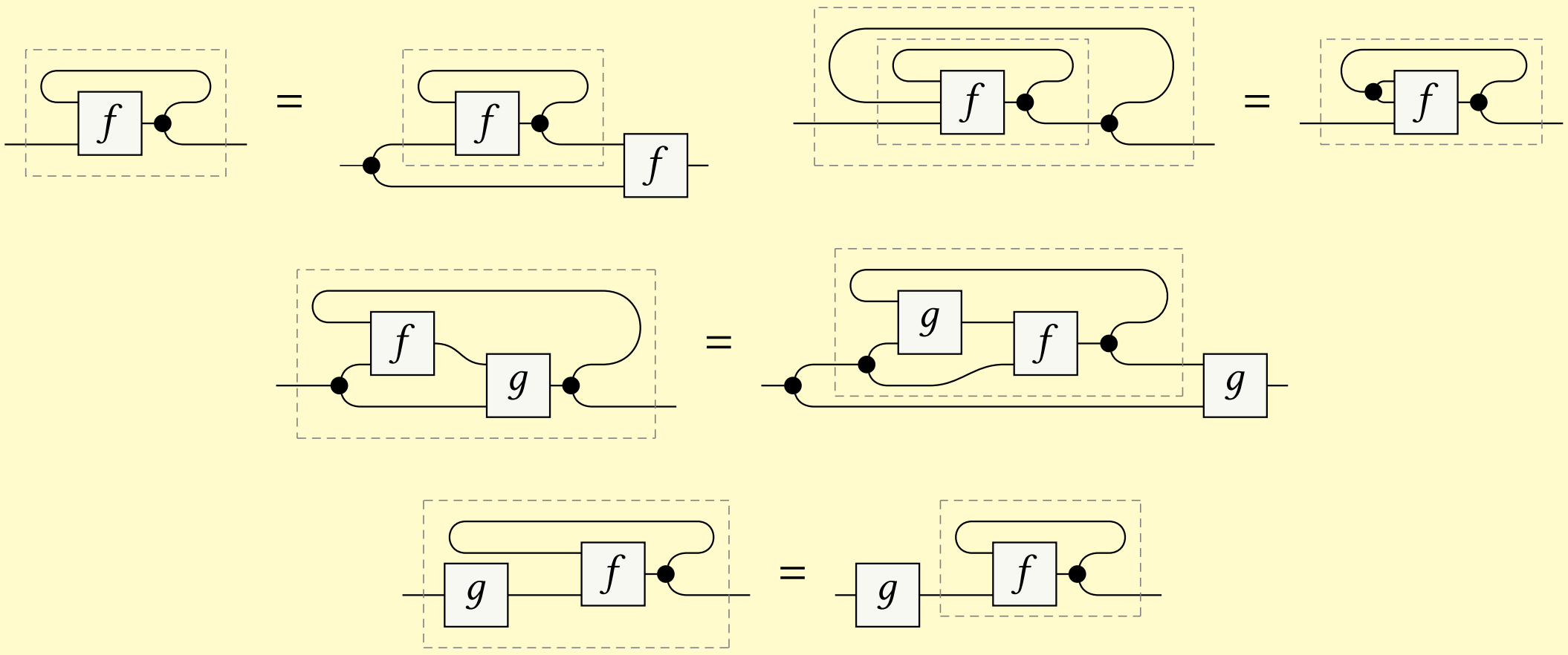

The notation we use for fixed point operators is already evocative of a trace, and the two are indeed equivalent. This was independently observed by Bloom and Ésik [BÉ93], Ştefănescu [Ste00], and Hyland, the former two before the notion of traced monoidal categories were formalised.

**Theorem 10.8** ([Has97, Thm. 3.1], [Has99]). A Cartesian category is traced if and only if it has a Conway operator.

*Proof.* This can be shown by constructing one operator using the other.

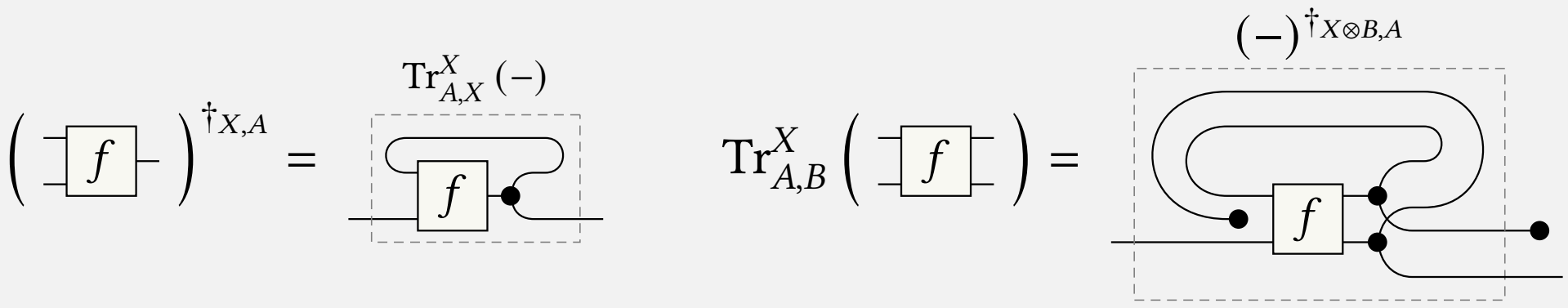

As these constructions are inverses, the conditions are equivalent. □

This means that we can reason about fixpoints using the same principles as reasoning with the trace and the comonoid structure, with the addition of the Cartesian equations.

**Example 10.9** (Unfolding). Reasoning about fixpoints in a traced category can be performed using the *unfolding* rule.

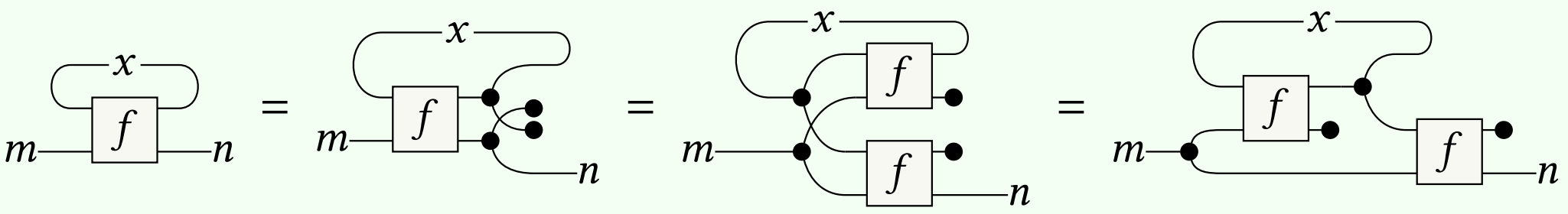

In the syntactic setting, this requires the application of multiple equations: the two counitality equations followed by the copy equation and optionally some axioms of STMCs for housekeeping. If we use the hypergraph interpretation, the comonoid equations are absorbed into the notation so only one rule needs to be applied.

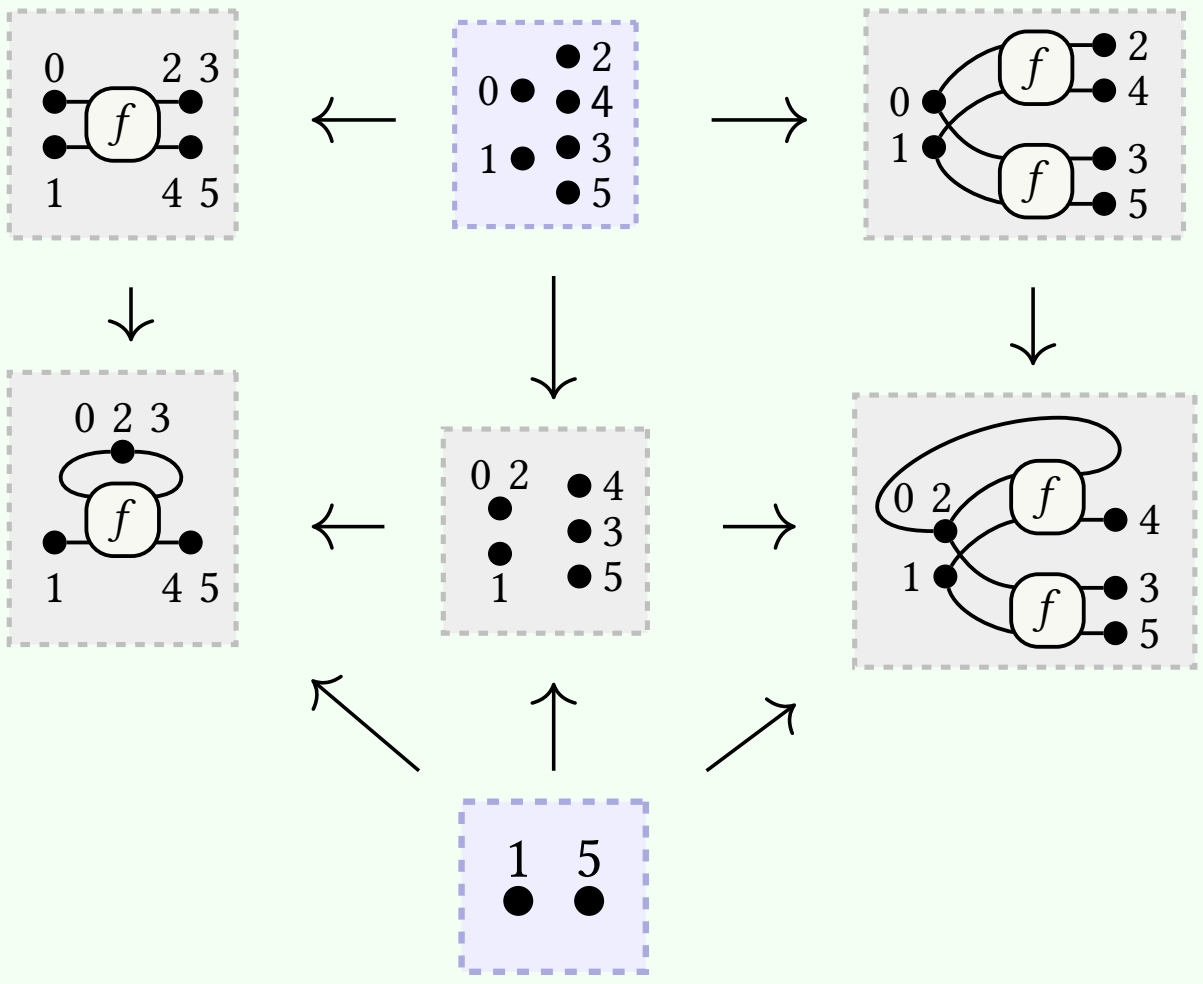

The dual notion of traced *cocartesian* categories [Bai76] are also important in computer science: a trace in a traced cocartesian category corresponds to *iteration* in the context of *control flow*. The details of this section could also be applied to the cocartesian case by flipping all the directions and working with partial *right*-monogamous cospans.

## 10.2 Sequential circuits

When performing traced graph rewriting using hypergraphs, any wires going from right-to-left are treated in exactly the same way as wires going from left-to-right: as tentacles connecting source and target vertices. In the operational semantics for sequential circuits, there are rules that operate especially on graphs of a certain form enclosed by a trace; while it would be sound to apply these arbitrarily (by virtue of the sliding axioms), for a *productive* procedure we need to be more guided. We will select particular tentacles as the 'chosen trace wires' (coloured red) for global transformations.

**Example 10.10.** Recall the SR NOR latch circuit from Example 3.18. This is interpreted as follows, where the $\wedge$, $\neg$ and $\delta$ edges respectively represent the OR gate, NOT gate, and delay.

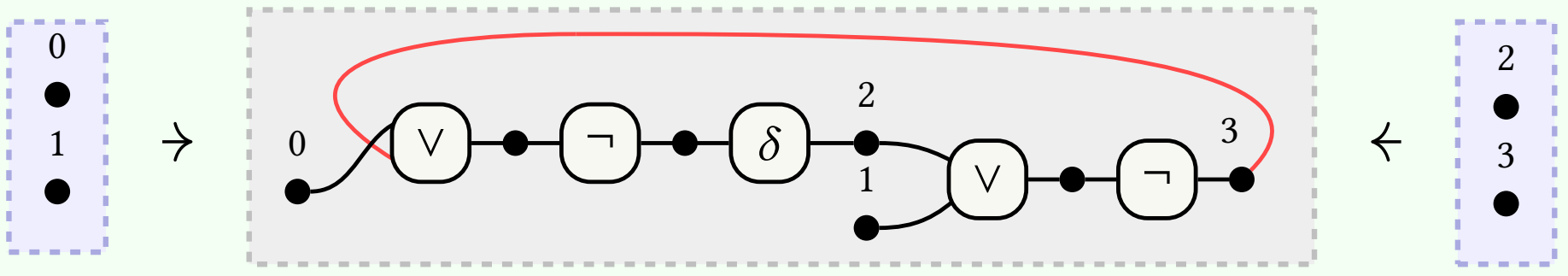

Since the transformation into global trace-delay form is through axioms of STMCs, as hypergraphs a circuit and its trace-delay form are isomorphic. The first rewrite that needs to be applied is the global Mealy reduction (Mealy).

**Example 10.11.** Applying the Mealy rewrite to Example 10.10 produces the following cospan of hypergraphs:

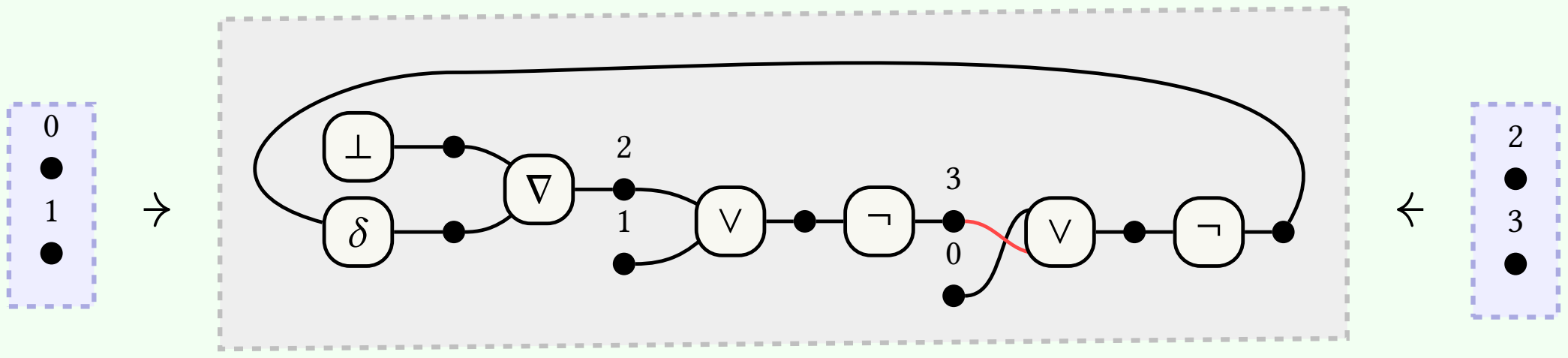

The instant feedback reduction can produce a complicated term with many forks; in the hypergraph representation, these forks are all absorbed into one.

**Example 10.12.** Below is an example showing how the instant feedback rewrite is applied to a circuit in Mealy form containing one generator $e$.

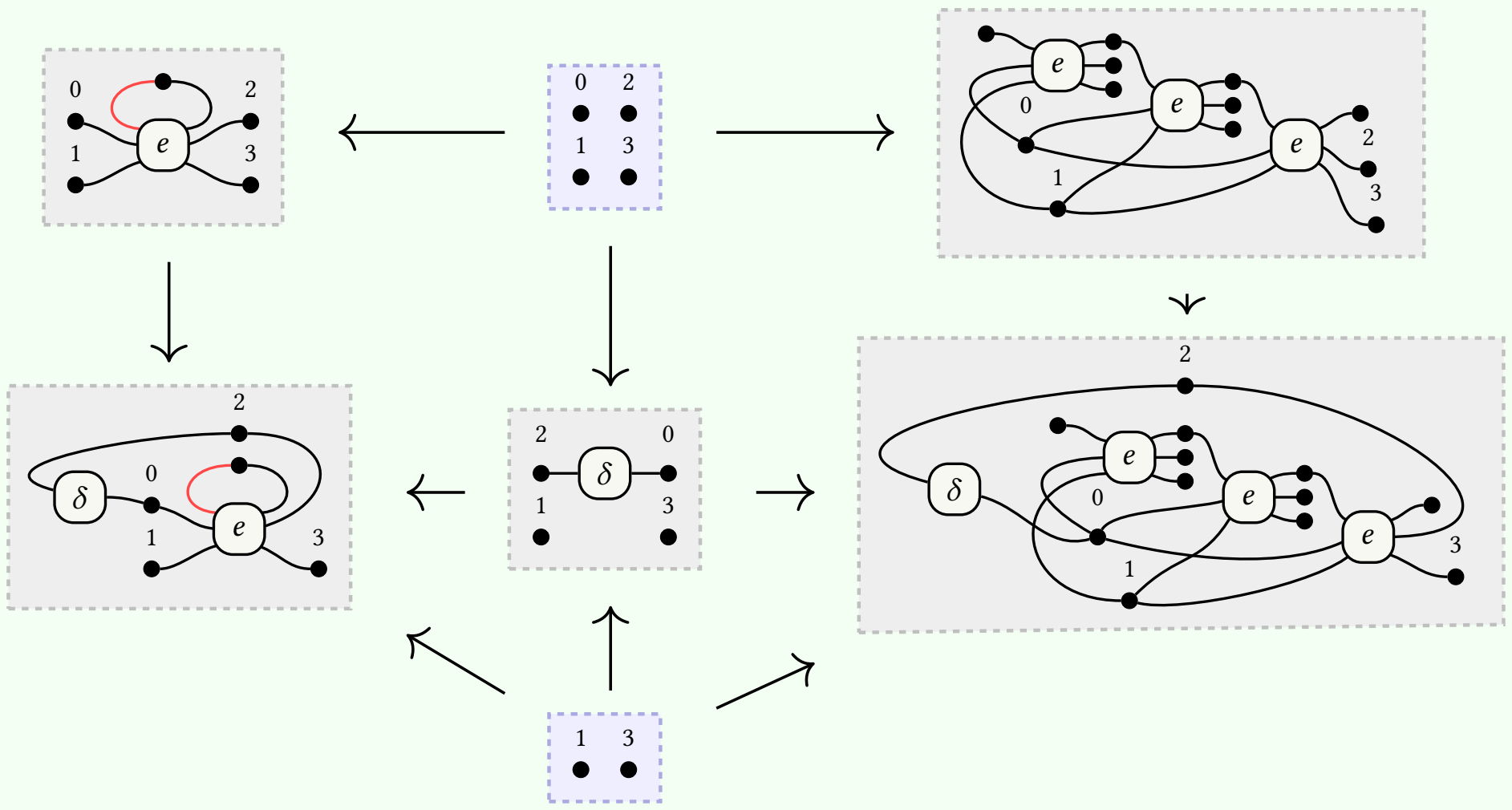

As the instant feedback rule eliminates non-delay-guarded feedback loops, there are no red tentacles in the right-hand side of the rule or the rewritten graph.

**Example 10.13.** The interpretation of the SR latch from Example 10.11 after being rewritten by the instant feedback rewrite is shown below.

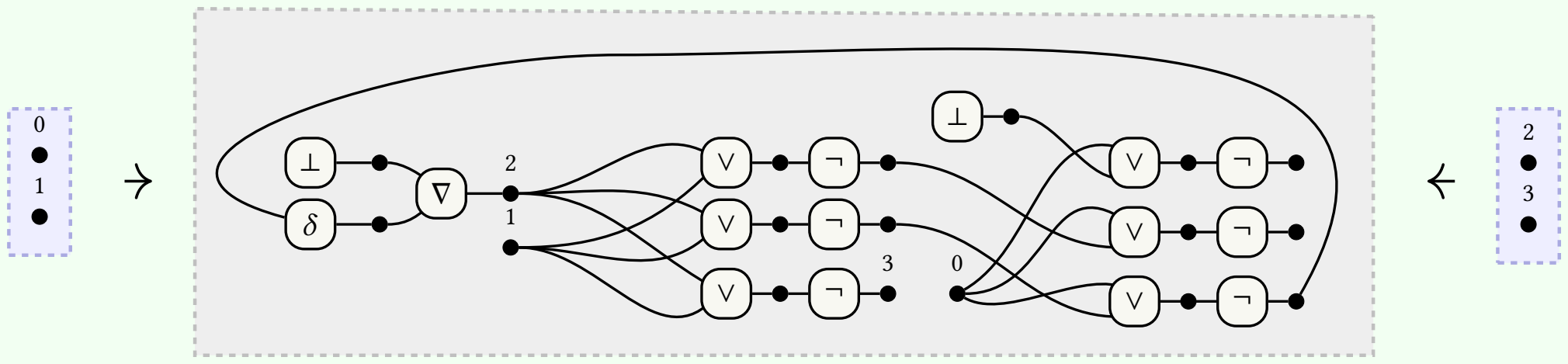

After performing the instant feedback rewrite, the graph is ready to receive inputs.

**Example 10.14.** We apply the inputs tf to the prepared SR latch hypergraph from Example 10.13 by precomposing some value registers.

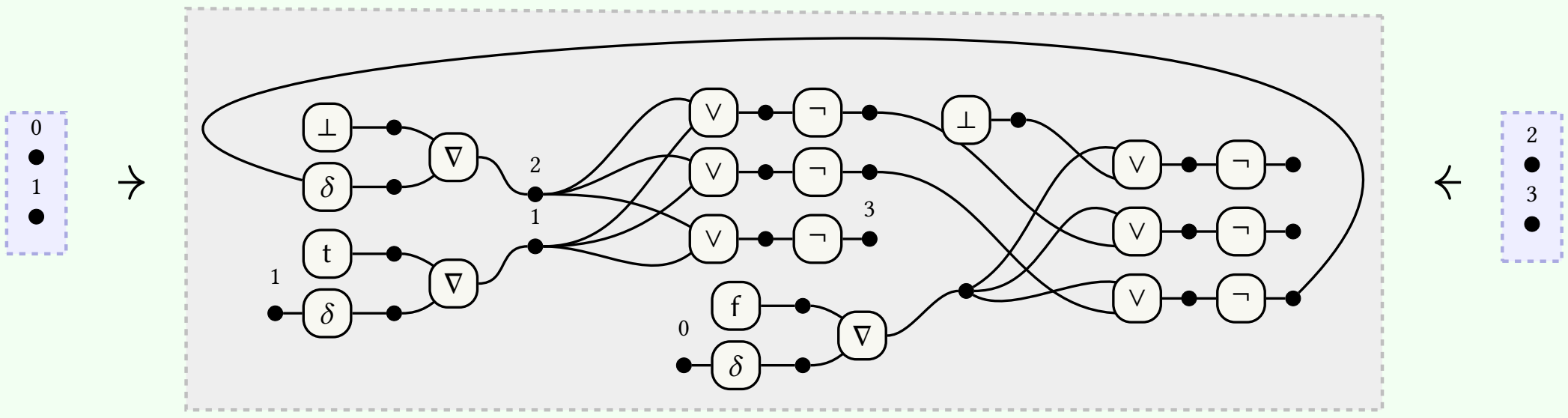

This graph is then rewritten by the streaming rule.

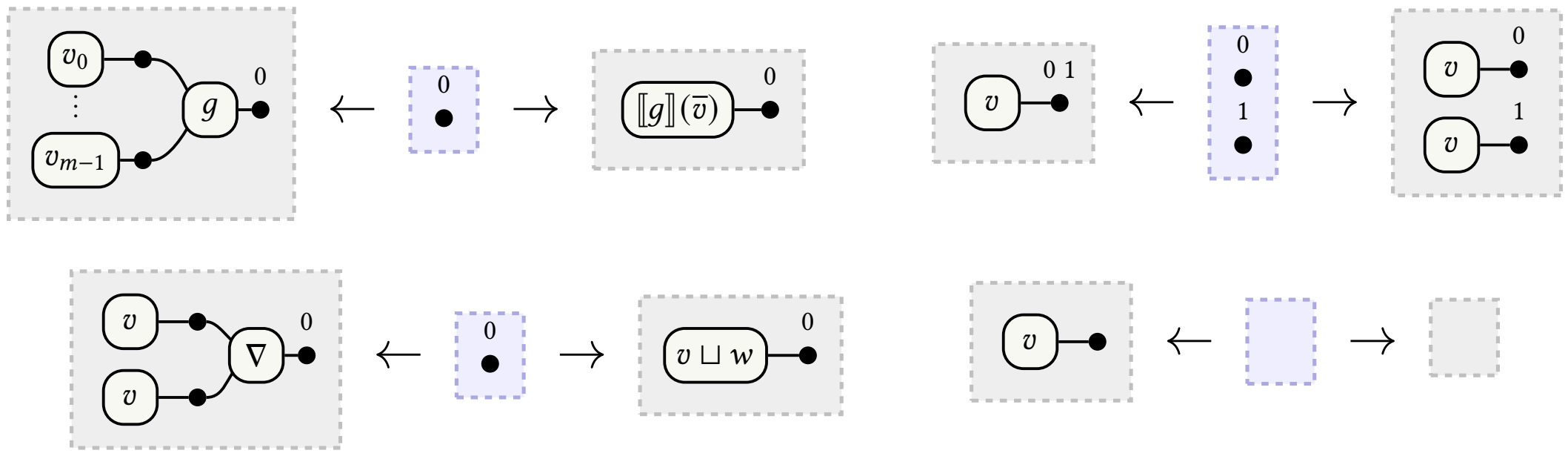

Figure 10.4: Hypergraph interpretations of the value rules

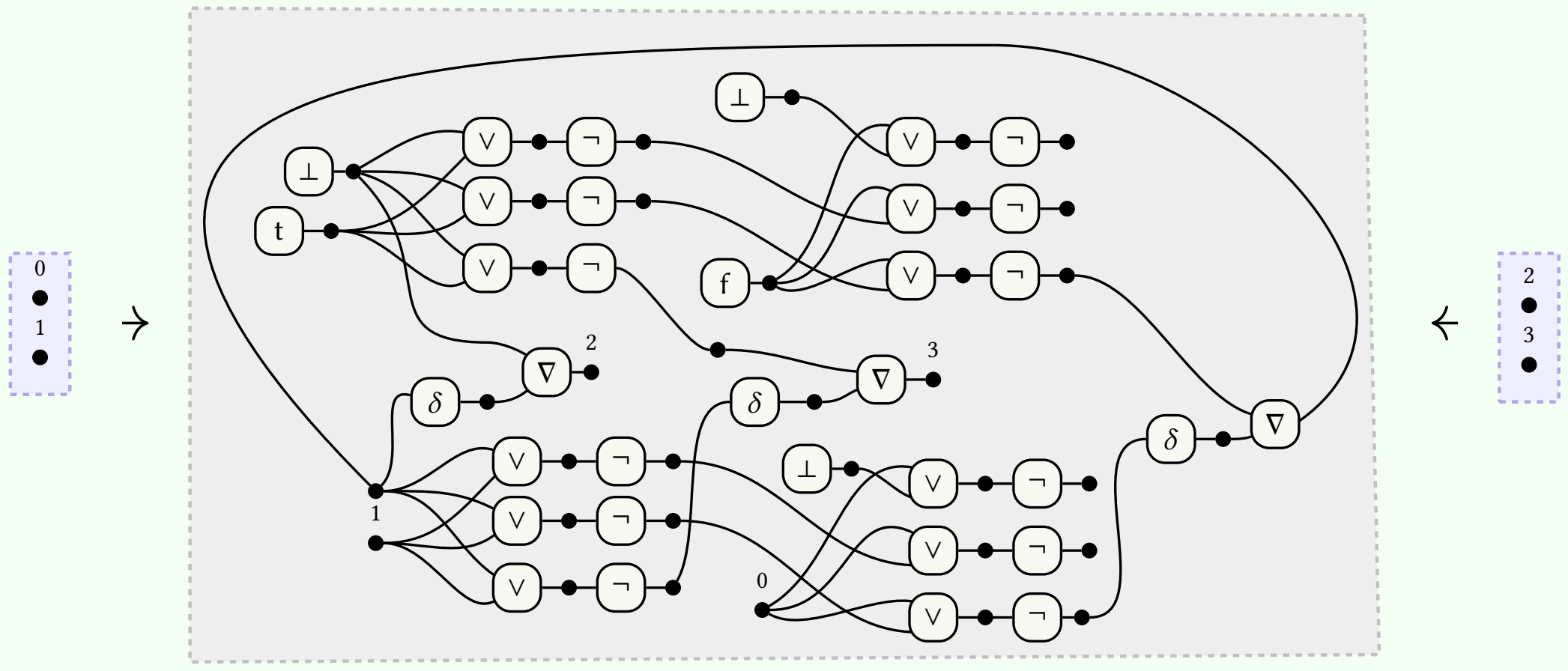

The final step is to propagate the values using the value rules. These have straightforward hypergraph interpretations, which are illustrated in Figure 10.4.

**Example 10.15.** When applying the value rules to the streamed circuit from Example 10.14, we apply the fork rules as much as possible to propagate the values:

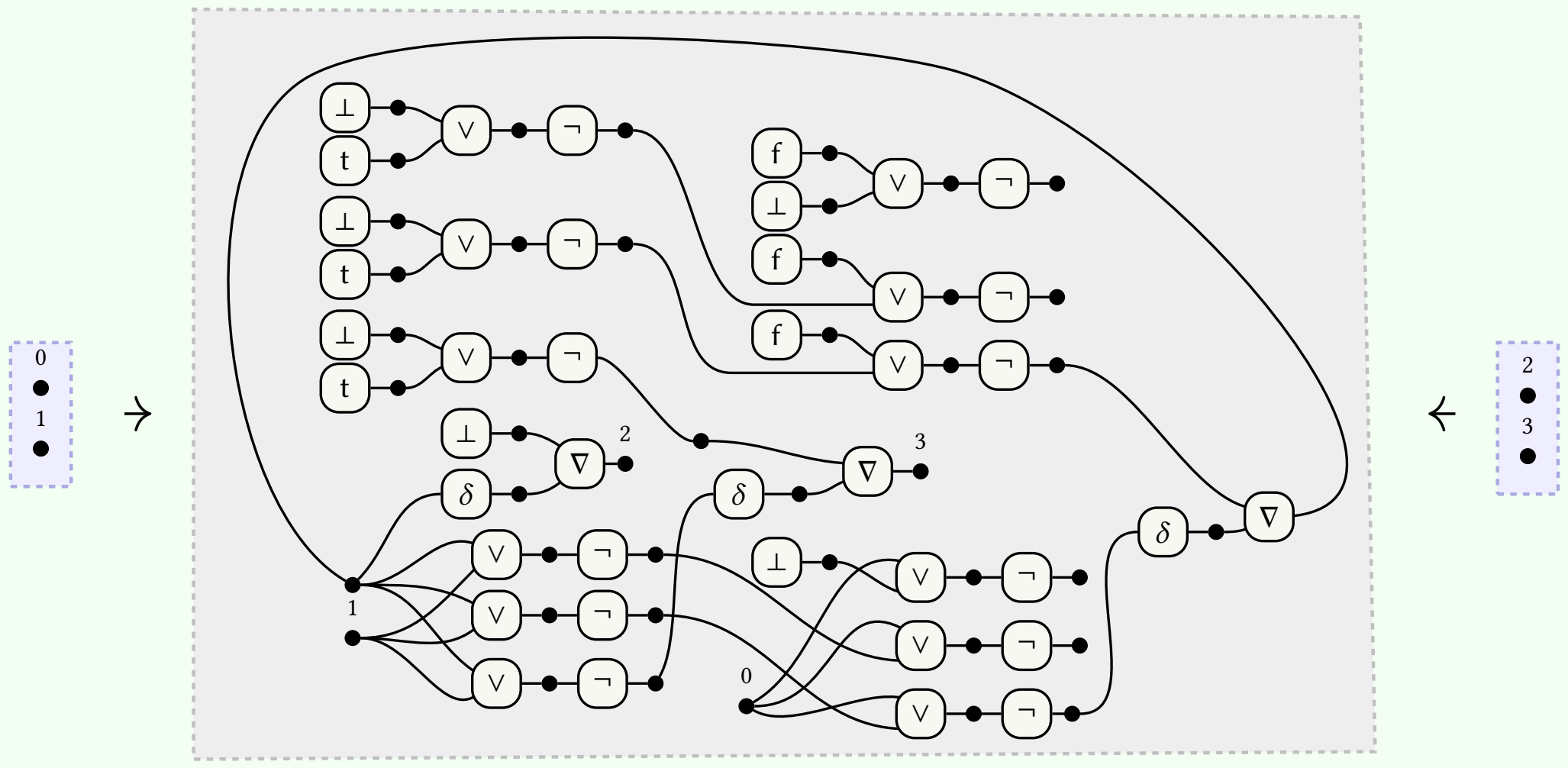

We can then repeatedly apply the gate and eliminate rule to obtain the outputs and next state, which can be seen below.

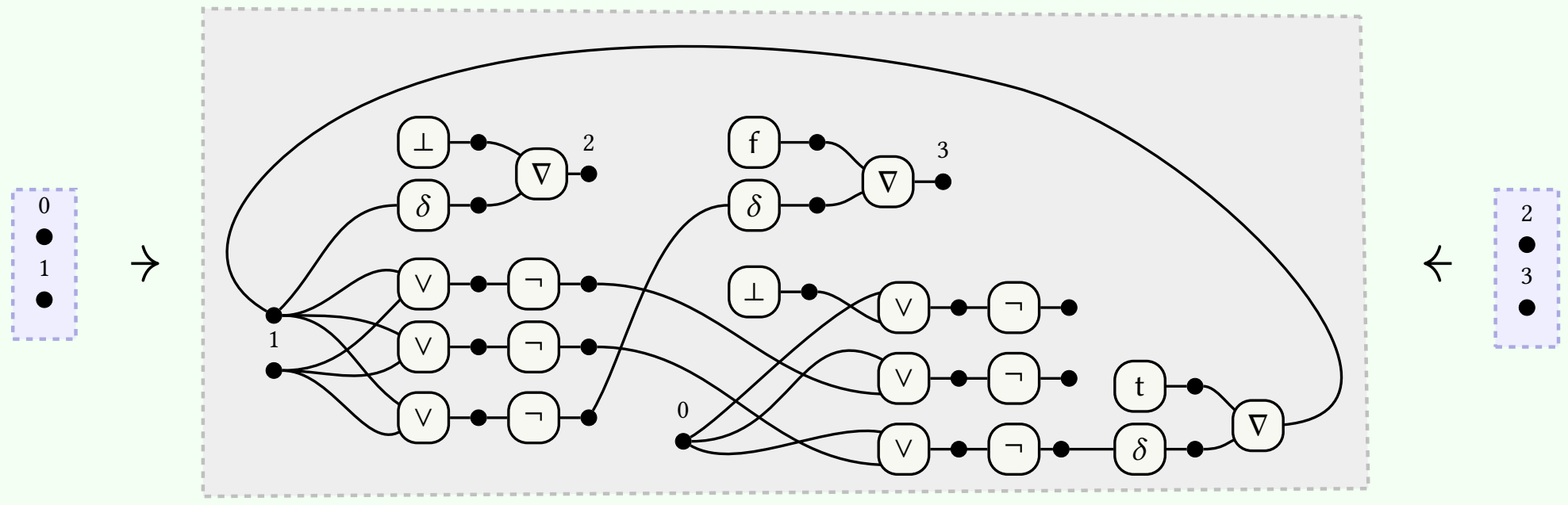

To identify the outputs, one simply needs to traverse from the outputs of the graph.

The initial global transformations do not have to be performed for each time-step; for subsequent inputs one only needs to use the streaming and value rules.

**Remark 10.16.** Note that the fork rule is not left-linear as it uses the comonoid structure. Consider the term 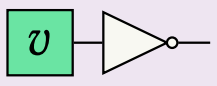 ; in the hypergraph interpretation it is possible to apply the fork rule to this term.

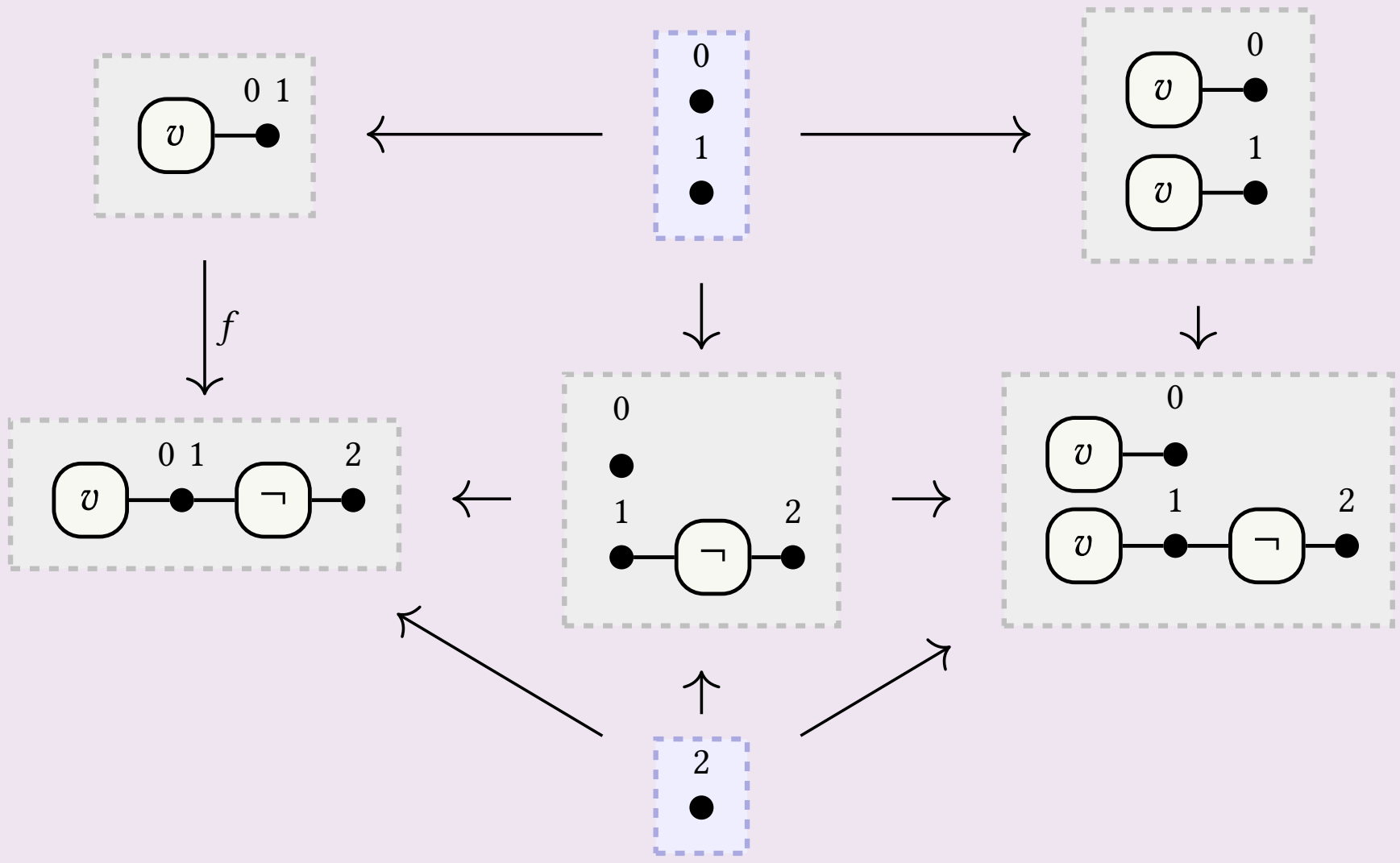

This reduction has arisen due to the counitality of the comonoid.

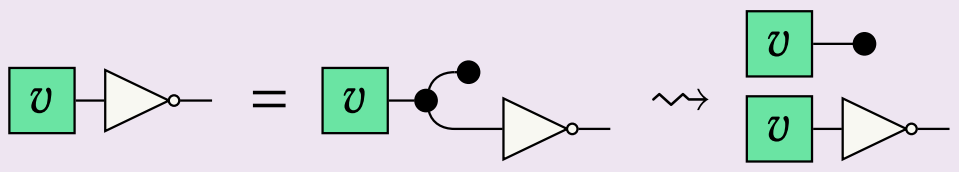

This means that a fork rewrite is only productive if the vertex in the image of $f$ has out-degree greater than 1.

## 10.3 Hardware description language

We motivated graph rewriting for digital circuits as an avenue for *automating* their reasoning. To this end, the operational semantics for sequential circuits has been implemented into a *hardware description language* (HDL) in Cangjie, a programming language by Huawei. The source code is at `https://github.com/georgejkaye/circuit-cj`.

### 10.3.1 Types of HDLs

Designing intricate circuits using pen and paper would be incredibly tedious and complicated. For this reason, *hardware description languages*, programming languages specialised for designing hardware, are often employed to design circuits on a computer.

A hardware description may be a completely bespoke language, such as VHDL [IEE88] and Verilog [IEE96]. As well as being able to construct circuits *structurally* by piecing components together, behavioural descriptions can be specified in a *dataflow* manner, in which constants are immediately propagated across the program upon updating, much like in a spreadsheet. This differs from the *control flow* style of execution in ordinary programming languages, in which each line of code runs sequentially.

Because they are so different to traditional programming languages, this can make VHDL and Verilog inpenetrable to outsiders. An alternative is to *embed* a HDL into an already existing language. Choices of parent language include Haskell (Lava [BCSS98], Bluespec [Nik04], Clash [Koo09; BKK+10]), OCaml (Hardcaml [RDQY23]), and Scala (Chisel [BVR+12]). Because these are primarily *functional*, these HDLs are more structural in nature; circuits are created by composing functions together; we opted to follow this approach when implementing our language.

### 10.3.2 Design

Rather than designing circuits using the categorical style of juxtaposing tiles in sequence and parallel, the tool uses a more conventional approach where the user manipulates wires and provides them as arguments to other components.

**Example 10.17.** We will first demonstrate how to define the combinational half adder circuit from Example 3.11. We begin by defining an XOR gate.

```
let xorA = sig.UseWire(1)
let xorB = sig.UseWire(1)
let xorOr = UseOr(xorA, xorB)
let xorNand = UseNot(UseAnd(xorA, xorB))
let xorAnd = UseAnd(xorOr, xorNand)
```

```
    let xor = MakeSubcircuit(
      [InterfaceWire(xorA, "A"), InterfaceWire(xorB, "B")],
      [InterfaceWire(xorAnd, "Z")],
      "XOR"
    )
```

Once a subcircuit has been defined, a specification in Dot can be generated and rendered using Graphviz.

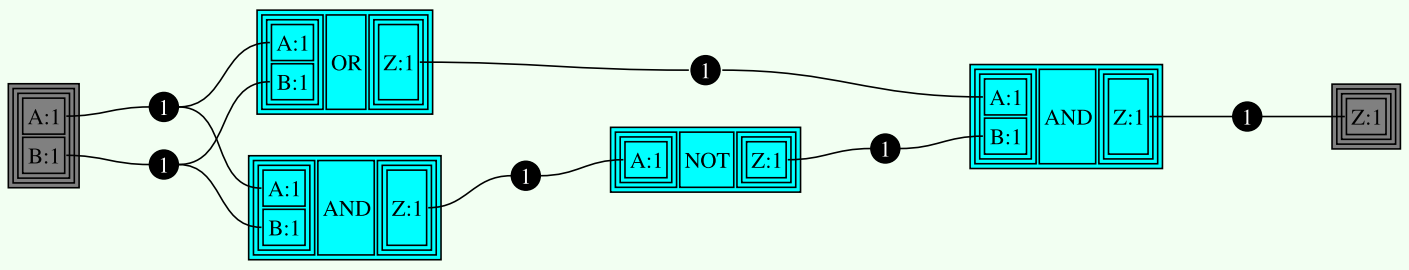

Using the XOR as a subcomponent, we can define the half adder.

```
let addA = sig.UseWire(1)
let addB = sig.UseWire(1)
let sum = UseSubcircuit(xor, [addA, addB])[0]
let carry = UseAnd(addA, addB)
let halfAdder = MakeSubcircuit(
  [InterfaceWire(addA, "A"), InterfaceWire(addB, "B")],
  [InterfaceWire(sum, "S"), InterfaceWire(carry, "C")],
  "half adder"
)
```

The generated graphs have a hierarchical structure: because we defined the XOR gate as a subcircuit, we can view it as a black box or expand it.

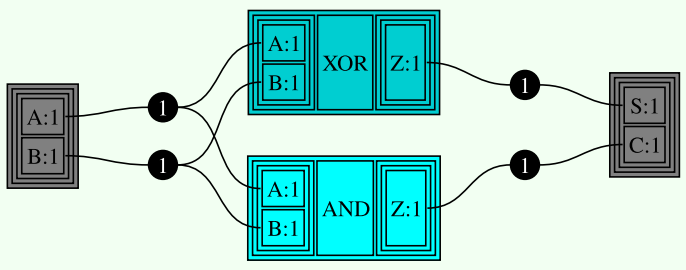

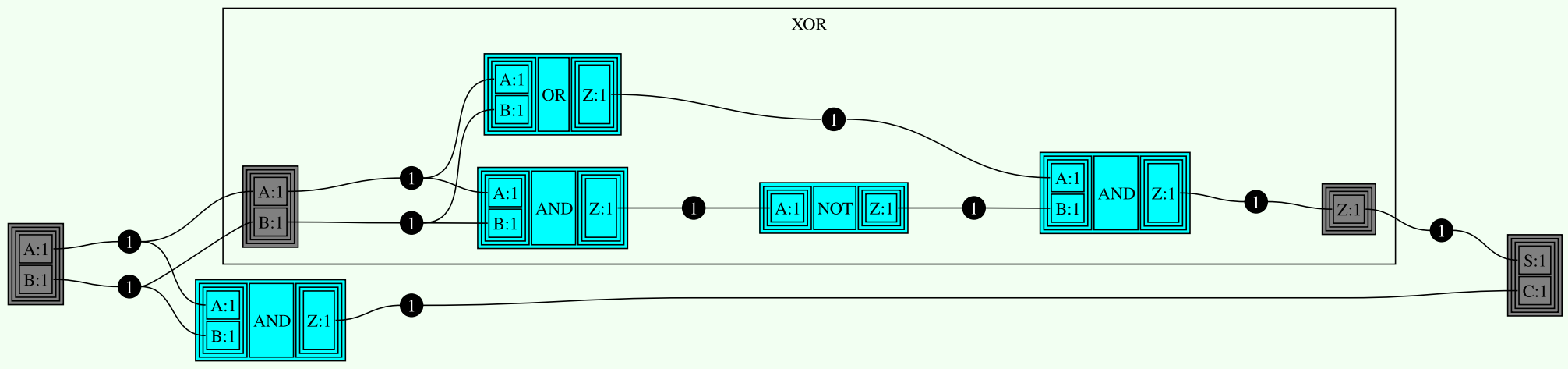

As usual, it is the *sequential* circuits which are the most interesting. The tool can be used to insert delays and feedback loops to circuits.

**Example 10.18.** The SR NOR latch from Example 3.18 can be created

```
let r = sig.UseWire(1)
let s = sig.UseWire(1)
let fb = sig.UseWire(1)
let or1 = UseOr(r, fb)
let not1 = UseNot(or1, delay: 1)
let or2 = UseOr(not1, s)
let not2 = UseNot(or2)
Feedback(not2, fb)
let latch = MakeSubcircuit(
  [InterfaceWire(r, "R"), InterfaceWire(s, "S")],
  [InterfaceWire(not1, "Q"), InterfaceWire(not2, "Q'")],
  "SR NOR Latch"
)
```

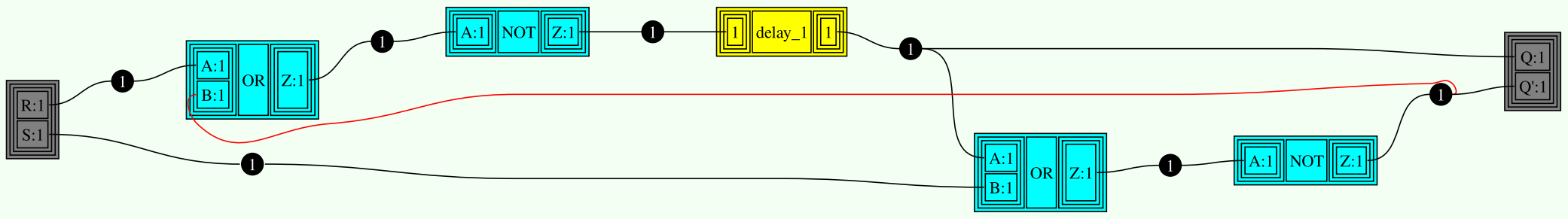

## 10.3.3 Evaluation

The tool can automatically *reduce* them using the operational semantics, by first assembling circuits into Mealy form and eliminating non-delay-guarded feedback.

**Example 10.19.** Before evaluation can be performed, the SR NOR latch defined in Example 3.18 is automatically translated into Mealy form.

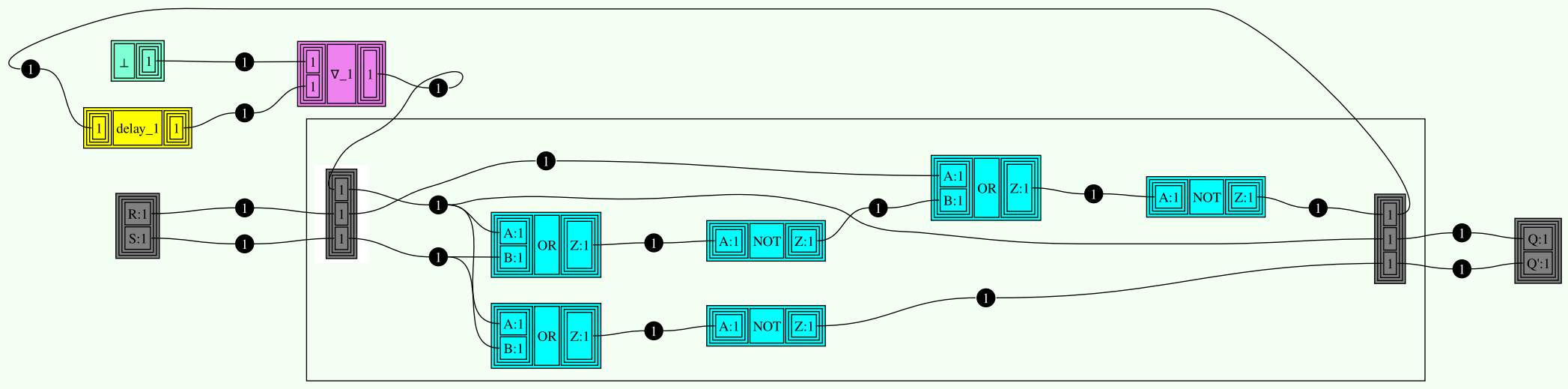

Note that this is a simpler circuit than the corresponding string diagram version in Example 5.20 because the tool automatically applies combinational reductions (in particular, the elimination rule) to tidy up the resulting circuit.

Once the circuit is translated into the correct form, inputs can be provided and the circuit evaluated step-by-step.

**Example 10.20.** We now provide inputs to the evaluator created in Example 10.19; recall that the first input is the Reset input and the second is the Set input.

```
let eval = Evaluator(latch)
eval.PerformCycle([FALSE, TRUE])
eval.PerformCycle([FALSE, FALSE])
eval.PerformCycle([TRUE, FALSE])
eval.PerformCycle([FALSE, FALSE])
```

This automatically applies the reduction rules to determine the output values over time. The inputs detailed above produce the output stream ⊥f::tf::tf::ft, illustrated below in the small boxes. This is the expected output as the delay causes the first tick of outputs to be underdefined.

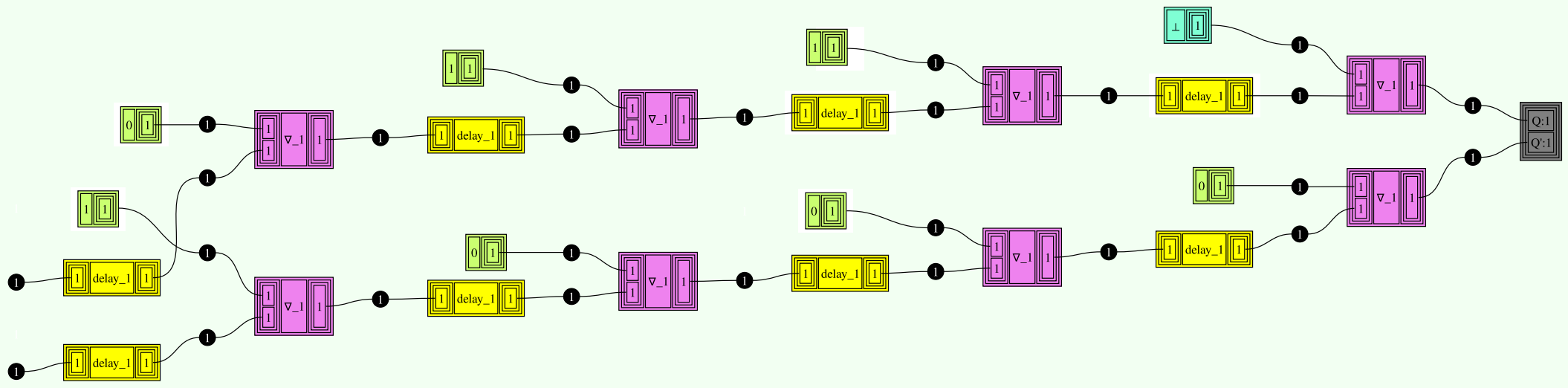

### 10.3.4 Cyclic combinational circuits

The use of the instant feedback rule means that the tool can also handle circuits with non-delay-guarded feedback that exhibit combinational behaviour.

**Example 10.21.** Recall the circuit from Example 5.15 containing blackboxes $f$ and $g$, in which the control signal dictates the order the circuits are applied.

```
// Input wires
let x = sig.UseWire(1)
let c = sig.UseWire(1)
// Wire from feedback
let feedback = sig.UseWire(1)
// Top half of the circuit
let muxa = UseMux2(s0: c, i0: x, i1: feedback)
let fbb = belnapSignature.AddBlackbox("f", [Port(1, "A")],
[Port(1, "Z")])
let f = UseBlackbox(fbb, [muxa])[0]
// Bottom half of the circuit
let muxb = UseMux2(s0: c, i0: f, i1: x)
let gbb = belnapSignature.AddBlackbox("g", [Port(1, "A")],
```

```
[Port(1, "Z")])
let g = UseBlackbox(gbb, [muxb])[0]
Feedback(g, feedback)
// Final multiplexer
let muxc = UseMux2(s0: c, i0: g, i1: f)
let cyclic = MakeSubcircuit(
  [InterfaceWire(X, "x"), InterfaceWire(C, "c")],
  [InterfaceWire(muxc, "Z")],
  "cyclic_combinational"
)
```

This generates the following circuit:

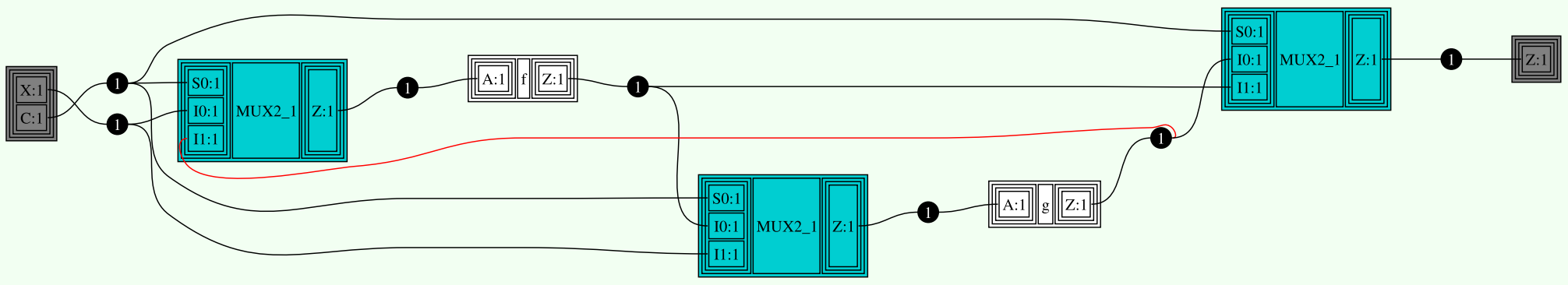

By providing input values we can verify that this circuit truly does have combinational behaviour.

```
let eval = Evaluator(sig, cyclic)
eval.PerformCycle([TRUE, TRUE])
eval.PerformCycle([TRUE, FALSE])
```

Due to the blackboxes, this circuit cannot be be reduced to a stream of output values, but it can be reduced to an expression in terms of $f$ and $g$.

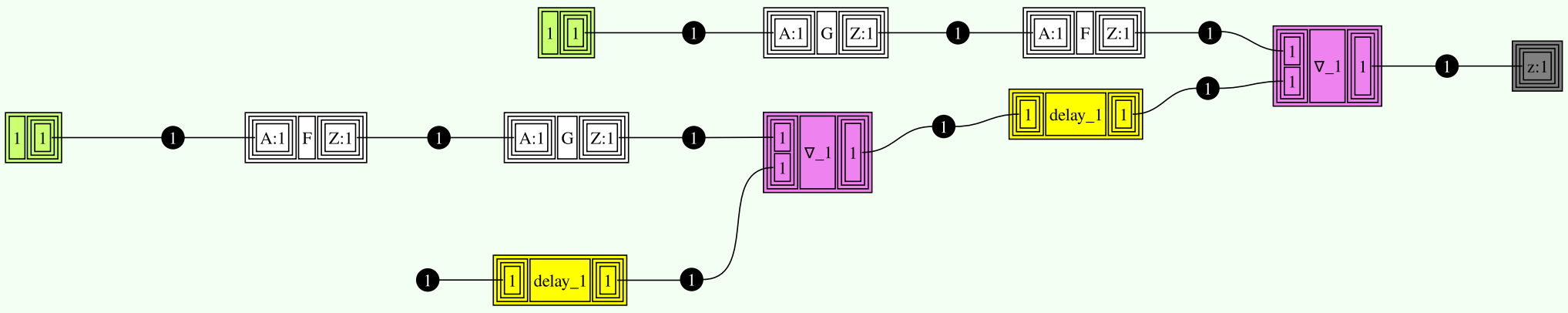

### 10.3.5 Partial evaluation

One of the benefits of the graph-rewrite-based evaluation style is that it allows for *partial evaluation*. The tool implements some of the strategies discussed in Section 7.2, such as tidying rules, shortcut rules, and uncertain values for reasoning with protocols.

**Example 10.22.** Recall the circuit from Example 7.9, which reduces to the identity when the first input is fixed as either true or false. This can be designed as follows:

```
let v = sig.UseWire(1)
let w = sig.UseWire(1)
let not = UseNot(v)
let or1 = UseOr(not, v)
let fb = sig.UseWire(1)
let or2 = UseOr(fb, or1)
let and = UseAnd(or2, w)
DelayGuardedFeedback(and, fb)
let circ = MakeSubcircuit(
  [InterfaceWire(v, "A"), InterfaceWire(w, "B")],
  [InterfaceWire(and, "Z")],
  "circuit"
)
```

This produces the following circuit:

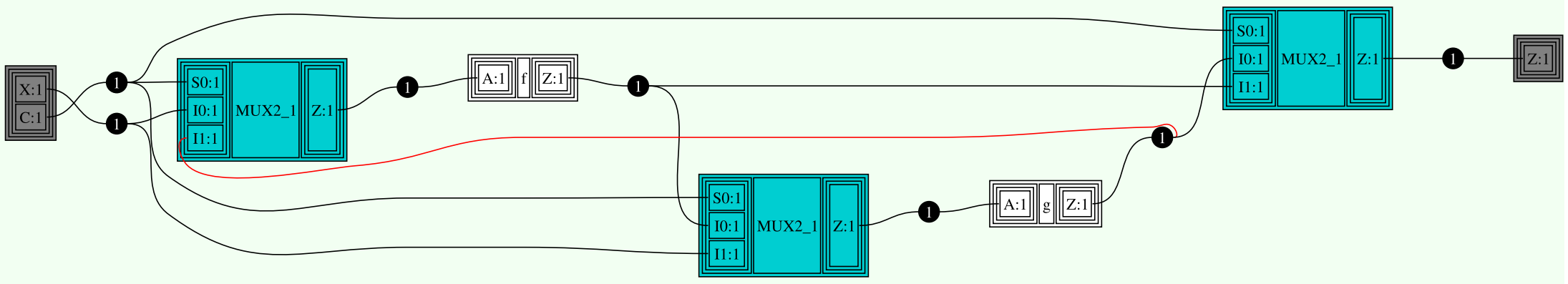

We apply some uncertain values and partially evaluate.

```
let eval = PartiallyEvaluate(
  sig, [GetVariable([TRUE, FALSE]), GetUnspecified(sig)], circ
)
```

For each input provided, the tool prepends infinite waveforms containing the (potentially uncertain) waveforms.

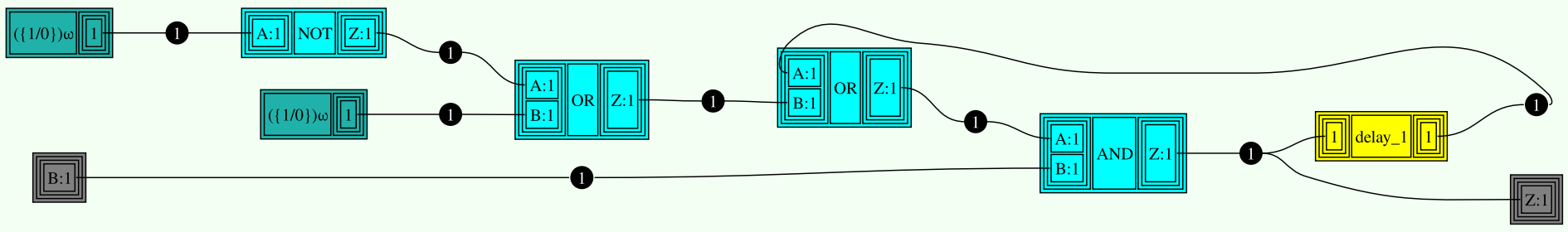

We can then visualise the reduction procedure step by step. First the waveform is propagated over the NOT gate:

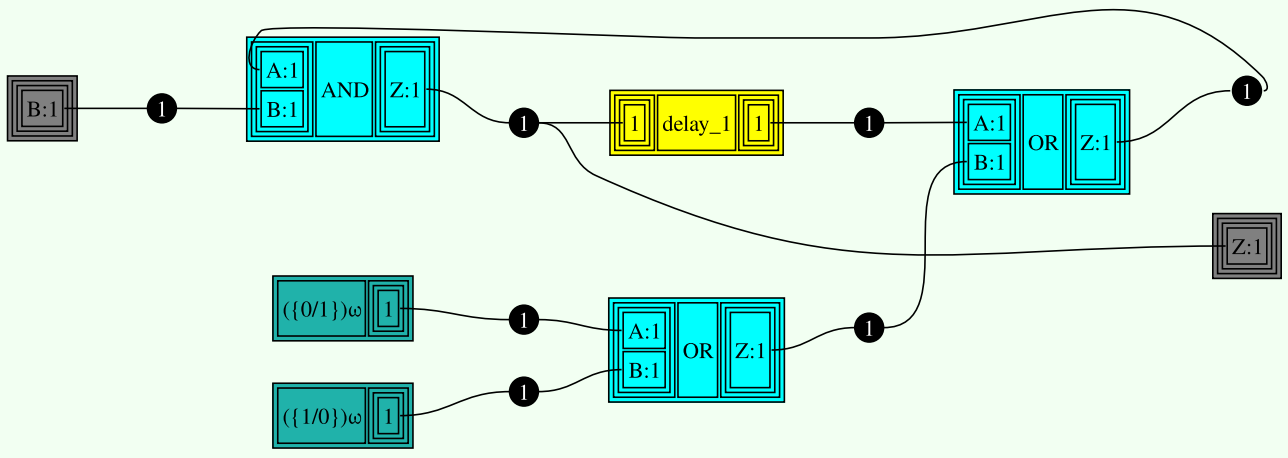

As the inputs to the OR gate can only ever produce true, this can be rewritten:

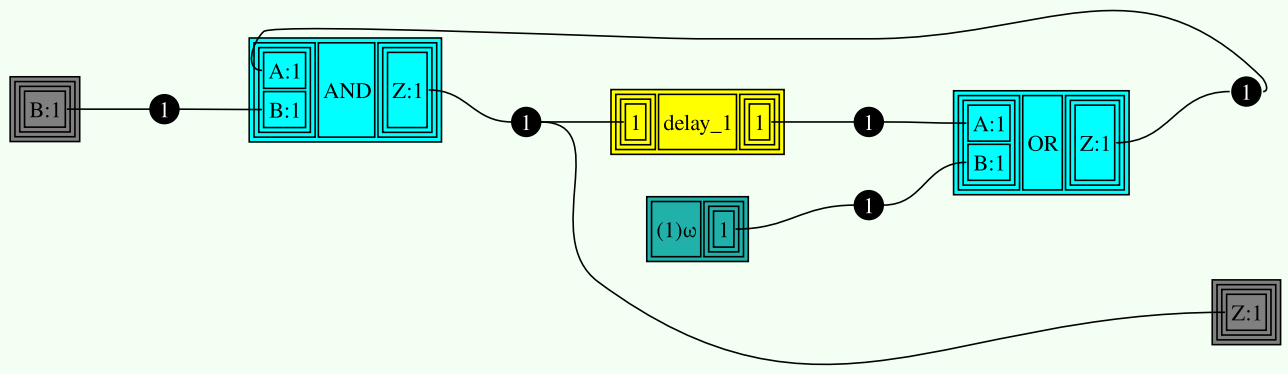

An OR gate with one input as true will always produce true, so this can be replaced and its other input eliminated:

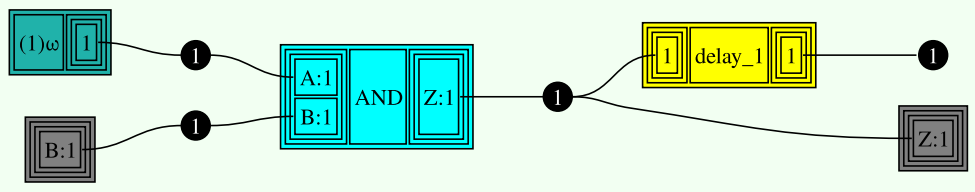

The delay can also be eliminated:

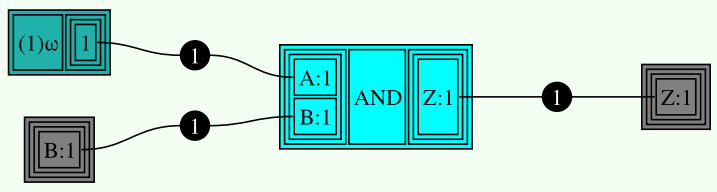

Finally, since an AND with a true input acts as the identity on the other input, this can also be replaced:

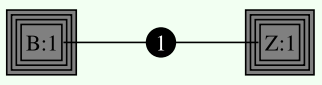

This shows how the entire circuit can be automatically reduced to the identity.

While this is just a toy example, the procedure can easily be applied to more complex circuits and potentially find optimisations.

Chapter 11

# Conclusion

This brings us to the conclusion of this work; in this final chapter we will sum up our contributions and look ahead to the future.

## 11.1 Summary of contributions

Our contributions can be divided into two main topics: the development of a fully compositional theory of sequential circuits, and the application of this framework to graph rewriting.

### 11.1.1 Semantics of Digital Circuits

The first major contribution of this thesis was to take the existing informal work on categorical semantics of sequential circuits [GJ16; GJL17a] and develop it into a rigorous mathematical theory.

#### Syntax of sequential circuits

In the original work on categorical semantics for digital circuits [GJ16], the semantics of circuits were defined as part of the base categories of circuits. Not only was this a confusing presentation, it also 'hardcoded' a particular approach to semantics, which made later developments more fiddly. In Chapter 3, we built the foundations for a different approach, in which circuits are first constructed as morphisms in a PROP $\mathbf{SCirc}_{\Sigma}$ of *syntactic* circuits with no associated behaviour. Behaviour of circuits can be assigned by quotienting this category by some semantic relation; this makes the framework as a whole more modular and 'semantic-agnostic'.

### Denotational semantics

The semantics of circuit components was previously defined as part of equations on the syntactic category; in Chapter 4 we defined the notion of an *interpretation* $\mathcal{I}$ of components as morphisms in a PROP of monotone functions $\mathbf{Func}_{\mathcal{I}}$ in order to keep syntax and semantics separate. This interpretation parameterises the PROP $\mathbf{Stream}_{\mathcal{I}}$ in which morphisms are causal, monotone, finitely specified stream functions; these are the denotations of sequential circuits. The major contribution of this section is that this PROP has a *trace*: the least fixed point.

In order to map from circuits in $\mathbf{SCirc}_{\Sigma}$ to stream functions in $\mathbf{Stream}_{\mathcal{I}}$, we used *Mealy machines* in Section 4.2. We defined a traced PROP $\mathbf{Mealy}_{\mathcal{I}}$ of *monotone* Mealy machines, and showed how we could use existing work on their *coalgebraic* properties to map between Mealy machines and stream functions using two PROP morphisms.

The most novel contributions of this chapter arise by relating Mealy machines and stream functions back to circuits in $\mathbf{SCirc}_{\Sigma}$. In Section 4.3, we defined PROP morphisms between circuits and Mealy machines, in one direction by freely mapping circuit components to primitive Mealy machines, and in the other by *encoding* states of a Mealy machine while preserving monotonicity. Using these PROP morphisms, we showed in Section 4.4 that $\mathbf{Stream}_{\mathcal{I}}$ is a sound and complete denotational semantics for sequential circuits.

### Operational semantics

Previous work had defined an restricted operational semantics for closed circuits without non-delay-guarded feedback [GJL17a]. The major contribution of Chapter 5 was the introduction of the 'instant feedback' rule in Section 5.1: eliminating non-delay-guarded feedback by iterating a circuit a certain number of times. This rule played a key part in the productive reduction strategy presented in Section 5.2, which is integral to the formalism of a notion of *observational equivalence* using a relation $\sim_{\mathcal{I}}$. We showed this relation is sound and complete with respect to the denotational semantics, and that it is the *largest adequate congruence* [Mor69; Gor80] on $\mathbf{SCirc}_{\Sigma}$.

### Algebraic semantics

In Chapter 6 we presented a sound and complete algebraic semantics for sequential circuits, which can be divided into three broad classes. The first class (Section 6.1) contains equations for *normalising* circuits into a Mealy form with a canonical (essentially combinational) core. The second class (Section 6.2) contains equations for *encoding* initial states of circuits in Mealy form using Mealy homomorphisms. The final class (Section 6.3) contains equations for translating between circuits which act the same way when *restricted* to the accessible internal states. In Section 6.4 we showed how these equations are enough to translate between any two denotationally equivalent circuits, thus exhibiting that this is a sound and complete algebraic semantics.

### 11.1.2 Graph rewriting for digital circuits

Viewing sequential circuits through the categorical lens opens up new ways of reasoning with circuits, such as by applying recent work on string diagram rewriting [BGK+22a; BGK+22b] using terms interpreted as morphisms in a category $\mathsf{Csp}_D(\mathbf{Hyp}_\Sigma)$ of cospans of hypergraphs. Our second major contribution is to extend this for settings with a *traced comonoid* structure, of which $\mathbf{SCirc}_\Sigma$ is an example.

#### String diagrams as hypergraphs

In Chapter 8 we defined two sub-PROPs of $\mathsf{Csp}_D(\mathbf{Hyp}_\Sigma)$: the PROP $\mathsf{PMCsp}_D(\mathbf{Hyp}_\Sigma)$ of *partial monogamous* cospans and the PROP $\mathsf{PLMCsp}_D(\mathbf{Hyp}_\Sigma)$ of *partial left-monogamous* cospans. We showed the former are in correspondence with *traced* terms and the latter are in correspondence with such terms equipped with a *cocommutative comonoid* structure: terms equal by equations of STMCs or cocommutative comonoids are mapped to isomorphic cospans of hypergraphs.

#### Graph rewriting

The primary reason for interpreting terms as hypergraphs is to perform automatic reasoning with them via *graph rewriting*. In Chapter 9 we characterised valid graph rewrites in a traced or traced comonoid setting using *traced boundary complements* and *traced left-boundary complements* respectively. While there may be multiple valid graph rewrites, we showed that every graph rewrite performed in this way corresponds to a valid term rewrite, so the rewriting system is sound and complete.

### 11.1.3 Case studies in Belnap logic

The framework presented in this thesis is parameterised over a *signature* specifying signals and components, and an *interpretation* mapping them to behaviour. Throughout the thesis we have considered a particular instantiation in terms of *Belnap's four-valued logic*. In Section 4.5 we showed how the Belnap interpretation is *functionally complete* in that all monotone functions between Belnap values can be expressed in terms of the three Belnap operations, and in Section 6.5 we defined the equations required to bring any essentially combinational Belnap circuit into a normal form. We envision that similar strategies could be applied for reasoning about transistor-level circuits to more abstract viewpoints.

### 11.1.4 Implementations

The developments of this thesis have been supported with some small toy examples, including a running example depicting an SR NOR latch. Unfortunately, any circuit larger than this quickly explodes in size and scope, and as such is not suitable for rendering in a book.

For experimenting with the Belnap interpretation, we developed a small tool (`https://belnap.georgejkaye.com`) which can generate the corresponding circuits given Belnap functions and truth tables. On a larger scale, we also developed a *hardware description language* for designing and evaluating with circuits using the operational semantics, presented in Section 10.3.

## 11.2 Future work

The work presented in this thesis acts as a major milestone in the project to develop a fully compositional theory of sequential circuits, but there are several ways in which the work can be continued to create an even more thorough package of categorical methods for reasoning with digital circuits.

### 11.2.1 Theoretical extensions

The categorical framework for digital circuits may be sound and complete, but it only models circuits that are fully specified with concrete values; in Chapter 7 we presented some ideas for how we could extend the framework. We proposed some alternate reduction rules to automate the tidying-up of circuits (Section 7.1) or to handle persistent inputs modelled as infinite waveforms (Section 7.2.1). We also examined how we could model inputs that follow protocols by adding new components modelling uncertain values (Section 7.2.3). At a meta-level, we also discussed how we could apply work on *layered PROPs* [LZ22] to view circuits at different levels of abstraction (Section 7.3), or implement new inequality relations to compare circuits that output the same signals but over different timespans (Section 7.4).

We are particularly keen to investigate the ways that our framework can be applied to the *partial evaluation* of sequential circuits; while this is a topic that has been examined to some extent [SM99; MS98; TM06], we believe our rigorous foundations will provide a new perspective on the way forward.

### 11.2.2 More applications

In Chapter 7 we presented some ideas for how the categorical framework could be applied to real-world digital circuits. It would be interesting to actually develop these ideas into proper industry-grade techniques for working with circuits. We could then compare *benchmarks* with existing procedures, allowing us to see whether our work has practical benefit in addition to bringing theoretical clarity.

On the topic of implementation, our hardware description language is still quite open-ended. While circuits can currently be designed and (partially) evaluated, it would be useful to add a native way of *verifying* circuits rather than merely inspecting the outputs of two circuits manually, perhaps with some built-in verification language. Moreover, *synthesis* of circuits to more traditional circuit design languages such as VHDL or Verilog, or even to a language suitable for printing on silicon (so-called 'netlists'), would allow for the benefits of our tool to be combined with the experienced power of the traditional methods.

### 11.2.3 Beyond the abstraction

Our categorical framework focuses on a commonly-used abstraction of *sequential synchronous circuits*. Crucially we operate in a *discrete* setting, both in terms of the signals used and the notion of time. A potential future research direction could be to see if it is possible to adapt the techniques used here to a more *continuous* setting.

A concept important in circuit design that is not present in our abstraction is that of *fan-out*, the idea that there are only so many times one can fork a wire before the signal it carries becomes unstable. To model this, one would need to work with continuous signals that degrade by some factor on each fork. One immediate issue that would arise is that the fork would no longer be coassociativity of the comonoid, as different branches would carry different 'strengths' of signals.

When working with potentially degraded signals, circuit designers use *amplifiers* to restore the strength of signals. Which signal is restored can be nondeterministic, so this could add an element of *probability* to digital circuits. Care would have to be taken, as nondeterministic computations breaks the naturality of the copy: rolling a die once and copying the outcome is not the same as duplicating the die and rolling each of them separately.

Our notion of delay is quite primitive, as it delays all inputs by one step. In reality, the *propagation delay* can differ depending on the change in signal; for example, the transition from high to low may be quicker than the transition from low to high. This could be implemented by parameterising the $\mathsf{shift}_\perp$ stream function with different delays for different inputs; how this might affect the rest of the framework remains to be seen. The propagation delay can also be affected by fan-out; the higher the fan-out, the higher the propagation delay.

The model of delay could be altered even further; rather than modelling it as a series of discrete timesteps, it could be modelled *continuously* to handle asynchronous circuits. This would be a significant change; since stream functions have discrete elements, what would the denotation of circuits over continuous time be? It is possible that modelling asynchronous circuits would require a completely different way of reasoning.

# Index of symbols

# Index